\documentstyle[12pt]{report}
  \def\singlespacing{\parskip 5 pt plus 1 pt \baselineskip 13 pt
      \lineskip 7 pt \normallineskip 7 pt}
  \def\doublespacing{\parskip 5 pt plus 1 pt \baselineskip 25 pt
      \lineskip 13 pt \normallineskip 13 pt}
\begin{document}
\input psfig 
\def\Figdir{./Figures}
\newcommand{\binom}[2]{\left( \matrix{{#1}\cr {#2}\cr} \right)}
\newcommand{\PRD}[3]{{\it Phys.~Rev.}~{\bf D~#1},~#2~(19#3)}
\newcommand{\PR}[3]{{\it Phys.~Rev.}~{\bf #1},~#2~(19#3)}
\newcommand{\PRL}[3]{{\it Phys.~Rev.~Lett.}~{\bf #1},~#2~(19#3)}
\newcommand{\PLB}[3]{{\it Phys.~Lett.}~{\bf B~#1},~#2~(19#3)}
\newcommand{\NPB}[3]{{\it Nucl.~Phys.}~{\bf B~#1},~#2~(19#3)}
\newcommand{\ZPC}[3]{{\it Z.~Phys.}~{\bf C#1},~#2~(19#3)}
\newcommand{\CPC}[3]{{\it Comput.~Phys.~Commun.}~{\bf #1},~#2~(19#3)}
\newcommand{\gtrsim}{{\ \lower-1.2pt\vbox{\hbox{\rlap{$>$}\lower5pt\vbox{\hbox{$\sim$}}}}\ }}
\newcommand{\text}[1]{\rm{#1}}
\newcommand{\lesim}{{\ \lower-1.2pt\vbox{\hbox{\rlap{$<$}\lower5pt\vbox{\hbox{$\sim$}}}}\ }}
\newcommand{\grsim}{{\ \lower-1.2pt\vbox{\hbox{\rlap{$>$}\lower5pt\vbox{\hbox{$\sim$}}}}\ }}
\newcommand{\qgen}{q}
\def\su{$SU(2)_{\em l} \times SU(2)_h\times U(1)_Y$\,}
\def\uem{$U(1)_{\rm{em}}$\,}
\def\suu{$SU(2)\times U(1)_Y$\,}
\def\beq{\begin{equation}}
\def\enq{\end{equation}}
\def\ra{\rightarrow}
\def\la{\leftarrow}
\def\del{\partial}
\def\gs{g_s}
\def\D0{D\O}
\def\ETslash{\not{\hbox{\kern-4pt $E_T$}}}
\def\slash{\not{}{\mskip-3.mu}}
\def\ra{\rightarrow}
\def\lra{\leftrightarrow}
\def\bea{\begin{eqnarray}}
\def\ena{\end{eqnarray}}
\def\beq{\begin{equation}}
\def\enq{\end{equation}}
\def\bec{\begin{center}}
\def\enc{\end{center}}
\def\cw{\cos \theta_W}
\def\sw{\sin \theta_W}
\def\tw{\tan \theta_W}
\def\eg{${\it e.g.}$}
\def\ie{${\it i.e.}$}
\def\etc{${\it etc}$}
\def\kln{\kappa_{L}^{NC}}
\def\krn{\kappa_{R}^{NC}}
\def\klc{\kappa_{L}^{CC}}
\def\krc{\kappa_{R}^{CC}}
\def\ttz{{\mbox {\,$t$-${t}$-$Z$}\,}}
\def\bbz{{\mbox {\,$b$-${b}$-$Z$}\,}}
\def\tta{{\mbox {\,$t$-${t}$-$A$}\,}}
\def\bba{{\mbox {\,$b$-${b}$-$A$}\,}}
\def\tbw{{\mbox {\,$t$-${b}$-$W$}\,}}
\def\tbW{{\mbox {\,$t$-${b}$-$W$}\,}}
\def\tltlz{{\mbox {\,$t_L$-$\overline{t_L}$-$Z$}\,}}
\def\blblz{{\mbox {\,$b_L$-$\overline{b_L}$-$Z$}\,}}
\def\brbrz{{\mbox {\,$b_R$-$\overline{b_R}$-$Z$}\,}}
\def\tlblw{{\mbox {\,$t_L$-$\overline{b_L}$-$W$}\,}}
\def\pbarp{ \bar{{\rm p}} {\rm p} }
\def\pp{ {\rm p} {\rm p} }
\def\ipb{ {\rm pb}^{-1} }
\def\ifb{ {\rm fb}^{-1} }
\def\stds{\strut\displaystyle}
\def\SST{\scriptscriptstyle}
\def\TT{\textstyle}
\def\ra{\rightarrow}
\def\cro{\cropen{12pt}}
\def\mf{m_f}
\def\mb{m_b}
\def\mt{m_t}
\def\MW2{M^2_W}
\def\MZ{M_Z}
\def\Cw{C_w}
\def\Sw{S_w}
\def\mHn{m_{\SST H^{\SST 0}} }
\def\mh0{m_{\SST h^{\SST 0}} }
\def\mHp{m_{\SST H^{\pm}} }
\def\mA0{m_{\SST A^{\SST 0}} }
\def\mH{m_{\SST H} }
\def\mHs{m^2_{\SST H} }
\def\qq{q_1 \bar{q}_2}
\def\jj{j_1 j_2}
\def\ee{e^+ e^-}
\def\ptW{P_{\SST T}^{\SST W} }
\def\DR{\Delta R}
\def\fL{f_{\SST L}}
\def\yW{y_{\SST W}}
\def\qs{\theta^\ast}
\def\MWW{M_{\SST WW}}
\def\EWQCD{(1.1)~}
\def\eg{${\it e.g.}$}
\def\ie{${\it i.e.}$}
\def\etc{${\it etc}$}
\def\etal{${\it et al.}$}
\def\hatt{ \hat {\rm T} }
\def\ETslash{\not{\hbox{\kern-4pt $E_T$}}}
\def\mynot#1{\not{}{\mskip-3.5mu}#1}
\def\sss{\scriptscriptstyle}
\def\rtS{\sqrt{S}}
\def\ra{\rightarrow}
\def\d{{\rm d}}
\def\M {{\cal M}}   
\def\qgtb{q' g \ra q t \bar b}
\def\Wgtb{q' g (W^+ g) \ra q t \bar b}
\def\ubdt{q' b \ra q t}
\def\udbt{q' \bar q \ra W^* \ra t \bar b}
\def\Wbt{W^+ b \ra t}
\def\ggtt{q \bar q, \, g g \ra t \bar t}
\def\ttb{t \bar t}
\def\Wt{W t}
\def\gbtW{g b \ra W^- t}
\def\width{\Gamma(t \ra b W^+)}
\def\tevs{Di-TeV}
\def\flong{f_{\rm Long}}
\def\ltap{\;\centeron{\raise.35ex\hbox{$<$}}{\lower.65ex\hbox{$\sim$}}\;}
\def\gtap{\;\centeron{\raise.35ex\hbox{$>$}}{\lower.65ex\hbox{$\sim$}}\;}
\def\gsim{\mathrel{\gtap}}
\def\lsim{\mathrel{\ltap}}
\def\del{\partial }
\def\D0{D\O~}

\def\bec{\begin{center}}
\def\enc{\end{center}}

\def\tbw{{\mbox {\,$t$-${b}$-$W$}\,}}
\def\tbW{{\mbox {\,$t$-${b}$-$W$}\,}}

\def\veps{{\varepsilon}}
\def\slash{\not{}{\mskip-3.mu}}
\newcommand{\myabs}[1]{{| {\vec{#1}} \, |}}
\newcommand{\br}[1]{{\langle {#1} |}}
\newcommand{\kt}[1]{{| {#1} \rangle}}
\newcommand{\bra}[2]{{\langle \, {\hat{#1}} \, {#2} |}}
\newcommand{\ket}[2]{{| \, {\hat{#1}} \, {#2} \rangle}}
\newcommand{\bk}[2]{{\langle \, {#1} | \, {#2} \rangle}}
\newcommand{\braket}[4]{{\langle \, {\hat{#1}} \, {#2} | \, {\hat{#3}} \, {#4} \rangle}}
\newcommand{\bsk}[5]{{\langle \, {\hat{#1}} \, {#2} | \slash {#3} \,
| \, {\hat{#4}} \, {#5} \rangle}}
\newcommand{\bssk}[6]{{\langle \, {\hat{#1}} \, {#2} | \slash {#3} \slash {#4} \, 
| \, {\hat{#5}} \, {#6} \rangle}}
\newcommand{\bsssk}[7]{{\langle \, {\hat{#1}} \, {#2} | \slash {#3} \slash {#4}\slash {#5} \, | \, {\hat{#6}} \, {#7} \rangle}}
\newcommand{\vect}[2]{\pmatrix{{#1}\cr{#2}\cr}}
\newcommand{\mat}[4]{\pmatrix{{#1}&{#2}\cr{#3}&{#4}\cr}}
\def\sone{\mat{0}{1}{1}{0}}
\def\stwo{\mat{0}{-i}{i}{0}}
\def\sthr{\mat{1}{0}{0}{-1}}
\def\gamz{\mat{0}{1}{1}{0}}
\def\gamj{\mat{0}{-\sigma_j}{\sigma_j}{0}}
\def\gam5{\mat{1}{0}{0}{-1}}
\def\Ppm{{1 \over 2} (1 {\pm} \gamma^5)}
\def\Pp{\mat{1}{0}{0}{0}}
\def\Pm{\mat{0}{0}{0}{1}}

\def\slash{\not{}{\mskip-3.mu}}
\def\rts{\sqrt{s}}
\def\sh{\hat{s}}
\def\rsh{\sqrt{\hat{s}}}

\def\phip{e^{i\phi}}
\def\phipsq{e^{2i\phi}}
\def\phim{e^{-i\phi}}
\def\phimsq{e^{-2i\phi}}
\def\vtwo{\sqrt{2}\, }
\def\vmt{\sqrt{m_t}\, }
\def\vmttwo{\sqrt{2m_t}\, }
\def\vbp{\sqrt{E_b+p}\, }
\def\vbm{\sqrt{E_b-p}\, }
\def\epplus{(E_b+p)}
\def\epminus{(E_b-p)}
\def\ra{\rightarrow}
\def\Wtb{W-t-b}
\def\Lag{L}
\def\MWW{M_{WW}}
\def\tow{{m_t \over M_W}}
\def\fol{f_1^L}
\def\for{f_1^R}
\def\cpyftl{f_2^L}
\def\cpyftr{f_2^R}
\def\ctt{\cos{\theta \over 2}}
\def\stt{\sin{\theta \over 2}}
\def\rt{\sqrt{2}}
\def\flong{f_{\rm Long}}
\def\veps{{\varepsilon}}
\def\slash{\not{}{\mskip-3.mu}}
\def\gamz{\mat{0}{1}{1}{0}}
\def\gamj{\mat{0}{-\sigma_j}{\sigma_j}{0}}
\def\gam5{\mat{1}{0}{0}{-1}}
\def\Ppm{{1 \over 2} (1 {\pm} \gamma^5)}
\def\Pp{\mat{1}{0}{0}{0}}
\def\Pm{\mat{0}{0}{0}{1}}

\def\ra{\rightarrow}
\def\st{{\sin\theta}}
\def\ct{{\cos\theta}}
\def\sp{{\sin\phi}}
\def\cp{{\cos\phi}}

\def\ugtb{q' g \ra q t \bar b}
\def\gutb{g q' \ra q t \bar b}
\def\ubdt{q' b \ra q t}
\def\budt{b q' \ra q t}
\def\veps{{\varepsilon}}
\def\slash{\not{}{\mskip-3.mu}}

\def\shat{\hat{s}}
\def\that{\hat{t}}
\def\uhat{\hat{u}}
\def\shat2{\hat{s}^2}
\def\that2{\hat{t}^2}
\def\uhat2{\hat{u}^2}

\newcommand{\xt}{{x_t=2p_t/\sqrt{s}}}


\def\bbar{\bar b}
\def\cbar{\bar c}
\def\qbarp{\bar{q}^{\prime}}
\def\alphas{\alpha _s}
\def\be{\begin{equation}}
\def\tanb{\tan\beta}
\def\CL{{\cal C}_L}
\def\CR{{\cal C}_R}
\def\dis{\displaystyle}
\def\f{\frac}
\def\tolr{\leftrightarrow}
\def\ba{\begin{array}}
\def\tauhat{\widehat{\tau}}
\def\ea{\end{array}}
\def\ee{\end{equation}}
\def\sq2{\sqrt{2}}
\def\qqbar{q\bar{q}}
\def\tbar{\bar t}
\def\bb{b\bar b}
\def\tbar{\bar t}

\newcommand{\lae}{\stackrel{<}{\sim}}
\newcommand{\gae}{\stackrel{>}{\sim}}


\def\ppbar{p\bar{p}}
\def\phibb{\phi b \bar{b}}
\def\bbbb{b \bar{b} b \bar{b}}
\def\Zbb{Z b \bar{b}}
\def\bbjj{b \bar{b} j j}
\def\ifb{ ${\rm fb}^{-1}$ }
\newcommand{\cotb}{\cot\beta}
\newcommand{\sinb}{\sin\beta}
\newcommand{\cosb}{\cos\beta}
\newcommand{\cosa}{\cos\alpha}
\newcommand{\sina}{\sin\alpha}
\newcommand{\cosba}{\cos (\beta -\alpha)}
\newcommand{\sinba}{\sin (\beta -\alpha)}

\input epsf
\pagestyle{plain}
\pagenumbering{roman}
\setcounter{page}{0}
 
\pagestyle{empty}

\begin{flushright}
CTEQ 906\\
MSUHEP-90615\\
hep-ph/9906420
\end{flushright}
\begin{center}
\singlespacing

\vspace*{.7in}

{\large SOFT GLUON EFFECTS ON ELECTROWEAK
BOSON PRODUCTION IN HADRON COLLISIONS}

\doublespacing

By

Csaba Bal\'azs

\vspace{2.5in}
A DISSERTATION \\
\vspace{0.3in}

Submitted to
\singlespacing

Michigan State University

in partial fulfillment of the requirements

for the degree of

\doublespacing
DOCTOR OF PHILOSOPHY

\vspace{.2in}
Department of Physics and Astronomy
\doublespacing

1999
\end{center}



\newpage
\doublespacing

\pagestyle{empty}
\begin{center}
ABSTRACT \\ [15pt]

{Soft Gluon Effects on Electroweak
Boson Production in Hadron Collisions} \\ [30pt]

By \\ [12pt]

Csaba Bal\'azs \\ [30pt]

\end{center}

Departures from the Standard Model (SM) are expected to emerge at
colliders, especially in the best understood precision electroweak (EW)
experiments, and in order to isolate signals of new physics we must
predict them and their backgrounds precisely. In the foreseeable future the 
two highest energy colliders operating will be the upgraded Fermilab
Tevatron and the CERN Large Hadron Collider (LHC), both hadron-hadron
machines. In hadronic collisions corrections from Quantum Chromodynamics
(QCD) tend to become large, and the sizable effects of the multiple
soft--gluon emission has to be included in the theoretical description.
This is achieved by the resummation of the large logarithmic
contributions due to the gluon radiation. In this work, extending the
Collins--Soper--Sterman (CSS) resummation formalism in a renormalization
group invariant manner, a uniform description of soft--gluon phenomena 
is presented in a wide variety of hadronic initiated EW
processes, ranging from Drell--Yan type lepton--pair production through
the single or pair production of colorless vector bosons (including the
standard, or the vector bosons of the extended, unified gauge theories)
to Higgs boson production.

One of the outstanding open questions of the SM, which initiates new
physics, is the underlying dynamics of the EW symmetry breaking (SB). It
is common to assume the existence of (pseudo-) scalar bosons, either
elementary or composite, associated with the EWSB, and the search for
the(se) Higgs boson(s) has the highest priorities at the next generation
of colliders. A SM like Higgs boson with a mass less than or about the
top quark mass can be detected at the upgraded Tevatron via
$p\bar{p}\to W^{\pm} (\to \ell^{\pm} \nu)~H(\to b\bar{b},\tau^+ \tau^-)
X$, or $p \bar p(gg) \to H(\to W^*W^* \to \ell \nu jj$ and $\ell \nu
\ell\nu$) X. Even before its detection the Tevatron is able constrain
the mass of the SM Higgs boson through the measurement of the top quark
and $W^\pm$ boson masses. The latter requires not only the detailed
knowledge of the leptonic distributions from $W^\pm$ decay, but also the
same for the $Z^0$ bosons. In this work the resummed distributions are
given and compared to the next-to-leading-order predictions in detail. At the 
LHC the extraction of the Higgs signal will also be challenging, and the
precise knowledge of the transverse momentum distributions of the Higgs
decay products will be vital. To this end, the resummed calculation of the 
Higgs boson background for the gold-plated $pp(gg)\to H^0 X\to Z^0Z^0$ and
$\gamma \gamma$ modes is also presented.

It was recently proposed that, due to enhanced Yukawa
coupling, the s-channel (pseudo-) scalar production via heavy
quark--anti-quark annihilation can be an important new mechanism for
discovering non-standard charged (and some neutral) scalar particles at
hadron colliders. 
To improve the theoretical prediction for the signal rates and distributions, 
the complete ${\cal O}(\alpha _s)$ QCD corrections to this s-channel 
production process are calculated for hadron collisions. 
In particular, the systematic QCD-improved production and decay rates of the 
charged top-pions of the topcolor models, and the charged Higgses of the 
generic two-Higgs doublet models are computed. 
The physics potential of the Tevatron and the
LHC for probing charged $s$-channel resonance via the single-top
production is analyzed. The extension to the $s$-channel
production of the neutral (pseudo-) scalars (such as in the Minimal
Supersymmetric Model with large $\tan \beta$ and the topcolor models) from the
$b\bar{b}$-annihilation is briefly discussed.

Higgs boson production associated with bottom quarks,
$p\bar{p}/pp \to \phi^0 b\bar{b} \to b\bar{b}b\bar{b}$, at the Tevatron
and the LHC was also studied. 
It was found that strong, model-independent constraints 
can be obtained on the size of the $\phi^0$-$b$-$\bar b$ coupling for a wide
range of Higgs boson masses. 
Based on these constraints it was showed that the small
mass of the bottom quark makes it an effective probe of new physics in
Higgs and top sectors of several different theories reaching beyond the
SM. The implications for supersymmetric models with large $\tan \beta$
were studied. We concluded that the Tevatron and the LHC can impose
stringent bounds on these models, if the $\phi^0 \to b \bar{b}$ signal is
not found.

The QCD fixed order, and resummed corrections were implemented in the Monte
Carlo event generator, called ResBos. Matching the regions described best 
by the resummed and fixed order
calculations, ResBos predicts the kinematic distributions (including the
transverse momentum distribution) of the electroweak bosons (and their decay 
products when applicable) in the {\it whole} kinematic range. ResBos is
currently used by both the CDF and \D0 collaborations at the analysis of
various distributions of $W^{\pm }$'s, $Z$'s and photons and their decay
products (e.g. in the asymmetry in the rapidity distribution of charged
leptons from $W^{\pm }$ decay, which constrains the error of the $W^\pm$ mass;
or in the lepton transverse momentum distribution from $W^{\pm }$ decay
which is essential at the extraction of the $W^\pm$ mass). Both collider 
and fixed target experiments at the Tevatron use ResBos in the analysis of 
their diphoton data. ResBos is also utilized by the LHC detector 
collaborations, ATLAS and CMS, in the study of the Higgs boson signal and its 
backgrounds.

\newpage


\pagestyle{plain}
\pagenumbering{roman}
\setcounter{page}{4}
{\tiny .}
\vspace{3.9in}
\begin{center}
To the Unknown Graduate Student, who is doing research by its definition: \\ 
being lost and learning the way as proceeding. 
\end{center}
\newpage

\begin{center} 
{\large ACKNOWLEDGMENTS}  
\end{center}

\bigskip

I would like to express my gratitude to my thesis advisors, Wu-Ki Tung
and C.-P. Yuan, who taught me almost everything that this work is based
on. I am grateful to Wu-Ki for encouraging my graduate application, for
his inspiration all through the years I spent at Michigan State
University, for the continuous financial support I received from him,
and for the freedom in research and thinking he provided me with. His
deep knowledge of physics and mathematics motivated, stimulated, and
sometimes provoked (in the good sense of the word) my work, teaching me
to be always rigorous, thorough, complete, accurate, and careful when
doing physics. 

I feel very fortunate to have C.-P. as my advisor as well. His
encouragement and extensive help started me on the work presented in this 
thesis. His exceptional patience in discussing details of projects and his 
wide knowledge helped me through obstacles I could not overcome without
him. His outstanding calculational and numerical abilities are woven
through every line of this thesis. I learned from him to always do
physics with enthusiasm and enjoyment. I am also indebted to him for his
help in finding my first job.
I also thank my thesis committee members, Chip Brock, Norman Birge, and
Pavel Danielewicz, for their careful reading of the manuscript.


While working on this thesis, I had extremely successful collaborations
with Ed Berger, Lorenzo Diaz-Cruz, Hong-Jian He, Steve Mrenna, Wayne
Repko, Carl Schmidt, Tim Tait, and Jianwei Qiu. 
I am thankful to them since they
literally contributed to this work. I am grateful to the high-energy
theorists at Michigan State University: Dan Stump, and Jon
Pumplin for insightful discussions on various topics. I thank Chip Brock
for encouraging me to participate in the $tev\_$2000 activities and Joey
Houston for his recognition and support of my work. High-energy
experimentalists at Michigan State University,
Maris Abolins, Jim Linemann, Harry Weerts, 
have their impact on my work; for this I am thankful to them.

I wish to thank Glenn Ladinsky who left his legacy of resummed codes to
me. I am grateful to the numerous users of the ResBos Monte Carlo event
generator for providing me with useful feedback and helping me
developing it, especially Michael Begel, Dylan Casey, Wei Chen, Mark
Lancaster, Andre Maul, Jim McKinley, Pavel Nadolsky, Fan Qun, Willis
Sakumoto, and John Wahl. I thank high-energy physicists outside of
Michigan State University, who have influenced my work. Among the many:
Uli Baur, who helped me to make my results on $\alpha _s$ public, Chris
Hill, who invited me as summer visitor to Fermilab, Tao Han, Gregory
Korchemsky, Steve Kuhlmann, Fred Olness, Alexander Pukhov, Xerxes Tata,
and Marek Zielinski with whom I had fruitful discussions. Many thanks to
Rolf Mertig, Hagen Eck, and Thomas Hahn, authors of the $FeynCalc$ and
$FeynArts$ $Mathemetica$ packages, of which I made extensive use while
calculating the results of this work.

My thanks also go to my colleagues in the Michigan State University
high-energy group: Hung-Liang Lai for providing the \LaTeX~thesis
format, Pankaj Agrawal, Jim Amundson, Doug Carlson, Dave Chao, Kate
Frame, Chris Glosser, Jim Hughes, Francisco Larios, Ehab Malkawi,
Xiaoning Wang, and Mike Wiest for useful discussions. I thank Julius
Kovacs for providing continuous support of my studies, Stephanie Holland
and Debbie Simmons, coordinators of student affairs, Lorie Neuman,
Jeanette Dubendorf, Lisa Ruess, Mary Curtis, the high-energy physics
secretaries, for all their help.

Finally, I thank my parents for their love and support, and my wife who
helped casting this work into English and has sacrificed so much for my
career.

\newpage

\singlespacing
\tableofcontents
\clearpage

\addcontentsline{toc}{chapter}{LIST OF TABLES}
\listoftables
\clearpage

\addcontentsline{toc}{chapter}{LIST OF FIGURES}
\listoffigures
\clearpage

\doublespacing

   \pagestyle{plain}
   \pagenumbering{arabic}
   \setcounter{page}{1}
   \makeatletter
   \def\@evenfoot{}
   \def\@evenhead{\hfil\thepage\hfil}
   \def\@oddhead{\@evenhead}
   \def\@oddfoot{\@evenfoot}
   \makeatother


%
%
\def\FigSMCoulings
{
\begin{figure}[tbh]
\epsfysize=4.0in
\centerline{\epsfbox{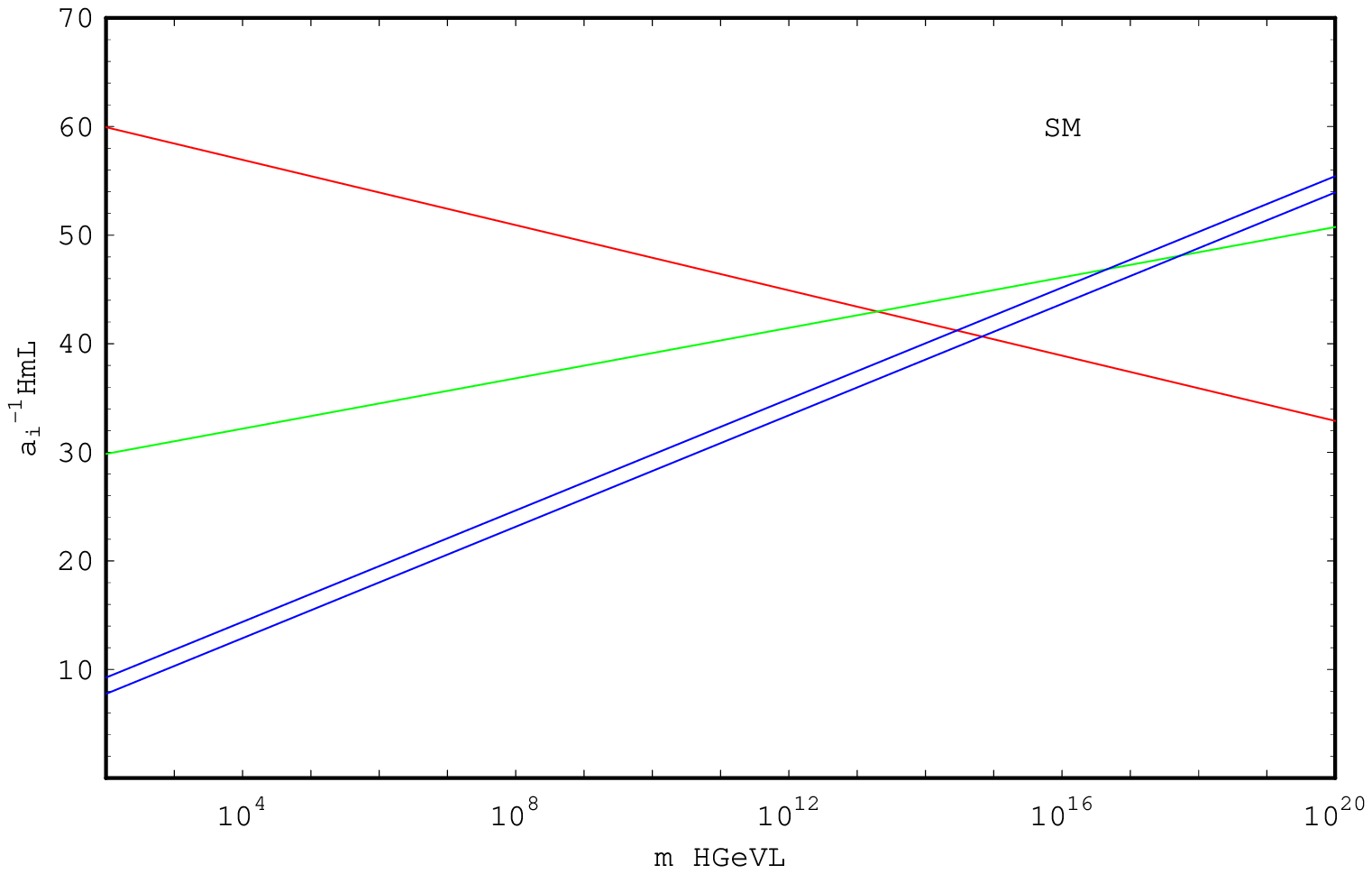}}
\caption{The inverse of the Standard Model gauge couplings as the function of 
the energy scale, extrapolated to high energies. The $U(1)_Y$ coupling (dark 
line) is normalized to .... The $SU(3)_C$  coupling (double line) is shown 
for two boundary values: $\alpha_s(m_Z) = 0.118 \pm 0.003$.}
\label{Fig:SMCoulings}
\end{figure}
}

\def\FigSMasRunning
{
\begin{figure}[t]
\epsfysize=8.5cm
\centerline{\epsfbox{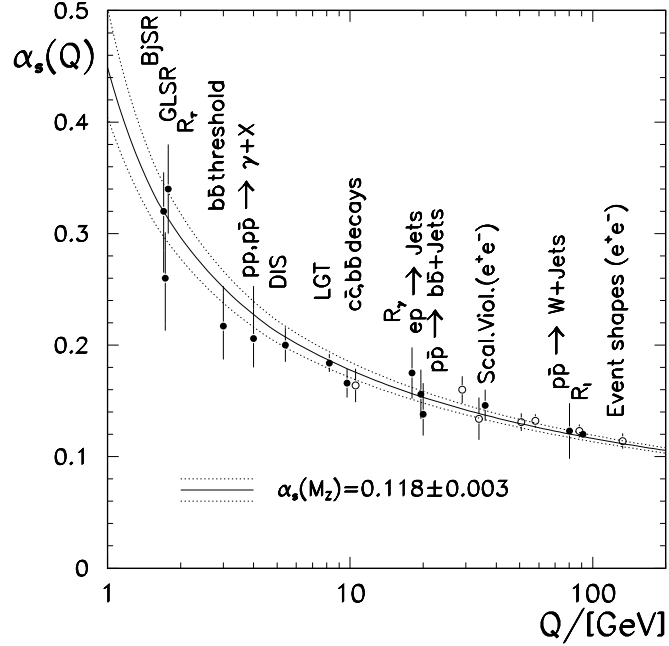}}
\caption{
Running of the QCD coupling constant $\alpha_s$, 
for different boundary values at $\mu = m_Z$.
The QCD prediction is compared to current experiments.
(Reproduced from Ref.~\cite{Wilczek98}%
.)}
\label{Fig:SMasRunning}
\end{figure}
}

\def\FigSMMasses
{
\begin{figure}[tbh]
\epsfysize=4.0in
\centerline{\epsfbox{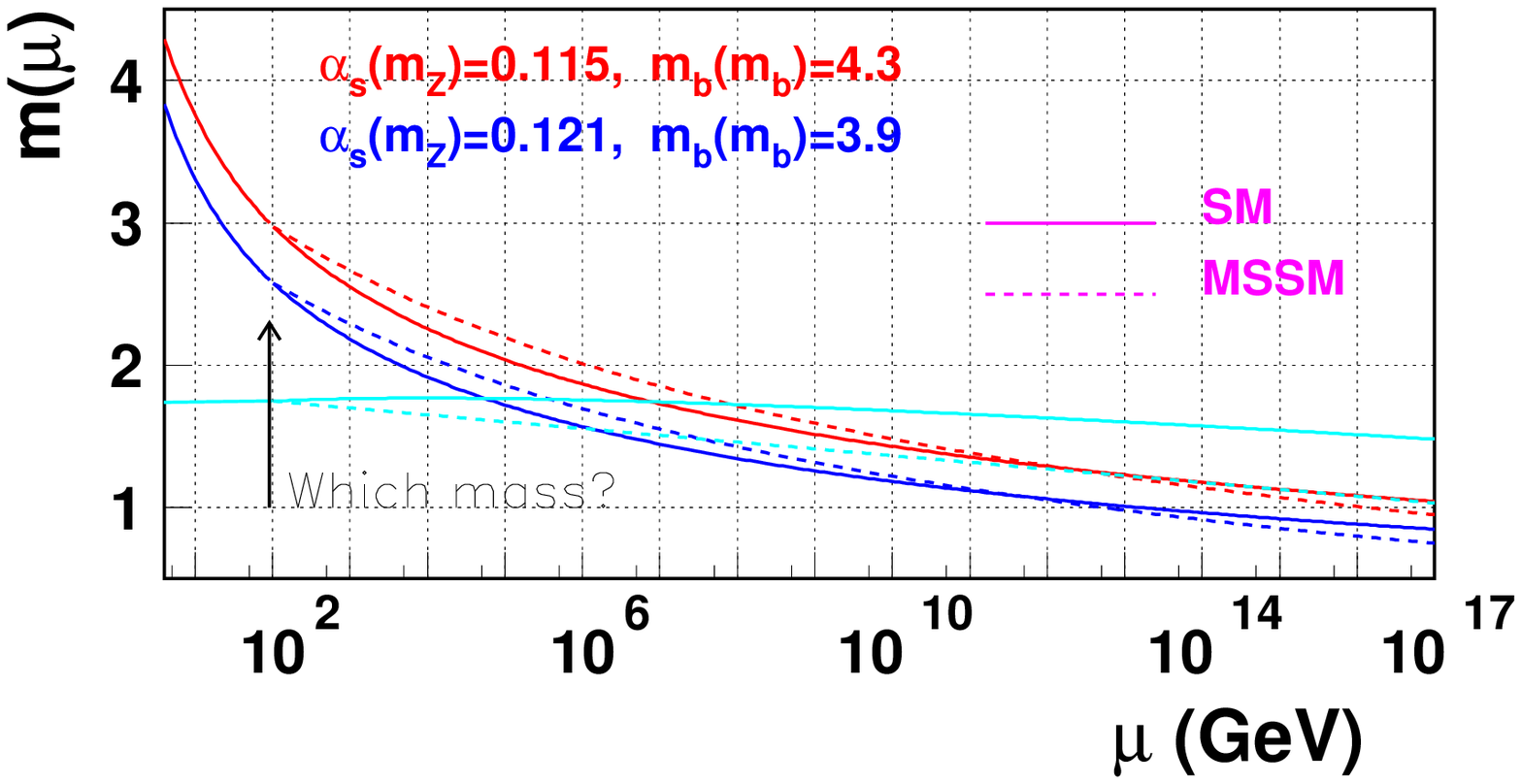}}
\caption{
Running of the $b$ quark and $\tau$ lepton masses within the SM and the MSSM.
The quark masses (dark curves) are shown for different values of the strong 
coupling constant $\alpha_s$.
All masses are in units of GeV's.
(Reproduced from A. Santamaria, G. Rodrigo,  
M. Bilenky, hep-ph/9802359 (1998).)}
\label{Fig:SMMasses}
\end{figure}
}


\chapter{ Introduction \label{ch:Introduction} }

\section{The Standard Model of Elementary Particles}

The Standard Model (SM) \cite{SM} of particle physics is the refined essence
of our wisdom of the microscopic world. It embodies most of our knowledge
about the smallest constituents of our universe. It unifies three of the
four, known fundamental forces: the electromagnetic, weak and strong
interactions, within a compact, economic framework. It describes a wide
range of the observed physical phenomena, most everything\footnote{%
The simplest component of the SM, Quantum Electrodynamics alone describes 
''all of chemistry, and most of physics'' as P.A.M. Dirac put it 
\cite{Dyson53}.} other than gravity. Its validity is tested
daily by numerous experiments with a precision of many decimal points \cite
{PDB}. Since it is the underlying theoretical structure of our work, in
this Section we highlight some of the most important features of the SM.

From the mathematical standpoint the SM is a relativistic, local, 
non-Abelian quantum field theory (QFT) \cite{QFT,StermanQFT}. 
The framework of the theory is based on the following physical
assumptions \cite{DHoker}:

-- space-time symmetry: the theory respects Poincare invariance,

-- point-particles: the elementary particles are point-like in space-time,

-- locality: local (point-like) interactions, i.e. no actions at a distance,

-- causality: commutativity of space-like separated observables,

-- unitarity: quantum mechanical evolution conserving probability, and

-- renormalizability: predictions of the theory to be free of infinities.
\newline
The above principles constrain the theory quite uniquely and grant the SM
substantial predicting power. Theories which surpass the SM either discard
one of these assumptions (string theory \cite{Green-Schwarz-Witten},
non-local QFT's \cite{Namsrai}, effective field theories \cite{Polchinski},
theories of gravity \cite{Weinberg72}), and/or try to improve on them
(supersymmetry \cite{SUSY}, supergravity \cite{SuGra}). 

Field theories are customarily expressed within the Lagrangian formalism 
\cite{Lagrange}, in which a single central quantity, the action density,
or Lagrangian, embodies all the dynamical information about the physical
system. The Lagrangian is constructed from quantum fields which represent the
fundamental particle types. In the SM the different types of particles fall
into two categories. The fermions, which are the building blocks of matter,
and the bosons which are the force mediators. The fermions of the SM are
further split into two branches: leptons and quarks. The former
participating only in the electroweak interaction, while the latter also
engaged by the strong force.

The essence of the SM is in its forces, that is, in the interactions between
the elementary fields. This, in short, is called the dynamics. The dynamics
of the SM is dictated by symmetries \cite{Weyl}, transformations of the
Lagrangian which leave it invariant. One of the central results of the field
theories, Noether's theorem \cite{Noether}, connects symmetries with
conserved physical quantities. This result enables us to relate the
mathematical description with the observed reality, uncovering the
symmetries of Nature through its invariants. Among the symmetries of the SM,
the most important ones are the space-time dependent, or local, so called:
gauge symmetries.

Symmetry transformations form groups \cite{GroupTheory,GeorgiGroupTheory},
and symmetries are usually referred to by their group. The symmetry of the
SM is given by its semi-simple group: $SO(3,1)\times SU(3)_C\times
SU(2)_L\times U(1)_Y$. Here $SO(3,1)$ represents the Poincare invariance and
the rest the gauge symmetries. The Poincare group is usually implicit, it is
understood that the SM to be Lorentz invariant. The structure of the
Lagrangian respecting the above symmetry is discussed in the following
subsections.

\subsection{Electroweak Interactions}

The electroweak (EW) sector of the SM is invariant under the transformations 
of the gauge group: $SU(2)_L\times U(1)_Y$ \cite{Weinberg-Salam,Glashow}. 
Its unbroken subgroup (cf. Section \ref{Sec:TheHiggsMech}), 
the $U(1)_{EM}$ group,
emerged historically in Quantum Electrodynamics (QED) \cite
{Feynman-Schwinger-Tomonaga}, where it represented the conservation of the
electric charge. Similarly, the $U(1)_Y$ symmetry of the SM, had remained
unbroken, would conserve the hypercharge $Y$. The $SU(2)$ group also
appeared empirically in the theory of the weak interactions \cite
{Fermi-Feynman-Gell-Mann-Sudarshan-Marshak}, originally describing a
fermionic flavor symmetry. The invariants of the $SU(2)_L$ group are the
square, $T^2$, and the third component, $T_3$, of the weak isospin
generator. 
These group invariants are connected by the Gell-Mann--Nishijima relation: 
\[
Q=T_3+\frac Y2, 
\]
where $Q$ is the electric charge.

The fermions of the SM are manifested by the fundamental representation of
the $SU(2)_L\times U(1)_Y$ gauge group. The left handed components of the
fermions are assumed to transform as doublets, and the right handed ones as
singlets under $SU(2)_L$. This is signified in the following notation: 
\[
\psi _L=\binom{{\varphi_L^{(+1/2)}}}{{\varphi_L^{(-1/2)}}}, ~~~ \psi
_R=\varphi_R^{(0)}. 
\]
Here 
\[
\varphi_L=\frac{1-\gamma _5}{2} \varphi {\rm ~~~ and ~~~} \varphi_R=\frac{%
1+\gamma _5}{2} \varphi 
\]
represent the left and right handed components of the fermion spinors $%
\varphi$, with $\gamma _5=i\gamma _0\gamma _1\gamma _2\gamma _3$ in four
space-time dimensions, and the Dirac matrices, $\gamma ^\mu $, defined
by their anti-commutation relation: 
\[
\{\gamma ^\mu ,\gamma ^\nu \}=2g^{\mu \nu }, 
\]
with $g^{\mu \nu }$ being the space-time metric tensor\footnote{%
Throughout this work we use the $g^{\mu \nu }=diag(1,-1,-1,-1)$ metric,
because with this choice the square of physical momenta are non-negative.}.
The weak isospin eigenvalues, $0$ and $\pm 1/2$, show the transformation
property of the given fermion field under $SU(2)_L$. In the first family $%
\varphi_L^{(+1/2)}=\nu_L^e,u_L$, $\varphi_L^{(-1/2)}=e_L,d_L$, and $%
\varphi_R^{(0)}=e_R,u_R,d_R$, where $\nu ^e$ and $e^{-}$ are the leptons,
and $u$ and $d$ are the quarks. Table \ref{TblSMQMNs} shows the most
important quantum numbers of the first family members. 
\begin{table}[t]
\begin{center}
\begin{tabular}{lrrr} 
\hline \hline \\[-0.2cm]
\vspace{.2cm} Fermion & $Q $ & $T_3$ & $Y $ \\ 
\hline \\[-0.2cm]
\vspace{.2cm} $\nu_L^e $ & $0 $ & $\frac{1}{2}$ & $-1 $ \\ 
\vspace{.2cm} $e_L$ & $-1 $ & $-\frac{1}{2}$ & $-1 $ \\ 
\vspace{.2cm} ${u}_L$ & $\frac{2}{3}$ & $\frac{1}{2}$ & $\frac{1}{3}$ \\ 
\vspace{.2cm} $d_L$ & $-\frac{1}{3}$ & $-\frac{1}{2}$ & $\frac{1}{3}$ \\ 
\vspace{.2cm} $e_R$ & $-1 $ & $0 $ & $-2 $ \\ 
\vspace{.2cm} $u_R$ & $\frac{2}{3}$ & $0 $ & $\frac{4}{3}$ \\ 
\vspace{.2cm} $d_R$ & $-\frac{1}{3}$ & $0 $ & $-\frac{2}{3}$ \\ 
\hline \hline
\end{tabular}
\end{center}
\caption{Electroweak quantum numbers of the first family of fermions.}
\label{TblSMQMNs}
\end{table}
In the SM the neutrinos are massless, which means that $\varphi_R^{(0)}=\nu
_R$ is decoupled from the theory. This pattern is repeated in the second and
third families of the leptons and quarks, where $e$ is replaced by $\mu $
and $\tau $, and $u,d$ by $c,s$ and $t,b$ respectively.

Since fermions are assigned to the fundamental representation of the gauge
group, their infinitesimal gauge transformation properties are: 
\[
\delta \psi _{L,R}=\left( ig\, T_j \theta _j(x)+ig{\prime }\frac Y2\theta
(x)\right) \psi _{L,R}. 
\]
Here $g$ and $g{\prime }$ are the gauge couplings, $T_j$ and $Y$ are the
generators of $SU(2)_L$ and $U(1)_Y$, respectively, $\theta _j(x)$ and $%
\theta (x)$ are space-time dependent transformation parameters, and $j=1,2,3$
is an isospin index. Since $SU(2)_L$ and $U(1)_Y$ are Lie groups \cite
{GeorgiGroupTheory}, the generators are defined by the Lie algebra of the $%
SU(2)_L \times U(1)_Y$ group, through their commutators 
\begin{eqnarray*}
\left[ T_i,T_j\right] &=&i\epsilon ^{ijk}T_k, \\
\left[ T_i,Y\right] &=&0 {\rm ~~for~any~} i.
\end{eqnarray*}
The fully anti-symmetric unit tensor $\epsilon ^{ijk}$ gives the structure
constants of the $SU(2)_L$ group ($\epsilon^{123}=1$). 
The generators can be represented by
two the dimensional Pauli matrices: $T_i = \sigma_i/2$.

Invariance requirement under the gauge and Poincare groups, 
and renormalizability constrain the fermionic part of the Lagrangian to: 
\[
{\cal L_F}=\overline{\psi }_L^f\gamma ^\mu iD_{L\mu }\psi _L^f+\overline{%
\psi }_R^f\gamma ^\mu iD_{R\mu }\psi _R^f. 
\]
Implicit summation is implied over the double fermion family indices $f$, as
well as the double Lorentz indices $\mu $. The gauge-covariant derivatives
are given by 
\begin{eqnarray*}
D_{L\mu } &=&\partial _\mu -ig\,T_iW_\mu ^i-ig{\prime }{\frac Y2}B_\mu , \\
D_{R\mu } &=&\partial _\mu -ig{\prime }QB_\mu .
\end{eqnarray*}
The gauge bosons $W^i$ and $B$ appear as the consequence of the gauge
invariance requirement, which also fixes their transformation properties: 
\begin{eqnarray*}
\delta W_\mu ^i &=&\partial _\mu \theta ^i(x)-g\,\epsilon ^{ijk}\theta
^j(x)W_\mu ^k, \\
\delta B_\mu &=&\partial _\mu \theta (x).
\end{eqnarray*}
From the above we can infer that the gauge bosons belong to the adjoint
representation of the gauge group. The kinetic term of the gauge bosons is 
written in the form
\[
{\cal L_V}=-{\frac 14}{B}_{\mu \nu }{B}^{\mu \nu }-{\frac 14}{W}_{\mu \nu }^i%
{W}^{i\mu \nu }, 
\]
where the field strength tensors defined as 
\begin{eqnarray*}
{W}_{\mu \nu }^i &=&\partial _\mu W_\nu ^i-\partial _\nu W_\mu
^i+g\,\epsilon ^{ijk}W_\mu ^jW_\nu ^k, \\
{B}_{\mu \nu } &=&\partial _\mu B_\nu -\partial _\nu B_\mu .
\end{eqnarray*}
The Yang-Mills \cite{Yang-Mills} nature of the electroweak gauge fields is
reflected in the fact that ${\cal L_V}$ contains the self-interactions of
the gauge bosons, due to the third term of the tensor ${W}_{\mu \nu }^i$.
This terms is required by invariance under the $SU(2)_L$, a non-Abelian
transformation group.

The $W_{\mu }^i$ and $B_{\mu }$ fields are the, so called, electroweak (or
interaction) eigenstates of the gauge bosons. Their physical counterparts,
the $W^{\pm }$, $Z^0$ particles and the photon ($A^0$), given by the
following transformations: 
\begin{eqnarray}
W_\mu ^{\pm } &=&\frac{W_\mu ^1\mp iW_\mu ^2}{\sqrt{2}}, \nonumber \\
Z_\mu ^0 &=&\cos \theta _w\,W_\mu ^3-\sin \theta _w\,B_\mu, \nonumber \\
A_\mu ^0 &=&\sin \theta _w\,W_\mu ^3+\cos \theta _w\,B_\mu, \label{Eq:DefWZA}
\end{eqnarray}
where $\theta _w$ is the weak mixing angle, and 
\[
\sin \theta _w=\frac{g{\prime }}{\sqrt{g^2+g{\prime}^2}}. 
\]
After expressing the Lagrangian in terms of the physical fields, we also
find that the charged fermions couple to the photon field by $Q_f e$, where 
\[
e=g\sin \theta _w, 
\]
which is the strength of the electromagnetic interactions.

It is remarkable that the gauge and Poincare symmetries, together with
the requirement of the renormalizability fully constrain the interactions
between the fermions and the gauge bosons and leave only two independent
free parameters in the electroweak sector. They are the group couplings $g$
and $g{\prime }$.

\subsection{Strong Interactions}

The strong interacting sector of the SM\ is called Quantum Chromodynamics,
QCD in short. QCD\ was developed along the lines of QED, except from
the beginning it was clear that the gauge group must be more involved. The $%
SU(3)$ symmetry was originally proposed to describe nuclear interactions as
a flavor symmetry, and was only later identified as a gauge (color) symmetry 
\cite{Fritzsch-Gell-Mann}.

It is straightforward to extend the gauge symmetry of the electroweak sector
of the SM Lagrangian to include the $SU(3)_C$ sector by requiring that the
quarks are $SU(3)_C$ triplets transforming as: 
\[
\delta \psi _{L,R}=\left( ig_st_a\theta _a(x)+ig\,T_i\theta _i(x)+ig{\prime }\frac Y2\theta
(x)\right) \psi _{L,R}.
\]
Above, $g_s$ is the gauge coupling and $t_a$ are the generators of the $%
SU(3)_C$ group, $\theta _a(x)$ are the gauge transformation parameters, and $%
a=1,...,8$ is a color index. The $SU(3)_C$ generators satisfy the algebra: 
\[
\left[ t_a,t_b\right] =if^{abc}t_c,
\]
and commute with the electroweak generators. The structure constants of the $%
SU(3)_C$ are denoted by $f^{abc}$. The three dimensional representation 
of the generators are the Gell-Mann matrices: $t_a = \lambda_i/2$.

Just like in the EW case, gauge and Poincare invariances and
renormalizability constrain the covariant derivatives of the quarks: 
\begin{eqnarray}
&& D_{L\mu } =\partial _\mu -ig_st_aG_\mu ^a -ig\,T_i W_\mu ^i-ig{\prime }
{\frac Y2}B_\mu, \nonumber \\
&& D_{R\mu } =\partial _\mu -ig_st_aG_\mu ^a -ig{\prime }QB_\mu .
\label{Eq:CovariantDerivatives}
\end{eqnarray}
The introduction of the gluon field $G_\mu ^a$ is the necessity of the gauge
invariance. It is assigned to the adjoint representation of the $SU(3)_C$ group,
and under the group transformations it transforms like 
\[
\delta G_\mu ^a=\partial _\mu \theta ^a(x)-g_sf^{abc}\theta ^b(x)G_\mu ^c. 
\]
The kinetic term of the Lagrangian 
is 
\[
{\cal L_G}=-{\frac 14}G_{\mu \nu}^a {G}_a^{\mu \nu }, 
\]
with the field strength tensor 
\[
{G}_{\mu \nu }^a=\partial _\mu G_\nu ^a-\partial _\nu G_\mu
^a+g_sf^{abc}G_\mu ^bG_\nu ^c. 
\]
The strong interactions also have self interacting gauge bosons due to the
non-Abelian nature of the symmetry group $SU(3)_C$.

QCD introduces only one additional parameter into the SM: $g_s$, the gauge
coupling strength between quarks and gluons. In principle it is possible to
introduce another, so called $\theta $ term, into the QCD Lagrangian which
satisfy all the symmetry requirements: 
\[
{\cal L_\theta }=
\frac{g_s^2}{64\pi^2}\theta \epsilon ^{\kappa \lambda \mu \nu }G_{\kappa
\lambda }^a{G}_{\mu \nu }^a,
\]
where $\epsilon ^{\kappa \lambda \mu \nu }$ is the fully antisymmetric unit
tensor. This term violates $P$ and $CP$ conservation, where $P$ stands for the 
parity and $C$ for the charge conjugation discrete transformation. 
From the measurement of the electric dipole moment of the neutron one concludes
that the $\theta $ parameter must be very small ($\theta < 10^{-9}$). 
This term has no significance for this work, therefore we simply ignore it.

\subsection{The Higgs Mechanism \label{Sec:TheHiggsMech}}

In the above discussion of the EW and QCD interactions all the fields,
representing different types of fundamental particles, are massless.
Since fermion mass terms describe transitions between left- and right-
handed chirality states, and the different handed fermions have
different transformation properties under $SU(2)_L$ the simplest fermion 
mass terms are forbidden by the SM gauge group. Gauge boson masses are
not allowed either, since even in the simplest gauge group $U(1)$ a term 
like $m_B^2 B_\mu B^\mu$ breaks the gauge symmetry.
This seemed to conflict the
observation that the weak interactions are very short ranged, which
originally hinted the existence of massive mediators \cite{Lee}. To resolve
this problem, the spontaneous breaking of the gauge symmetry was proposed 
\cite{Higgs}. The essence is that while the Lagrangian can be kept invariant
under the $SU(2)_L\times U(1)_Y$ transformations, one can break the symmetry
of the vacuum, a result of which the gauge bosons can be rendered massive.
This maneuver is called the Higgs mechanism. The spontaneous symmetry
breaking (SSB) is implemented in the SM by the introduction of an elementary
scalar which has a non-vanishing vacuum expectation value (vev) \cite
{Weinberg-Salam}. As an added bonus the same mechanism offers a possibility
to generate the fermion masses \cite{Yukawa}.

The idea of the Higgs mechanism is based on the Goldstone theorem \cite
{Goldstone}. In order to achieve the symmetry breaking, customarily, a 
$SU(2)_L$ doublet, complex scalar field is introduced: 
\[
\Phi =
\binom{i\varphi ^{+}}{\displaystyle{\frac{v+H-i\varphi _0}{\sqrt{2}}}}. 
\]
The Nambu-Goldstone fields $\varphi ^{\pm}$ and $\varphi _0$ are
non-physical since their degrees of freedom will be absorbed by the $W^\pm$
and $Z^0$ gauge bosons as longitudinal polarizations. The physical Higgs
field $H$ has the following electroweak quantum numbers: 
\begin{eqnarray*}
\begin{tabular}{lrr}
\vspace{.2cm} $Q $ & $T_3$ & $Y $ \\ 
\vspace{.2cm} $0 $ & $-\frac{1}{2}$ & $1 $%
\end{tabular}
\end{eqnarray*}

It is assumed that the vev of the $\Phi$ field is 
\[
\left\langle \Phi \right\rangle =\frac 1{\sqrt{2}}\binom 0v, 
\]
where $v$ is a real parameter. This implies that the vev's of the $SU(2)_L$, 
and $U(1)_Y$ generators, and the charge operator acting on the $\Phi $ field 
are: 
\begin{eqnarray*}
\left\langle T_i\;\Phi \right\rangle = 
\frac{\sigma_i}{2} \left\langle \Phi \right\rangle \neq 0, ~~~
\left\langle Y\;\Phi \right\rangle   = \frac v{\sqrt{2}}\neq 0, ~~~
\left\langle Q\;\Phi \right\rangle   = \left\langle \left( T_3+\frac
Y2\right) \Phi \right\rangle =0,
\end{eqnarray*}
where $\sigma_i$ are the Pauli matrices.
Namely, $SU(2)_L \times U(1)_Y$ is broken by the Higgs vacuum but the
electromagnetic symmetry $U(1)_{EM}$ is intact.

In order to show that simultaneously with the breaking of $SU(2)_L$ and $%
U(1)_Y$, the gauge bosons will acquire a mass, the Lagrangian of the scalar
sector is written in an $SU(2)_L\times U(1)_Y$ invariant form: 
\[
{\cal L_H}=\left( D_{\mu }\Phi \right) ^{\dagger }\left( D^\mu \Phi \right)
-\lambda \left( \Phi ^{\dagger }\Phi -\frac{v^2}2\right) ^2, 
\]
where $D^\mu = D_L^\mu$ of Eq.(\ref{Eq:CovariantDerivatives}). 
The quadratic terms for the gauge bosons emerge from the first term. 
After diagonalizing the bosonic mass matrices and using Eq.~(\ref{Eq:DefWZA}) 
we arrive at the mass relations: 
\begin{eqnarray*}
m_W =\frac{gv}2, ~~~ m_Z =\frac{\sqrt{g^2+g\prime ^2}\;v}2, ~~~ 
m_H =v\sqrt{2\lambda },
\end{eqnarray*}
and $m_A=0$, as expected.

Fermion masses can be generated trough Yukawa interaction terms between
the fermions and the $\Phi$ field. Utilizing that the scalar field is an
$SU(2)_L$ doublet, the fermion Yukawa terms (shown here only for the first 
generation of fermions) can be written in a gauge invariant form:
\[
{\cal L_Y}=-\lambda _u(\overline{u}_L~\overline{d}_L)\,\widetilde{\Phi }\,
u_R-\lambda _d(\overline{u}_L~\overline{d}_L)\,\Phi\,d_R-\lambda _e(\overline{%
\nu }_{eL}~\overline{e}_L)\,\Phi\,e_R+{\rm hermitian~conjugate},
\]
where 
$\widetilde{\Phi }=2 i T_2 \Phi^*$ is the charge conjugate of $\Phi$,
and we assumed that neutrinos are massless.
Yukawa mass terms induce the possibility of physically observable mixing 
between fermions, and there is empirical evidence that the fermions of 
different families mix. This mixing can be represented as follows.
The electroweak (gauge) eigenstates are expressed in terms of the mass 
eigenstates through a unitary rotation as 
\[
\left( \matrix{
u_L\cr
c_L\cr
t_L\cr
}\right) =U_L^u\left( \matrix{
{\bf u}_L\cr
{\bf c}_L\cr
{\bf t}_L\cr
}\right) ,~~~\left( \matrix{
d_L\cr
s_L\cr
b_L\cr
}\right) =U_L^d\left( \matrix{
{\bf d}_L\cr
{\bf s}_L\cr
{\bf b}_L\cr
}\right) ,~~~etc,
\]
where we denote the spinors of the mass eigenstates by bold letters. After
the diagonalization of the fermion mass matrices the mixing among the
leptons (assuming zero neutrino masses) or the right handed quarks can be 
absorbed into the definition of the fields.
After redefining the left-handed up type quarks
such that their mass matrix is diagonal,
the remaining effect can be described by the
Cabibbo--Kobayashi--Maskawa (CKM) weak mixing matrix \cite
{Cabibbo-Kobayashi-Maskawa}, $V=U_L^{u\dagger }U_L^d$, relating the mass
and electroweak eigenstates of the left handed down type quarks as 
$(q_{L~mass}^{(-1/2)})_f = V_{ff^{\prime }}\;(q_L^{(-1/2)})_{f^{\prime}}$. 
The indices $f,f^{\prime }=1,2,3$ run over the three families.

The free parameters associated with the spontaneous symmetry breaking are
the Higgs boson mass $m_H$, and the vev $v$. Besides, the Yukawa couplings 
$\lambda_{flavor}$ are not restricted by the symmetries of the SM, so
none of the fermion masses are predicted, which increases the number of the
free parameters by 9. After the re-phasing of the quark fields 3 independent
angles and a phase parametrizes the unitary CKM matrix.
In total, there are 19 free parameters in the SM, including the QCD $\theta$ parameter.
While the gauge sector of the SM is well established, the symmetry breaking
mechanism is awaiting confirmation from the experiments. The Higgs boson, up
to date, is a hypothetical particle, and the Higgs potential has hardly any
experimental constraints \cite{Langacker}. There are also alternative
formulations of the spontaneous symmetry breaking which avoid the
introduction of an elementary scalar \cite{Chivukula-Spira-Zerwas-Others}.

\section{The Quantum Nature of Gauge Fields}

The detailed description of the quantization process of gauge fields is
laid out in many textbooks \cite{QFT,StermanQFT}. Here we only
highlight those results which are necessary for the understanding the
rest of this work. Because we want to focus on the partial summation of
the perturbative series, to introduce the basic definitions and to set
up a simple example, in this Section first we examine the strong
coupling constant, as the running coupling between quantum fields. Then
we explain the basic ideas of factorization which connects the hadronic
level cross sections to the parton cross sections which are calculable
from the SM Lagrangian.

\subsection{Renormalization}

In the pioneering days of QED it became evident that beyond the lowest order 
calculations individual terms appear to contain infinities. Ultraviolet (UV)
singularities arise because of the local nature of the field theory,
combined with the assumption that the theory is valid at all energy scales.
When particle loops are shrunk to a single space-time point, that is the
momentum of the particle in the loop is taken to be infinity, the
corresponding integral in the Feynman rules is divergent. Similarly, when a
massless particle, e.g. a photon or a gluon, forms a loop and its momentum
vanishes, the Feynman integral over the loop momentum is singular. This
latter is a type of the infrared (IR) singularities.

In a field theory we encounter two typical attributes when calculating
physical observables using the method of the perturbative expansion. The
first is the systematic redefinition of the parameters of the theory,
order by order in the perturbation series. This feature is the
consequence of the perturbative method, and already present in quantum
mechanics. The second property, specific to field theories, are the
appearance of the infinities. These two features are connected in
renormalization, when the redefinition of the parameters is tied to the
removal of the infinities. Renormalization is only possible if the
infinities are universal, which is the remarkable case in the SM.

In order to execute the renormalization procedure in a quantum field theory,
first we have to regularize its infinities. There were several regularization
methods proposed in the literature \cite{Regularization}. Today the most
commonly used regularization method is dimensional regularization \cite
{DimensionalRegularization} , which preserves the symmetries of the SM
Lagrangian, most importantly, the Lorentz and gauge invariances. The idea of
dimensional regularization is based on the simple fact that integrals which
are singular in a given number of dimension, can be finite in another. In the
framework of the dimensional regularization we assume that the number of
dimensions in the loop-integrals $D$, i.e. the number of the space-time
dimensions, differs from four. We then calculate all integrals in $%
unspecified $ dimensions $D$, where all the results are finite. After the
removal of the singular terms from the results we can safely take the $%
D\rightarrow 4$ limit, and obtain meaningful physical predictions.

The procedure of relating the measurable parameters to the parameters of
the initial Lagrangian, while systematically removing the UV
divergences, by the introduction of suitable counter terms into the
Lagrangian order by order in the perturbation theory, is called
renormalization \cite{Collins:Renormalization}. If
the counter terms can be absorbed in the original terms of the Lagrangian by
multiplicative redefinition of the couplings ($g$), masses ($m$), gauge fixing
parameters ($\xi$) or the normalization of the fields themselves, 
then we say that
the theory is renormalizable. If an infinite number of counter terms needed
to define all measurables (in all perturbative orders), then the theory is not 
renormalizable. 
Historically, it was the proof of the
renormalizability of the Yang-Mills theories with spontaneous symmetry
breaking \cite{'tHooft71} which opened the door in front of the $%
SU(2)_L\times U(1)_Y$ gauge theory toward today's SM. 

After the renormalization procedure the parameters, $g$, $m$, $\xi $, etc.,
and the fields of the original, unrenormalized Lagrangian are referred to as
bare ones, and the redefined parameters 
\[
g_r=Z_g^{-1}g,\;\;\;m_r=Z_m^{-1/2}m,\;\;\;\xi _r=Z_\xi ^{-1}\xi ,\;\;\;\;etc., 
\]
and the fields as renormalized ones. There is a finite arbitrariness in the
definition of the singular terms which are removed by renormalization. This
is fixed by the renormalization scheme. Throughout this work we use the $%
\overline{MS}$ scheme, unless it is stated otherwise.

\subsection{Asymptotic Freedom and Confinement}

It is a generic feature of the parameters of a quantum field theory that
they exhibit a dependence on the (energy or distance) scale at which the
theory is applied. This dependence leads to crucial consequences in
Yang-Mills theories, for it results in asymptotic freedom and hints
confinement in QCD. In this Section we outline these features, which
play important roles in our resummation calculations.

When calculating a physical quantity beyond the lowest order, we
encounter infinities. After regularization these infinities can 
be isolated either in additive or multiplicative fashion. Taking the
QCD coupling constant as an example, we can write:
\begin{equation}
g = \mu ^{-\epsilon} Z_g(\mu) \; g_r(\mu).
\label{Eq:BareAndRenorm}
\end{equation}
The factor $\mu ^\epsilon $ is introduced to preserve the mass dimension
of the 4 dimensional bare coupling constant $g$, and $\epsilon
=(4-D)/2$. The above equation also reflects that in any process of
regularization we inevitably introduce a scale into the theory. In the
case of dimensional regularization this energy scale is the
renormalization scale $\mu$. The renormalization constant $Z_g$ embodies
the infinities, while renormalized coupling $g_r$ is
finite. Renormalizability of the SM exhibits itself in that $Z_g$ is process
independent. The evolution of the renormalized parameter as a function
of the renormalization scale can be described by the transformation 
\[
g_r(\mu )=z_g(\mu ,\mu ^{\prime })\;g_r(\mu ^{\prime }),
\]
where 
\[
z_g(\mu ,\mu ^{\prime })=Z_g^{-1}(\mu )\;Z_g(\mu ^{\prime }).
\]
These transformations form an Abelian group which is called the
renormalization group (RG).

The trivial requirement that the bare parameters of the Lagrangian are
independent of the renormalization scale is expressed, for the coupling, as: 
\[
\frac{dg}{d\mu }=0.
\]
This and the relation between the bare and renormalized coupling lead to
the differential equation 
\begin{equation}
\mu \frac{dg_r(\mu )}{d\mu }=\beta (g_r).  \label{Eq:RGEg}
\end{equation}
This equation is the renormalization group equation (RGE) for the coupling 
\cite{'tHooft73}. From the RGE and the relation of the bare and
renormalized coupling we deduce the relation between the $\beta$ function
and the renormalization constant $Z_g$: 
\[
\beta(g_r) = 
-g_r \left(\epsilon + \frac{\mu}{Z_g(\mu)} \frac{dZ_g(\mu)}{d\mu} \right).
\]
The above equation is the RGE for the renormalization constant
$Z_g$ and, in general, it is written as
\[
\gamma_g = \frac{\mu}{Z_g(\mu)} \frac{dZ_g(\mu)}{d\mu},
\]
where $\gamma_g$ is called the anomalous dimension (of the coupling, in this
specific case).
The renormalization constant is calculable order by order in the
perturbation theory by the requirement that the counter terms must cancel
the singular terms arising in the given order. In the lowest non-trivial
order in the coupling, for the $SU(N)$ gauge coupling one finds 
\cite{StermanQFT}: 
\[
Z_g(\mu )=1-\frac{g_r^2(\mu )}{4\pi ^2}\frac{11N-4T_RN_f}{24}\frac 1\epsilon
+{\cal O}(g_r^4).
\]
Here $N$ is the number of dimensions of the fundamental representation, $T_R$
is the normalization of the trace of the $SU(N)$ generators ($%
Tr(t^at^b)=\delta ^{ab}T_R$), and $N_f$ is the number of the light fermionic
flavors (i.e. $m_f<\mu $). With the aid of the last two equations we can
determine the lowest order coefficient in the perturbative expansion of the $%
\beta $ function
\[
\beta (g_r)=-g_r\frac \alpha \pi \beta _1-g_r\left( \frac \alpha \pi \right)
^2\beta _2-g_r\left( \frac \alpha \pi \right) ^3\beta _3+
{\cal O}(g_r\alpha ^4),
\]
where we introduced the running coupling $\alpha (\mu )=g_r^2(\mu )/(4\pi )$%
, in the fashion of the QED\ fine structure constant. The lowest order
coefficient is identified to be: 
\[
\beta _1=\frac{11N-4T_RN_f}{12}.
\]
The higher coefficients, $\beta _2$ and $\beta _3$, in the expansion of the 
$\beta (g_r)$ function were also calculated 
\cite{Tarasov-Vladimirov-Zharkov}.

By solving the RGE, using the lowest order truncation of the $\beta $
function, the running coupling can be expressed in terms of $\beta _1$%
\[
\alpha (\mu )=
\frac{\alpha _0}{1+\displaystyle{\frac{\alpha _0}\pi} \beta _1\ln 
\displaystyle{\frac{\mu ^2}{\mu _0^2}}},
\]
where $\alpha _0$ and $\mu _0$ are boundary value parameters. It is
customary to introduce a single parameter 
\[
\Lambda =\mu _0e^{-\pi /(2\beta _1\alpha _0)},
\]
and to write 
\[
\alpha (\mu )=
\frac \pi {\beta _1\ln \displaystyle{\frac{\mu ^2}{\Lambda ^2}}}.
\]

Equation (\ref{Eq:RGEg}) and its solution for the coupling has a striking
physical meaning: the coupling depends on the energy scale of the
interaction. From the solution we immediately see that as the scale $\mu $
increases the coupling strength decreases: 
\[
\lim_{\mu \rightarrow \infty }\alpha (\mu )=0.
\]
This is the phenomenon of asymptotic freedom, a necessity of a Yang-Mills 
type interaction. In QCD it means that the higher the energy we probe the 
quarks at, the least they interact. Asymptotically they are free at high 
energies, that is at short distances. On the other hand, as they separate, or 
at lower energies, their interaction becomes strong. This implies that when
quarks separate the color field becomes so strong between them that it will
be possible spontaneously create a new quark pair. This way quarks can never
be separated by more that a few femto-meters. They are confined within
hadronic bound states.

\FigSMasRunning
In order to determine the coupling constant we have to calculate at
least one physical quantity, e.g. a total scattering cross section, and
compare the calculation with the measurement of the cross section. In
practice, the calculation is necessarily truncated at a certain finite
order. For the truncated series to be a good approximation of the full
sum, the expansion variable, the coupling constant, has to be small. We
also want to keep the logarithms of the renormalization scale in the
coefficients of the expansion small. These two conditions constrain the
value of the renormalization scale close to the typical physical scale
of the process at hand. 
Fig.~\ref{Fig:SMasRunning} shows the running of the QCD coupling, 
$\alpha_s(Q )=g_s^2(Q )/(4\pi )$, as the function of the physical energy scale 
$Q = \mu$, in the experimentally accessible region for 
the boundary values $\alpha _s(m_Z)=0.118\pm 0.003$.
The running, predicted by QCD, is in agreement with the experiments within
the uncertainties. 
From our argument it is non-trivial that $\alpha _s(\mu )$ is a
smooth function of the energy, since the value of the coefficient $\beta _1$
is changing abruptly as the function of $N_f$ at every quark mass threshold.
The continuous definition of $\alpha _s(\mu )$ is achieved by requiring a
non-continuous $\Lambda $ parameter. The $\Lambda $ parameter in QCD, $%
\Lambda _{QCD}(N_f,N_{order})$, becomes the function of the number of active
($m_q<\mu $) quark flavors $N_f$, and the perturbative order $N_{order}$,
up to which the $\beta $ function was calculated, to ensure a smooth
matching at the quark thresholds. As an example we show a set of $\Lambda
_{QCD}(N_f,N_{order})/$GeV values which is recently in use \cite{CTEQ4}: 
\[
\begin{array}{rrrrr}
\vspace{0.2 cm}
& N_f=3 & N_f=4 & N_f=5 & N_f=6 \\ 
\vspace{0.2 cm}
N_{order}=1 & 0.272 & 0.236 & 0.181 & 0.094 \\ 
N_{order}=2 & 0.354 & 0.298 & 0.202 & 0.081
\end{array}
\]

The running coupling is one of the simplest examples of a resummed
quantity. It is calculated by the reorganization of the perturbative
series, such that the largest all order contributions are included in
the new leading terms. The partial summation of the perturbative series
implemented in this fashion is called resummation, which is the main
theme of this work. Since the running coupling is a resummed quantity,
when used in a fixed order expression it automatically includes certain
{\it all order} effects in the calculation, and as such, it improves the
quality of the prediction.

\subsection{Factorization and Infrared Safety \label{Sec:Factorization}}

The QCD Lagrangian describes the interaction of quarks and gluons. On the
other hand, because of confinement, we observe only hadrons. According to
the parton model \cite{PartonModel}, hadrons are composed from partons:
quarks and gluons. Although, numerical implementations of QCD suggest that
the theory describes confinement,
we do not have a mechanism to
derive low energy (long distance) properties from the QCD Lagrangian. 
In order to describe interactions involving hadrons at 
high energies we use the tool of factorization. 
Factorization is the idea of
the separation of the short distance physics from the long distance physics
in high energy collisions. The short distance physics is described by
perturbative QCD. The long distance effects are showed to be universal,
parametrized, and extracted from measurements.

In order to illustrate the plausibility of factorization in $hadronic$
scattering cross sections, we first highlight the calculation of
$partonic$ cross sections. When calculating a partonic cross section
beyond the leading order, after UV renormalization, and cancellation of
the IR singularities between real emission and virtual diagrams the
cross section still contains collinear singularities. These collinear
singularities, on the other hand, can be isolated as multiplicative factors, 
just like in the case of UV renormalization. Similarly to 
Eq.~\ref{Eq:BareAndRenorm}, the partonic scattering cross section can be 
written as
\begin{equation}
d\sigma_{a_1a_2\to X}(\{s\}) = 
\int_0^1dx_1\int_0^1dx_2 
\,\phi_{b_1/a_1}(x_1,\mu )\,
d\hat{\sigma}_{b_1b_2\to X}(\{s\},x_1,x_2,\mu )\,
\phi_{b_2/a_2}(x_2,\mu ),
\label{Eq:Factorization}
\end{equation}
where summation over the double partonic indices $b_i$ is implied,
$\hat{\sigma}_{b_1b_2\to X}$ is the hard scattering cross section 
of two partons $b_1$ and $b_2$, and $\phi_{b/a}(x,\mu )$ can be interpreted as
the probability of the collinear emission of parton $b$
from parton $a$ with a momentum fraction $x$ at energy $\mu$, and $\{s\}$
represents the collection of the kinematical variables relevant to
$d\hat{\sigma}$ (e.g. the center of mass energy of the collision, etc.),
and $X$ is some arbitrary final state.
The hard cross section $\hat{\sigma}_{b_1b_2\to X}$ and the distributions
$\phi_{b/a}$ are calculable in perturbative QCD order by
order.\footnote{ In the lowest order of the perturbation theory the
distributions are just delta functions: 
$\phi _{b/a}(x,\mu )=\delta _{ba}\delta(1-x)+{\cal O}(\alpha_s(\mu),\mu)$.}
The above form is useful because, as it can be shown by direct calculation
for a given process, all the non-cancelling infrared
singularities associated with the emission of collinear partons are universal
and can be defined into the distributions $\phi_{b/a}$ and the hard 
scattering cross section $\hat{\sigma}_{b_1b_2\to X}$ is infrared safe 
(i.e. finite).
When calculating $\hat{\sigma}_{b_1b_2\to X}$ at higher orders of the 
perturbation theory, its infrared safety can be maintained by absorbing 
its collinear singularities into the distributions $\phi_{b/a}$ 
\cite{Altarelli-Parisi77}.
In this procedure there is a freedom of shifting part of the hard cross
section into the distributions. This introduces the factorization scale, 
$\mu$, dependence into the definition of both $\hat{\sigma}_{b_1b_2\to X}$ 
and $\phi_{b/a}$. Just like in the case of the UV renormalization, there are
different scheme choices for the parton densities, depending on the amount 
of the finite contributions arbitrarily defined into $\phi_{b/a}$.
This factorization of the infrared behavior is a factorization of the
short and long-distance physics, because the fact that 
$\hat{\sigma}_{b_1b_2\to X}$ is infrared safe implies that it cannot
depend on physics which is associated by long distances.
The long distance physics is described by the universal distributions 
$\phi_{b/a}$, which can be shown to be independent of the short distance
features captured by $\hat{\sigma}_{b_1b_2\to X}$.

We can show by explicit calculation that Eq.~(\ref{Eq:Factorization})
generalizes to hadronic cross sections:
\begin{equation}
d\sigma _{h_1h_2\to X}(\{s\})=\int_0^1dx_1\int_0^1dx_2\,
f_{b_1/h_1}(x_1,\mu )\,d%
\hat{\sigma}_{b_1b_2\to X}(\{s\},x_1,x_2,\mu )\,f_{b_2/h_2}(x_2,\mu ),
\label{Eq:FactTheo}
\end{equation}
where $f_{a/h}(x,\mu )$ is a universal (process independent)
function which gives the probability of finding parton $a$ in a $hadron$ $h$
with a momentum fraction of $x$ at the scale $\mu$. (While the introduction
of the partonic distributions is somewhat heuristic in the parton model,
they are defined formally in the operator product expansion in terms of
expectation values of non-local operators.) The factorization theorem \cite
{FactTheo}, generally formulated in Eqs.~(\ref{Eq:Factorization}) and 
(\ref{Eq:FactTheo}) is an
assumption, until it is proven process by process, and order by order
of the perturbation theory that the infrared singularities of the 
hard cross section can be factorized into the parton distributions, and an 
infrared safe hard cross section can be defined. 
The proofs are given in the literature for the most important processes
as deep inelastic scattering or the Drell-Yan lepton pair production.
After it is proven, the 
factorization theorem is used to predict cross sections involving hadrons. 
Comparing these predictions to the experiments, the parton distributions then 
can be extracted.

Although, presently we do not know how to calculate the parton distribution
functions of hadrons, $f_{a/h}(x,\mu)$, from the first principles, 
since their scale dependence is governed by perturbative QCD, based on the RG
technique we can calculate their evolution with the energy scale $\mu$:
\[
\mu \frac{df_{a/h}(x,\mu )}{d\mu }=\int_0^1\frac{d\xi }\xi P_{a\leftarrow
b}\left( \frac x\xi ,\alpha _s(\mu )\right) f_{b/h}(\xi ,\mu ),
\]
where the splitting functions, $P_{a\leftarrow b}(x,\alpha _s(\mu ))$,
are calculable order by order of the perturbation theory \cite{DGLAP}, 
since they describe the short distance physics of partonic splitting. 
(The analog of this equation can be written for partonic distributions
$\phi_{b/a}$, where the splitting functions are calculated by perturbative
methods.)
For the case of gluon radiation their ${\cal O}(\alpha_s)$ expressions are:
\begin{eqnarray}
P_{{j\leftarrow k}}^{(1)}(z) &=&
C_F\left( {\frac{1+z^2}{{1-z}}}\right) _{+}, \nonumber \\
P_{{j\leftarrow G}}^{(1)}(z) &=&
{\frac 12}\left[ z^2+(1-z)^2\right],
\label{Eq:DGLAP} 
\end{eqnarray}
where the ``+'' prescription for arbitrary functions $f(z)$ and $g(z)$ 
is defined as 
\[
\int_{x}^{1} dz \left( f(z) \right)_+ g(z) =
 \int_{0}^{1} dz \; f(z) \left[\Theta(z - x) g(z) - g(1) \right] , 
\]
with
\[
\Theta(x) = \cases{~0, & {\rm if} $x < 0$, \cr ~1, & {\rm if} $x \geq 0$,} 
\]
the unit step function. 
(It is assumed that $f(z)$ is less singular than $1/(1-z)^2$ at $z=1$.)
Similarly to the case of the running coupling, the
parton distributions, being solutions of the above RGE, include resummed
collinear contributions.

The renormalization of the ultraviolet
singularities and the factorization of the infrared divergences can be
viewed in parallel \cite{Tung}. The ultraviolet
singularities absorbed into the renormalization constants $Z_i$, 
just like the collinear divergences absorbed into the parton
distributions $f_{a/b}$, that is
in the factorization the parton distributions play the role 
of the renormalization constants, with the added complication that they
also depend on the momentum fraction $x$, besides the energy scale $\mu$.
The evolution equations, describing the scale dependence of the 
renormalization constants and the parton distributions can also be
casted into analogous forms, with the anomalous dimensions $\gamma_i$ 
identified with the splitting functions $P_{a\leftarrow b}$.




\chapter{Soft Gluon Resummation \label{ch:Resummation}}

\section{Lepton Pair Production at Fixed Order in $\alpha _s$ \label{sec:R1}}

In this work we apply and extend the resummation formalism developed by
Collins, Soper and Sterman (CSS) \cite{CS,CSS,Collins-Soper-Sterman}, which
resums large logarithmic contributions, arising as the consequence of the
initial state multiple soft gluon emission. We calculate these resummed
corrections to processes of the type $h_1h_2\rightarrow v_1v_2X$, where $h_i$
are hadrons, $v_1$ and $v_2$ is a pair of leptons, photons or $Z^0$ bosons, 
and $X$
is anything else produced in the collision undetected. More specifically, we
consider lepton pair production through electroweak vector boson production
and decay: $h_1h_2\rightarrow V(\rightarrow \ell _1{\bar{\ell}_2)}X$ ($V$ is 
$W^{\pm }$, $Z^0$ or a virtual photon), and vector boson pair production $%
h_1h_2\rightarrow VVX$ with $V$ being photon or $Z^0$ boson. To resum the
contributions from the soft gluon radiation, we use the results of the
various fixed order calculations of these processes, which are found in the
literature \cite
{Ellis-Martinelli-Petronsio,Aurenche-Lindfors,Mirkes,Bailey-Owens-Ohnemus,
Owens-Ohnemus}. In this Chapter we review the lepton pair production through 
$W^{\pm }$, $Z^0$ or virtual photon ($\gamma^*$) production and decay in some 
detail, to introduce a pedagogical example identifying technical details
which are most important for the resummation calculation.

\subsection{The Collins-Soper Frame}

Since the Collins-Soper-Sterman resummation formalism gives the cross
sections in a special frame, called the Collins-Soper ($CS$) frame~\cite
{CSFrame}, we give the detailed form of the Lorentz transformation between
the $CS$ and the laboratory ($lab$) frames. The $lab$ frame is defined as
the center-of-mass frame of the colliding hadrons $h_1$ and $h_2$. In the $%
lab$ frame, the cartesian coordinates of the hadrons are: 
\[
p_{h_1,h_2}^\mu (lab)=\frac{\sqrt{S}}2\,(1,0,0,\pm 1),
\]
where $\sqrt{S}$ is the center-of-mass energy of the collider. The $CS$
frame is the special rest frame of the vector boson in which the $z$ axis
bisects the angle between the $h_1$ hadron momentum $p_{h_1}(CS)$ and the
negative $h_2$ hadron momentum $-p_{h_2}(CS)$~\cite{lamtung}.

To derive the Lorentz transformation $\Lambda _{~\nu }^\mu (lab\rightarrow
CS)$ that connects the $lab$ and $CS$ frames (in the active view point): 
\[
p^\mu (CS)=\Lambda _{~\nu }^\mu (lab\rightarrow CS)\;p^\nu (lab), 
\]
we follow the definition of the $CS$ frame. Since the invariant amplitude is
independent of the azimuthal angle of the vector boson ($\phi _V$), without
loosing generality we start from a $lab$ frame in which $\phi _V$ is zero. 
First, we find the boost into a vector boson rest frame. Then, in the vector
boson rest frame we find the rotation which brings the hadron momentum $%
p_{h_1}(CS)$ and negative hadron momentum $-p_{h_2}(CS)$ into the desired
directions.

A boost by $\vec{\beta}=-\vec{q}(lab)/q^0$ brings four vectors from the lab
frame (with $\phi _V = 0$) into a vector boson rest frame ($rest$). The
matrix of the Lorentz boost from the $lab$ frame to the $rest$ frame,
expressed explicitly in terms of $q^\mu$ is 
\begin{eqnarray*}
&&\Lambda _{~\nu }^\mu (lab\rightarrow rest)= \\
&&{\frac 1Q}\left( 
\begin{tabular}{rrrr}
$q^0$ \quad & $-Q_T$ \quad & $0$ \quad & $-q^3$ \\ 
$-Q_T$ \quad & $Q+{\frac{Q_T^2}{q^0+Q}}$\quad & $0$ \quad & ${\frac{Q_Tq^3}{%
q^0+Q}}$ \\ 
$0$ \quad & ${0}$ \quad & $Q$ \quad & ${0}$ \\ 
$-q^3$ \quad & ${\frac{Q_Tq^3}{q^0+Q}}$ \quad & $0$ \quad & $Q+\frac{(q^3)^2%
}{q^0+Q}$%
\end{tabular}
\right) .
\end{eqnarray*}
where $Q=\sqrt{(q^0)^2-Q_T^2-(q^3)^2}$ is the vector boson invariant mass,
and the transverse mass is defined as $M_T=\sqrt{Q^2+Q_T^2}$. 

After boosting the lab frame hadron momenta into this rest frame, we obtain 
\begin{eqnarray*}
p_{h_1,h_2}^\mu (rest) &=&\Lambda _{~\nu }^\mu (lab\rightarrow
rest)\;p_{h_1,h_2}^\nu (lab)= \\
&&\frac{\sqrt{S}}2\left( \frac{q^0\mp q^3}Q,-\frac{Q_T}Q\frac{q^0+Q\mp q^3}{%
q^0+Q},0,\frac{(\pm Q-q^3)(q^0+Q)\pm (q^3)^2}{Q(q^0+Q)}\right) ,
\end{eqnarray*}
and the polar angles of $p_{h_1}^\mu (rest)$ and $-p_{h_2}^\mu (rest)$ are
not equal unless $Q_T=0$. (In the above expressions the upper signs refers
to $h_1$ and the lower signs to $h_2$.) In the general $Q_T\neq 0$ case we
have to apply an additional rotation in the $rest$ frame so that the $z$%
-axis bisects the angle between the hadron momentum $p_{h_1}(CS)$ and the
negative hadron momentum $-p_{h_2}(CS)$. It is easy to verify that to keep $%
\vec{p}_{h_1,h_2}$ in the $xz$ plane, this rotation should be a rotation
around the $y$ axis by an angle $\alpha =\arccos {[Q(q^0+M_T)/(M_T(q^0+Q))]}$%
.

Thus the Lorentz transformation from the $lab$ frame to the $CS$ frame is $%
\Lambda _{~\nu} ^\mu (lab\rightarrow CS)=\Lambda _{~\lambda} ^\mu
(rest\rightarrow CS)\;\Lambda _{~\nu} ^\lambda (lab\rightarrow rest)$.
Indeed, this transformation results in equal polar angles $\theta
_{h_1,-h_2}=\arctan (Q_T/Q)$. The inverse of this transformation takes
vectors from the $CS$ frame to the $lab$ frame is: 
\begin{eqnarray*}
\Lambda _{~\nu }^\mu (CS &\rightarrow &lab)=\left( \Lambda _{~\nu }^\mu
(lab\rightarrow CS)\right) ^{-1}= \\
&& {\frac{1}{Q M_T}} \left( 
\begin{tabular}{rrrr}
$q^0 M_T$ \quad & $q^0 Q_T$ \quad & $0$ \quad & $q^3 Q$ \\ 
$Q_T M_T$ \quad & $M_T^2$ \quad & $0$ \quad & $0$ \\ 
$0$ \quad & $0$ \quad & $Q M_T$ \quad & $0$ \\ 
$q^3 M_T$ \quad & $q^3 Q_T$ \quad & $0$ \quad & $q^0 Q$%
\end{tabular}
\right) .
\end{eqnarray*}

The kinematics of the leptons from the decay of the vector boson can be
described by the polar angle $\theta $ and the azimuthal angle $\phi $,
defined in the Collins-Soper frame. The above transformation formulae lead
to the four-momentum of the decay product fermion (and anti-fermion) in the
lab frame as 
\begin{eqnarray*}
p^\mu &=&{\frac Q2}\left( {\frac{q^\mu }Q}+\sin {\theta }\cos {\phi }~X^\mu
+\sin {\theta }\sin {\phi }~Y^\mu +\cos {\theta }~Z^\mu \right) , \\
\overline{p}^\mu &=&q^\mu -p^\mu ,
\end{eqnarray*}
where 
\begin{eqnarray}
&&q^\mu =(M_T\cosh {y},\,Q_T\cos {\phi _V},\,Q_T\sin {\phi _V},\,M_T\sinh {y}%
),  \nonumber \\
\ &&X^\mu =-{\frac Q{Q_TM_T}}\left( q_{+}n^\mu +q_{-}{\bar{n}}^\mu -{\frac{%
M_T^2}{Q^2}}q^\mu \right) ,  \nonumber \\
\ &&Z^\mu ={\frac 1{M_T}}\left( q_{+}n^\mu -q_{-}{\bar{n}}^\mu \right) , 
\nonumber \\
\ &&Y^\mu =\varepsilon ^{\mu \nu \alpha \beta }{\frac{q_\nu }Q}Z_\alpha
X_\beta .  \label{eq:QXYZ}
\end{eqnarray}
Here, $q_{\pm }={\frac 1{\sqrt{2}}}(q^0\pm q^3)$, $y=\frac 12\ln
(q_{+}/q_{-})$, $n^\nu =\frac 1{\sqrt{2}}(1,0,0,1)$, ${\bar{n}}^\nu ={\frac
1{\sqrt{2}}}(1,0,0,-1)$ and the totally anti-symmetric unit tensor is
defined as $\varepsilon ^{0123}=-1$.

\subsection{General Properties of the Cross Section \label{GeneralProp}}

Along the lines of the factorization theorem (cf. Eq.~\ref{Eq:FactTheo}), 
the inclusive differential
cross section of the lepton pair production in a hadron collision process $%
h_1h_2\to V(\to \ell _1{\bar{\ell}_2})X$ is written as 
\begin{eqnarray*}
&&d\sigma _{h_1h_2}(Q,Q_T,y,\theta ,\phi ,\alpha _s(Q))= \\
&&~~~\int_{x_1}^1d\xi _1\int_{x_2}^1d\xi _2\sum_{a,b} \;\,
f_{a/h_1}(\xi_1,\mu)\,\;
d\hat{\sigma}_{ab}(Q,Q_T,y,\theta ,\phi ,z_1,z_2,\mu)\;\,
f_{b/h_2}(\xi_2,\mu ).
\end{eqnarray*}
Here $d\hat{\sigma}_{ab}$ is the differential cross section of the parton
level process $ab\to V(\to \ell _1{\bar{\ell}_2})X$, and $f_{a/h}$ are the
parton distribution functions. The variable $Q$ denotes the invariant mass, $%
y$ the rapidity and $Q_T$ the transverse momentum of the lepton pair, while $%
\theta $ and $\phi $ are the polar and azimuthal angles of the lepton $\ell
_1$ in the rest frame of $V$.
The indices $a$ and $b$ denote different parton flavors (including
gluon), and $V$ may stand for $W^{\pm }$, $Z^0$ or $\gamma ^{*}$. The
partonic momentum fractions $x_i$ are related to these independent 
variables as 
\[
x_1=\frac{M_T}{\sqrt{S}}e^y,\;\;\;x_2=\frac{M_T}{\sqrt{S}}e^{-y}, 
\]
and 
\[
z_{1,2}={\frac{x_{1,2}}{\xi _{1,2}}.} 
\]

The partonic cross section $\hat{\sigma}_{ab}$ is calculated as 
\[
d\hat{\sigma}_{ab}(Q,Q_T,y,\theta ,\phi ,x_1,x_2,\mu )=\frac 1{2s}\overline{%
\left| {\cal M}_{ab}(s,t,u,\mu )\right| ^2}d\Phi . 
\]
where $s$, $t$, and $u$ are the Mandelstam invariants 
\[
s=(p_{a}+p_{b})^2,\;\;\;t=(p_{a}-Q)^2,\;\;\;u=(p_{b}-Q)^2 
\]
with $p_{a}$ and $p_{b}$ being the four momenta of the incoming partons, and 
$Q$ is the four momentum of the lepton pair. Momentum conservation, $%
p_{a}+p_{b}=Q+p_X$, can be expressed in the form of the Mandelstam relation 
\[
s+t+u=Q^2, 
\]
where we used $p_X^2=0$. The phase space factor $d\Phi $, assuming massless
leptons, in terms of the relevant kinematical variables, is 
\[
d\Phi =\frac 1{2^9\pi ^4}\;\delta (s+t+u-Q^2)\;dQ^2dQ_T^2dyd\cos \theta
d\phi . 
\]
The averaged square of the invariant amplitude 
\[
\overline{\left| {\cal M}_{ab}(s,t,u,\mu)\right| ^2}=\frac
1{n_a^cn_b^cn_a^sn_b^s}\sum_{colors}\left| {\cal M}_{ab}(s,t,u,\mu
_f)\right| ^2, 
\]
where $n_{a,b}^s$ and $n_{a,b}^c$ are the number of spin and color degrees
of freedom of partons $a$ and $b$.

Using the relation $s=\xi _1\xi _2S$, where $\sqrt{S}$ is the center of mass
energy of the collider, the differential cross section of the lepton pair
production in a fixed order of the strong coupling is written as 
\begin{eqnarray}
&&{\left( {\frac{d\sigma (h_1h_2\rightarrow V(\rightarrow \ell _1{\bar{\ell}%
_2})X)}{dQ^2dQ_T^2dyd\cos {\theta }d\phi }}\right) }_{{\rm Fixed}}=\frac
1{2^{10}\pi ^4S}\sum_{a,b}\int_{x_1}^1{\frac{d\xi _1}{\xi _1}}\int_{x_2}^1{%
\frac{d\xi _2}{\xi _2}}  \nonumber  \label{eq:PerCroSec} \\
&&~~~\times ~\,f_{a/h_1}(\xi _1,\mu ) \overline{\left| {\cal M}%
_{ab}(s,t,u,\mu )\right| ^2}\delta (s+t+u-Q^2) \,f_{b/h_2}(\xi _2,\mu ). 
\nonumber
\end{eqnarray}
In the above equation both the invariant amplitude ${\cal M}_{ab}(s,t,u,\mu )
$ and the parton distribution functions $f_{a/h}(x,\mu )$ depend on the
renormalization scale 
\[
\mu =C_4Q 
\]
and the corresponding value, $g_s(\mu )=\sqrt{4\pi \alpha _s(\mu )}$, of the
QCD coupling. Normally one sets the constant $C_4$ to be of order 1, to
avoid large logarithms $\ln (Q/\mu )$, that would otherwise spoil the
usefulness of a low--order perturbative approximation of ${\cal M}_{ab}$.

The squared invariant amplitude of the partonic level process is calculated
contracting the hadronic and leptonic tensors $H^{\mu \nu }$ and $L_{\mu \nu
}$: 
\[
\left| {\cal M}_{ab}(s,t,u,Q^2)\right| ^2=H_{ab}^{\mu \nu }(k,l,Q)L_{\mu \nu
}(l_1,l_2,Q) 
\]
where, for the $q\bar{q}\rightarrow V(\rightarrow \ell _1{\bar{\ell}_2})g$
process $k,l,l_1,l_2$ are the quark, anti-quark, lepton and anti-lepton
four-momenta, respectively. The detailed from of the hard cross section is
found in the next Section. The hadronic tensor $H_{ab}$ is calculated by
evaluating the ${\cal O}(\alpha _s)$ real and virtual gluon emission
diagrams and adding together these contributions. In order to keep track of
the vector boson polarization one has to evaluate the symmetric and
anti-symmetric parts of both the hadronic and leptonic tensors. Since the
calculation of the leptonic tensor is trivial we focus on the hadronic
tensor.

The hadronic tensor can be split into a sum of a symmetric and an
anti-symmetric tensor 
\[
H_{ab}^{\mu \nu }(k,l,Q)=H_{ab,S}^{\mu \nu }(k,l,Q)+H_{ab,A}^{\mu \nu
}(k,l,Q). 
\]
Here $H_{ab,S}^{\mu \nu }(k,l,Q)$ depends on symmetric combinations of the
four-momenta and the metric tensor. The calculation of the symmetric part of
the hadronic tensor $H_{ab,S}^{\mu \nu }(k,l,Q)$ is straightforward and does
not depend on the prescription of the $\gamma _5$. This is because one does
not have to evaluate a trace that contains $\gamma _5$ to calculate $%
H_{ab,S}^{\mu \nu }(k,l,Q)$. But the anti-symmetric part of the hadronic
tensor involves traces containing single $\gamma _5$'s. These traces are, by
definition, proportional to the fully anti-symmetric tensor $\varepsilon
^{\mu \nu \alpha \beta }$, so that the calculation of $H_{ab,A}^{\mu \nu
}(k,l,Q)$ lead to the form 
\[
H_{ab,A}^{\mu \nu }(k,l,Q)=h_1(s,t,u,Q^2)\varepsilon ^{\mu \nu
kQ}+h_2(s,t,u,Q^2)\varepsilon ^{\mu \nu lQ}+
\epsilon \; h_3(s,t,u,Q^2)\varepsilon ^{\mu \nu kl}. 
\]
The last term in this equation would violate the Ward identities in four
dimension. Although proportional to $\epsilon =(4-D)/2$, 
after the phase-space integration it contributes to the four-dimensional cross
section if $h_3$ does not vanish. 

\subsection{The Cross Section at ${\cal O}(\alpha _s)$ \label{TheCrossS}}

To correctly extract the distributions of the leptons, we have to calculate
the production and the decay of a polarized vector boson. The ${\cal O}%
(\alpha _s)$ QCD corrections to the production and decay of a polarized
vector boson can be found in the literature \cite{Aurenche-Lindfors}, in
which both the symmetric and the anti-symmetric parts of the hadronic tensor
were calculated. Such a calculation was, as usual, carried out in general
number ($D$) of space-time dimensions, and dimensional regularization 
was used to regulate infrared (IR) divergences because it preserves the
gauge and the Lorentz invariances. Since the anti-symmetric part of the
hadronic tensor contains traces with an odd number of $\gamma _5$'s, one has
to choose a definition (prescription) of $\gamma _5$ in $D$ dimensions. It
was shown in a series of papers \cite{gamma5Def,gamma5c} 
that in $D\neq 4$ dimension,
the consistent $\gamma _5$ prescription to use is the t'Hooft-Veltman
prescription \cite{gamma5Def}. 
Since in Ref.~\cite{Aurenche-Lindfors} a different
prescription~\cite{gamma5n} was used, we give below the results of our
calculation in the t'Hooft-Veltman $\gamma _5$ prescription.

For calculating the virtual corrections, we follow the argument of Ref.~\cite
{Korner-Schuler} and impose the chiral invariance relation, which is
necessary to eliminate ultraviolet anomalies of the one loop axial vector
current when calculating the structure function. Applying this relation for
the virtual corrections we obtain the same result as that in Refs.~\cite
{Aurenche-Lindfors} and~\cite{AEM79}. The final result of the virtual
corrections gives 
\begin{eqnarray}
&&{\cal M}_{Born}^{\dagger }{\cal M}_{Virt}+{\cal M}_{Virt}^{\dagger }{\cal %
M}_{Born}=  \nonumber \\
&&~~~~~~~~~~ 
C_F\frac{\alpha _s}{2\pi }\left( \frac{4\pi \mu ^2}{Q^2}\right)^\epsilon
\frac 1{\Gamma (1-\epsilon )}\left( -\frac 2{\epsilon ^2}-\frac 3\epsilon
+\pi ^2-8\right) \left| {\cal M}_{Born}\right| ^2,  \label{Eq:VirtCorr}
\end{eqnarray}
where $\epsilon =(4-D)/2$, $\mu $ is the t'Hooft mass (renormalization)
scale, and $C_F=4/3$ in QCD. The four-dimensional Born level amplitude is 
\begin{eqnarray}
\left| {\cal M}_{Born}\right| ^2 &=&{\frac{16Q^4}{(Q^2-M_V^2)^2+Q^4\Gamma
_V^2/M_V^2}}  \nonumber \\
&&\times \left[ (g_L^2+g_R^2)(f_L^2+f_R^2){\cal L}%
_0+(g_L^2-g_R^2)(f_L^2-f_R^2){\cal A}_3\right] ,  \label{eq:MBorn}
\end{eqnarray}
where we have used the LEP prescription for the vector boson resonance with
mass $M_V$ and width $\Gamma _V$. The angular functions are ${\cal L}%
_0=1+\cos ^2{\theta }$ and ${\cal A}_3=2\cos {\theta }$. The initial state
spin average (1/4), and color average (1/9) factors are not yet included in
Eq.~(\ref{eq:MBorn}).

When calculating the real emission diagrams, we use the same
(t'Hooft-Veltman) $\gamma _5$ prescription. It is customary to organize the $%
{\cal O}(\alpha _s^n)$ corrections by separating the lepton degrees of
freedom from the hadronic ones, so that 
\begin{eqnarray*}
&&{\left( {\frac{d\sigma (h_1h_2\rightarrow V(\rightarrow \ell _1{\bar{\ell}%
_2})X)}{dQ^2dydQ_T^2d\cos {\theta }d\phi }}\right) }_{{\cal O}(\alpha _s)}^{%
{\rm real~emission}}={\frac{\alpha _s(Q)C_F}{(2\pi )^3S}}{\frac{Q^2}{%
(Q^2-M_V^2)^2+Q^4\Gamma _V^2/M_V^2}} \\
&&~~~\times \sum_{a,b,i}\int_{x_1}^1{\frac{d\xi _1}{\xi _1}}\int_{x_2}^1{%
\frac{d\xi _2}{\xi _2}}~{\cal G}^{i\;}{\cal L}_{ab}(\xi _1,\xi _2,Q^2)\;%
{\cal T}_{ab}^i(Q_T,Q,z_1,z_2)\ {\cal A}_i(\theta ,\phi ),
\end{eqnarray*}
with $z_1=x_1/\xi _1$ and $z_2=x_2/\xi _2$. The dependence on the lepton
kinematics is carried by the angular functions 
\begin{eqnarray*}
\ &&{\cal L}_0=1+\cos ^2{\theta },~{\cal A}_0={\frac 12}(1-3\cos ^2{\theta }%
),~{\cal A}_1=\sin {2\theta }\cos {\phi },~{\cal A}_2={\frac 12}\sin ^2{%
\theta }\cos {2\phi },~ \\
\ &&{\cal A}_3=2\cos {\theta },~{\cal A}_4=\sin {\theta }\cos {\phi }.
\end{eqnarray*}
In the above differential cross section, $i=-1,...,4$ with ${\cal A}%
_{-1}\equiv {\cal L}_0$; and ${\cal G}^i=(g_L^2+g_R^2)(f_L^2+f_R^2)$ for $%
i=-1,0,1,2;$ ${\cal G}^i=(g_L^2-g_R^2)(f_L^2-f_R^2)$ for $i=3,4$. The parton
level cross sections are summed for the parton indices $a$, $b$ in the
following fashion 
\[
\sum_{a,b}{\cal L}_{ab}\;{\cal T}_{ab}^i=\sum_{q=u,d,s,c,b}\left( {\cal L}_{q%
\bar{q}}\;{\cal T}_{q\bar{q}}^i+{\cal L}_{\bar{q}q}\;{\cal T}_{\bar{q}q}^i+%
{\cal L}_{qG}\;{\cal T}_{qG}^i+{\cal L}_{\bar{q}G}\;{\cal T}_{\bar{q}G}^i+%
{\cal L}_{Gq}\;{\cal T}_{Gq}^i+{\cal L}_{G\bar{q}}\;{\cal T}_{G\bar{q}%
}^i\right). 
\]
The partonic luminosity functions ${\cal L}_{ab}$\ are defined as 
\[
{\cal L}_{q\bar{q}}(\xi _1,\xi _2,Q^2)=f_{q/h_1}(\xi _1,Q^2)\;f_{\bar{q}%
/h_2}(\xi _2,Q^2),\;etc., 
\]
where $f_{q/h_1}$ is the parton probability density of parton $q$ in hadron $%
h_1$, etc. The squared matrix elements for the annihilation sub-process $q{%
\bar{q}}\rightarrow VG$ in the $CS$ frame, including the $\epsilon $
dependent terms, are as follows: 
\begin{eqnarray*}
\ {\cal T}_{q\bar{q}}^{-1} &=&\frac 1{ut}\left( T_{+}(u,t)-(t+u)^2\epsilon
\right) ,\;\; \\
{\cal T}_{q\bar{q}}^0\;\; &=&{\cal T}_{q\bar{q}}^2=\frac 1{ut}\frac{Q_T^2}{%
M_T^2}\left( T_{+}(u,t)-(Q^2+s)^2\epsilon \right) ,\; \\
{\cal T}_{q\bar{q}}^1\;\; &=&\frac 1{ut}\frac{Q_TQ}{M_T^2}%
T_{-}(u,t)(1-\epsilon ),\; \\
{\cal T}_{q\bar{q}}^3\;\; &=&\frac 1{ut}\frac Q{M_T}\left( T_{+}(u,t)-{\frac{%
(Q^2-u)t^2+(Q^2-t)u^2}{Q^2}}\epsilon \right) ,\; \\
{\cal T}_{q\bar{q}}^4\;\; &=&\frac 2{ut}\frac{Q_T}{M_T}\left(
T_{-}(u,t)+Q^2(u-t)\epsilon \right) .
\end{eqnarray*}
For the Compton sub-process $qG\rightarrow Vq$, we obtain 
\begin{eqnarray*}
\ {\cal T}_{qG}^{-1} &=&\frac{-1}{su}\left( T_{+}(s,u)-{(s+u)^2}\epsilon
\right) ,\;\; \\
{\cal T}_{qG}^0\;\; &=&{\cal T}_{qG}^2=\frac{-1}{su}\frac{Q_T^2}{M_T^2}%
\left( (Q^2-u)^2+(Q^2+s)^2-{(s+u)^2}\epsilon \right) , \\
\;{\cal T}_{qG}^1\;\; &=&\frac{-1}{su}\frac{Q_TQ}{M_T^2}\left(
2(Q^2-u)^2-(Q^2-t)^2+{(s+u)^2}\epsilon \right) ,\; \\
{\cal T}_{qG}^3\;\; &=&\frac{-1}{su}\frac Q{M_T}\left( T_{+}(s,u)-2u(Q^2-s)+%
\frac{(Q^2-t)(Q^2s-su-u^2)}{Q^2}\epsilon \right) , \\
{\cal T}_{qG}^4\;\; &=&\frac{-2}{su}\frac{Q_T}{M_T}\left(
2s(Q^2-s)+T_{+}(s,u)-(Q^2-t)u\epsilon \right) .
\end{eqnarray*}
In the above equations, the Mandelstam variables: $s=(k+l)^2$, $t=(k-Q)^2$,
and $u=(l-Q)^2$ where $k$, $l$ and $q$ are the four momenta of the partons
from hadrons $h_1$, $h_2$ and that of the vector boson, respectively, and $%
T_{\pm }(t,u)=(Q^2-t)^2\pm (Q^2-u)^2$. All other necessary parton level
cross sections can be obtained from the above as summarized by the following
crossing rules:

\begin{center}
\begin{tabular}{ll}
\hline\hline \\[-.2cm]
\vspace{.2cm} $i=-1,0,1,2$ & $i=3,4$ \\ \hline \\[-.2cm]
\vspace{.2cm} ${\cal T}_{\bar{q}q}^i={\cal T}_{q\bar{q}}^i$ & ${\cal T}_{%
\bar{q}q}^i= -{\cal T}_{q\bar{q}}^i$ \\ 
\vspace{.2cm} ${\cal T}_{Gq}^i={\cal T}_{qG}^i\;(u\leftrightarrow
t)\;\;\;\;\;\;\;$ & ${\cal T}_{Gq}^i=-{\cal T}_{qG}^i\;(u\leftrightarrow t)$
\\ 
\vspace{.2cm} ${\cal T}_{\bar{q}G}^i={\cal T}_{qG}^i$ & ${\cal T}_{\bar{q}%
G}^i=-{\cal T}_{qG}^i$ \\ 
\vspace{.2cm} ${\cal T}_{G\bar{q}}^i={\cal T}_{Gq}$ & ${\cal T}_{G\bar{q}%
}^i=-{\cal T}_{Gq}^i,$ \\ \hline\hline
\end{tabular}
\end{center}
\vspace{0.1cm} 
with the only exceptions that ${\cal T}_{Gq}^1=-{\cal T}%
_{qG}^1\;(u\leftrightarrow t)$\ and ${\cal T}_{Gq}^4={\cal T}%
_{qG}^4\;(u\leftrightarrow t)$. These results are consistent with the
regular pieces of the $Y$ term given in Section \ref
{Sec:RegularContributions} and with~those in Ref. \cite{Mirkes}.

In the above matrix elements, only the coefficients of ${\cal L}_0$ and $%
{\cal A}_3$ are not suppressed by $Q_T$ or $Q_T^2$, so they contribute to
the singular pieces which are resummed in the CSS formalism. By definition
we call a term singular if it diverges as 
$Q_T^{-2}$ $\times$ [1 or $\ln (Q^2/Q_T^2)$]
as $Q_T\rightarrow 0$. Using the t'Hooft-Veltman prescription
of $\gamma _5$ we conclude that the singular pieces of the symmetric (${\cal %
L}_0$) and anti-symmetric (${\cal A}_3$) parts are the same, and 
\begin{eqnarray*}
\lim\limits_{Q_T\rightarrow 0}\ {\cal T}_{q\bar{q}}^{-1}\;\delta (s+t+u-Q^2)
&=&\lim\limits_{Q_T\rightarrow 0}\ {\cal T}_{q\bar{q}}^3\;\delta (s+t+u-Q^2)=%
{s}_{q\bar{q}}, \\
\lim\limits_{Q_T\rightarrow 0}\ {\cal T}_{Gq}^{-1}\;\delta (s+t+u-Q^2)
&=&\lim\limits_{Q_T\rightarrow 0}\ {\cal T}_{Gq}^3\;\delta (s+t+u-Q^2)={s}%
_{Gq},
\end{eqnarray*}
where 
\begin{eqnarray*}
{s}_{q\bar{q}} &=&\frac 1{Q_T^2}\left[ 2\;\delta (1-z_1)\;\delta
(1-z_2)\left( \ln \frac{Q^2}{Q_T^2}-\frac 32\right) \right. \\
&&\left. +\delta (1-z_1)\left( \frac{1+z_2{}^2}{1-z_2}\right) _{+}+\delta
(1-z_2)\left( \frac{1+z_1{}^2}{1-z_1}\right) _{+}\right. \\
&&\left. -\left( (1-z_1)\;\delta (1-z_2)+(1-z_2)\;\delta (1-z_1)\right)
\epsilon \right] +{\cal O}\left( {\frac 1{Q_T}}\right) , \\
{s}_{Gq} &=&\frac 1{Q_T^2}\left[ (z_1^2+\ (1-z_1^2))\;\delta (1-z_2)-\delta
(1-z_2)\epsilon \right] +{\cal O}\left( {\frac 1{Q_T}}\right) .
\end{eqnarray*}
As $Q_T\rightarrow 0$, only the ${\cal L}_0$ and ${\cal A}_3$ helicity cross
sections survive as expected, since the ${\cal O}(\alpha _s^0)$ differential
cross section contains only these angular functions [cf. Eq.~(\ref{eq:MBorn}%
)].

\section{The Resummation Formalism \label{sec:22}}

\subsection{Renormalization Group Analysis \label{sec:221}}

In hard scattering processes, the dynamics of the multiple
soft--gluon radiation is predicted by resummation \cite{resum}--\cite
{resum4}.
Since our work is based on the Collins-Soper-Sterman resummation formalism 
\cite{CS,CSS,Collins-Soper-Sterman}, in this Section we review the most
important details of the proof of this formalism. 
After the pioneering work
of Dokshitser, D'Yakonov, and Troyan \cite{DDT}, it was proven by Collins
and Soper in Ref.~\cite{CS} that besides the leading logarithms~\cite
{Parisi-Petronsio} all the large logarithms, including the sub-logs in the
order-by-order calculations, can be expressed in a closed form (resummed)
for the energy correlation in $e^{+}e^{-}$ collisions. This proof was
generalized to the single vector boson $V$ production process $%
h_1h_2\rightarrow VX$, which we consider as an example. Generalization of
the results to other processes is discussed later.

As a first step, using the factorization theorem, the cross section of the
process is written in a general form 
\begin{eqnarray}
&&{{\frac{d\sigma (h_1h_2\rightarrow VX)}{dQ^2dQ_T^2dy}}}=  \nonumber \\
&&~~~\int_{x_1}^1{\frac{d\xi _1}{\xi _1}}\int_{x_2}^1{\frac{d\xi _2}{\xi _2}}
\sum_{a,b}f_{a/h_1}(\xi _1,\mu )\;T_{ab}(Q,Q_T,z_1,z_2,\mu
_f)\;f_{b/h_2}(\xi _2,\mu ).  \nonumber
\end{eqnarray}
The hard part of the cross section is calculated order by order in the
strong coupling 
\[
T_{ab}(Q,Q_T,z_1,z_2,\mu )=\sum_{n=0}^\infty \left( \frac{\alpha _s}\pi
\right) ^nT_{ab}^{(n)}(Q,Q_T,z_1,z_2,\mu ). 
\]
At some fixed order in $\alpha _s$ the coefficients $T_{ab}^{(n)}$ can be
split into singular and regular parts 
\begin{eqnarray*}
&&T_{ab}^{(n)}(Q,Q_T,z_1,z_2,\mu )=T_{ab}^{(n,\delta )}(Q,Q_T,z_1,z_2,\mu
)\delta (Q_T)+ \\
&&~~~\sum_{m=0}^{2n-1}T_{ab}^{(n,m)}(Q,Q_T,z_1,z_2,\mu )\frac 1{Q_T^2}\ln
^m\left( \frac{Q^2}{Q_T^2}\right) +R_{ab}^{(n)}(Q,Q_T,z_1,z_2,\mu )
\end{eqnarray*}
where $R_{ab}^{(n)}$ contains terms which are less singular than $Q_T^{-2}$
and $\delta (Q_T)$ as $Q_T\rightarrow 0$. The perturbative expansion of the
cross section defined in this manner is not useful when 
$\alpha_s^n Q_T^{-2}\ln ^m\left( Q^2/Q_T^2\right) $ 
is large, i.e. in the $%
Q_T\rightarrow 0$ limit. This is expressed in saying that the large log's
are not under control when the perturbative series is calculated in terms of 
$T_{ab}^{(n)}$.

To gain control over these large log's, that is to express the cross section
in terms of coefficients which do not contain large log's, the summation of
the perturbative terms is reorganized. The cross section is rewritten in the
form 
\[
{\frac{d\sigma (h_1h_2\rightarrow VX)}{dQ^2\,dQ_T^2dy\,\,}}={\frac 1{(2\pi
)^2}}\int d^2b\,e^{i{\vec{Q}_T}\cdot {\vec{b}}}{\widetilde{W}(b,Q,x_1,x_2)}%
+~Y(Q_T,Q,x_1,x_2) 
\]
where ${\widetilde{W}(b,Q,x_1,x_2)}$ accommodates the singular logarithmic
terms in the form of $\ln (b^2Q^2)$. The ''regular'' terms are included in
the $Y$ piece 
\begin{eqnarray}
&&Y(Q_T,Q,x_1,x_2)=  \nonumber \\
&&~~~\int_{x_1}^1{\frac{d\xi _1}{\xi _1}}\int_{x_2}^1{\frac{d\xi _2}{\xi _2}}%
~\,f_{a/h_1}(\xi _1,\mu )\;\sum_{n=0}^\infty \left( \frac{\alpha _s}\pi
\right) ^nR_{ab}^{(n)}(Q,Q_T,z_1,z_2,\mu )\;\,f_{b/h_2}(\xi _2,\mu ). 
\nonumber
\end{eqnarray}
The Fourier integral form was introduced to explicitly conserve ${\vec{Q}_T}$ 
\cite{Parisi-Petronsio}.

Based on the form of the singular pieces in the $Q_T\rightarrow 0$ 
limit \cite{Collins-Soper-Sterman}, it is assumed that the $x_1$
and $x_2$ dependence of $\widetilde{W}$ factorizes 
\[
{\widetilde{W}(Q,b,x_1,x_2)}=
\sum_j{\cal C}_{jh_1}(Q,b,x_1)\;{\cal C}_{jh_2}(Q,b,x_2)\;Q_j^2, 
\]
where ${\cal C}_{jh}$ is a convolution of the parton distribution with a
calculable Wilson coefficient, called $C$ function 
\[
{\cal C}_{jh}(Q,b,x)=\sum_{a}\int_x^1{\frac{d\xi }\xi }\;C_{ja}(\frac{x}{\xi}%
,b,\mu,Q)\;f_{a/h}(\xi ,\mu ), 
\]
where $a$ sums over incoming partons, and $j$ denotes the quark flavors with
charge $Q_j$ in the units of the positron charge.

It is argued 
, based on the work in Ref. \cite{CSS}, 
that $\widetilde{W}$ obeys the evolution equation 
\[
\frac \partial {\partial \ln Q^2}{\widetilde{W}(Q,b,x_1,x_2)=}\left[ {K}%
\left( {b\mu ,g}_s(\mu )\right) {+G}\left( {Q/\mu ,g}_s(\mu )\right) \right] 
{\widetilde{W}(Q,b,x_1,x_2)}, 
\]
where ${K}\left( {b\mu ,g}_s(\mu )\right) ${\ and }${G}\left( {Q/\mu ,g}%
_s(\mu )\right) $ satisfy the renormalization group equations (RGE's) 
\begin{eqnarray*}
\mu \frac d{d\mu }{K}\left( {b\mu ,g}_s(\mu )\right) =-\gamma _K\left( {g}%
_s(\mu )\right) , \\
\mu \frac d{d\mu }{G}\left( {b\mu ,g}_s(\mu )\right) =+\gamma _K\left( {g}%
_s(\mu )\right) .
\end{eqnarray*}
The anomalous dimension $\gamma _K$ can be determined from the singular part
of the cross section \cite{Davies-Stirling}. These RGE's allow the control
of the log's, since using them\ one can change the scale ${\mu }$ in ${K}%
\left( {b\mu ,g}_s(\mu )\right) ${\ and }${G}\left( {Q/\mu ,g}_s(\mu
)\right) $ independently, such that it is the order of $1/b$ and $Q$
respectively, so they do not contain large log's. The solution of the RGE's
constructed in this fashion contain the functions $A$ and $B$, which one
uses to rewrite the evolution equation of ${\widetilde{W}}$ 
\begin{eqnarray}
&&\frac \partial {\partial \ln Q^2}{\widetilde{W}(Q,b,,x_1,x_2)}=  \nonumber
\\
&&~~~\left\{ \int_{C_1/b}^{C_2Q}\frac{d\overline{\mu }^2}{\overline{\mu }^2}%
\left[ {A}\left( {g}_s(\overline{\mu }),C_1\right) {+B}\left( {g}%
_s(C_2Q),C_1,C_2\right) \right] \right\} {\widetilde{W}(Q,b,x_1,x_2)}. 
\nonumber
\end{eqnarray}
The $A$ and $B$ functions can be approximated well, calculating them order
by order in the perturbation theory, because they are free from the large
log's.

The only remaining task is to relate the $A$ and $B$ functions to the cross
section. This is done by solving the evolution equation of ${\widetilde{W}}$%
. The solution is of the form 
\[
{\widetilde{W}(Q,b,x_1,x_2)}=e^{-{\cal S}(Q,b{,C_1,C_2})}{\widetilde{W}(%
\frac{C_1}{C_2b},b,x_1,x_2)}, 
\]
where the Sudakov exponent is defined as 
\[
{\cal S}(Q,b{,C_1,C_2})=\int_{C_1^2/b^2}^{C_2^2 Q^2}\frac{d\overline{\mu }^2}{%
\overline{\mu }^2}\left[ {A}\left( {g}_s(\overline{\mu }),C_1\right) \ln
\left( \frac{C_2Q^2}{\overline{\mu }^2}\right) {+B}\left( {g}_s(\overline{%
\mu }),C_1,C_2\right) \right] . 
\]
From this we can write the cross section in terms of the $A$, $B$ and $C$
functions 
\begin{eqnarray*}
&&{\frac{d\sigma (h_1h_2\rightarrow W^{\pm }X)}{dQ^2\,dQ_T^2dy\,\,}}= \\
&&~~~{\frac 1{(2\pi )^2}}\int d^2b\,e^{i{\vec{Q}_T}\cdot {\vec{b}}}\,e^{-%
{\cal S}(Q{,b,C_1,C_2})}~\sum_j{\cal C}_{jh_1}(\frac{C_1}{C_2b},b,x_1)\;%
{\cal C}_{jh_2}(\frac{C_1}{C_2b},b,x_2)\;Q_j^2 \\
&&~~~~+ Y(Q_T,Q,x_1,x_2).
\end{eqnarray*}
This important expression is the resummation formula which we make vital use
in the rest of this work.

The remaining slight modification of the resummation formula ensures that
the impact parameter $b$ does not extend into the $b\grsim 1/\Lambda _{QCD}$
region, where perturbation theory is invalid. To achieve this the $b$
dependence of ${\widetilde{W}}$ is replaced by 
\[
b_{*}={\frac b{\sqrt{1+(b/b_{max})^2}}}\,,
\]
which is always smaller than $b_{max}$. This arbitrary cutoff of the $b$
integration is compensated by the parametrization of the non-perturbative
region with the introduction of a non-perturbative function for each quark
and anti-quark flavors $j$ and $\overline{k}$%
\[
W_{jk}^{NP}(b,Q,Q_0,x_1,x_2)=\exp \left[ -F_1(b)\ln \left( \frac{Q^2}{Q_0^2}%
\right) -F_{j/{h_1}}(x_1,b)-F_{\overline{{k}}/{h_2}}(x_2,b)\right] ,
\]
where the functions $F_1$, $F_{j/{h_1}}$ and $F_{{\bar{k}}/{h_2}}$ have to
be determined using experimental data. The $\ln \left( Q^2/Q_0^2\right) $
term is introduced to match the logarithmic term of the Sudakov exponent and
its coefficient is expected to be process independent.

For the production of vector bosons in hadron collisions another formalism
was presented in the literature to resum the large contributions due to
multiple soft gluon radiation by Altarelli, Ellis, Greco, Martinelli (AEGM)~%
\cite{AEGM}. The detailed differences between the CSS\ and AEGM formulations
were discussed in Ref.~\cite{Arnold-Kauffman}. It was shown that the two are
equivalent up to the few highest power of $\ln (Q^2/Q_T^2)$ at every order
in $\alpha _s$ for terms proportional to ${Q}_T^{-2}$, provided $\alpha _s\,$
in the AEGM\ formalism is evaluated at $b_0^2/b^2$ rather than at ${Q^2}$. A
more noticeable difference, except the additional contributions of order ${Q}%
^{-2}$ included in the AEGM formula, is caused by different ways of
parametrizing the non-perturbative contribution in the low $Q_T$ regime.
Since, the CSS formalism was proven to sum over not just the leading logs
but also all the sub-logs, and the piece including the Sudakov factor was
shown to be renormalization group invariant \cite{CSS}, we only discuss the
results of CSS formalism in the rest of this work.

\subsection{Resummation Formula for Lepton Pair Production}

Due to the increasing accuracy of the experimental data on the properties of
vector bosons at hadron colliders, it is no longer sufficient to consider the
effects of multiple soft gluon radiation for an on-shell vector boson and
ignore the effects coming from the decay width and the polarization of the
massive vector boson to the distributions of the decay leptons. Hence, it is
desirable to have an equivalent resummation formalism~\cite{Balazs-Qui-Yuan}
for calculating the distributions of the decay leptons. This formalism
should correctly include the off-shellness of the vector boson (i.e. the
effect of the width ) and the polarization information of the produced
vector boson which determines the angular distributions of the decay
leptons. In this Section, we give our analytical results for such a
formalism that correctly takes into account the effects of the multiple soft
gluon radiation on the distributions of the decay leptons from the vector
boson. 

To derive the building blocks of the resummation formula, we use the
dimensional regularization scheme to regulate the infrared divergencies, and
adopt the canonical-$\gamma _5$ prescription to calculate the anti-symmetric
part of the matrix element in $D$-dimensional space-time.\footnote{%
In this prescription, $\gamma_5$ anti-commutes with other $\gamma$'s in the
first four dimensions and commutes in the others \cite{gamma5Def,gamma5c,twoloopf3}.}
The infrared-anomalous contribution arising from using the canonical-$\gamma
_5$ prescription was carefully handled by applying the procedures outlined
in Ref.~\cite{Korner-Mirkes} for calculating both the virtual and the real
diagrams.\footnote{%
In Ref.~\cite{Korner-Mirkes}, the authors calculated the anti-symmetric
structure function $F_3$ for deep-inelastic scattering.}

The resummation formula for the differential cross section of lepton pair
production is given in Ref.~\cite{Balazs-Qui-Yuan}:
\begin{eqnarray}
&&\left( {\frac{d\sigma (h_1h_2\rightarrow V(\rightarrow \ell _1{\bar{\ell}_2%
})X)}{dQ^2\,dy\,dQ_T^2\,d\cos {\theta }\,d\phi }}\right) _{res}={\frac
1{48\pi S}}\,{\frac{Q^2}{(Q^2-M_V^2)^2+Q^4\Gamma _V^2/M_V^2}}  \nonumber \\
&&~~\times \left\{ {\frac 1{(2\pi )^2}}\int d^2b\,e^{i{\vec{Q}_T}\cdot {\vec{%
b}}}\,\sum_{j,k}{\widetilde{W}_{j{\bar{k}}}(b_{*},Q,x_1,x_2,\theta ,\phi
,C_1,C_2,C_3)}\,\widetilde{W}_{j{\bar{k}}}^{NP}(b,Q,x_1,x_2)\right.  
\nonumber \\
&&~~~~\left. +~Y(Q_T,Q,x_1,x_2,\theta ,\phi ,{C_4})\right\} .
\label{eq:ResFor}
\end{eqnarray}
In the above equation the parton momentum fractions are defined as $x_1=e^yQ/%
\sqrt{S}$ and $x_2=e^{-y}Q/\sqrt{S}$, where $\sqrt{S}$ is the center-of-mass
(CM) energy of the hadrons $h_1$ and $h_2$. For $V=W^{\pm }$ or $Z^0$, we
adopt the LEP line-shape prescription of the resonance behavior, with $M_V$
and $\Gamma _V$ being the mass and the width of the vector boson. The
renormalization group invariant quantity $\widetilde{W}_{j{\bar{k}}}(b)$,
which sums to all orders in $\alpha _s$ all the singular terms that behave as
$Q_T^{-2}$ $\times$ [1 or $\ln (Q^2/Q_T^2)$] 
for $Q_T\rightarrow 0$, is
\begin{eqnarray}
&&\widetilde{W}_{j{\bar{k}}}(b,Q,x_1,x_2,\theta ,\phi {,C_1,C_2,C_3})=\exp
\left\{ -{\cal S}(b,Q{,C_1,C_2})\right\} \mid V_{jk}\mid ^2  \nonumber \\
&&~\times \left\{ \left[ {\cal C}_{j/h_1}(x_1)~{\cal C}_{{\bar{k}}/h_2}(x_2)+%
{\cal C}_{{\bar{k}}/h_1}(x_1)~{\cal C}_{j/h_2}(x_2)\right] {\cal F}%
_{ij}^{(+)}(\alpha _{em}(C_2Q),\alpha _s(C_2Q),\theta )\right.   \nonumber \\
&&~~~\left. +\left[ {\cal C}_{j/h_1}(x_1)~{\cal C}_{{\bar{k}}/h_2}(x_2)-%
{\cal C}_{{\bar{k}}/h_1}(x_1)~{\cal C}_{j/h_2}(x_2)\right] {\cal F}%
_{ij}^{(-)}(\alpha _{em}(C_2Q),\alpha _s(C_2Q),\theta )\right\} . \nonumber \\
&&~~~
\label{eq:WTwi}
\end{eqnarray}
Here the Sudakov exponent ${\cal S}(b,Q{,C_1,C_2})$ is defined as
\begin{equation}
{\cal S}(b,Q{,C_1,C_2})=\int_{C_1^2/b^2}^{C_2^2Q^2}{\frac{d{\bar{\mu}}^2}{{%
\bar{\mu}}^2}}\left[ A\left( \alpha _s({\bar{\mu}}),C_1\right) \ln \left( {%
\frac{C_2^2Q^2}{{\bar{\mu}}^2}}\right) +B\left( \alpha _s({\bar{\mu}}%
),C_1,C_2\right) \right] .  \label{eq:SudExp}
\end{equation}
The explicit forms of the $A$, $B$ and $C$ functions and the renormalization
constants $C_i$ ($i$=1,2,3) are summarized in Section \ref{Sec:OASExpansion}. 
The $V_{jk}$ coefficients are given by
\begin{eqnarray}
V_{jk}=\cases{ {\rm Cabibbo-Kobayashi-Maskawa~matrix~elements} & for $V =
W^\pm$, \cr $$\delta_{jk}$$ & for $V = Z^0,\gamma^*$.}
\label{eq:Vjk}
\end{eqnarray}
In Eq.~(\ref{eq:WTwi}) ${\cal C}_{i/h}$ denotes the convolution of the 
$C_{ia}$
Wilson coefficients with the parton distributions $f_{a/h}$%
\begin{equation}
{\cal C}_{i/h}(x)=\sum_a\int_x^1{\frac{d\xi }\xi }\,C_{ia}
\left( {\frac x\xi },b,\mu =
\frac{C_3}b,C_1,C_2\right) \;f_{a/h}\left( \xi ,\mu =%
\frac{C_3}b\right).  \label{Eq:DefCalC}
\end{equation}
In this notation we suppressed the $b$ and $\mu$ dependences of ${\cal
C}_{i/h}$, which play the role of generalized parton distributions, including 
(through their $b$ dependence) transverse effects of soft--gluon emission.
In the above expressions $j$ represents quark flavors and $\bar{k}$ stands
for anti-quark flavors. The indices $a$ and $b$ are meant to sum over quarks
and anti-quarks or gluons. Summation on these double indices is implied. In
Eq.~(\ref{eq:WTwi}) ${\cal F}_{ij}^{(\pm )}$ are kinematic factors that
depend on the coupling constants and the polar angle of the lepton $\ell _1$
\begin{eqnarray}
&&{\cal F}_{ij}^{(+)}(\alpha _{em}(C_2Q),\alpha _s(C_2Q),\theta
)=(g_L^2+g_R^2)(f_L^2+f_R^2)(1+\cos ^2\theta )  \nonumber \\
&&~{\cal F}_{ij}^{(-)}(\alpha _{em}(C_2Q),\alpha _s(C_2Q),\theta
)=(g_L^2-g_R^2)(f_L^2-f_R^2)(2\cos \theta )  \nonumber
\end{eqnarray}
The couplings $f_{L,R}$ and $g_{L,R}$ are defined through the $\ell _1{\bar{%
\ell}_2}V$ and the $q{\bar{q}^{\prime }}V$ vertices, which are written
respectively, as 
\[
i\gamma _\mu \left[ f_L(1-\gamma _5)+f_R(1+\gamma _5)\right] ~~~{\rm and}%
~~~i\gamma _\mu \left[ g_L(1-\gamma _5)+g_R(1+\gamma _5)\right] .
\]
For example, for $V=W^{+},q=u$, ${\bar{q}^{\prime }}={\bar{d}}$, $\ell
_1=\nu _e$, and ${\bar{\ell}_2}=e^{+}$, the couplings $g_L^2=f_L^2=G_FM_W^2/%
\sqrt{2}$ and $g_R^2=f_R^2=0$, where $G_F$ is the Fermi constant. The
detailed information on the values of the parameters used in Eqs.~(\ref
{eq:ResFor}) and (\ref{eq:WTwi}) is given in Table~{\ref{tbl:parameters}}.

\begin{table}[tbp]
\begin{center}
\begin{tabular}{crrrr}
\hline\hline \\[-.2cm]
\vspace{.2cm} $V$ & $M_V$ (GeV) & $\Gamma_V$ (GeV) & $g_L$ & $g_R$ \\ 
\hline \\[-.2cm]
\vspace{.2cm} $\gamma$ & 0.00 & 0.00 & $g Q_f s_w/2$ & $g Q_f s_w/2$ \\ 
\vspace{.2cm} $W^\pm$ & 80.36 & 2.07 & $g/( 2\sqrt{2})$ & 0 \\ 
\vspace{.2cm} $Z^0$ & 91.19 & 2.49 & $g (T_3 - Q_f s_w^2)/(2 c_w)$ & $-g Q_f
s_w^2/(2 c_w)$ \\ \hline\hline
\end{tabular}
\end{center}
\caption{ Vector boson parameters and couplings to fermions. The $f{\bar{f}%
^{\prime }}V$ vertex is defined as $i\gamma _\mu [g_L(1-\gamma
_5)+g_R(1+\gamma _5)]$ and $s_w=\sin \theta _w$ ($c_w=\cos \theta _w$) is
the sine (cosine) of the weak mixing angle: $\sin ^2(\theta _w(M_{Z^0}))_{%
\overline{MS}}=0.2315$. $Q_f$ is the fermion charge ($Q_u=2/3,Q_d=-1/3,Q_\nu
=0,Q_{e^{-}}=-1$), and $T_3$ is the eigenvalue of the third component of the 
$SU(2)_L$ generator ($T_3^u=1/2$, $T_3^d=-1/2$, $T_3^\nu =1/2$, $%
T_3^{e^{-}}=-1/2$).}
\label{tbl:parameters}
\end{table}

In Eq.~(\ref{eq:ResFor}) the magnitude of the impact parameter $b$ is
integrated from 0 to $\infty $. However, in the region where $b\gg 1/\Lambda
_{QCD}$, the Sudakov exponent ${\cal S}(b,Q,C_1,C_2)$ diverges as the result
of the Landau pole of the QCD coupling $\alpha _s(\mu )$ at $\mu =\Lambda
_{QCD}$, and the perturbative calculation is no longer reliable. As
discussed in the previous section, in this region of the impact parameter
space (i.e. large $b$), a prescription for parametrizing the
non-perturbative physics in the low $Q_T$ region is necessary. Following the
idea of Collins and Soper~\cite{Collins}, the renormalization group
invariant quantity $\widetilde{W}_{j{\bar{k}}}(b)$ is written as 
\[
\widetilde{W}_{j{\bar{k}}}(b)=
\widetilde{W}_{j{\bar{k}}}(b_{*})
\widetilde{W}_{j{\bar{k}}}^{NP}(b)\,,
\]
where
\[
b_{*}={\frac b{\sqrt{1+(b/b_{max})^2}}}\,.
\]
Here $\widetilde{W}_{j{\bar{k}}}(b_{*})$ is the perturbative part of $%
\widetilde{W}_{j{\bar{k}}}(b)$ and can be reliably calculated by
perturbative expansions, while $\widetilde{W}_{j{\bar{k}}}^{NP}(b)$ is the
non-perturbative part of $\widetilde{W}_{j{\bar{k}}}(b)$ that cannot be
calculated by perturbative methods and has to be determined from
experimental data. To test this assumption, one should verify that there
exists a universal functional form for this non-perturbative function $%
\widetilde{W}_{j{\bar{k}}}^{NP}(b)$. This expectation is based on the 
general feature that there exists a universal set of parton distribution
functions (PDF's) that can be used in any perturbative QCD calculation to
compare it with experimental data. 
The non-perturbative function was parametrized by (cf. Ref.~\cite
{Collins-Soper-Sterman}) 
\begin{equation}
\widetilde{W}_{j\bar{k}}^{NP}(b,Q,Q_0,x_1,x_2)=\exp \left[ -F_1(b)\ln \left( 
\frac{Q^2}{Q_0^2}\right) -F_{j/{h_1}}(x_1,b)-F_{{\bar{k}}/{h_2}%
}(x_2,b)\right],  \label{Eq:WNonPert}
\end{equation}
where $F_1$, $F_{j/{h_1}}$ and $F_{{\bar{k}}/{h_2}}$ have to be first
determined using some sets of data, and later can be used to predict the
other sets of data to test the dynamics of multiple gluon radiation
predicted by this model of the QCD theory calculation. As noted in Ref.~\cite
{Collins-Soper-Sterman}, $F_1$ does not depend on the momentum fraction
variables $x_1$ or $x_2$, while $F_{j/{h_1}}$ and $F_{{\bar{k}}/{h_2}}$ in
general depend on those kinematic variables.\footnote{%
Here, and throughout this work, the flavor dependence of the
non-perturbative functions is ignored, as it is postulated in Ref.~\cite
{Collins-Soper-Sterman}.} The $\ln (Q^2/Q_0^2)$ dependence, associated with
the $F_1$ function, was predicted by the renormalization group analysis 
described in Section~\ref{sec:221}. It is necessary to balance the 
$\ln (Q^2)$ dependence of the Sudakov exponent.  
Furthermore, $F_1$ was shown to be universal, and
its leading behavior ($\sim b^2$) can be described by renormalon physics 
\cite{Korchemsky-Sterman}. Various sets of fits to these non-perturbative
functions can be found in Refs.~\cite{Davies} and \cite{Ladinsky-Yuan}.

In our numerical results we use the Ladinsky-Yuan parametrization of the
non-perturbative function (cf. Ref.~\cite{Ladinsky-Yuan}): 
\begin{eqnarray}
\widetilde{W}_{j\bar k}^{NP}(b,Q,Q_0,x_1,x_2) = {\rm exp} \left[- g_1 b^2 -
g_2 b^2 \ln\left( {\frac{Q }{2 Q_0}} \right) - g_1 g_3 b \ln{(100 x_1 x_2)}
\right],  \label{eq:WNPLY}
\end{eqnarray}
where $g_1 = 0.11^{+0.04}_{-0.03}~{\rm GeV}^2$, $g_2 = 0.58^{+0.1}_{-0.2}~%
{\rm GeV}^2$, $g_3 =-1.5^{+0.1}_{-0.1}~{\rm GeV}^{-1}$ and $Q_0 = 1.6~{\rm %
GeV}$. (The value $b_{max}=0.5~{\rm GeV}^{-1}$ was used in determining the
above $g_i$'s and in our numerical results.) These values were fit for
CTEQ2M PDF with the canonical choice of the renormalization constants, i.e. $%
C_1=C_3=2e^{-\gamma _E}$ ($\gamma _E$ is the Euler constant) and $C_2=1$. In
principle, for a calculation using a more update PDF, these non-perturbative
parameters should be refit using a data set that should also include the
recent high statistics $Z^0$ data from the Tevatron. This is however beyond
the scope of this work.

In Eq.~(\ref{eq:ResFor}), $\widetilde{W}_{j{\bar{k}}}$ sums over the soft
gluon contributions that grow as 
$Q_T^{-2}$ $\times$ [1 or $\ln (Q^2/Q_T^2)$] 
to all orders in $\alpha _s$.
Contributions less singular than those included in $\widetilde{W}_{j{\bar{k}}%
}$ should be calculated order-by-order in $\alpha _s$ and included in the $Y$
term, introduced in Eq.~(\ref{eq:ResFor}). This would, in principle, extend
the applicability of the CSS resummation formalism to all values of $Q_T$.
However, as to be shown below, since the $A$, $B$, $C$, and $Y$ functions
are only calculated to some finite order in $\alpha _s$, the CSS resummed
formula as described above will cease to be adequate for describing data
when the value of $Q_T$ is in the vicinity of $Q$. Hence, in practice, one
has to switch from the resummed prediction to the fixed order perturbative
calculation as $Q_T \gtrsim Q$. The $Y$ term, which is defined as the
difference between the fixed order perturbative contribution and those
obtained by expanding the perturbative part of $\widetilde{W}_{j{\bar{k}}}$
to the same order, is given by 
\begin{eqnarray}
&& Y(Q_T,Q,x_1,x_2,\theta ,\phi ,C_4)=\int_{x_1}^1{\frac{d\xi _1}{\xi _1}}
\int_{x_2}^1{\frac{d\xi_2}{\xi _2}}\sum_{n=1}^\infty \left( {\frac{\alpha
_s(C_4Q)}\pi }\right) ^n  \nonumber \\
&& ~~~\times f_{a/h_1}(\xi _1,C_4Q)\,R_{ab}^{(n)}(Q_T,Q,\frac{x_1}{\xi _1}, 
\frac{x_2}{\xi _2},\theta ,\phi )\,f_{b/h_2}(\xi _2,C_4Q),  
\label{eq:RegPc}
\end{eqnarray}
where the functions $R_{ab}^{(n)}$ contain contributions which are less
singular than $Q_T^{-2}$ $\times$ [1 or $\ln (Q^2/Q_T^2)$] 
as $Q_T\rightarrow 0$. Their
explicit expressions and the choice of the scale $C_4$ are summarized in
Section \ref{Sec:RegularContributions}.

Within the Collins-Soper-Sterman resummation formalism $\widetilde{W}_{j{%
\bar{k}}}(b)$ sums all the singular terms which grow as 
$\alpha _s^nQ_T^{-2}\ln ^m(Q^2/Q_T^2)$ 
for all $n$ and $0\leq m\leq 2n-1$ provided
that all the $A^{(n)}$, $B^{(n)}$ and $C^{(n-1)}$ coefficients are included
in the perturbative expansion of the $A$, $B$ and $C$ functions,
respectively. This was illustrated in Eqs.~(A.12) and (A.13) of Ref.~\cite
{Arnold-Kauffman}. In our numerical results we included $A^{(1)}$, $A^{(2)}$%
, $B^{(1)}$, $B^{(2)}$, $C^{(0)}$ and $C^{(1)}$, which means we resummed the
following singular pieces \cite{Arnold-Kauffman}: 
\begin{eqnarray}
\frac{d\sigma }{dQ_T^2}\sim \frac{(L+1)}{Q_T^2}\left\{ {}\right. \alpha _s
&+&\alpha _s^2L^2+\alpha _s^3L^4+\alpha _s^4L^6+~...  \nonumber \\
&+&\alpha _s^{2\;\;}\;\;\;+\alpha _s^3L^2+\alpha _s^4L^4+~...\left.
{}\right\} ,  \label{eq:logs}
\end{eqnarray}
where $L$ denotes $\ln (Q^2/Q_T^2)$ and the explicit coefficients
multiplying the logs are suppressed. The lowest order singular terms that
were not included are $Q_T^{-2}(L+1)(\alpha _s^3+\alpha _s^4 L^2+...)$~.
Also, in the $Y$ term we included $R^{(1)}$ and $R^{(2)}$ (cf. Eq.(\ref
{eq:RegPc})), which are derived from the fixed order $\alpha _s$ and $\alpha
_s^2$ calculations \cite{Arnold-Reno,Arnold-Kauffman}.

The results of the usual ${\cal O}(\alpha _s)$ calculation can be obtained
by expanding the above CSS resummation formula to the $\alpha _s$ order,
which includes both the singular piece and the $Y$ term. Details are given
in Sections \ref{Sec:OASExpansion} and \ref{Sec:RegularContributions},
respectively.

\subsection{${\cal O}$($\alpha _s$) Expansion \label{Sec:OASExpansion}}

In this section we expand the resummation formula, as given in Eq.~(\ref
{eq:ResFor}), up to ${\cal O}(\alpha _s)$, and calculate the ${Q_T}$
singular piece as well as the integral of the ${\cal O}(\alpha _s)$
corrections from 0 to ${P_T}$. These are the ingredients, together with the
regular pieces to be given in Section \ref{Sec:RegularContributions}, needed
to construct our NLO calculation.

First we calculate the singular part at the ${\cal O}(\alpha _s)$. By
definition, this consist of terms which are at least as singular as 
$Q_T^{-2}$ $\times$ [1 or $\ln (Q^2/Q_T^2)$]. 
We use the perturbative
expansion of the $A,\;B$ and $C$ functions in the strong coupling constant $%
\alpha _s$ as: 
\begin{eqnarray}
A\left( \alpha _s({\bar{\mu}}),C_1\right) &=& \sum_{n=1}^\infty \left( \frac{%
\alpha _s({\bar{\mu}})}\pi \right) ^nA^{(n)}(C_1),  \nonumber \\
B\left( \alpha _s({\bar{\mu}}),C_1,C_2\right) &=&\sum_{n=1}^\infty \left( 
\frac{\alpha _s({\bar{\mu}})}\pi \right) ^nB^{(n)}(C_1,C_2),
\label{eq:ABCExp} \\
C_{ja}(z,b,\mu ,C_1,C_2) &=& \sum_{n=0}^\infty \left( \frac{\alpha _s({\mu })%
}\pi \right)^n C_{ja}^{(n)}(z,b,\mu ,\frac{C_1}{C_2}).  \nonumber
\end{eqnarray}
The explicit expressions of the $A^{(n)},\;B^{(n)}$ and $C^{(n)}$
coefficients are given in Section \ref{sec:ABC}. After integrating over the
lepton variables and the angle between ${\vec{b}}$ and ${\vec{Q}_T}$, and
dropping the regular ($Y$) piece in Eq.~(\ref{eq:ResFor}), we obtain 
\begin{eqnarray*}
\lim_{Q_T\rightarrow 0} \frac{d\sigma }{dQ^2dydQ_T^2}&=& \frac{\sigma _0}%
S\delta (Q^2-M_V^2)\left\{ \frac 1{2\pi Q_T^2}\int_0^\infty d\eta \;\eta
\;J_0(\eta )\;{\text e}^{-{\cal S}(\eta /Q{_T,Q,C}_{{1}},{C}_{{2}})}\right. \\
&&\left. \times \;f_{j/h_1}\left( x_1,\frac{C_3^2Q_T^2}{\eta ^2}\right) \;f_{%
\overline{k}/h_2}\left( x_2,\frac{C_3^2Q_T^2}{\eta ^2}\right)
+\;j\leftrightarrow \overline{k}\right\} +{\cal O}(Q_T^{-1}),
\end{eqnarray*}
where we have substituted the resonance behavior by a fixed mass for
simplicity, and defined $\sigma _0$ as\footnote{%
For our numerical calculation (within the ResBos Monte Carlo package), we
have consistently used the on-shell scheme for all the electroweak
parameters in the improved Born level formula for including large
electroweak radiative corrections. In the $V=Z^0$ case, they are the same as
those used in studying the $Z^0$-pole physics at LEP~\cite{Peccei}.}~\cite
{CSS} 
\begin{eqnarray*}
\sigma _0 &=&\frac{4\pi \alpha ^2}{9Q^2},~~~{\text for\ }V=\gamma ^{*}, \\
\sigma _0 &=&\frac{\pi ^2\alpha }{3 s_W^2}\sum_{jk}|V_{jk}|^2,~~~{\text for\ }%
V=W^{\pm }, \\
\sigma _0 &=&\frac{\pi ^2\alpha }{12 s_W^2 c_W^2} \sum_{jk}[(1-4|Q_j|
s_W^2)^2+1]|V_{jk}|^2,~~~{\text for\ }V=Z^0.
\end{eqnarray*}
Here $\alpha $ is the fine structure constant, $s_W$ ($c_W$) is the sine
(cosine) of the weak mixing angle $\theta _W$, $Q_j$ is the electric charge
of the incoming quark in the units of the charge of the positron (e.g. $%
Q_{up}=2/3$, $Q_{down}=-1/3$, etc.), and $V_{jk}$ is defined by Eq.~(\ref
{eq:Vjk}). To evaluate the integral over $\eta =bQ_T$, we use the following
property of the Bessel functions: 
\[
\int_0^\infty d\eta \;\eta \;J_0(\eta )\;F(\eta )=-\int_0^\infty d\eta
\;\eta \;J_1(\eta )\;\frac{dF(\eta )}{d\eta }, 
\]
which holds for any function $F(\eta )$ satisfying $\left[ \eta \;J_1(\eta
)\;F(\eta )\right] _0^\infty =0$. Using the expansion of the Sudakov
exponent ${\cal S}(b,Q,C_1,C_2)={\cal S}^{(1)}(b,Q,C_1,C_2)+{\cal O}(\alpha
_S^2)$ with%
\[
{\cal S}^{(1)}(b,Q,C_1,C_2)=\frac{\alpha _s(Q^2)}\pi \left[ \frac
12A^{(1)}(C_1)\ln ^2\left( \frac{C_2^2Q^2}{C_1^2/b^2}\right)
+B^{(1)}(C_1,C_2)\ln \left( \frac{C_2^2Q^2}{C_1^2/b^2}\right) \right] , 
\]
and the evolution equation of the parton distribution functions 
\[
\frac{d\;f_{j/h}(x,\mu ^2)}{d\ln \mu ^2}=\frac{\alpha _s(\mu ^2)}{2\pi }%
\left( P_{j\leftarrow a}^{(1)}\otimes f_{a/h}\right) (x,\mu ^2)+{\cal O}%
(\alpha _s^2), 
\]
we can calculate the derivatives of the Sudakov form factor and the parton
distributions with respect to $\eta $: 
\[
\frac d{d\eta }{\text e}^{-{\cal S}(\eta /Q_T,Q,C_1,C_2)} = \frac{-2}\eta 
\frac{\alpha _s(Q^2)}\pi \left[ A^{(1)}(C_1)\ln \left( \frac{C_2^2Q^2\eta ^2%
}{C_1^2Q_T^2}\right) +B^{(1)}(C_1,C_2)\right] +{\cal O}(\alpha _s^2),
\]
and
\begin{eqnarray*}
\frac d{d\eta }f_{j/h}(x,\frac{C_3^2Q_T^2}{\eta ^2}) &=&\frac{-2}\eta \frac{%
\alpha _s(Q^2)}{2\pi }\left( P_{j\leftarrow a}^{(1)}\otimes f_{a/h}\right)
(x,Q^2)+{\cal O}(\alpha _s^2).
\end{eqnarray*}
Note that $\alpha _s$ itself is expanded as 
\[
\frac{\alpha _s(\mu ^2)}{2\pi} =\frac{\alpha _s(Q^2)}{2\pi} -2 \beta _1
\left( \frac{\alpha _s(Q^2)}{2\pi} \right) ^2\ln \left( \frac{\mu ^2}{Q^2}%
\right) +{\cal O}(\alpha _s^3(Q^2)), 
\]
with $\beta _1=(11N_C-2N_f)/12$, where $N_C$ is the number of colors (3 in
QCD) and $N_f$ is the number of light quark flavors with masses less than $Q$%
. In the evolution equation of the parton distributions %
$P_{{j\leftarrow k}}^{(1)}(z)$ and $P_{{j\leftarrow G}}^{(1)}(z)$
are the leading order DGLAP splitting kernels~\cite{DGLAP} 
given by Eq.~(\ref{Eq:DGLAP}), and $\otimes $
denotes the convolution defined by 
\begin{eqnarray*}
\left( P_{j\leftarrow a}^{(1)}\otimes f_{a/h}\right) (x,\mu ^2)=\int_x^1{%
\frac{d\xi }\xi }\,P_{j\leftarrow a}^{(1)}\left( {\frac x\xi }\right)
\;f_{a/h}\left( \xi ,\mu ^2\right) ,
\end{eqnarray*}
and the double parton index $a$ is running over all light quark flavors and
the gluon. 

After utilizing the Bessel function property and substituting the
derivatives into the resummation formula above, the integral over $\eta $
can be evaluated using 
\begin{equation}
\int_0^\infty d\eta \;J_1(\eta )\;\ln ^m\left( \frac \eta {b_0}\right) = %
\cases{~1, & {\rm if} $m = 0$, \cr ~0, & {\rm if} $m=1,2$ {\rm and}
$b_0=2e^{-\gamma_E}$, }  \label{eq:IntBess}
\end{equation}
where $\gamma _E$ is the Euler constant. The singular piece up to ${\cal O}%
(\alpha _s)$ is found to be 
\begin{eqnarray}
&& \left. \frac{d\sigma }{dQ^2dydQ_T^2}\right| _{Q_T\rightarrow 0} = 
\nonumber \\ && ~~~
\frac{%
\sigma _0}S\delta (Q^2-M_V^2)\frac 1{2\pi Q_T^2}\frac{\alpha _s (Q^2)} \pi
\left\{ \left[ \;f_{j/h_1}(x_1,Q^2)\left( P_{\overline{k}\leftarrow
b}\otimes f_{b/h_2}\right) (x_2,Q^2)\right. \right.  
\nonumber \\ && ~~~ ~~~~~~~~~~~~~~~~~~~~~~~~~~~~~~~~~~~~~
+\left. \left( P_{j\leftarrow a}\otimes f_{a/h_1}\right) (x_1,Q^2)\;f_{%
\overline{k}/h_2}(x_2,Q^2)\right]  
\nonumber \\ && ~~~
+ \left[ A^{(1)}(C_1) \ln \left( \frac{C_2^2}{C_1^2/b_0^2} \frac{Q^2}{%
Q_T^2}\right) + B^{(1)}(C_1,C_2) \right] \;f_{j/h_1}(x_1,Q^2)\;f_{\overline{k%
}/h_2}(x_2,Q^2)  
\nonumber \\ && ~~~
\left. + \;j\leftrightarrow \overline{k}\right\} +~{\cal O}(\alpha _s^2,{%
\frac{1}{Q_T}}),  \label{eq:SingPartCi}
\end{eqnarray}
for arbitrary $C_1$ and $C_2$ constants. If $C_1$ is not equal to $C_2 b_0$
then, when $Q_T$ is of the order of $Q$, the arbitrary log terms $\ln
(C_1^2/(C_2^2 b_0^2))$ can potentially be larger than $\ln (Q^2/Q_T^2)$.
Therefore, to properly describe the $Q_T$ distribution of the vector boson
in the matching region, i.e. for $Q_T \sim Q$, Eq.~(\ref{eq:SingPartCi}) has
to be used to define the asymptotic piece at ${\cal O}(\alpha _s)$. This
asymptotic piece is different from the singular contribution derived from a
fixed order perturbative calculation at ${\cal O}(\alpha _s)$ which is given
by 
\begin{eqnarray}
&& \left. \frac{d\sigma }{dQ^2dydQ_T^2}\right| _{Q_T\rightarrow 0} =
\nonumber \\ && ~~~
\frac{\sigma _0}S\delta (Q^2-M_V^2)\frac 1{2\pi Q_T^2}\frac{\alpha _s (Q^2)}\pi
\left\{ \left[ \;f_{j/h_1}(x_1,Q^2)\left( P_{\overline{k}\leftarrow
b}\otimes f_{b/h_2}\right) (x_2,Q^2)\right. \right.  
\nonumber \\ && ~~~
+\left. \left( P_{j\leftarrow a}\otimes f_{a/h_1}\right) (x_1,Q^2)\;f_{%
\overline{k}/h_2}(x_2,Q^2)\right]  
\nonumber \\ && ~~~
+\left. \left[ A^{(1)}\ln \left( \frac{Q^2}{Q_T^2}\right) +B^{(1)}\right]
\;f_{j/h_1}(x_1,Q^2)\;f_{\overline{k}/h_2}(x_2,Q^2)+\;j\leftrightarrow 
\overline{k}\right\}  
\nonumber \\ && ~~~
+~{\cal O}(\alpha _s^2,{\frac{1}{Q_T}}),  \label{eq:SingPart}
\end{eqnarray}
where $A^{(1)}=C_F$ and $B^{(1)}=-3 C_F/2$. Compared to the general results
for $A^{(1)} (C_1)$ and $B^{(1)} (C_1,C_2)$, as listed in Section \ref
{sec:ABC}, the above results correspond to the special case of $C_1=C_2 b_0$%
. The choice of $C_1=C_2 b_0=C_3=b_0=2 e^{-\gamma_E}$ is usually referred to
as the canonical choice. Throughout this work, we use the canonical choice
in our numerical calculations.

To derive the integral of the ${\cal O}(\alpha _s)$ corrections over $Q_T$,
we start again from the resummation formula [Eq.~(\ref{eq:ResFor})] and the
expansion of the $A,\;B$ and $C$ functions [Eq.~(\ref{eq:ABCExp})]. This
time the evolution of parton distributions is expressed as 
\begin{eqnarray*}
f_{j/h}(x,\mu ^2) &=&f_{j/h}(x,Q^2)+f_{j/h}^{(1)}(x,\mu ^2)+{\cal O}(\alpha
_S^2),~{\text with} \\
f_{j/h}^{(1)}(x,\mu ^2) &=&\frac{\alpha _s(Q^2)}{2\pi }\ln \left( \frac{\mu
^2}{Q^2}\right) \left( P_{j\leftarrow a}^{(1)}\otimes f_{a/h}\right) (x,Q^2),
\end{eqnarray*}
where summation over the partonic index $a$ is implied. Substituting these
expansions in the resummation formula Eq.~(\ref{eq:ResFor}) and integrating
over both sides with respect to $Q_T^2$. We use the integral formula, valid
for an arbitrary function $F(b)$: 
\[
\frac 1{(2\pi )^2}\int_0^{P_T^2}dQ_T^2\int d^2b\;{\text e}^{i{\vec{Q}_T}\cdot 
{\vec{b}}}F(b)=\frac 1{2\pi }\int_0^\infty db\;P_TJ_1(bP_T)F(b), 
\]
together with Eq.~(\ref{eq:IntBess}) to derive 
\begin{eqnarray}
&& \int_0^{P_T^2}dQ_T^2\;\frac{d\sigma }{dQ^2dydQ_T^2}=\frac{\sigma _0}S
\delta (Q^2-M_V^2)  \nonumber \\
&&\times \left\{ \left( 1-\frac{\alpha _s (Q^2)}\pi \left[ \frac
12A^{(1)}\ln ^2\left( \frac{Q^2}{P_T^2}\right) +B^{(1)}\ln \left( \frac{Q^2}{%
P_T^2} \right) \right] \right) \;f_{j/h_1}(x_1,Q^2)\;f_{\overline{k}%
/h_2}(x_2,Q^2)\right.  \nonumber \\
&&-\frac{\alpha _s (Q^2)}{2\pi }\ln \left( \frac{Q^2}{P_T^2} \right) \left[
\left( P_{j\leftarrow a}\otimes f_{a/h_1}\right) (x_1,Q^2)\;f_{\overline{k}%
/h_2}(x_2,Q^2) \right.  
\nonumber \\ && ~~~~~~~~~~~~~~~~~~~~~~~~~~ 
\left. - f_{j/h_1}(x_1,Q^2)
\left( P_{\overline{k}\leftarrow b}\otimes f_{b/h_2}\right) (x_2,Q^2)\right]  
\nonumber \\ &&
+\frac{\alpha _s (Q^2)}\pi \left[ \left( C_{ja}^{(1)}\otimes
f_{a/h_1}\right) (x_1,Q^2)\;\;f_{\overline{k}/h_2}(x_2,Q^2) \right.
\nonumber \\ && ~~~~~~~~~~~~~~~~~~~~~~~~~~ 
\left. \left. +f_{j/h_1}(x_1,Q^2)\;\left( 
C_{\overline{k}b}^{(1)}\otimes f_{b/h_2}\right)
(x_2,Q^2)\right] \right\} 
\nonumber \\ &&
+j\leftrightarrow \overline{k}+\int_0^{P_T^2}dQ_T^2%
\;Y(Q_T,Q,x_1,x_2),  \label{Eq:DelSig}
\end{eqnarray}
where $x_1 = {\rm e}^y Q/\sqrt{S}$ and $x_2 = {\rm e}^{-y} Q/\sqrt{S}$.
Equations (\ref{eq:SingPart}) and (\ref{Eq:DelSig}) (together with the
regular pieces, discussed in Section \ref{Sec:RegularContributions}) are
used to program the ${\cal O}(\alpha _s)$ results.

\subsection{$A$, $B$ and $C$ functions \label{sec:ABC}}

For completeness, we give here the coefficients $A$, $B$ and $C$ utilized in
our numerical calculations. The coefficients in the Sudakov exponent are~%
\cite{Collins-Soper-Sterman,Davies}. 
\begin{eqnarray*}
A^{(1)}(C_1) &=&C_F, \\
A^{(2)}(C_1) &=&C_F\left[ \left( {\frac{67}{36}}-{\frac{\pi ^2}{12}}\right)
N_C-{\frac 5{18}}N_f-2 \beta _1\ln \left( {\frac{b_0}{C_1}}\right) \right] ,
\\
B^{(1)}(C_1,C_2) &=&C_F\left[ -{\frac 32}-2\ln \left( {\frac{C_2b_0}{C_1}}%
\right) \right] , \\
B^{(2)}(C_1,C_2) &=&C_F\left\{ C_F\left( {\frac{\pi ^2}4}-{\frac 3{16}}%
-3\zeta (3)\right) +N_C\left( {\frac{11}{36}}\pi ^2-{\frac{193}{48}}+{\frac
32}\zeta (3)\right) \right. + \\
&&\frac{N_F}{2}\left( -{\frac 19}\pi ^2+{\frac{17}{12}}\right) - \left[
\left( {\frac{67}{18}}-{\frac{\pi ^2}6}\right) N_C-{\frac{5}{9}}N_f\right]
\ln \left( {\frac{C_2b_0}{C_1}}\right) + \\
&&\left. 2\beta _1\left[ \ln ^2\left( {\frac{b_0}{C_1}}\right) -\ln ^2(C_2)-{%
\frac 32}\ln (C_2)\right] \right\} ,
\end{eqnarray*}
where $N_f$ is the number of light quark flavors ($m_q<Q_V$, e.g. $N_f=5$
for $W^{\pm }$ or $Z^0$ production), $C_F={\rm tr}(t_at_a)$ is the second
order Casimir of the quark representation (with $t_a$ being the SU(N$_C$)
generators in the fundamental representation), $\beta _1=(11N_C-2N_f)/12$
and $\zeta (x)=\sum_{n=1}^{\infty} n^{-x}$ is the Riemann zeta function, and 
$\zeta (3) \approx 1.202$. For QCD, $N_C = 3$ and $C_F = 4/3$.

The $C_{jk}^{(n)}$ coefficients up to $n=1$ are: 
\begin{eqnarray*}
C_{jk}^{(0)}(z,b,\mu ,{\frac{C_1}{C_2}}) &=&\delta _{jk}\delta ({1-z}), \\
C_{jG}^{(0)}(z,b,\mu ,{\frac{C_1}{C_2}}) &=&0, \\
C_{jk}^{(1)}(z,b,\mu ,{\frac{C_1}{C_2}}) &=&\delta _{jk}C_F\left\{ {\frac 12}%
(1-z)-{\frac 1{C_F}}\ln \left( {\frac{\mu b}{b_0}}\right) P_{j\leftarrow
k}^{(1)}(z)\right. \\
&&\left. +\delta (1-z)\left[ -\ln ^2\left( {\frac{C_1}{{b_0C_2}}}%
e^{-3/4}\right) +{\frac{\pi ^2}4}-{\frac{23}{16}}\right] \right\} , \\
C_{jG}^{(1)}(z,b,\mu ,{\frac{C_1}{C_2}}) &=&{\frac 12}z(1-z)-\ln \left( {%
\frac{\mu b}{b_0}}\right) P_{j\leftarrow G}^{(1)}(z),
\end{eqnarray*}
where $P_{j\leftarrow a}^{(1)}(z)$ are the leading order DGLAP splitting
kernels \cite{DGLAP} given in Section \ref{Sec:Factorization}, and $j$ and 
$k$ represent quark or anti-quark flavors.

The constants $C_1,\;C_2$ and $C_3\equiv \mu b$ were introduced when solving
the renormalization group equation for $\widetilde{W}_{jk}$. $C_1$ enters
the lower limit $\overline{\mu }=C_1/b$ in the integral of the Sudakov
exponent [cf. Eq.~(\ref{eq:SudExp})], and determines the onset of the
non-perturbative physics. The renormalization constant $C_2,$ in the upper
limit $\overline{\mu }=C_2 Q$ of the Sudakov integral, specifies the scale
of the hard scattering process. The scale $\mu = C_3/b$ is the scale at
which the $C$ functions are evaluated. The canonical choice of these
renormalization constants is $C_1=C_3=2e^{-\gamma _E}\equiv b_0$ and $%
C_2=C_4=1$ \cite{CSS}. We adopt these choices of the renormalization
constants in the numerical results of this work, because they eliminate
large constant factors within the $A,\;B$ and $C$ functions.

After fixing the renormalization constants to the canonical values, we
obtain much simpler expressions of $A^{(1)}$, $B^{(1)}$, $A^{(2)}$ and $%
B^{(2)}$. The first order coefficients in the Sudakov exponent become 
\[
A^{(1)}(C_1)=C_F,~~~{\rm and}~~~B^{(1)}(C_1=b_0,C_2=1)=-3C_F/2. 
\]
The second order coefficients in the Sudakov exponent simplify to 
\begin{eqnarray*}
\ A^{(2)}(C_1=b_0) &=&C_F\left[ \left( {\frac{67}{36}}-{\frac{\pi ^2}{12}}%
\right) N_C-{\frac 5{18}}N_f\right] , \\
B^{(2)}(C_1=b_0,C_2=1) &=&C_F^2\left( {\frac{\pi ^2}4}-{\frac 3{16}}-3\zeta
(3)\right) +C_FN_C\left( {\frac{11}{36}}\pi ^2-{\frac{193}{48}}+{\frac 32}%
\zeta (3)\right) \\
&&+C_FN_f\left( -{\frac 1{18}}\pi ^2+{\frac{17}{24}}\right).
\end{eqnarray*}

The Wilson coefficients $C_{ja}^{(i)}$ for the parity-conserving part of the
resummed result are also greatly simplified under the canonical definition
of the renormalization constants. Their explicit forms are 
\begin{eqnarray*}
C_{jk}^{(1)}(z,b,\mu =\frac{b_0}b,\frac{C_1}{C_2}=b_0) &=&\delta
_{jk}\left\{ {\frac 23}(1-z)+{\frac 13}\,\delta (1-z)(\pi ^2-8)\right\} ~~~%
{\rm and}~~~ \\
C_{jG}^{(1)}(z,b,\mu =\frac{b_0}b,\frac{C_1}{C_2}=b_0) &=&{\frac 12}z(1-z).
\end{eqnarray*}
The same Wilson coefficient functions $C_{ja}$ also apply to the parity
violating part which is multiplied by the angular function ${\cal A}_3 = 2
\cos{\theta}$.

\subsection{Regular Contributions \label{Sec:RegularContributions}}

The $Y$ piece in Eq.~(\ref{eq:ResFor}), which is the difference of the fixed
order perturbative result and their singular part, is given by the
expression 
\begin{eqnarray}
&& Y(Q_T,Q,x_1,x_2,\theta ,\phi ,C_4) =\int_{x_1}^1{\frac{d\xi _1}{\xi _1}}%
\int_{x_2}^1{\frac{d\xi _2}{\xi _2}}\sum_{n=1}^\infty \left( {\frac{\alpha
_S(C_4Q)}\pi }\right) ^n  \nonumber \\
&&~~~\times f_{a/h_1}(\xi _1,C_4Q)\,R_{ab}^{(n)}(Q_T,Q,z_1,z_2,\theta ,\phi
)\,f_{b/h_2}(\xi _2,C_4Q),  \label{eq:YPc}
\end{eqnarray}
where $z_i=x_i/\xi _i$ $(i=1,2)$. The regular functions $R_{ab}^{(n)}$ only
contain contributions which are less singular than  
$Q_T^{-2}$ $\times$ [1 or $\ln (Q^2/Q_T^2)$]
as $Q_T\rightarrow 0$. Their explicit expressions for $%
h_1h_2\rightarrow V(\rightarrow \ell _1{\bar{\ell}_2})X$ are given below.
The scale for evaluating the regular pieces is $C_4Q$. To minimize the
contribution of large logarithmic terms from higher order corrections, we
choose $C_4=1$ when calculating the $Y$ piece.

We define the $q\bar{q^{\prime }}V$ and the $\ell _1{\bar \ell _2}V$
vertices, respectively, as 
\[
i\gamma _\mu \left[ g_L(1-\gamma _5)+g_R(1+\gamma _5)\right] ~~~{\rm and}~~~
i\gamma _\mu \left[ f_L(1-\gamma _5)+f_R(1+\gamma _5)\right] . 
\]
For example, for $V=W^{+},q=u$, ${\bar{q}^{\prime }}={\bar{d}}$, $\ell
_1=\nu _e$, and ${\bar \ell_2}=e^{+}$, the couplings $g_L^2=f_L^2=G_FM_W^2/%
\sqrt{2} $ and $g_R^2=f_R^2=0$, where $G_F$ is the Fermi constant. Table \ref
{tbl:parameters} shows all the couplings for the general case. In Eq.~(\ref
{eq:YPc}), 
\[
R_{ab}^{(1)}={\frac{16\mid V_{ab}\mid ^2}{\pi Q^2}}\left[
(g_L^2+g_R^2)(f_L^2+f_R^2)R_1^{ab}+(g_L^2-g_R^2)(f_L^2-f_R^2)R_2^{ab}
\right], 
\]
where the coefficient functions $R_i^{ab}$ are given as follows\footnote{%
Note that in Ref.~\cite{Balazs-Qui-Yuan} there were typos in $R_1^{j{\bar{k}}%
}$ and $R_2^{Gj}$.}: 
\begin{eqnarray*}
&&R_1^{j{\bar{k}}}=r^{j{\bar{k}}}{\cal L}_0+{\frac{T_{+}(t,u)}s}\,\delta
(s+t+u-Q^2)\left[ {\cal A}_0+{\cal A}_2+{\frac Q{Q_T}} {\frac{T_{-}(u,t)}{%
T_{+}(t,u)}}{\cal A}_1\right] {\frac{Q^2}{M_T^2}}, \\
&&R_2^{j{\bar{k}}}=r^{j{\bar{k}}}{\cal A}_3+{\frac{T_{+}(t,u)}s}\,\delta
(s+t+u-Q^2) \\
&&~~~~~~~~~~~~\times \left\{ {\frac{Q^2}{Q_T^2}}\left( {\frac Q{M_T}}%
-1\right) {\cal A}_3-{\frac{2Q^2}{Q_TM_T}}{\frac{T_{-}(t,u)} {T_{+}(t,u)}}%
{\cal A}_4\right\} , \\
&&R_1^{Gj}=r^{Gj}{\cal L}_0-{\frac{Q^2Q_T^2}{uM_T^2}}\, {\frac{T_{+}(u,s)}s}%
\,\delta (s+t+u-Q^2) \\
&&~~~~~~~~~~~~\times \left\{ {\frac{T_{+}(u,-s)}{T_{+}(u,s)}} \left[ {\cal A}%
_0+{\cal A}_2\right] +{\frac Q{Q_T}}{\frac{(Q^2-u)^2+ T_{-}(u,t)}{T_{+}(u,s)}%
}{\cal A}_1\right\} , \\
&&R_2^{Gj}=-r^{Gj}{\cal A}_3-{\frac{Q_T^2}u}~{\frac{T_{+}(u,s)}s} \,\delta
(s+t+u-Q^2) \\
&&~~~~~~~~~~~~\times \left\{ {\frac{Q^2}{Q_T^2}}\left[ {\frac Q{M_T}}\left( {%
\frac{2u(Q^2-s)}{T_{+}(u,s)}}-1\right) +1\right] {\cal A}_3\right. \\
&&~~~~~~~~~~~~~~~~~~~~\left. -{\frac{2Q^2}{Q_TM_T}}\left[ {\frac{2s(Q^2-s)} {%
T_{+}(u,s)}}+1\right] {\cal A}_4\right\} ,
\end{eqnarray*}
with 
\begin{eqnarray*}
&&r^{j{\bar{k}}}={\frac{Q^2}{Q_T^2}}\left\{ {\frac{T_{+}(t,u)}s}~\delta
(s+t+u-Q^2)-2~\delta (1-z_1)~\delta (1-z_2)\left[ \ln {\left( {\frac{Q^2}{%
Q_T^2}}\right) }-{\frac 32}\right] \right. \\
&&~~~~~~~~~~~~~~~\left. -\delta (1-z_1)\left( {\frac{1+z_2^2}{1-z_2}}\right)
_{+}-\delta (1-z_2)\left( {\frac{1+z_1^2}{1-z_1}}\right) _{+}\right\} ,
\end{eqnarray*}
and 
\[
r^{Gj}={\frac{Q^2}{Q_T^2}}\left\{ -{\frac{Q_T^2}u}~{\frac{T_{+}(u,s)}s}%
\,\delta (s+t+u-Q^2)-\left[ z_1^2+(1-z_1)^2\right] \,\delta (1-z_2)\right\}
, 
\]
where $T_{\pm }(t,u)=(Q^2-t)^2\pm (Q^2-u)^2$. The Mandelstam variables 
$s,t,u $ and the angular functions ${\cal L}_0,{\cal A}_i$ are defined in
Sections \ref{GeneralProp} and \ref{TheCrossS}.
The $V_{jk}$ coefficients are defined by Eq.~(\ref{eq:Vjk}). For 
$a=j$ and $b = G$: $\displaystyle |V_{jG}|^2 = \sum_k |V_{jk}|^2$ where $j$
and $k$ are light quark flavors with opposite weak isospin quantum numbers.
Up to this order, there is no contribution from gluon-gluon initial state,
i.e. $R_{GG}^{(1)} = 0$. The remaining coefficient functions with all
possible combinations of the quark and gluon indices (for example $R^{{\bar k%
}j}$, $R^{G{\bar j}}$ or $R^{jG}$, etc.) are obtained by the same crossing
rules summarized in Section \ref{Sec:OASExpansion}. 

Having both the singular and the regular pieces expanded up to ${\cal O}%
(\alpha _s)$, we can construct the NLO Monte Carlo calculation by first
including the contribution from Eq.~(\ref{Eq:DelSig}), with $P_T=Q_T^{Sep}$,
for $Q_T<Q_T^{Sep}$. Second, for $Q_T>Q_T^{Sep}$, we include the ${\cal O}%
(\alpha _s)$ perturbative results, which is equal to the sum of the singular
[Eq.~(\ref{eq:SingPart})] and the regular [Eq.~(\ref{eq:YPc})] pieces up to $%
{\cal O}(\alpha _s)$. (Needless to say that the relevant angular functions
for using Eqs.(\ref{eq:SingPart}) and (\ref{Eq:DelSig}) are ${\cal L}_0=1 +
\cos^2{\theta}$ and ${\cal A}_3=2 \cos{\theta}$, cf. Eq.~(\ref{eq:MBorn}).)
Hence, the NLO total rate is given by the sum of the contributions from both
the $Q_T<Q_T^{Sep}$ and the $Q_T>Q_T^{Sep}$ regions.




\chapter{ Vector Boson Production and Decay in Hadron Collisions 
\label{ch:VBP}}

\section{Vector Boson Distributions \label{secV1}}

At the Fermilab Tevatron, about ninety percent of the production cross
section of the $W^{\pm}$, $Z^0$ bosons and virtual photons 
is in the small transverse
momentum region, where $Q_T\lesim 20$ GeV (hence $Q_T^2\ll Q^2$). In this
region the higher order perturbative corrections, dominated by soft and
collinear gluon radiation and of the form $Q_T^{-2}$ $\sum_{n=1}^\infty$
$\sum_{m=0}^{2n-1}$ ${}_nv_m\,$ $\alpha _s^n$ $\ln ^m({Q_T^2/Q^2})$, 
are substantial
because of the logarithmic enhancement \cite{CSS}. (${}_nv_m$ are the
coefficient functions for a given $n$ and $m$.) As we discussed, these
corrections are divergent in the $Q_T\rightarrow 0$ limit at any fixed order
of the perturbation theory. After applying the renormalization group
analysis, these singular contributions in the low $Q_T$ region can be
resummed to derive a finite prediction for the $Q_T$ distribution to compare
with experimental data.

In this Chapter we discuss the phenomenology predicted by the resummation
formalism. To illustrate the effects of multiple soft gluon radiation, we
also give results predicted by a next-to-leading order (NLO, ${\cal O}%
(\alpha _s)$) calculation. As expected, the resummed and the NLO predictions
of observables that are directly related to the transverse momentum of the 
vector boson will exhibit large differences. These observables,
are for example, the transverse momentum of the leptons from vector boson
decay, the back-to-back correlations of the leptons from $Z^0$ decay, etc.
The observables that are not directly related to the transverse momentum of
the vector boson can also show noticeable differences between the resummed
and the NLO calculations if the kinematic cuts applied to select the signal
events are strongly correlated to the transverse momentum of the vector
boson.

Due to the increasing precision of the experimental data at hadron
colliders, it is necessary to improve the theoretical prediction of the QCD
theory by including the effects of the multiple soft gluon emission to all
orders in $\alpha _s$. To justify the importance of such an improved QCD
calculation, we compare various distributions predicted by the resummed and
the NLO calculations. For this purpose we categorize measurables into two
groups. We call an observable to be {\it directly sensitive} to the soft
gluon resummation effect if it is sensitive to the transverse momentum of
the vector boson. The best example of such observable is the transverse
momentum distribution of the vector boson ($d\sigma /dQ_T$). Likewise, the
transverse momentum distribution of the decay lepton ($d\sigma /dp_T^\ell $)
is also directly sensitive to resummation effects. The other examples are
the azimuthal angle correlation of the two decay leptons $(\Delta \phi
^{\ell _1{\bar{\ell}_2}})$, the balance in the transverse momentum of the
two decay leptons $(p_T^{\ell _1}-p_T^{\bar{\ell}_2})$, or the correlation
parameter $z=-{\vec{p}_T^{~\ell _1}}\cdot {\vec{p}_T^{~{\bar{\ell}_2}}}%
/[\max ({{p_T^{~\ell _1}},{p_T^{~{\bar{\ell}_2}})}}]^2$. These distributions
typically show large differences between the NLO and the resummed
calculations. The differences are the most dramatic near the boundary of the
kinematic phase space, such as the $Q_T$ distribution in the low $Q_T$
region and the $\Delta \phi ^{\ell _1{\bar{\ell}_2}}$ distribution near $\pi 
$. Another group of observables is formed by those which are {\it indirectly
sensitive} to the resummation of the multiple soft gluon radiation. The
predicted distributions for these observables are usually the same in either
the resummed or the NLO calculations, provided that the $Q_T$ is fully
integrated out in both cases. Examples of indirectly sensitive quantities
are the total cross section $\sigma $, the invariant mass $Q$, the rapidity $%
y$, and $x_F\left( =2q^3/\sqrt{S}\right) $ of the vector boson\footnote{%
Here $q^3$ is the longitudinal-component of the vector boson momentum $q^\mu$
[cf. Eq.~(\ref{eq:QXYZ})].}, and the rapidity $y^\ell $ of the decay lepton.
However, in practice, to extract signal events from the experimental data
some kinematic cuts have to be imposed to suppress the background events. It
is important to note that imposing the necessary kinematic cuts usually
truncate the range of the $Q_T$ integration, and causes different
predictions from the resummed and the NLO calculations. We demonstrate such
an effect in the distributions of the lepton charge asymmetry $A(y^\ell )$
predicted by the resummed and the NLO calculations. We show that they are
the same as long as there are no kinematic cuts imposed, and different when
some kinematic cuts are included. They differ the most in the large rapidity
region which is near the boundary of the phase space.

To systematically analyze the differences between the results of the NLO and
the resummed calculations we implemented the ${\cal O}(\alpha _s^0)$ (LO),
the ${\cal O}(\alpha _s)$ (NLO), and the resummed calculations in a unified
Monte Carlo package: ResBos (the acronym stands for $Res$ummed Vector $Bos$%
on production). The code calculates distributions for the hadronic
production and decay of a vector bosons via $h_1h_2\rightarrow V(\rightarrow
\ell _1 {\bar \ell_2})X$, where $h_1$ is a proton and $h_2$ can be a proton,
anti-proton, neutron, an arbitrary nucleus or a pion. Presently, $V$ can be
a virtual photon $\gamma^*$ (for Drell-Yan production), $W^{\pm }$ or $Z^0$.
The effects of the initial state soft gluon radiation are included using the
QCD soft gluon resummation formula, given in Eq.~(\ref{eq:ResFor}). This
code also correctly takes into account the effects of the polarization and
the decay width of the massive vector boson.

It is important to distinguish ResBos from the parton shower Monte Carlo
programs like HERWIG \cite{herwig}, ISAJET \cite{isajet}, PYTHIA \cite{pythia}, 
etc., which use the backward radiation technique~\cite{Sjostrand}
to simulate the physics of the initial state soft gluon radiation. They are
frequently shown to describe reasonably well the shape of the vector boson
distribution. On the other hand, these codes do not have the full
resummation formula implemented and include only the leading logs and some
of the sub-logs of the Sudakov factor. The finite part of the higher order
virtual corrections which leads to the Wilson coefficient ($C$) functions is
missing from these event generators. ResBos contains not only the physics
from the multiple soft gluon emission, but also the higher order matrix
elements for the production and the decay of the vector boson with large $Q_T
$, so that it can correctly predict both the event rates and the
distributions of the decay leptons.

\begin{figure}[t]
\ifx\nopictures Y \else{
\centerline{\epsfysize=8.5cm \epsfbox{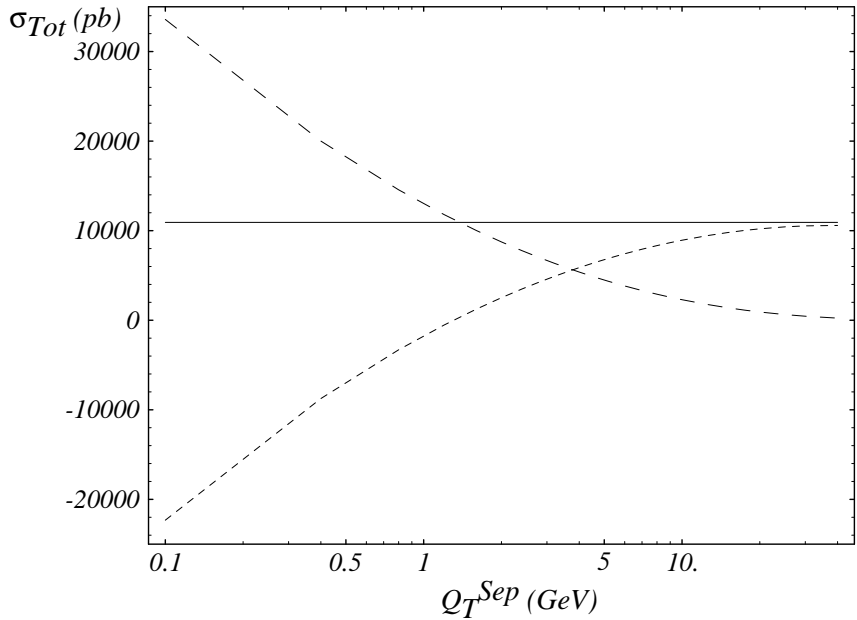}}} \fi
\caption{ Total $W^+$ production cross section as a function of the
parameter $Q_T^{Sep}$ (solid curve). The long dashed curve is the part of
the ${\cal {O}(\alpha _s)}$ cross section integrated from $Q_T^{Sep}$ to the
kinematical boundary, and the short dashed curve is the integral from $Q_T =
0$ to $Q_T^{Sep}$ at ${\cal O}(\alpha _s)$. The total cross section is
constant within $10^{-5}$ \% through more than two order of magnitude of $%
Q_T^{Sep}$.}
\label{fig:QTSep}
\end{figure}
In a NLO Monte Carlo calculation, it is ambiguous to treat the singularity
of the vector boson transverse momentum distribution near $Q_T=0$. There are
different ways to deal with this singularity. Usually one separates the
singular region of the phase space from the rest (which is calculated
numerically) and handles it analytically. We choose to divide the $Q_T$
phase space with a separation scale $Q_T^{Sep}$. We treat the $Q_T$ singular
parts of the real emission and the virtual correction diagrams analytically,
and integrate the sum of their contributions up to $Q_T^{Sep}$. If $%
Q_T<Q_T^{Sep}$ we assign a weight to the event based on the above integrated
result and assign it to the $Q_T=0$ bin. If $Q_T>Q_T^{Sep}$, the event
weight is given by the usual NLO calculation. The above procedure not only
ensures a stable numerical result but also agrees well with the logic of the
resummation calculation. In Fig.~\ref{fig:QTSep} we demonstrate that the
total cross section, as expected, is independent of the separation scale $%
Q_T^{Sep}$ in a wide range. As explained above, in the $Q_T<Q_T^{Sep}$
region we approximate the $Q_T$ of the vector boson to be zero. For this
reason, we choose $Q_T^{Sep}$ as small as possible. We use $Q_T^{Sep}=0.1$
GeV in our numerical calculations, unless otherwise indicated. This division
of the transverse momentum phase space gives us practically the same results
as the invariant mass phase space slicing technique. This was precisely
checked by the lepton charge asymmetry results predicted by DYRAD \cite
{Giele}, and the NLO [up to ${\cal O}(\alpha _s)$] calculation within the
ResBos Monte Carlo package.

To facilitate our comparison, we calculate the NLO and the resummed
distributions using the same parton luminosities and parton distribution
functions, EW and QCD parameters, and renormalization and factorization
scales so that any difference found in the distributions is clearly due to
the different QCD physics included in the theoretical calculations. (Recall
that they are different models of calculations based upon the same QCD
theory, and the resummed calculation contains the dynamics of the multiple
soft gluon radiation.) This way we compare the resummed and the NLO results
on completely equal footing. The parton distributions used in the different
order calculations are listed in Table~\ref{tbl:PDF}. In Table~\ref{tbl:PDF}
and the rest of this work, we denote by resummed (2,1,2) the result of the
resummed calculation with $A$ and $B$ calculated to $\alpha _s^2$ order, $C$
to $\alpha _s$, and $R$ to $\alpha _s^2$ order, that is with $A^{(1,2)}$, $%
B^{(1,2)}$, $C^{(0,1)}$ and, $R^{(1,2)}$ included [cf. Section \ref{sec:ABC}%
]. Similarly, resummed (1,0,1) includes $A^{(1)}$, $B^{(1)}$, $C^{(0)}$, and 
$R^{(1)}$, and resummed (1,0,0) includes $A^{(1)}$, $B^{(1)}$, $C^{(0)}$
without the $Y$ piece. Unless specified otherwise, hereafter we use $%
A^{(1,2)}$, $B^{(1,2)}$, $C^{(0,1)}$ and, $R^{(1,2)}$ in our resummed
calculation. In the following, we discuss the relevant experimental
observables predicted by these models of calculations using the ResBos code.
Our numerical results are given for the Tevatron, a $p{\bar{p}}$ collider
with $\sqrt{S}=1.8$ TeV, and CTEQ4 PDF's unless specified otherwise.

\begin{table}[t]
\begin{center}
\begin{tabular}{lcccccc}
\hline \hline \\[-0.2cm]
& \multicolumn{3}{c}{Fixed order} & \multicolumn{3}{c}{Resummed} \\ 
& ${\cal O}(\alpha _s^0)$ & ${\cal O}(\alpha _s)$ & ${\cal O}(\alpha _s^2)$
& (1,0,0) & (1,0,1) & (2,1,2) \\ 
\hline \\[-0.2cm]
PDF & CTEQ4L & CTEQ4M & CTEQ4M & CTEQ4L & CTEQ4M & CTEQ4M 
\\ \hline \hline
\end{tabular}
\end{center}
\caption{List of PDF's used at the different models of calculations. The
values of the strong coupling constants used with the CTEQ4L and CTEQ4M
PDF's are $\alpha _s^{(1)}(M_{Z^0}) = 0.132$ and $\alpha _s^{(2)}(M_{Z^0}) =
0.116$ respectively.}
\label{tbl:PDF}
\end{table}


\subsection{Vector Boson Transverse Momentum Distribution}

According to the parton model the primordial transverse momenta of partons
entering into the hard scattering are zero. This implies that a $\gamma ^{*}$%
, $W^{\pm }$ or $Z^0$ boson produced in the Born level process has no
transverse momentum, so that the LO $Q_T$ distribution is a Dirac-delta
function peaking at $Q_T=0$. In order to have a vector boson produced with a
non-zero $Q_T$, an additional parton has to be emitted from the initial
state partons. This happens in the QCD process. However the singularity at $%
Q_T=0$ prevails up to {\it any fixed order} in $\alpha _s$ of the
perturbation theory, and the transverse momentum distribution $d\sigma
/dQ_T^2 $ is proportional to $Q_T^{-2} \times [1~{\rm {or}~}\ln (Q_T^2/Q^2)]$
at small enough transverse momenta. The most important feature of the
transverse momentum resummation formalism is to correct this unphysical
behavior and render $d\sigma /dQ_T^2$ finite at zero $Q_T$ by exponentiating
the $Q_T$ singular logs.

The CSS formalism itself is constructed to do even more than that. By
including the regular $Y$ contribution, it interpolates between the low and
the high $Q_T$ regions smoothly, provided that the $A$, $B$, $C$ functions
and the $Y$ contribution are evaluated to all orders in $\alpha _s$.\footnote{%
Strictly speaking, this is only true when the energy of the collider is much
larger than $Q$, because the resummed and the perturbative pieces are
evaluated at different $x$ values. The former depends on $x_1$ and $x_2$
defined in Eq.~(\ref{eq:ResFor}), however the latter depends on $\xi_1$ and $%
\xi_2$ (cf. Eq.~(\ref{eq:RegPc})) in which the energy carried away by the
emitted gluon is also included.} The $Y$ piece is defined as the difference
of the fixed order perturbative result and its $Q_T$ singular (asymptotic)
part which grows as $Q_T^{-2} \times$ [1 or $\ln (Q_T^2/Q^2)$] when $Q_T \to
0$. In the $Q_T \ll Q$ region, the $\ln (Q/Q_T)$ terms are large and the
perturbative distribution is dominated by these singular logs, that is the
perturbative and the asymptotic parts are about the same. Consequently, for
low $Q_T$, the exponentiated asymptotic pieces, i.e. the CSS piece,
dominates over the $Y$ piece. In the $Q_T \sim Q$ region the $\ln (Q_T/Q)$
terms are small, and the perturbative part is dominated by its other terms.
The CSS and the asymptotic terms cancel each other leaving the perturbative
piece to dominate the high $Q_T$ region. The cancellation between the
perturbative and asymptotic pieces is always exact (by definition) in the
low $Q_T$ region order by order in $\alpha _s$, and the formalism is well
defined for low $Q_T$, no matter in which order the $A$, $B$, $C$ functions
and $Y$ are known. On the other hand, the cancellation between the CSS and
the asymptotic pieces in the high $Q_T$ region becomes better if the
asymptotic piece is calculated in higher order in $\alpha _s$. This is
because the CSS piece contains logs in all order (cf. Eq.(\ref{eq:logs}))
while the asymptotic part only up to a fixed order in $\alpha _s$. The above
will be clearly illustrated later in Fig.~\ref{fig:Matching}. Consequently,
the CSS formalism must break down for $Q_T {\ \lower-1.2pt%
\vbox{\hbox{\rlap{$>$}\lower5pt\vbox{\hbox{$\sim$}}}}\ } Q$, since $A$, $B$, 
$C$ and $Y$ are known only up to a finite order.

Although the matching is built into the formalism, in practice it is still
necessary to specify a matching prescription which provides a smooth
transition between the resummed and the fixed order perturbative results. In
Fig.~\ref{fig:Matching} we show the resummed (1,1,1) (resummation with $%
A^{(1)},B^{(1)}$, $C^{(0,1)}$, and $R^{(1)}$ included) and the fixed order $%
{\cal O}(\alpha _s)$ $Q_T$ distributions for $W^+$ and $Z^0$ bosons. As
shown, the resummed (1,1,1) and the fixed order curves are close to each
other for $Q/2<Q_T<Q$, and they cross around $Q_T\sim Q/2$. Based on this
observation we adopt the following procedure for calculating the fully
differential cross section $d\sigma/dQ^2dQ_Tdy$. For $Q_T$ values below the
crossing points $Q_T^{match}(Q,y)$ of the resummed and the fixed order $Q_T$
distributions, as the function of $Q$ and $y$, we use the resummed cross
section, and above it we use the fixed order perturbative cross section. The
resulting $d\sigma/dQ{^2}dQ_Tdy$ distribution is continuous, although not
differentiable with respect to $Q_T$ right at the matching points $%
Q_T^{match}(Q,y)$. The differential cross section $d\sigma/dQ_T$, on the
other hand, is completely smooth since it has no specific matching point.
Most importantly, the above prescription does not alter either the resummed
or the fixed order perturbative distributions in the kinematic regions where
they are proven to be valid. 
\begin{figure*}[t]
\vspace{-.75cm}
\begin{center}
\begin{tabular}{cc}
\ifx\nopictures Y \else{ \epsfysize=5.2cm $\!\!\!\!\!\!\!\!\!\!\!\!$
\epsffile{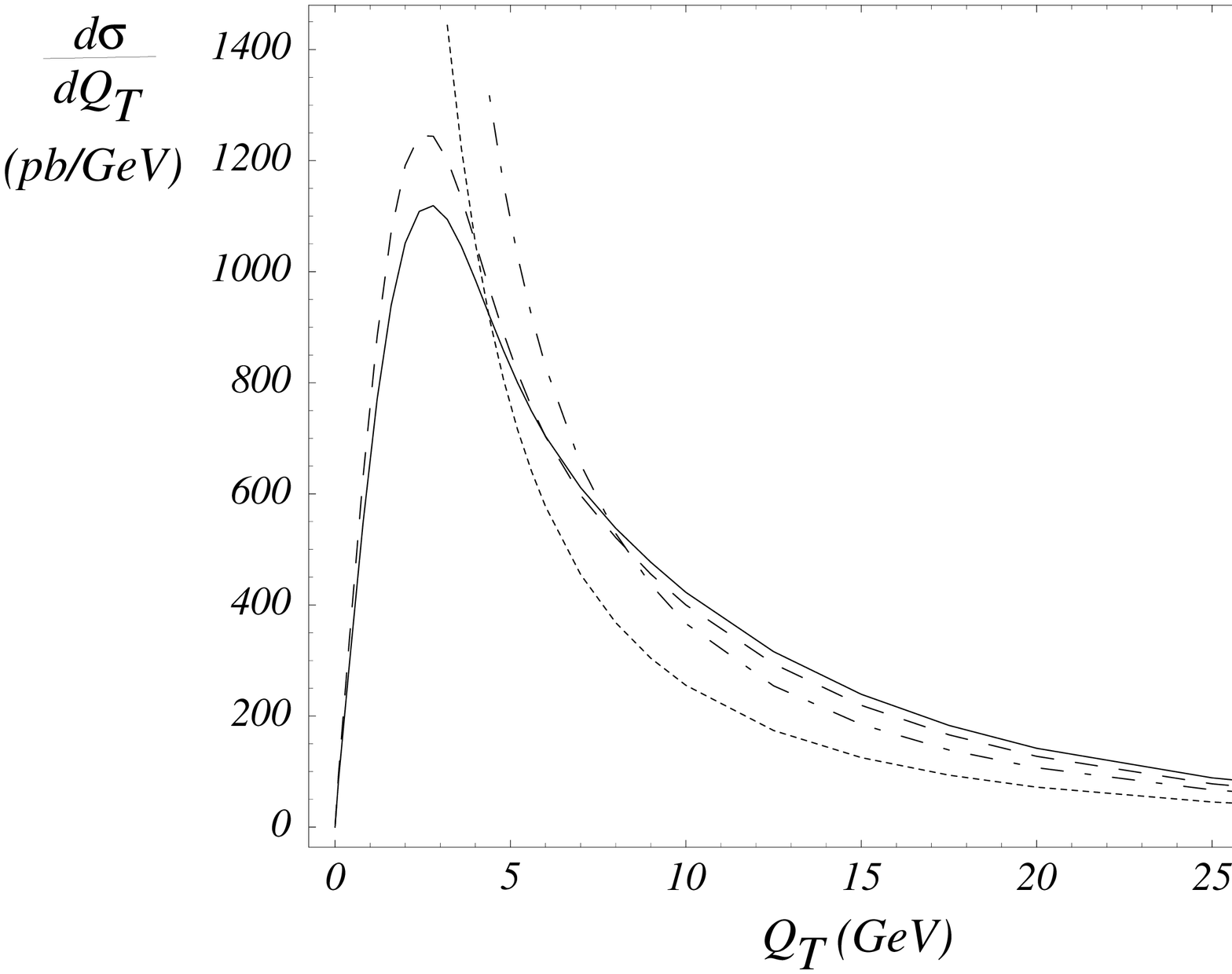} \epsfysize=5.2cm
\epsffile{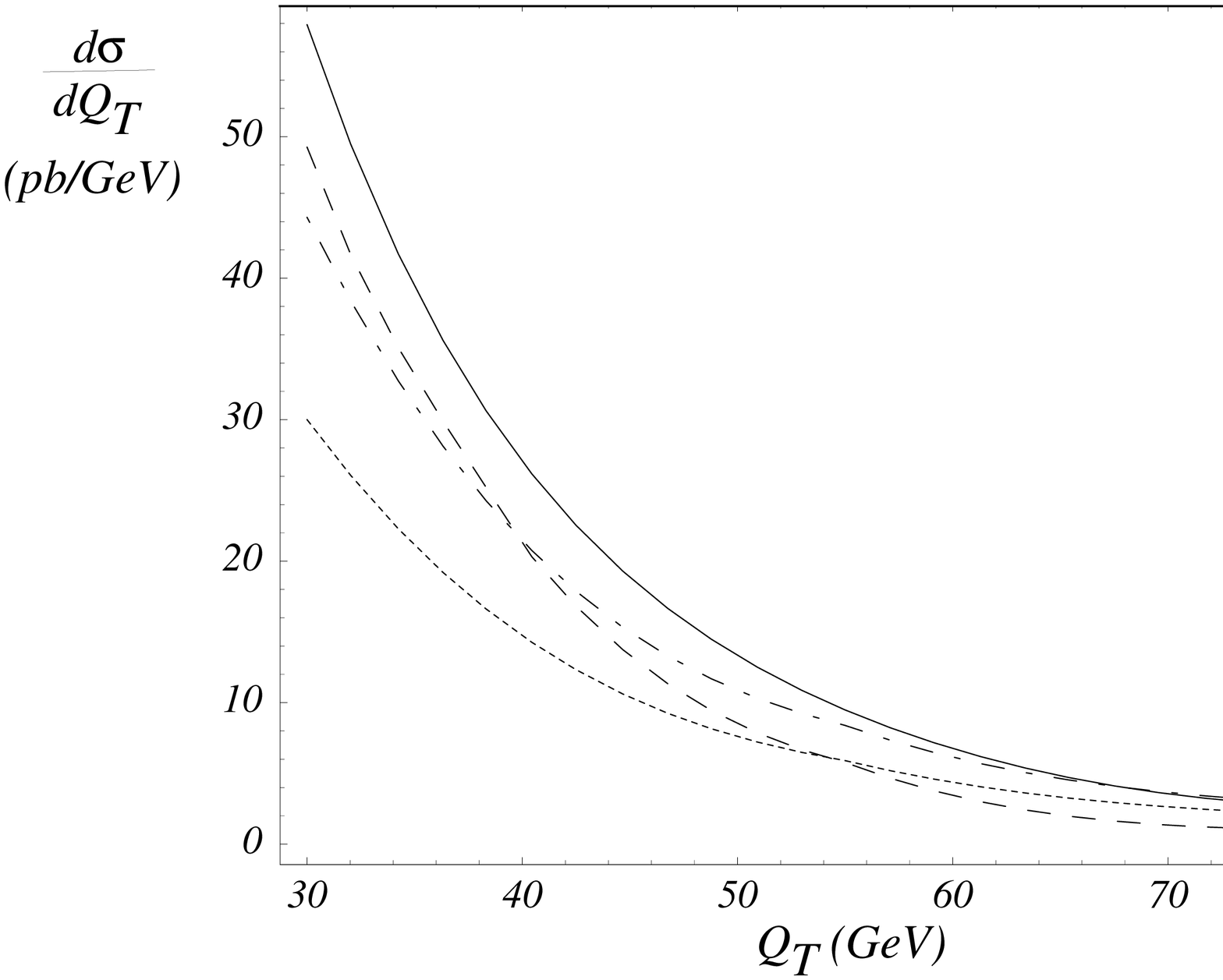} } \fi &  \\ 
\ifx\nopictures Y \else{ \epsfysize=5.2cm $\!\!\!\!\!\!\!\!\!\!\!\!$
\epsffile{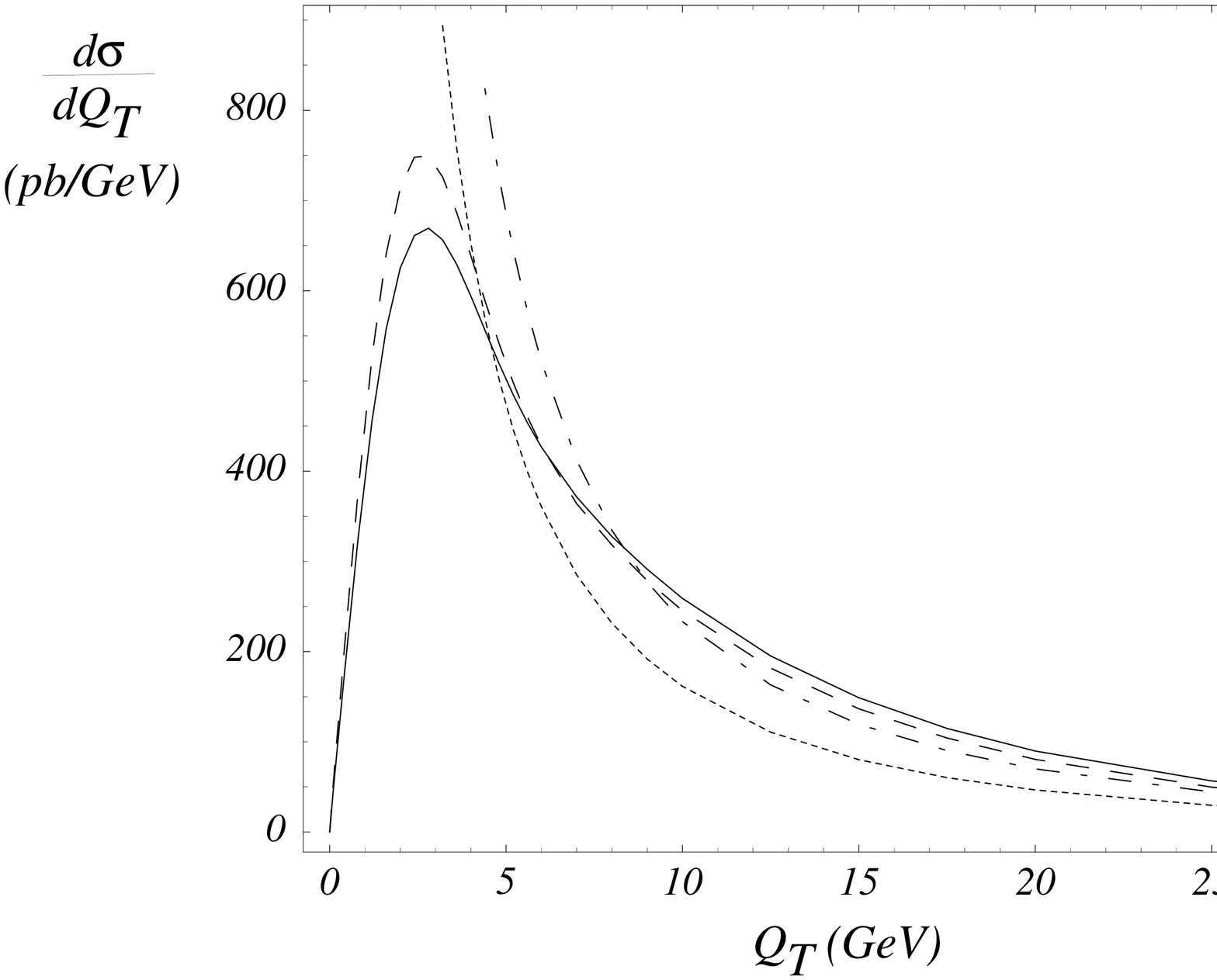} \epsfysize=5.2cm
\epsffile{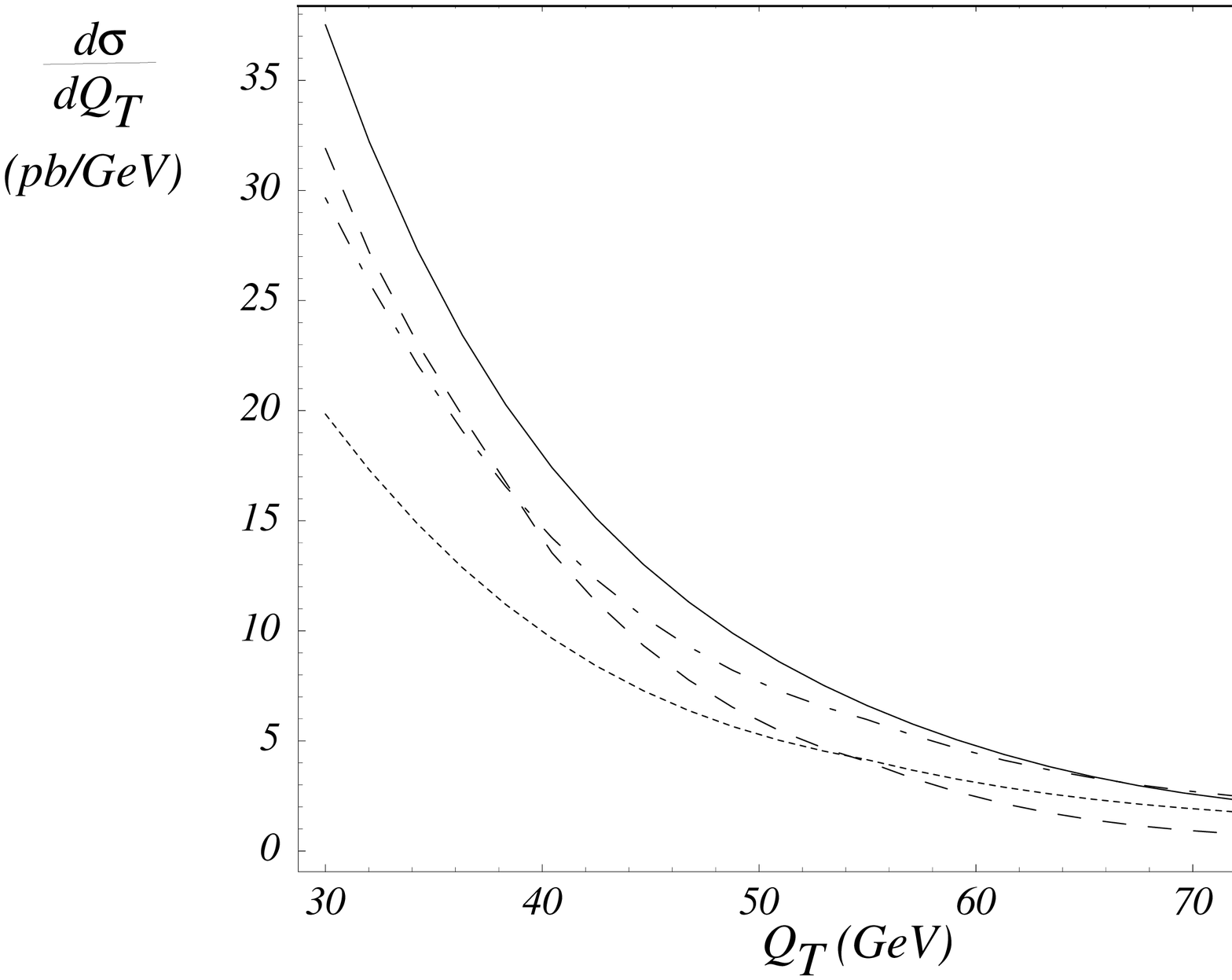} } \fi & 
\end{tabular}
\end{center}
\vspace{-1cm}
\caption{The low and intermediate $Q_T$ regions of the $W^+$ and $Z^0$
distributions at the Tevatron, calculated in fixed order ${\cal {O}(\alpha _s)%
}$ (dotted) and ${\cal O}(\alpha _s^2)$ (dash-dotted), and resummed (1,1,1)
(dashed) and (2,1,2) (solid) [cf. Table~III]. The cross-over occurs at 54
GeV for the ${\cal O}(\alpha _s)$, and at 69 GeV for the ${\cal O}(\alpha _s^2)
$ $W^\pm$ distributions. The matching between the resummed and the fixed
order disrtibutions becomes much smoother at ${\cal O}(\alpha _s^2)$ than at $%
{\cal O}(\alpha _s)$. The situation is very similar for the $Z^0$ boson. }
\label{fig:Matching}
\end{figure*}

To improve the theory prediction for the $Q_T$ distribution, we also include
the effect of some known higher order (at ${\cal O}(\alpha _s^2)$)
corrections to the Sudakov factor~\cite{Davies}, the $Y$ 
piece~ \cite{Arnold-Kauffman}, 
and the fixed order perturbative cross section~\cite{Arnold-Reno}. 
As we described, the $Y$ piece plays an essential role in the
matching between the resummed and the fixed order $Q_T$ distributions which
are dominated by the $Y$ piece when $Q_T$ is in the matching region. To
emphasize this in Fig.~\ref{fig:YtoTotal} we show the ratio of the $Y$ piece
to the total resummed (2,1,1) cross section. 
\begin{figure*}[t]
\vspace{-.5cm}
\begin{center}
\begin{tabular}{cc}
\ifx\nopictures Y \else{ \epsfysize=6.2cm \epsffile{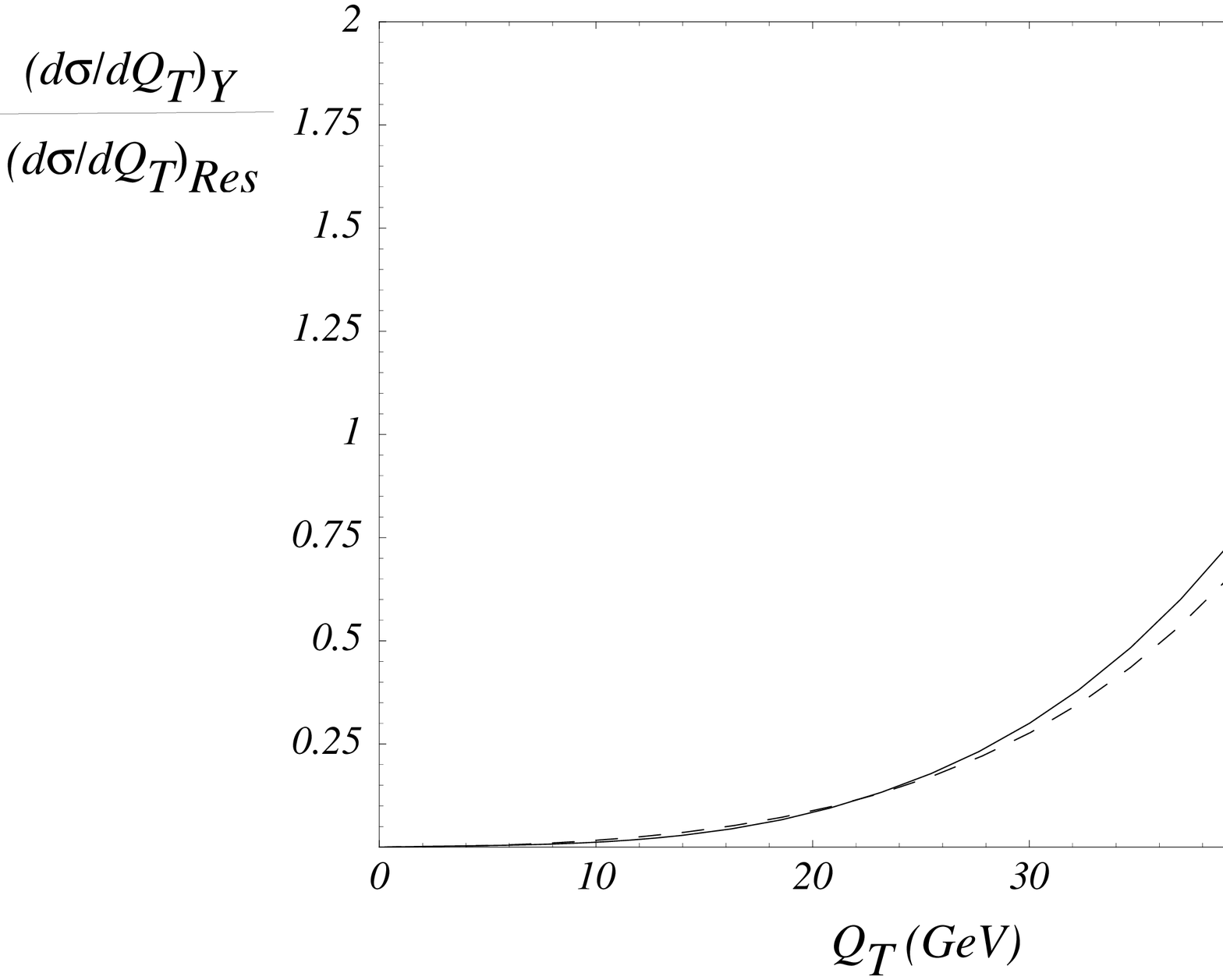} } %
\fi &  \\ 
\ifx\nopictures Y \else{ \epsfysize=6.2cm \epsffile{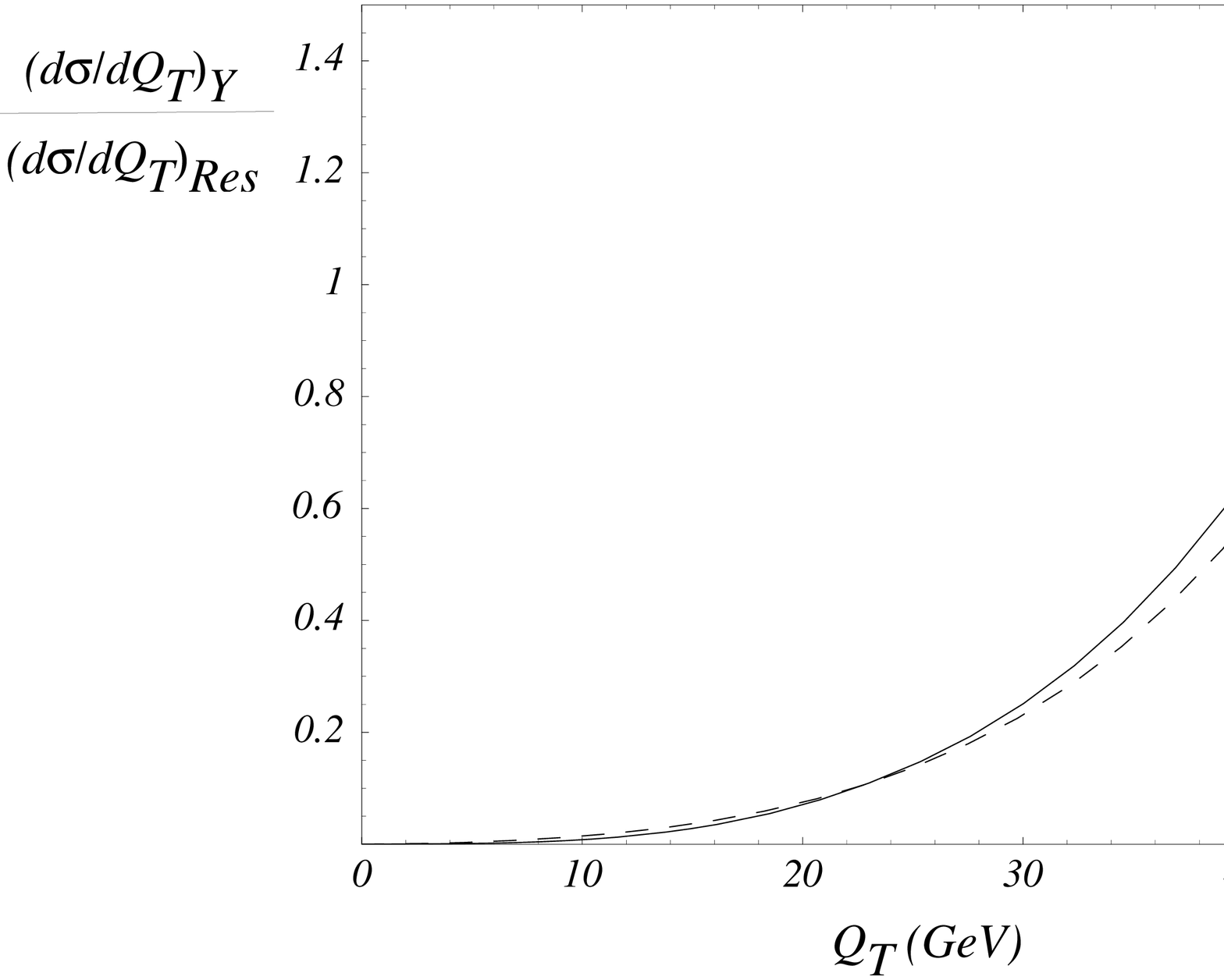} } %
\fi & 
\end{tabular}
\end{center}
\vspace{-1cm}
\caption{ The ratio of the ${\cal O}(\alpha _s+\alpha _s^2)$ $Y$ piece
(solid curve, $R^{(1)}$ and $R^{(2)}$ included), and the ${\cal O}(\alpha _s)
$ $Y$ piece (dashed curve, $R^{(1)}$ included) to the resummed (2,1,1)
distribution for $W^+$ and $Z^0$ bosons. }
\label{fig:YtoTotal}
\end{figure*}
This ratio can be larger than one because the CSS piece, which is the
difference between the total resummed cross section and the $Y$ piece, can
be negative for large enough $Q_T ({\lower-1.2pt\vbox{\hbox{\rlap{$>$}%
\lower5pt\vbox{\hbox{$\sim$}}}}\ } Q/2$). As indicated, the $Y$ contribution
is small for $Q_T < 30$ GeV. At $Q_T = 30$ GeV it only contributes by about
25\% to $d\sigma/dQ_T$. The total contribution of the $Y$-term to $%
\int_0^{30~GeV} dQ_T~(d\sigma/dQ_T)$ is less than a percent. Therefore, in
the region of $Q_T < 30$ GeV, the CSS piece dominates. We can also define
the $K_Y$-factor as the ratio of the $Y$ pieces calculated at the ${\cal O}%
(\alpha _s + \alpha _s^2)$ to that at the ${\cal O}(\alpha _s)$. The $K_Y$%
-factor is plotted in Fig.~\ref{fig:Y2toY1} as the function of $Q_T$ and $y$
for $W^\pm$ and $Z^0$ bosons (for $Q = M_V$). 
\begin{figure*}[t]
\vspace{-.5cm}
\begin{center}
\begin{tabular}{cc}
\ifx\nopictures Y \else{ \epsfysize=6.2cm \epsffile{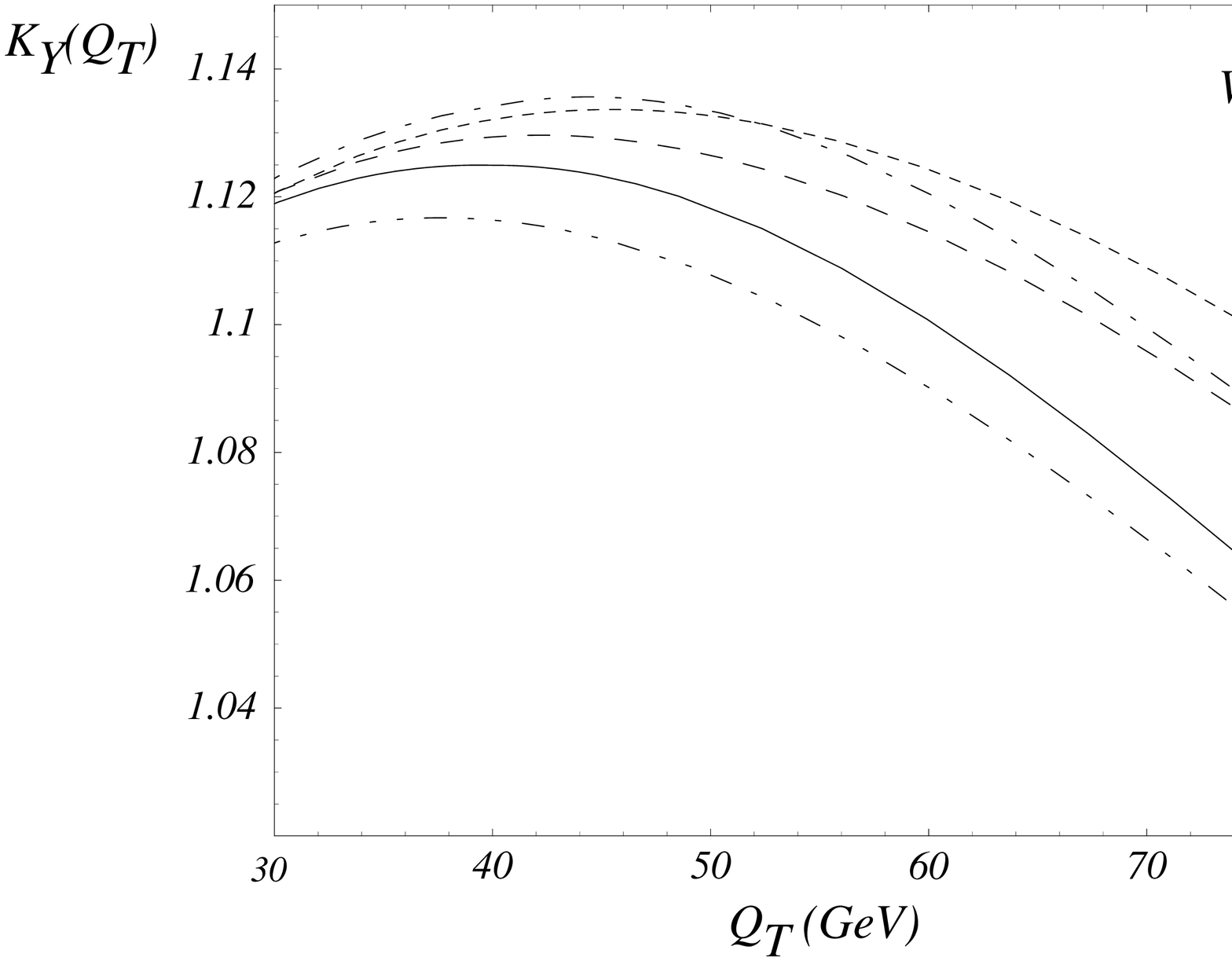} } %
\fi &  \\ 
\ifx\nopictures Y \else{ \epsfysize=6.2cm \epsffile{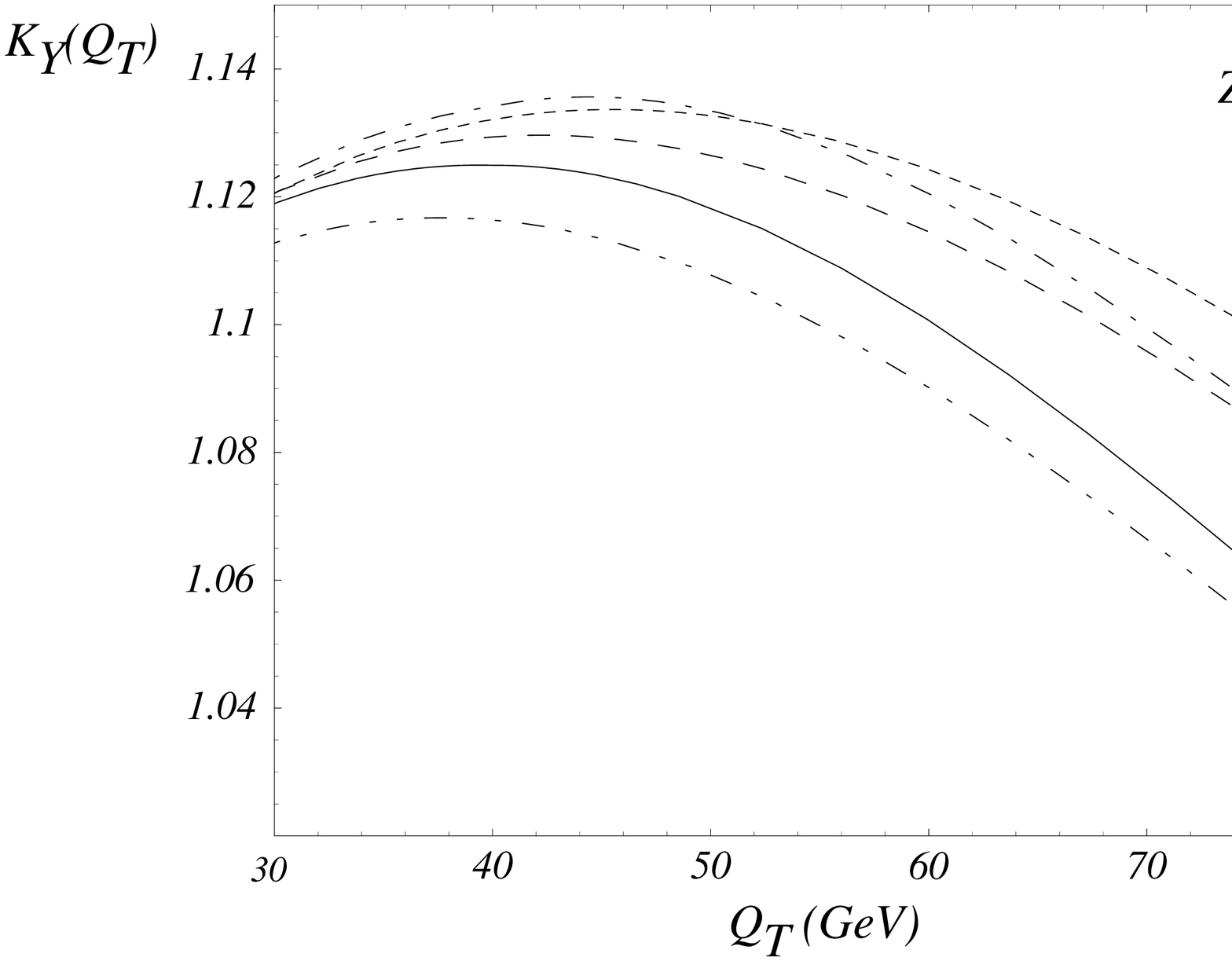} } %
\fi & 
\end{tabular}
\end{center}
\vspace{-1cm}
\caption{ The $K_Y$-factor: ratio of the ${\cal O}(\alpha _s + \alpha _s^2)$ 
$Y$ piece ($R^{(1)}$ and $R^{(2)}$ included) to the ${\cal O}(\alpha _s)$ $Y$
piece ($R^{(1)}$ included). The curves are plotted for $Q=M_V$ and $y = -$%
2.0 (solid), $-$1.0 (long dash), 0.0 (short dash), 1.0 (dash-dot) and $y = $%
2.0 (dash-double-dot). }
\label{fig:Y2toY1}
\end{figure*}
As shown, when $Q_T$ is between 30 to 80 GeV, the $K_Y$ factor is about 10\%
unless the rapidity of the $W$ and $Z$ bosons become large (i.e. $|y| > 2$).
Similarly, in Fig.~\ref{fig:P2toP1}, we show the $K_P$\ factor, which is
defined as the ratio of the fixed order differential cross sections
calculated at the ${\cal O}(\alpha _s+\alpha _s^2)$ to that at the ${\cal O}%
(\alpha _s)$, as a function of $Q_T$, and $y$ for $Q = M_V$. As expected,
when $|y|$ is large, i.e. near the boundary of the available phase space,
the $K_P$ factor can be large. The variation as a function of $Q_T$ for $Q_T
> 50$ GeV is small, of the order of 10\%. 
\begin{figure*}[t]
\vspace{-.5cm}
\begin{center}
\begin{tabular}{cc}
\ifx\nopictures Y \else{ \epsfysize=6.2cm \epsffile{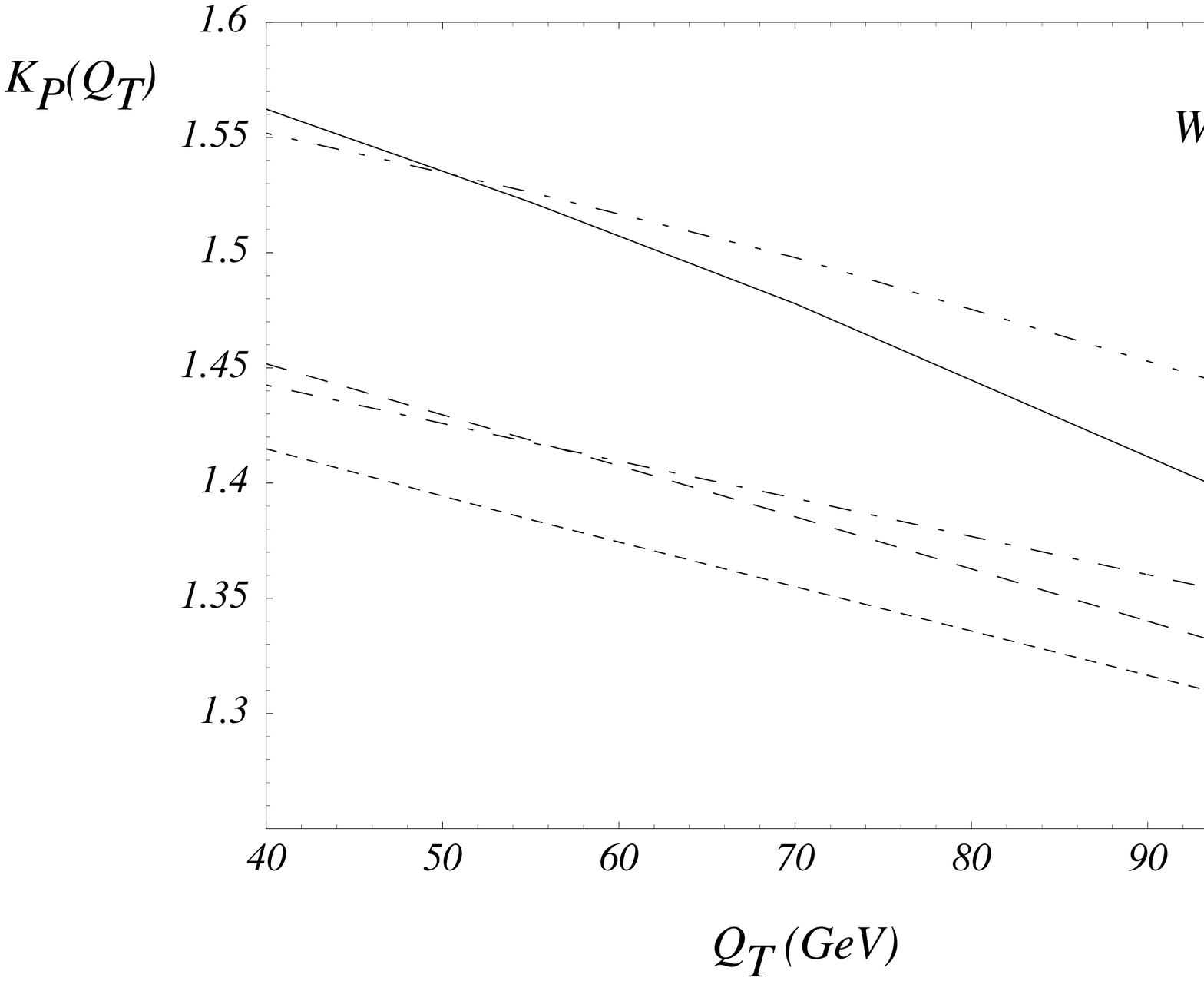} } %
\fi &  \\ 
\ifx\nopictures Y \else{ \epsfysize=6.2cm \epsffile{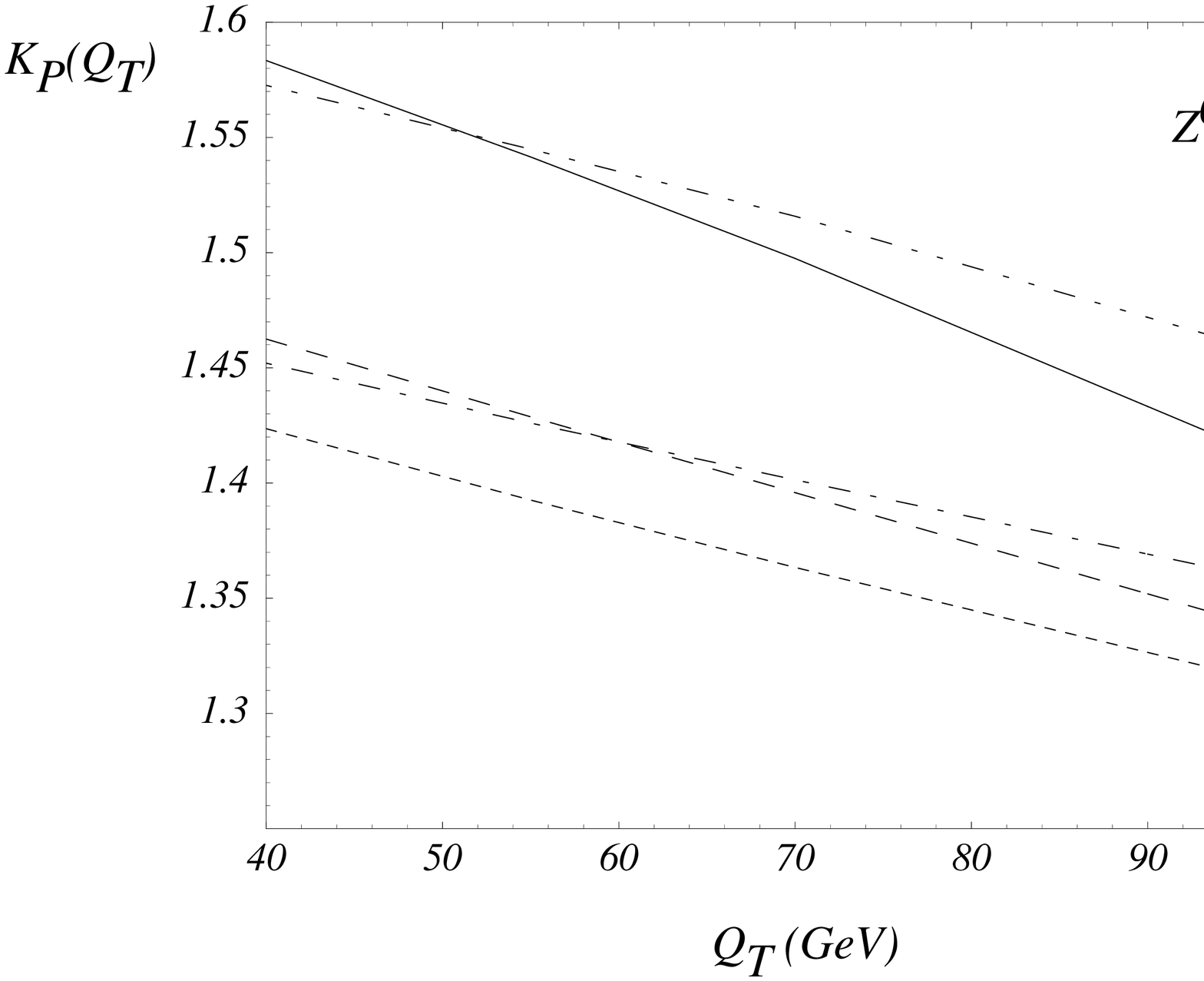} } %
\fi & 
\end{tabular}
\end{center}
\vspace{-1cm}
\caption{ The fixed order perturbative ${\cal O}(\alpha _s^2)$ to ${\cal O}%
(\alpha _s)$ $K$-factor as the function of $Q_T$. The curves are plotted for 
$Q=M_V$ and $y = -$2.0 (solid), $-$1.0 (long dash), 0.0 (short dash), 1.0
(dash-dot) and $y = $2.0 (dash-double-dot). }
\label{fig:P2toP1}
\end{figure*}

In Fig.~\ref{fig:Matching} we also show the resummed (2,1,2) (with $%
A^{(1,2)},B^{(1,2)}$, $C^{(0,1)}$, and $R^{(1,2)}$ included) and the fixed
order ${\cal O}(\alpha _s^2)$ $Q_T$ distributions. Joining these
distributions at the triple differential cross section level defines the
resummed ${\cal O}(\alpha _s^2)$ distribution in the whole $Q_T$ region. As
shown, while in ${\cal O}(\alpha _s)$ the matching takes place at lower $Q_T$
values leaving a noticeable kink in the cross section, in ${\cal O}(\alpha
_S^2)$ the matching occurs at larger $Q_T$ values and is much smoother. This
happens because, as discussed above, the cancellation between the CSS and
the asymptotic pieces becomes better.

In summary, the CSS resummation formalism is constructed in such a manner
that if the $A$, $B$, $C$ functions and the $R$ coefficients were calculated
to all order then the matching would be completely natural in the sense that
the resummed cross section would blend into the fixed order one smoothly and
no additional matching prescription would be necessary. However, in a
practical calculation, because $A$, $B$, $C$ and $Y$ are only known to some
finite order in $\alpha _s$, the matching prescription described above is
necessary. Using this procedure, we discuss below a few other interesting
results calculated from the resummation formalism.

We find that in the resummed calculation, after taking out the resonance
weighting factor $Q^2/((Q^2-M_V^2)^2+Q^4\Gamma _V^2/M_V^2)$ in Eq.~(\ref
{eq:ResFor}), the shape of the transverse momentum distribution of the
vector boson $V$ for various $Q$ values in the vicinity of $M_V$ is
remarkably constant for $Q_T$ between 0 and 20 GeV. Fixing the rapidity $y$
of the vector boson $V$ at some value $y_0$ and taking the ratio 
\begin{eqnarray*}
R(Q_T,Q_0)= {\frac {\left. \displaystyle \frac{d\sigma}{d Q^2d Q_T^2dy}%
\right| _{Q=Q_0,y=y_0}} {\left. \displaystyle \frac{d\sigma}{d Q^2d Q_T^2dy}%
\right| _{Q=M_V,y=y_0}}},
\end{eqnarray*}
we obtain almost constant curves (within 3 percent) for $Q = M_V \pm 10$ GeV
(cf. Fig.~\ref{fig:RqTQ}) for $V = W^+$ and $Z^0$. 
The fact that the shape of the transverse momentum distribution shows such a
weak dependence on the invariant mass $Q$ in the vicinity of the vector
boson mass can be used to make the Monte Carlo implementation of the
resummation calculation faster. This weak dependence was also used in the \D%
0 $W$ mass analysis when assuming that the mass dependence of the fully
differential $W$ boson production cross section factorizes as a
multiplicative term~\cite{Flattum}. 
\begin{figure*}[t]
\vspace{-.5cm}
\begin{center}
\begin{tabular}{cc}
\ifx\nopictures Y \else{ \epsfysize=6.0cm \epsffile{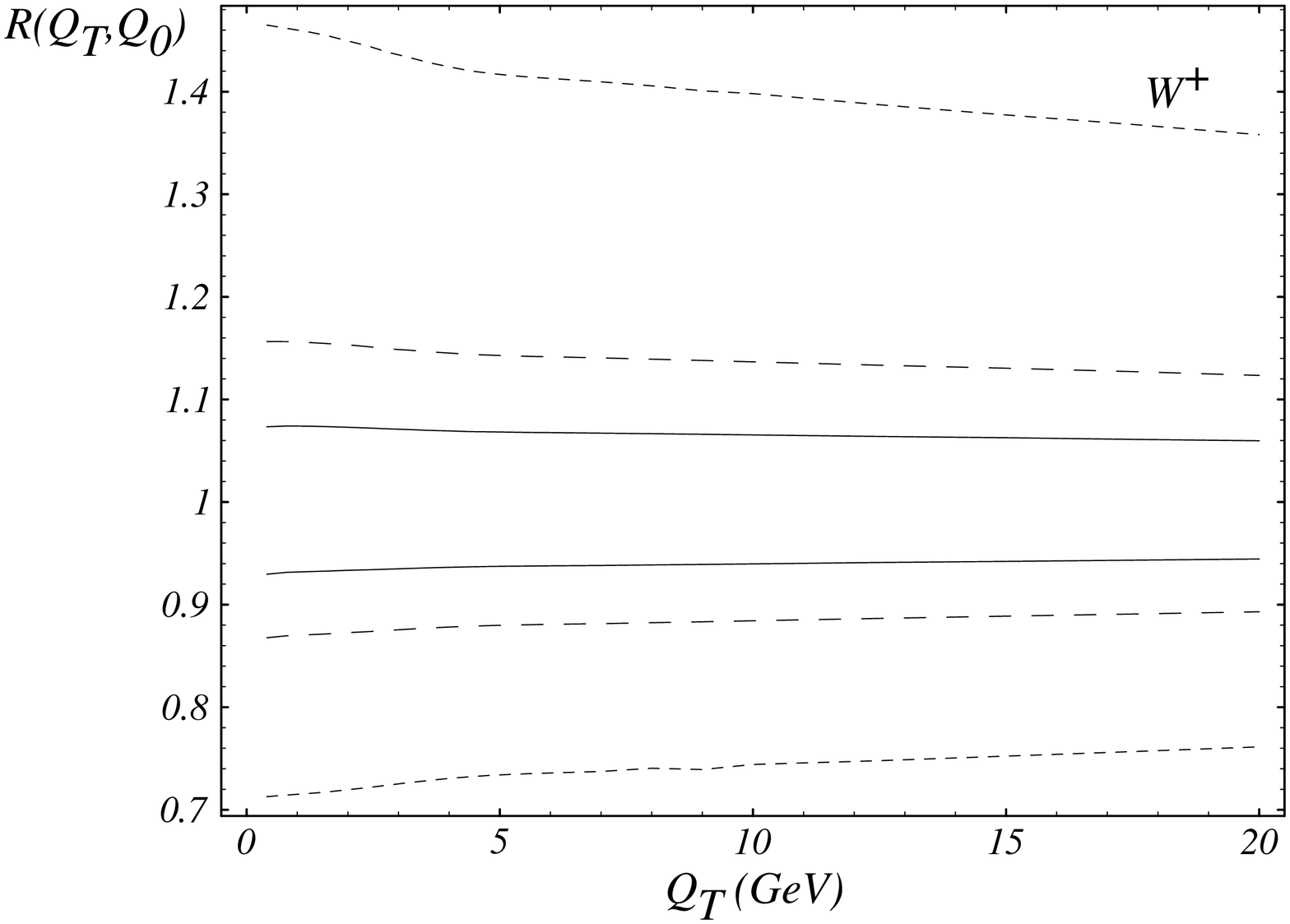}} \fi
&  \\ 
\ifx\nopictures Y \else{ \epsfysize=6.0cm \epsffile{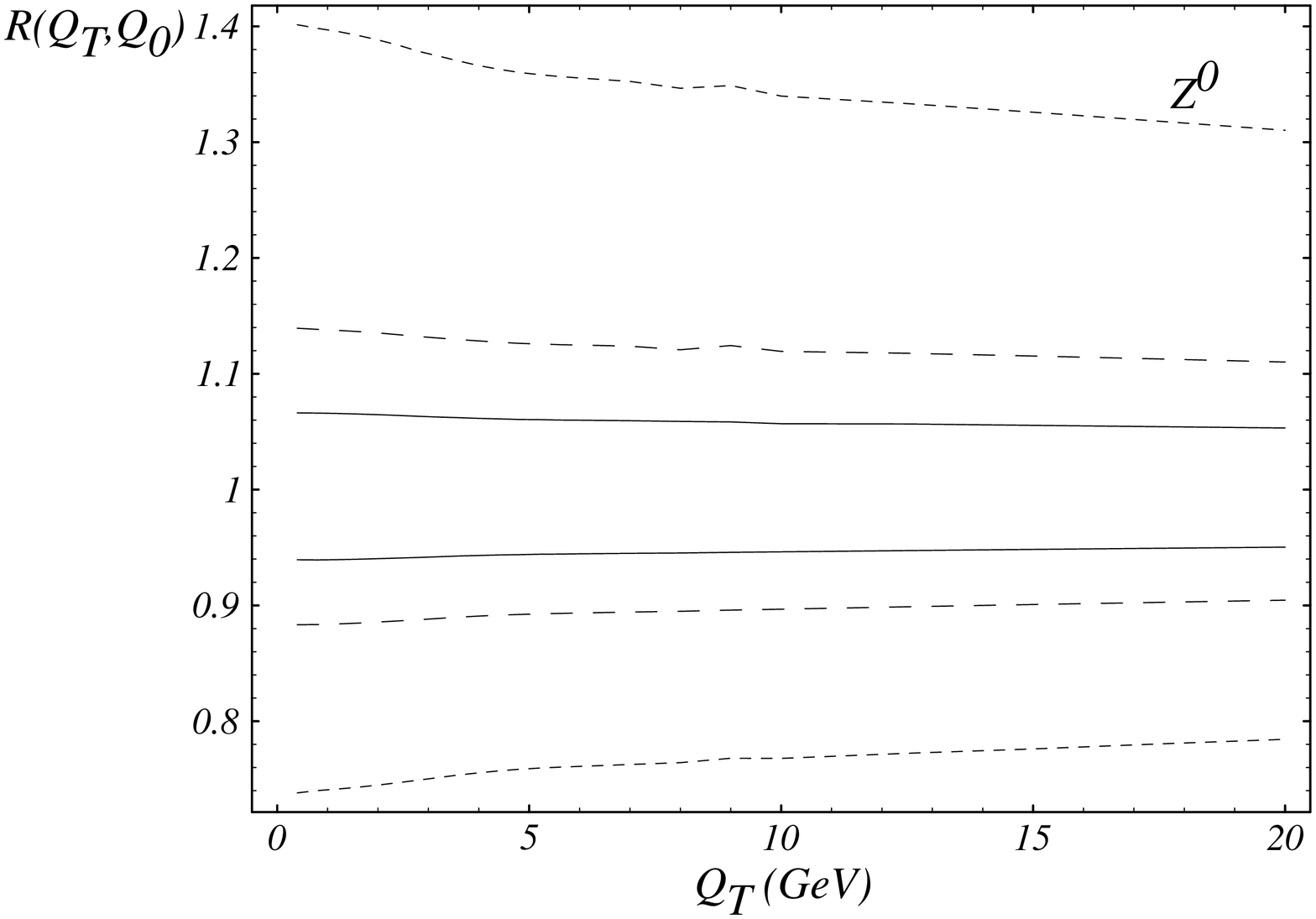}} \fi
& 
\end{tabular}
\end{center}
\vspace{-1cm}
\caption{ The ratio $R(Q_T,Q_0)$, with $y_0 = 0$, for $W^+$ and $Z^0$ bosons
as a function of $Q_T$. For $W^+$, solid lines are: $Q_0 = 78$ GeV (upper)
and 82 GeV (lower), dashed: $Q_0 = 76$ GeV (upper) and 84 GeV (lower),
dotted: $Q_0 = 70$ GeV (upper) and 90 GeV (lower). For $Z^0$ bosons, solid
lines: $Q_0 = 88$ GeV (upper) and 92 GeV (lower), dashed: $Q_0 = 86$ GeV
(upper) and 94 GeV (lower), dotted: $Q_0 = 80$ GeV (upper) and 100 GeV
(lower).}
\label{fig:RqTQ}
\end{figure*}
Similarly, we define the ratio 
\begin{eqnarray*}
R(Q_T,y_0)= {\frac{\left. \displaystyle \frac{d\sigma}{dQ^2dQ_T^2dy}\right|
_{Q=M_V,y=y_0}} {\left. \displaystyle \frac{d\sigma}{dQ^2dQ_T^2dy}\right|
_{Q=M_V,y=0}}},
\end{eqnarray*}
to study the $Q_T$ shape variation as a function of the vector boson
rapidity. Our results are shown in Fig.~\ref{fig:RqTy}. Unlike the ratio $%
R(Q_T,Q_0)$ shown in Fig.~\ref{fig:RqTQ}, the distributions of $R(Q_T,y_0)$
for the $W^\pm$ and $Z^0$ bosons are clearly different for any value of the
rapidity $y_0$.

To utilize the information on the transverse momentum of the $W^+$ boson in
Monte Carlo simulations to reconstruct the mass of the $W^+$ boson, it was
suggested in Ref.~\cite{Reno} to predict $Q_T(W^+)$ distribution from the
measured $Q_T(Z^0)$ distribution and the calculated ratio of $Q_T(W^+)$ and $%
Q_T(Z^0)$ predicted by the resummation calculations \cite
{CSS,Arnold-Kauffman}, in which the vector boson is assumed to be on its
mass-shell. 
Unfortunately, this idea will not work with a good precision because, as
clearly shown in Fig.~\ref{fig:RqTy}, the ratio of the $W^+$ and $Z^0$
transverse momentum distributions depends on the rapidities of the vector
bosons. Since the rapidity of the $W^+$ boson cannot be accurately
reconstructed without knowing the longitudinal momentum (along the beam pipe
direction) of the neutrino, which is in the form of missing energy carried
away by the neutrino, this dependence cannot be incorporated in data
analysis and the above ansatz cannot be realized in practice for a precision
measurement of $M_W$.\footnote{%
If a high precision measurement were not required, then one could choose
from the two-fold solutions for the neutrino longitudinal momentum to
calculate the longitudinal momentum of the $W^\pm$ boson.} Only the Monte
Carlo implementation of the exact matrix element calculation (included in
ResBos) can correctly predict the distributions of the decay leptons, such
as the transverse mass of the $W^\pm$ boson, and the transverse momentum of
the charged lepton, so that they can be directly compared with experimental
data to extract the value of $M_W$. We comment on these results later in
this section. 
\begin{figure*}[t]
\vspace{-.5cm}
\begin{center}
\begin{tabular}{cc}
\ifx\nopictures Y \else{ \epsfysize=6.2cm \epsffile{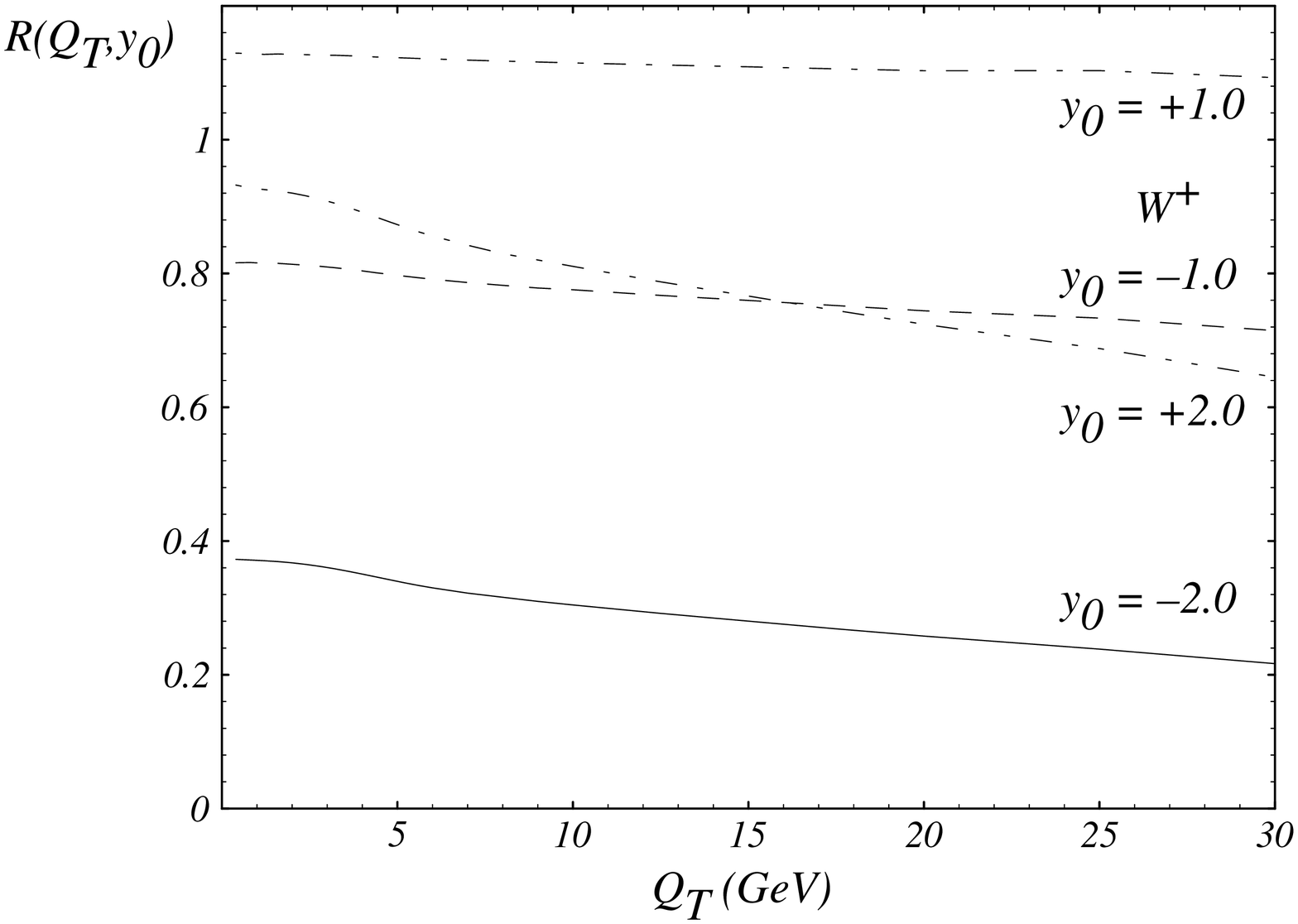}} \fi
&  \\ 
\ifx\nopictures Y \else{ \epsfysize=6.2cm \epsffile{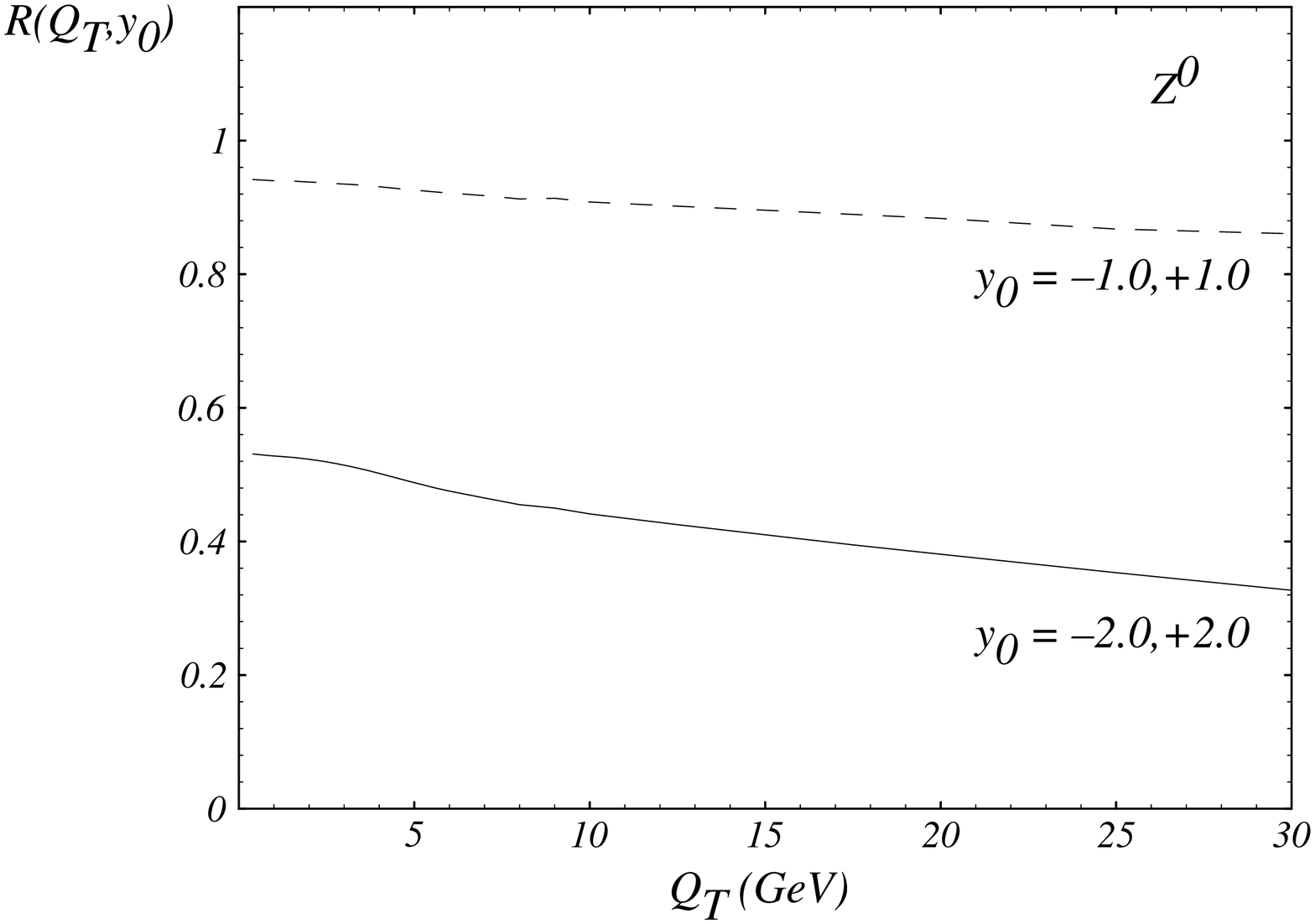}} \fi
& 
\end{tabular}
\end{center}
\vspace{-1cm}
\caption{The ratio $R(Q_T,y_0)$, with $Q_0 = M_V$, for $W^+$ and $Z^0$
bosons as a function of $Q_T$.}
\label{fig:RqTy}
\end{figure*}

Another way to compare the results of the resummed and the NLO calculations
is given by the distributions of $\sigma (Q_T>Q_T^{\min })/\sigma _{Total}$,
as shown in Fig.~\ref{fig:Integrated}. We defined the ratio as 
\[
R_{CSS} \equiv \frac{\sigma (Q_T>Q_T^{\min })}{\sigma _{Total}}=\frac
1{\sigma _{Total}}\int_{Q_T^{\min }}^{Q_T^{\max}}dQ_T\;\frac{d\sigma
(h_1h_2\rightarrow V)}{dQ_T}, 
\]
where $Q_T^{\max}$ is the largest $Q_T$ allowed by the phase space. In the
NLO calculation, $\sigma(Q_T>Q_T^{\min })$ grows without bound near $%
Q_T^{\min }=0$, as the result of the singular behavior $1/Q_T^2$ in the
matrix element. The NLO curve runs well under the resummed one in the 2 GeV $%
<Q_T^{\min }<$ 30 GeV region, and the $Q_T$ distributions from the NLO and
the resummed calculations have different shapes even in the region where $Q_T
$ is of the order 15 GeV.

With large number of fully reconstructed $Z^0$ events at the Tevatron, one
should be able to use data to discriminate these two theory calculations. In
view of this result it is not surprising that the \D0 analysis of the $%
\alpha _s$ measurement \cite{D0alphaS} based on the measurement of $\sigma
(W+1\;jet)/\sigma (W+0\;jet)$ does not support the NLO\ calculation in which
the effects of the multiple gluon radiation are not included. We expect that
if this measurement was performed by demanding the transverse momentum of
the jet to be larger than about 50 GeV, at which scale the resummed and the
NLO distributions in Fig.~\ref{fig:Matching} cross, the NLO calculation
would adequately describe the data. 
\begin{figure*}[t]
\vspace{-.5cm}
\begin{center}
\begin{tabular}{cc}
\ifx\nopictures Y \else{ \epsfysize=6.2cm \epsffile{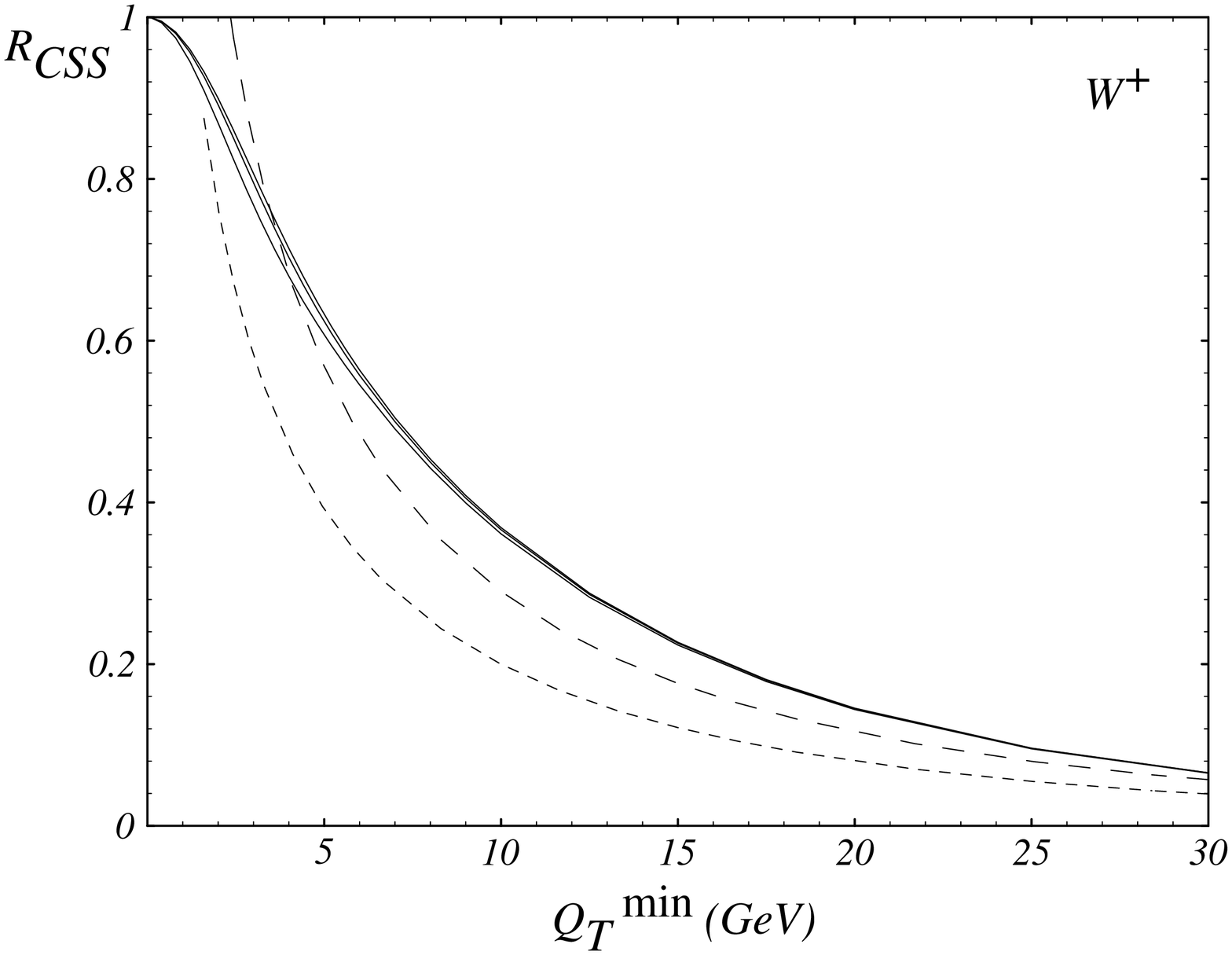}
} \fi &  \\ 
\ifx\nopictures Y \else{ \epsfysize=6.2cm \epsffile{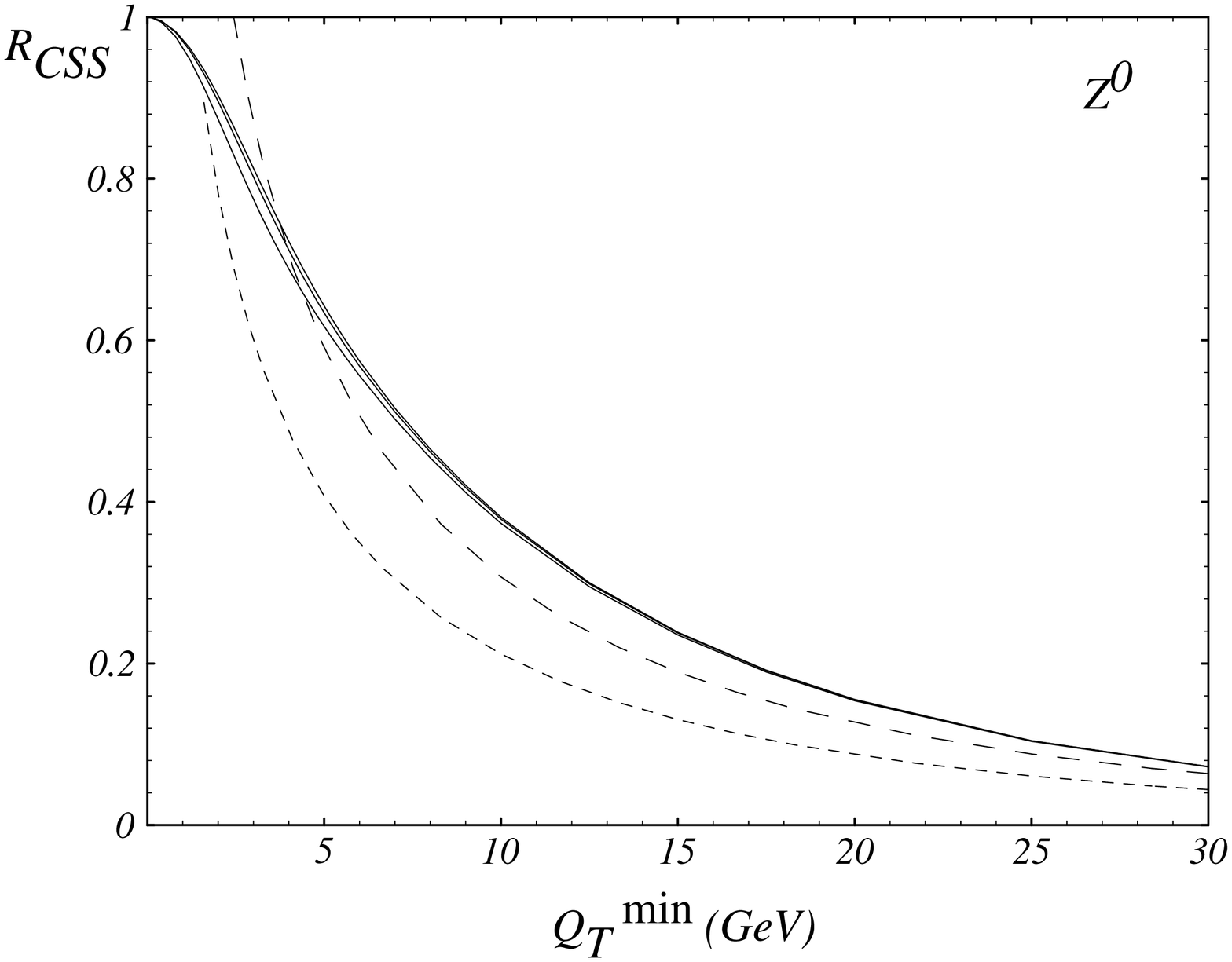}
} \fi & 
\end{tabular}
\end{center}
\vspace{-1cm}
\caption{ The ratio $R_{CSS}$ as a function of $Q_T^{\min}$ for $W^+$ and $%
Z^0$ bosons. The fixed order (${\cal O}(\alpha _s)$ short dashed, ${\cal O}%
(\alpha _s^2)$ dashed) curves are ill-defined in the low $Q_T$ region. The
resummed (solid) curves are calculated for $g_2$ = 0.38 (low), 0.58 (middle)
and 0.68 (high) GeV$^2$ values. }
\label{fig:Integrated}
\end{figure*}

To show that for $Q_T$ below 30 GeV, the QCD multiple soft gluon radiation
is important to explain the \D0 data \cite{D0alphaS}, we also include in
Fig.~\ref{fig:Integrated} the prediction for the $Q_T$ distribution at the
order of $\alpha^2_S$. As shown in the figure, the $\alpha^2_S$ curve is
closer to the resummed curve which proves that for this range of $Q_T$ the
soft gluon effect included in the $\alpha^2_S$ calculation is important for
predicting the vector boson $Q_T$ distribution. In other words, in this
range of $Q_T$, it is more likely that soft gluons accompany the $W^\pm$
boson than just a single hard jet associated with the vector boson
production. For large $Q_T$, it becomes more likely to have hard jet(s)
produced with the vector boson.

Measuring $R_{CSS}$ in the low $Q_T$ region (for $Q_T \lesim Q/2$) provides
a stringent test of the dynamics of the multiple soft gluon radiation
predicted by the QCD theory. The same measurement of $R_{CSS}$ can also
provide information about some part of the non-perturbative physics
associated with the initial state hadrons. As shown in Fig.~\ref
{fig:W_qT_W_g2} and in Ref.~\cite{Ladinsky-Yuan}, the effect of the
non-perturbative physics on the $Q_T$ distributions of the $W^\pm$ and $Z^0$
bosons produced at the Tevatron is important for $Q_T$ less than about 10
GeV. This is evident by observing that different parametrizations of the
non-perturbative functions do not change the $Q_T$ distribution for $Q_T > 10
$ GeV, although they do dramatically change the shape of $Q_T$ for $Q_T < 10$
GeV. Since for $W^\pm$ and $Z^0$ production, the $\ln(Q^2/Q^2_0)$ term is
large, the non-perturbative function, as defined in Eq.~(\ref{Eq:WNonPert}),
is dominated by the $F_1(b)$ term (or the $g_2$ parameter) which is supposed
to be universal for all Drell-Yan type processes and related to the physics
of the renormalon \cite{Korchemsky-Sterman}. Hence, the measurement of $%
R_{CSS}$ cannot only be used to test the dynamics of the QCD multiple soft
gluon radiation, in the $10 \, {\rm GeV} \, < Q_T < 40$ GeV region, but may
also be used to probe this part of non-perturbative physics for $Q_T < 10$
GeV. It is therefore important to measure $R_{CSS}$ at the Tevatron. With a
large sample of $Z^0$ data at Run 2, it is possible to determine the
dominant non-perturbative function which can then be used to calculate the $%
W^\pm$ boson $Q_T$ distribution to improve the accuracy of the $M_W$ and the
charged lepton rapidity asymmetry measurements. 
\begin{figure*}[t]
\vspace{-.5cm}
\begin{center}
\begin{tabular}{cc}
\ifx\nopictures Y \else{ \epsfysize=6.2cm \epsffile{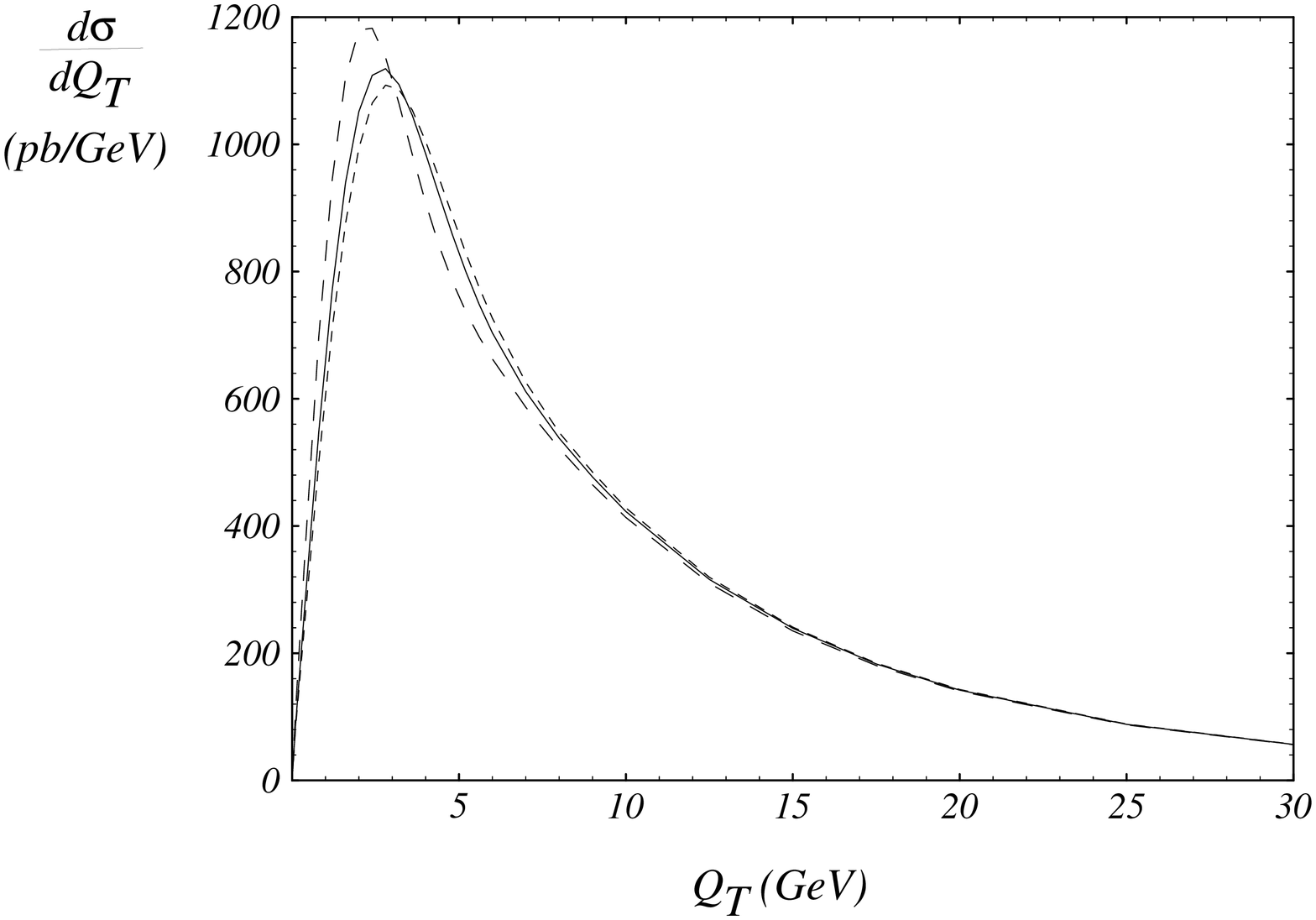} } %
\fi &  \\ 
\ifx\nopictures Y \else{ \epsfysize=6.2cm \epsffile{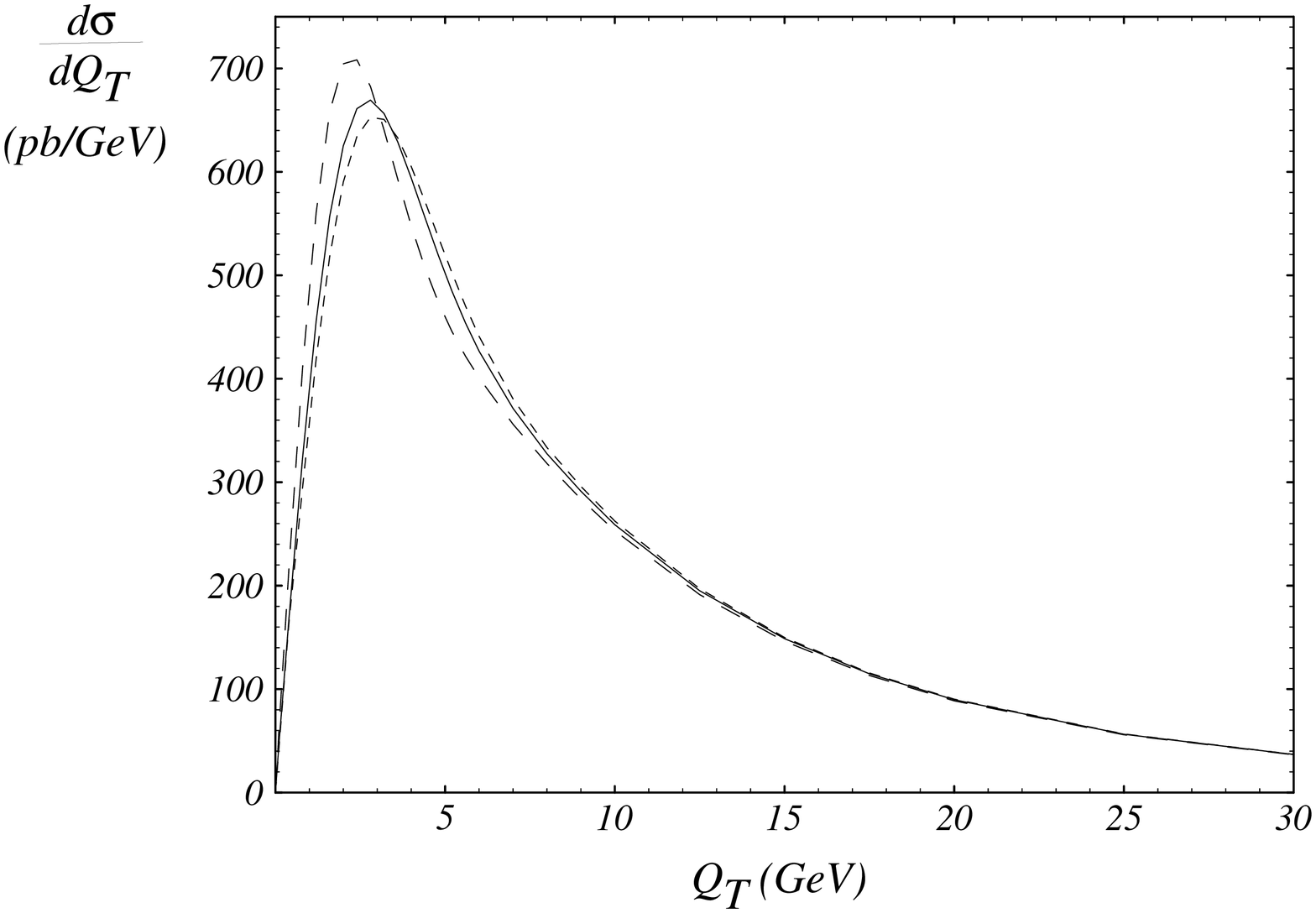} } %
\fi & 
\end{tabular}
\end{center}
\vspace{-1cm}
\caption{ Transverse momentum distributions of $W^+$ and $Z^0$ bosons
calculated with low (long dash, $g_2$ = 0.38 GeV$^2$), nominal (solid, $g_2$
= 0.58 GeV$^2$) and high (short dash, $g_2$ = 0.68 GeV$^2$) $g_2$
non-perturbative parameter values. The low and high excursions in $g_2$ are
the present one standard deviations from the nominal value in the
Ladinsky-Yuan parametrization. }
\label{fig:W_qT_W_g2}
\end{figure*}

The transverse momentum distribution of $W^\pm$ bosons within a modified 
CSS formalism was also calculated by in Ref. \cite{Ellis-Ross-Veseli}.
The conclusions and results, when they overlap, agree with ours.


\subsection{Vector Boson Longitudinal Distributions}

The resummation of the logs involving the transverse momentum of the vector
boson does not directly affect the shape of the longitudinal distributions
of the vector bosons. 
A good example of this is the distribution of the longitudinal momentum of
the $Z^0$ boson which can be measured at the Tevatron with high precision,
and can be used to extract information on the parton distributions. It is
customary to plot the rescaled quantity $x_F=2q^3/\sqrt{S}$, where $q^3 =
\sinh (y) \sqrt{Q^2 + Q_T^2}$ is the longitudinal momentum of the $Z^0$
boson measured in the laboratory frame. In Fig.~\ref{fig:Z_qZ_Z}, we plot
the distributions predicted in the resummed and the NLO calculations. As
shown, their total event rates are different in the presence of kinematic
cuts. (Although they are the same if no kinematic cuts imposed.) This
conclusion is similar to that of the $y^\ell $ distributions, as discussed
in Sections~\ref{subsec:Total} and~\ref{subsec:LCA}. 
\begin{figure*}[t]
\vspace{-.5cm}
\begin{center}
\begin{tabular}{cc}
\ifx\nopictures Y \else{ \epsfysize=8.5cm \epsffile{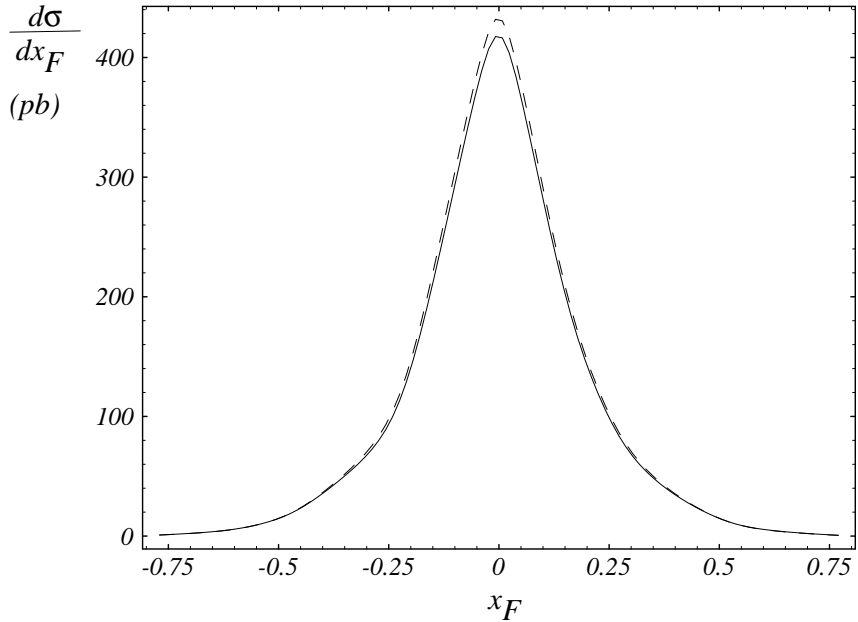} } \fi
& 
\end{tabular}
\end{center}
\vspace{-1cm}
\caption{ Longitudinal $x_F$ distributions of $Z^0$ bosons produced at the
Tevatron. The NLO (dashed) curves overestimate the rate compared to the
resummed (solid) ones, because kinematic cuts enhance the low $Q_T$ region
where the NLO and resummed distributions are qualitatively different.
Without cuts, the NLO and the resummed $x_F$ distributions are the same. }
\label{fig:Z_qZ_Z}
\end{figure*}

Without any kinematic cuts, the vector boson rapidity distributions are also
the same in the resummed and the NLO calculations. This is so because when
calculating the $y$ distribution the transverse momentum $Q_T$ is integrated
out so that the integral has the same value in the NLO and the resummed
calculations. On the other hand, experimental cuts on the final state
leptons restrict the phase space, so the difference between the NLO and the
resummed $Q_T$ distributions affects the vector boson rapidity
distributions. This shape difference is very small at the vector boson
level, as shown in Fig.~\ref{fig:Z_y_Z}. 
\begin{figure*}[t]
\vspace{-.5cm}
\begin{center}
\begin{tabular}{cc}
\ifx\nopictures Y \else{ \epsfysize=8.5cm \epsffile{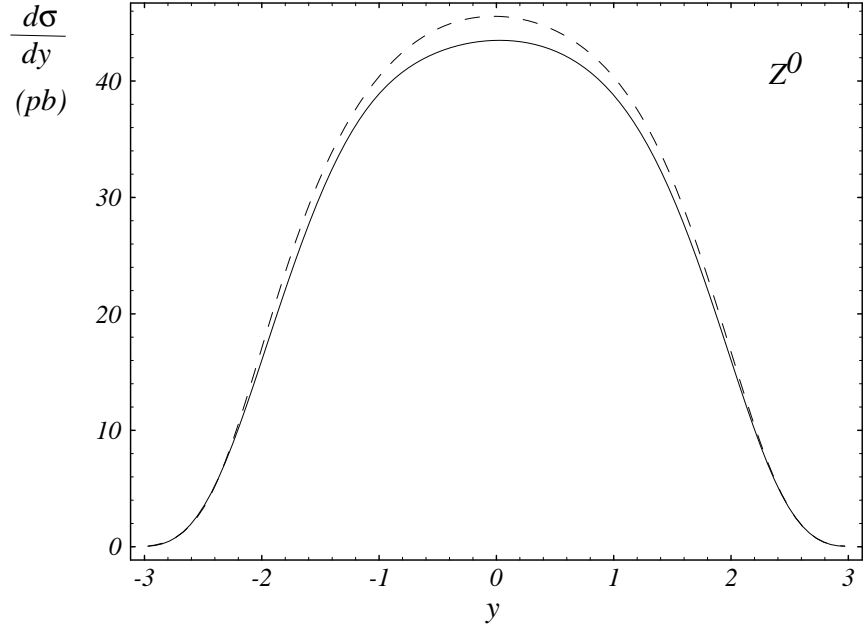} } \fi
& 
\end{tabular}
\end{center}
\vspace{-1cm}
\caption{Rapidity distributions (resummed: solid, NLO: dashed) of $Z^0$
bosons produced at the Tevatron with the kinematic cuts given in the text.}
\label{fig:Z_y_Z}
\end{figure*}

\subsection{The Total Cross Section}

\label{subsec:Total}

Before we compare the distributions of the decay leptons, we examine the
question whether or not the $Q_T$ resummation formalism changes the
prediction for the total cross section. In Ref.~\cite{AEM85} it was shown
that in the AEGM formalism, which differs from the CSS formalism, the ${\cal %
O}(\alpha _s)$ total cross section is obtained after integrating their
resummation formula over the whole range of the phase space.

In the CSS formalism, without including the $C$ and $Y$ functions, the fully
integrated resummed result recovers the ${\cal O} (\alpha _s^0)$ cross
section, provided that $Q_T$ is integrated from zero to $Q$. This can be
easily verified by expanding the resummation formula up to ${\cal O}(\alpha
_S)$, dropping the $C^{(1)}$ and the $Y$ pieces (which are of order ${\cal O}%
(\alpha _s)$), and integrating over the lepton variables. It yields 
\begin{eqnarray}
\ &&\int_0^{P_T^2}dQ_T^2\;\frac{d\sigma }{dQ^2dydQ_T^2}=\frac{\sigma _0}S
\delta(Q^2 - M_V^2)  \nonumber \\
&&\ \ \ \ \ \times \left\{ \left( 1-\frac{\alpha _s(Q)}\pi \left[ \frac 12
A^{(1)}\ln ^2\left( \frac{Q^2}{P_T^2}\right) +B^{(1)}\ln \left( \frac{Q^2}{
P_T^2}\right) \right] \right) \right. 
\nonumber \\ 
&& ~~~~~~~~~~
\left.\times \;f_{j/h_1}(x_1,Q^2)\;f_{\overline{k}/h_2}(x_2,Q^2)\right. 
\nonumber \\
&&\ \ \ \ \ -\frac{\alpha _s(Q)}{2\pi }\ln \left( \frac{Q^2}{P_T^2}\right)
\left[ \left( P_{j\leftarrow a}\otimes f_{a/h_1}\right) (x_1,Q^2)\;f_{%
\overline{k}/h_2}(x_2,Q^2) \right.  \nonumber \\
&& ~~~~~~~~~~~~~~~~~~~~~~~~ \left. \left. + \;f_{j/h_1}(x_1,Q^2) \left(
P_{\overline{k}\leftarrow b}\otimes f_{b/h_2}\right) (x_2,Q^2)\right] +
j\leftrightarrow \overline{k}\right\} ,  \label{eq:Total}
\end{eqnarray}
where $P_T$ is the upper limit of the $Q_T$ integral and we fixed the mass
of the vector boson for simplicity. To derive the above result we have used
the canonical set of the $C_i$ ($i=1,2,3$) coefficients 
(cf. Section \ref{sec:ABC}).
When the upper limit $P_T$ is taken to be $Q$, all the logs in the above
equation vanish and Eq.~(\ref{eq:Total}) reproduces the Born level (${\cal O}%
(\alpha _s^0)$) cross section. Similar conclusion holds for higher order
terms from the expansion of the resummed piece when $C$ and $Y$ are not
included. This is evident because the singular pieces from the expansion are
given by 
\begin{eqnarray*}
\left. \frac{d\sigma }{dQ_T^2}\right| _{singular}=\frac
1{Q_T^2}\sum_{n=1}^\infty \sum_{m=0}^{2n-1}{}_nv_m\alpha _s^n\ln ^m\left( 
\frac{Q_T^2} {Q^2}\right)
\end{eqnarray*}
The integral of these singular terms will be proportional to $\ln
(Q^2/P_T^2) $ raised to some power. Again, for $P_T=Q$ all the logs vanish
and the tree level result is obtained.

Including $C^{(1)}$ and the $Y$ contribution changes the above conclusion
and leads to a different total cross section, because $C^{(1)}$ contains the
hard part virtual corrections and $Y$ contains the hard gluon radiation.
Combining the resummed (1,1,1) and the fixed order ${\cal O}(\alpha _s)$
distributions, the above described matching prescription provides us with an 
${\cal O}(\alpha _s)$ resummed total cross section with an error of ${\cal O}%
(\alpha_s^2)$, as shown in Ref.~\cite{Arnold-Kauffman}. In practice this
translates into less than a percent deviation between the resummed ${\cal O}%
(\alpha _s)$ total cross section and the inclusive NLO (${\cal O}(\alpha _s)$%
) calculation. This can be understood from the earlier discussion that if
the matching were done at $Q_T$ equal to $Q$, then the total cross section
calculated from the CSS resummation formalism should be the same as that
predicted by the NLO calculation, provided that $C^{(1)}$ and $Y^{(1)}$ are
included. However, this matching prescription would not result in a smooth
curve for the $Q_T$ distribution at $Q_T=Q$. The matching procedure
described above causes a small (about a percent) difference between the $%
{\cal O}(\alpha _s)$ resummed and the NLO total cross sections. This
difference indicates the typical size of the higher order corrections not
included in the NLO total cross section calculation.

The total cross section predicted from the various theory calculations are
listed in Table \ref{tbl:Total}. 
\begin{table}[tbp]
\begin{center}
\begin{tabular}{l c cc c c c r}
\\ \hline \hline \\[-.2cm]
& $E_{cm}$ & \multicolumn{2}{c}{Fixed Order} & \multicolumn{3}{c}{Resummed} & 
Experiment \\
$$ & (TeV) & ${\cal O}(\alpha _s^0)$ & ${{\cal O} (\alpha _s)}$ &
(1,1,1) & (2,1,1) & (2,1,2) & (Ref.~\cite{CDFTotal}) \\ 
&  &  &  & $\oplus~
{\cal O}(\alpha _s )$ & $\oplus~{\cal O}(\alpha _s )$ & $%
\oplus~{\cal O}(\alpha _s^2)$ &  
\\ \hline \\[-.2cm]
W$^{+}$ & 1.8 & 8.81 & 11.1 & 11.3 & 11.3 & 11.4 & 11.5 $\pm $ 0.7 \\ 
W$^{+}$ & 2.0 & 9.71 & 12.5 & 12.6 & 12.6 & 12.7 &  \\ 
Z$^0$ & 1.8 & 5.23 & 6.69 & 6.79 & 6.79 & 6.82 & 6.86 $\pm $ 0.36 \\ 
Z$^0$ & 2.0 & 6.11 & 7.47 & 7.52 & 7.52 & 7.57 & 
\\ \hline \hline
\end{tabular}
\end{center}
\caption{Total cross sections of $p {\bar p} \rightarrow (W^+ {\rm or}~ Z^0)
X$ at the present and upgraded Tevatron, calculated in different
prescriptions, in units of nb. The finite order total cross section results
are based on the calculations in Ref.~\cite{AEM85}. 
The ${\cal O}(\alpha _s^2)$ results were obtained from Ref.~\cite{Arnold-Reno}. 
The ``$\oplus$'' signs refer to the
matching prescription discussed in the text.}
\label{tbl:Total}
\end{table}
As shown in Table~\ref{tbl:Total}, as far as the total rate is concerned,
there is hardly any observable difference between the predicted results from
the resummed (2,1,1) matched to the fixed order perturbative ${\cal O}%
(\alpha _s)$ and the resummed (2,1,2) matched to the fixed order
perturbative ${\cal O}(\alpha _s^2)$, although the latter gives a smoother $%
Q_T$ curve, as shown in Fig.~\ref{fig:Matching}.

Kinematic cuts affect the total cross section in a subtle manner. It is
obvious from our matching prescription that the resummed ${\cal O}(\alpha
_S^2)$ and the fixed order ${\cal O}(\alpha _s)$ curves in Fig.~\ref
{fig:Matching} will never cross in the high $Q_T$ region. On the other hand,
the resummed ${\cal O}(\alpha _s^2)$ total cross section is about the same
as the fixed order ${\cal O}(\alpha _s)$ cross section when integrating $Q_T$
from 0 to $Q$. These two facts imply that when kinematic cuts are made on
the $Q_T$ distribution with $Q_T < Q$, we will obtain a higher total cross
section in the fixed order ${\cal O}(\alpha _s)$ than in the resummed ${\cal %
O}(\alpha _s^2)$ calculation. In this Chapter we follow the CDF cuts (for the 
$W^+$ boson mass analysis) and demand $Q_T < 30$ GeV~\cite{CDFMW}.
Consequently, in many of our figures, to be shown below, the fixed order $%
{\cal O}(\alpha _s)$ curves give about 3\% higher total cross section than
the resummed ones.


\section{Lepton Distributions}

\label{sec:LeptonDistributions}

\subsection{Lepton Charge Asymmetry}

\label{subsec:LCA}

The CDF lepton charge asymmetry measurement~\cite{CDFLCA} played a crucial
role in constraining the slope of the $u/d$ ratio in recent parton
distribution functions. It was shown that one of the largest theoretical
uncertainty in the $W^\pm$ mass measurement comes from the parton
distributions~\cite{TEV2000}, and the lepton charge asymmetry was shown to
be correlated with the transverse mass distribution~\cite{Stirling-Martin}.
Among others, the lepton charge asymmetry is studied to decrease the errors
in the measurement of $M_W$ coming from the parton distributions. Here we
investigate the effect of the resummation on the lepton rapidity
distribution, although it is not one of those observables which are most
sensitive to the $Q_T$ resummation, i.e. to the effect of multiple soft
gluon radiation.

The definition of the charge asymmetry is 
\begin{eqnarray*}
A(y)=\frac{\displaystyle \frac{d\sigma}{dy_+}-\frac{d\sigma}{dy_-}} {%
\displaystyle \frac{d\sigma}{dy_+}+\frac{d\sigma}{dy_-}},
\end{eqnarray*}
where $y_{+}$ ($y_{-}$) is the rapidity of the positively (negatively)
charged particle (either vector boson or decay lepton). Assuming CP
invariance,\footnote{%
Here we ignore the small CP violating effect due to the CKM matrix elements
in the SM.} the following relation holds: 
\begin{eqnarray*}
{\frac{d\sigma }{dy_{+}}}(y)={\frac{d\sigma }{dy_{-}}}(-y).
\end{eqnarray*}
Hence, the charge asymmetry is frequently written as 
\begin{eqnarray*}
A(y)=\frac{\displaystyle \frac{d\sigma}{dy}(y>0)-\frac{d\sigma}{dy}(y<0)} {%
\displaystyle \frac{d\sigma}{dy}(y>0)+\frac{d\sigma}{dy}(y<0)}.
\end{eqnarray*}
For the charge asymmetry of the vector boson ($W^{\pm }$) or the charged
decay lepton ($\ell ^{\pm }$), the fixed order and the resummed ${\cal O}%
(\alpha _s)$ [or ${\cal O}(\alpha _s^2)$] results are the same, provided
that there are no kinematic cuts imposed. This is because the shape
difference in the vector boson transverse momentum has been integrated out
and the total cross sections are the same up to higher order corrections in $%
\alpha _s$. In Fig.~\ref{fig:W_y_e_Asy}(a) we show the lepton charge
asymmetry without cuts for CTEQ4M PDF. The NLO and the resummed curves
overlap, although they differ from the ${\cal O}(\alpha _s^0)$ prediction. 
\begin{figure}[t]
\epsfysize=6.0cm \centerline{\epsfbox{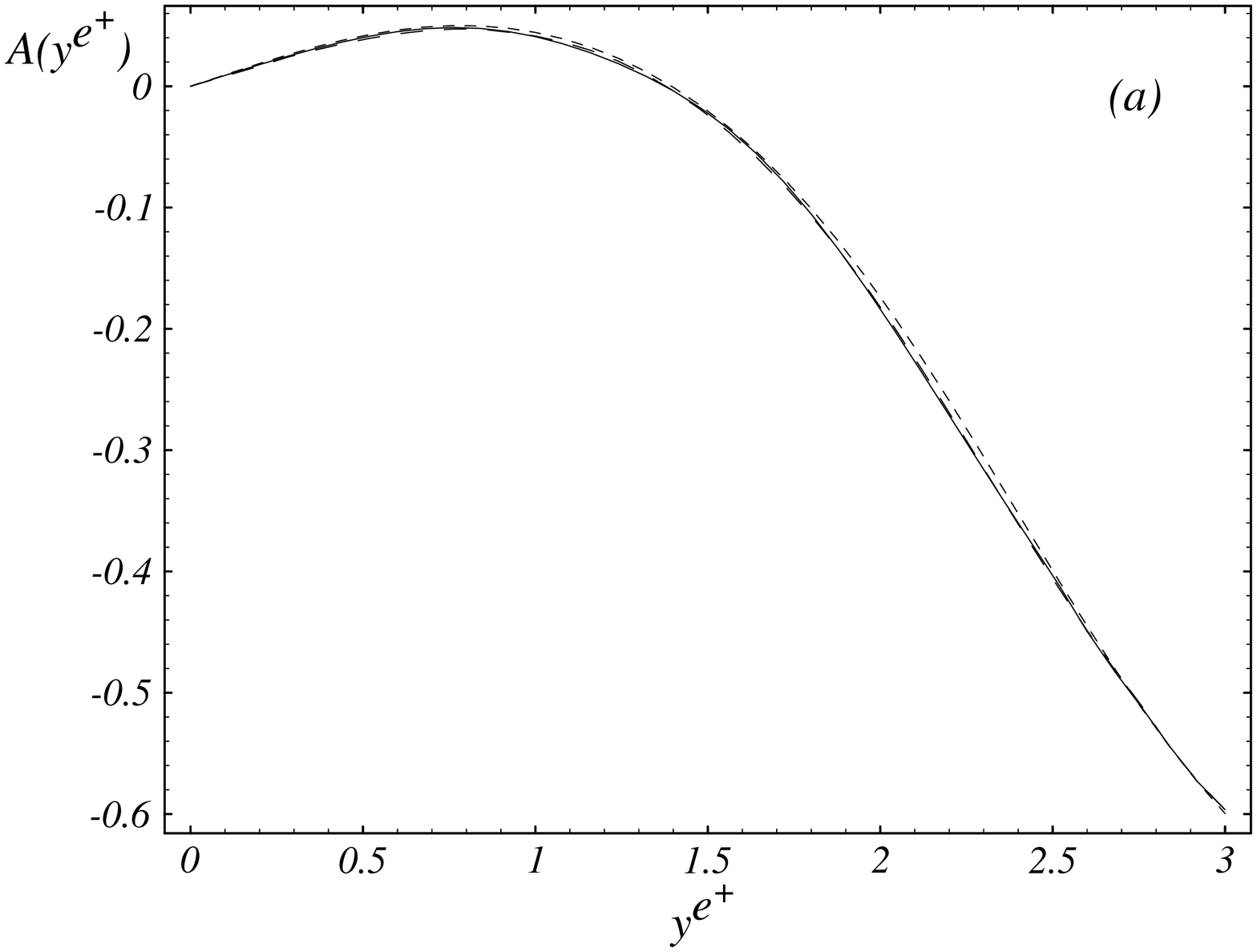}}
\epsfysize=6.0cm \centerline{\epsfbox{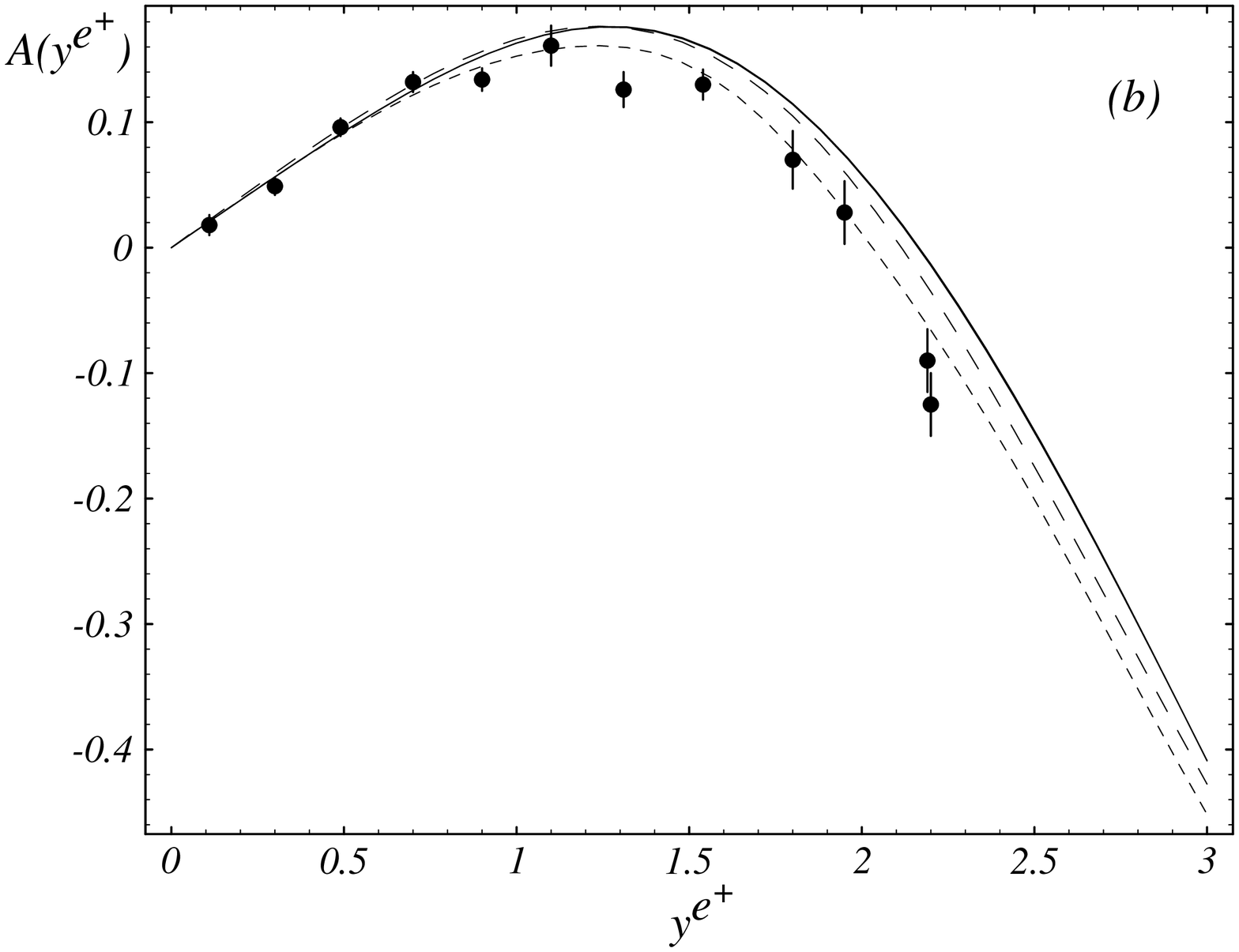}}
\caption{Lepton charge asymmetry distributions.
(a) Without any kinematic
cuts, the NLO (long dashed) and the resummed ${\cal O}(\alpha _s)$ (solid)
curves overlap and the LO (short dashed) curve differs somewhat from them.
(b) With cuts ($Q_T<30$ GeV, $p_T^{e,\nu}>25$ GeV), the effect
of the different $Q_T$ distributions renders the lepton rapidity asymmetry
distributions different. The two resummed curves calculated with $g_2$ =
0.58 and 0.78 GeV$^2$ cannot be distinguished on this plot.
}
\label{fig:W_y_e_Asy}
\end{figure}

On the other hand, when kinematic cuts are applied to the decay leptons, the
rapidity distributions of the vector bosons or the leptons in the fixed
order and the resummed calculations are different. Restriction of the phase
space implies that only part of the vector boson transverse momentum
distribution is sampled. The difference in the resummed and the fixed order $%
Q_T$ distributions will prevail as a difference in the rapidity
distributions of the charged lepton. We can view this phenomenon in a
different (a Monte Carlo) way. In the rest frame of the $W^\pm$, the decay
kinematics is the same, whether it is calculated up to ${\cal O}(\alpha _s)$
or within the resummation formalism. On the other hand, the $W^\pm$ rest
frame is different for each individual Monte Carlo event depending on the
order of the calculation. This difference is caused by the fact that the $Q_T
$ distribution of the $W^\pm$ is different in the ${\cal O}(\alpha _s)$ and
the resummed calculations, and the kinematic cuts select different events in
these two cases. Hence, even though the $Q_T$ distribution of the $W^\pm$ is
integrated out, when calculating the lepton rapidity distribution, we obtain
slightly different predictions in the two calculations. 
The difference is larger for larger $|y^{\ell}|$, being closer to the edge
of the phase space, because the soft gluon radiation gives high corrections
there and this effect up to all order in $\alpha _s$ is contained in the
resummed but only up to order of $\alpha _s$ in the NLO calculation. Because
the rapidity of the lepton and that of the vector boson are highly
correlated, large rapidity leptons mostly come from large rapidity vector
bosons. Also, a vector boson with large rapidity tends to have low
transverse momentum, because the available phase space is limited to low $Q_T
$ for a $W^\pm$ boson with large $|y|$. Hence, the difference in the low $Q_T
$ distributions of the NLO and the resummed calculations yields the
difference in the $y^{\ell}$ distribution for leptons with high rapidities.

Asymmetry distributions of the charged lepton with cuts using the CTEQ4M PDF
are shown in Fig.~\ref{fig:W_y_e_Asy}(b). The applied kinematic cuts are: $%
Q_T<30$~GeV, $p_T^{e^+,\nu}>25$. These are the cuts that CDF used when
extracted the lepton rapidity distribution from their data~\cite{CDFLCA}. We
have checked that the ResBos fixed order ${\cal O}(\alpha _s)$ curve agrees
well with the DYRAD~\cite{Giele} result. As anticipated, the ${\cal O}%
(\alpha _s^0)$, ${\cal O}(\alpha _s)$ and resummed results deviate at higher
rapidities ($|y^e|>1.5$).\footnote{%
As indicated before, here and henceforth, unless specified otherwise, by a
resummed calculation we mean our resummed ${\cal O}(\alpha _s^2)$ result.}
The deviation between the NLO and the resummed curves indicates that to
extract information on the PDF in the large rapidity region, the resummed
calculation, in principle, has to be used if the precision of the data is
high enough to distinguish these predictions. Fig.~\ref{fig:W_y_e_Asy}(b)
also shows the negligible dependence of the resummed curves on the
non-perturbative parameter $g_2$. We plot the result of the resummed
calculations with the nominal $g_2$ = 0.58 GeV$^2$, and with $g_2$ = 0.78 GeV%
$^2$ which is two standard deviations higher. The deviation between these
two curves (which is hardly observable on the figure) is much smaller than
the deviation between the resummed and the NLO ones.

There is yet another reason why the lepton charge asymmetry can be reliably
predicted only by the resummed calculation. When calculating the lepton
distributions in a numerical ${\cal O}(\alpha _s)$ code, one has to
artificially divide the vector boson phase space into hard and soft regions,
depending on -- for example -- the energy or the $Q_T$ of the emitted gluon
(e.g. $q\overline{q} \rightarrow W+\;$hard or soft gluon). The observables
calculated with this phase space slicing technique acquire a dependence on
the scale which separates the hard from the soft regions. For example, when
the phase space is divided by the $Q_T$ separation, the dependence of the
asymmetry on the scale $Q_T^{Sep}$ can be comparable to the difference in
the ${\cal O}(\alpha _s)$ and the resummed results. This means that there is
no definite prediction from the NLO calculation for the lepton rapidity
distribution. Only the resummed calculation can give an unambiguous
prediction for the lepton charge asymmetry.

Before closing this section, we also note that although in the lepton
asymmetry distribution the NLO and resummed results are about the same for $%
|y^{e^+}| < 1$, it does not imply that the rapidity distributions of the
leptons predicted by those two theory models are the same. As shown in Fig.~%
\ref{fig:W_y_e}, this difference can in principle be observable with a large
statistics data sample and a good knowledge of the luminosity of the
colliding beams. 
\begin{figure*}[t]
\vspace{-.5cm}
\begin{center}
\begin{tabular}{cc}
\ifx\nopictures Y \else{ \epsfysize=8.5cm \epsffile{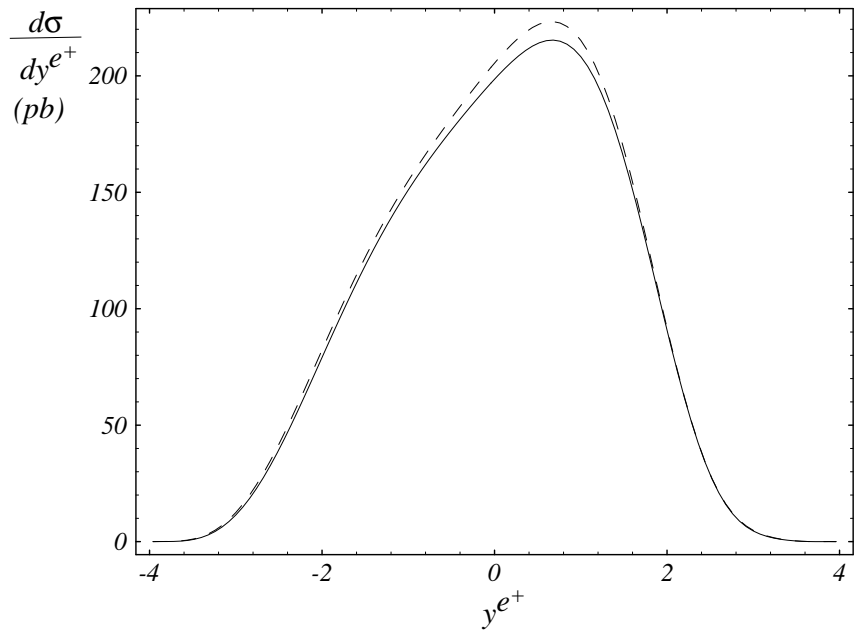} } \fi
& 
\end{tabular}
\end{center}
\vspace{-1cm}
\caption{Distributions of positron rapidities from the decays of $W^+$'s
produced at the Tevatron, predicted by the resummed (solid) and the NLO
(dashed) calculations with the same kinematic cuts as for the asymmetry plot.
}
\label{fig:W_y_e}
\end{figure*}


\subsection{Transverse Mass Distribution}


Since the invariant mass of the $W^{\pm }$ boson cannot be reconstructed
without knowing the longitudinal momentum of the neutrino, one has to find a
quantity that allows an indirect determination of the mass of the $W^\pm$
boson. In the discovery stage of the $W^{\pm }$ bosons at the CERN\ ${\rm S
p \overline{p} S}$ collider, the mass and width were measured using the
transverse mass distribution of the charged lepton-neutrino pair from the $%
W^\pm$ boson decay. Ever since the early eighties, the transverse mass
distribution, $m_T = \sqrt{2p_T^{e}p_T^\nu (1-\cos \Delta \phi_{e \nu })}$,
has been known as the best measurable for the extraction of both $M_W$ and $%
\Gamma _W$, for it is insensitive to the transverse momentum of the $W^\pm$
boson. The effect of the non-vanishing vector boson transverse momentum on
the $m_T$ distribution was analyzed ~\cite{Barger,Smith} well before the $Q_T
$ distribution of the $W^\pm$ boson was correctly calculated by taking into
account the multiple soft gluon radiation. Giving an average transverse
boost to the vector boson, the authors of Ref.~\cite{Barger} concluded that
for the fictive case of $\Gamma _W$ = 0, the end points of the transverse
mass distribution are fixed at zero: $d\sigma /dm_T^2(m_T^2=0)=d\sigma
/dm_T^2(m_T^2=M_W^2)=0$. The sensitivity of the $m_T$ shape to a non-zero $%
Q_T$ is in the order of $\left\langle (Q_T/M_W)^2\right\rangle \approx $1\%
without affecting the end points of the $m_T$ distribution. 
Including the effect of the finite width of the $W^\pm$ boson, the authors
in Ref.~\cite{Smith} showed that the shape and the location of the Jacobian
peak are not sensitive to the $Q_T$ of the $W^\pm$ boson either. The
non-vanishing transverse momentum of the $W^{\pm }$ boson only significantly
modifies the $m_T$ distribution around $m_T=0$. 

Our results confirm that the shape of the Jacobian peak is quite insensitive
to the order of the calculation. We show the NLO and the resummed transverse
mass distributions in Fig.~\ref{fig:W_mt} for $W^\pm$ bosons produced at the
Tevatron with the kinematic cuts: $Q_T < 30$ GeV, $p_T^{e^+,\nu} > 25$ GeV
and $|y^{e^+}|<3.0$. 
Fig.~\ref{fig:W_mt}(a) covers the full (experimentally interesting) $m_T$
range while Fig.~\ref{fig:W_mt}(b) focuses on the $m_T$ range which contains
most of the information about the $W^\pm$ mass. There is little visible
difference between the $shapes$ of the NLO and the resummed $m_T$
distributions. On the other hand, the right shoulder of the curve appears to
be ``shifted'' by about 50 MeV, because, as noted in Section~\ref
{subsec:Total}, the total cross sections are different after the above cuts
imposed in the NLO and the resummed calculations. At Run 2 of the Tevatron,
with large integrated luminosity ($\sim 2~{\rm fb}^{-1}$), the goal is to
extract the $W^\pm$ boson mass with a precision of 30-50 MeV from the $m_T$
distribution \cite{TEV2000}. 
Since $M_W$ is sensitive to the position of the Jacobian peak \cite{Smith},
the high precision measurement of the $W^\pm$ mass has to rely on the
resummed calculations. 
\begin{figure*}[t]
\vspace{-.5cm}
\begin{center}
\begin{tabular}{cc}
\ifx\nopictures Y \else{ \epsfysize=6.2cm \epsffile{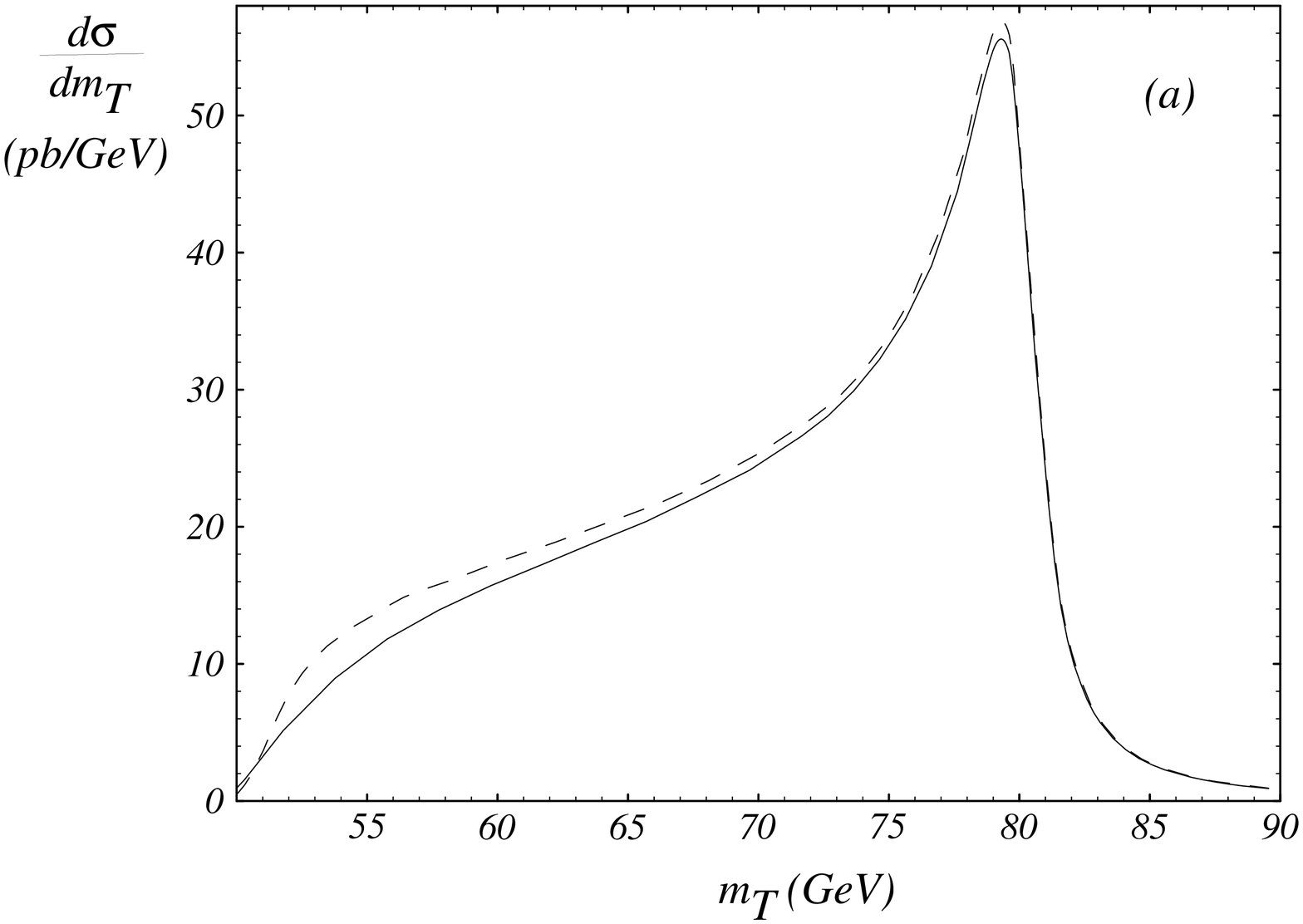} } \fi
&  \\ 
\ifx\nopictures Y \else{ \epsfysize=6.2cm \epsffile{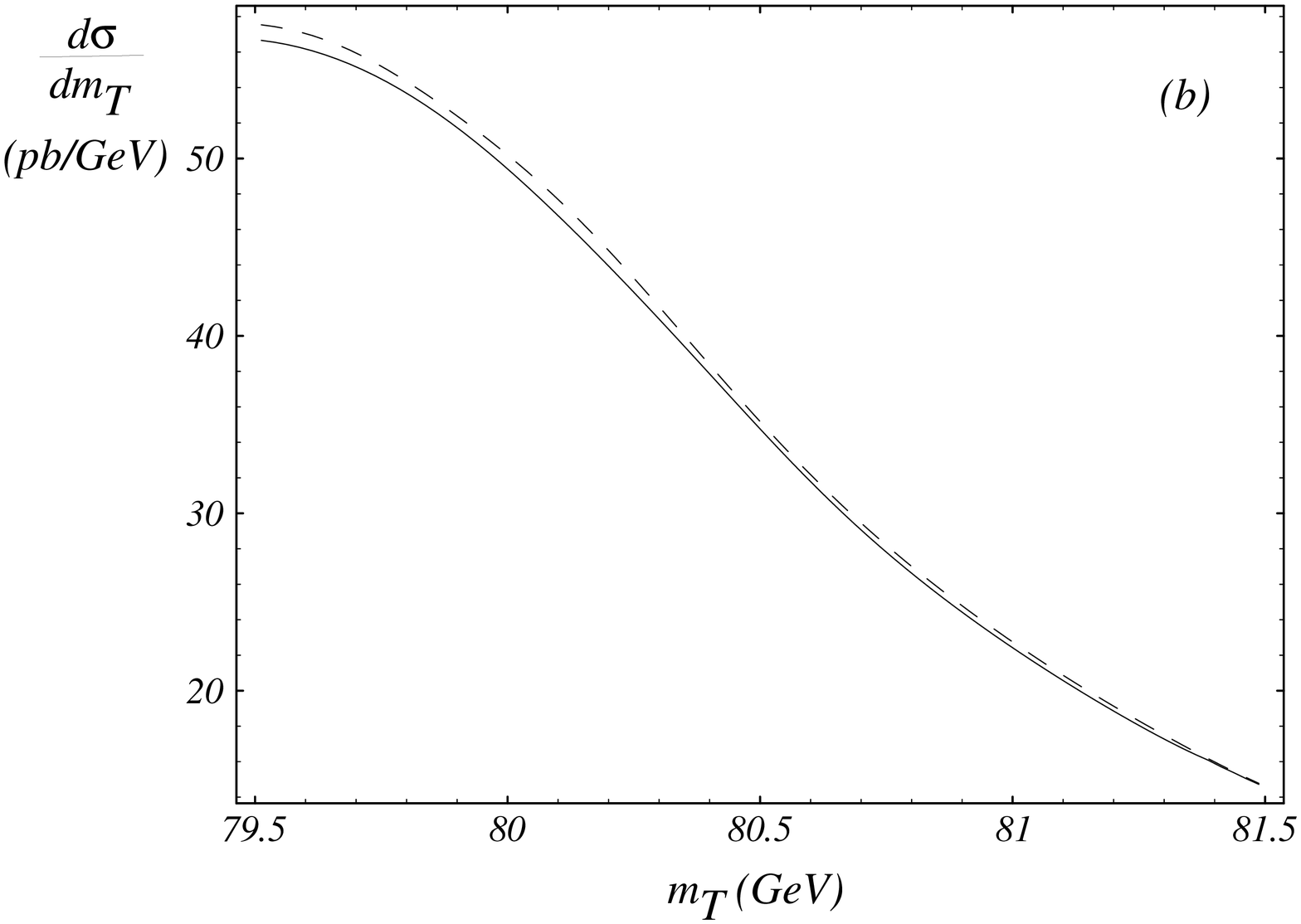} } \fi
& 
\end{tabular}
\end{center}
\vspace{-1cm}
\caption{ Transverse mass distribution for $W^{+}$ production and decay at
the 1.8 TeV Tevatron. }
\label{fig:W_mt}
\end{figure*}

The extraction of $M_W$ from the transverse mass distribution has some
drawbacks. The reconstruction of the transverse momentum $p_T^\nu $ of the
neutrino involves the measurement of the underlying event transverse
momentum: $\vec{p}_T^{~\nu }=-\vec{p}_T^{~\ell }-\vec{p}_T^{~recoil}-\vec{p}%
_T^{~underlying~event}$. This resolution degrades by the number of
interactions per crossing ($N_{I_c}$)~\cite{TEV2000}. 
With a high luminosity ($\sim 100\,{\rm fb}^{-1}$) at the 2 TeV Tevatron
(TEV33)~\cite{TEV33}, $N_{I_c}$ can be as large as 10, so that the Jacobian
peak is badly smeared. 
This will lead to a large uncertainty in the measurement of $M_W$. For this
reason the systematic precision of the $m_T$ reconstruction will be less at
the high luminosity Tevatron, and an $M_W$ measurement that relies on the
lepton transverse momentum distribution alone could be more promising. We
discuss this further in the next section. 

The theoretical limitation on the $M_W$ measurement using the $m_T$
distribution comes from the dependence on the non-perturbative sector, i.e.
from the PDF's and the non-perturbative parameters in the resummed
formalism. Assuming the PDF's and these non-perturbative parameters to be
independent variables, the uncertainties introduced are estimated to be less
than 50 MeV and 10 MeV, respectively, at the TEV33 \cite{Flattum,SM96EW}. It
is clear that the main theoretical uncertainty comes from the PDF's. As to
the uncertainty due to the non-perturbative parameters (e.g. $g_2$) in the
CSS resummation formalism, it can be greatly reduced by carefully study the $%
Q_T$ distribution of the $Z^0$ boson which is expected to be copiously
produced at Run 2 and beyond.

The $M_W$ measurement at the LHC may also be promising. Both ATLAS and CMS
detectors are well optimized for measuring the leptons and the missing $E_T$~%
\cite{SM96EW}. The cross section of the $W^+$ boson production is about four
times larger than that at the Tevatron, and in one year of running with 20 fb%
$^{-1}$ luminosity yields a few times $10^7$ $W\rightarrow \ell \nu $ events
after imposing similar cuts to those made at the Tevatron. 
Since the number of interactions per crossing may be significantly lower (in
average $N_{I_c}$ = 2) at the same or higher luminosity than that at the
TEV33~\cite{SM96EW}, the Jacobian peak in the $m_T$ distribution will be
less smeared at the LHC than at the TEV33. 
Furthermore, the non-perturbative effects are relatively smaller at the LHC
because the perturbative Sudakov factor dominates. On the other hand, the
probed region of the PDF's at the LHC has a lower value of the average $%
x~(\sim 10^{-3})$ than that at the Tevatron $(\sim 10^{-2})$, hence the
uncertainty from the PDF's might be somewhat larger. A more detailed study
of this subject is desirable.

\subsection{Lepton Transverse Momentum}

Due to the limitations mentioned above, the transverse mass method may not
be the only and the most promising way for the precision measurement of $M_W$
at some future hadron colliders. As discussed above, the observable $m_T$
was used because of its insensitivity to the high order QCD corrections. In
contrast, the lepton transverse momentum ($p_T^\ell$) distribution receives
a large, ${\cal O}\left( \left\langle Q_T/M_W\right\rangle \right) \sim 10\%$%
, perturbative QCD correction at the order $\alpha _s$, as compared to the
Born process. With the resummed results in hand it becomes possible to
calculate the $p_T^\ell $ distribution precisely within the perturbative
framework, and to extract the $W^\pm$ mass straightly from the transverse
momentum distributions of the decay leptons.

Just like in the $m_T$ distribution, the mass of the $W^\pm$ boson is mainly
determined by the shape of the distribution near the Jacobian peak. The
location of the maximum of the peak is directly related to the $W^\pm$ boson
mass, while the theoretical width of the peak varies with its decay width $%
\Gamma _W$. Since the Jacobian peak is modified by effects of both $Q_T$ and 
$\Gamma _W$, it is important to take into account both of these effects
correctly. In our calculation (and in ResBos) we have properly included both
effects. 

The effect of resummation on the transverse momentum distribution of the
charged lepton from $W^{+}$ and $Z^0$ decays is shown in Fig.~\ref
{fig:V_pT_e}. The NLO and the resummed distributions differ a great amount
even without imposing any kinematic cuts. The clear and sharp Jacobian peak
of the NLO distribution is strongly smeared by the finite transverse
momentum of the vector boson introduced by multiple gluon radiation. This
higher order effect cannot be correctly calculated in any finite order of
the perturbation theory and the resummation formalism has to be used. 
\begin{figure*}[t]
\vspace{-.5cm}
\begin{center}
\begin{tabular}{cc}
\ifx\nopictures Y \else{ \epsfysize=6.2cm \epsffile{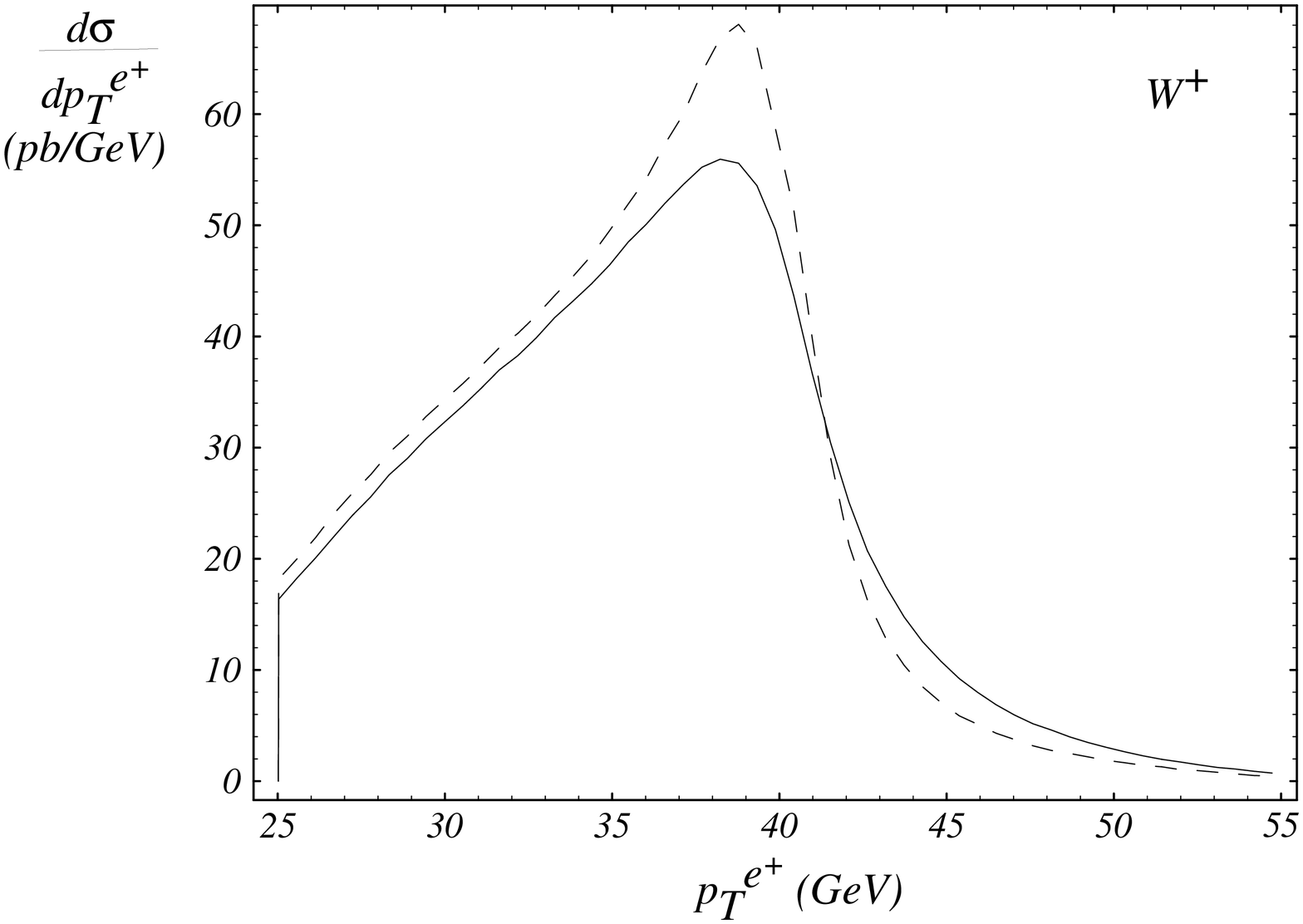} } \fi
&  \\ 
\ifx\nopictures Y \else{ \epsfysize=6.2cm \epsffile{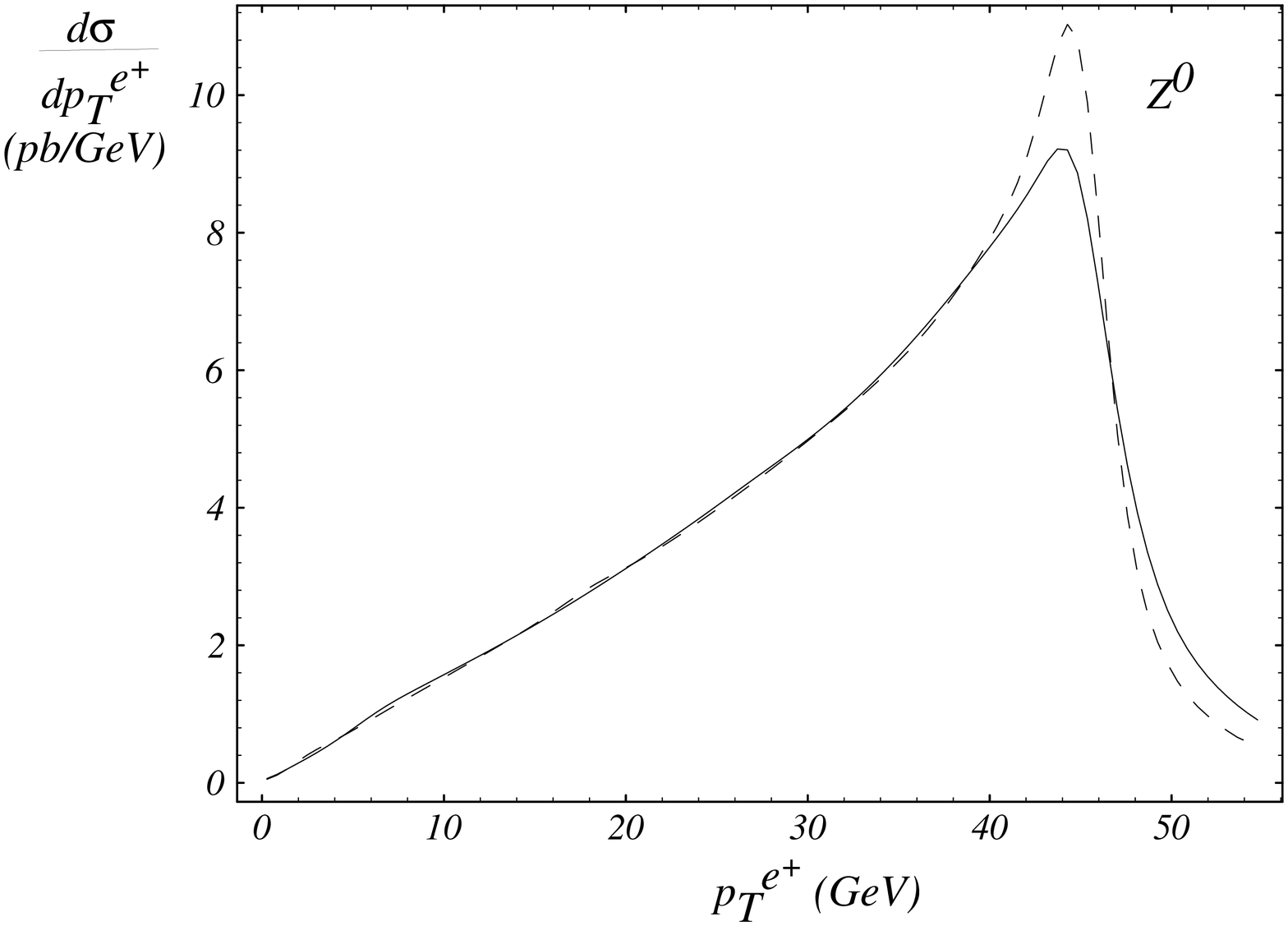} } \fi
& 
\end{tabular}
\end{center}
\vspace{-1cm}
\caption{Transverse momentum distributions of $p_T^{e^{+}}$ from $W^{+}$ and 
$Z^0$ decays for the NLO (dashed) and the resummed ${\cal O}(\alpha _s)$
(solid) calculations. Resumming the initial state multiple soft-gluon
emission has the typical effect of smoothening and broadening the Jacobian
peak (at $p_T^{e^{+}}=M_V/2$). The CDF cuts are imposed on the $W^+$
distributions, but there are no cuts on the $Z^0$ distributions.}
\label{fig:V_pT_e}
\end{figure*}

One of the advantages of using the $p_T^\ell $ distribution to determine $%
M_W $ is that there is no need to reconstruct the $p_T^\nu $ distribution
which potentially limits the precision of the $m_T$ method. From the
theoretical side, the limitation is in the knowledge of the non-perturbative
sector. 
Studies at \D0 \cite{Adam} show that the $p_T^\ell $ distribution is most
sensitive to the PDF's and the value of the non-perturbative parameter $g_2$%
. The $p_T^\ell $ distribution is more sensitive to the PDF choice, than the 
$m_T$ distribution is. The uncertainty in the PDF causes an uncertainty in $%
M_W$ of about 150 MeV, which is about three times as large as that using the 
$m_T$ method~\cite{Adam}. A 0.1 GeV$^2$ uncertainty in $g_2$ leads to about $%
\Delta M_W=30$ MeV uncertainty from the $p_T^{\ell}$ fit, which is about
five times worse than that from the $m_T$ measurement~\cite{Adam}. 
Therefore, to improve the $M_W$ measurement, it is necessary to include the $%
Z^0$ data sample at the high luminosity Tevatron to refit the $g_i$'s and
obtain a tighter constrain on them from the $Q_T$ distribution of the $Z^0$
boson. The \D0 study showed that an accuracy of $\Delta g_2=0.01$ GeV$^2$
can be achieved with Run 2 and TeV33 data, 
which would contribute an error of $\Delta M_W<5$ MeV from the $p_T^\ell $ 
\cite{Adam}. In this case the uncertainty coming from the PDF's remains to
be the major theoretical limitation. 
At the LHC, the $p_T^\ell$ distribution can be predicted with an even
smaller theoretical error coming from the non-perturbative part, because at
higher energies the perturbative Sudakov factor dominates over the
non-perturbative function.

It was recently suggested to extract $M_W$ from the ratios of the transverse
momenta of leptons produced in $W^{\pm }$ and $Z^0$ decay~\cite{Giele-Keller}%
. The theoretical advantage is that the non-perturbative uncertainties are
decreased in such a ratio. On the other hand, it is not enough that the
ratio of cross sections is calculated with small theoretical errors. For a
precision extraction of the $W^\pm$ mass the theoretical calculation must be
capable of reproducing the individually observed transverse momentum
distributions themselves. The $W^\pm$ mass measurement requires a detailed
event modeling, understanding of detector resolution, kinematical acceptance
and efficiency effects, which are different for the $W^{\pm }$ and $Z^0$
events, as illustrated above. Therefore, the ratio of cross sections can
only provide a useful check for the $W^\pm$ mass measurement.

For Drell-Yan events or lepton pairs from $Z^0$ decays, additional
measurable quantities can be constructed from the lepton transverse momenta.
They are the distributions in the balance of the transverse momenta $(\Delta 
{p}_T=|{\vec{p}}_T^{\,\ell _1}|-|{\vec{p}}_T^{\,{\bar \ell_2}}|)$ and the
angular correlation of the two lepton momenta $({z=-\vec{p}}%
_T^{\,\ell_1}\cdot {\vec{p}}_T^{\,{\bar \ell_2}}/[\max (p_T^{\,\ell
_1},p_T^{\,{\bar \ell_2}})]^2)$. It is expected that these quantities are
also sensitive to the effects of the multiple soft gluon radiation. These
distributions are shown in Figure~\ref{fig:Z_pTCorr}. As shown, the resummed
distributions significantly differ from the NLO ones. In these, and the
following figures for $Z^0$ decay distributions, it is understood that the
following kinematic cuts are imposed: $Q_T^{Z^0}<30$~GeV, $%
p_T^{e^{+},e^{-}}>25$ GeV and $|y^{e^{+},e^{-}}|<3.0$, unless indicated
otherwise. 
\begin{figure*}[t]
\vspace{-.5cm}
\begin{center}
\begin{tabular}{cc}
\ifx\nopictures Y \else{ \epsfysize=6.2cm \epsffile{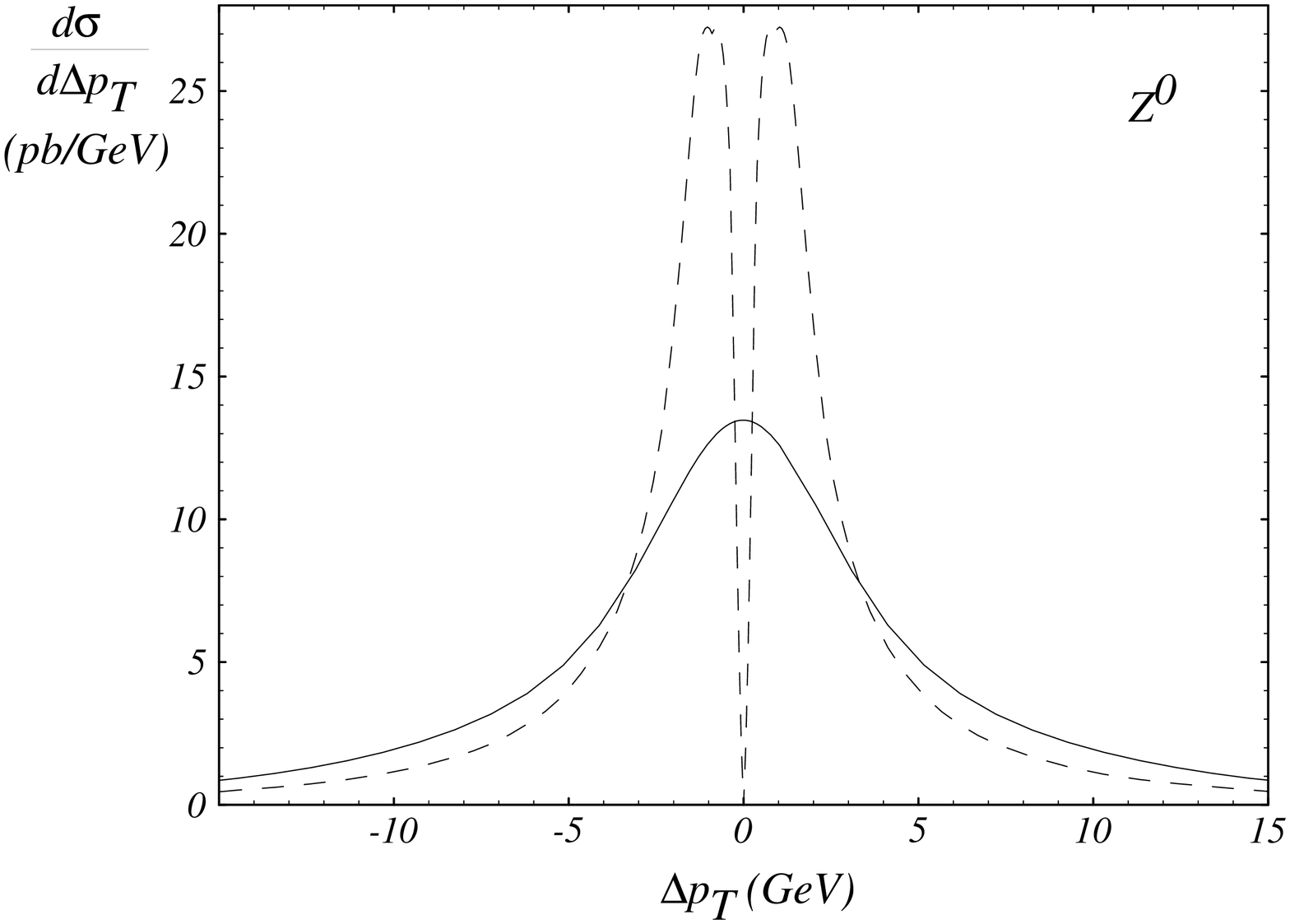} } \fi
&  \\ 
\ifx\nopictures Y \else{ \epsfysize=6.2cm \epsffile{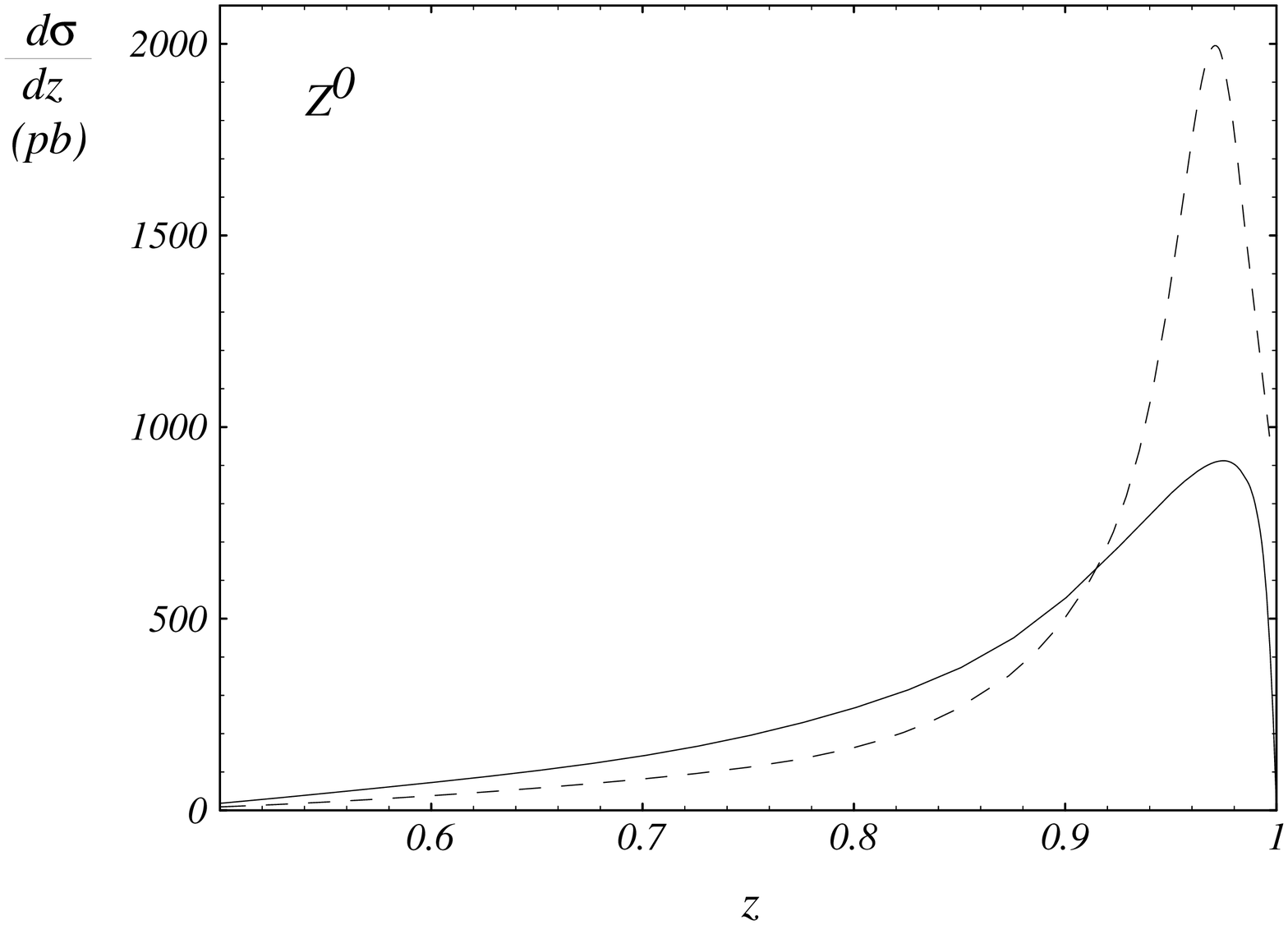} } %
\fi & 
\end{tabular}
\end{center}
\vspace{-1cm}
\caption{Balance in transverse momentum $\Delta {p}_T=|{\vec p}_T^{\,\ell
_1}|-|{\vec p}_T^{\,{\bar \ell_2}}|$ and angular correlation ${z=-\vec p}
_T^{\,\ell _1}\cdot {\vec p}_T^{\,{\bar \ell_2}}/[\max (p_T^{\ell_1},p_T^{{%
\bar \ell_2}})]^2$ of the decay leptons from $Z^0$ bosons produced at the
Tevatron.}
\label{fig:Z_pTCorr}
\end{figure*}


\subsection{Lepton Angular Correlations}

Another observable that can serve to test the QCD theory beyond the
fixed-order perturbative calculation is the difference in the azimuthal
angles of the leptons $\ell _1$ and ${\bar \ell_2}$ from the decay of a
vector boson $V$. In practice, this can be measured for $\gamma ^{*}\text{
or }Z^0\rightarrow \ell_1 {\bar \ell_2}$. We show in Fig.~\ref
{fig:Z_DelPhi_ee} the difference in the azimuthal angles of $e^{+}$ and $%
e^{-}$ ($\Delta \phi ^{e^{+}e^{-}}$), measured in the laboratory frame for $%
Z^0\rightarrow e^{+}e^{-}$, calculated in the NLO and the resummed
approaches. As indicated, the NLO result is ill-defined in the vicinity of $%
\Delta \phi \sim \pi $, where the multiple soft-gluon radiation has to be
resummed to obtain physical predictions. 
\begin{figure*}[t]
\vspace{-.5cm}
\begin{center}
\begin{tabular}{cc}
\ifx\nopictures Y \else{ \epsfysize=8.5cm \epsffile{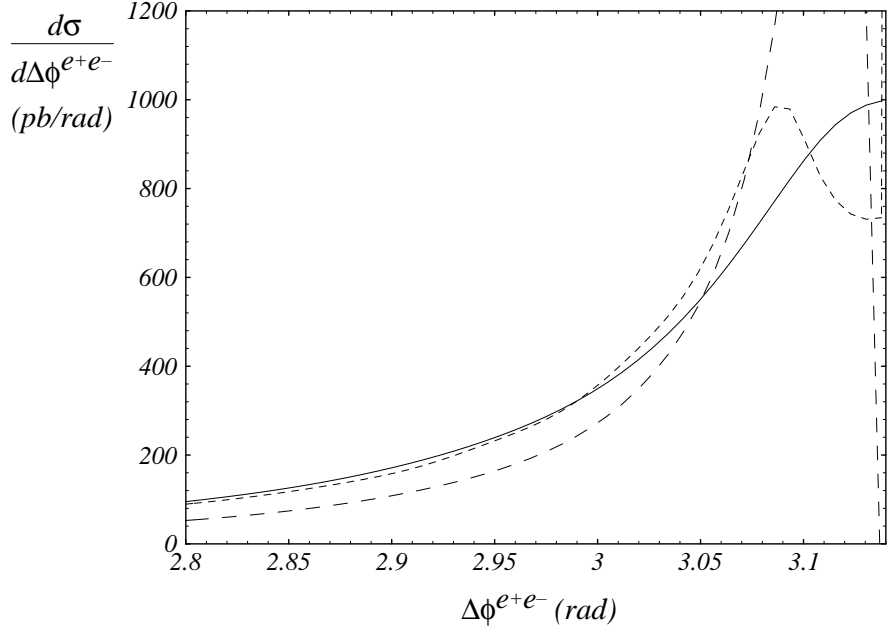}}
\fi & 
\end{tabular}
\end{center}
\vspace{-1cm}
\caption{The correlation between the lepton azimuthal angles near the region 
$\Delta \phi \sim \pi$ for $p{\bar p}\rightarrow (Z^0 \rightarrow
e^{+}e^{-})X$. The resummed (solid) distribution gives the correct angular
correlation of the lepton pair. The NLO (dashed lines) distribution near $%
\Delta \phi =\pi $ is ill-defined and depends on $Q_T^{Sep}$ (the scale for
separating soft and hard gluons in the NLO calculation). The two NLO
distributions were calculated with $Q_T^{Sep} = 1.2$ GeV (long dash) and $%
Q_T^{Sep} = 2.0$ GeV (short dash). }
\label{fig:Z_DelPhi_ee}
\end{figure*}

Another interesting angular variable is the lepton polar angle distribution $%
\cos \theta ^{*\ell }$ in the Collins-Soper frame. It can be calculated for
the $Z^0$ decay and used to extract $\sin ^2\theta _w$ at the Tevatron~\cite
{CDFAFB}. The asymmetry in the polar angle distribution is essentially the
same as the forward-backward asymmetry $A_{FB}$ measured at LEP. Since $%
A_{FB}$ depends on the invariant mass $Q$ and around the energy of the $Z^0$
peak $A_{FB}$ happens to be very small, the measurement is quite
challenging. At the hadron collider, on the other hand, the invariant mass
of the incoming partons is distributed over a range so the asymmetry is
enhanced~\cite{Rosner}. The potentials of the measurement deserve a
separated study. In Fig.~\ref{fig:Z_CTS} we show the distributions of $\cos
\theta ^{*\ell }$ predicted from the NLO and the resummed results. 
\begin{figure*}[t]
\vspace{-.5cm}
\begin{center}
\begin{tabular}{cc}
\ifx\nopictures Y \else{ \epsfysize=8.5cm \epsffile{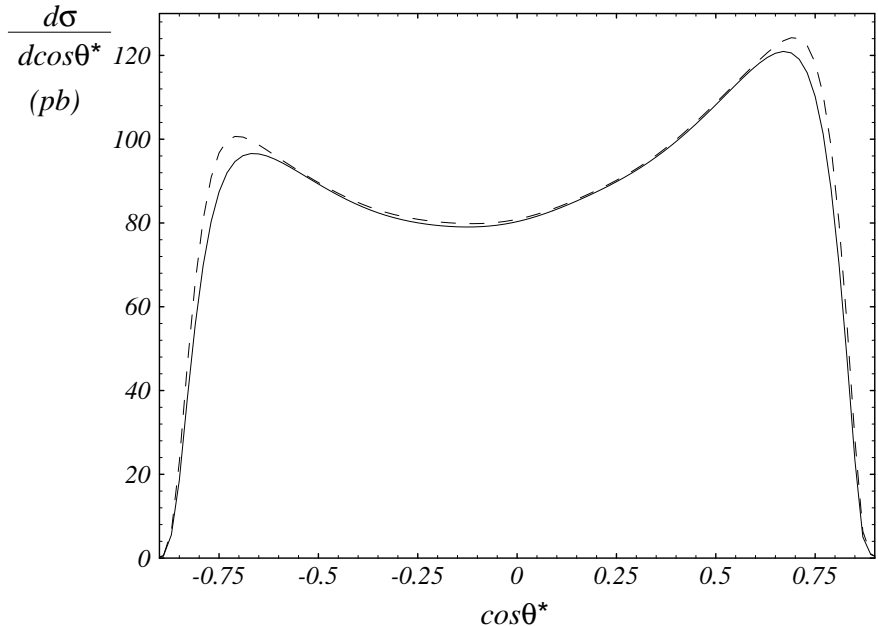} } \fi
& 
\end{tabular}
\end{center}
\vspace{-1cm}
\caption{Distribution of the $e^{+}$ polar angle $\cos (\theta ^{*})$ in the
Collins-Soper frame from $Z^0$ decays at the Tevatron with cuts indicated in
the text.}
\label{fig:Z_CTS}
\end{figure*}


\section{Conclusions}

With a $100\,{\rm pb}^{-1}$ luminosity at the Tevatron, around $2 \times 10^6
$ $W^\pm$ and $6 \times 10^5$ $Z^0$ bosons are produced, and the data sample
will increase by a factor of 20 in the Run 2 era. In view of this large
event rate, a careful study of the distributions of leptons from the decay
of the vector bosons can provide a stringent test of the rich dynamics of
the multiple soft gluon emission predicted by the QCD theory. Since an
accurate determination of the mass of the $W^\pm$ boson and the test of
parton distribution functions demand a highly precise knowledge of the
kinematical acceptance and the detection efficiency of $W^\pm$ or $Z^0$
bosons, the effects of the multiple gluon radiation have to be taken into
account. In this work, we have extended the formalism introduced by Collins,
Soper and Sterman for calculating an on-shell vector boson to include the
effects of the polarization and the decay width of the vector boson on the
distributions of the decay leptons. Our resummation formalism can be applied
to any vector boson $V$ where $V=\gamma^*, \, W^\pm, \, Z^0, \, W^{\prime},
\, Z^{\prime}$, etc., with either vector or axial-vector couplings to
fermions (leptons or quarks). To illustrate how the multiple gluon radiation
can affect the distributions of the decay leptons, we studied in detail
various distributions for the production and the decay of the vector bosons
at the Tevatron.

One of the methods to test the rich dynamics of the multiple soft gluon
radiation predicted by the QCD theory is to measure the ratio $R_{CSS}
\equiv \frac{\sigma (Q_T>Q_T^{\min })}{\sigma _{Total}}$ for the $W^\pm$ and 
$Z^0$ bosons. We found that, for the vector boson transverse momentum less
than about 30\,GeV, the difference between the resummed and the fixed order
predictions (either at the $\alpha _s$ or $\alpha^2_S$ order) can be
distinguished by experimental data. This suggests that in this kinematic
region, the effects of the multiple soft gluon radiation are important,
hence, the $Q_T$ distribution of the vector boson provides an ideal
opportunity to test this aspect of the QCD dynamics. For $Q_T$ less than
about 10 GeV, the distribution of $Q_T$ is largely determined by the
non-perturbative sector of QCD. At the Tevatron this non-perturbative
physics, when parametrized by Eq.~(\ref{eq:WNPLY}) for $W^\pm$ and $Z^0$
production, is dominated by the parameter $g_2$ which was shown to be
related to properties of the QCD vacuum~\cite{Korchemsky-Sterman}.
Therefore, precisely measuring the $Q_T$ distribution of the vector boson in
the low $Q_T$ region, e.g. from the ample $Z^0$ events, can advance our
knowledge of the non-perturbative QCD physics.

Although the rapidity distributions of the leptons are not directly related
to the transverse momentum of the vector boson, they are predicted to be
different in the resummed and the fixed order calculations. This is because
to compare the theoretical predictions with the experimental data, some
kinematic cuts have to be imposed so that the signal events can be observed
over the backgrounds. We showed that the difference is the largest when the
rapidity of the lepton is near the boundary of the phase space (i.e. in the
large rapidity region), and the difference diminishes when no kinematic cuts
are imposed. When kinematic cuts are imposed another important difference
between the results of the resummed and the NLO calculations is the
prediction of the event rate. These two calculations predict different
normalizations of various distributions. For example, the rapidity
distributions of charged leptons ($y^{\ell^\pm}$) from the decays of $W^\pm$
bosons are different. They even differ in the central rapidity region in
which the lepton charge asymmetry distributions are about the same (cf.
Figs.~\ref{fig:W_y_e_Asy} and~\ref{fig:W_y_e}). As noted in Ref.~\cite
{Stirling-Martin}, with kinematic cuts, the measurement of $M_W$ is
correlated to that of the rapidity and its asymmetry through the transverse
momentum of the decay lepton. Since the resummed and the NLO results are
different and the former includes the multiple soft gluon emission dynamics,
the resummed calculation should be used for a precision measurement of $M_W$.

In addition to the rapidity distribution, we have also shown various
distributions of the leptons which are either directly or indirectly related
to the transverse momentum of the vector boson. For those which are directly
related to the transverse momentum of the vector boson, such as the
transverse momentum of the lepton and the azimuthal correlation of the
leptons, the resummation formalism predicts significant differences from the
fixed order perturbation calculations in some kinematic regions. The details
were discussed in Section \ref{sec:LeptonDistributions}.

\begin{figure*}[t]
\vspace{-.5cm}
\begin{center}
\begin{tabular}{cc}
\ifx\nopictures Y \else{ \epsfysize=8.5cm \epsffile{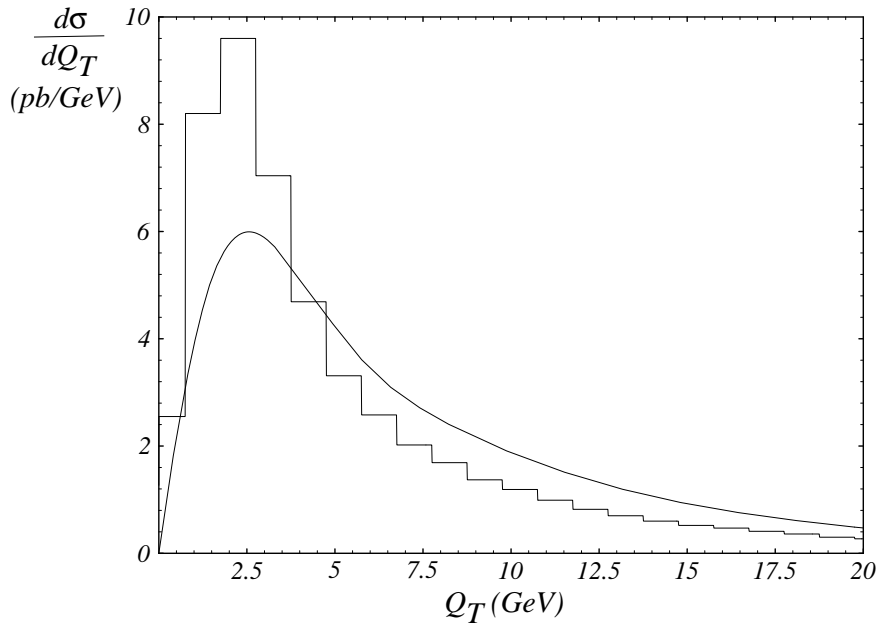}} \fi
& 
\end{tabular}
\end{center}
\vspace{-1cm}
\caption{ Transverse momentum distribution of virtual photons in $p {\bar p}
\rightarrow \gamma^* \rightarrow e^+ e^-$ events predicted by ResBos (solid
curve) and ISAJET (histogram), calculated for the invariant mass range 30
GeV $< Q <$ 60 GeV at the 1.8 TeV Tevatron. }
\label{fig:QTph}
\end{figure*}
As noted in the Introduction, a full event generator, such as ISAJET, can
predict a reasonable shape for various distributions because it contains the
backward radiation algorithm~\cite{Sjostrand}, which effectively includes
part of the Sudakov factor, i.e. effects of the multiple gluon radiation.
However, the total event rate predicted by the full event generator is
usually only accurate at the tree level, as the short distance part of the
virtual corrections cannot yet be consistently implemented in this type of
Monte Carlo program. To illustrate the effects of the high order corrections
coming from the virtual corrections, which contribute to the Wilson
coefficients $C$ in our resummation formalism, we showed in Fig.~\ref
{fig:QTph} the predicted distributions of the transverse momentum of the
Drell-Yan pairs by ISAJET and by ResBos (our resummed calculation). In this
figure we have rescaled the ISAJET prediction to have the same total rate as
the ResBos result, so that the shape of the distributions can be directly
compared. We restrict the invariant mass of the virtual photons $Q$ to be
between 30 and 60 GeV without any kinematic cuts on the leptons. If
additional kinematic cuts on the leptons are imposed, then the difference is
expected to be enhanced, as discussed in Section~\ref{subsec:LCA}. As
clearly shown, with a large data sample in the future, it will be possible
to experimentally distinguish between these two predictions, and, more
interestingly, to start probing the non-perturbative sector of the QCD
physics.



%
%
\def\FigFeynman
{
\begin{figure}[t]
\vspace{-1cm}
\ifx\nopictures Y 
\else{\centerline{\epsfysize=10.0cm \epsfbox{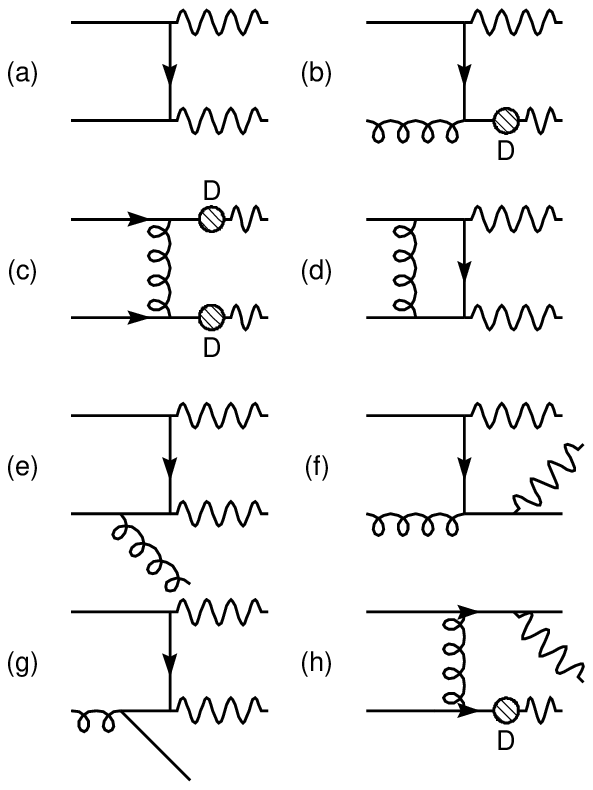}}} 
\fi
\vspace{-1.5cm}
\caption{ Feynman diagrams representing the leading order and
next--to--leading order contributions to photon pair production in hadron
collisions. The shaded circles signify the production of long--distance
fragmentation photons, which are described by the fragmentation function 
$D_{\gamma \leftarrow q}$. }
\label{Fig:Feynman}
\end{figure}
}
\def\FigFrag
{
\begin{figure}[p]
\vspace{-1cm}
\ifx\nopictures Y \else{
\centerline{\epsfysize=7.5cm \epsfbox{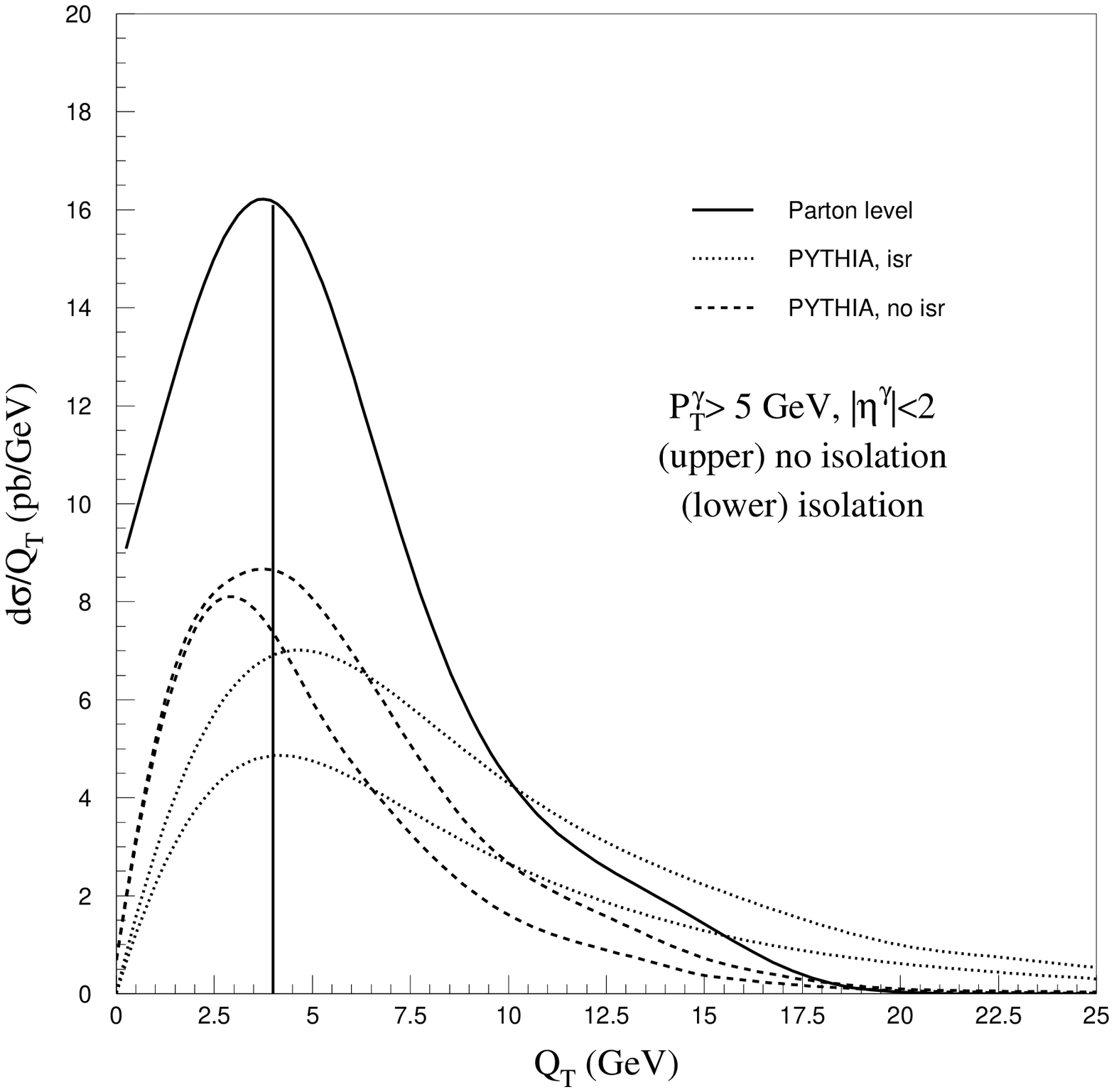}}
\vspace{-1cm}
\centerline{\epsfysize=7.5cm \epsfbox{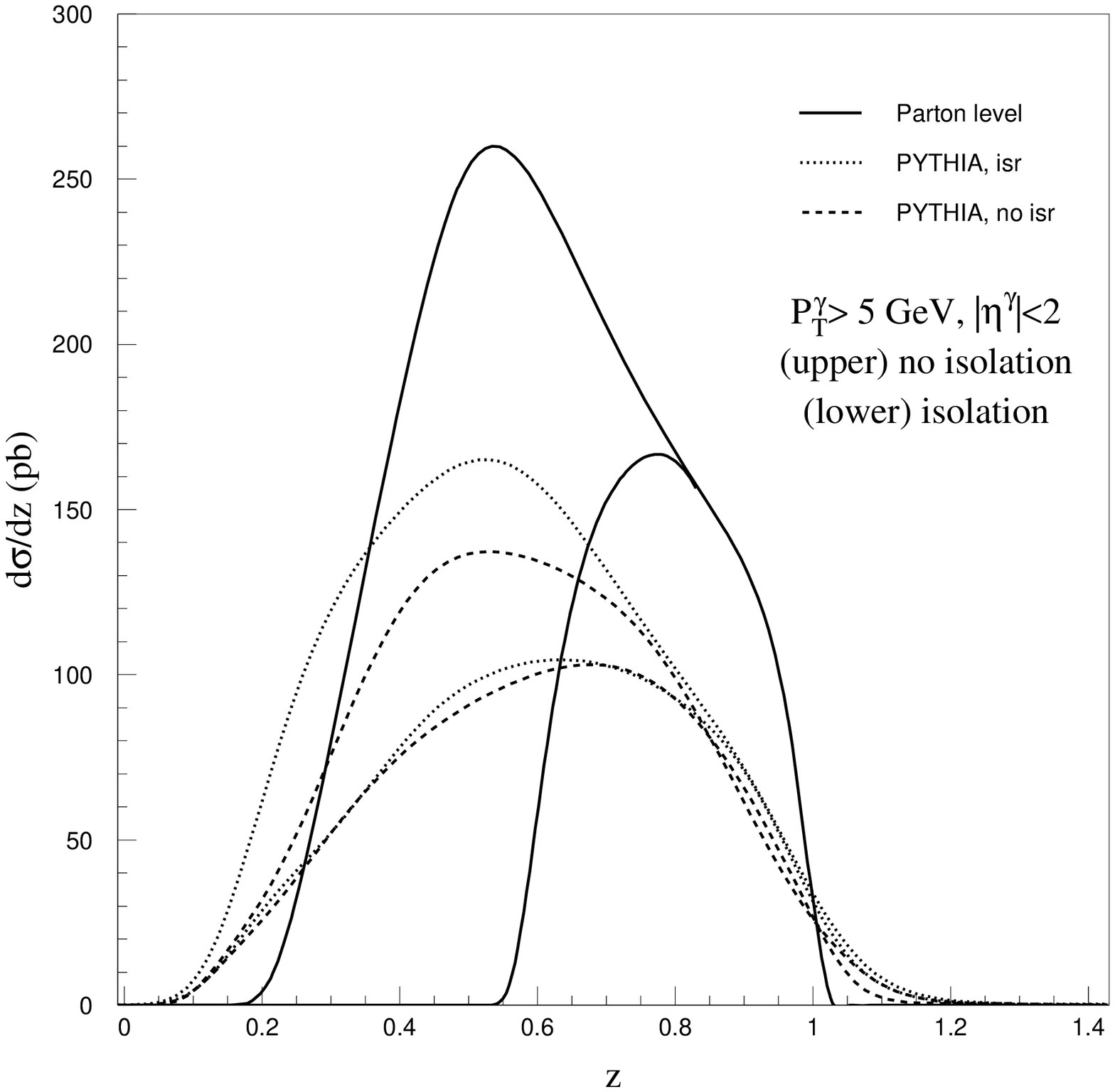}}
} \fi
\vspace{-.25cm}
\caption{ Comparison of the parton--level and Monte Carlo fragmentation
contributions at the Tevatron. The upper and lower curves of the same type
show the contribution before and after an isolation cut. The left figure
shows the transverse momentum of the photon pair $Q_T$. The right figure
shows the light--cone momentum fraction carried by the fragmentation photon. 
}
\label{Fig:Frag}
\end{figure}
}
\def\FigCDFa
{
\begin{figure}[t]
\vspace{-1cm}
\ifx\nopictures Y 
\else{\centerline{\epsfysize=10cm \epsfbox{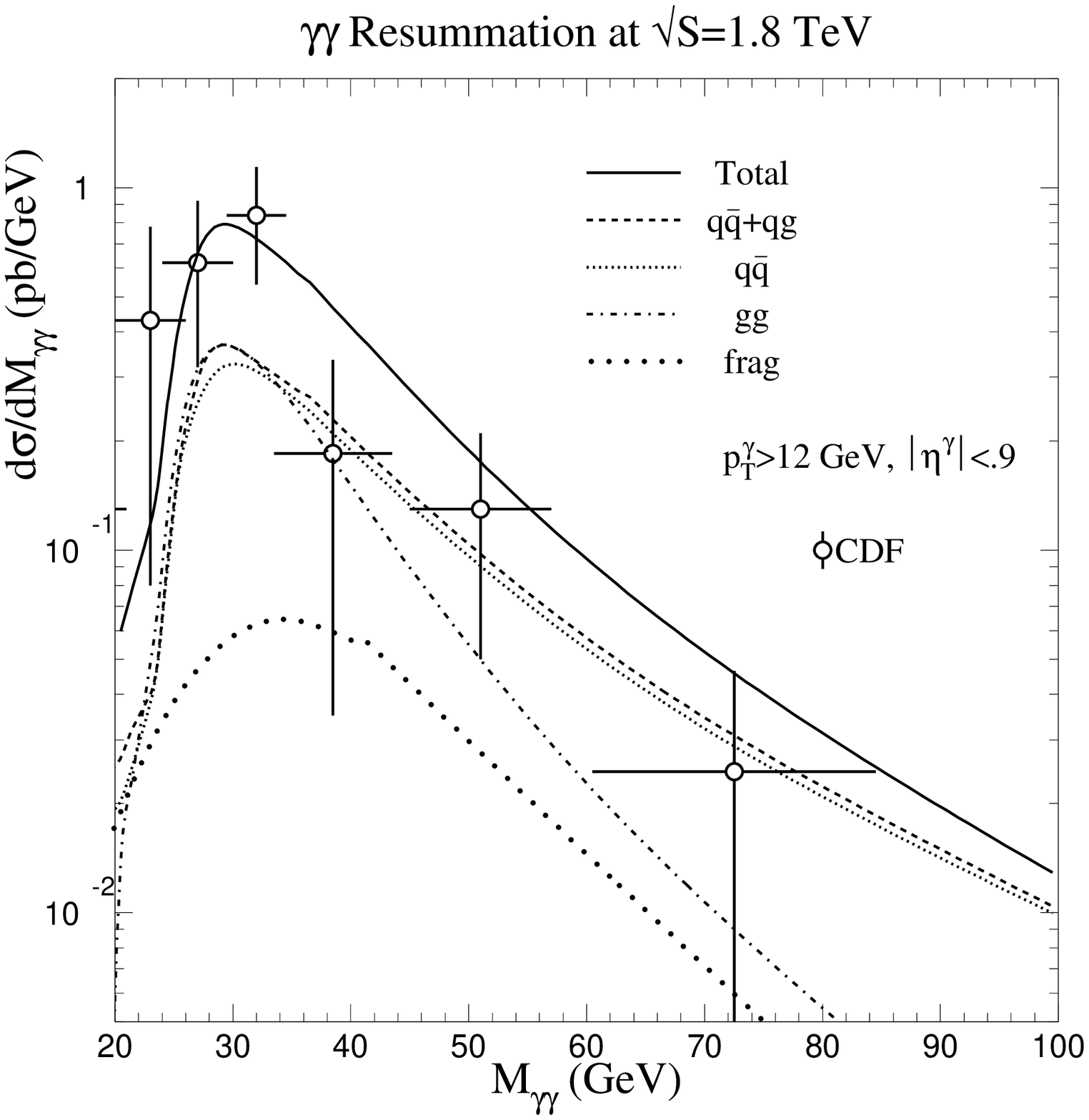}}} 
\fi
\vspace{-.25cm}
\caption{The predicted distribution for the invariant mass of the photon
pair $M_{\gamma \gamma }$ from the resummed calculation compared to the CDF
data, with the CDF cuts imposed in the calculation.}
\label{Fig:CDF0}
\end{figure}
}
\def\FigCDFb
{
\begin{figure}[t]
\vspace{-1cm}
\ifx\nopictures Y 
\else{\centerline{\epsfysize=10cm \epsfbox{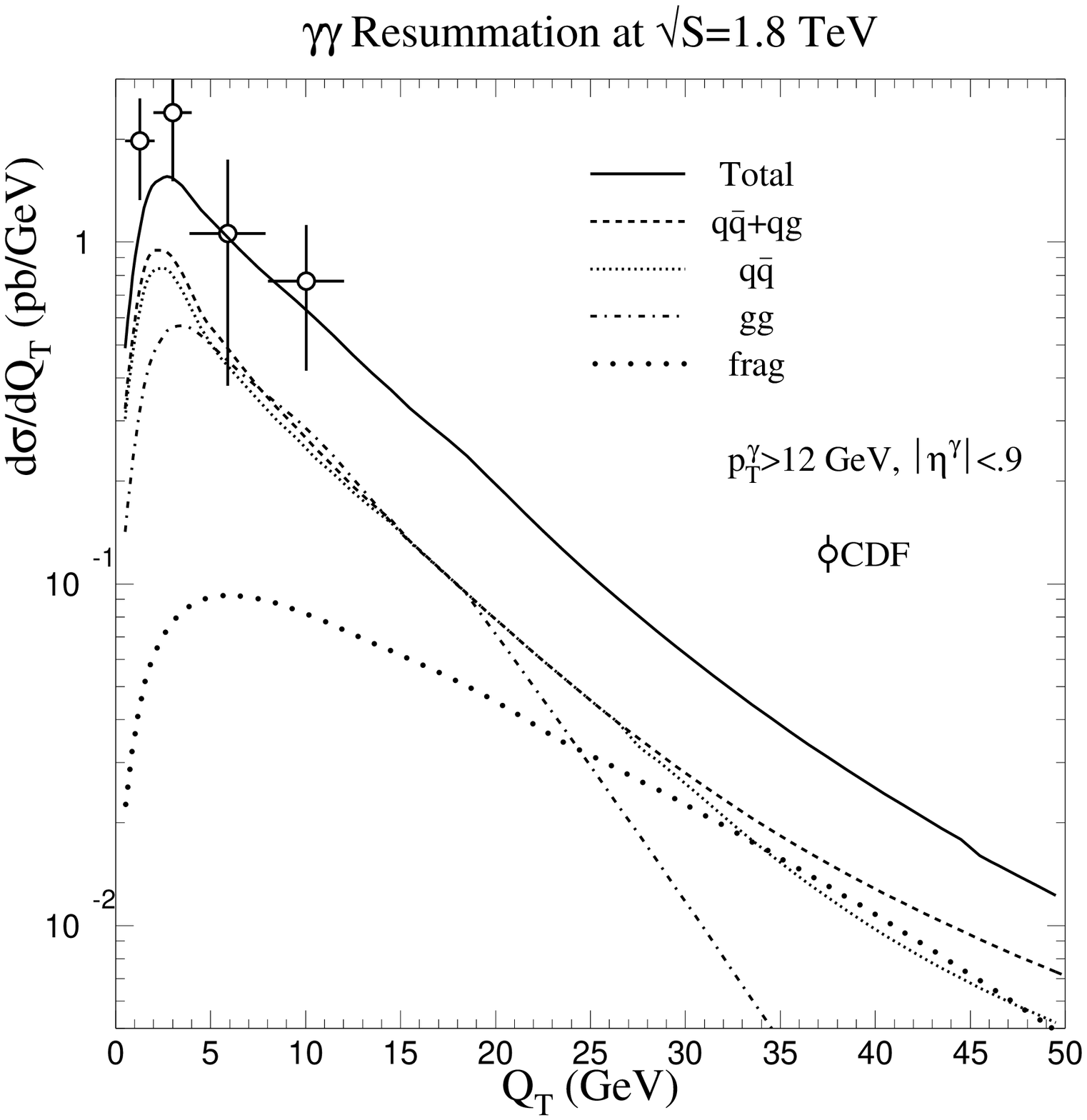}}} 
\fi
\vspace{-.25cm}
\caption{The predicted distribution for the transverse momentum of the
photon pair $Q_T$ from the resummed calculation compared to the CDF data,
with the CDF cuts imposed in the calculation.}
\label{Fig:CDF1}
\end{figure}
}
\def\FigCDFc
{
\begin{figure}[t]
\vspace{-1cm}
\ifx\nopictures Y 
\else{\centerline{\epsfysize=10cm \epsfbox{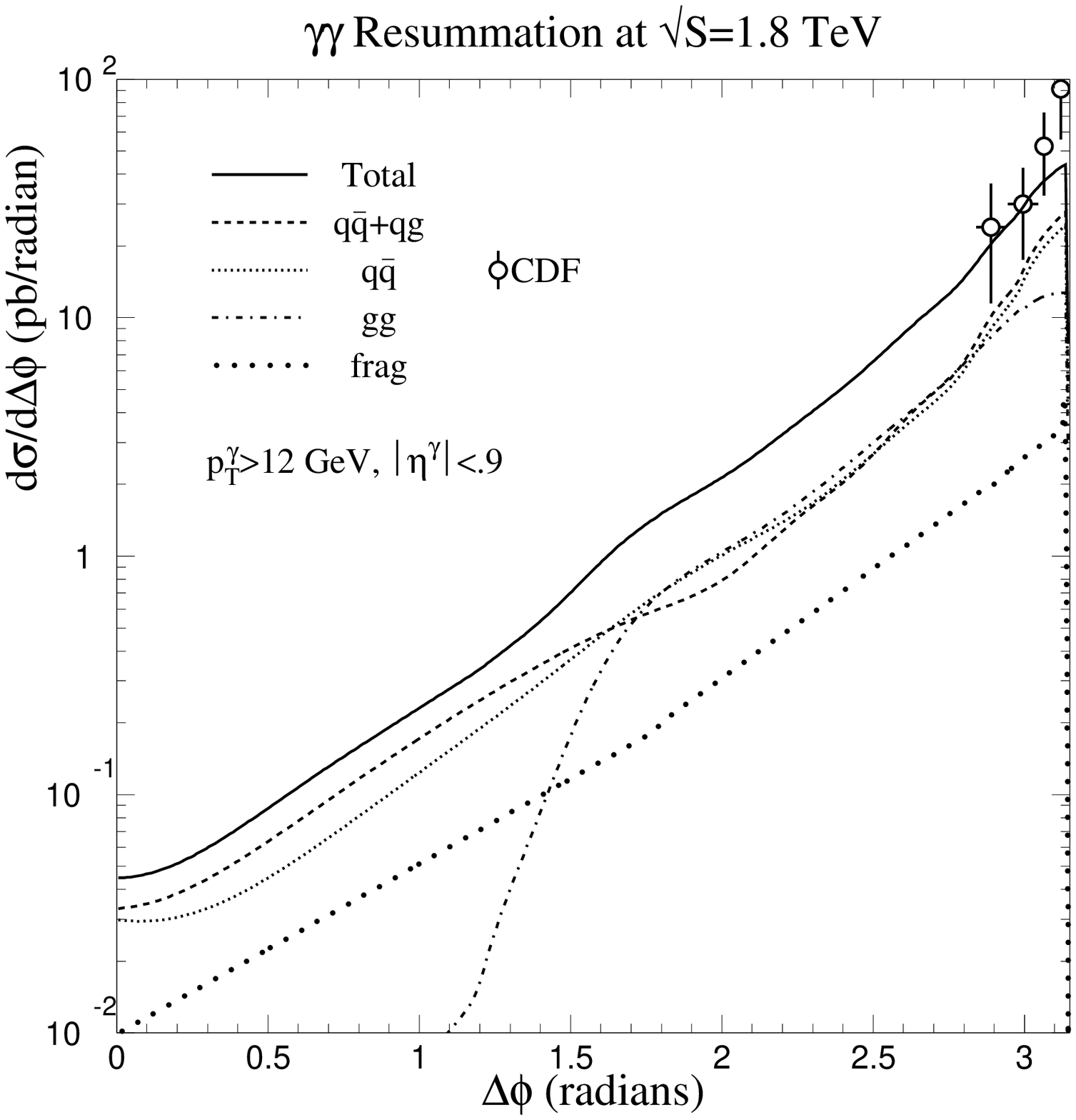}}} 
\fi
\vspace{-.25cm}
\caption{The predicted distribution for the difference between the azimuthal
angles of the photons $\Delta \phi _{\gamma \gamma }$ from the resummed
calculation compared to the CDF data, with the CDF cuts imposed in the
calculation.}
\label{Fig:CDF2}
\end{figure}
}
\def\FigDa
{
\begin{figure}[t]
\vspace{-1cm}
\ifx\nopictures Y 
\else{\centerline{\epsfysize=10cm \epsfbox{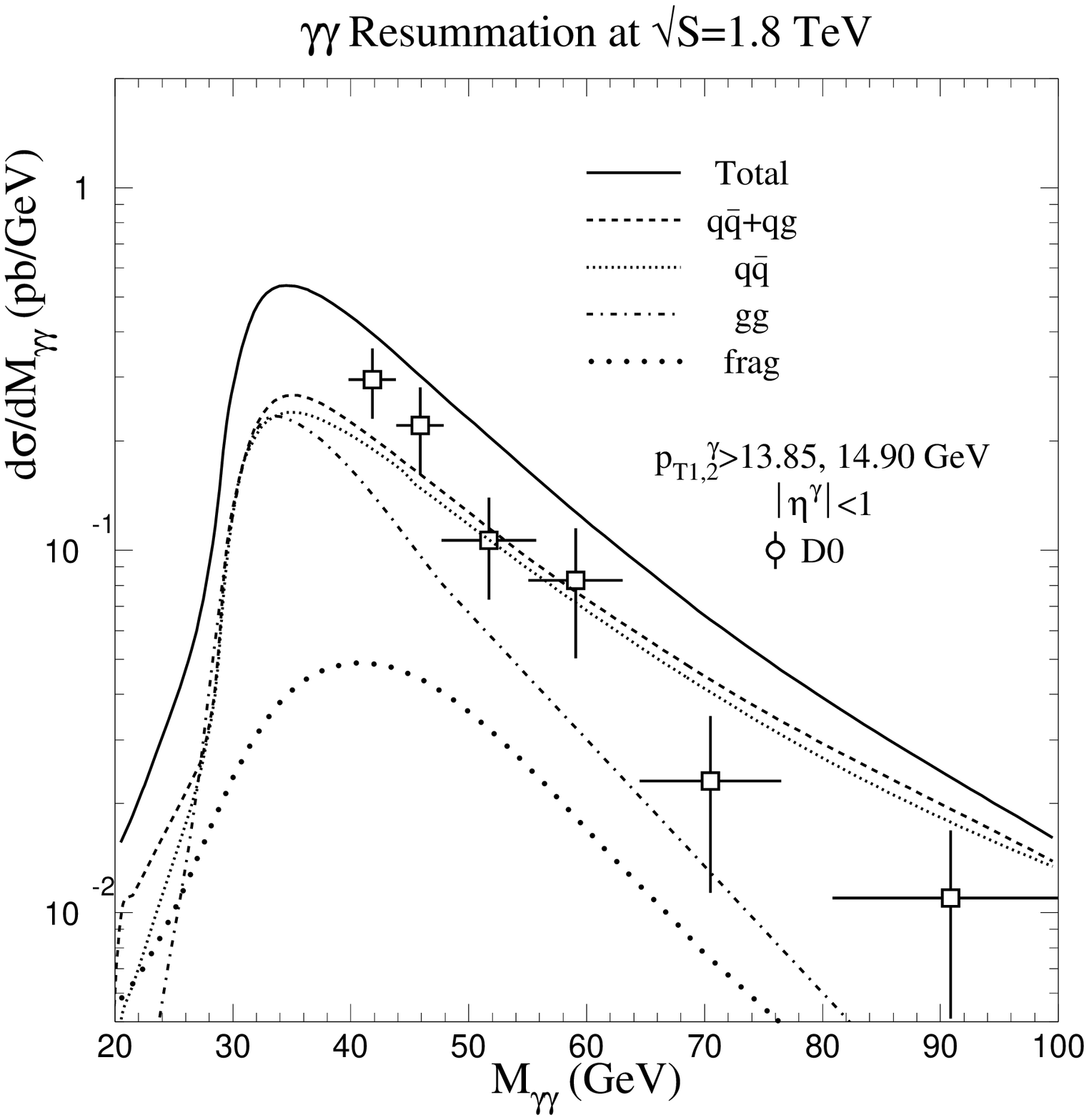}}} 
\fi
\vspace{-.25cm}
\caption{The predicted distribution for the invariant mass of the photon
pair $M_{\gamma \gamma }$ from the resummed calculation compared to the \D%
0~data, with the \D0~cuts imposed in the calculation.}
\label{Fig:D00}
\end{figure}
}
\def\FigDb
{
\begin{figure}[t]
\vspace{-1cm}
\ifx\nopictures Y 
\else{\centerline{\epsfysize=10cm \epsfbox{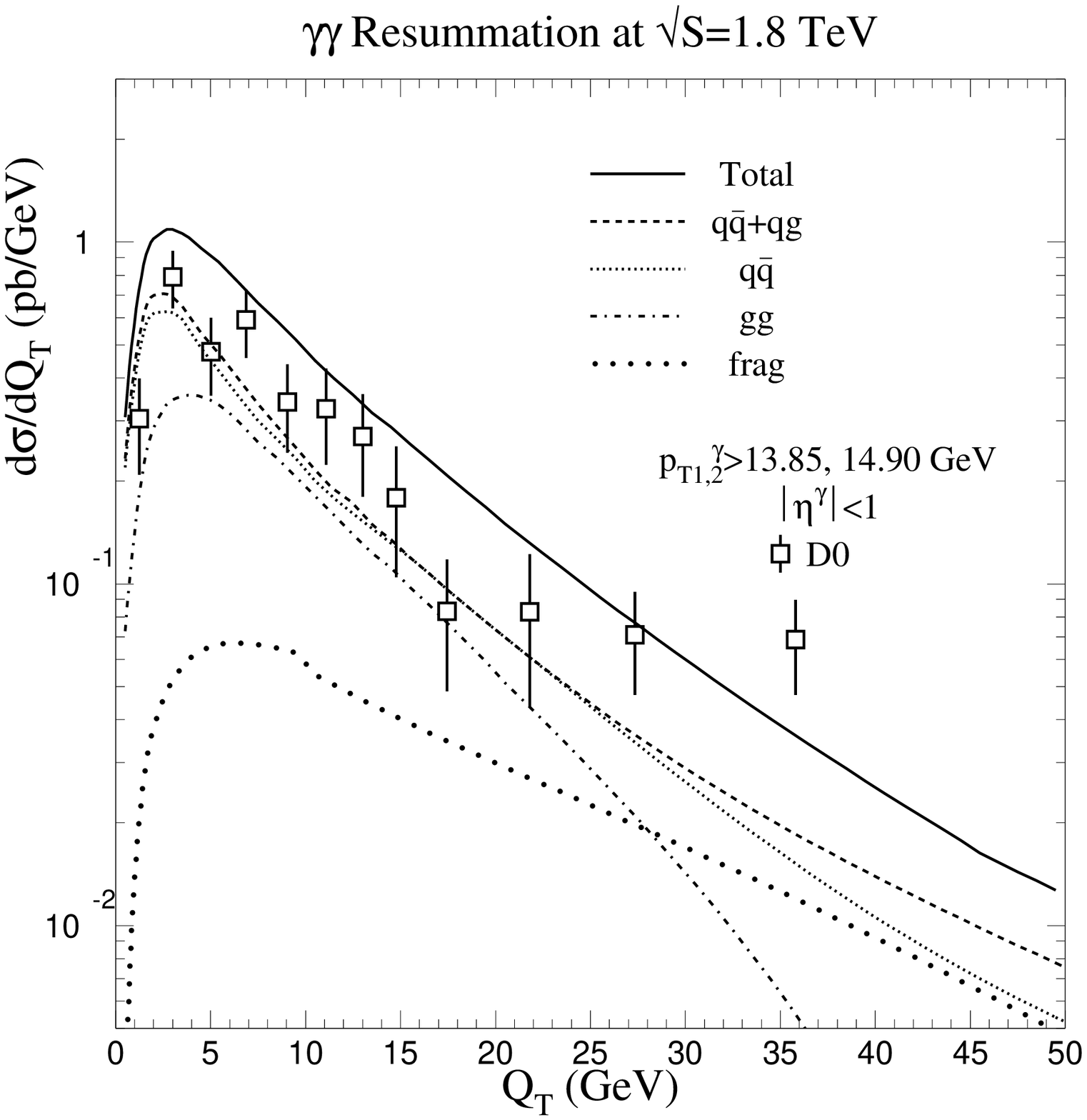}}} 
\fi
\vspace{-.25cm}
\caption{The predicted distribution for the transverse momentum of the
photon pair $Q_T$ from the resummed calculation compared to the \D0~data,
with the \D0~cuts imposed in the calculation.}
\label{Fig:D01}
\end{figure}
}
\def\FigDc
{
\begin{figure}[t]
\vspace{-1cm}
\ifx\nopictures Y 
\else{\centerline{\epsfysize=10cm \epsfbox{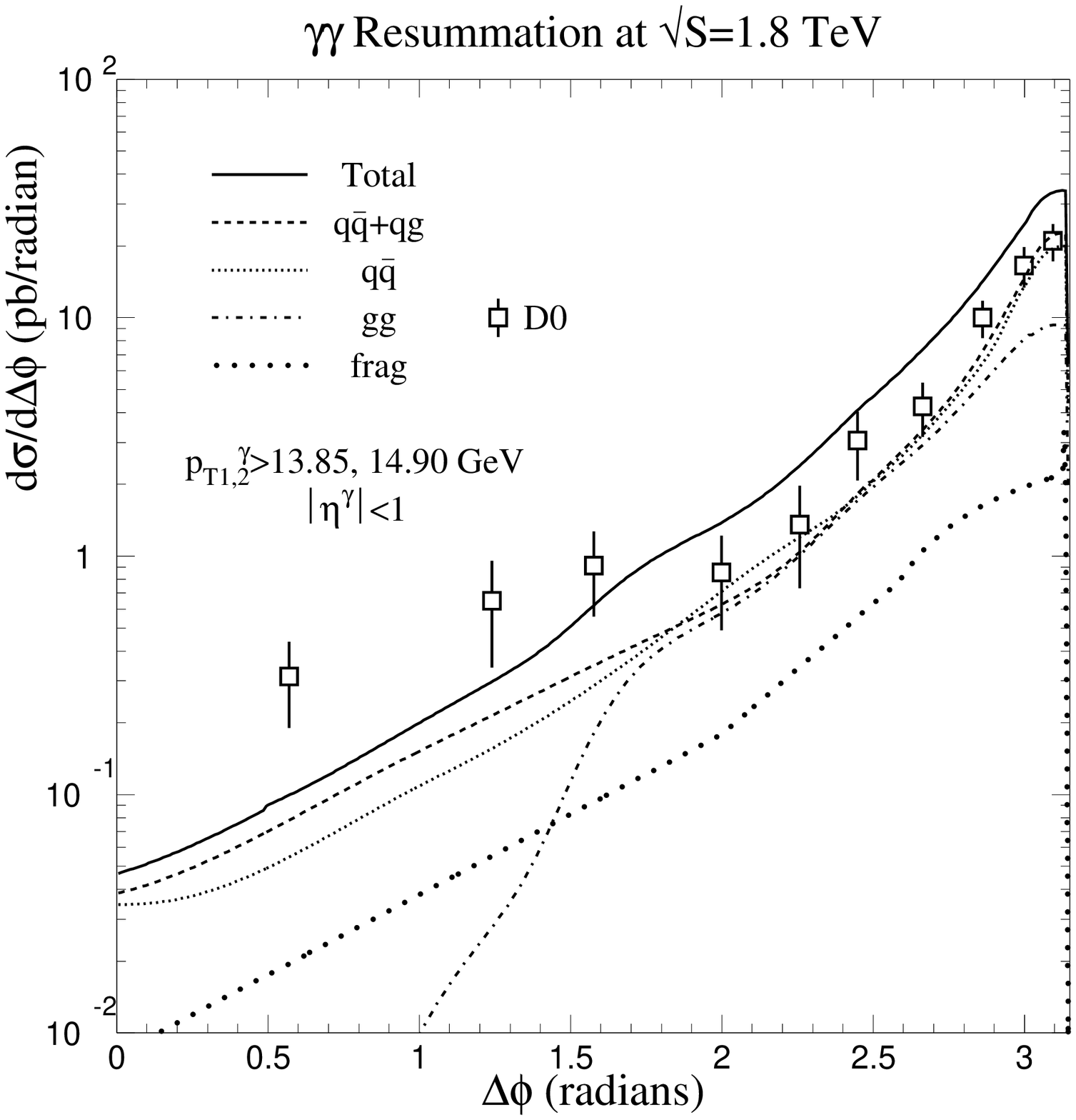}}} 
\fi
\vspace{-.25cm}
\caption{The predicted distribution for the difference between the azimuthal
angles of the photons $\Delta \phi _{\gamma \gamma }$ from the resummed
calculation compared to the \D0~data, with the \D0~cuts imposed in the
calculation.}
\label{Fig:D02}
\end{figure}
}
\def\Figqtcuta
{
\begin{figure}[t]
\vspace{-1cm}
\ifx\nopictures Y \else{
\centerline{\epsfysize=10cm \epsfbox{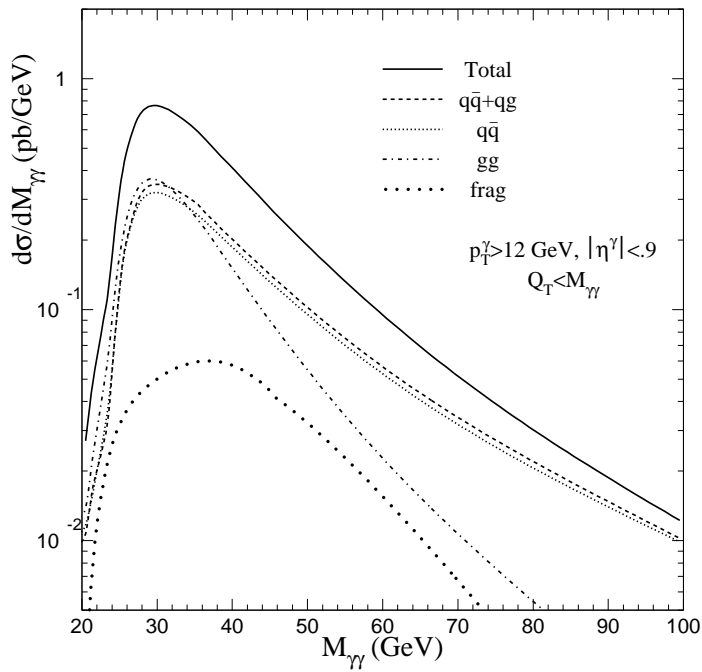} } 
} \fi
\vspace{-.25cm}
\caption{The predicted distribution for the invariant mass of the photon
pair $M_{\gamma \gamma }$ from the resummed calculation. The additional cut $%
Q_T<M_{\gamma \gamma }$ has been applied to reduce the theoretical
uncertainty.}
\label{Fig:qtcut0}
\end{figure}
}
\def\Figqtcutb
{
\begin{figure}[t]
\vspace{-1cm}
\ifx\nopictures Y \else{
\centerline{\epsfysize=10cm \epsfbox{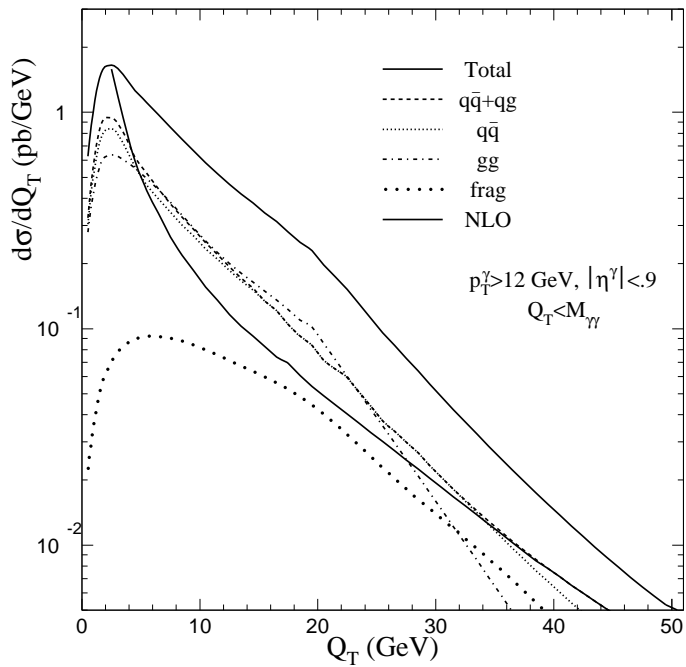} }
} \fi
\vspace{-.25cm}
\caption{The predicted distribution for the transverse momentum of the
photon pair $Q_T$ from the resummed calculation. The additional cut $%
Q_T<M_{\gamma \gamma }$ has been applied to reduce the theoretical
uncertainty. The lower solid curve shows the prediction of the pure NLO
(fixed--order) calculation for the $q\bar{q}$ and $qg$ subprocesses, but
without fragmentation contributions.}
\label{Fig:qtcut1}
\end{figure}
}
\def\Figqtcutc
{
\begin{figure}[t]
\vspace{-1cm}
\ifx\nopictures Y \else{
\centerline{\epsfysize=10cm \epsfbox{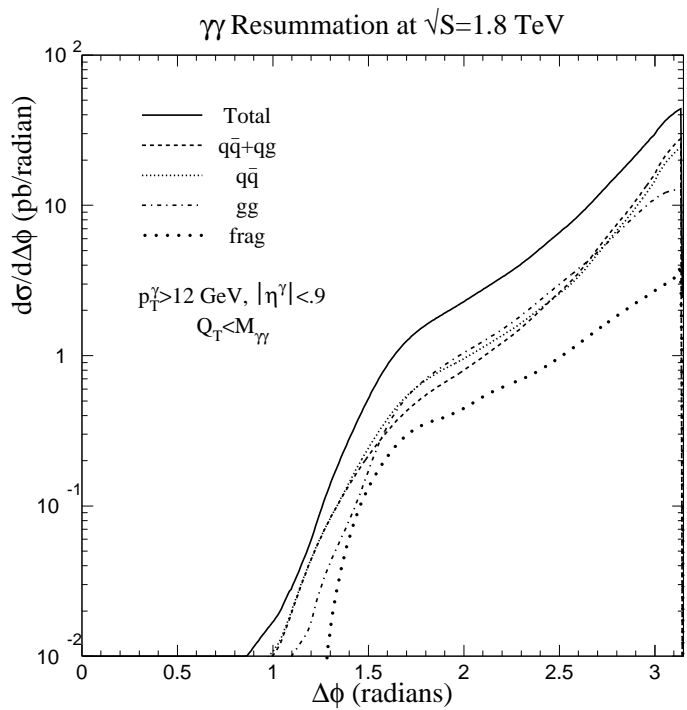} } } \fi
\vspace{-.25cm}
\caption{The predicted distribution for the difference between the azimuthal
angles of the photons $\Delta \phi _{\gamma \gamma }$ from the resummed
calculation. The additional cut $Q_T<M_{\gamma \gamma }$ has been applied to
reduce the theoretical uncertainty.}
\label{Fig:qtcut2}
\end{figure}
}
\def\FigEa
{
\begin{figure}[t]
\vspace{-1cm}
\ifx\nopictures Y \else{
\centerline{\epsfysize=10cm \epsfbox{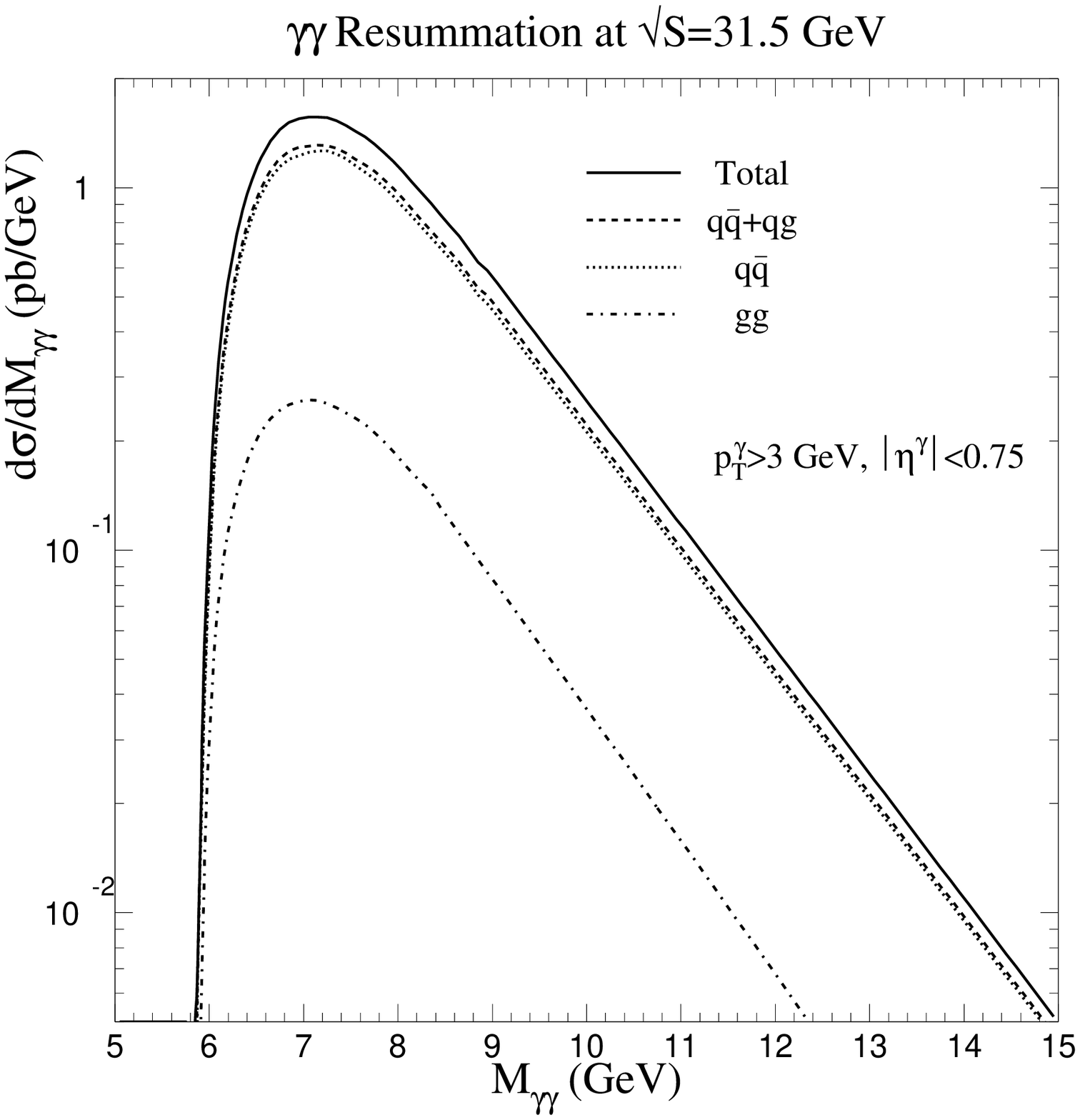} } } \fi
\vspace{-.25cm}
\caption{The predicted distribution for the invariant mass of the photon
pair $M_{\gamma \gamma }$ from the resummed calculation appropriate for $%
pN\to \gamma \gamma X$ at ${\protect\sqrt{S}}$=31.5 GeV.}
\label{Fig:E7060}
\end{figure}
}
\def\FigEb
{
\begin{figure}[t]
\vspace{-1cm}
\ifx\nopictures Y \else{
\centerline{\epsfysize=10cm \epsfbox{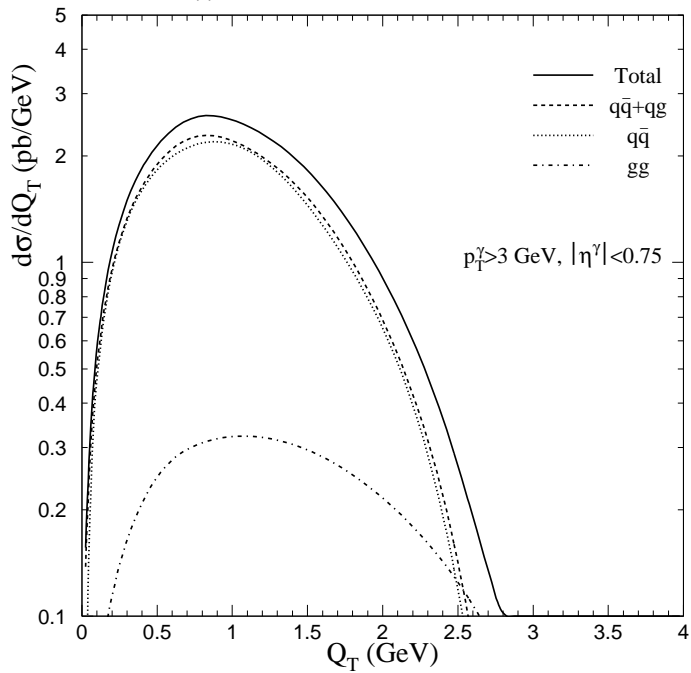} } 
} \fi
\vspace{-.25cm}
\caption{The predicted distribution for the transverse momentum of the
photon pair $Q_T$ from the resummed calculation appropriate for $pN\to
\gamma \gamma X$ at ${\protect\sqrt{S}}$=31.5 GeV.}
\label{Fig:E7061}
\end{figure}
}
\def\FigEc
{
\begin{figure}[t]
\vspace{-1cm}
\ifx\nopictures Y \else{
\centerline{\epsfysize=10cm \epsfbox{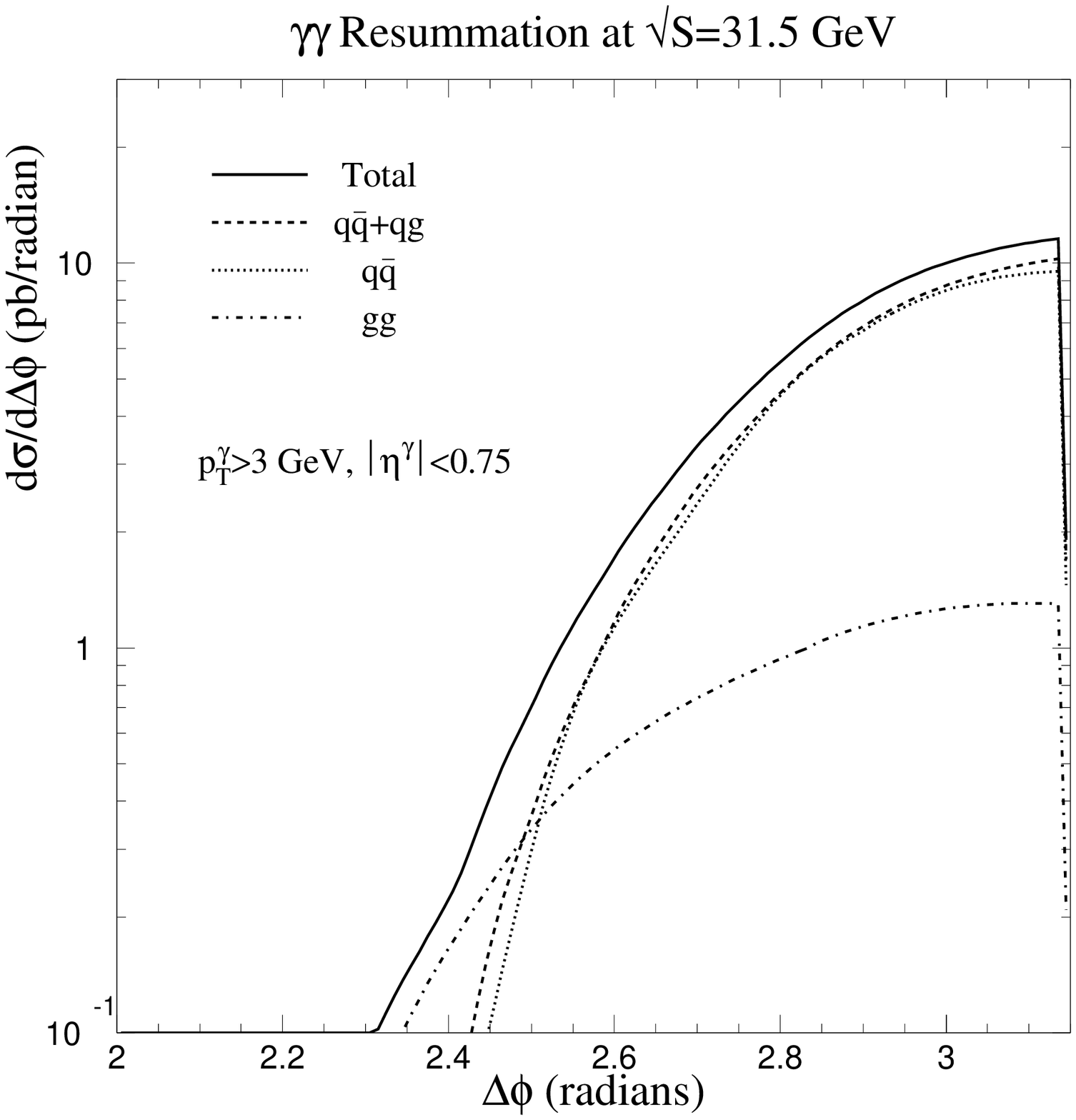}} } \fi
\vspace{-.25cm}
\caption{The predicted distribution for the difference between the azimuthal
angles of the photons $\Delta \phi _{\gamma \gamma }$ from the resummed
calculation appropriate for $pN\to \gamma \gamma X$ at ${\protect\sqrt{S}}$%
=31.5 GeV.}
\label{Fig:E7062}
\end{figure}
}


\chapter{Photon Pair Production In Hadronic Interactions \label{ch:PhotonPair}}

An increasing amount of prompt diphoton data is becoming available from the
Tevatron collider and the fixed--target experiments at Fermilab \cite
{cdfdata,d0data,e706data}. After the upgrade of the Fermilab Tevatron the
amount of the di-gauge boson data increases by a factor of 20, and the LHC\
by three order of magnitude. A comparison of the data to the calculation of
the diphoton production rate and kinematic distributions provides a test of
many aspects of perturbative quantum chromodynamics (pQCD). Furthermore,
understanding the diphoton data is important for new physics searches. For
example, diphoton production is an irreducible background to the light Higgs
boson decay mode $h\to \gamma \gamma $. The next--to--leading order (NLO)
cross section for the $p\bar{p}\to \gamma \gamma X$ process \cite{nloqq} was
shown to describe well the invariant mass distribution of the diphoton pair
after the leading order (LO) $gg\to \gamma \gamma $ contribution (from
one--loop box diagrams) was included \cite{box,Bailey-Owens-Ohnemus}. 
However, to accurately
describe the distribution of the transverse momentum of the photon pair and
the kinematical correlation of the two photons, a calculation has to be
performed that includes the effects of initial--state multiple soft--gluon
emission. 

In this Section, we extend the Collins--Soper--Sterman (CSS) soft gluon
resummation formalism, to describe the production of photon pairs. This
extension is similar to the formalism developed for describing the
distribution of the leptons from vector boson decays in Chapter 2, because the
final state of the diphoton process is also a color singlet state at LO.
Initial--state multiple soft--gluon emission in the scattering subprocesses $%
q{\bar{q}},qg$ and $gg\to \gamma \gamma X$ is resummed by treating the
photon pair $\gamma \gamma $ similarly to the Drell--Yan photon $\gamma ^{*}$%
. In addition, there are contributions in which one of the final photons is
produced through a long--distance fragmentation process. An example is $%
qg\to \gamma q$ followed by the fragmentation of the final state quark $q\to
\gamma X$. An earlier study of soft--gluon resummation effects in photon
pair production may be found in Ref.~\cite{qqres}.

The results of the calculation are compared with CDF \cite{cdfdata}
and \D0 \cite{d0data} data taken at the collider energy $\sqrt{S}=1.8$ TeV.
A prediction for the production rate and kinematic distributions of the
diphoton pair in proton--nucleon interactions at the fixed--target energy $%
\sqrt{S}=31.5$ GeV, appropriate for the E706 experiment at Fermilab \cite
{e706data}, is also presented.

\section{Diphoton Production at Fixed Order \label{sec:DiphotonProduction}}

\FigFeynman
The leading order (LO) subprocesses for diphoton production in hadron
interactions are of order $\alpha_{em}^2$, where $\alpha_{em}$ denotes the
electromagnetic coupling strength. There are three classes of LO partonic
contributions to the reaction $h_1h_2\to \gamma \gamma X$, where $h_1$ and $%
h_2$ are hadrons, illustrated in 
Fig.~(\ref{Fig:Feynman}a)--(\ref{Fig:Feynman}c). 
The first (\ref{Fig:Feynman}a) is the
short--distance $q\bar q\to\gamma\gamma$ subprocess. 
The second (\ref{Fig:Feynman}b) is the
convolution of the short--distance $qg\to\gamma q$ subprocess with the
long--distance fragmentation of the final state quark $q\to \gamma X$. This
is a LO contribution since the hard scattering is of order $\alpha_{em}
\alpha_s$, while fragmentation is effectively of order $\alpha_{em}/\alpha_s$. 
Here, $\alpha_s$ denotes the QCD coupling strength. Class (\ref{Fig:Feynman}b) also
includes the subprocess $q\bar q\to \gamma g$ convoluted with the
fragmentation $g\to \gamma X$. Finally, there are LO contributions 
(\ref{Fig:Feynman}c)
involving subprocesses like $qq\to qq$, where both final state quarks
fragment $q\to\gamma X$. The transverse momenta of the photons are denoted $%
\vec{p}_{T_1}$ and $\vec{p}_{T_2}$, and the transverse momentum of the pair
is $\vec{Q}_T = \vec{p}_{T_1} + \vec{p}_{T_2}$. In the absence of transverse
momentum carried by the incident partons, the LO process (\ref{Fig:Feynman}a) 
provides $\vec{%
Q}_T=0$. With the added assumption of collinear final--state fragmentation,
(\ref{Fig:Feynman}b) provides $\vec{Q}_T = \vec{p}_{T_1} + \vec{p}_{T_2} = (1-z) \vec{p}%
_{T_1}$, where photon 2 carries a fraction $z$ of the momentum of the
final--state quark. Given a lower limit on the magnitude of the transverse
momentum $p_T^\gamma$ of each photon, the total cross section at LO is
finite.

The next--to--leading order (NLO) subprocesses for diphoton production are
of order $\alpha_{em}^2\alpha_s$. One class of one--loop Feynman diagrams
(\ref{Fig:Feynman}d) contributes by interfering with the tree level diagram (\ref{Fig:Feynman}a). Real gluon
emission (\ref{Fig:Feynman}e) is also present at NLO. The subprocess $qg\to\gamma\gamma q$
contains a singular piece (\ref{Fig:Feynman}f) that renormalizes the lower order
fragmentation (\ref{Fig:Feynman}b) and a piece (\ref{Fig:Feynman}g) that is free of final--state collinear
singularities. Finally, subprocesses like $qq\to qq\gamma\gamma$ contain a
regular piece involving photon emission convoluted with a fragmentation
function (\ref{Fig:Feynman}h) and pieces that renormalize the double fragmentation process
(\ref{Fig:Feynman}c). The regular 3--body final state contributions from (\ref{Fig:Feynman}e), (\ref{Fig:Feynman}f), and
(\ref{Fig:Feynman}g) provide $\vec{Q}_T = -\vec{p}_{T_j}$, where $j$ represents the
final--state quark or gluon. The full set of NLO contributions just
described is free of final--state singularities, and the total integrated
cross section at NLO is finite for a finite lower limit on each $p_T^\gamma$.

Higher order calculations in $\alpha_s$ improve the accuracy of predictions
for total cross sections involving quarks or gluons when only one hard scale 
$Q$ is relevant. For $h_1h_2\to \gamma \gamma X$, this scale can be chosen
proportional to the invariant mass of the photon pair, $Q=M_{\gamma \gamma }$%
, which is about equal to $2 p_T^\gamma$ for two well separated photons in
the central rapidity region. For kinematic distributions that depend on more
than one scale, a NLO calculation may be less reliable. 
This is the case for the distribution of the transverse momentum of the photon pair, $Q_T=|\vec{Q}_{T}|$. 
Similarly to the Drell-Yan pair, 
at fixed $Q$, the behavior of the fixed order differential 
cross section for small $Q_T$ is
\begin{eqnarray}
\lim_{Q_T \to 0} \frac{d\sigma}{dQ_T^2}=
\sum_{n=0}^{\infty}\sum_{m=0}^{2n-1} 
\left( \frac{\alpha_s}{\pi} \right)^n
\frac{_n v_m}{Q_T^2}
\ln^m \left( \frac{Q^2}{Q_T^2}\right) + 
{\cal O} \left( \frac{1}{Q_T},\delta(Q_T) \right),
\label{Eq:NLO}
\end{eqnarray}
where $_n v_m$ are calculable perturbatively. The structure of
Eq.~({\ref{Eq:NLO}) indicates that the fixed order QCD prediction is
reliable when $Q_T\simeq Q$, but becomes less reliable when $Q_T\ll Q$,
where $\ln(Q^2/Q_T^2)$ becomes large. In the region $Q_T\ll Q$, the photon
pair is accompanied by soft and/or collinear gluon radiation. To calculate
distributions like $\displaystyle\frac{d{\sigma}}{dQ_T^2}$ reliably in the
region $Q_T \ll Q$, effects of multiple soft gluon emission must be taken
into account. The contributions (\ref{Fig:Feynman}e)
and (\ref{Fig:Feynman}g) exhibit singular behavior that can be tamed by resummation of the
effects of initial--state multiple soft--gluon radiation to all orders in $%
\alpha_s$. Other contributions that do not become singular as $Q_T\to 0$ do
not need to be treated in this manner. Fragmentation contributions like (\ref{Fig:Feynman}b)
are found to be small in magnitude after isolation restrictions are imposed
on the energy of the hadronic remnant from the fragmentation. Therefore,
contributions like (\ref{Fig:Feynman}c) and (\ref{Fig:Feynman}h) are ignored in this work. Gluon
fragmentation to a photon can be ignored, since its magnitude is small. }

The subprocess $gg\to\gamma\gamma$, involving a quark box diagram, is of
order $\alpha_{em}^2\alpha_s^2$. While formally of even higher order than
the NLO contributions considered so far, this LO $gg$ contribution is
enhanced by the size of the gluon parton distribution function.
Consideration of the order $\alpha_{em}^2\alpha_s^3$ correction leads to
resummation of the $gg$ subprocess in a manner analogous to the $q\bar q$
resummation.

\section{Extension of the Resummation Formalism}

\subsection{The Resummation Formula}

To improve upon the prediction of Eq.~(\ref{Eq:NLO}) for the region $Q_T\ll Q
$, perturbation theory can be applied using an expansion parameter $%
\alpha_s^m\ln ^n(Q^2/Q_T^2)$, with $n=0,\ldots ,2m-1,$ instead of $\alpha_s^m
$. The terms $\alpha_s^m\ln^n(Q^2/Q_T^2)$ represent the effects of soft
gluon emission at order $\alpha_s^m$. Resummation of the singular part of
the perturbative series to all orders in $\alpha_s$ by Sudakov
exponentiation yields a regular differential cross section as $Q_T\to 0$.

The differential cross section in the CSS resummation formalism for the
production of photon pairs in hadron collisions is given by 
Eq.~(\ref{eq:ResFor}), with the changes of $M_V=0$ and $\Gamma _V=0$:
\begin{eqnarray}
&&{\frac{d\sigma (h_1h_2\to \gamma _1\gamma _2X)}{dQ^2\,dy\,dQ_T^2\,d\cos {%
\theta }\,d\phi }}={\frac 1{48\pi S}}\,{\frac 1{Q^2}}  \nonumber \\
&&~~\times \left\{ {\frac 1{(2\pi )^2}}\int d^2b\,e^{i{\vec{Q}_T}\cdot {\vec{%
b}}}\,\sum_{i,j}{\widetilde{W}_{ij}(b_{*},Q,x_1,x_2,\theta ,\phi
,C_1,C_2,C_3)}\,\widetilde{W}_{ij}^{NP}(b,Q,x_1,x_2)\right.   \nonumber \\
&&~~~~\left. +~Y(Q_T,Q,x_1,x_2,\theta ,\phi ,{C_4})\right\} .  
\label{master}
\end{eqnarray}
The variables $Q$, $y$, and $Q_T$ here denote the invariant mass, rapidity,
and transverse momentum of the photon pair in the laboratory frame, while $%
\theta $ and $\phi $ are the polar and azimuthal angle of one of the photons
in the Collins--Soper frame \cite{CSFrame}. The initial--state parton
momentum fractions are defined as $x_1=e^yQ/\sqrt{S}$, and $x_2=e^{-y}Q/%
\sqrt{S}$, and $\sqrt{S}$ is the center--of--mass (CM) energy of the hadrons 
$h_1$ and $h_2$.

The renormalization group invariant quantity $\widetilde{W}_{ij}(b)$ sums
the large logarithmic terms $\alpha _s^m\ln ^n(b^2Q^2)$ to all orders in $%
\alpha _s$. For a hard scattering process initiated by the partons $i$ and $j
$, 
\begin{eqnarray}
&&\widetilde{W}_{ij}(b,Q,x_1,x_2,\theta ,\phi ,C_1,C_2,C_3)=\exp \left\{ -%
{\cal S}_{ij}(b,Q{,C_1,C_2})\right\}   \nonumber \\
&&~\times \left[ {\cal C}_{i/h_1}(x_1){\cal C}_{j/h_2}(x_2)+{\cal C}%
_{j/h_1}(x_1){\cal C}_{i/h_2}(x_2)\right] {\cal F}_{ij}(\alpha
_{em}(C_2Q),\alpha _s(C_2Q),\theta ,\phi ),  \label{Eq:WTwiDiPhoton}
\end{eqnarray}
where the Sudakov exponent ${\cal S}_{ij}(b,Q{,C_1,C_2})$ is defined by 
Eq.~(\ref{eq:SudExp}), and ${\cal C}_{i/h}(x)$ denotes the convolution of the
perturbative Wilson coefficient functions $C_{ij}^{(n)}$ with parton
distribution functions $f_{a/h}$ as given by Eq.~(\ref{Eq:DefCalC}). The $%
A_{ij}$ and $B_{ij}$ coefficients of the Sudakov exponent and the functions $%
C_{ij}$ are calculated perturbatively in powers of $\alpha _s/\pi $ as in
Eq.~(\ref{eq:ABCExp}). The kinematic factor ${\cal F}_{ij}$ will be defined
later for each particular partonic process.

The dimensionless constants $C_1,\;C_2$ and $C_3\equiv \mu b$ were
introduced in the solution of the renormalization group equations for $%
\widetilde{W}_{ij}$. The constant $C_1$ determines the onset of
non-perturbative physics, $C_2$ specifies the scale of the hard scattering
process, and $\mu =C_3/b$ is the factorization scale at which the $%
C_{ij}^{(n)}$ functions are evaluated. We use the conventional choice of the
renormalization constants: $C_1=C_3=2e^{-\gamma _E}\equiv b_0$ and $C_2=C_4=1
$ \cite{Collins-Soper-Sterman}, where $\gamma _E$ is the Euler constant.

As in the case of lepton pair production, in Eq.~(\ref{master}), the impact
parameter $b$ is to be integrated from 0 to $\infty $. However, for $b\ge b_{%
{\rm max}}$, which corresponds to an energy scale less than $1/b_{{\rm max}}$%
, the QCD coupling $\alpha _s(\bar{\mu}\sim 1/b)$ becomes so large that a
perturbative calculation is no longer reliable, and non-perturbative physics
must set in. The non-perturbative physics in this region is described by the
empirically fit function $\widetilde{W}_{ij}^{NP}$ 
\cite{npfit,Ladinsky-Yuan}, and 
$\widetilde{W}_{ij}$ is evaluated at a revised value of $b$, $\displaystyle %
b_{*}={\frac b{\sqrt{1+(b/b_{{\rm max}})^2}}},$ where $b_{{\rm max}}$ is a
phenomenological parameter used to separate long and short distance physics.
With this change of variable, $b_{*}$ never exceeds $b_{{\rm max}}$; $b_{%
{\rm max}}$ is a free parameter of the formalism \cite{css} that can be
constrained by other data (e.g. Drell--Yan).

The function $Y$ in Eq.~(\ref{master}) contains contributions in the full
NLO perturbative calculation that are less singular than ${Q_T^{-2}}$ or $%
Q_T^{-2}\ln({Q^2 /Q_T^2})$ as $Q_T \to 0$ (both the factorization and the
renormalization scales are chosen to be $C_4 Q$). It is the difference
between the exact perturbative result to a given order and the result from $%
\widetilde{W}_{ij}$ expanded to the same fixed order (called the asymptotic
piece). The function $Y$ restores the regular contribution in the fixed
order perturbative calculation that is not included in the resummed piece $%
\widetilde{W}_{ij}$. It does not contain a contribution from final--state
fragmentation, which is included separately as described in
Section \ref{sec:Fragmentation}.

The CSS formula Eq.~(\ref{master}) contains many higher--order logarithmic
terms, such that when $Q_T \sim Q$, the resummed differential cross section
can become negative in some regions of phase space. In this calculation, the
fixed--order prediction for the differential cross section is used for $Q_T %
\grsim Q$ whenever it is larger than the prediction from Eq.~(\ref{master}).
The detailed properties of this matching prescription can be found in Ref. 
\cite{wres}.

\subsection{Resummation for the $q\bar{q}\to \gamma \gamma $ subprocess}

For the $q\bar{q}\to \gamma \gamma $ subprocess, the application of the CSS
resummation formalism is similar to the Drell--Yan case $q{\bar{q}}%
^{~(^{\prime })}\to V^{*}\to \ell _1{\bar{\ell}_2}$, where $\ell _1$ and $%
\ell _2$ are leptons produced through a gauge boson $V^{*}$ \cite{wres}.
Since both processes are initiated by $q{\bar{q}}^{~(^{\prime })}$ color
singlet states, the $A^{(1)},A^{(2)}$ and $B^{(1)}$ coefficients in the Sudakov
form factor are identical to those of the Drell--Yan case when each photon
is in the central rapidity region with large transverse momentum and is well
separated from the other photon. This universality can be understood as
follows. The invariants $\hat{s}$, $\hat{t}$ and $\hat{u}$ are defined for
the $q(p_1)\bar{q}(p_2)\to \gamma (p_3)\gamma (p_4)$ subprocess as 
\[
\hat{s}=(p_1+p_2)^2,~~~~~~\hat{t}=(p_1-p_3)^2,~~~~~~\hat{u}=(p_2-p_3)^2.
\]
The transverse momentum of each photon can be written as $p_T^\gamma =\sqrt{%
\hat{t}\hat{u}/\hat{s}}$. When $p_T^\gamma $ is large, $\hat{t}$ and $\hat{u}
$ must also be large, so the virtual--quark line connecting the two photons
is far off the mass shell, and the leading logarithms due to soft gluon
emission beyond the leading order can be generated only from the diagrams in
which soft gluons are connected to the incoming (anti--)quark. To obtain the 
$B^{(2)}$ function, it is necessary to calculate beyond NLO, so it is not
included in this calculation. However, the Sudakov form factor becomes more
accurate when more terms are included in $A_{ij}$ and $B_{ij}$. Since the
universal functions $A_{ij}^{(n)}$ depend only on the flavor of the incoming
partons (quarks or gluons), $A_{q\bar{q}}^{(2)}$ can be appropriated from
Drell--Yan studies, and its contribution {\it is} included in this work.

To describe the effects of multiple soft--gluon emission, Eq.~(\ref{master})
can be applied, where $i$ and $j$ represent quark and anti--quark flavors,
respectively, and
\[
{\cal F}_{ij}=2\delta _{ij}(g_L^2+g_R^2)^2(1+\cos ^2\theta )/(1-\cos
^2\theta ).
\]
The couplings $g_{L,R}$ are defined through the $q{\bar{q}}\gamma $ vertex,
written as 
\[ 
i\gamma _\mu \left[ g_L(1-\gamma _5)+g_R(1+\gamma _5)\right],
\]
with $g_L=g_R=eQ_f/2$, and $eQ_f$ is the electric charge of the incoming
quark ($Q_u=2/3,Q_d=-1/3$). The explicit forms of the $A$ and $B$ coefficients
used in the numerical calculations are: 
\begin{eqnarray*}
A_{q\bar{q}}^{(1)}(C_1) &=&C_F, \\
A_{q\bar{q}}^{(2)}(C_1) &=&C_F\left[ \left( \frac{67}{36}-\frac{\pi ^2}{12}%
\right) N_C-\frac 5{18}N_f-2\beta _1\ln \left( \frac{b_0}{C_1}\right)
\right] , \\
B_{q\bar{q}}^{(1)}(C_1,C_2) &=&C_F\left[ -\frac 32-2\ln \left( \frac{C_2b_0}{%
C_1}\right) \right] ,
\end{eqnarray*}
where $N_f$ is the number of light quark flavors, $N_C=3$, $C_F=4/3$, and $%
\beta _1=(11N_C-2N_f)/12$.

To obtain the value of the total cross section to NLO, it is necessary to
include the Wilson coefficients $C_{ij}^{(0)}$ and $C_{ij}^{(1)}$. These can
be derived from the full set of LO contributions and NLO corrections to 
$\gamma \gamma $ production \cite{Bailey-Owens-Ohnemus}. 
After the leading order and the
one--loop virtual corrections to $q\bar{q}\to \gamma \gamma $ and the tree
level contribution from $q\bar{q}\to \gamma \gamma g$ are included, the
coefficients are:
\begin{eqnarray}
C_{jk}^{(0)}(z,b,\mu ,{\frac{C_1}{C_2}}) &=&\delta _{jk}\delta ({1-z}), 
\nonumber \\
C_{jG}^{(0)}(z,b,\mu ,{\frac{C_1}{C_2}}) &=&0,  \nonumber \\
C_{jk}^{(1)}(z,b,\mu ,{\frac{C_1}{C_2}}) &=&\delta _{jk}C_F\left\{ \frac
12(1-z)-\frac 1{C_F}\ln \left( \frac{\mu b}{b_0}\right) P_{j\leftarrow
k}^{(1)}(z)\right.   \nonumber \\
&&\left. +\delta (1-z)\left[ -\ln ^2\left( {\frac{C_1}{{b_0C_2}}}%
e^{-3/4}\right) +{\frac{{\cal V}}4}+{\frac 9{16}}\right] \right\} .
\label{eq:quarks}
\end{eqnarray}
After factorization of the final--state collinear singularity, as described
below, 
the real emission subprocess $qg\to \gamma \gamma q$ yields: 
\[
C_{jG}^{(1)}(z,b,\mu ,{\frac{C_1}{C_2}})={\frac 12}z(1-z)-\ln \left( \frac{%
\mu b}{b_0}\right) P_{j\leftarrow G}^{(1)}(z).
\label{eq:gluons}
\]
In the above expressions, the splitting kernels \cite{DGLAP} are 
given by Eq.~\ref{eq:DGLAP}.
For photon pair production, the function ${\cal V}$ is 
\begin{eqnarray}
{\cal V}_{\gamma \gamma } &=&-4+{\frac{\pi ^2}3}+{\frac{\hat{u}\hat{t}}{\hat{%
u}^2+\hat{t}^2}}\left( F^{virt}(v)-2\right) ,  \nonumber  \label{eq:Vgg} \\
F^{virt}(v) &=&\left( 2+{\frac v{1-v}}\right) \ln ^2(v)+\left( 2+{\frac{1-v}v%
}\right) \ln ^2(1-v)  \nonumber \\
&&+\left( {\frac v{1-v}}+{\frac{1-v}v}\right) \left( \ln ^2(v)+\ln ^2(1-v)-3+%
{\frac{2\pi ^2}3}\right)   \nonumber \\
&&+2\left( \ln (v)+\ln (1-v)+1\right) +3\left( {\frac v{1-v}}\ln (1-v)+{%
\frac{1-v}v}\ln (v)\right) ,
\end{eqnarray}
where $v=-\hat{u}/\hat{s}$, and $\hat{u}=-\hat{s}(1+\cos \theta )/2$ in the $%
q\bar{q}$ center--of--mass frame. Because of Bose symmetry, $%
F^{virt}(v)=F^{virt}(1-v)$. A major difference from the Drell--Yan case ($%
{\cal V}_{DY}=-8+\pi ^2$) is that ${\cal V}_{\gamma \gamma }$ depends on the
kinematic correlation between the initial and final states through its
dependence on $\hat{u}$ and $\hat{t}$.

The non-perturbative function used in this study is the empirical fit 
\cite{Ladinsky-Yuan} 
\[
\widetilde{W}_{q\overline{q}}^{NP}(b,Q,Q_0,x_1,x_2)={\rm exp}\left[
-g_1b^2-g_2b^2\ln \left( {\frac Q{2Q_0}}\right) -g_1g_3b\ln {(100x_1x_2)}%
\right] ,
\]
where $g_1=0.11_{-0.03}^{+0.04}~{\rm GeV}^2$, $g_2=0.58_{-0.2}^{+0.1}~{\rm %
GeV}^2$, $g_3=-1.5_{-0.1}^{+0.1}~{\rm GeV}^{-1}$, and $Q_0=1.6~{\rm GeV}$.
(The value $b_{max}=0.5~{\rm GeV}^{-1}$ was used in determining the above $%
g_i$'s and for the numerical results presented in this work.) These values
were fit for the CTEQ2M parton distribution function, with the
conventional choice of the renormalization constants, i.e. $C_1=C_3=b_0$ and 
$C_2=1$. In principle, these coefficients should be refit for the CTEQ4M
distributions \cite{cteq} used in this study. The parameters of 
Eq.~(\ref{Eq:WNonPert}) 
were determined from Drell--Yan data. It is assumed that the same
values should be applicable for the $\gamma \gamma $ final state.

\subsection{Contributions From $qg$ Subprocesses}

As described in Section \ref{sec:DiphotonProduction}, 
the complete NLO calculation of diphoton production
in hadron collisions includes photons from long--distance fragmentation
processes like (\ref{Fig:Feynman}b) and short--distance processes like (\ref{Fig:Feynman}f) and (\ref{Fig:Feynman}g). The
latter processes yield a regular 3--body final state contribution, while the
former describes a photon recoiling against a collinear quark and photon.

The singular part of the squared amplitude of the $q(p_1)g(p_2)\to \gamma
(p_3)\gamma (p_4)q(p_5)$ subprocess can be factored into a product of the
squared amplitude of $q(p_1)g(p_2)\to \gamma (p_3)q(p_{4+5})$ and the
splitting kernel for $q(p_{4+5})\to \gamma (p_4)q(p_5)$. In the limit that
the emitted photon $\gamma (p_4)$ is collinear with the final state quark $%
q(p_5)$: 
\begin{eqnarray}
\lim_{p_4\parallel p_5} &&\left| {\cal M}\left( q(p_1)g(p_2)\to \gamma
(p_3)\gamma (p_4)q(p_5)\right) \right| ^2=  \nonumber \\
&&{\frac{e^2}{p_4\cdot p_5}}P_{\gamma \leftarrow q}^{(1)}\left( z\right)
\left| {\cal M}\left( q(p_1)g(p_2)\to \gamma (p_3)q(p_{4+5})\right) \right|
^2.  \label{subtraction}
\end{eqnarray}
A similar result holds when $p_3$ and $p_5$ become collinear and/or the
quark is replaced with an anti--quark. Conventionally, the splitting
variable $z$ is the light--cone momentum fraction of the emitted photon with
respect to the fragmenting quark, $z=p_4^{+}/(p_4^{+}+p_5^{+})$, where $%
p_i^{+}=\left( p_i^{(0)}+p_i^{(3)}\right) /\sqrt{2}$. ($p_i^{(0)}$ is the
energy and $p_i^{(3)}$ is the longitudinal momentum component along the
moving direction of the fragmenting quark in the $qg$ center-of-mass frame.)
Alternatively, since the final state under consideration contains only a
fragmenting quark and a spectator, a Lorentz invariant splitting variable
can be defined as: \cite{Catani-Seymour} 
\[
\tilde{z}=1-\frac{p_i\cdot p_k}{p_j\cdot p_k+p_i\cdot p_k+p_i\cdot p_j}.
\]
In this notation, $i=5$ is the fragmentation quark, $j$ is the fragmentation
photon, and $k$ is the prompt spectator photon. When the pair $ij$ becomes
collinear, $\tilde{z}$ becomes the same as the light--cone momentum fraction 
$z$ carried by the photon. Aside from the color factor, $P_{\gamma
\leftarrow q}^{(1)}(z)$ in Eq.~(\ref{subtraction}) is the usual DGLAP
splitting kernel for $q\to gq$ 
\[
P_{\gamma \leftarrow q}^{(1)}(z)=\left( \frac{1+(1-z)^2}z\right) .
\]

The regular contribution $qg\to\gamma\gamma q$ (\ref{Fig:Feynman}f) is defined by removing
the final--state, collinear singularity from the full amplitude of the
partonic subprocess. The matrix element squared for (\ref{Fig:Feynman}f) can be written \cite
{Catani-Seymour}: 
\begin{eqnarray}
\left| {\cal M}\left( qg\to\gamma\gamma q \right) \right|^2_{reg} &=& 
\nonumber \\
\left| {\cal M}\left( qg\to\gamma\gamma q \right) \right|^2_{full} &-& {{%
\frac{{e^2}}{{p_4 \cdot p_5}}}}P_{\gamma\leftarrow q}^{(1)} \left(\tilde
z\right) \left| {\cal M}\left( q(p_1)g(p_2)\to \gamma (p_3)q(p_{4+5})\right)
\right| ^2.  \label{regular}
\end{eqnarray}
After the final--state collinear singularity is subtracted, the remainder
expresses the regular 3--body final state contribution $\gamma\gamma q$.
This remainder, as shown in (\ref{Fig:Feynman}g), contains terms that diverge when $Q_T\to 0$
which should be regulated by renormalizing the parton distribution at the
NLO. The contribution from this divergent part is included in the resummed $%
q\bar q$ cross section in $C^{(1)}_{jG}$, as shown in Eq.~(\ref{eq:gluons}).
The part that is finite as $Q_T\to 0$ is included in the function $Y$. When $%
Q_T \grsim Q$, Eq.~(\ref{regular}) describes the NLO contribution from the $%
qg\to\gamma\gamma q$ subprocess to the $Q_T$ distribution of the photon
pair. The subtracted final--state collinear singularity from the NLO $%
qg\to\gamma\gamma q$ subprocess is absorbed into the fragmentation process
(\ref{Fig:Feynman}b).

\subsection{Fragmentation Contributions \label{sec:Fragmentation}}

Final--state photon fragmentation functions $D_{\gamma /i}(z,\mu _F^2)$ are
introduced in an analogous manner to initial--state parton distribution
functions $f_{i/h_1}(x,\mu _I^2)$. Here, $z(x)$ is the light--cone momentum
fraction of the fragmenting quark (incident hadron) carried by the photon
(initial--state parton), and $\mu _F(\mu _I)$ is the final state (initial
state) fragmentation (factorization) scale. The parton--level cross section
for the fragmentation contribution (\ref{Fig:Feynman}b) is evaluated from the general
expression for a hard scattering to a parton $m$, which then fragments to a
photon: 
\[
d\hat{\sigma}=\frac 1{2\hat{s}}|{\cal M}(p_1p_2\to p_3\ldots
p_m)|^2d^{(m-2)}[PS]dzD_{\gamma /m}(z,\mu _F^2).
\]
Here, ${\cal M}$ is the matrix element for the hard scattering subprocess, $%
d^{(m-2)}[PS]$ is the $m-2$--body phase space, and an integral is performed
over the photon momentum fraction $z$ weighted by the fragmentation function 
$D_{\gamma /m}(z,\mu _F^2)$. Since fragmentation is computed here to LO
only, the infrared divergences discussed by Berger, Guo and Qiu are not an
issue \cite{BGQiu}.

The fragmentation function $D_{\gamma \leftarrow q}$ obeys an evolution
equation, and the leading--logarithm, asymptotic solution 
$D_{\gamma /q}^{LL}$ is \cite{Bailey-Owens-Ohnemus}:
\begin{eqnarray}
D_{\gamma /q}^{LL}(z,\mu _F^2) &=&{\frac{\alpha _{em}}{2\pi }}\ln \left( {%
\frac{\mu _F^2}{\Lambda _{QCD}^2}}\right) D_{\gamma \leftarrow q}^{(1)}(z), 
\nonumber \\
zD_{\gamma \leftarrow q}^{(1)}(z) &=&\frac{Q_q^2(2.21-1.28z+1.29z^2)z^{0.049}%
}{1-1.63\ln (1-z)}+0.0020(1-z)^{2.0}z^{-1.54},  
\label{Eq:DefFrag}
\end{eqnarray}
where $\Lambda _{QCD}$ is the QCD scale for four light quark flavors. As
shown in Fig. \ref{Fig:Frag}, the collinear approximation made in defining $%
D_{\gamma \leftarrow q}$ leads to kinematic distributions with an
unrealistic sensitivity to kinematic cuts, such as cuts to define an
isolated photon.
\FigFrag

The Monte Carlo showering method goes beyond the collinear approximation
used in solving the evolution equation for the fragmentation function $%
D_{\gamma\leftarrow q}$. In Monte Carlo calculations, the probability for
photon emission is determined from the splitting function $%
P_{\gamma\leftarrow q}(z)$, which is a collinear approximation. However, the
kinematics are treated by assigning a virtuality to the fragmenting quark
whose value lies between the hard scale of the process and a
phenomenological cutoff $\sim 1$ GeV. This cutoff replaces the parameter $%
\Lambda_{QCD}$ in Eq.~(\ref{Eq:DefFrag}). Most importantly, gluon emission
can be incorporated into the description of final state fragmentation.
Because there is no collinear approximation in the kinematics, kinematic
distributions do not exhibit the unrealistic behavior of the parton--level
calculation. The ``correctness'' of either approach can be judged only after
a careful comparison of their respective predictions.

The collinear approximation becomes an issue because of the experimental
definition of isolated photons. Experimentally, an isolation cut is
necessary to separate prompt photons from various hadronic backgrounds,
including $\pi^\circ$ and $\eta$ meson decays. The separation between a
particle $j$ and the photon is expressed as $R_j = \sqrt{(\eta-\eta_j)^2 +
(\phi-\phi_j)^2}$, where the coordinates $\eta (\eta_j)$ and $\phi (\phi_j)$
are the pseudorapidity and azimuthal angle of the photon (particle $j$). At
hadron colliders, the standard isolation criterion is that the sum of excess
transverse energy $E_T$ contained inside a cone of size $R_0$ centered on
the photon candidate is below a cutoff $E_T^{iso}$, $\displaystyle 
\sum_{R_j<R_0} E_T^j < E_T^{iso}$. The sum is over each particle $j$. Since
the resummed CSS piece of the final state cross section describes the
radiation of multiple soft gluons approximately collinear with the incident
partons, it produces only isolated photons. For NLO $\gamma\gamma j$ final
states (\ref{Fig:Feynman}e), (\ref{Fig:Feynman}f), and (\ref{Fig:Feynman}g), where there is only one extra parton $j=q$ or $g
$, isolation enforces a separation $R_{j} \ge R_0$, provided that $p_{T_j} >
E_T^{iso}$. Above $Q_T=E_T^{iso}$, the perturbative corrections contained in
the function $Y$ are affected by isolation. On the other hand, because of
the collinear approximation, the parton--level fragmentation calculation
based on Eq.~(\ref{Eq:DefFrag}) does not depend on the isolation cone $R_0$;
the hadronic remnant of the fragmentation (\ref{Fig:Feynman}b) {\it always} satisfies $R<R_0$%
. Hence, for this case, $\vec{Q}_T = (1-z) \vec{p}_{T_1}$, and the isolation
cut reduces to a step function requirement $\theta(E_T^{iso}-Q_T)$.

The parton--level calculation of the fragmentation contribution at the
Tevatron based on the fragmentation function $D_{\gamma \leftarrow q}(z,\mu
_F^2)$ has been compared with a Monte Carlo estimate based on PYTHIA \cite
{pythia}. For the parton--level calculation, the scale $\mu _F=M_{\gamma
\gamma }$ is used. For the PYTHIA calculation, the scale is $\mu _F=%
\sqrt{\hat{s}}$, and hadronization is not performed, so that no photons
arise from $\pi ^{\circ }$ or $\eta $ meson decays, for example. For this
comparison, the invariant mass $\sqrt{\hat{s}}$ of the hard--scattering
subprocess is limited to $20<\sqrt{\hat{s}}<50$ GeV in both approaches, and
the photons are required to satisfy $p_T^\gamma >5$ GeV and $|\eta ^\gamma
|<2$. These kinematic cuts are chosen to increase the statistics of the
PYTHIA calculation, while reflecting the kinematic region of interest for a
comparison with data. PYTHIA can simulate the QED and QCD showering of the
final--state quark as well as the QCD showering of the initial--state quark
and gluon. To isolate the effect of initial--state gluon radiation, PYTHIA
calculations were performed with and without the QCD initial--state
radiation (i.e. by preventing space--like showering). In neither case is
initial--state QED radiation simulated. It is possible for the partons
produced in initial--state showering to develop time--like showering. Any
photons produced from this mechanism are discarded, since they are formally
of higher--order than the contributions considered here. Such contributions,
however, might be necessary to understand photon pairs with small invariant
mass and small $Q_T$.

Figure~\ref{Fig:Frag} is a comparison of kinematic quantities from the
parton--level and Monte Carlo calculations. The left--side of Fig.~\ref
{Fig:Frag} shows the $Q_T$--distribution for the parton--level (solid),
PYTHIA with initial--state radiation of gluons (short--dashed), and PYTHIA
without initial--state radiation (long--dashed) calculations. Each curve is
plotted twice, with and without an isolation cut $E_T^{iso}=4$ GeV and $%
R_0=0.7$. Before the isolation cut, the total parton--level fragmentation
cross section is approximately 50\% higher than the Monte Carlo cross
section. After isolation, the total cross sections are in good agreement,
even though the parton--level calculation is discontinuous at $Q_T=E_T^{iso}$%
. The effect of initial--state gluon radiation in the PYTHIA calculation is
to increase $Q_T$ without compromising the isolation of the photons.

The right--side of Fig.~\ref{Fig:Frag} shows the distribution of the
light--cone momentum fraction $z$ of the quark carried by the fragmentation
photon (for this figure, $z$ is defined in the laboratory frame). After
isolation, the parton--level contribution is limited to $z>0.55$ by
kinematics, whereas the Monte Carlo contribution is more uniformly
distributed between 0 and 1. For the PYTHIA result, $z$ is calculated with
respect to the final state quark {\it before} showering. In the showering
process, some energy--momentum can be exchanged between the final state
prompt photon and the fragmenting quark, since the quark is assigned a
virtuality. As a result, the effective $z$ value can extend beyond the naive
limit $z=1$.

The conclusions of this comparison are: (1) after isolation, the total cross
sections from the parton--level and Monte Carlo fragmentation calculations
are in good agreement, and (2) the Monte Carlo kinematic distributions (e.g. 
$Q_T$ and $z$) are not very sensitive to the isolation cut. For these
reasons, the Monte Carlo estimate with initial--state radiation is used to
account for the (\ref{Fig:Feynman}b) contribution in the final results. Furthermore, with
initial--state radiation, the PYTHIA calculation includes the leading
effects of a full resummation calculation of the $qg\to \gamma q$ process.
It is approximately equivalent to performing a resummation calculation in
the CSS formalism with coefficients $A^{(1)}$ and $B^{(1)}$ calculated for a $%
qg$ initial state and the LO Wilson function.

One final comparison was made with the Monte Carlo calculation by treating
the subtracted term in Eq. (\ref{regular}), with $P^{(1)}$ replaced by $%
D^{(1)}$ defined in Eq. (\ref{Eq:DefFrag}), as a 3--body matrix element. The
collinear divergence was regulated by requiring a separation $R_0$ between
the photon and quark remnant for all $Q_T$. This calculation agrees with
PYTHIA in the shape and normalization of various distributions, except when $%
Q_T<E_T^{iso}$, where there is a substantial difference.

\subsection{Resummation for the $gg\to \gamma \gamma $ subprocess}

A resummation calculation for the $gg\to \gamma \gamma $ subprocess is
included in the theoretical prediction. The LO contribution comes from
one--loop box diagrams of order $\alpha _{em}^2\alpha_s^2$ in perturbative
QCD. At present, a full NLO calculation, of ${\cal O}(\alpha_{em}^2%
\alpha_s^3)$,  for this process is not available. Nevertheless, the
resummation technique can be applied to resum part of the higher order
contributions and improve the theoretical prediction. The exact NLO $gg\to
\gamma \gamma g$ calculation must include gluon emission from the internal
quark lines of the box diagram, thus generating pentagon diagrams. However,
such diagrams do not generate large logarithms when the final state photons
have large transverse momentum, are in the central rapidity region, and are
well separated from each other. All the large logarithms originate from the
diagrams with soft gluons coupling to the initial--state gluons. Similarly,
the exact NLO $qg\to\gamma\gamma q$ calculation, of ${\cal O}%
(\alpha_{em}^2\alpha_s^3)$, must include contributions involving a box
diagram with one incoming gluon off shell. Large logarithms only arise from
soft gluon emission off the initial--state quark or gluon. The leading
logarithms due to initial--state radiation are universal, and the $A^{(1)}$
and $A^{(2)}$ coefficients calculated for the resummed $gg\to H$ process \cite
{higgsres,cpyhiggs} or the color singlet part of the $gg\to Q\bar Q$ process 
\cite{resum4} can be applied directly to the resummed $gg\to\gamma\gamma$
calculation, since these subprocesses have the same QCD\ color structure.

When the transverse momentum of the photon pair is much smaller than its
invariant mass, i.e. $Q_T\ll Q$, and each photon has large transverse
momentum, then the box diagram of the hard scattering subprocess $gg\to
\gamma \gamma $ can be approximated as a point--like interaction (multiplied
by a form factor which depends on $\hat{s},\hat{t}$ and $\hat{u}$). This
approximation ignores pentagon diagrams in the $gg\to\gamma\gamma g$
subprocess and the virtuality of intermediate quarks in the $qg\to
\gamma\gamma q$ subprocess. It does not have the complete structure of the
hard process, but it does contain the most important logarithmic terms from
initial state gluon radiation. Under such an approximation, the sub-leading
logarithmic terms associated with $B^{(1)}$, and $C^{(1)}$ of
Eqs.~(4) and (5) can be included in the resummation calculation. These
functions were calculated for the $gg\to H$ process \cite{higgsres,cpyhiggs}%
. Without a complete ${\cal O}(\alpha_{em}^2\alpha_s^3)$ calculation, the
exact Wilson coefficient function $C^{(1)}$ is not known. Since part of the
exact $C^{(1)}$ function must include the piece for the $gg\to H$ process,
it is included to estimate the possible NLO enhancement to the production
rate of the $gg$ subprocess. After these ingredients are incorporated into
Eq. (\ref{master}), the resummed kinematics of the photon pair from the $%
gg\to \gamma \gamma $ subprocess can be obtained. The distribution of the
individual photons can be calculated approximately from the LO\ angular
dependence of the box diagram.

The above approximation certainly fails when $Q_T$ is of the order of $Q$.
In the absence of a complete ${\cal O}(\alpha_{em}^2\alpha_s^3)$ 
calculation of the $gg\to \gamma \gamma g$ and $qg\to\gamma\gamma q$
subprocesses, it is not possible to estimate the uncertainties introduced by
the approximation. In the limit of $Q_T\ll Q$, the approximation should be
reliable, since the soft gluon approximation is applicable. In the same
spirit, the approximate function $Y$ for photon pair production is taken
from the results of the perturbative piece for the $gg\to Hg$ and $g\qgen\to
H\qgen$ processes \cite{higgsres,cpyhiggs}.

\smallskip In summary, the resummed distributions of the photon pair from
the $gg$ subprocess in the region of $Q_T\ll Q$ can be described by Eq. (\ref
{master}), with $i=j=g$, and ${\cal F}_{gg}=N_C\left| {\cal M}_{gg\to \gamma
\gamma }(s,t,u)\right| ^2/2^{11}$. Here, $\left| {\cal M}_{gg\to \gamma
\gamma }(s,t,u)\right| ^2$ is the absolute square of the invariant amplitude
of the $gg\to \gamma \gamma $ subprocess \cite{box} summed over spins,
colors, and the fermion flavors in the box loop, but without the
initial--state color $(1/8^2)$, spin $(1/2^2)$ average, and the final--state
identical particle $(1/2)$ factors. The $A$ and $B$ functions used in the
calculation for the $gg$ initial state are 
\begin{eqnarray*}
A_{gg}^{(1)}(C_1) &=&C_A=3, \\
A_{gg}^{(2)}(C_1) &=&\frac{C_A}{C_F}A_{q\bar{q}}^{(2)}(C_1), \\
B_{gg}^{(1)}(C_1,C_2) &=&2\left[ 3\ln \left( \frac{C_1}{C_2b_0}\right)
-\beta _1\right] .
\end{eqnarray*}
The LO and NLO Wilson coefficients, extracted from the $gg\to H$ subprocess,
are: 
\begin{eqnarray*}
C_{gg}^{(0)}\left( z,b;\frac{C_1}{C_2};\mu \right)  &=&\delta (1-z), \\
C_{qg}^{(0)}\left( z,b;\frac{C_1}{C_2};\mu \right)  &=&0 \\
C_{gg}^{(1)}\left( z,b;\frac{C_1}{C_2};\mu \right)  &=&-\ln \!\left( \frac{%
\mu b}{b_0}\right) P_{g\leftarrow g}(z)+\delta (1-z)\left\{ \frac{11}4+\frac{%
3\pi ^2}4\right.  \\
&&\left. -3\ln ^2\left( \frac{C_1}{C_2b_0}\right) +3\ln \left( \frac{C_1}{%
C_2b_0}\right) +(2\beta _1-3)\ln \left( \frac{\mu b}{b_0}\right) \right\} ,
\\
C_{qg}^{(1)}\left( z,b;{\frac{C_1}{C_2}};\mu \right)  &=&-\ln \left( \frac{%
\mu b}{b_0}\right) P_{g\leftarrow q}(z)+\frac 23z.
\end{eqnarray*}

Since the NLO pentagon and off--shell box diagram calculations are not
included, the Wilson coefficients $C_{ij}^{(1)}$ are expected to predict
accurately the total cross section only when $Q_T\ll Q$, the transverse
momenta of the individual photons are large, and their rapidities are small.
Under the {\it approximation} made above, the resummed $gg$ result increases
the integrated rate by about a factor of 2, for kinematic cuts typical of
the Tevatron, as compared to the lowest order (one--loop calculation)
perturbative result. This comparison suggests that the full NLO contribution
of the $gg$ initiated subprocess is large. Because it is necessary to impose
the condition $Q_T<Q$ to make the above approximations valid, the $gg$
resummed result presented in this work probably underestimates the rate when 
$Q_T$ is large or the separation of the azimuthal angle ($\Delta\phi$)
between the two photons is small. This deficiency can be improved only by a
complete ${\cal O}(\alpha_{em}^2\alpha_s^3)$ calculation.

At the Tevatron, the $gg$ contribution is important when the invariant mass (%
$M_{\gamma \gamma }=Q$) of the two photon pair is small. Because of the
approximation made in the $gg$ calculation beyond the LO, the prediction
will be more reliable for the data with larger $Q$. A more detailed
discussion is presented in the next section.

The full calculation of the $gg$ contribution in the CSS formalism depends
also upon the choice of non-perturbative functions. However, the best fits to
the parametrizations are performed for $q\bar{q}$ processes \cite
{npfit,Ladinsky-Yuan}. 
Two assumptions were studied: $(i)$ the non-perturbative
functions are truly universal for $q\bar{q}$ and $gg$ processes, and $(ii)$
the non-perturbative functions obey the same renormalization group properties
as the $A$ functions for each type of process (which are universal for all $q%
\bar{q}$ or $gg$ subprocesses), so the coefficient of the $\ln \left( \frac
Q{2Q_0}\right) $ term in the non-perturbative function Eq.~(\ref{Eq:WNonPert}) 
is scaled by $C_A/C_F$ relative to that of the $q\bar{q}$ process.
Specifically, the different assumptions are: 
\begin{eqnarray*}
(i)\;\widetilde{W}_{gg}^{NP}(b,Q,Q_0,x_1,x_2) &=&\widetilde{W}_{q\bar{q}%
}^{NP}(b,Q,Q_0,x_1,x_2) \\
(ii)\;\widetilde{W}_{gg}^{NP}(b,Q,Q_0,x_1,x_2) &=&\widetilde{W}_{q\overline{q%
}}^{NP}(b,Q,Q_0,x_1,x_2)\;\;\text{with}\;\;g_2\to {\frac{C_A}{C_F}}g_2.
\label{Eq:ggnp}
\end{eqnarray*}
The numerical values of $g_1,g_2,$ and $g_3$ are listed following Eq.~(\ref
{Eq:WNonPert}). These two assumptions do not exhaust all possibilities, but 
ought to be representative of reasonable choices. Choice $(ii)$ is used for 
the results presented in this work. The effect of different choices is 
discussed in the next Section.

\section{Numerical Results}

\subsection{Tevatron Collider Energies}

Two experimental collaborations at the Tevatron $p\bar p$ collider have
collected diphoton data at $\sqrt{S}=1.8$ TeV: CDF \cite{cdfdata}, with 84 pb%
$^{-1}$, and \D0~\cite{d0data}, with 81 pb$^{-1}$. The kinematic cuts
applied to the resummed prediction for comparison with the CDF data are $%
p_T^\gamma > 12$ GeV and $|\eta^{\gamma}|<0.9$. For \D0, the kinematic cuts
are $p_T^{\gamma_1} > 14$ GeV and $p_T^{\gamma_2} > 13$ GeV, and $%
|\eta^{\gamma}|<1$. For CDF, an isolation cut for each photon of $R_0=0.7$
and $E_T^{iso}=4$ GeV is applied; for \D0, the cut is $R_0=0.4$ and $%
E_T^{iso}=2$ GeV.

Other ingredients of the calculation are: $(i)$ the CTEQ4M parton
distribution functions, $(ii)$ the NLO expression for $\alpha _s$, $(iii)$
the NLO expression for $\alpha _{em}$, and $(iv)$ the non-perturbative
coefficients of Ladinsky--Yuan \cite{Ladinsky-Yuan}.

\FigCDFa
The predictions for the CDF cuts and a comparison to the data are shown in
Figs.~\ref{Fig:CDF0}--\ref{Fig:CDF2}. Figure~\ref{Fig:CDF0} shows the
distribution of the photon pair invariant mass, $d\sigma/dM_{\gamma\gamma}$
vs. $M_{\gamma\gamma}$.  The dashed--dot curve represents the resummation of
the $gg$ subprocess, which is the largest contribution for $%
M_{\gamma\gamma}\lesim 30$ GeV. The long--dashed curve represents the full 
$q\bar q$ resummation, while the short--dashed curve is a similar
calculation with the gluon parton distribution function artificially set to
zero. Schematically, there are contributions to the resummed calculation
that behave like $q\to g q_1\otimes q_1\bar q\to\gamma\gamma$ and $g\to \bar
q q_1 \otimes q_1\bar q\to\gamma\gamma$. These contributions are contained
in the terms proportional to $P_{j\leftarrow k}^{(1)}(z)$ in Eq.~(\ref
{eq:quarks}) and $P_{j\leftarrow G}^{(1)}(z)$ in Eq.~(\ref{eq:gluons}),
respectively. The full $q\bar q$ resummation contains both the $q\bar q$ and 
$qg$ contributions. The short--dashed curve is calculated by setting $%
C_{jG}^{(1)}=0$ and retaining only the $q\bar q$ contribution in the
function $Y$. Since the short--dashed curve almost saturates the full $q\bar
q+qg$ contribution, except at large $Q_T$ or small $\Delta\phi$, the $qg$
initiated subprocess is not important at the Tevatron in most of phase space
for the cuts used. The fragmentation contribution is denoted by the dotted
line. The sum of all contributions including fragmentation is denoted by the
solid line. After isolation, the fragmentation contribution is much smaller
than ``direct'' ones, but contributes $\simeq 10\%$ near the peak. The
uncertainty in the contribution of the fragmentation process can be
estimated by comparing the Monte Carlo result with a parton--level
calculation, as shown in Fig.~\ref{Fig:Frag}.

\FigCDFb
Figure \ref{Fig:CDF1} shows the distribution of the transverse momentum of
the photon pair, $d\sigma/dQ_T$ vs. $Q_T$. Over the interval $5\lesim Q_T
\lesim 25$ GeV, the contribution from the $gg$ subprocess is comparable to
the $q\bar q+qg$ subprocess. The change in slope near $Q_T=20$ GeV arises
from the $gg$ subprocess (dot--dashed line) for which $Q_T \stackrel{%
\scriptscriptstyle<}{\scriptscriptstyle\sim} M_{\gamma\gamma}$ is required
in our approximate calculation. The peak near $Q_T\simeq 1.5$ GeV is
provided mostly by the $q\bar q+qg$ (long--dashed line) subprocess. In
general, the height and breadth of the peak in the $Q_T$ distribution
depends on the details of the non-perturbative function in Eq.~(\ref{master}). 
The effect of different non-perturbative contributions may be estimated if
the parameter $g_2$ is varied by $\pm 2\sigma$. For $Q_T > 10$ GeV, the
distribution is not sensitive to this variation. The height and the width
(half--maximum) of the peak change by approximately 20\% and 35\%,
respectively, but the integrated rate from 0 to 10 GeV is almost constant.
The peak of the distribution (which is below 5 GeV), shifts approximately $%
+0.5$ GeV or $-0.6$ GeV for a $+2\sigma$ or $-2\sigma$ variation. The mean $%
Q_T$ for $Q_T<10$ GeV shifts at most by $0.4$ GeV. For $gg$ resummation, it
is not clear which parametrization of the non-perturbative physics should be
used. However, the final effect of the two different parametrizations
outlined in Eq.~(\ref{Eq:ggnp}) is minimal, shifting the mean $Q_T$ for $%
Q_T<40$ GeV by about 0.4 GeV. The parametrization $(ii)$ is used in the
final results, so that the coefficient $g_2$ is scaled by $C_A/C_F$ relative
to the $q\bar q$ non-perturbative function.

\FigCDFc
Figure~\ref{Fig:CDF2} shows $d\sigma/d\Delta\phi$ vs. $\Delta\phi$, where $%
\Delta\phi$ is the azimuthal opening angle between the two photons. The
change in slope near $\Delta\phi=\pi/2$ is another manifestation of the
approximations made in the treatment of the $gg$ contribution (dot--dashed
line). The height of the distribution near $\Delta\phi\simeq\pi$ is also
sensitive to the details of the non-perturbative function.

In the absence of resummation or NLO effects, the $gg$ box contribution
supplies $\vec{Q}_T=0$ and $\Delta\phi=\pi$. In this calculation, as
explained earlier, the NLO contribution for the $gg$ subprocess is handled
in an approximate fashion. For the cuts listed above, the total cross
section from the complete $gg$ resummed calculation, including the function $%
Y$, is 6.28 pb. If the resummed CSS piece is used alone, the contribution is
reduced to 4.73 pb. This answer can be compared to the contribution at LO.
For the same structure functions, the LO $gg$ cross section for the CDF cuts
is 3.18 pb for the scale choice $Q=M_{\gamma\gamma}$. Therefore, the effect
of including part of the NLO contribution to the $\gamma\gamma$ process is
to approximately double the LO $gg$ box contribution to the cross section.
This increase indicates that the exact NLO correction can be large for the $%
gg$ subprocess and motivates a full calculation.

The predictions for the \D0~cuts and a comparison to data are shown in Figs.~%
\ref{Fig:D00}--\ref{Fig:D02}. Because of the steep distribution in the
transverse momentum of the individual photons, the higher $p_T^\gamma$
threshold in the \D0~case significantly reduces the total cross section.
Otherwise, the behavior is similar to the resummed calculation shown for the
CDF cuts. The \D0~data plotted in the figures are not corrected for
experimental resolution. To compare with the uncorrected \D0~data with the
kinematic cuts $p^{\gamma_1}_T >14$\,GeV, $p^{\gamma_2}_T >13$\,GeV and $%
\eta^\gamma<1.0$, an ``equivalent'' set of cuts is used in the theoretical
calculation: $p^{\gamma_1}_T >14.9$\,GeV, $p^{\gamma_2}_T >13.85$\,GeV, and $%
\eta^\gamma<1.0$ \cite{d0www}. The effect of this ``equivalent'' set is to
reduce the theoretical rate in the small $M_{\gamma \gamma}$ region.
\FigDa
\FigDb
\FigDc

While the agreement in both shapes and absolute rates is generally good,
there are some discrepancies between the resummed prediction and the data as
presented in these plots. At small $Q_T$ (Fig.~\ref{Fig:CDF1}) and large $%
\Delta\phi$ (Fig.~\ref{Fig:CDF2}), where the CDF cross section is large, the
theoretical results are beneath the data. Since this is the kinematic region
in which the non-perturbative physics is important, better agreement can be
obtained if the non-perturbative function is altered. In Fig.~\ref{Fig:D00},
the calculated $M_{\gamma\gamma}$ distribution  is larger than the \D0~data
at large $M_{\gamma\gamma}$, while the calculation appears to agree with the
CDF data in Fig.~\ref{Fig:CDF0}. The small discrepancy in Fig.~\ref{Fig:D00}
at large values of $M_{\gamma\gamma}$ is not understood. (The systematic
errors of the data, which are about 25\% \cite{d0www}, are not included in
this plot.) On the other hand, Figs.~\ref{Fig:D01} and \ref{Fig:D02} show
that the resummed calculation is {\it beneath} the data at large $Q_T$ or
small $\Delta\phi$. The discrepancies in Figs.~\ref{Fig:D01} and \ref
{Fig:D02} may result from the approximations made in the $gg$ process
(notice the kinks in the dot--dashed curves). A complete NLO calculation for
the $gg$ subprocess is needed, and may improve the comparison with data for
small $\Delta\phi$.

Because of the uncertainty in the prediction for the $gg$ contribution of
the resummed calculation, the distributions in $M_{\gamma\gamma}, Q_T$ and $%
\Delta\phi$ are shown in Figs.~\ref{Fig:qtcut0}--\ref{Fig:qtcut2} for the
CDF cuts and the additional requirement that $Q_T < M_{\gamma\gamma}$. This
additional requirement should significantly reduce the theoretical
uncertainty for large $Q_T$ and small $\Delta\phi$.
\Figqtcuta
\Figqtcutb
\Figqtcutc

In Fig.~\ref{Fig:qtcut1}, the lower of the two solid curves in the $Q_T$
distribution shows the prediction of the pure NLO ${\cal O}(\alpha_s)$
(fixed--order) calculation, without resummation, for the $q\bar q$ and $qg$
subprocesses, excluding fragmentation. For $Q_T \grsim 25$ GeV, the lower
solid curve is very close to the long--dashed ($q\bar q+qg$) curve obtained
after resummation, as is expected. As $Q_T$ decreases below $Q_T\simeq 25$
GeV, all--orders resummation produces significant changes. Most apparent,
perhaps, is that the $Q_T\to 0$ divergence in the fixed--order calculation
is removed. However, there is also a marked difference in shape over the
interval $5<Q_T<25$ GeV between the fixed--order $q\bar q+qg$ result and its
resummed counterpart. These are general features in a comparison of resummed
and NLO calculations \cite{resum2}--\cite{qqres}.

\subsection{Fixed--Target Energy}

The fixed--target experiment E706 \cite{e706data} at Fermilab has collected
diphoton data from the collision of a $p$ beam on a $Be$ ($A=9.01, Z=4$)
target at $\sqrt{S}=31.5$ GeV. The kinematic cuts applied to the resummed
prediction in the center--of--mass frame of the beam and target are $%
p_T^\gamma > 3$ GeV and $|\eta^{\gamma}|<0.75$. No photon isolation is
required. The same phenomenological inputs are used for this calculation as
for the calculation at collider energies. The $Be$ nucleon target is treated
as having an admixture of $4/9.01$ proton and $5.01/9.01$ neutron parton
distribution functions. The $A$ dependence effect appears to be small in the
prompt photon data (the effect is parametrized as $A^\alpha$ and the
measured dependence is $\alpha\simeq 1$), so it is ignored \cite{sorrell}.

Figures \ref{Fig:E7060}--\ref{Fig:E7062} show the same distributions
discussed previously. Because of the kinematic cuts, the relative
contribution of gluon initiated processes is highly suppressed, except at
large $Q_T$, where the $gg$ box contribution is seen to dominate, and at
large $M_{\gamma\gamma}$ where the $qg$ contribution is dominant. The
fragmentation contribution (not shown) is minimal (of a few percent). The
dominance of $gg$ resummation over the $q\bar q$ resummation at large $Q_T$
in Fig.~\ref{Fig:E7061} occurs because it is more likely (enhanced by the
ratio $C_A/C_F=9/4$) for a gluon to be radiated from a gluon than a quark
line. The exact height of the distribution is sensitive to the form of the
non-perturbative function (in the low $Q_T$ region) and to the approximation
made in calculating the NLO corrections (of ${\cal O}(\alpha_{em}^2%
\alpha_s^3)$) to the hard scattering. However, since $Q_T < Q$ is satisfied
for the set of kinematic cuts, the final answer with complete NLO
corrections should not differ significantly from the result reported here.
\FigEa
\FigEb
\FigEc

The scale dependence of the calculation was checked by comparing with the
result obtained with $C_2=C_1/b_0=0.5$, $C_3=b_0$, and $C_4=1$. The $q {\bar
q}$ rate is not sensitive to the scale choice, and the $gg$ rate increases
by less than about 20\%. This insensitivity can be understood as follows.
For the E706 data, the non-perturbative physics completely dominates the $Q_T$
distribution. The perturbative Sudakov resummation is not important over the
entire $Q_T$ region, and the NLO $Y$ piece is sizable only for $Q_T > 3$%
\,GeV where the event rate is small. Since the LO $q \bar q$ rate does
depend on $\alpha_s$, and the LO $gg$ rate is proportional to $%
\alpha_s^2(C_2 M_{\gamma \gamma})$, the $gg$ rate increases for a smaller $%
C_2$ value, but the $q \bar q$ rate remains about the same. In conclusion,
the E706 data can be used to constrain the non-perturbative functions
associated with the $q {\bar q}$ and $gg$ hard processes in hadron
collisions.

\section{Conclusions}

Prompt photon pair production at fixed target and collider energies is of
interest in its own right as a means of probing the dynamics of strong
interactions. The process is of substantial interest also in searches for
new phenomena, notably the Higgs boson.

In this Chapter, a calculation is presented of the production rate and
kinematic distributions of photon pairs in hadronic collisions. This
calculation incorporates the full content of the next--to--leading order
(NLO) contributions from the $q{\bar{q}}$ and $qg$ initial--state
subprocesses, supplemented by resummation of contributions to these
subprocesses from initial state radiation of soft gluons to all orders in
the strong coupling strength. The computation also includes important
contributions from the $gg$ box diagram. The $gg$ contributions from
initial--state multiple soft gluons are resummed to all orders, but the NLO
contribution, of ${\cal O}(\alpha _{em}^2\alpha _s^3)$, to the hard
scattering subprocess is handled in an approximate fashion. The
approximation should be reliable at relatively small values of the pair
transverse momentum $Q_T$ as compared to the invariant mass of the photon
pair $M_{\gamma \gamma }$. At collider energies, the $gg$ contribution is
comparable to that of the $q{\bar{q}}$ and $qg$ contributions over a
significant part of phase space where $M_{\gamma \gamma }$ is not large, and
its inclusion is essential. The exact ${\cal O}(\alpha _{em}^2\alpha _s^3)$
corrections to the $gg$ box diagram should be calculated to test the
validity of the approximations made in this calculation. Finally, the
calculation also includes long--distance fragmentation contributions at
leading order from the subprocess $qg\rightarrow \gamma q$, followed by
fragmentation of the final quark, $q\rightarrow \gamma X$. After photon
isolation, fragmentation plays a relatively minor role. The fragmentation
contribution is computed in two ways: first, in the standard parton model
collinear approximation and second, with a Monte Carlo shower simulation.
This overall calculation is the most complete treatment to date of photon
pair production in hadronic collisions. Resummation plays a very important
role particularly in the description of the behavior of the $Q_T$
distribution at small to moderate values of this variable, where the cross
section takes on its largest values.

The resummed calculation is necessary for a reliable prediction of kinematic
distributions that depend on correlations between the photons. It is a
significant improvement over fixed--order NLO calculations that do not
include the effects of initial--state multiple soft--gluon radiation.
Furthermore, even though the  hard scattering $q \bar q$ and $q g$
subprocesses are computed to the same order in the resummed and fixed--order
NLO calculations, the cross sections from the two calculations can differ
after kinematic cuts are imposed \cite{wres}.

The results of the calculation are compared with data from the CDF and \D%
0~collaborations, and the agreement is generally good in both absolute
normalization and shapes of the distributions in the invariant mass $%
M_{\gamma\gamma}$ of the diphoton system, the pair transverse momentum $Q_T$%
, and  the difference in the azimuthal angles $\Delta\phi$. Discrepancies
with CDF results at the smallest values of $Q_T$ and $\Delta \phi$ near $\pi$
might originate from the strong dependence on the non-perturbative functions
in this kinematic region. In comparison with the \D0~data, there is also
evidence for disagreement at intermediate and small values of $\Delta \phi$.
The region of intermediate $\Delta \phi$, where the two photons are not in a
back--to--back configuration, is one in which the full treatment of three
body final--state contributions of the type $\gamma \gamma j$ are important,
with $j = q$ or $g$. The distributions in Figs. 5 and 8 suggest that an
exact calculation of the NLO contribution associated with the $g g$ initial
channel would ameliorate the situation and will be necessary to describe
data at future high energy hadron colliders.

Predictions are also presented in the work for $pN\rightarrow \gamma \gamma X
$ at the center--of--mass energy 31.5 GeV, appropriate for the E706
fixed--target experiment at Fermilab. The large $Q_T$ and small $\Delta \phi 
$ behavior of the kinematic distributions is dominated by the resummation of
the $gg$ initial state. Non-perturbative physics controls the $Q_T$
distribution, and neither the perturbative Sudakov nor the regular NLO
contribution plays an important role, except in the very large $Q_T$ region
where the event rate is small. For the E706 kinematics, the requirement $%
Q_T<Q$ is generally satisfied. Therefore, the approximate $gg$ calculation
presented in this work should be reliable.

In this calculation, the incident partons are assumed to be collinear with
the incident hadrons. A recurring question in the literature is the extent
to which finite ``intrinsic'' $k_T$ may be required for a quantitative
description of data \cite{grabbag,e706data}. An important related issue is
the proper theoretical specification of the intrinsic component \cite
{kt_intrinsic}. In the CSS resummation formalism, this physics is included
by properly parametrizing the non-perturbative function ${\widetilde{W}}%
^{NP}(b)$, which can be measured in Drell--Yan, $W$, and $Z$ production.
Because photons participate directly in the hard scattering, because their
momenta can be measured with greater precision than that of hadronic jets or
heavy quarks, and because the $\gamma \gamma $ final state is a color
singlet, the reaction $p{\bar{p}}\rightarrow \gamma \gamma X$ may serve as a
particularly attractive laboratory for the understanding of the role of
intrinsic transverse momentum. The agreement with data on the $Q_T$
distributions in Figs.~\ref{Fig:CDF1} and \ref{Fig:D01} is suggestive that
the CSS formalism is adequate. However, the separate roles of gluon
resummation and the assumed non-perturbative function in the successful
description of the $Q_T$ distributions are not disentangled. In the
non--perturbative function of Eq.~(\ref{Eq:WNonPert}), the dependence on $b$
(and, thus, the behavior of $d\sigma /dQ_T$ at small $Q_T$) is predicted to
change with both $Q$ and the values of the parton momentum fractions $x_i$.
At fixed $Q$, dependence on the values of the $x_i$ translates into
dependence on the overall center--of--mass energy of the reaction. As data
with greater statistics become available, it should be possible to verify
these expectations. In combination with similar studies with data on massive
lepton--pair production (the Drell--Yan process), it will be possible to
determine whether the same non--perturbative function is applicable in the
two cases, as is assumed in this work.

The diphoton data may allow a study of the non-perturbative as well as the
perturbative physics associated with multiple gluon radiation from the {\it %
gluon}--initiated hard processes, which cannot be accessed from Drell--Yan, $%
W^{\pm }$, and $Z$ data. With this knowledge, it may be possible to improve
calculations of single photon production and other reactions sensitive to
gluon--initiated subprocesses. In the \D0~data analysis \cite{d0data}, an
asymmetric cut is applied on the transverse momenta ($p_T^\gamma $) of the
two photons in the diphoton event. This cut reduces the effect of multiple
gluon radiation in the event. To make the best use of the data for probing
the interesting multiple gluon dynamics predicted by the QCD theory, a
symmetric $p_T^\gamma $ cut should be applied. 



\chapter{ Gauge Boson Pair Production at the Upgraded Tevatron and at the LHC 
\label{ch:ZPair}}

\ifx\nopictures Y \else \input epsf.tex \fi
\input psfig 
\def\TblTotal{ 
\begin{table}[t]
\begin{center}
\begin{tabular}{r r r r r r}
\hline \hline \\[-.2cm]
Di-boson & 
Collison &
$E_{cm}$ & 
\multicolumn{2}{c}{Fixed Order ${\cal O}(\alpha_s^0)$} & 
Resummed \\ 
produced &
    type &
   (TeV) & 
  CTEQ4L &
  CTEQ4M & 
$\oplus$ ${\cal O}(\alpha _s)$ 
\\ \hline \\[-.2cm]
$Z^0 Z^0$       & $      pp$ &  14 & 9.14 & 10.3 & 14.8 \\
$Z^0 Z^0$       & $p\bar{p}$ &   2 & 0.91 & 1.01 & 1.64 \\ 
$\gamma \gamma$ & $      pp$ &  14 & 22.1 & 24.5 & 60.8 \\ 
$\gamma \gamma$ & $p\bar{p}$ &   2 & 8.48 & 9.62 & 22.8 \\ 
$\gamma \gamma$ & $p\bar{p}$ & 1.8 & 6.30 & 7.15 & 17.0 \\
\hline \hline
\end{tabular}
\end{center}
\par
\caption{Total cross sections of diphoton and $Z^0$ boson pair production
at the LHC and the upgraded Tevatron, in units of pb.
The kinematic cuts are described in the text.
The ``$\oplus$'' sign refers to the matching prescription discussed in the text.
}
\label{tbl:TotalZPair}
\end{table}
}
\def\TblSubTotalZZ{
\begin{table}[t]
\begin{center}
\begin{tabular}{r r r r}
\hline \hline \\[-.2cm]
%
$E_{cm}$ & 
Collison &
$q\bar{q}\to  Z^0 Z^0 X$ &
$      qg\to  Z^0 Z^0 X$ 
\\ 
(TeV) &
type &    
 &
\\ \hline \\[-.2cm]
 14 &       $pp$ & 10.9 & 3.91 \\ 
  2 & $p\bar{p}$ & 1.62 & 0.02 \\ 
\hline \hline
\end{tabular}
\end{center}
\par
\caption{
Resummed cross sections of the subprocesses for $Z^0$ boson pair production
at the LHC and the upgraded Tevatron, in units of pb.
The kinematic cuts are described in the text.
}
\label{tbl:SubTotalZZ}
\end{table}
}
\def\TblSubTotalAA{
\begin{table}[t]
\begin{center}
\begin{tabular}{r c c c c c c c}
\hline \hline \\[-.2cm]
$E_{cm}$ & 
Collison &
$q\bar{q}\to \gamma \gamma X$ &
$      qg\to \gamma \gamma X$ &         
\multicolumn{2}{c}{$gg \to \gamma \gamma$} & 
$      gg\to \gamma \gamma g$ & 
Fragmen- \\ 
(TeV) &
type &        
  &
  &
$4L$& 
$4M$& 
  & tation
\\
\hline \\[-.2cm]
  14 &       $pp$ & 20.5 & 16.6 & 22.3 & 14.4 & 23.9 & 6.76 \\ 
   2 & $p\bar{p}$ & 9.68 & 4.81 & 6.02 & 4.34 & 8.26 & 2.15 \\ 
\hline \hline
\end{tabular}
\end{center}
\par
\caption{
Cross sections of the subprocesses for diphoton production
at the LHC and the upgraded Tevatron, in units of pb.
The resummed $qg \to \gamma \gamma X$ rate includes
the fragmentation contribution.
The ${\cal O}(\alpha_s^2)$ $gg\to \gamma \gamma$ rates were calculated using
both the CTEQ4L and CTEQ4M PDF's.
The kinematic cuts are described in the text.
}
\label{tbl:SubTotalAA}
\end{table}
}
\def\FigDiagrams
{
\begin{figure*}[t]
\vspace{-.5cm}
\begin{center}
\begin{tabular}{c}
\ifx\nopictures Y \else{ \epsfysize=8.5cm \epsffile{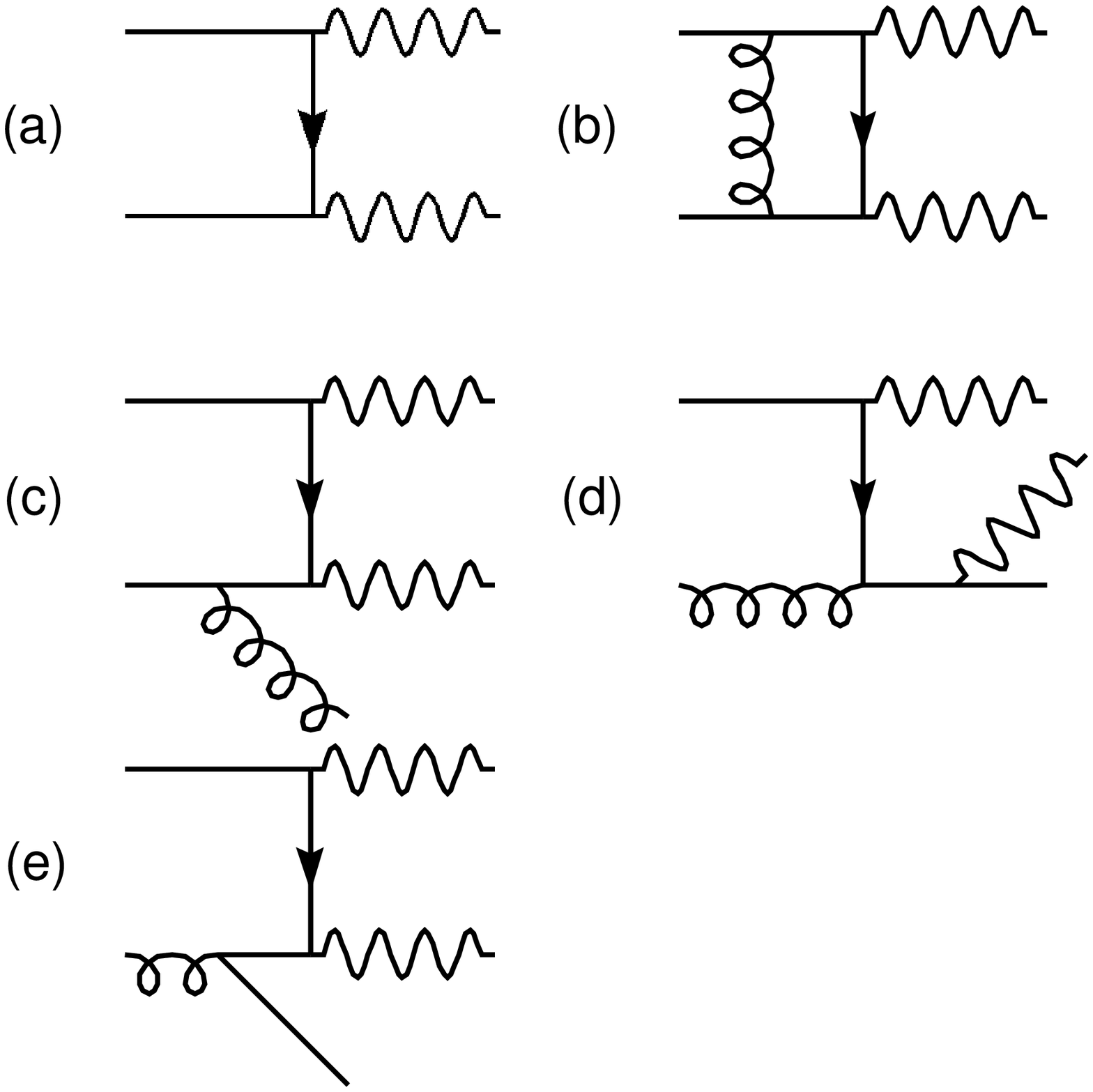}}
\fi 
\end{tabular}
\end{center}
\vspace{-1cm}
\caption{ A representative set of Feynman diagrams included in the NLO
 calculation of $Z^0$ pair production. }
\label{fig:Diagrams}
\end{figure*}
}
\def\FigZZTevQypT
{
\begin{figure*}[p]
\vspace{-1.5cm}
\begin{center}
\ifx\nopictures Y \else{
\begin{tabular}{cc}
\epsfysize=7.0cm \epsffile{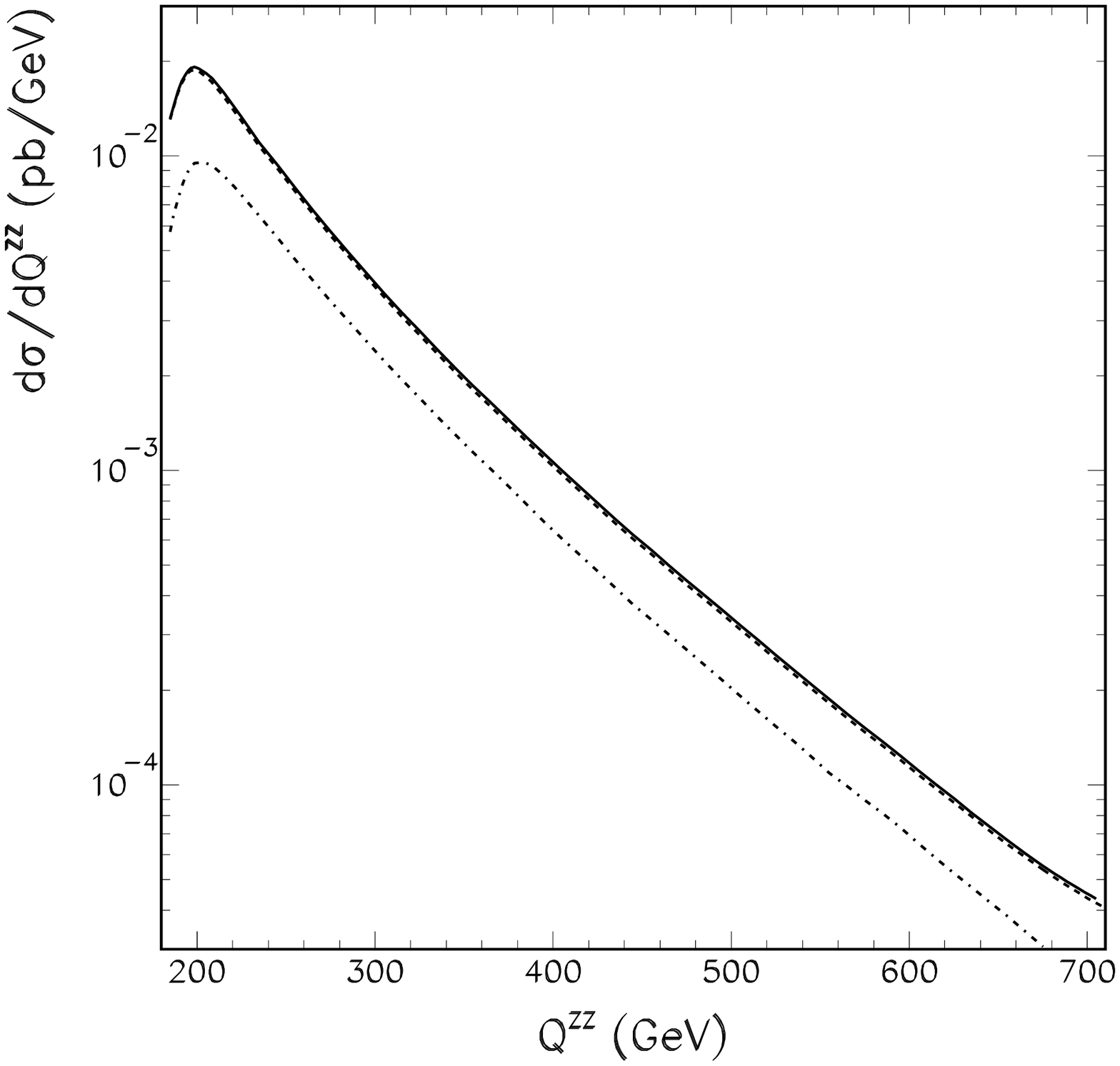} &
\epsfysize=7.0cm \epsffile{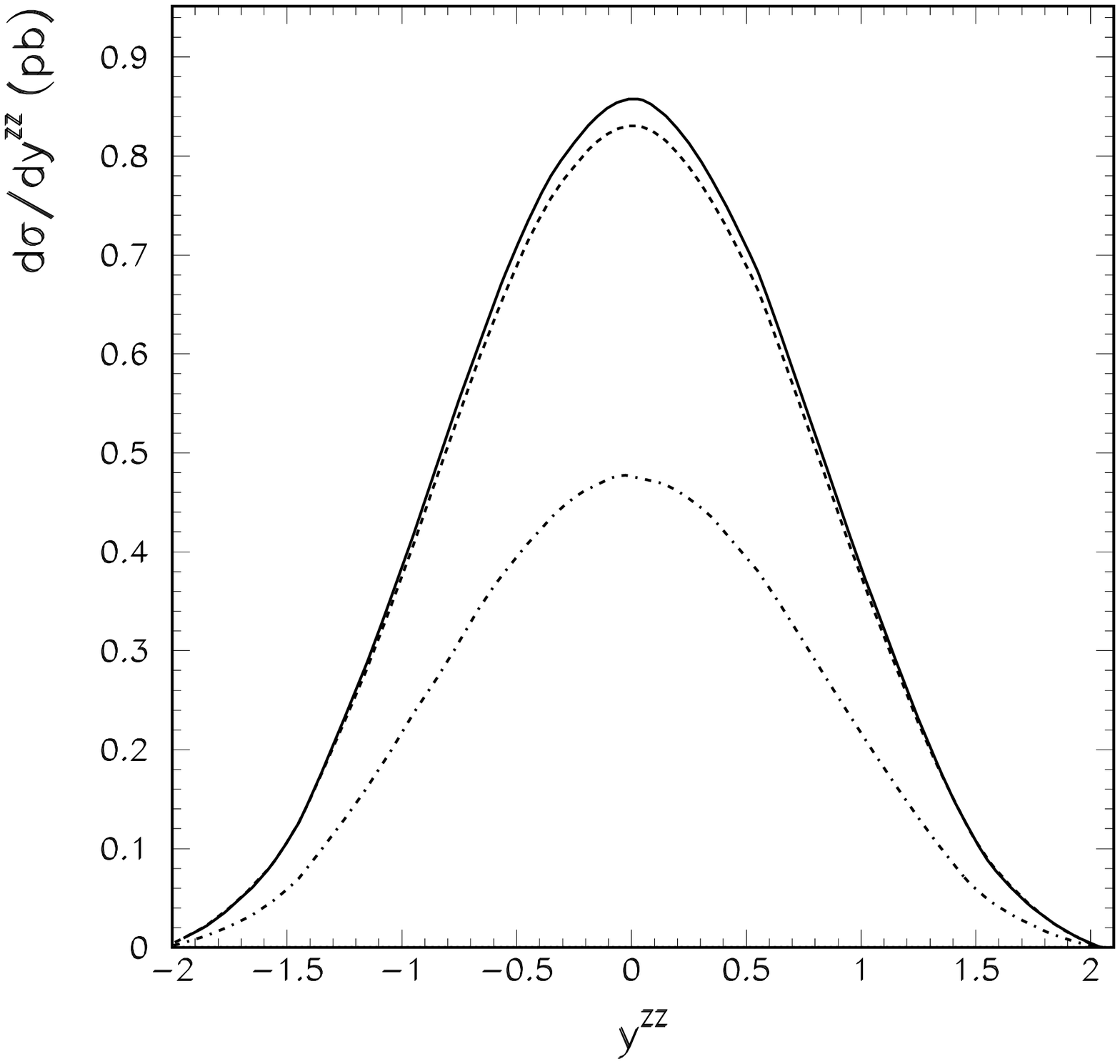} \\
\multicolumn{2}{c}{\epsfysize=7.0cm \epsffile{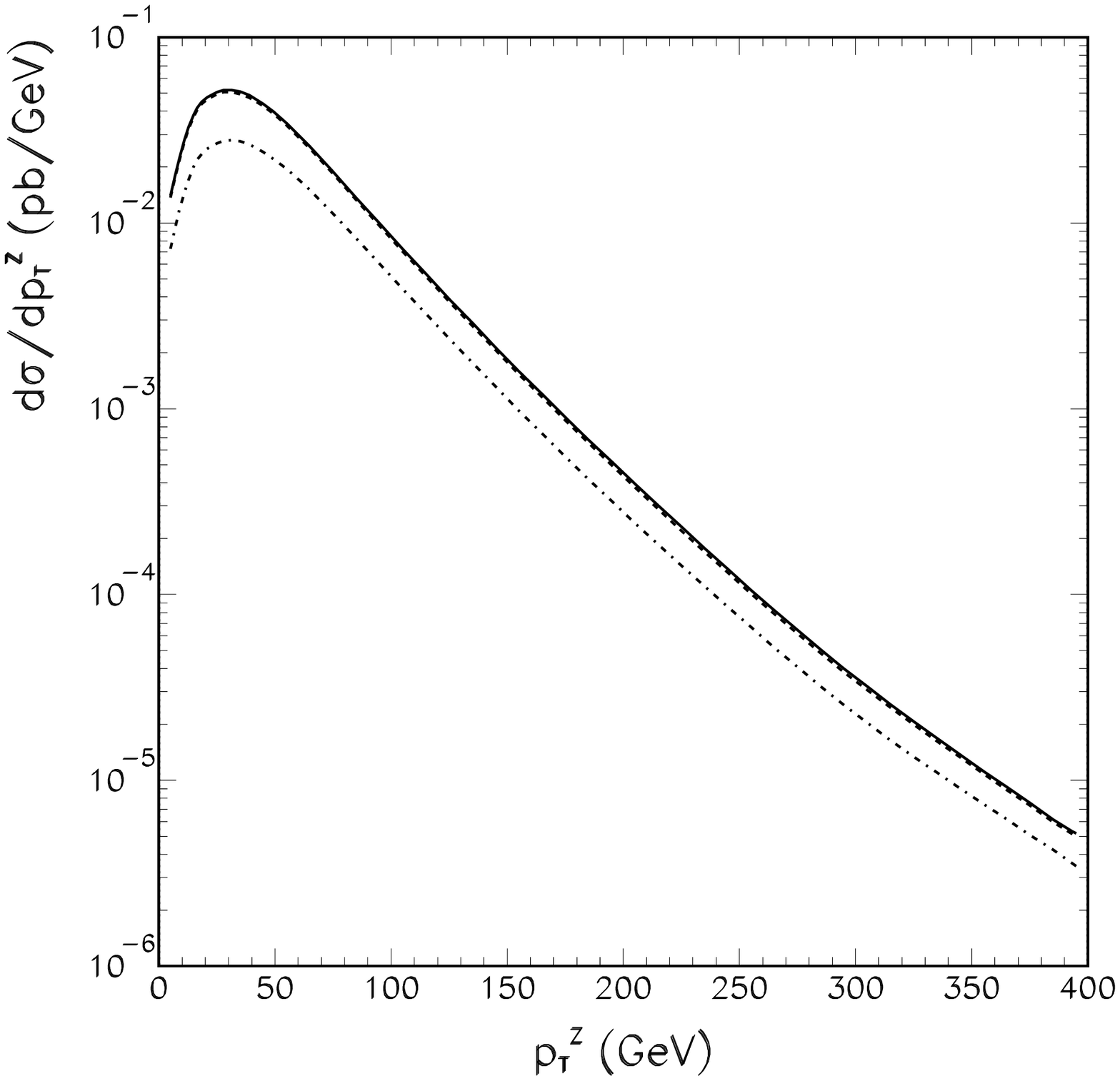} }
\end{tabular}
} \fi
\end{center}
\vspace{-1cm}
\caption{
Same as Fig.~\ref{fig:ZZLHCQypT} except for the upgraded Tevatron.
}
\label{fig:ZZTevQypT}
\end{figure*}
}
\def\FigZZTevQT
{
\begin{figure*}[t]
\vspace{-1.5cm}
\begin{center}
\begin{tabular}{c}
\ifx\nopictures Y \else{ 
\epsfysize=10.0cm \epsffile{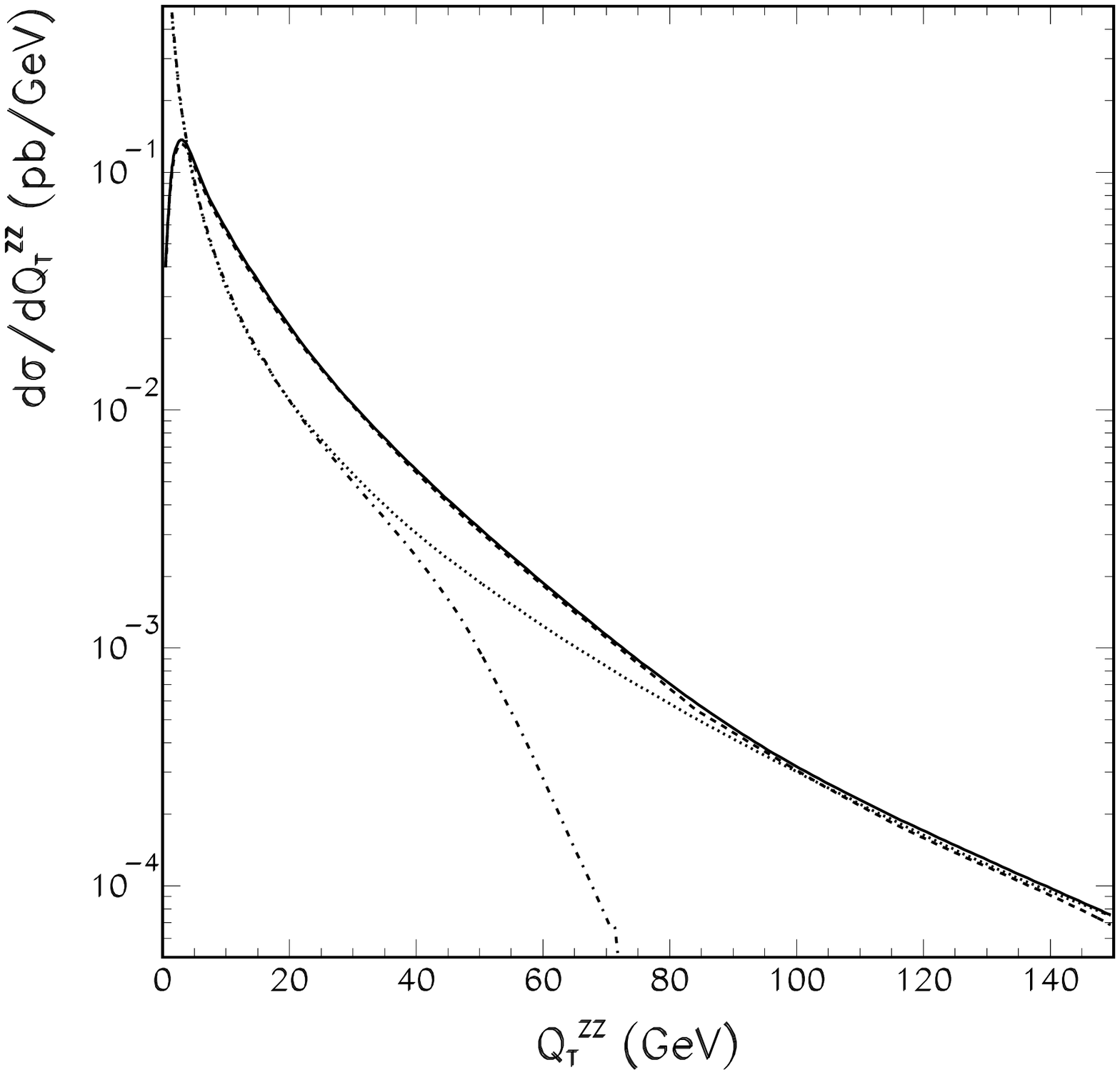}} 
\fi 
\end{tabular}
\end{center}
\vspace{-1cm}
\caption{
Same as Fig.~\ref{fig:ZZLHCQT} except for the upgraded Tevatron.
}
\label{fig:ZZTevQT}
\end{figure*}
}
\def\FigZZTevInt
{
\begin{figure*}[t]
\vspace{-1.5cm}
\begin{center}
\begin{tabular}{c}
\ifx\nopictures Y \else{ \epsfysize=10.0cm \epsffile{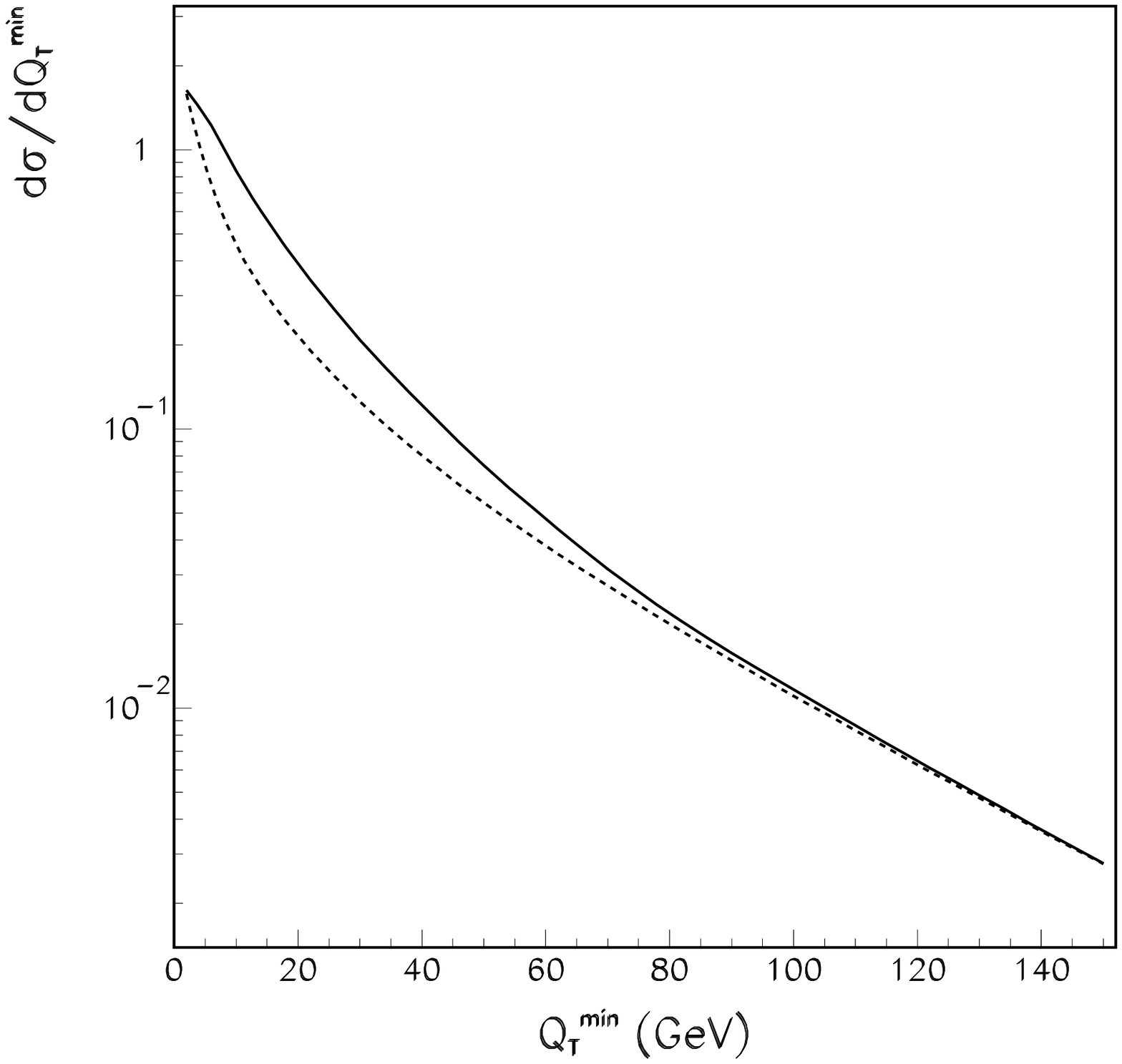}} \fi
\end{tabular}
\end{center}
\vspace{-1cm}
\caption{
Same as Fig.~\ref{fig:ZZLHCInt} except for the upgraded Tevatron.
}
\label{fig:ZZTevInt}
\end{figure*}
}
\def\FigZZLHCQT
{
\begin{figure*}[t]
\vspace{-1.5cm}
\begin{center}
\begin{tabular}{c}
\ifx\nopictures Y \else{ \epsfysize=10.0cm \epsffile{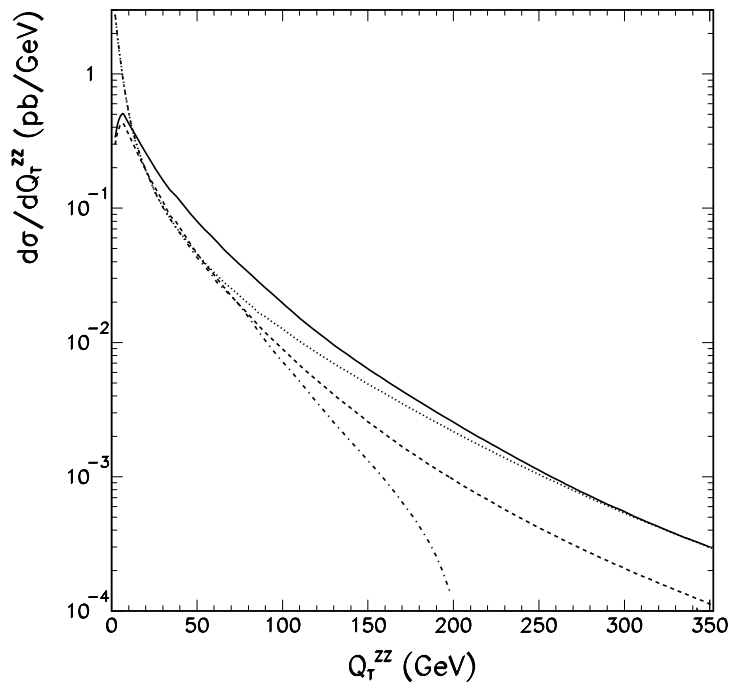}} \fi
\end{tabular}
\end{center}
\vspace{-1cm}
\caption{
Transverse momentum distribution of $Z^0$ pairs from $q {\bar q} + q g$
partonic initial states at the LHC. The ${\cal O}(\alpha_s)$
(dotted) and the asymptotic (dash-dotted) pieces are coincide and
diverge as $Q_T \to 0$. 
The resummed (solid) curve matches the ${\cal O}(\alpha_s)$ curve at
about $Q_T = 320$ GeV. 
The resummed $q {\bar q}$ contribution (excluding the $qg$ contribution) 
is shown as dashed line.
}
\label{fig:ZZLHCQT}
\end{figure*}
}
\def\FigZZLHCInt
{
\begin{figure*}[t]
\vspace{-1.5cm}
\begin{center}
\begin{tabular}{c}
\ifx\nopictures Y \else{ \epsfysize=10.0cm \epsffile{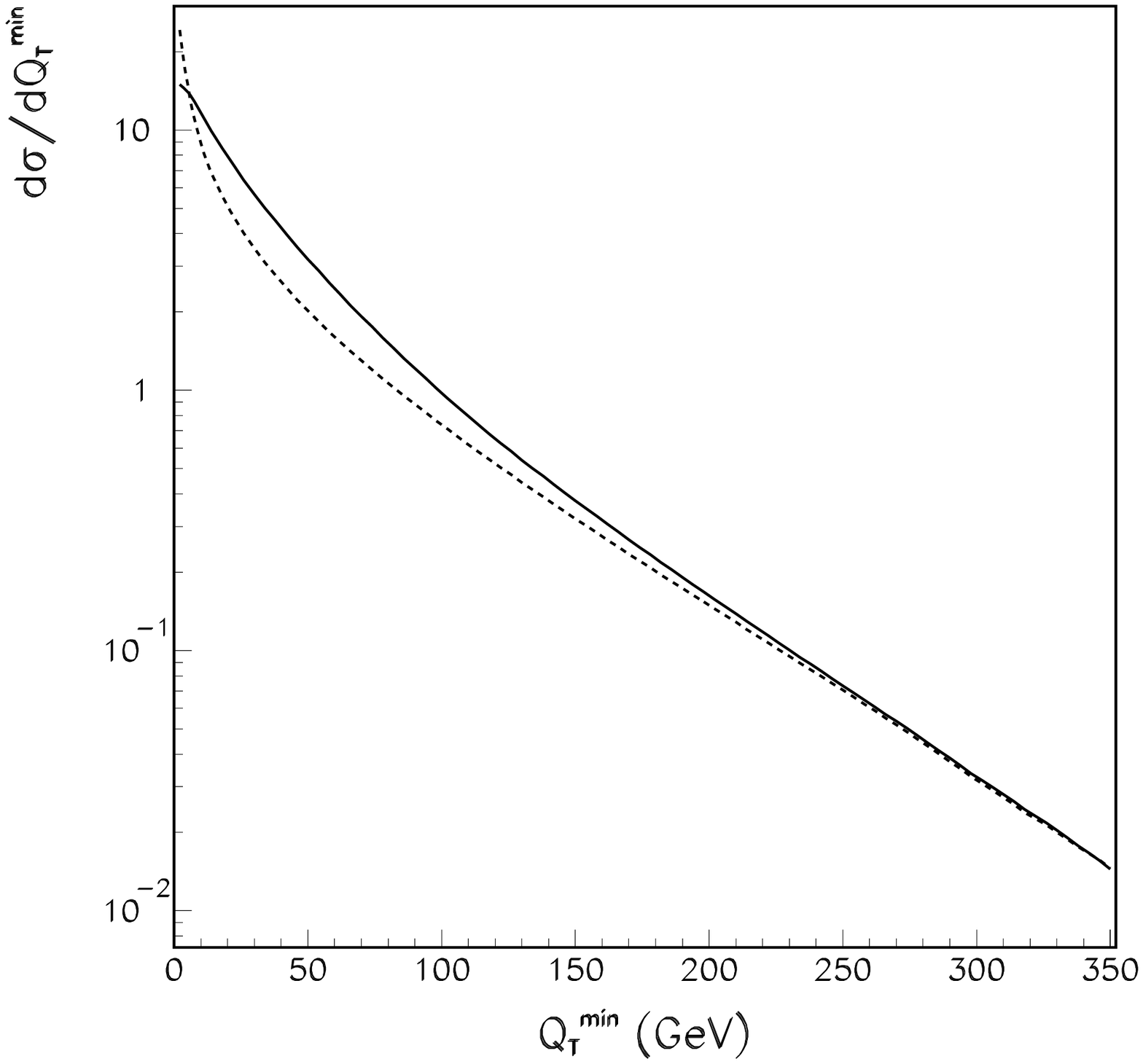}} \fi
\end{tabular}
\end{center}
\vspace{-1cm}
\caption{
The integrated cross section for $Z^0$ boson pair production at the LHC. 
The resummed and the 
${\cal O}(\alpha_s)$ distributions are shown in solid and dashed lines, 
respectively.
 }
\label{fig:ZZLHCInt}
\end{figure*}
}
\def\FigZZLHCQypT
{
\begin{figure*}[p]
\vspace{-1.5cm}
\begin{center}
\ifx\nopictures Y \else{
\begin{tabular}{cc}
\epsfysize=7.0cm \epsffile{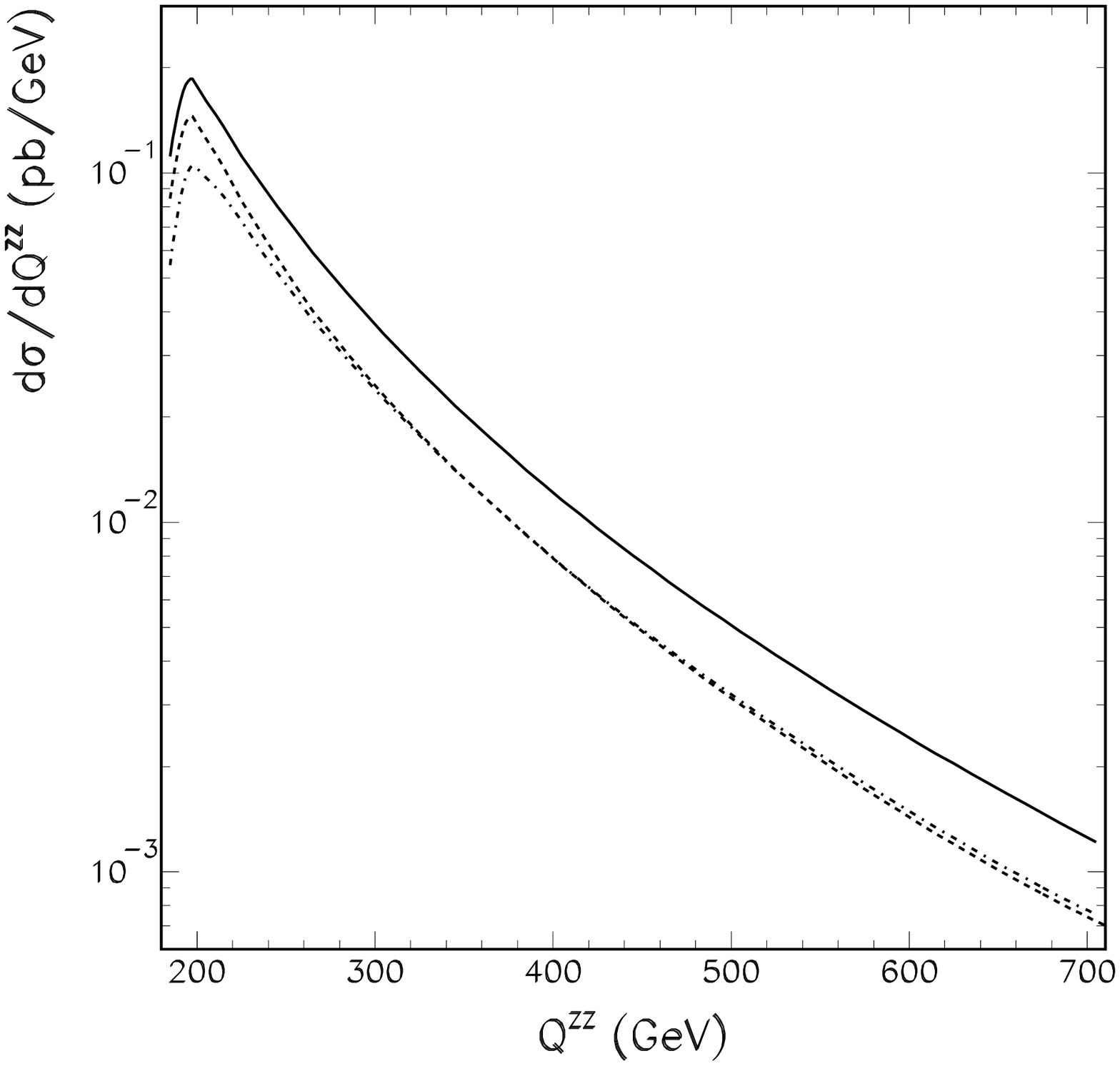} &
\epsfysize=7.0cm \epsffile{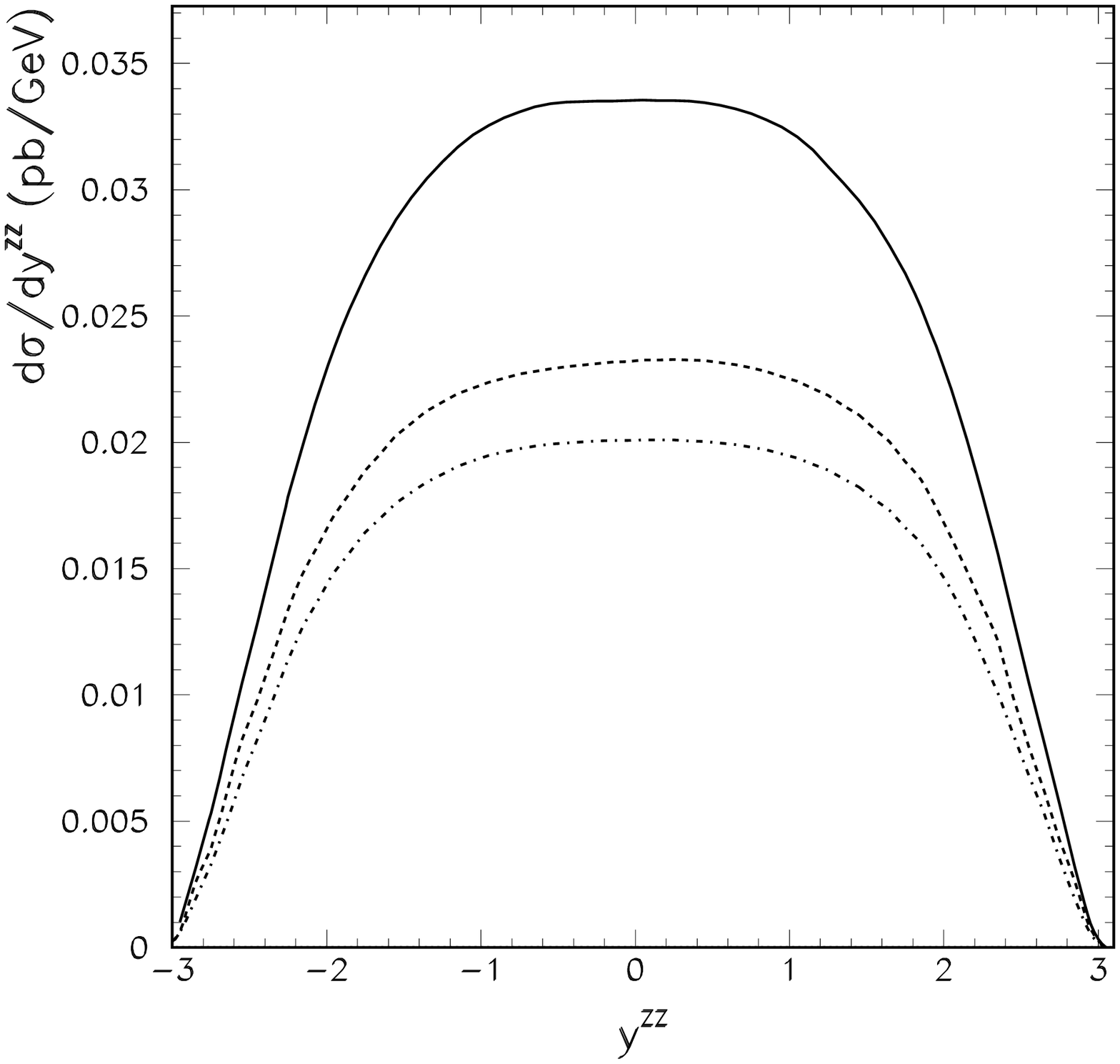} \\
\multicolumn{2}{c}{\epsfysize=7.0cm \epsffile{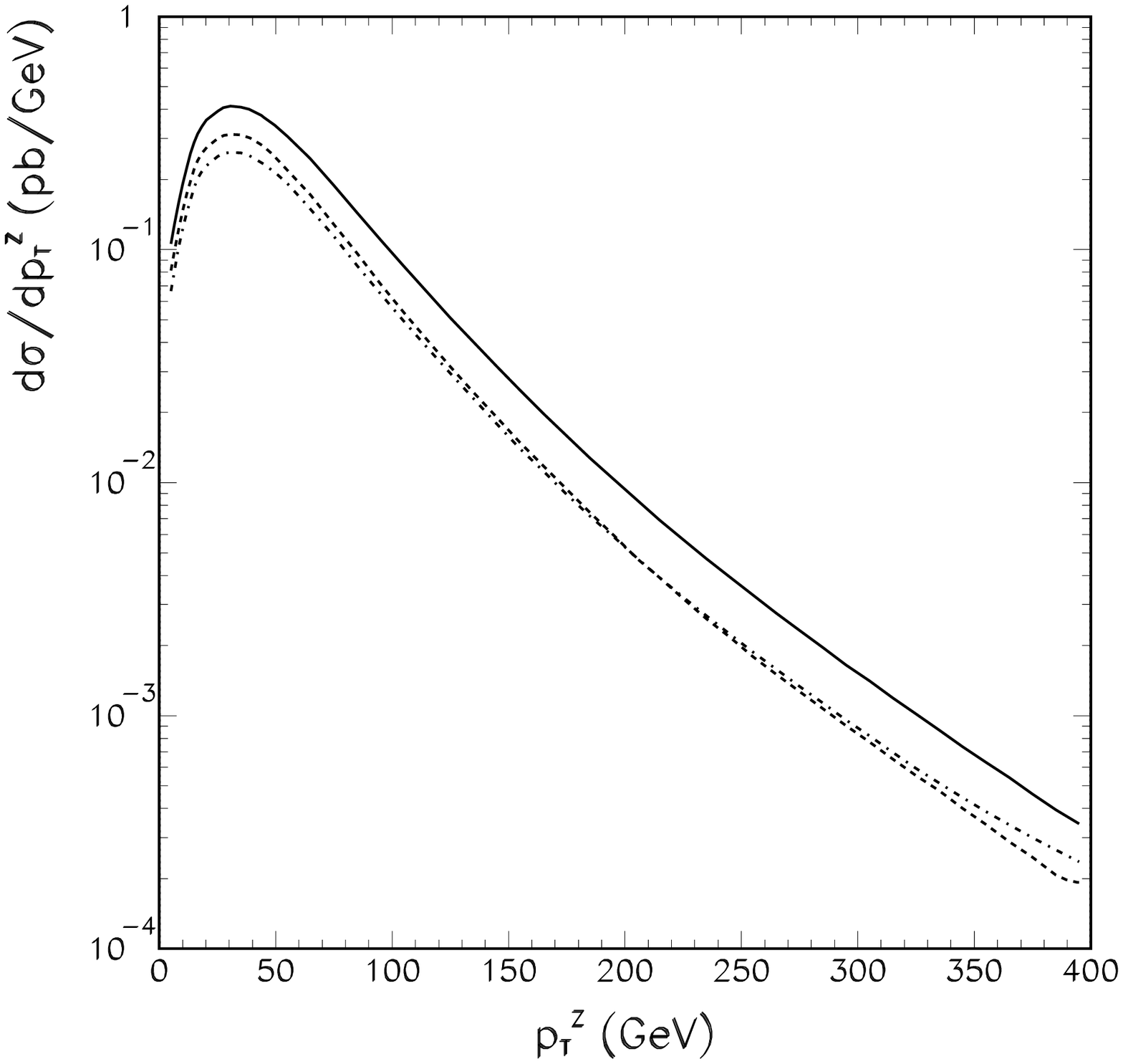} }
\end{tabular}
} \fi
\end{center}
\vspace{-1cm}
\caption{
Invariant mass and rapidity distributions of $Z^0$ boson pairs, and 
transverse momentum distributions of the individual $Z^0$ bosons at the LHC. 
The resummed contribution of the $q {\bar q} + q g
\to Z^0 Z^0 X$ subprocess is shown by the solid curve, and of the 
$q \bar{q} \to Z^0 Z^0 X$ subprocess by the dashed curve. 
The leading order distribution of $q \bar{q} \to Z^0 Z^0$ is 
shown by the dash-dotted curve.
}
\label{fig:ZZLHCQypT}
\end{figure*}
}
\def\FigAATevQypT
{
\begin{figure*}[p]
\vspace{-1.5cm}
\begin{center}
\ifx\nopictures Y \else{
\begin{tabular}{cc}
\epsfysize=7.0cm \epsffile{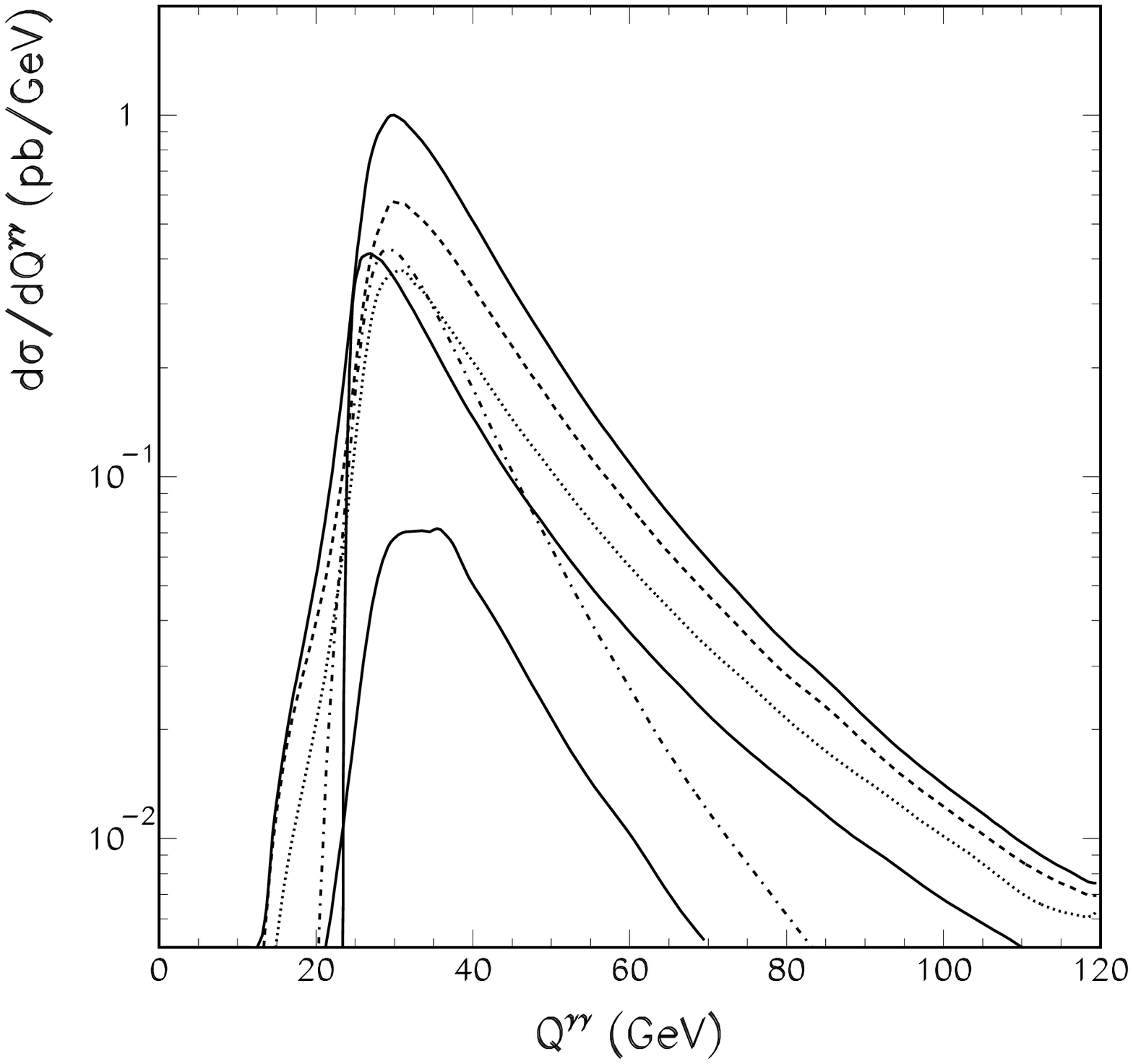} &
\epsfysize=7.0cm \epsffile{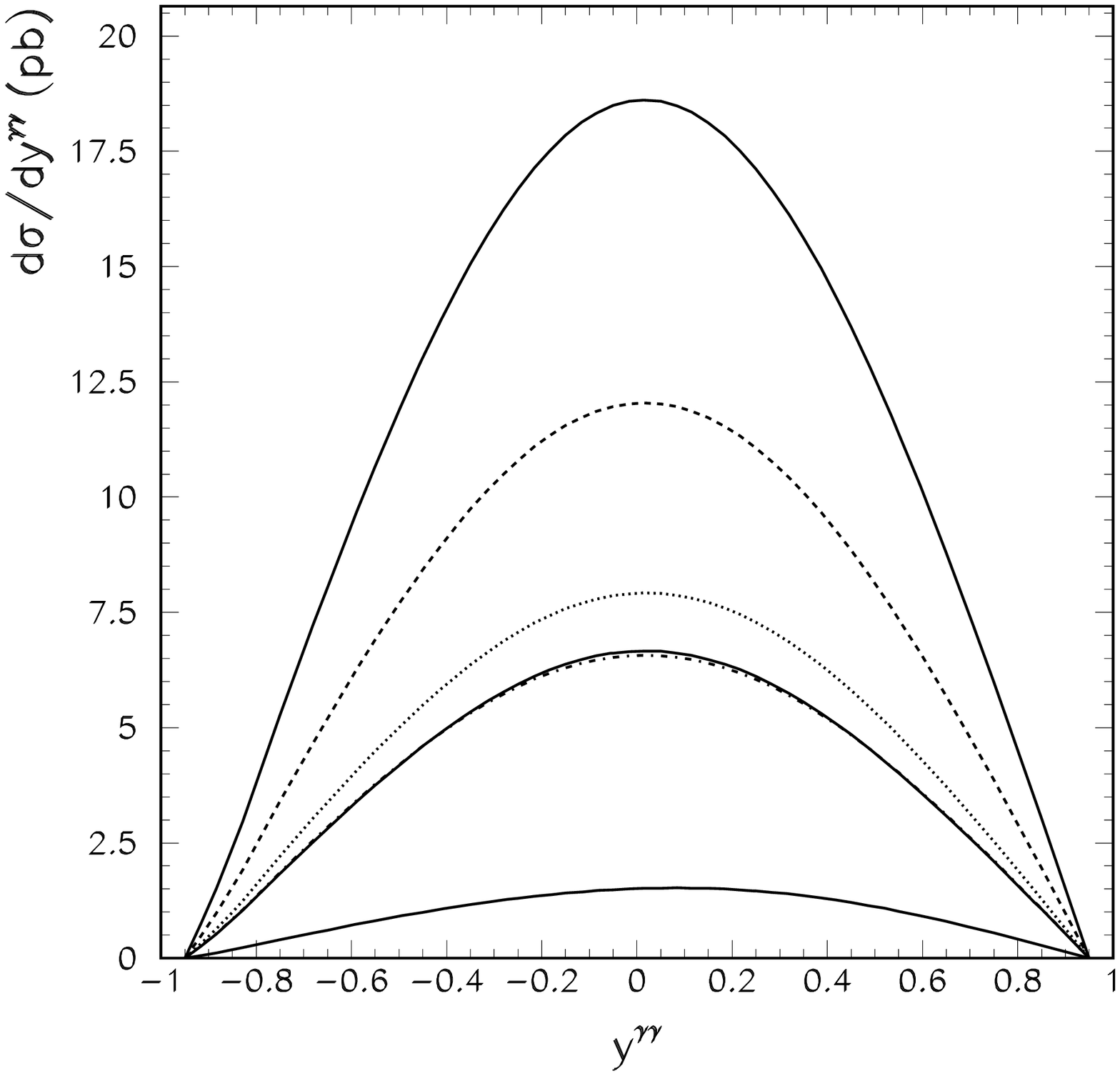} \\
\multicolumn{2}{c}{\epsfysize=7.0cm \epsffile{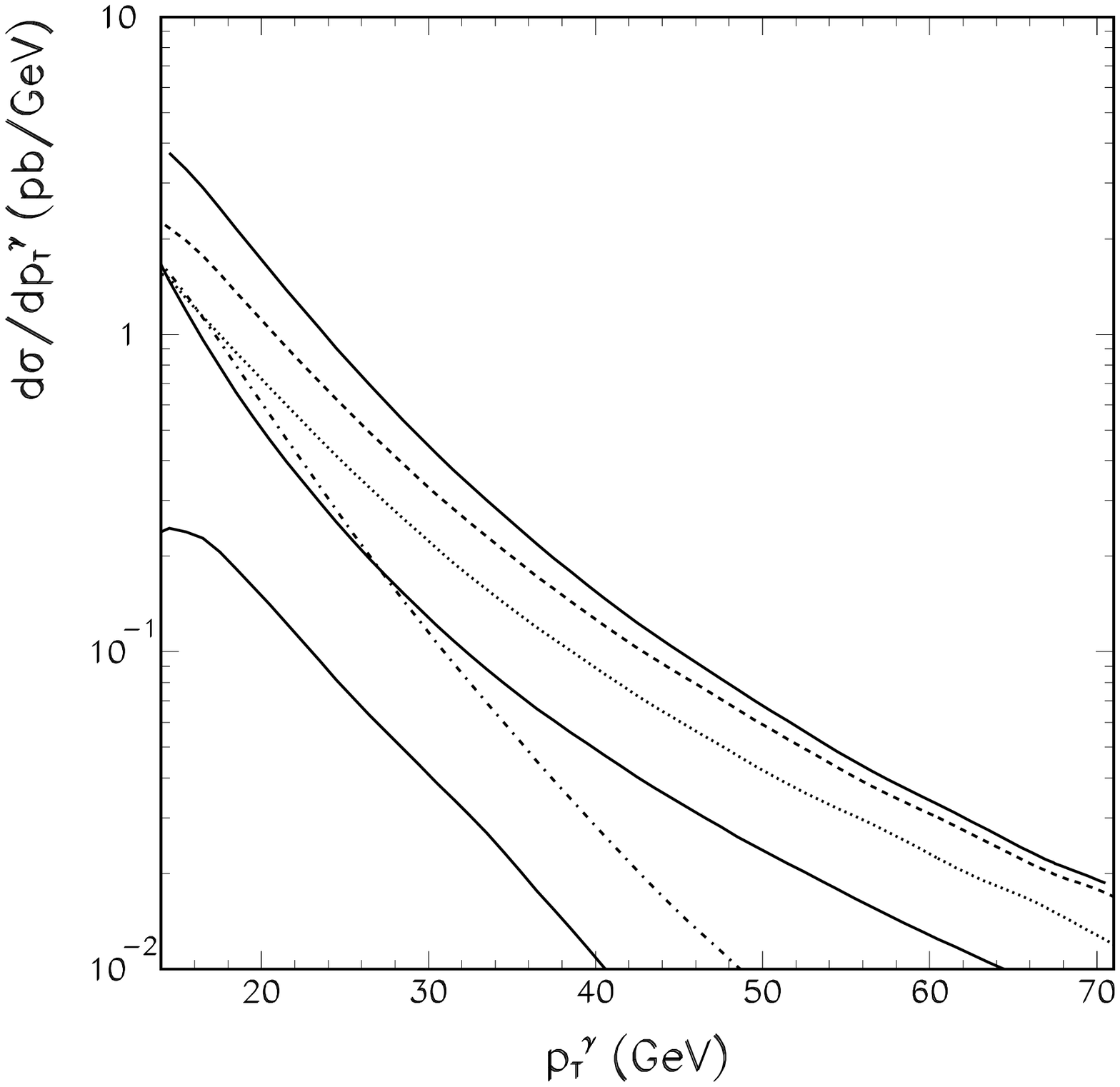} }
\end{tabular}
} \fi
\end{center}
\vspace{-1cm}
\caption{
Same as Fig.~\ref{fig:AALHCQypT} but for the upgraded Tevatron.
}
\label{fig:AATevQypT}
\end{figure*}
}
\def\FigAATevQT
{
\begin{figure*}[t]
\vspace{-1.5cm}
\begin{center}
\begin{tabular}{cc}
\ifx\nopictures Y \else{ \epsfysize=10.0cm \epsffile{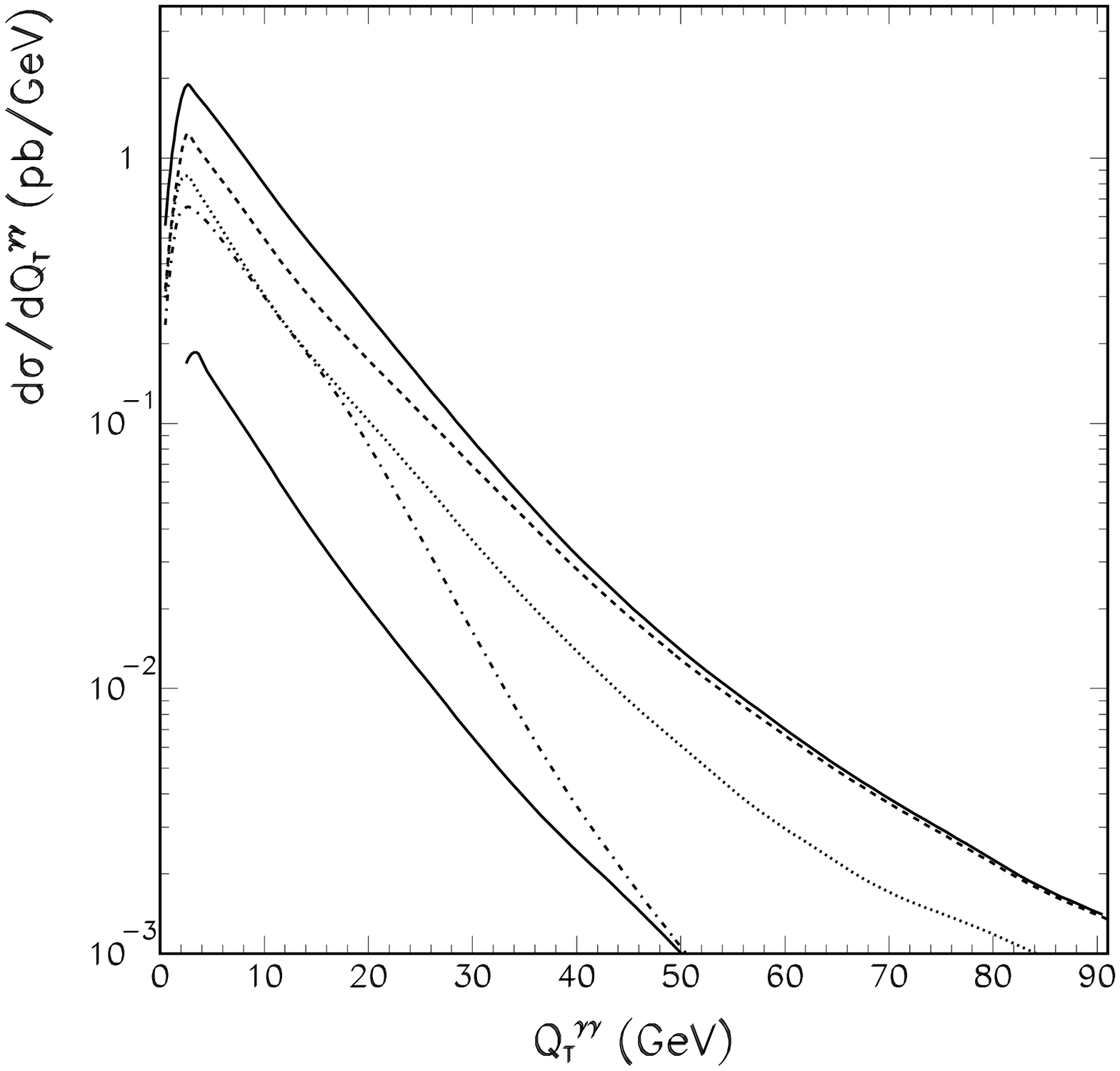}}
\fi & 
\end{tabular}
\end{center}
\vspace{-1cm}
\caption{
Same as Fig.~\ref{fig:AALHCQT} but for the upgraded Tevatron.
}
\label{fig:AATevQT}
\end{figure*}
}
\def\FigAATevInt
{
\begin{figure*}[t]
\vspace{-1.5cm}
\begin{center}
\begin{tabular}{c}
\ifx\nopictures Y \else{ \epsfysize=10.0cm \epsffile{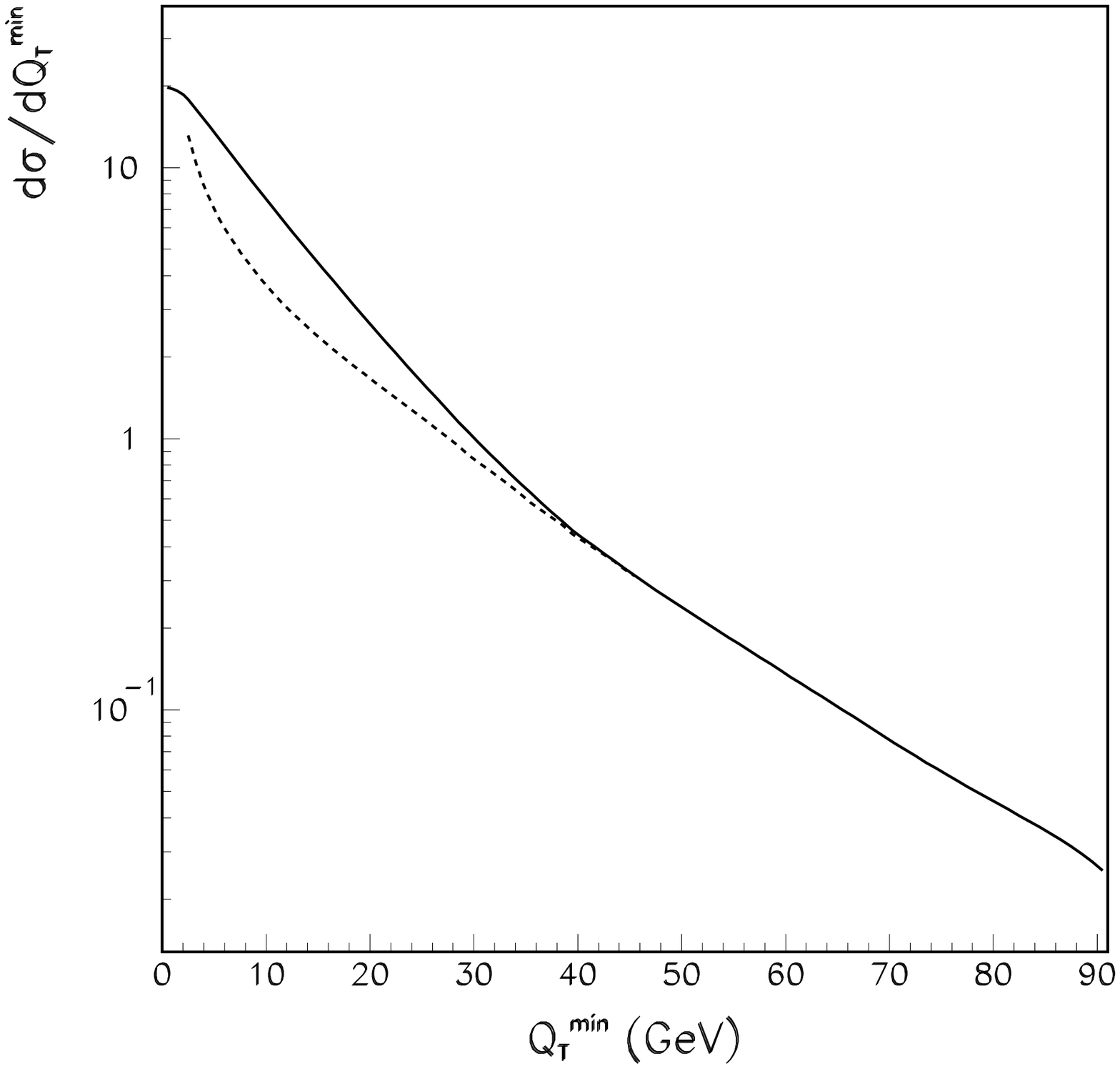}} \fi
\end{tabular}
\end{center}
\vspace{-1cm}
\caption{
Same as Fig.~\ref{fig:AALHCInt} but for the upgraded Tevatron.
}
\label{fig:AATevInt}
\end{figure*}
}
\def\FigAALHCQypT
{
\begin{figure*}[p]
\vspace{-1.5cm}
\begin{center}
\ifx\nopictures Y \else{
\begin{tabular}{cc}
\epsfysize=7.0cm \epsffile{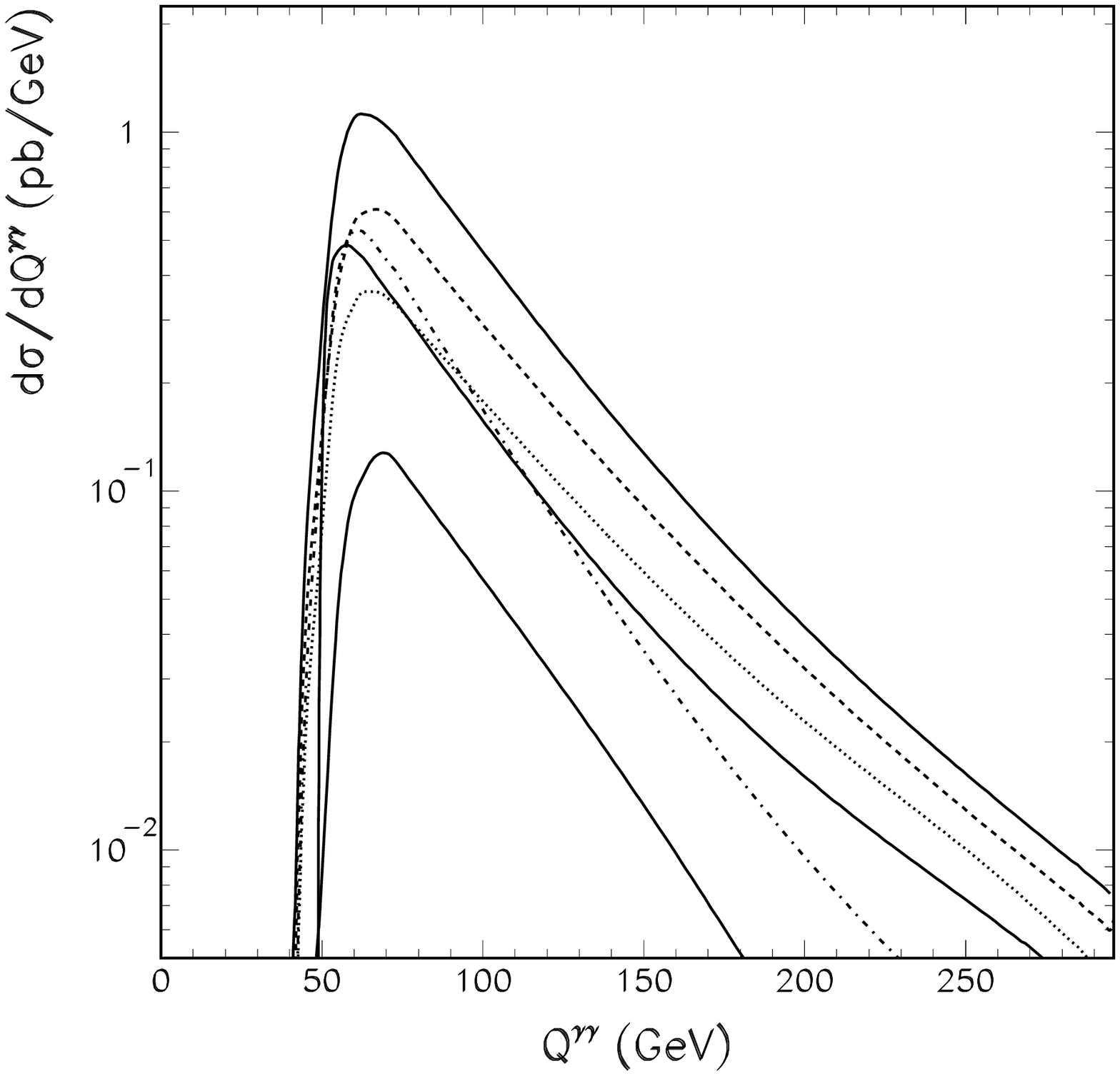} &
\epsfysize=7.0cm \epsffile{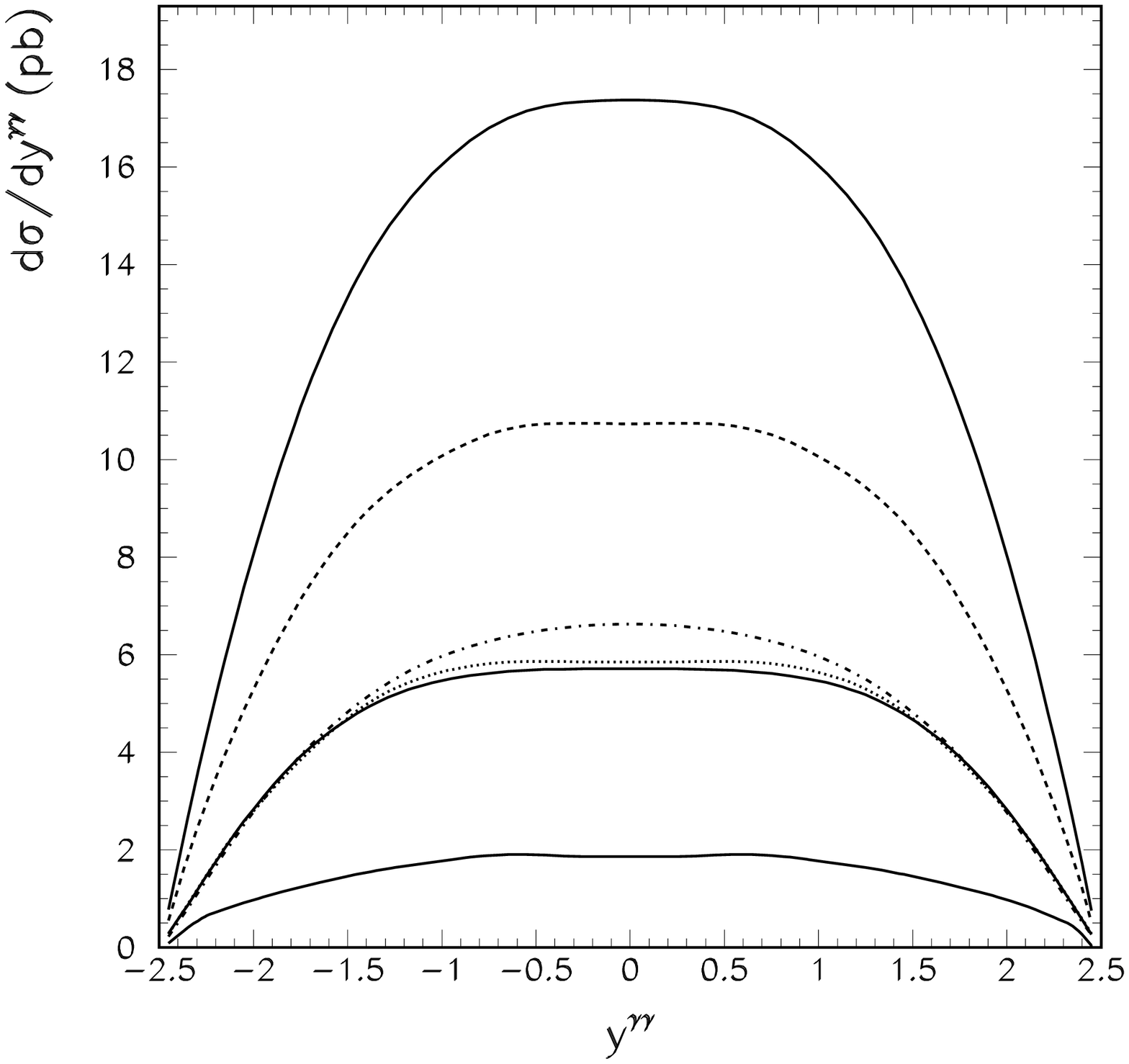} \\
\multicolumn{2}{c}{\epsfysize=7.0cm \epsffile{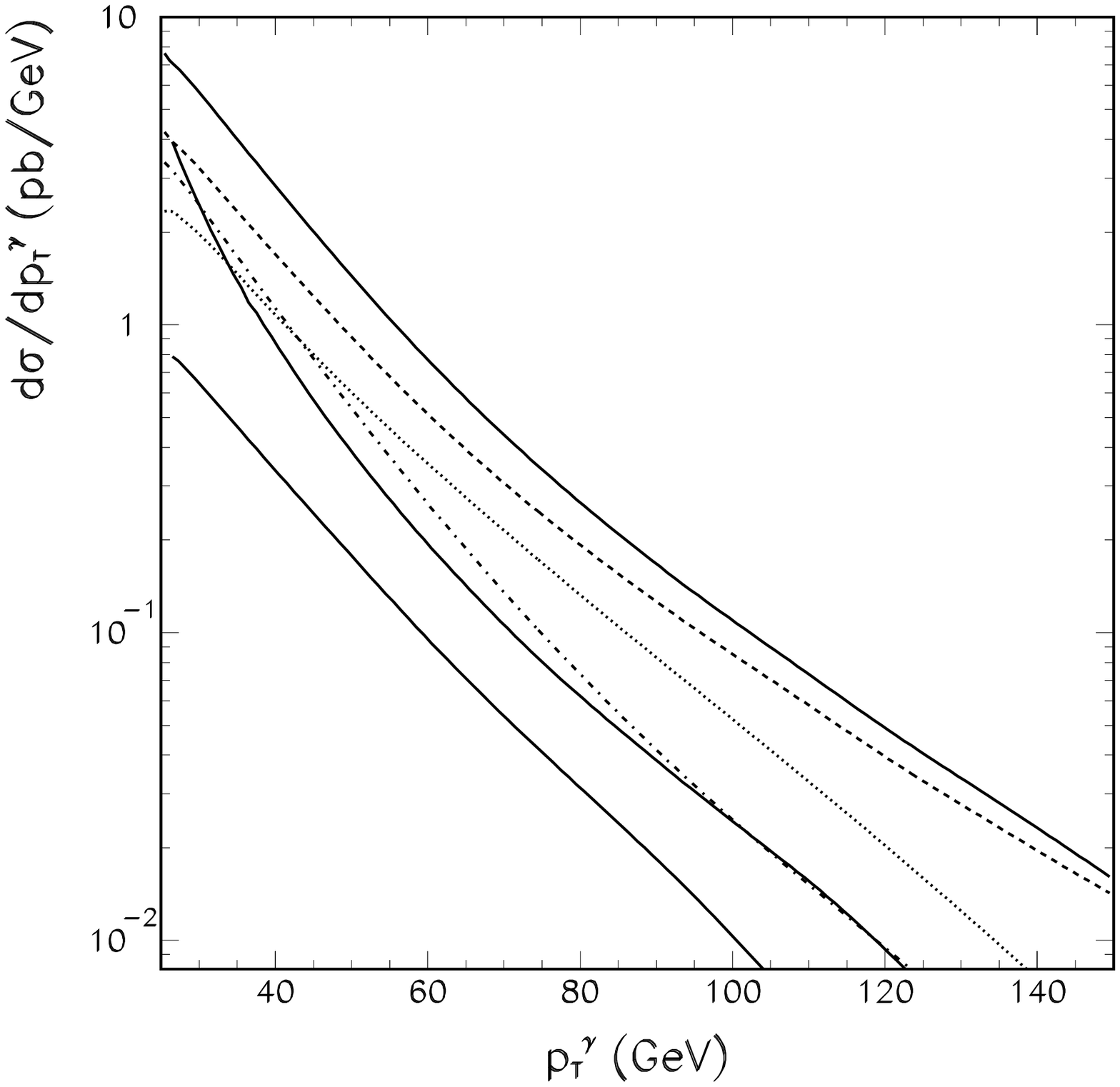} }
\end{tabular}
} \fi
\end{center}
\vspace{-1cm}
\caption{
Invariant mass and rapidity distributions of photon pairs, and
transverse momentum distributions of the individual photons  
at the LHC. 
The total resummed contribution (upper solid), and the resummed 
$q{\bar q} + q g \to \gamma \gamma X$ (dashed), 
$q{\bar q}\to \gamma \gamma X$ (dotted), 
$g g \to \gamma \gamma g$ (dash-dotted), as well as the 
fragmentation (lower solid)
contributions are shown separately. 
The $q{\bar q}\to \gamma \gamma$ ${\cal O}(\alpha_s^0)$ 
distribution is shown in the middle solid curve. 
}
\label{fig:AALHCQypT}
\end{figure*}
}
\def\FigAALHCQT
{
\begin{figure*}[t]
\vspace{-1.5cm}
\begin{center}
\begin{tabular}{c}
\ifx\nopictures Y \else{ \epsfysize=10.0cm \epsffile{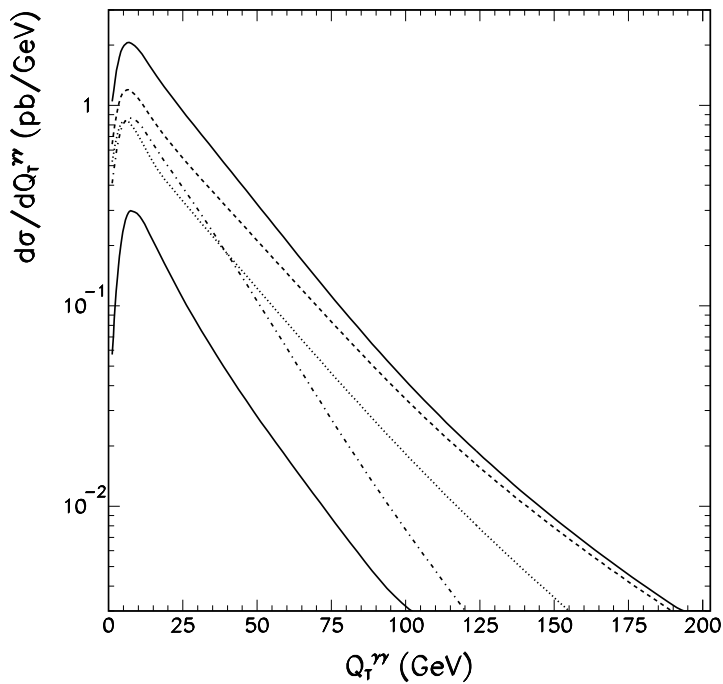}} \fi
\end{tabular}
\end{center}
\vspace{-1cm}
\caption{Transverse momentum distribution of photon pairs at the LHC. 
The total resummed contribution (upper solid), the resummed
$q{\bar q} + q g \to \gamma \gamma X$ (dashed), 
$q{\bar q}\to \gamma \gamma X$ (dotted),
$g g \to \gamma \gamma X$ (dash-dotted), as well as the
fragmentation (lower solid) contributions are shown separately. }
\label{fig:AALHCQT}
\end{figure*}
}
\def\FigAALHCInt
{
\begin{figure*}[t]
\vspace{-1.5cm}
\begin{center}
\begin{tabular}{c}
\ifx\nopictures Y \else{ \epsfysize=10.0cm \epsffile{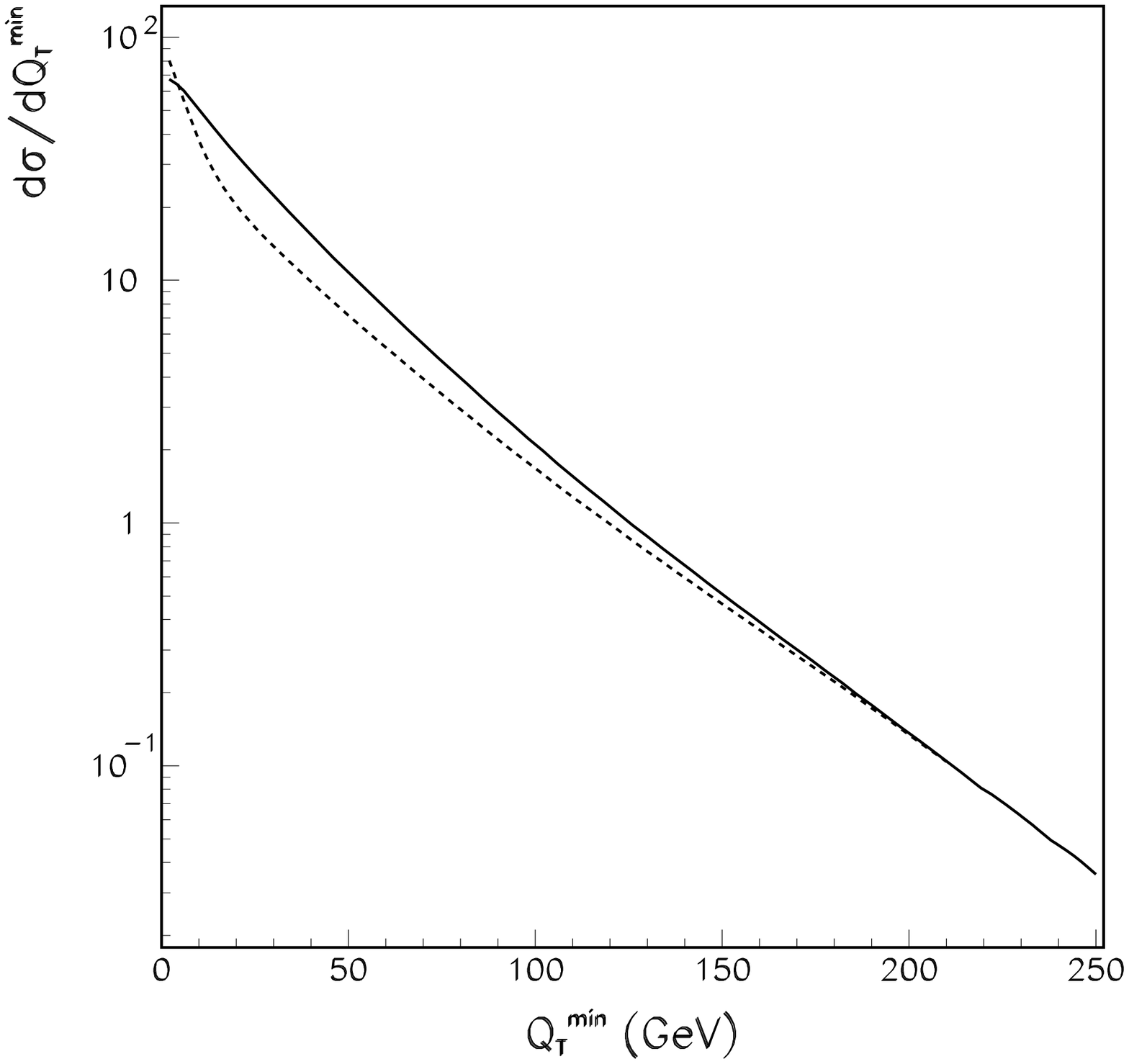}} 
\fi
\end{tabular}
\end{center}
\vspace{-1cm}
\caption{The integrated cross section for photon pair production at the LHC. 
The resummed and the ${\cal O}(\alpha_s)$ distributions are shown in solid and
dashed lines, respectively.
 }
\label{fig:AALHCInt}
\end{figure*}
}



\section{Introduction}

The underlying dynamics of the electroweak symmetry breaking sector of
the Standard Model (SM) awaits understanding. The principal
goal of the CERN Large Hadron Collider (LHC) is to shed light on this open
question. 
The direct searches at the CERN Large Electron Positron (LEP) collider 
have constrained the mass of the SM Higgs boson to be higher than 
90 GeV \cite{Moriond98}. 
Furthermore, 
global analyses of electroweak data \cite{Langacker} and the
 values of the top
quark and the $W^{\pm }$ boson masses \cite{Parke} suggest 
that the SM Higgs boson is light, less than a few hundred GeV. 
Arguments based on supersymmetry (SUSY) also indicate that the
lightest Higgs boson is lighter than the top quark \cite{SUSYHiggsMass}.
Hence, the existence of a light Higgs boson is highly possible.
 
It has been shown in the literature that a SM like Higgs boson 
with a mass less than or about 180 GeV can be detected 
at the upgraded Fermilab Tevatron via 
$p\bar{p}\to W^{\pm} (\to \ell^{\pm} \nu)~H(\to b\bar{b},\tau^+ \tau^-)$
\cite{TEV2000}, or
$p \bar p(gg) \to H(\to W^*W^* \to \ell \nu jj$ and $\ell \nu \ell \nu$)
\cite{Han-Zhang},
and a SUSY Higgs boson can be detected 
in the $W^{\pm} h$ and $hb\bar{b}$ modes 
\cite{Balazs-Diaz-He-Tait-Yuan}.
To observe a light ($m_H<120$ GeV) neutral Higgs
boson at the LHC, the most promising detection mode is the diphoton channel 
$H\to \gamma \gamma$ \cite{AtlasTechProposal}
via the production process $pp(gg)\to HX$.  
In the intermediate mass ($120\;{\rm GeV}<m_H<2~m_Z$)
region, the $Z^{0*}Z^0$ channel is also useful 
in addition to the $\gamma \gamma$ channel \cite{AtlasTechProposal}.
If the Higgs boson is heavier than twice the mass of the $Z^0$
boson, the gold--plated decay mode into two $Z^0$ bosons 
(which sequentially decay into leptons) \cite{AtlasTechProposal}
is the best way to detect it. 
At the LHC, as at any hadron--hadron collider, initial--state 
radiative corrections from
Quantum Chromodynamics (QCD) interaction to electroweak 
processes can be large. 
Some fixed order QCD corrections have been calculated
to the Higgs signal and to its most important backgrounds 
\cite{HiggsFixO,Ohnemus-Owens,Mele-Nason-Ridolfi,DiphotonFixO}. 
The next-to-leading order (NLO) corrections to the total cross section 
of $pp(gg)\to HX$ have been found to be large (50-100 \%) \cite{HiggsFixO}, 
and the largest contribution in the fixed order corrections results from 
soft gluon emission \cite{Kramer-Laenen-Spira}.
This signals the slow convergence of the perturbative series,
and the importance of still higher order corrections.
Furthermore, the
fixed order corrections fail to predict the transverse momentum
distributions of the
Higgs boson and its decay products correctly. The knowledge of these
distributions is necessary to precisely predict the signal and the
background in the presence of various kinematic cuts,
in order to deduce the accurate event rates to compare with theory 
predictions \cite{ADIKSS}. 
To predict the correct distribution of the transverse
momentum of the photon or $Z^0$ pair and the individual 
vector bosons, or the
kinematical correlation of the two vector bosons produced 
at hadron colliders, it is necessary to include
the effects from the initial--state multiple soft--gluon emission.
In this Section, we present the results
of our calculation for the most important continuum backgrounds to the 
Higgs boson signal detected at hadron colliders. 
The distributions of the Higgs boson signal for the 
$h_1 h_2 \to H\to \gamma \gamma X$ and $Z^0 Z^0 X$ processes,
including the soft--gluon effects, 
will be discussed in our future work \cite{Balazs-Yuan_Higgs}.

In scattering processes involving hadrons, the dynamics of the multiple 
soft--gluon radiation can be described by the resummation technique.
We extend the Collins--Soper--Sterman (CSS) resummation formalism 
\cite{CS,CSS,Collins-Soper-Sterman} to describe the production of photon
and $Z^0$ pairs. 
This extension is analogous to our recent resummed calculation of
the hadronic production of photon pairs \cite{Balazs-Berger-Mrenna-Yuan}. 
In comparison, an 
earlier work \cite{Han-Meng-Ohnemus} on the soft--gluon resummation for the 
$q{\bar{q}}\to Z^0 Z^0 X$ process did not include the 
complete NLO corrections.
In the present Section, the effect of 
initial--state multiple soft--gluon emission in 
$q{\bar{q}}\to Z^0 Z^0X$ is resummed with the inclusion of the full NLO 
contributions, so that the inclusive rate of the $Z^0$ boson pair 
production agrees with the NLO result presented 
in Ref.~\cite{Ohnemus-Owens}. 
Furthermore, we also include part of the higher order contributions 
in our results by using the CSS resummation formalism.

The collected diphoton data at the Tevatron, 
a $p\bar{p}$ collider with center-of-mass energy 
$\sqrt{S}=1.8$ TeV with 84 and 81 pb$^{-1}$ 
integrated luminosity (for CDF and \D0), 
are in the order of $10^3$ events per experiment 
\cite{CDFDiPhoton,D0DiPhoton}.
After the upgrade of the Tevatron, with $\sqrt{S}=2$ TeV
and a 2 fb$^{-1}$ integrated luminosity, about $4\times 10^4$ photon 
pairs can be detected, 
and more than $3\times 10^3$ $Z^0$ boson pairs can be produced. 
At the LHC, a $\sqrt{S}=14$ TeV $pp$ collider with 
a 100 fb$^{-1}$ integrated luminosity, 
we expect about $6\times 10^6$ photon 
and $1.5\times 10^6$ $Z^0$ pairs to be produced,
after imposing the kinematic cuts described later in the text.
This large data sample will play an important role
in the search for the Higgs boson(s) and new physics 
that modifies the production of the vector boson pairs
(e.g., by altering the vector boson tri-linear couplings
\cite{www}).

The rest of this Chapter is organized as follows. Section \ref{Analytical}
briefly summarizes the extension of the CSS resummation formalism to the $Z^0
Z^0$ pair production. 
In Section \ref{Numerical} the numerical results of the
resummed and fixed order calculations are compared for various
distributions of the photon and $Z^0$ pairs produced at the LHC and the
upgraded Tevatron. 

\section{Analytical Results \label{Analytical}}

\subsection{The CSS Resummation Formalism for $Z^0$ Pair Production}

When QCD corrections to the $Z^0$ boson pair production cross section are
calculated order by order in the strong coupling constant 
$\alpha _s$, 
the emission of potentially soft gluons spoils the convergence of the
perturbative series for small transverse momenta ($Q_T$) of the $Z^0$ boson
pair. In the $Q_T\ll Q$ region, the cross section can be written as \cite
{Collins-Soper-Sterman} 
\[
\lim_{{Q_T\to 0}}\frac{d\sigma }{dQ_T^2}=\sum_{n=1}^\infty
\sum_{m=0}^{2n-1}\alpha _s^n~\frac{_nv_m}{Q_T^2}~\ln ^m\left( 
\frac{Q^2}{Q_T^2}
\right) +{\cal O}\left (\frac 1{Q_T} \right), 
\]
where $Q$ is the invariant mass of the $Z^0$ boson pair, and 
the coefficients $_nv_m\,$are perturbatively calculable. 
At each order of the strong coupling the emitted
gluon(s) can be soft and/or collinear, which yields a small $Q_T$.
When the two scales $Q$ and $Q_T$ are very different, the
logarithmic terms $\ln ^m (Q^2/Q_T^2)$ are large, 
and for $Q_T\ll Q$ the perturbative series is
dominated by these terms. 
It was shown in Refs.~\cite{CS,CSS,Collins-Soper-Sterman} that 
these logarithmic contributions can be summed up to all order in $\alpha_s$, 
resulting in a well behaved cross section in the full $Q_T$ region.

The resummed differential cross section of the $Z^0$ boson pair production
in hadron collisions is written, similarly to the cross sections of the
lepton pair production \cite{Balazs-YuanPRD}, or photon pair production \cite
{Balazs-Berger-Mrenna-Yuan}, in the form: 
\begin{eqnarray}
&&{\frac{d\sigma (h_1h_2\to Z^0Z^0X)}{dQ^2\,dy\,dQ_T^2\,d\cos {\theta }%
\,d\phi }}={\frac 1{48\pi S}}\,{\frac {\beta}{Q^2}}  \nonumber \\
&&~~\times \left\{ {\frac 1{(2\pi )^2}}\int d^2b\,e^{i{\vec{Q}_T}\cdot
 {\vec{%
b}}}\,\sum_{i,j}{\widetilde{W}_{ij}(b_{*},Q,x_1,x_2,\theta ,\phi
,C_{1,2,3})}\,\widetilde{W}_{ij}^{NP}(b,Q,x_1,x_2)\right.  \nonumber \\
&&~~~~\left. +~Y(Q_T,Q,x_1,x_2,\theta ,\phi ,{C_4})\right\} .
\label{Eq:ResFor}
\end{eqnarray}
In this case, the variables $Q$, $y$, and $Q_T$ are the invariant mass,
rapidity, and transverse momentum of the $Z^0$ boson pair in the laboratory 
frame, while $\theta $ and $\phi $ are the polar and azimuthal angle
 of one of 
the $Z^0$ bosons in the Collins-Soper frame \cite{CSFrame}. The factor 
\[
\beta = \sqrt{1 -\frac{4 m_Z^2}{Q^2}} 
\]
originates from the phase space for producing the massive $Z^0$ boson pair. 
The parton
momentum fractions are defined as $x_1=e^yQ/\sqrt{S}$, and $x_2=e^{-y}Q/%
\sqrt{S}$, and $\sqrt{S}$ is the center--of--mass (CM) energy of the hadrons 
$h_1$ and $h_2$.

The renormalization group invariant function $\widetilde{W}_{ij}(b)$ sums
the large logarithmic terms $\alpha _s^n \ln ^m(b^2Q^2)$ to all orders in $%
\alpha _s$. For a scattering process initiated by the partons $i$ and $j$, 
\begin{eqnarray}
&&\widetilde{W}_{ij}(b,Q,x_1,x_2,\theta ,\phi ,C_{1,2,3})= \exp
\left\{ -{\cal S}_{ij}(b,Q,C_{1,2})\right\}  \nonumber \\
&& ~\times \left[ {\cal C}_{i/h_1}(x_1,b,C_{1,2,3},t,u)\, {\cal C}%
_{j/h_2}(x_2,b,C_{1,2,3},t,u) \right.  \nonumber \\
&& ~~ + \left. {\cal C}_{j/h_1}(x_1,b,C_{1,2,3},u,t)\, {\cal C}%
_{i/h_2}(x_2,b,C_{1,2,3},u,t)\right]  \nonumber \\
&& ~\times {\cal F}_{ij}(\alpha(C_2Q),\alpha _s(C_2Q),\theta
,\phi).  \label{Eq:WTwiZPair}
\end{eqnarray}
Here the Sudakov exponent ${\cal S}_{ij}(b,Q,C_{1,2})$ is defined in 
Eq.(\ref{eq:SudExp}).
In Eq.~(\ref{Eq:WTwiZPair}), ${\cal F}_{ij}$ originates from the hard 
scattering process, and will be
given later for specific initial state partons.
Eq.~(\ref{Eq:DefCalC}) defines ${\cal C}_{i/h}(x)$ which denotes
the convolution of the perturbative Wilson coefficient functions 
$C_{ia}$ with parton distribution functions (PDF) 
$f_{a/h}(\xi)$ (describing
the probability density of parton $a$ inside hadron $h$ 
with momentum fraction $\xi$).
The invariants $s$, $t$ and $u$ are defined for the $q(p_1)\bar{q}%
(p_2)\to Z^0(p_3)Z^0(p_4)$ subprocess as 
\begin{equation}
s=(p_1+p_2)^2,~~~~~~t=(p_1-p_3)^2,~~~~~~u=(p_2-p_3)^2,
\end{equation}
with $s+t+u=2m_Z^2$.

The functions $A_{ij}$, $B_{ij}$ and $C_{ij}$ are calculated
perturbatively in powers of $\alpha _s/\pi$ as indicated by
Eq.~(\ref{eq:ABCExp}).
The dimensionless constants $C_1,\;C_2$ and $C_3\equiv \mu b$ were
introduced in the solution of the renormalization group equations for $%
\widetilde{W}_{ij}$. Their canonical choice is $C_1=C_3=2e^{-\gamma
_E}\equiv b_0$, $C_2=C_1/b_0=1$, and $C_4=C_2=1$ 
\cite{Collins-Soper-Sterman}, where $\gamma _E$ is the Euler constant.

For large $b$, which is relevant for small $Q_T$, the 
perturbative evaluation 
of Eq.~(\ref{Eq:WTwiZPair}) is questionable. 
Thus in Eq.~(\ref{Eq:ResFor}), $\widetilde{W}_{ij}$ is 
evaluated at $b_{*}={b/\sqrt{1+(b/b_{{\rm max}})^2}}$, 
so that the perturbative calculation is reliable.
Here $b_{{\rm max}}$ is a free parameter of the
formalism \cite{Collins-Soper-Sterman} 
that has to be constrained by other data (e.g. Drell--Yan),
along with the non-perturbative function $\widetilde{W}_{ij}^{NP}(b)$
which is introduced in Eq.~(\ref{Eq:ResFor}) to parametrize the 
incalculable long distance effects.
Since the $q{\bar q}\to \gamma \gamma$ or $Z^0 Z^0$, and the 
$q{\bar q}\to V \to \ell \ell '$ processes have the same initial state 
as well as the same QCD color structure, 
in this work we assume that the 
non-perturbative function $\widetilde{W}_{ij}^{NP}(b)$, fitted to existing 
low energy Drell-Yan data \cite{Ladinsky-Yuan}, 
also describes the non-perturbative 
effects in the $q{\bar q}\to \gamma \gamma$ and $Z^0 Z^0$ processes.
Needless to say, this assumption has to be tested by experimental data.

The function $Y$ in Eq.~(\ref{Eq:ResFor}) contains contributions from the
NLO calculation that are less singular than $1/Q_T^2$ or 
$\ln (Q^2/Q_T^2)/Q_T^2$ as $Q_T\to 0$. 
This function restores the regular contribution
in the fixed order perturbative calculation that is not included in the
resummed piece $\widetilde{W}_{ij}$. In the $Y$ function, both the
factorization and the renormalization scales are chosen to be $C_4Q$. The
detailed description of the matching (or ``switching'')
between the resummed and the fixed
order cross sections for $Q_T\sim Q$ can be found in
Ref.~\cite{Balazs-YuanPRD}.

\subsection{The $q\bar{q}$, $qg$ and $\bar{q}g\to Z^0Z^0X$ subprocesses}

\FigDiagrams
The largest background to the Higgs boson signal 
in the $Z^0 Z^0$ channel is the continuum production
of $Z^0$ boson pairs via the $q{\bar{q}}\to Z^0Z^0X$ partonic subprocess 
\cite{Glover-Bij}. The next--to--leading order calculations of this process
are given in Refs. \cite{Ohnemus-Owens,Mele-Nason-Ridolfi}. 
A representative set of Feynman diagrams, 
included in the NLO calculations,
is shown in Fig.~\ref{fig:Diagrams}.
The application
of the CSS resummation formalism for the $q\bar{q}\to Z^0Z^0X $ subprocess
is the same as that for the case of $q\bar{q}\to \gamma \gamma X $ \cite
{Balazs-Berger-Mrenna-Yuan}. The $A^{(1)}$, $A^{(2)}$ and $B^{(1)}$
coefficients in the Sudakov exponent are identical to those of the
Drell--Yan case. This follows from the observation that to produce a heavy 
$Z^0$ boson pair, 
the virtual--quark line connecting the two $Z^0$ bosons in 
Fig.~\ref{fig:Diagrams} is far off the mass shell, and the leading logarithms
due to soft gluon emission beyond the leading order can only be generated
from the diagrams in which soft gluons are connected to the incoming
(anti--)quark. This situation was described in more detail for di-photon
production \cite{Balazs-Berger-Mrenna-Yuan}.

The resummed cross section is given by Eq.~(\ref{Eq:ResFor}), with 
$i$ and $%
j $ representing quark and anti--quark flavors, respectively, and 
\[
{\cal F}_{ij} (g,g_s,\theta,\phi) = 
2\delta _{ij}(g_L^2+g_R^2)^2 \, \frac{1+\cos ^2\theta}{1-\cos
^2\theta}. 
\]
The left- and right-handed couplings $g_{L,R}$ are defined through the $q{%
\bar{q}Z}^0$ vertex, which is written as $i\gamma _\mu \left[ g_L(1-\gamma
_5)+g_R(1+\gamma _5)\right]$, with 
\begin{equation}
g_L= g \, \frac{T_3-s_w^2Q_f}{2c_w} ~~~ {\rm and} ~~~ g_R=-g \, \frac{%
s_w^2Q_f}{2c_w}.  
\label{Eq:DefgLgR}
\end{equation}
Here $g$ is the weak coupling constant, $T_3$ is the third component of the
SU(2)$_L$ generator 
($T_3 = 1/2$ for the up quark $Q_u$, and $ -1/2$ for the down 
quark $Q_d$), 
$s_w$ ($c_w$) is the sine (cosine) of the weak mixing
angle, and $Q_f$ is the electric charge of the incoming quark in the units
of the positron charge ($Q_u=2/3$ and $Q_d=-1/3$). 
The values of these parameters will be given in the next section.

The explicit forms of the $A$ and $B$ coefficients, 
used in the numerical calculations are: 
\begin{eqnarray}
A_{q\bar{q}}^{(1)}(C_1) &=&C_F,  \nonumber \\
A_{q\bar{q}}^{(2)}(C_1) &=&C_F\left[ \left( \frac{67}{36}-\frac{\pi ^2}{12}%
\right) N_C-\frac 5{18}N_f-2\beta _1\ln \left( \frac{b_0}{C_1}\right)
\right] ,  \nonumber \\
B_{q\bar{q}}^{(1)}(C_1,C_2) &=&C_F\left[ -\frac 32-2\ln \left( \frac{C_2b_0}{%
C_1}\right) \right] ,
\label{Eq:Cjs}
\end{eqnarray}
where $N_f$ is the number of light quark flavors, $N_C=3$ is the number of
colors in QCD, $C_F=4/3$, and $\beta _1=(11N_C-2N_f)/12$.

To obtain the value of the total cross section to NLO, it is necessary to
include the Wilson coefficients $C_{ij}^{(0)}$ and $C_{ij}^{(1)}$, 
which can be derived similarly to those for diphoton production in 
Chapter~\ref{ch:PhotonPair}.
The results are: 
\begin{eqnarray}
C_{jk}^{(0)}(z,b,\mu ,{\frac{C_1}{C_2}},t,u) &=&\delta _{jk}\delta ({1-z}),
\nonumber \\
C_{jG}^{(0)}(z,b,\mu ,{\frac{C_1}{C_2}},t,u) &=&0,  \nonumber \\
C_{jk}^{(1)}(z,b,\mu ,{\frac{C_1}{C_2}},t,u) &=&\delta _{jk}C_F\left\{
\frac 12(1-z)-\frac 1{C_F}\ln \left( \frac{\mu b}{b_0}\right) P_{j\leftarrow
k}^{(1)}(z)\right.  \nonumber \\
&&\left. +\delta (1-z)\left[ -\ln ^2\left( {\frac{C_1}{{b_0C_2}}}%
e^{-3/4}\right) +{\frac{{\cal V}(t,u)}4}+{\frac 9{16}}\right] \right\} , 
\nonumber \\
C_{jG}^{(1)}(z,b,\mu ,{\frac{C_1}{C_2}},t,u) &=&{\frac 12}z(1-z)-\ln
\left( \frac{\mu b}{b_0}\right) P_{j\leftarrow G}^{(1)}(z).
\end{eqnarray}
In the above expressions, the splitting kernels are \cite{DGLAP}  
\begin{eqnarray}
P_{{j\leftarrow k}}^{(1)}(z) &=&C_F\left( {\frac{1+z^2}{{1-z}}}\right)_+ ~~~%
{\rm and}  \nonumber \\
P_{{j\leftarrow G}}^{(1)}(z) &=&{\frac 12}\left[ z^2+(1-z)^2\right] .
\label{eq:DGLAP}
\end{eqnarray}
For $Z^0$ boson pair production the function ${\cal V}$ in Eq.~(\ref{Eq:Cjs})
is given by 
\[
{\cal V}(t,u)={\cal V}_{Z^0Z^0}(t,u)=-4+{\frac{\pi ^2}3}+{\frac{tu}{%
t^2+u^2}}\left( F(t,u)+F(u,t)-2\right) . 
\]
The definition of the function $F(t,u)$ is somewhat lengthy and can be
found in Appendix C of Ref.~\cite{Ohnemus-Owens} (cf. Eqs.~(C1) and (C2)).
The function ${\cal V}(t,u)$ depends on the kinematic
correlation between the initial and final states through its dependence 
on $t
$ and $u$. In the $m_Z \to 0$ limit, $F(t,u)+F(u,t)$ reduces to $%
F^{virt}(t,u)$ of the diphoton case which is given in Ref.~\cite
{Balazs-Berger-Mrenna-Yuan}.\footnote{%
This is connected to the fact that as $m_Z \to 0$
the virtual corrections of the $Z^0$ pair
and diphoton productions are the same (up to the couplings), which is
apparent when comparing Eq.(11) of Ref.~\cite{Bailey-Ohnemus-Owens} and 
Eq.(12) of Ref.~\cite{Ohnemus-Owens}, after including a missing factor 
of $1/(16 \pi s)$ in the latter equation.}

The non-perturbative function used in this study is \cite{Ladinsky-Yuan} 
\[
\widetilde{W}_{q\overline{q}}^{NP}(b,Q,Q_0,x_1,x_2)={\rm exp}\left[
-g_1b^2-g_2b^2\ln \left( {\frac Q{2Q_0}}\right) -g_1g_3b\ln {(100x_1x_2)}%
\right] , 
\]
with $g_1=0.11~{\rm GeV}^2$, $g_2=0.58~{\rm GeV}^2$, 
$g_3=-1.5~{\rm GeV}^{-1}$, and $Q_0=1.6~{\rm GeV}$. 
These values were
fit for the CTEQ2M parton distribution function, with the canonical
choice of the renormalization constants, i.e. $C_1=C_3=b_0$ and $C_2=1$,
and $b_{max}=0.5~{\rm GeV}^{-1}$ was used.  
In principle, these coefficients should be refit for CTEQ4M distributions 
\cite{CTEQ4} used in this study.
We have checked that using the updated fit in Ref.~\cite{Yuan_Moriond} 
does not change largely our conclusion because at the LHC and Tevatron
energies the perturbative Sudakov contribution is more important
compared to that in the low energy fixed target experiments.

Before concluding this section we note that for the diphoton
production, we use the formalism described in 
Ref.~\cite{Balazs-Berger-Mrenna-Yuan} 
to include the $gg\to \gamma \gamma X$ contribution, in which part of
the higher order corrections has been included via resummation. 
Since a gauge invariant calculation of the $gg\to Z^0 Z^0 g$ cross section in
the SM involves diagrams with the Higgs particle, 
we shall defer its discussion to a separate work ~\cite{Balazs-Yuan_Higgs}.

\section{Numerical Results \label{Numerical}}

We implemented our analytic results in the ResBos Monte Carlo event
generator \cite{Balazs-YuanPRD}. 
As an input we use the following electroweak parameters \cite{PDB}: 
\begin{eqnarray}
&& G_F=1.16639\times 10^{-5}~{\rm GeV}^{-2},~~m_Z=91.187~{\rm GeV}, \\
&& m_W=80.41~{\rm GeV},~~\alpha (m_Z)=\frac 1{128.88}.
\nonumber
\end{eqnarray}
In the on-shell renormalization scheme we define the effective weak mixing 
angle
\begin{eqnarray}
\sin^2 \theta_w^{eff} = 1 - \frac{m_W^2}{\rho m_Z^2},
\nonumber
\end{eqnarray}
with
\begin{eqnarray}
\rho = \frac{m_W^2}{m_Z^2}
\left( 1 - \frac{\pi \alpha (m_Z)}{\sqrt{2} G_F m_W^2} \right)^{-1}.
\nonumber
\end{eqnarray}

In Eq.(\ref{Eq:DefgLgR}), the coupling of the $Z^0$ boson to fermions, $g$,
is defined using the improved Born approximation: 
\[
g^2=4\sqrt{2}G_F(c_w^{eff})^2m_Z^2\rho,
\]
with $c_w^{eff} = \sqrt{1 - \sin^2 \theta_w^{eff}}$, 
the cosine of the effective weak mixing angle.
(In Eq.~(\ref{Eq:DefgLgR}) $c_w$ is identified with $c_w^{eff}$.)
We use the NLO expression for the running strong and electroweak couplings
$\alpha _s(\mu)$ and $\alpha(\mu)$, as well as the NLO 
parton distribution function CTEQ4M (defined in the modified
minimal subtraction, i.e. ${\overline {\rm MS}}$, scheme), 
unless stated otherwise. 
Furthermore, in all cases we set the renormalization scale equal to the
factorization scale: $\mu_R=\mu_F=Q$.

\TblTotal
Table~\ref{tbl:TotalZPair} summarizes the total rates for the leading order 
(LO), i.e. ${\cal O}(\alpha_s^0)$,
and resummed photon- and $Z^0$-pair production cross sections for the 
LHC and the Tevatron.
For the lowest order calculation we show results using LO (CTEQ4L) and 
NLO (CTEQ4M) parton distributions, 
because there is a noticeable difference
due to the PDF choice.
As it was discussed in Ref.~\cite{Balazs-YuanPRD}, 
the resummed total rate
is expected to reproduce the ${\cal O}(\alpha_s)$ rate, 
provided that in the resummed calculation the $A^{(1)}$, $B^{(1)}$ and 
$C^{(1)}$ coefficients and the ${\cal O}(\alpha_s)$ $Y$ piece are included, 
and the $Q_T$ distribution is described by the resummed result for 
$Q_T \leq Q$ and by the ${\cal O}(\alpha_s)$ result for $Q_T > Q$.
In our present calculation we added the $A^{(2)}$ coefficient to include
the most important higher order corrections in the Sudakov exponent.
Our matching prescription (cf. Ref.~\cite{Balazs-YuanPRD})
is to switch from the resummed prediction to the 
fixed-order perturbative calculation as they cross around $Q_T \sim Q$.
This switch is performed for any given $Q$ 
and $y$ of the photon or $Z^0$ boson pairs.
In the end, the total cross section predicted by our resummed 
calculation is about the same as that predicted by the NLO calculation. 
The small difference of those two predictions  
can be interpreted as an estimate of the contribution beyond
the NLO.

\subsection{$Z^0$ pair production at the LHC}

In the LHC experiments the $H \to Z^0 Z^0 $ channel can be identified 
through the decay products of the $Z^0$ bosons. The detailed experimental 
kinematic cuts for this process are given in Ref.~\cite{AtlasTechProposal}.
Since the aim of this work is not to analyze the decay kinematics of the 
background, rather to present the effects of the initial--state soft--gluon 
radiation, following Ref.~\cite{Ohnemus-Owens} 
for the LHC energies we restrict the rapidities of each $Z^0$ bosons as: 
$|y^Z|<3.0$. We do not apply any other kinematic cuts.
The total rates are given in Table~\ref{tbl:TotalZPair}.
Our ${\cal O}(\alpha_s^0)$ rates are in agreement with that of 
Ref.~\cite{Ohnemus-Owens} when calculated using the same PDF.
We expect the resummed rate to be higher than the 
${\cal O}(\alpha_s)$ rate due to the inclusion of the $A^{(2)}$ term.
Indeed, our $K$ factor, defined as the ratio of the resummed to the LO rate
using the same PDF in both calculations, is higher than the naive soft gluon
$K$-factor ($K_{\rm naive} = 1 + 8\pi \alpha_s(Q)/9 \sim 1.3$) of 
Ref.~\cite{Barger-Lopez-Putikka}, which estimates the NLO 
corrections to the production rate of $q\bar{q}\to Z^0 Z^0 X$ 
in the DIS (deep-inelastic scattering) scheme.
Our $K$-factor approaches the naive one with the increase of the 
center-of-mass energy, as expected.

\TblSubTotalZZ
The rates for the different subprocesses of the $Z^0$ boson pair production
are given in Table~\ref{tbl:SubTotalZZ}.
At the LHC the $q g \to Z^0 Z^0 X$ subprocess contributes
about 25\% of the $q q + q g \to Z^0 Z^0 X$ rate.
The $K$-factor is defined as the ratio $\sigma(q\bar{q}+qg\to Z^0 Z^0 X)
/\sigma(q\bar{q}\to Z^0 Z^0)$, which is about 1.4 for using CTEQ4M PDF.

\FigZZLHCQT
\FigZZLHCQypT
Figs.~\ref{fig:ZZLHCQT}--\ref{fig:ZZLHCQypT} show our results for proton-proton  
collisions at the LHC energy, $\sqrt{S}=14$ GeV. 
Fig.~\ref{fig:ZZLHCQT} shows the transverse momentum distribution of $Z^0$ 
pairs. 
The NLO (${\cal O}(\alpha_s)$) prediction for the 
$q{\bar q} + q g \to Z^0 Z^0 X$ 
subprocesses, shown by the dotted curve, is singular as $Q_T \to 0$.
This singular behavior originates from the contribution of terms which
grow at least as fast as $1/Q_T^2$ or $ln(Q^2/Q_T^2)/Q_T^2$.
This, so-called asymptotic part, is shown by the dash-dotted curve,
which coincides with the ${\cal O}(\alpha_s)$ distribution as $Q_T \to 0$.
After exponentiating these terms, the distribution
is well behaved in the low $Q_T$ region, as shown by the solid curve. 
The resummed curve matches the ${\cal O}(\alpha_s)$ curve at about 
$Q_T=320$ GeV.   
Following our matching prescription described in the previous section, 
we find that this matching takes place around $Q_T = 300$ GeV, 
depending on the actual values of $Q$ and $y$.
Fig.~\ref{fig:ZZLHCQT} also shows that at the LHC there
is a substantial contribution from $q g$ scattering,
which is evident from the difference between the solid and 
dashed curves, where the dashed curve is the resummed contribution from 
the $q \bar q \to Z^0 Z^0 X$ subprocesses.

\FigZZLHCInt
In Fig.~\ref{fig:ZZLHCInt} we give the integrated distributions, defined as
\begin{equation}
\frac{d\sigma}{dQ_T^{\min }} =
\int_{Q_T^{\min }}^{Q_T^{\max}}dQ_T\;
\frac{d\sigma}{dQ_T}, 
\label{eq:Integrated}
\end{equation}
where $Q_T^{\max}$ is the largest $Q_T$ allowed by the phase space. 
In the NLO calculation, this distribution grows
without bound near $Q_T^{\min }=0$, as a result of the singular behavior 
of the scattering amplitude when $Q_T \rightarrow 0$. 
It is clearly shown by Fig.~\ref{fig:ZZLHCInt} that the $Q_T$ 
distribution of the resummed calculation is different from that of 
the NLO calculation.
The different shapes of the two curves in Fig.~\ref{fig:ZZLHCInt}
indicates that the predicted $Z^0$ pair production rates, with a minimal value
of the transverse momentum $Q_T$, are different in the two calculations. This
is important at the determination of the background for the detection of a
Higgs boson even with moderately large transverse momentum.
For $Q_T^{min}=50$ GeV, the resummed cross section is 
about 1.5 times of the NLO cross section.

The invariant mass and the rapidity distributions of the $Z^0$ boson pairs, 
and the transverse momentum distribution of the individual $Z^0$ bosons
are shown in Fig~\ref{fig:ZZLHCQypT}.
When calculating the production rate as the function of the $Z^0$ pair 
invariant mass,
we integrate the $Q_T$ distribution for any $Q$, and $y$. 
When plotting the transverse momentum distributions of the individual 
$Z^0$ bosons, we include both of the $Z^0$ bosons per event.
 In the shape of the invariant mass and 
rapidity distributions we do not expect large
deviations from the NLO results. 
Indeed, the shape of our invariant mass distribution agrees with that in 
Ref.~\cite{Ohnemus-Owens}.
However, the resummed transverse momentum distribution 
$P_T^Z$ of the individual $Z^0$ bosons
is slightly broader than the NLO distribution 
(not shown in Fig~\ref{fig:ZZLHCQypT}, cf. Ref.~\cite{Ohnemus-Owens}).
This is expected because, in contrast with the NLO distribution, the resummed
transverse momentum distribution of the $Z^0$ boson pair is finite as 
$Q_T \rightarrow 0$ so that the $P_T^Z$ 
distribution is less peaked.

\subsection{$Z^0$ Pair Production at the upgraded Tevatron}

After the upgrade of the Fermilab Tevatron, there are
more than $3\times 10^3$ $Z^0$ boson pairs to be produced.
Since this data sample can be used to test
the tri-linear gauge boson couplings \cite{www}, we also give our 
results for  
the upgraded Tevatron with proton--anti-proton collisions
at a center-of-mass energy of $2$ TeV. 
Our kinematic cuts constrain the rapidity of both of the $Z^0$
bosons such that $|y^Z| < 3$.
Both the LO and resummed total rates are listed in Table~\ref{tbl:TotalZPair}.
The ratio $\sigma(q\bar{q}+qg\to Z^0 Z^0 X)/\sigma(q\bar{q}\to Z^0 Z^0)$ 
is about 1.6, which is larger than the naive soft gluon $K$-factor of 
1.3. 
Table~\ref{tbl:SubTotalZZ} shows that  
$q g \to Z^0 Z^0 X$ partonic subprocess contributes only a 
small amount (about 3\%) at this energy, in contrast to
25\% at the LHC.

\FigZZTevQypT
\FigZZTevInt
Figs.~\ref{fig:ZZTevQypT}--\ref{fig:ZZTevInt} show the resummed
predictions for the upgraded Tevatron. 
The invariant mass and rapidity distributions of $Z^0$ boson pairs, 
and the transverse momentum distribution of the individual $Z^0$ bosons
are shown in Fig.~\ref{fig:ZZTevQypT}. 
The solid curve shows the resummed contributions from the 
$q {\bar q} + q g \to Z^0 Z^0 X$ subprocess. 
The resummed contribution from the $q {\bar q} \to Z^0 Z^0 X$ subprocess
is shown by the dashed curve.
The leading order $q {\bar q} \to Z^0 Z^0$ cross section is also 
shown, by the dash-dotted curve.
The invariant mass distribution of the $q {\bar q} + q g \to Z^0 Z^0 X$
subprocess is in agreement with the NLO result of Ref.~\cite
{Ohnemus-Owens}, when calculated for $\sqrt{S}=1.8$ TeV.
From this figure we also find that the
contribution from the $q g \to Z^0 Z^0 X$ subprocess at the energy of the 
Tevatron is very small.

\FigZZTevQT
In Fig.~\ref{fig:ZZTevQT} we compare the NLO and resummed 
 distributions of the transverse momentum of the $Z^0$ pair.
The figure is qualitatively similar to that at the LHC, 
as shown in Fig.~\ref{fig:ZZLHCQT}. 
The resummed and the NLO curves merge at about 100 GeV.
The resummed contribution from the $q {\bar q} \to Z^0 Z^0 X$
subprocess is shown by the dashed curve, which clearly 
dominates the total rate.

In Fig.~\ref{fig:ZZTevInt} we show the integrated distribution 
$d\sigma/dQ_T^{min}$ for $Z^0$ boson pair production at the 
upgraded Tevatron. 
The figure is qualitatively the same as that for the LHC 
(cf. Fig.~\ref{fig:ZZLHCInt}).
The NLO curve runs well under the resummed one in the $Q_T^{\min }<$ 
80 GeV region, and the $Q_T$ distributions from the NLO and the resummed 
calculations have different shapes even in the region where $Q_T$ is of the 
order 60 GeV. 
For $Q_T^{min}=30$ GeV, the resummed rate is about 1.5 times of the NLO rate.

\subsection{Diphoton production at the LHC}

\TblSubTotalAA
Photon pairs from the decay process $H \to \gamma \gamma $ can
be directly detected at the LHC. 
When calculating its most important background rates, 
we impose the kinematic cuts on the final state photons
that reflect the optimal detection capabilities 
of the ATLAS detector \cite{AtlasTechProposal}:

$p_T^\gamma >25$ GeV, for the transverse momentum of each photons,

$\left| y^\gamma \right| <2.5$, for the rapidity of each photons, and

$p_T^1/(p_T^1+p_T^2)<0.7$, to suppress the fragmentation contribution,
where $p_T^1$ is the transverse momentum of the photon 
with the higher $p_T$ value.\\
We also apply a $\Delta R=0.4$ separation cut on the photons, but our results
are not sensitive to this cut. 
(This conclusion is similar to that in 
Ref.~\cite{Balazs-Berger-Mrenna-Yuan}.)
The total rates and cross sections from the different partonic subprocesses 
are presented in Tables~\ref{tbl:TotalZPair} and ~\ref{tbl:SubTotalAA}.
We have incorporated part of the higher order contributions to this process
by including $A^{(2)}$ in the Sudakov factor and $C^{(1)}_{gg}$ in 
the Wilson coefficient functions (cf. Ref.~\cite{Balazs-Berger-Mrenna-Yuan}).
Within this ansatz, up to ${\cal O}(\alpha_s^3)$, the 
$gg\to \gamma \gamma X$ rate is about 24 pb, which increases the total 
$K$-factor by almost 1.0.
The leading order $gg\to \gamma \gamma$ rate, via the box diagram, is about 
22 pb and 14 pb for using the LO PDF CTEQ4L and the NLO PDF CTEQ4M, 
respectively. The large difference mainly due to the differences in the
strong coupling constants used in the two calculations%
\footnote{When using the CTEQ4L PDF, we consistently use the LO running 
coupling constant $\alpha_s$.}:
CTEQ4L requires $\alpha_s(m_Z)=0.132$, while for CTEQ4M $\alpha_s(m_Z)=0.116$.
The ratio 
$\sigma(q\bar{q}+qg\to \gamma \gamma X)/\sigma(q\bar{q}\to \gamma \gamma)$ 
is 1.5, and 
$\sigma(gg\to \gamma \gamma X)/\sigma(q\bar{q}\to \gamma \gamma)$ 
is about 1. 
Hence, the ratio of the resummed and the ${\cal O}(\alpha_s^0)$ rates, 
is quite substantial.

\FigAALHCQypT
\FigAALHCInt
Figs.~\ref{fig:AALHCQypT}--\ref{fig:AALHCInt} show our predictions
for distributions of diphotons produced at the LHC. 
In Fig.~\ref{fig:AALHCQypT} we plot
the invariant mass and rapidity distribution of the photon pairs, 
and the transverse momentum distribution of the individual photons.
When plotting the transverse momentum distributions of the 
individual photons
we include both photons per event.
The total (upper solid) and the resummed 
$q{\bar q} + q g \to \gamma \gamma X$ (dashed), 
$q{\bar q}\to \gamma \gamma X$ (dotted), 
$g g \to \gamma \gamma X$ (dash-dotted), 
and the fragmentation (lower solid), 
as well as the leading order
 $q{\bar q} \to \gamma \gamma$ (middle solid)
contributions are shown separately.
The ratio of the resummed and the LO distributions is about 2.5 
which is consistent with the result in Table~\ref{tbl:TotalZPair}. 
The relative values of the contributions from each subprocesses reflect 
the summary given in Table~\ref{tbl:SubTotalAA}.

\FigAALHCQT
Fig.~\ref{fig:AALHCQT} shows various
contributions to the transverse momentum of the photon pair.
At low $Q_T$ values ($Q_T \ll Q$), the $q{\bar q} \to \gamma \gamma X$ 
contribution is larger than the
$qg \to \gamma \gamma X$ contribution,
while at high $Q_T$ values ($Q_T>Q$), the 
$qg \to \gamma \gamma X$ subprocess becomes more important.
The $gg$ contribution dominates the total rate in low $Q_T$ region, and
the kink in the $gg$ curve at about 50 GeV indicates the need for the 
inclusion of the complete ${\cal O}(\alpha_s^3)$ $gg\to\gamma\gamma g$ 
contribution.
(Recall that our prediction for the $gg$ contribution 
at ${\cal O}(\alpha_s^3)$ only holds for small $Q_T$, where the 
effect of the initial-state soft-gluon radiation is relatively
more important for a fixed $Q$.)
 
In Fig.~\ref{fig:AALHCInt} we give the integrated cross section 
as the function of the transverse momentum of the 
photon pair produced at the LHC. 
Similarly to the $Z^0$ pair production, 
there is a significant shape difference 
between the resummed and the NLO curves in the low to mid $Q_T$ region.
For $Q_T^{min}=50$ GeV, the resummed rate is about 1.5 times of the NLO
rate.

\subsection{Diphoton production at the upgraded Tevatron}

In Ref.~\cite{Balazs-Berger-Mrenna-Yuan}, we have presented the 
predictions of the CSS resummation formalism for the diphoton production
at the Tevatron with $\sqrt{S}=1.8$ TeV, and compared with the data 
\cite{CDFDiPhoton,D0DiPhoton}.
In this Section, we show the results for the upgraded Tevatron with 
$\sqrt{S}=2.0$ TeV.
We use the same kinematic cuts which were used in 
Ref.~\cite{Balazs-Berger-Mrenna-Yuan}:

$p_T^\gamma > 12$ GeV, for the transverse momentum of each photons,

$\left| y^\gamma \right| < 0.9$, for the rapidity of each photons. \\
An isolation cut of $\Delta R=0.7$ is also applied.
The total cross sections and the rates of the different subprocesses 
are given by Tables~\ref{tbl:TotalZPair} and ~\ref{tbl:SubTotalAA}.
The ratio of the $q\bar{q}+qg\to \gamma \gamma X$ and 
$q\bar{q}\to \gamma \gamma$ rates is about 1.5, similar to
that at the LHC.
The leading order rate for the $gg\to \gamma \gamma$ subprocess is
about 6.0 pb and 4.3 pb for using CTEQ4L and CTEQ4M PDF, respectively.
The NLO rate for $gg\to \gamma \gamma g$ is estimated to be 8.3 pb, using 
the approximation described in Ref.~\cite{Balazs-Berger-Mrenna-Yuan}, which 
is about the same magnitude 
as the leading order $q\bar{q} \to \gamma \gamma$ rate.
From our estimate of the NLO $gg$ rate, we expect that the 
complete ${\cal O}(\alpha_s^3)$ contribution will be important
for photon pair production at the Tevatron.

\FigAATevQypT
\FigAATevQT
\FigAATevInt
Figs.~\ref{fig:AATevQypT}--\ref{fig:AATevInt} show our results
for photon pairs produced at the upgraded Tevatron. 
The resummed predictions for the 
invariant mass and rapidity distributions of the photon pairs, and the
transverse momentum distribution of the individual photons are 
shown in Fig.~\ref{fig:AATevQypT}.
In Fig.~\ref{fig:AATevQT} we also plot
the contributions to the transverse momentum of 
the photon pair from the 
$q{\bar q} + q g \to \gamma \gamma X$ (dashed), 
$q{\bar q}\to \gamma \gamma X$ (dotted), 
$gg\to\gamma\gamma g$ (dash-dotted), 
and the fragmentation (lower solid) subprocesses, separately. 
The leading order $q{\bar q}\to\gamma\gamma$ 
cross section (middle solid) is also plotted.
In the low $Q_T$ region, the $gg$ and the $q\bar{q}$ rates
are about the same, and the $qg$ rate becomes more important in the large
$Q_T$ region. 
Furthermore, after imposing the above kinematic cuts,
the fragmentation contribution is found to be unimportant.

Fig.~\ref{fig:AATevInt} shows the integrated $Q_T$ distribution. 
The qualitative features of these distributions are the same as 
those predicted for the LHC.
For $Q_T^{min}=10$ GeV, the resummed cross section is 
about twice of the NLO cross section.

\section{Conclusions}

In this Section we studied the effects of the initial--state multiple 
soft--gluon emission on the total rates and various distributions of the 
most important background processes 
($pp, p{\bar p} \to \gamma \gamma X, Z^0Z^0 X$)
to the detection of the Higgs boson at the LHC.
We applied the extended CSS formalism to resum the large
logarithms induced by the soft--gluon radiation.
We found that for the $q {\bar q}$ and $q g$ initiated processes, 
the total cross sections and the invariant mass distributions
of the photon and $Z^0$ boson pairs are in agreement with the fixed
order calculations.
From our estimate of the NLO rate of the $gg$ initiated 
process, we expect that the 
complete ${\cal O}(\alpha_s^3)$ contribution will be important
for photon pair production at the Tevatron.
We showed that the resummed and the NLO transverse momentum distributions
of the $Z^0$ and photon pairs are substantially different for 
$Q_T\lesim Q/2$.
In terms of the integrated cross section above a given $Q_T^{\rm min}$,
this difference can be as large as 50\% in the low to mid-range of 
$Q_T^{\rm min}$.
Using the resummation calculation, we are able to give
a reliable prediction of the $Q_T$ and any other distribution 
in the full kinematical region at the LHC and the Tevatron, 
even in the presence of kinematic cuts.
Since the bulk of the signal is in the low transverse momentum region, 
we conclude that the difference between the NLO and resummed predictions 
of the background rates will be essential when extracting the signal of 
the Higgs boson at hadron colliders.



\chapter{ Charged Scalar Production at Hadron Colliders 
\label{ch:ChargedHiggs} }

\section{Introduction}
\vspace*{0.3cm}

The top quark ($t$), among the three generations of fermions, is the only
one with a large mass as high as the electroweak scale. This makes
the top the most likely place to discover new physics beyond the
Standard Model (SM).
In a recent study~\cite{HY}, it was proposed that, due to the top-mass
enhanced flavor mixing Yukawa coupling of the charm ($c$) and bottom ($b$)
with a charged scalar or pseudo-scalar ($\phi^\pm$), the $s$-channel partonic
process $c\bbar , \cbar b\to\phi^\pm$, can be an important mechanism for the
production of $\phi^\pm$ at various colliders. From the 
leading order (LO) calculation \cite{HY}, it was demonstrated that the 
Fermilab Tevatron Run-II has the potential to explore the mass range of the 
charged top-pions up to about 300--350\,GeV in the 
topcolor (TopC) models \cite{TopC,TopCrev}.
In this Chapter, we compute the complete next-to-leading order (NLO)
QCD corrections to the process $~q\qbarp\to\phi^\pm$,~
which includes the one-loop virtual corrections and the contributions
from the additional ${\cal O}(\alphas )$ processes,
\be
q\qbarp\to\phi^\pm g~~~{\rm and}~~~ qg\to q'\phi^\pm .
\label{eq:alphas}
\ee
The decay width and branching ratio (BR) of such a (pseudo-)scalar
are also included up to NLO to estimate the
event rates. The QCD resummation of multiple soft-gluon radiation
is also carried out,
which provides a better prediction of the
transverse momentum distribution of the (pseudo-)scalar particle.
We shall choose the TopC model \cite{TopC} 
as a benchmark of our analysis. 
The generalization to the generic 
type-III two-Higgs doublet model (2HDM) \cite{ansatz,2HDM3}
is straightforward since the QCD-corrections are 
universal.\footnote{We note that the finite part of the counter term
to the $q$-$\qbarp$-$\phi^{0,\pm}$ Yukawa vertex is 
renormalization scheme- and model-dependent.}
The direct extension to the production of neutral
(pseudo-)scalars via $b\bbar$ fusion is studied in  
the Minimal Supersymmetric SM (MSSM) \cite{SUSY0,SUSY} 
with large $\tanb$ and
in the TopC models with $U(1)$-tilted large bottom
Yukawa coupling \cite{TopC,TopCrev}.

\section{Charged (Pseudo-)Scalar Production via \\ Charm-Bottom Fusion }

\subsection{Fixed-Order Calculation}

We study charged (pseudo-)scalar production via the 
top-mass-enhanced flavor mixing vertex $c$-$b$-$\phi^\pm$ \cite{HY}.
The corresponding Yukawa coupling can be generally defined as
$~\CL\widehat{L}+\CR\widehat{R}~$
in which 
$~\widehat{L}=(1-\gamma_5)/2~$ and
$~\widehat{R}=(1+\gamma_5)/2$  .
The total cross sections for the $\phi^+$ production 
at hadron colliders (cf. Fig~1) can be generally expressed as
{
\begin{eqnarray}
&& \sigma \left(h_1h_2\to \phi^+X\right) =
\nonumber \\
&& \dis\sum_{\alpha ,\beta}
\dis\int^1_{\tau_0}dx_1\int^1_{\f{\tau_0}{x_1}}dx_2
\left[f_{\alpha/h_1}(x_1,Q^2)f_{\beta /h_2}(x_2,Q^2)
      +(\alpha\tolr\beta )\right] 
\widehat{\sigma}^{\alpha\beta}(\alpha\beta\to \phi^+X),
\nonumber \\ 
\label{eq:crosstot}
\end{eqnarray}
}
\hspace*{-1.5mm}where  
$\tau_0=m_\phi^2/S ,~
x_{1,2}\hspace*{-1mm}=\hspace*{-1.2mm}\sqrt{\tau_0}
\hspace*{1mm}\dis e^{\pm y}$,\, $m_\phi$ is the mass of $\phi^\pm$,
$\sqrt{S}$ is the center-of-mass energy of the $h_1h_2$ collider, 
and $f_{\alpha /h}(x,Q^2)$ is the 
parton distribution function (PDF) of a parton $\alpha$ with
the factorization scale $Q$.
The quantity $\widehat{\sigma}^{\alpha\beta}$ is the
partonic cross section and has the following LO contribution
for $c\bbar \to \phi^+$ 
(cf. Fig.~1a)~\cite{HY}:
\be
\dis\widehat{\sigma}^{\alpha\beta}_{\rm LO}
~=~\dis\delta_{\alpha c}\delta_{\beta\bbar}
\delta (1-\widehat{\tau})\widehat{\sigma}_0~,~~~~~
\widehat{\sigma}_0 \equiv
\f{\pi}{12\widehat{s}}\left(|\CL|^2+|\CR|^2\right),
\label{eq:sigmaLO}
\ee
where ~$\widehat{\tau}=m_\phi^2/\widehat{s}$~ with  $\widehat{s}$ the
center-of-mass energy of the sub-process, and 
the terms suppressed by the small mass ratio
$(m_{c,b}/m_\phi )^2$ have been ignored.
Since we are interested in the inclusive production of the scalar
$\phi$, it is natural to choose the factorization scale $Q$ to be its
mass $m_\phi$, which is of ${\cal O}(10^{2-3})$\,GeV and much larger than
the mass of charm or bottom quark. Hence, in this work, we will treat
$c$ and $b$ as massless partons inside proton or anti-proton and perform 
a NLO QCD calculation with consistent sets of 
PDFs \cite{Haber,ACOT,Collins98}.

The NLO contributions are of ${\cal O}(\alphas )$, which contain three
parts: (i)~the one-loop Yukawa vertex and quark self-energy corrections
(cf. Fig.~1b-d); (ii)~the real gluon emission in the $q\qbarp$-annihilations
(cf. Fig.~1e); (iii)~$s$- and $t$-channel gluon-quark fusions
(cf. Fig.~1f-g). The Feynman diagrams coming from permutations 
are not shown in Fig.~1. 
Unlike the usual Drell-Yan type of processes (where the
sum of the one-loop quark-wavefunction renormalization and vertex correction
gives the ultraviolet finite result), we need to include
the renormalization for the Yukawa coupling ($y_j$) which usually 
relates to the relevant quark mass ($m_{q_j}$), 
i.e., we have to add the counter term  
at the NLO (cf. Fig.1d)
besides the contribution from the usual wavefunction renormalization 
$~Z_{q_1q_2\phi}=\f{1}{2}(Z_{q_1}+Z_{q_2})$  (cf. Fig.1c).
This applies to the
Yukawa interactions of the SM and MSSM Higgs bosons 
as well as the top-pions in the TopC models. 
It is clear that, for flavor-mixing vertex 
$c$-$b$-$\phi^\pm$ in the TopC model [cf. Eq.~(\ref{eq:Ltoppi}) below],   
the counter-term of the Yukawa
coupling is equal to the top quark mass counter-term $\delta m_t/m_t$, 
which we determine from the top-quark mass renormalization
in the on-shell scheme so that $m_t$ is the pole mass of the top quark. 
In other cases such as in the general 2HDM (type-III) \cite{2HDM3} 
and the TopC models (with $b$-Higgs or $b$-pions) \cite{TopC2}, 
some of their Yukawa couplings are not related to 
quark masses or not of the above simple one-to-one correspondence, 
and thus have their independent counter terms ($\delta y_j/y_j$).
In addition to the virtual QCD-loop corrections, 
the contributions of the real gluon emission
from the initial state quarks have to be included (cf. Fig.~1e).
The soft and collinear singularities appeared in these diagrams
are regularized by the dimensional regularization prescription
at $D=4-2\epsilon$ dimensions. After summing up the contributions of
virtual gluon-loop and real gluon-radiation (cf. Fig.~1b-e),  the 
ultraviolet and soft singularities separately cancel. 
The collinear singularities
are still left over and should be absorbed into the renormalization
of the PDF \cite{AEM79}. (The $\overline{\rm MS}$ renormalization scheme
is used in our calculation.)
Finally, the gluon-quark fusion sub-processes 
(cf. Fig.~1f-g) should also be taken into account and computed at general
dimension-$D$. All these results are separately summarized in
Section \ref{sec:SummaryOfAnalytic}.

\begin{figure}[t]
\vspace{-.5cm}
\begin{center}
\begin{tabular}{c}
\epsfysize=8.5cm
\epsfbox{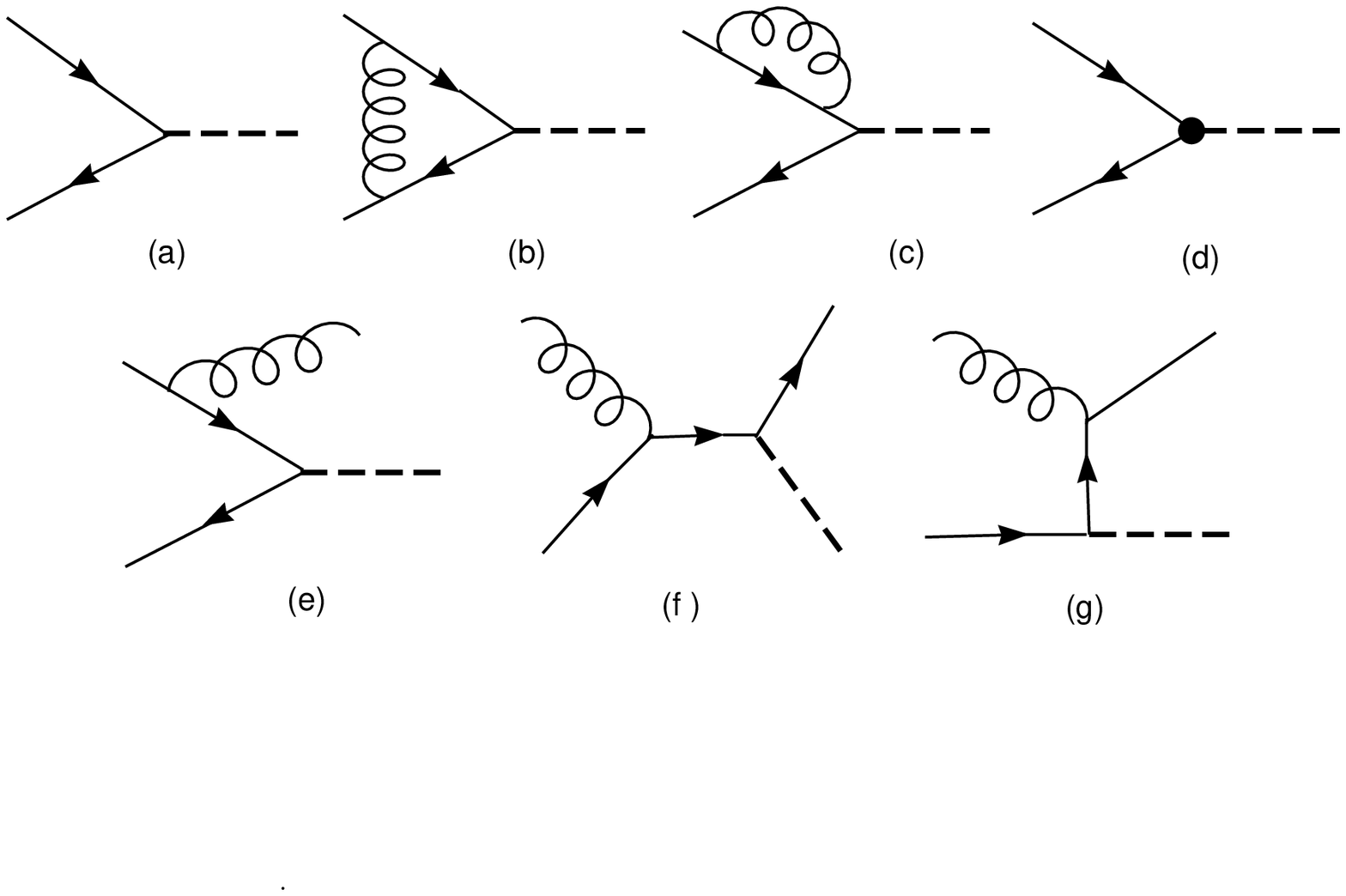}
\end{tabular}
\end{center}
\vspace{-3.5cm}
\caption{ 
Representative diagrams for charged or neutral (pseudo-)scalar
(dashed line) production from quark-antiquark and quark-gluon collisons at 
${\cal O}(\alphas^0)$ and ${\cal O}(\alphas^1)$:
(a)~leading order contribution; 
(b-d)~self-energy and vertex corrections (with counter term);
(e)~real gluon radiation in $q\qbarp$-fusion;
(f-g)~$s$- and $t$-channel gluon-quark fusions.
}
\label{fig:graphs}
\end{figure}

The hadron cross sections become regular
after renormalizing the Yukawa coupling and
the PDFs in (\ref{eq:crosstot}), which 
are functions of the renormalization scale $\mu$ and the
factorization scale $\mu_F(=\sqrt{Q^2})$. 
The partonic NLO cross section $\widehat{\sigma}^{\alpha\beta}_{\rm NLO}
\left(\alpha\beta\to\phi^+X\right)$ contains the contributions
$\Delta\widehat{\sigma}_{q\qbarp}(q\qbarp\to\phi^+,\,\phi^+g)$, 
$\Delta\widehat{\sigma}_{qg}(qg\to\phi^+q' )$, and 
$\Delta\widehat{\sigma}_{\bar{q}g}(\bar{q}g\to\phi^+\qbarp)$: 
\be
\ba{l}
  \left(\Delta\widehat{\sigma}_{q\qbarp},~
    \Delta\widehat{\sigma}_{qg},~
    \Delta\widehat{\sigma}_{\bar{q}g}\right)
=\dis\widehat{\sigma}_0\times\f{\alphas}{2\pi}
\left( \delta_{qc}\delta_{\qbarp \bbar}
\Delta\overline{\sigma}_{c\bbar},~
\delta_{qc}\Delta\overline{\sigma}_{cg},~
\delta_{\bar{q}\bbar}\Delta\overline{\sigma}_{\bbar g}
\right),\\[4mm]
\Delta\overline{\sigma}_{c\bbar}
=\dis C_F
\left[
 4\left(1+\tauhat^2\right)\left(\f{\ln (1-\tauhat )}{1-\tauhat}\right)_+
-2\f{1+\tauhat^2}{1-\tauhat}\ln\tauhat
\right. \\[4mm] \left. \dis
~~~~~~~~~~~~~~~+\left(\f{2\pi^2}{3}-2-\Omega\right)\delta (1-\tauhat )
+2(1-\tauhat )      \right] 
+\dis 2P^{(1)}_{q\leftarrow q}(\tauhat )\ln\f{m_\phi^2}{Q^2}~,  
\\[4mm]
\Delta\overline{\sigma}_{cg,\bbar g}=
\dis P^{(1)}_{q\leftarrow g}(\tauhat )
\left[\ln\f{\left(1-\tauhat\right)^2}{\tauhat} +\ln\f{m_\phi^2}{Q^2}\right]
-\f{1}{4}(1-\tauhat )\left(3-7\tauhat\right), \\[6mm]
\dis P^{(1)}_{q\leftarrow q}(\tauhat )=
\dis C_F\left(\f{1+\tauhat^2}{1-\tauhat}\right)_+, ~~~~
\dis P^{(1)}_{q\leftarrow g}(\tauhat )=
\f{1}{2}\left[\tauhat^2+(1-\tauhat )^2\right],
\ea
\label{eq:NLO}
\ee
where  $~\tauhat =m_\phi^2/\widehat{s}$ and $C_F=4/3$.
The mass counter term for the Yukawa vertex renormalization 
is determined in the on-shell scheme, i.e.,
\be
\dis\f{\delta m_t}{m_t}
=-\f{C_F\alpha_s}{4\pi}\left[3\left(\f{1}{\epsilon}-\gamma_E+\ln 4\pi \right)
+\Omega \right],
\label{eq:delta_mt}
\ee
in the TopC model.
Here, the bare mass $m_{t0}$ and the renormalized mass $m_{t}$ 
are related by $\,m_{t0}=m_t+\delta m_t$\, and  
$m_t\simeq 175$\,GeV is taken to be the top-quark pole mass. 
The finite part of the mass counter term is
$~\dis\Omega = 3\ln \left[\mu^2/m_t^2\right]+4~$ in the TopC model, 
where $\Omega \geq 0$ for $\mu \geq m_te^{-2/3}\simeq 90$~GeV.
In the following, we shall choose the QCD factorization scale $\mu_F$ 
(set as the invariant mass ${\sqrt{Q^2}}$) and the 
renormalization scale $\mu$ to be the same as the scalar mass,
i.e.,   ${\sqrt{Q^2}}=\mu=m_\phi$, 
which means that in (\ref{eq:NLO})  
the factor $\ln\left(m_\phi^2/Q^2\right)$ vanishes and 
the quantity $\Omega$ becomes 
\be
\dis\Omega = 3\ln \left[m_\phi^2/m_t^2\right]+4~.
\label{eq:omega_final}
\ee

For the case of $m_\phi\gg m_t$, 
the logarithmic term $\ln \left(m_\phi^2/m_t^2\right)$
becomes larger for $m_\phi \gg m_t$, and its contributions to all orders
in $\alphas\ln\left(m_\phi^2/m_t^2\right)$ may be resummed by introducing
the running Yukawa coupling $y_t(\mu )$, or correspondingly, the
running mass $m_t(\mu )$. 
In the above formula, $m_t$ is the pole mass ($m_t^{\rm pol}\simeq 175$\,GeV)
and is related to the one-loop running mass via the relation \cite{mtrun}
\be
\dis m_t(\mu )=m_t( m_t^{\rm pol})
\left[1-\f{3C_F}{4\pi}\alpha_s(\mu )\ln\f{\mu^2}{m_t^{\rm pol}}\right],~~~~~
m_t( m_t^{\rm pol})
=m_t^{\rm pol}\left[1+\f{C_F}{\pi}\alpha_s(m_t^{\rm pol})\right]^{-1} .
\label{eq:mtrun1loop}
\ee
Using the renormalization group equation, one can resum the leading
logarithms to all orders in $\alphas$ \cite{peskin} and obtains 
\be
\dis m_t(\mu )=m_t( m_t^{\rm pol})
\left[\f{\alpha_s (\mu )}{\alpha_s(m_t^{\rm pol})}\right]^{\f{9C_F}{33-2n_f}} ,
\label{eq:mtrunsum}
\ee
with $n_f=6$ for $\mu >m_t$ .
Thus, to include the running effect of the Yukawa coupling,  
we can replace the $(m_t^{\rm pol})^2$-factor (from the Yukawa
coupling) inside the square of the $S$-matrix element
[up to ${\cal O}(\alpha_s)$] by the running factor
\be
\dis m_t^2(\mu )\left\{1+2\f{C_F\alpha_s(\mu )}{\pi}\left[1+\f{3}{4}
\ln\left(\f{\mu}{m_t^{\rm pol}}\right)^2 \right] \right\}
\,=\, m_t^2(\mu ) \left[1+\f{C_F\alpha_s(\mu )}{2\pi}\Omega\right] ,
\label{eq:runfactor}
\ee
where the logarithmic term in the bracket $[\cdots ]$ 
is added to avoid double-counting with the resummed logarithms 
inside $m_t^2(\mu )$.  It is clear
that this  $~\dis\left[1+\left(C_F\alpha_s(\mu )/2\pi\right)\Omega\right]~$ 
factor will cancel the $\Omega$-term inside the NLO hard cross section 
$\Delta\widehat{\sigma}_{c\bbar}$ in Eq.~(\ref{eq:NLO}) at 
${\cal O}(\alphas )$, 
so that the net effect of the Yukawa vertex renormalization
(after the resummation of leading logarithms) is to replace the relevant 
tree-level on-shell quark mass (related to the Yukawa coupling) by its 
$\overline{\rm  MS}$ running mass [cf. Eq.~(\ref{eq:mtrunsum})] and  remove
the $\Omega$-term in Eq.~(\ref{eq:NLO}).
When the physical scale $\mu$ (chosen as the scalar mass 
$m_\phi$) is not much larger than $m_t$, 
the above running effect is small since the $\ln (\mu/m_t)$ factor in
the Yukawa counter-term $\delta m_t/m_t$ is small.  
However, the case 
for the neutral scalar production via the $b\bbar$ annihilation can be
different. When the loop correction to the $\phi^0$-$b$-$\bbar$ 
Yukawa coupling contains the logarithm $\ln (\mu /m_b)$, which is much
larger than $\ln (\mu /m_t)$,
these large logarithms should be resummed into the running coupling,
as we will do in Section~4.

In the TopC model, there are three pseudo-scalars, called top-pions,
which are predicted to be light, with a mass around of ${\cal O}(100 \sim 300)$
GeV. The relevant Yukawa interactions for top-pions,
including the large $t_R$-$c_R$ flavor-mixing, can be written 
as\footnote{As pointed out in Ref.~\cite{HY},
an important feature deduced from (\ref{eq:Ltoppi}) 
is that the charged top-pion 
$\pi_t^\pm$ mainly couples to the right-handed top 
($t_R$) or charm ($c_R$) but not
the left-handed top ($t_L$) or charm ($c_L$), 
in contrast to the standard $W$-$t$-$b$
coupling which involves only $t_L$. This makes the top-polarization
measurement very useful for further discriminating the signal from the 
background events.}
\cite{HY}
\be
\ba{ll}
{\cal L}_Y^{\pi_t}~=&
-\dis\f{m_t\tanb}{v}\left[
iK_{UR}^{tt}{K_{UL}^{tt}}^{\hspace*{-1.3mm}\ast}\overline{t_L}t_R\pi_t^0
+\sq2
{K_{UR}^{tt}}^{\hspace*{-1.3mm}\ast}K_{DL}^{bb}\overline{t_R}b_L\pi_t^+ 
+ \right. \\[3mm]
&~~~~~~~~~~~~~~\left.
iK_{UR}^{tc}{K_{UL}^{tt}}^{\hspace*{-1.3mm}\ast}\overline{t_L}c_R\pi_t^0
+\sq2
{K_{UR}^{tc}}^{\hspace*{-1.3mm}\ast}K_{DL}^{bb}\overline{c_R}b_L\pi_t^+ 
+{\rm h.c.}       \right],
\ea
\label{eq:Ltoppi}
\ee
where {$\tanb = \sqrt{(v/v_t)^2-1}$}$\,\sim\! {\cal O}(4\! -\! 1.3)$ 
with the top-pion decay constant $v_t\!\sim\! {\cal O}(60\!-\!150$ GeV), 
and the full vacuum expectation value (vev) 
$v\simeq 246$~GeV (determined by the Fermi constant).
The analysis from top-quark decay in the Tevatron $t\bar t$ events
sets a direct lower bound on the charged top-pion mass 
to be larger than about 150\,GeV \cite{bound,TopC}. 
The existing low energy LEP/SLD measurement of $R_b$, 
which slightly lies above the SM value 
by about 0.9$\sigma$ \cite{hollik}, also provides an indirect 
constraint on the top-pion Yukawa coupling 
${\cal C}_R^{tb}=(\sqrt{2}m_t/v)\tanb$ due to the one-loop contribution
of charged top-pions to $R_b$. 
However, given the crude approximation in estimating the top-pion
loops (with all higher loop orders ignored)
and the existence of many other sources of 
contributions associated with the strong dynamics, the indirect
$R_b$ constraint is not conclusive \cite{TopC}.
For instance, it was shown that the 3$\sigma$ $R_b$ bound from 
the one-loop top-pion correction can be fully
removed if the top-pion decay
constant $v_t$ is increased by a factor of 2
(which can be the typical uncertainty of the Pagels-Stokar 
estimate)\,\cite{TopC,Rb}; also, the non-perturbative contributions of 
coloron-exchanges can shift the $R_b$ above its SM 
value\,\cite{TopC} and tend to cancel the negative top-pion corrections.
Due to these reasons, it is clear that the inconclusive $R_b$-bound in TopC
models should not be taken seriously. 
Nevertheless, to be on the safe side, we will uniformally
impose the roughly estimated $R_b$-constraint in our current analysis of
the TopC model, by including {\it only}
the (negative) one-loop top-pion contribution as in 
Ref.\,\cite{Rb}.\footnote{
However, it is important to keep in mind 
that such a rough $R_b$-bound is likely to
over-constrain the top-pion Yukawa coupling since only the negative 
one-loop top-pion correction (but nothing else) is included in this 
estimate. A weaker $R_b$-bound will less reduce the top-pion Yukawa
coupling and thus allow larger production rates of charged top-pions at
colliders which can be obtained from our current analysis by simple
re-scaling.}~
\begin{figure}[t]
\vspace{-.2cm}
\begin{center}
\begin{tabular}{c}
\epsfysize=10cm
\epsfbox{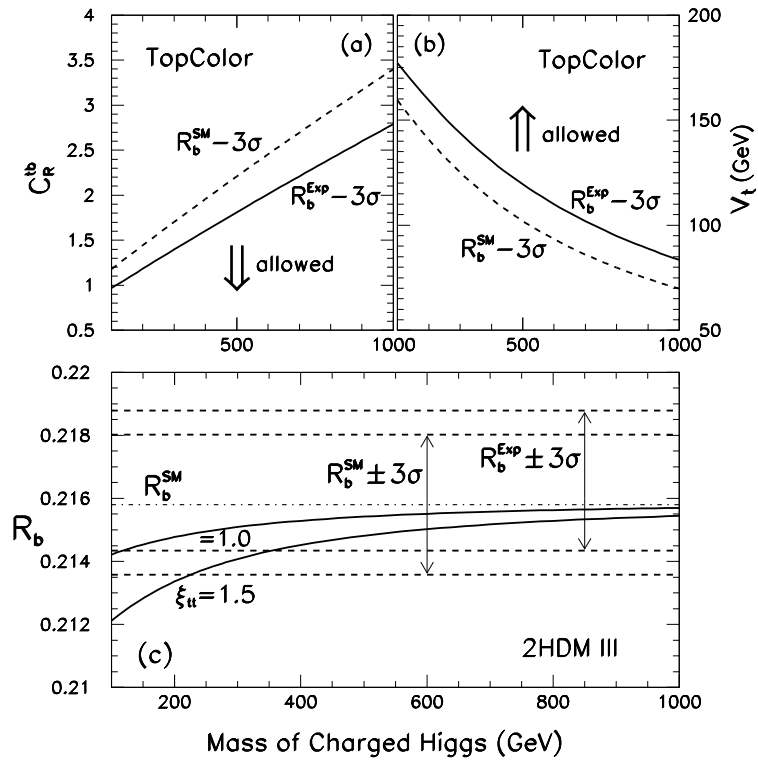}
\end{tabular}
\end{center}
\vspace{-.5cm}
\caption{ 
Estimated current $3\sigma$-bounds in the TopC model and 2HDM-III:
(a) the $3\sigma$ upper bound on the top-pion Yukawa coupling 
${\cal C}_R^{tb}$; (b) the $3\sigma$ lower bound on the top-pion
decay constant; (Here, in (a) and (b), the solid curves are derived
from the combined LEP/SLD data of 
$R_b^{\rm Exp}=0.21656\pm 0.00074$ while dashed
curves are from the same $3\sigma$ combined experimental error but with
the central $R_b$-value equal to $R_b^{\rm SM}=0.2158$.)
(c) the $R_b$-predictions of 2HDM-III with coupling $\xi_{tt}=1.0$ and
$1.5$ (solid curves) and the $3\sigma$ $R_b$-bounds (dashed lines).
}
\label{fig:Rb}
\end{figure}
As shown in Fig.\,\ref{fig:Rb}a, the current 3$\sigma$ $R_b$-bound
requires a smaller top-pion Yukawa coupling,
${\cal C}_R^{tb}\!\sim\! 1.3-2$ (or, $\tanb \!\sim\! 1.3-2$), 
for the low mass region of $m_{\pi_t^\pm}\!\sim\! 200-500$\,GeV.
Since the top-pion decay constant $v_t$ is related to $\tanb$,
this also results in a constraint of $v_t$ to be around 
$150\!\sim\!100$\,GeV when $m_{\pi_t^\pm}\!\sim\! 200-500$\,GeV
(cf. Fig.\,\ref{fig:Rb}b). 
The usual Pagels-Stokar estimate of $v_t$ (by keeping only the
leading logarithm but not constant terms)
gives $v_t\!\sim\! 64\!-\!97$\,GeV for the topC breaking scale 
$\Lambda\!\sim\! 1\!-\!10$\,TeV, where a typical factor of $2\!\sim\!3$
uncertainty in the calculation of $v_t^2$
is expected \,\cite{PS,TopC}.
This estimate is slightly lower than the $R_b$-constrained $v_t$-values in 
Fig.\,\ref{fig:Rb}b, but is still in reasonable consistency
(given the typical factor of $\sqrt{2}\!\sim\!\sqrt{3}$
error in the leading logarithmic Pagels-Stokar estimate of $v_t$).

In (\ref{eq:Ltoppi}), $K_{UL,R}$ and $K_{DL,R}$ are defined from 
diagonalizing the up- and down-type quark mass matrices $M_U$ and $M_D$:
{~$K_{UL}^\dag M_U K_{UR} = M_U^{\rm dia},~
K_{DL}^\dag M_D K_{DR} = M_D^{\rm dia}~,$}
with {\small $M_U^{\rm dia}$}$={\rm diag}(m_u,m_c,m_t)$ and
     {\small $M_D^{\rm dia}$}$={\rm diag}(m_d,m_s,m_b)$.
For the class-I TopC models \cite{TopC2}, 
we have constructed \cite{HY}  a realistic and attractive
pattern of $K_{UL}$ and $K_{DL}$ so that the well-constrained 
Cabibbo-Kobayashi-Maskawa (CKM) matrix
$V~(=K_{UL}^\dag K_{DL})$ can be reproduced 
in the Wolfenstein-parametrization
\cite{Wolfen} and all potentially large contributions 
to the low energy data (such as the
$K$-$\bar K$, $D$-$\bar D$ and $B$-$\bar B$ mixings 
and the $b\to s\gamma$ rate) can be avoided~\cite{HY}. 
We then found that the right-handed rotation matrix $K_{UR}$ is 
constrained such that its 33 and 32 elements take the values as~\cite{HY}
\be          
K_{UR}^{tt}\simeq 0.99\hspace*{-0.7mm}-\hspace*{-0.7mm}0.94,~~~
K_{UR}^{tc}\leq \sqrt{1-{K_{UR}^{tt}}^2} 
\simeq 0.11\hspace*{-0.7mm}-\hspace*{-0.7mm}0.33,
\label{eq:KURtc}
\ee
which show that the $t_R$-$c_R$ flavor mixing 
can be naturally around $10-30\%$.

For the current numerical analysis 
we make the benchmark choice \cite{HY} based upon the above TopC model
and consider
\be
{\cal C}^{tb}_R= {\cal C}_R^{tb}(R_b{\rm \,constrained}), ~~~~
{\cal C}^{cb}_R={\cal C}^{tb}_R{K_{UR}^{tc}}
               \simeq {\cal C}^{tb}_R\times 0.2,~~~~
{\cal C}_L^{tb}={\cal C}_L^{cb}=0.
\label{eq:Coupling0}
\ee
It is trivial to scale the numerical results presented in this Chapter to
any other values of ${\cal C}_{L,R}$ when needed. Unless specified otherwise, 
we use CTEQ4M PDF \cite{CTEQ4} to calculate the rates. Note that
CTEQ4M PDFs are consistent with the scheme used in the current study
which treats the initial state quarks as massless partons in computing
the Wilson coefficient functions. The only effect of the heavy quark mass is
to determine at which scale $Q$ this heavy quark parton becomes 
active.\footnote{This is the Collins-Wilczek-Zee (CWZ) scheme
\cite{CWZ}.} In our case, the scale $Q=m_\phi\gg m_c,\,m_b$.

In Fig.~\ref{fig:FigSigma_FixMt.eps}, we present the total cross sections for
the charged top-pion production as functions of its mass,
at the Tevatron 
(a $p\bar p$ collider at 1.8 and 2\,TeV) and 
the LHC (a $pp$ collider at 14\,TeV).
We compare the improvements by including the complete NLO results
[cf. (\ref{eq:NLO})] and by including the resummed running Yukawa coupling
or running mass [cf. (\ref{eq:mtrunsum})].
For this purpose, we first plot the LO total cross sections with the 
tree-level Yukawa coupling [dash-dotted curves, 
cf. (\ref{eq:sigmaLO}) and (\ref{eq:Coupling0})] and 
with the resummed running Yukawa coupling 
or running mass [dotted curves, cf. (\ref{eq:sigmaLO}) and (\ref{eq:mtrunsum})]; 
then we plot the NLO cross sections with the 
one-loop Yukawa coupling [dashed curves, cf.
(\ref{eq:NLO})] and with the resummed running Yukawa coupling or running mass 
[solid curves, cf. 
(\ref{eq:NLO}), (\ref{eq:mtrunsum}) and (\ref{eq:runfactor})] .
We see that at the LHC there is a visible difference between the
pure LO results with tree-level Yukawa coupling (dash-dotted curves) and
other NLO and/or running-coupling improved results. But at the Tevatron,
the LO results with running Yukawa coupling (dotted curves) are visibly
smaller than the results in all other cases for $m_\phi >300$\,GeV. 
This shows that without the complete NLO calculation, including only the 
running Yukawa coupling in a LO result may not always warrant a 
better improvement. 
Finally, the comparison in Fig.~\ref{fig:FigSigma_FixMt.eps}
shows that the resummed running Yukawa coupling or top-mass
[cf. Eq.~(\ref{eq:mtrunsum})]
does not generate any significant improvement from the one-loop running.
This is because the top-mass is large and
$\alphas\ln\left(m_\phi^2/m_t^2\right)$ is small for $m_\phi$ up to 1\,TeV. 
Thus, the
improvement of the resummation in (\ref{eq:mtrunsum}) has to come from
higher order effects of $\alphas\ln\left(m_\phi^2/m_t^2\right)$. 
However, as to be shown in Section~\ref{sec:Generalization}
the situation for summing over powers of 
$\alphas\ln\left(m_\phi^2/m_b^2\right)$ is different due to 
$m_b\ll m_t,\, m_\phi$\,. 

Fig.~\ref{fig:Sigma1} is to examine the individual NLO contributions 
to the charged top-pion production via the $\qqbar '$ and $qg$
sub-processes,
in comparison with the full NLO contributions.\footnote{
Unless specified, $qg$ includes both $qg$ and ${\bar q}g$ contributions.}
The LO contributions are also shown as a reference.\footnote{
With the exception of Figs.~\ref{fig:FigSigma_FixMt.eps}, ~\ref{fig:GammaBr},
and ~\ref{fig:Sigma_MSSMbb}, we only show our numerical results with the 
resummed running Yukawa coupling or running mass.} 
[Here $q$ denotes the heavy charm or bottom quark.] 
In this figure, there are
three sets of curves for the charged top-pion production
cross sections: the highest set 
is for the LHC ($\sqrt{S}=14$\,TeV), the middle set is 
for the upgraded Tevatron ($\sqrt{S}=2$\,TeV),
and the lowest set is for the Tevatron Run~I ($\sqrt{S}=1.8$\,TeV).
The LO cross sections are plotted as dotted lines while the NLO cross sections
as solid ones. The dashed lines show the contributions from the
$q\bar{q}'$-fusion sub-processes, 
and the dash-dotted lines describe the contributions
from the $qg$-fusion sub-processes. 
The $qg$-fusion cross sections are negative
and are plotted by multiplying a factor of $-1$, for convenience. 
For a quantitative comparison 
of the individual NLO contributions versus the full NLO
results, we further plot, in  Fig.~\ref{fig:InitKFac}, 
the ratios (called $K$-factors) of the different NLO contributions 
to the LO cross section by using the same set of CTEQ4M PDFs.
The solid lines of Fig. \ref{fig:InitKFac} show
that the overall NLO corrections to the $pp,p\bar{p} \to \phi^\pm X$
processes are positive for $m_\phi$ above $\sim$150\,(200)~GeV
and lie below $\sim$15\,(10)\% for the
Tevatron (LHC) in the relevant mass region. This is in contrast with
the NLO corrections to the $W^\pm$ boson production at hadron
colliders, which are always positive and as large as about
25\% at the Tevatron \cite{Balazs-YuanPRD}.   
The reason
of this difference originates from the differences in the $\Delta \sigma_ 
{q\bar{q}'}$ and $\Delta \sigma _{qg,g\bar{q}}$ for $\phi^\pm$ and $W^\pm$
production. While in the case of $W^\pm$ production the positive 
$\Delta \sigma_{q\bar{q}'}$ piece dominates, in the case of  $\phi^\pm$
production the size of negative $\Delta \sigma _{qg,g\bar{q}}$ piece becomes
comparable with that of the positive $\Delta \sigma_{q\bar{q}'}$\,
such that a non-trivial cancellation occurs.

While it is reasonable to take the renormalization and the factorization
scales to be $m_\phi$ for predicting the inclusive production rate of
$\phi^+$, it is desirable to estimate the uncertainty in the rates due to
different choices of PDFs. For that purpose, we examine a few typical
sets of PDFs from CTEQ4, which predict different shapes of charm, bottom
and gluon distributions. As shown in Table~\ref{tb:PDFS} and 
Fig.~\ref{fig:PDF}, the uncertainties due to the choice of PDF set are 
generally within $\pm 20\%$ for the relevant scalar mass ranges 
at both the Tevatron and the LHC.
\begin{table}[p]
\vspace{-5mm}
\begin{center}
\begin{tabular}{c  r r r r  r r r r}
\hline\hline
&&&&&&&&\\[-0.2cm]
Collider & 
\multicolumn{4}{c}{Upgraded Tevatron (2\,TeV)} & 
\multicolumn{4}{c}{LHC (14\,TeV)}  \\
[0.15cm] \cline{1-9}
&&&&&&&&\\[-0.2cm]
Process$\backslash$PDF 
                    &   4A1 &    4M &   4A5 &   4HJ
                    &   4A1 &    4M &   4A5 &   4HJ \\
[0.15cm]\hline\hline 
&&&&&&&&\\[-0.2cm]
                    &   367 &   382 &   376 &   387
                    &  5380 &  5800 &  6060 & 5890 \\
&&&&&&&&\\[-0.3cm]
LO                  &  42.6 &  43.7 &  41.5 &  46.6
                    &   863 &   901 &   896 &   906 \\
&&&&&&&&\\[-0.3cm]
                    &  6.88 &  7.05 &  6.56 &  8.38
                    &   235 &   240 &   323 &   241 \\
[0.2cm]\hline 
&&&&&&&&\\[-0.2cm] 
                    &   370 &   402 &   412 &   407
                    &  5430 &  6080 &  6510 &  6170 \\
&&&&&&&&\\[-0.3cm]
NLO                 &  45.6 &  48.6 &  47.9 &  51.6
                    &   912 &   976 &   997 &   981 \\
&&&&&&&&\\[-0.3cm]
                    &  7.70 &  8.21 &  7.89 &  9.56   
                    &   255 &   266 &   264 &   268 \\
[0.2cm]\hline 
&&&&&&&&\\[-0.2cm]
                    &   551 &   584 &   585 &   590
                    &  7530 &  8290 &  8740 &  8400 \\
&&&&&&&&\\[-0.3cm]
$q\bar{q}\to\phi^+X$&  64.5 &  67.4 &  65.5 &  71.7
                    &  1210 &  1280 &  1290 &  1290 \\
&&&&&&&&\\[-0.3cm]
                    &  10.6 &  11.1 &  10.5 &  13.0
                    &   331 &   341 &   335 &   343 \\
[0.2cm]\hline 
&&&&&&&&\\[-0.2cm]
                    &$-$ 180&$-$ 181&$-$ 174&$-$ 183
                    &$-$2100&$-$2200&$-$2240&$-$2230\\
&&&&&&&&\\[-0.3cm]
$qg\to\phi^+X$      &$-$19.2&$-$18.9&$-$17.5&$-$19.9
                    & $-$299& $-$302& $-$293& $-$303\\
&&&&&&&&\\[-0.3cm]
                    &$-$2.94&$-$2.86&$-$2.59&$-$3.34
                    &$-$76.0&$-$74.7&$-$70.6&$-$75.0\\
[0.2cm]\hline\hline 
\end{tabular}
\end{center}
\caption{
Cross sections in fb for charged top-pion production in the TopC model at
the upgraded Tevatron and the LHC are shown, by using four different CTEQ4 
PDFs. They are separately given for the LO and NLO processes, 
and for the $q\bar{q}\to\phi^+X$ and $qg\to\phi^+X$ sub-processes. 
At the upgraded Tevatron the top number is for $m_{\phi} = 200$ GeV, 
the middle is for $m_{\phi} = 300$ GeV, and
the lowest is for $m_{\phi} = 400$ GeV.
At the LHC the top number is for $m_{\phi} = 400$ GeV, 
the middle is for $m_{\phi} = 700$ GeV, and
the lowest is for $m_{\phi} = 1$ TeV.
}
\label{tb:PDFS}
\end{table}

\section{Soft-Gluon Resummation}

The ${\cal O}(\alpha _s)$ corrections to the (pseudo-)scalar production
involve the contributions from the emission of real and virtual
gluons, as shown in Figs.~1(e), (b), and (c). As the result of the real gluon
radiation, the (pseudo-)scalar particle will acquire a non-vanishing
transverse momentum ($Q_T$). When the emitted gluons are soft, they generate
large logarithmic contributions of the form: 
$\alpha _s$$\ln^m\left(Q^2/Q_T^2\right)$$/Q_T^2$ (in the lowest order), 
where $Q$ is the invariant mass of the
(pseudo-)scalar, and $m=0,1$. These large logarithms spoil the convergence of
the perturbative series, and falsify the ${\cal O}(\alpha _s) 
$ prediction of the transverse momentum when $Q_T\ll Q$.

To predict the transverse momentum distribution of the produced 
(pseudo-)scalar, we utilize the Collins--Soper--Sterman (CSS) 
formalism \cite{CS,CSS,Collins-Soper-Sterman}, 
resumming the logarithms of the type 
$\alpha _s^n$ $\ln ^m\left( Q^2/Q_T^2\right)/Q_T^2$, 
to all orders $n$ in $\alpha _s$ ($m=0,...,2n-1$). 
The resummation calculation is
performed along the same line as for vector boson production in
Chapter~\ref{ch:VBP}. Here we only give the differences from 
that given in Chapter~\ref{ch:VBP}. 
But for convenience, 
we also list the $A^{(1)}$, $A^{(2)}$, and $B^{(1)}$ coefficients of the 
Sudakov exponent, which have been used in the current analysis:
\be
\ba{l}
A^{(1)}\left(C_1\right) \,=\, C_F\,,~~~~~\,
\dis B^{(1)}\left(C_1 =b_0,C_2=1\right)\,=\,-\frac{3}{2}C_F\,,  \\[3mm]
\dis A^{(2)}\left(C_1 =b_0\right)\,=\,
C_F\left[ \left( {\frac{67}{36}}-{\frac{\pi ^2}{12}}
\right) N_C-\f{5}{18}n_f\right],
\ea
\label{eq:A12B1}
\ee
where $C_F=4/3$ is the Casimir of the fundamental representation of SU(3), 
$N_C=3$ is the number of SU(3) colors, and $n_f$ is the
number of light quark flavors with masses less than $Q$. 
In the above we used the
canonical values of the renormalization constants $C_1=b_0$, and $\ C_2=1$. 

To recover the $O\left( \alpha _s\right) $
total cross section, we also include the Wilson coefficients $C_{i\alpha
}^{(1)}$, among which $C_{ij}^{(1)}$ differs from the vector boson
production (here $i$ denotes quark or antiquark flavors, and 
$\alpha = q_i$ or gluon $g$). Explicitly,
\begin{eqnarray}
C_{jk}^{(0)}(z,b,\mu ,{{C_1}/{C_2}}) & \hspace*{-8mm}
=~\dis\delta _{jk}\delta ({1-z}), 
~~~~
C_{jg}^{(0)}(z,b,\mu ,{C_1}/{C_2}) ~=~0, 
\nonumber \\[2mm]
C_{jk}^{(1)}(z,b,\mu ,{{C_1}/{C_2}}) &
\hspace*{-8mm}=~ \dis\delta _{jk}C_F\left\{ \frac
12(1-z)-\frac 1{C_F}\ln \left( \frac{\mu b}{b_0}\right) P_{j\leftarrow
k}^{(1)}(z)\right.  
\nonumber \\[3mm]
& ~~~~~\dis\left. +\,\delta (1-z)\left[ -\ln ^2\left( {\frac{C_1}{{b_0C_2}}}
e^{-3/4}\right) +{\frac{\mathcal{V}}4}+{\frac 9{16}}\right] \right\}, 
\nonumber \\[3mm]
C_{jg}^{(1)}(z,b,\mu ,{{C_1}/{C_2}}) &
\hspace*{-27mm}=~\dis\f{1}{2}z(1-z)-\ln \left( 
\frac{\mu b}{b_0}\right) P_{j\leftarrow G}^{(1)}(z),
\label{eq:C01}
\end{eqnarray}
where $P_{j\leftarrow g}^{(1)}$ is the ${\cal O}(\alphas )$ 
gluon splitting kernels \cite{DGLAP} given in Section \ref{Sec:Factorization}.
In the above expressions, 
${\mathcal V}={\mathcal V}_{DY}=-8+\pi^2$ for the vector boson
production \cite{Balazs-YuanPRD}, and 
${\mathcal V}={\mathcal V}_\Phi =\pi^2$ for the (pseudo-)scalar
production, when using the running mass given in Eq.~(\ref{eq:mtrunsum}) 
for the Yukawa coupling. Using the canonical values of the
renormalization constants, $\ln (\mu b/b_0)$ vanishes, because 
$\mu=C_1/b=b_0/b$.

The only remaining difference between the resummed formulae of the vector
boson and (pseudo-)scalar production is in the regular ($Y$) terms, which
comes from the difference of the  $O\left( \alpha_s\right) $ 
real emission amplitude squares (cf., the definitions of
${\cal T}_{q\overline{q}}^{-1}$ and ${\cal T}_{qg}^{-1}$ in
Section \ref{TheCrossS} 
and Eqs.~(\ref{eq:M2qq}) and (\ref{eq:M2gq}) of this Chapter). 
The non-perturbative sector of the CSS resummation 
(the non-perturbative function and the related parameters) 
is assumed to be the same as that in Ref.~\cite{Balazs-YuanPRD}.

As described in Ref.~\cite{Balazs-YuanPRD}, the resummed total rate
is the same as the ${\cal O}(\alpha _s)$ rate, when we include 
$C^{(1)}_{i\alpha}$ and $Y^{(1)}$, 
and switch from the resummed distribution to the fixed order one
at $Q_T=Q$. When calculating the total rate, we have applied this matching
prescription. In the case of the (pseudo-)scalar production, 
the matching takes place at high $Q_T \sim Q$ values, 
and the above matching prescription is irrelevant 
when calculating the total rate
because the cross sections there are negligible. 
Thus, as expected, the resummed total rate differs from the 
${\cal O}(\alpha _s)$ rate only by a few percent. 
Since the difference of the resummed and fixed order rate 
indicates the size of the higher order corrections, 
we conclude that for inclusive (pseudo-)scalar production 
the ${\cal O}(\alpha _s^2)$ corrections are likely much smaller than
the uncertainty  from the parton distribution functions 
(cf. Fig.\ref{fig:InitKFac}).  

In Fig.~\ref{fig:QTDistn}, we present the numerical results for the
transverse momentum distributions of the charged top-pions (in TopC model)
and the charged Higgs bosons (in 2HDM) produced at the
upgraded Tevatron and the LHC. The solid curves show the resummation
prediction for the typical values of $m_\phi$. The dashed curves, from 
the ${\cal O}(\alpha _s)$ prediction, are irregular as $Q_T \to 0$. The large
difference of the transverse momentum distributions
between the results from the resummation and fixed-order analyses 
throughout a wide range of $Q_T$ shows the importance of using
the resummation prediction when extracting the top-pion and Higgs boson 
signals. 
We also note that the average value of $Q_T$ varies slowly as $m_\phi$
increases and it ranges from 
35 to 51\,GeV for $m_\phi$ between 250 and 550 GeV at the 14 TeV LHC, and from 
23 to 45\,GeV for $m_\phi$ between 200 and 300 GeV at the  2 TeV Tevatron.

\begin{figure}[t]
\vspace{-1cm}
\begin{center}
\begin{tabular}{c}
\epsfysize=10cm
\epsfbox{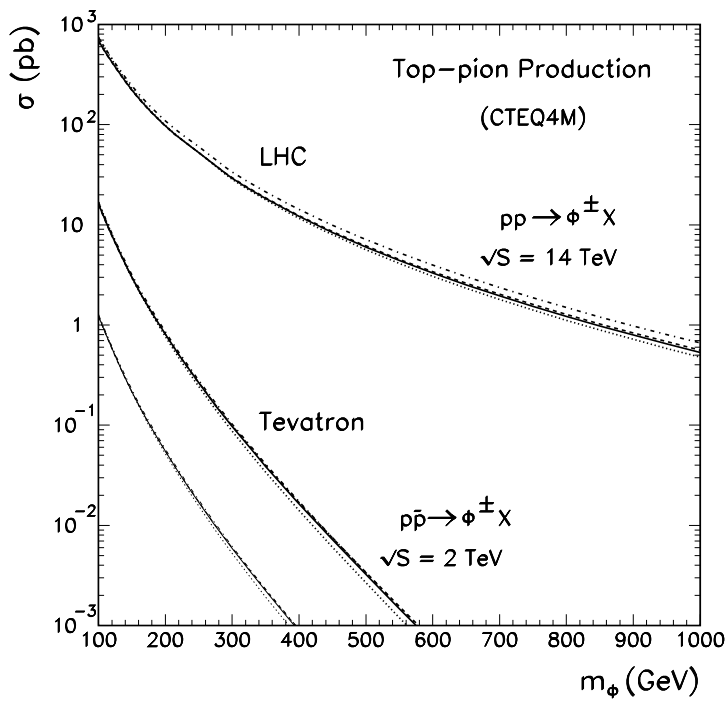}
\end{tabular}
\end{center}
\vspace{-0.5cm}
\caption{ 
Top-pion production cross sections at the present Tevatron, upgraded
Tevatron, and the LHC. For each collider we show the NLO cross section with
the resummed running Yukawa coupling (solid), and with one-loop
Yukawa coupling (dashed), as well as the
LO cross section with resummed running Yukawa coupling (dotted) and 
with tree-level (dash-dotted) Yukawa coupling.
The cross sections at $\sqrt{S} = 1.8$~TeV 
(thin set of lowest curves)
are multiplied by 0.1 to avoid overlap with the $\sqrt{S} = 2$ TeV curves. 
}
\label{fig:FigSigma_FixMt.eps}
\end{figure}
\begin{figure}[t]
\vspace{-1cm}
\begin{center}
\begin{tabular}{c}
\epsfysize=10cm
\epsfbox{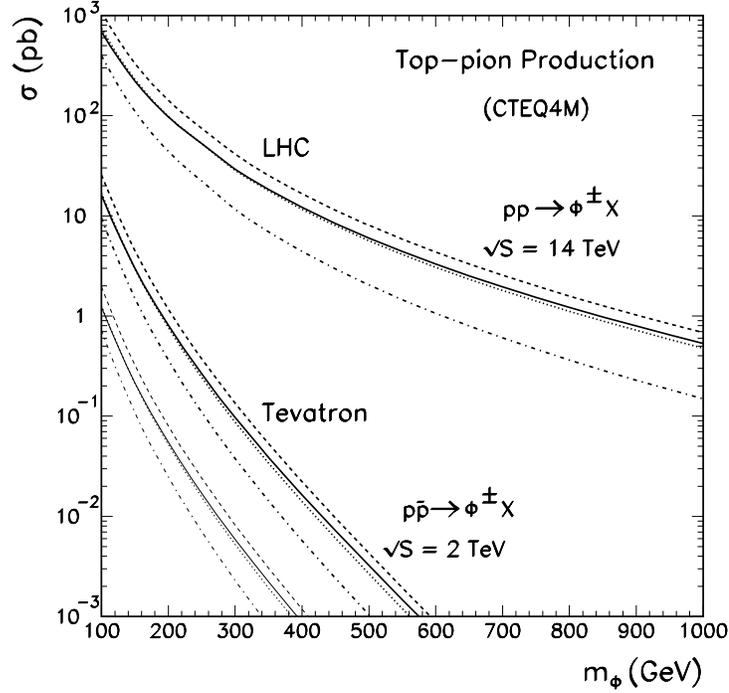}
\end{tabular}
\end{center}
\vspace{-0.5cm}
\caption{ 
Cross sections for the
charged top-pion production in the TopC model at the present Tevatron,
upgraded Tevatron and the LHC. 
The NLO (solid), the $q\bar{q}'$ (dashed) and $qg$
(dash-dotted) sub-contributions, and the LO (dotted) contributions are shown.
Since the $qg$ cross sections are negative, they are multiplied by
$-1$ in the plot. The cross sections at $\sqrt{S} = 1.8$~TeV 
(thin set of lowest curves)
are multiplied by 0.1 to avoid overlap with the $\sqrt{S} = 2$ TeV curves.   
}
\label{fig:Sigma1}
\end{figure}
\begin{figure}[t]
\vspace{-1cm}
\begin{center}
\begin{tabular}{c}
\epsfysize=10cm
\epsfbox{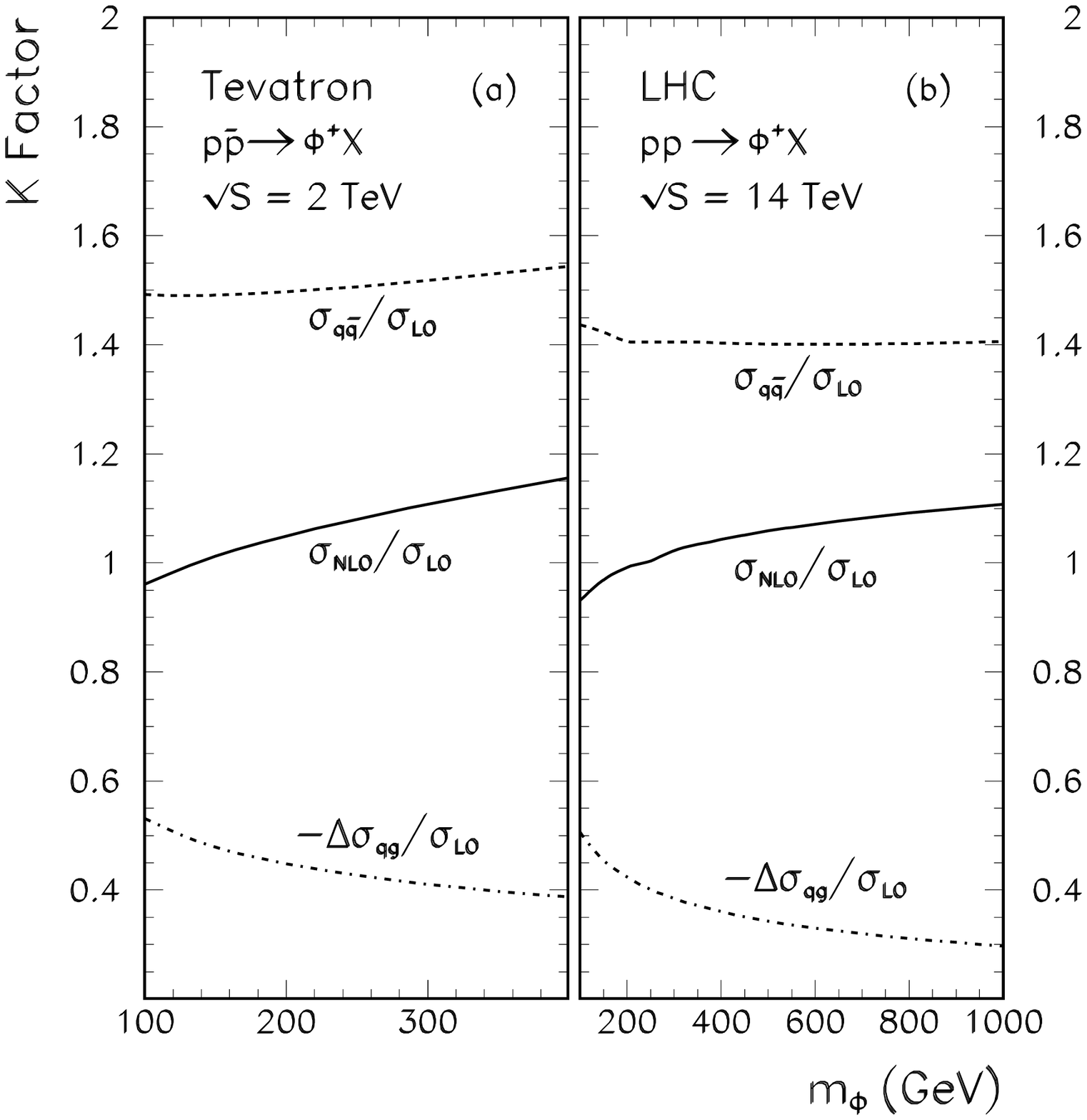}
\end{tabular}
\end{center}
\vspace{-0.5cm}
\caption{ 
The $K$-factors for the $\phi^+$ production in the TopC model are shown for
the NLO ($K=\sigma_{\rm NLO}/\sigma_{\rm LO}$, solid lines),
$q\bar{q}'$ ($K=\sigma_{\qqbar '}/\sigma_{\rm LO}=
(\sigma_{\rm LO}+\Delta\sigma_{\qqbar '})/\sigma_{\rm LO}$, 
dashed lines), and $qg$ ($K=-\Delta\sigma_{qg}/\sigma_{\rm LO}$, 
dash-dotted lines) contributions, 
at the upgraded Tevatron (a) and the LHC (b).
}
\label{fig:InitKFac}
\end{figure}
\begin{figure}[t]
\vspace{-1cm}
\begin{center}
\begin{tabular}{c}
\epsfysize=10cm
\epsfbox{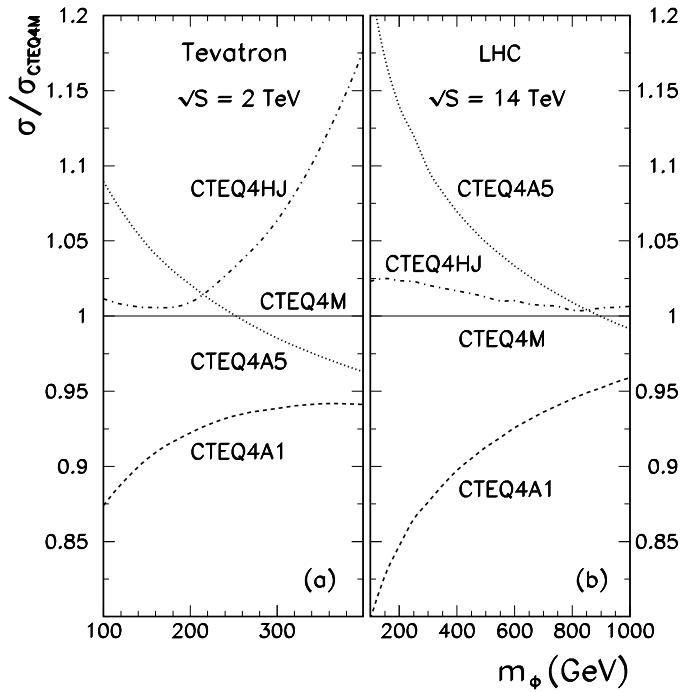}
\end{tabular}
\end{center}
\vspace{-0.5cm}
\caption{
The ratios of NLO cross sections computed by four different sets of 
CTEQ4 PDFs relative to that by the CTEQ4M
for charged top-pion production 
at the upgraded Tevatron (a) and the LHC (b). 
}
\label{fig:PDF}
\end{figure}
\begin{figure}[t]
\vspace{-1cm}
\begin{center}
\begin{tabular}{c}
\epsfysize=10cm
\epsfbox{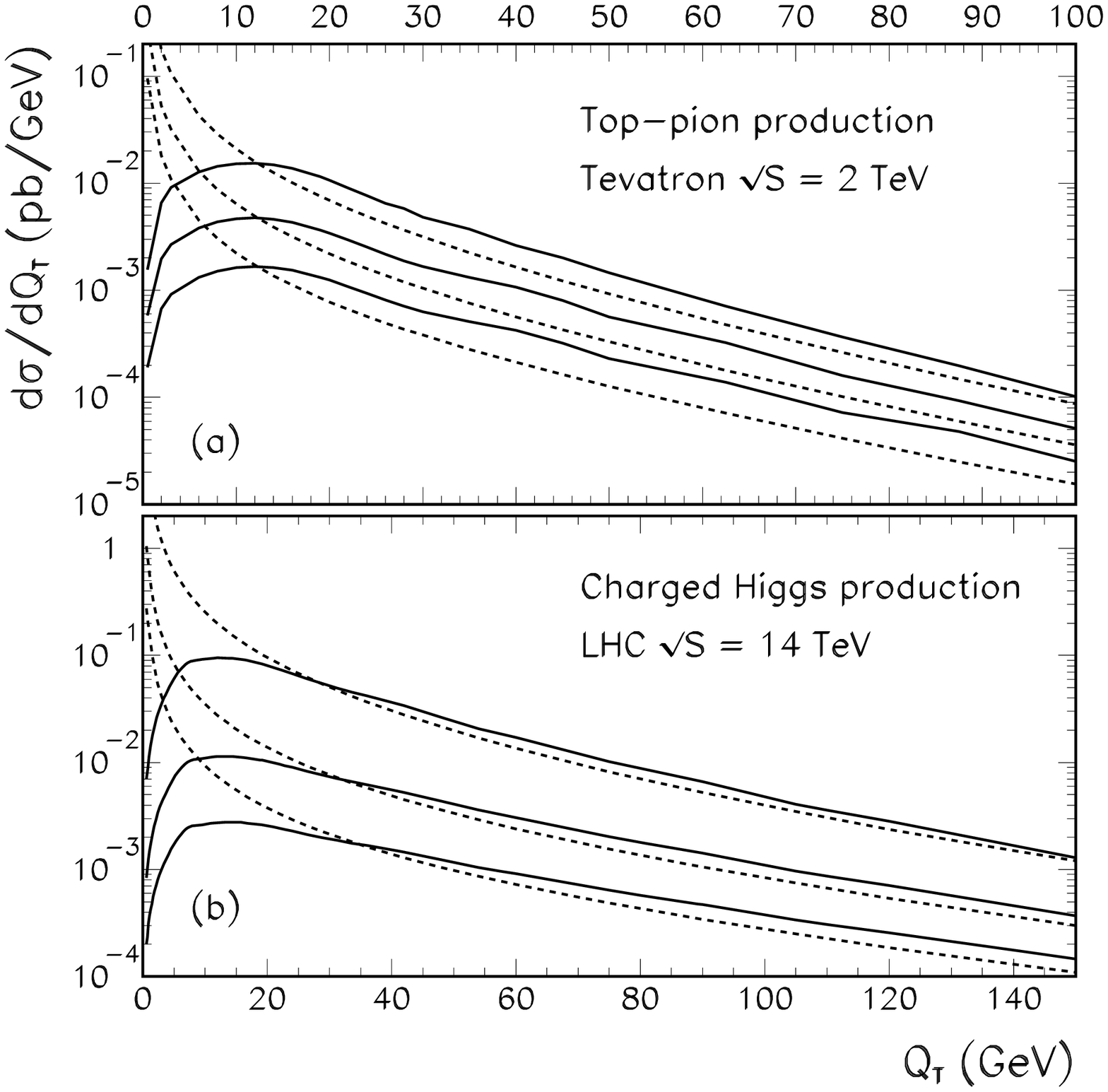}
\end{tabular}
\end{center}
\vspace{-0.5cm}
\caption{ 
Transverse momentum distributions of charged top-pions produced in hadronic
collisions. The resummed (solid) and ${\cal O}(\alpha _s)$ (dashed)
curves are calculated for $m_\phi =$ 200, 250, and 300 GeV at the
upgraded Tevatron (a), 
and for $m_\phi =$ 250, 400, and 550 GeV at the LHC (b). 
}
\label{fig:QTDistn}
\end{figure}

\section{Hadronic Decays of Charged (Pseudo-)Scalars to ${\cal O}(\alpha_s)$}

In the TopC models, the current Tevatron data from the top quark decay into
charged top-pion ($\pi_t^\pm$) and $b$-quark already requires the
mass of $\pi_t^\pm$ to be above $\sim $150~GeV \cite{TopC,bound}. 
In the current analysis, we shall consider $m_{\pi_t} > m_t+m_b$, 
so that its dominant decay channels are $\pi_t^\pm\to tb,cb$\,.

The decay width of $\pi_t^\pm$ ($=\phi^\pm$), 
including the ${\cal O}(\alphas )$ QCD corrections, 
is given by \cite{Htb-NLO,CSL}:

\be
\ba{l}
\dis\Gamma_{\rm NLO}(Q)~=~\dis \Gamma_{\rm LO}(Q)
\left[1+\f{\alphas C_F}{2\pi}{\cal R}\right],~~~~~\\[3mm]
\Gamma_{\rm LO}(Q) ~= 
\dis\f{3}{16\pi}Q\left( |{\cal C}_L|^2+|{\cal C}_R|^2\right)
(1-r)^2,\\[2.8mm]
{\cal R} ~=~\dis\f{9}{2}(1-r)^2+(1-r)\left(3-7r+2r^2\right)\ln\f{r}{1-r}
 +\left[\dis 3\ln\f{Q^2}{m_t^2}+4-\Omega\right]\\[2.3mm]
 ~~~~~~~~\dis -2(1-r)^2\left[\f{\ln (1-r)}{1-r}-2{\rm Li}_2\left(\f{r}{1-r}\right)
-\ln (1-r)\ln\f{r}{1-r} \right],  
\ea
\label{eq:widNLO}
\ee
in which $Q=\sqrt{Q^2}$ is the invariant mass of $\phi^\pm$.
The small bottom and charm masses are ignored so that
~$r\equiv \left( m_t/m_{\phi}\right)^2$ for $tb$ final state and $~r=0~$
for $cb$ final state.  Thus, for $~\phi^\pm \to cb$,~ the quantity
${\cal R}$ reduces to $\,{\cal R} = {17}/{2}-\Omega$\,.
In Fig.~\ref{fig:GammaBr}, we present the results for
total decay widths of $\phi^+$ and branching ratios of $\phi^+ \to t\bar{b}$
in the TopC model and the 2HDM. For the 2HDM, we also show the branching 
ratios of the $W^+h^0$ channel, which is complementary to the $t\bbar$
channel.  The NLO (solid) and LO (dashed) curves differ only by a small
amount. In the same figure,
the $K$-factor, defined as the ratio of the NLO to the LO partial decay 
widths, is plotted for the  $\phi^+\to t\bar{b}$ (solid) 
and $\to c\bar{b}$ (dashed) channels. Here,
the sample results for the 2HDM  are derived for the parameter 
choice: $\alpha =0$ and $(M_h,\,M_A)=(100,\,1200)$\,GeV.

With the decay width given above, we can study the invariant
mass distribution of $\,tb\,$ for the $s$-channel $\phi^+$-production:
\be
\dis\f{d\sigma}{dQ^2}
\left[h_1h_2\to\hspace*{-0.7mm}(\phi^+X)\hspace*{-0.7mm}\to t\bbar X\right]
=\dis\sigma\left[h_1h_2\hspace*{-1.5mm}\to\hspace*{-0.7mm}\phi^+(Q)X\right] 
\f{\left(Q^2\Gamma_\phi/m_\phi\right){\rm Br}\left[\phi^+\hspace*{-1.5mm}\to t\bbar\right]}
{\pi\left[\left(Q^2\hspace*{-0.7mm}-m_\phi^2\right)^2+\left(Q^2\Gamma_\phi
/m_\phi\right)^2\right]},
\label{eq:M_tb}
\ee
where $\Gamma_\phi$ and ${\rm Br}\left[\phi^+\hspace*{-1.5mm}\to t\bbar\right]$
are the total decay width of $\phi^+$ and the branching ratio of
$\phi^+\to t\bbar$, respectively, which are calculated up to the NLO.
We note that the one-loop box diagrams 
with a virtual gluon connecting the initial state quark and
final state quark (from the hadronic decay of $\phi$)
have vanishing contribution at ${\cal O}(\alphas)$
because the scalar $\phi$ is color-neutral.
In Fig.~\ref{fig:DsDQ}a and Fig.~\ref{fig:DsDQ-LHC}a,
we plot the invariant mass distribution for 
$t$-$\bar b$ and $\bar t$-$b$ pairs from $\phi^\pm$ (top-pion signal) and
$W^{\pm\ast}$ (background) decays in the TopC model. In these plots,
we have included the NLO contributions, as a function of $Q$, 
to the $W^{\pm \ast}$ background rate at the Tevatron and the LHC. 
The overall $K$-factor (after averaging over the invariant mass $Q$) 
including both the initial and final state radiations is about 1.4~(1.34) 
for the Tevatron (LHC)\,\cite{Tim-CP}.
The total rate of $W^{\pm \ast}$ up to the NLO is about 0.70 [0.86]~pb and 
11.0~pb at the 1.8 [2]~TeV Tevatron and the 14~TeV LHC, respectively.
\begin{figure}[t]
\vspace{-1cm}
\begin{center}
\begin{tabular}{c}
\epsfysize=10cm
\epsfbox{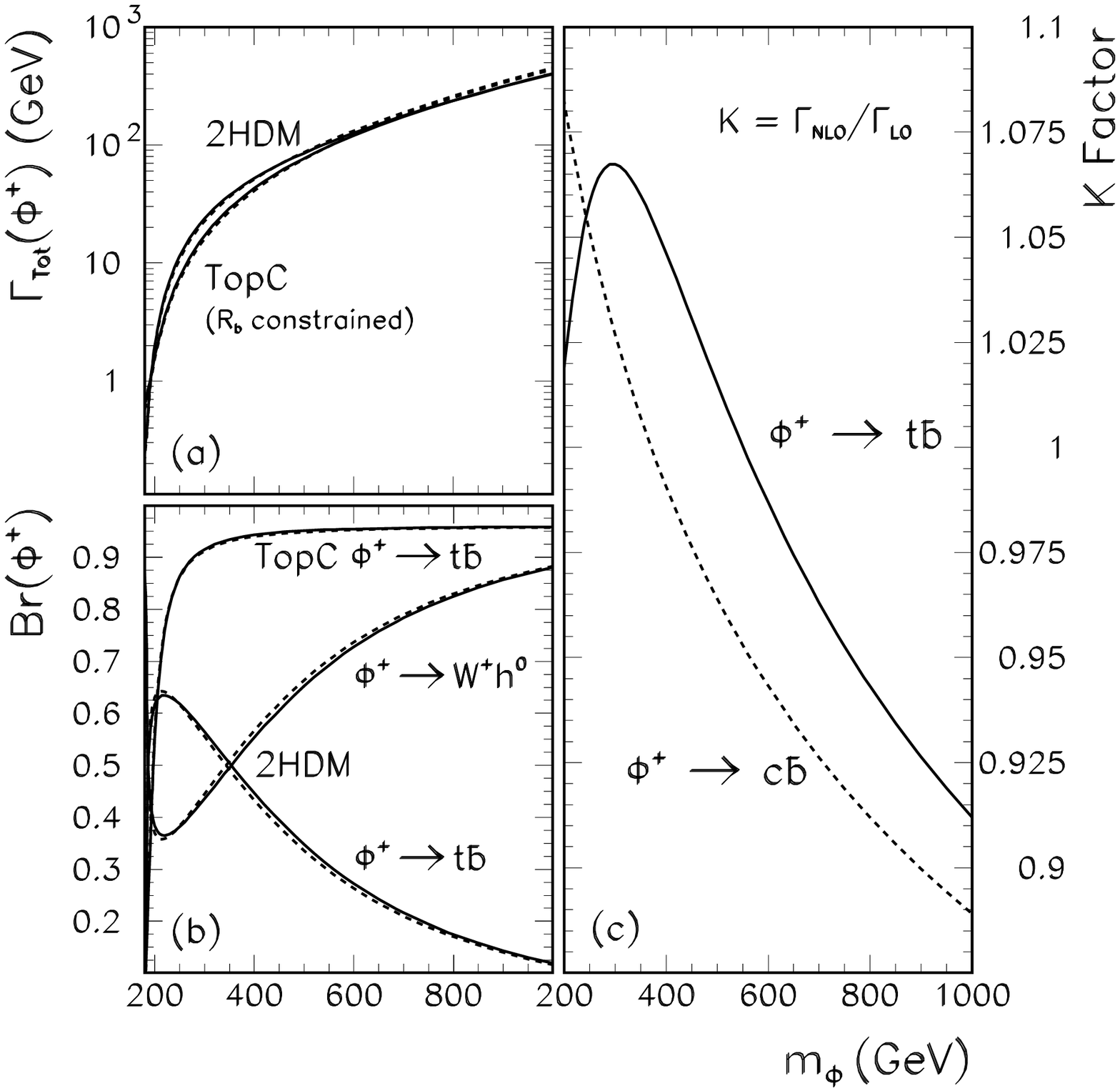}
\end{tabular}
\end{center}
\vspace{-0.5cm}
\caption{ 
Total decay widths of $\phi^+$ and BRs of $\phi^+ \to t\bar{b}$ in the
TopC model and 2HDM. (For the 2HDM, the BR of the $W^+h^0$ channel is
also shown, which is complementary to the $t\bbar$ channel.) In Fig. (a)
and (b), the NLO (solid) and LO (dashed) curves differ only by a small
amount. In Fig. (c), the $K$-factor, which is defined as the ratio of the
NLO to the LO partial decay widths, is shown for the $\phi^+\to
t\bar{b}$ (solid) and $\to c\bar{b}$ (dashed) channels. The sample
results for the 2HDM in this figure are derived for the parameter choice
$(\xi_{tt}^U,\,\xi_{tc}^U)=(1.5,\,1.5)$, 
$\alpha =0$, and $(m_h,\,m_A)=(120,\,1200)$\,GeV.
}
\label{fig:GammaBr}
\end{figure}

\begin{figure}[t]
\vspace{-1cm}
\begin{center}
\begin{tabular}{c}
\epsfysize=10cm
\epsfbox{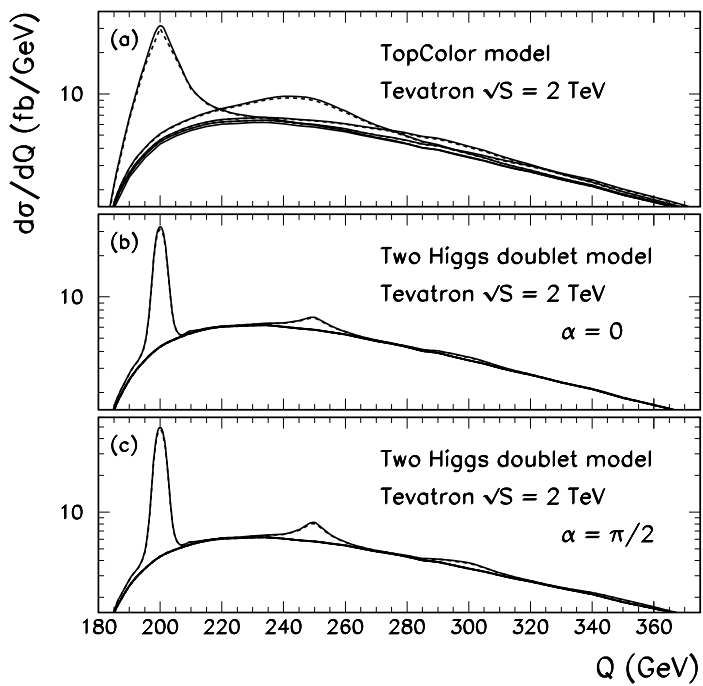}
\end{tabular}
\end{center}
\vspace{-0.5cm}
\caption{
Invariant mass distribution of $t$-$\bar{b}$ and  $\bar{t}$-$b$
pairs from  $\phi^\pm$ (signal) and $W^{\pm *}$ (background) 
decays at the Tevatron Run-II for the
TopC model (a), and 2HDM with Higgs mixing angles $\alpha =0$ (b),
and $\alpha =\pi /2$ (c). 
We show the signal for $m_{\phi}=200,~250$, 300 and 350\,GeV. The solid
curves show the results from the NLO calculation,
and the dashed ones from the LO analysis. 
}
\label{fig:DsDQ}
\end{figure}
%
\begin{figure}[t]
\vspace{-1cm}
\begin{center}
\begin{tabular}{c}
\epsfysize=10cm
\epsfbox{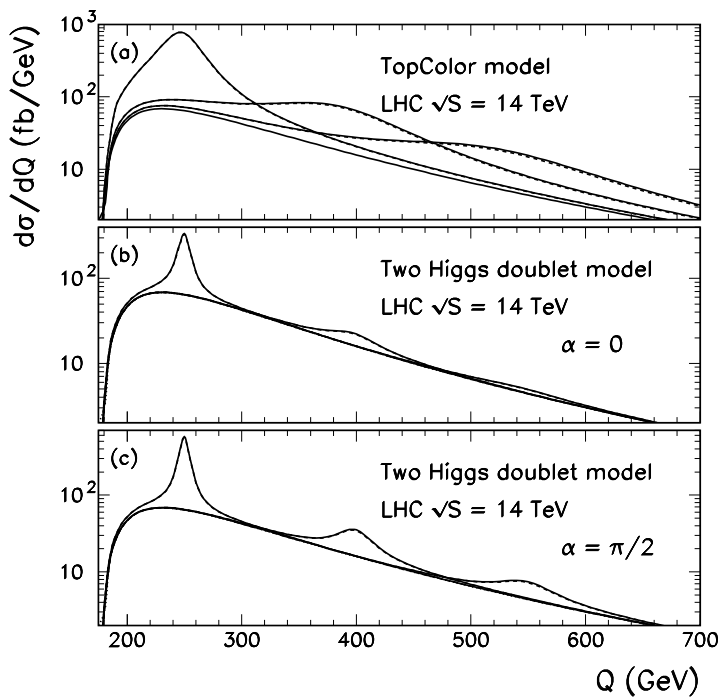}
\end{tabular}
\end{center}
\vspace{-0.5cm}
\caption{
Invariant mass distribution of $t$-$\bar{b}$ and $\bar{t}$-$b$ pairs
from $\phi^\pm$ (signal) and $W^{\pm *}$ (background) decays at the LHC
for the TopC model (a), and for the 2HDM with the Higgs mixing angles
$\alpha =0$ in (b), and $\alpha =\pi/2$ in (c). Here the charged
pseudo-scalar or scalar mass are chosen as the typical values of
$m_{\phi}=250,~400$ and 550\,GeV. The solid curves show the results by
the NLO calculation, while the dashed ones come from the LO analysis. 
}
\label{fig:DsDQ-LHC}
\end{figure}

\begin{figure}[t]
\vspace{-1cm}
\begin{center}
\begin{tabular}{c}
\epsfysize=10cm
\epsfbox{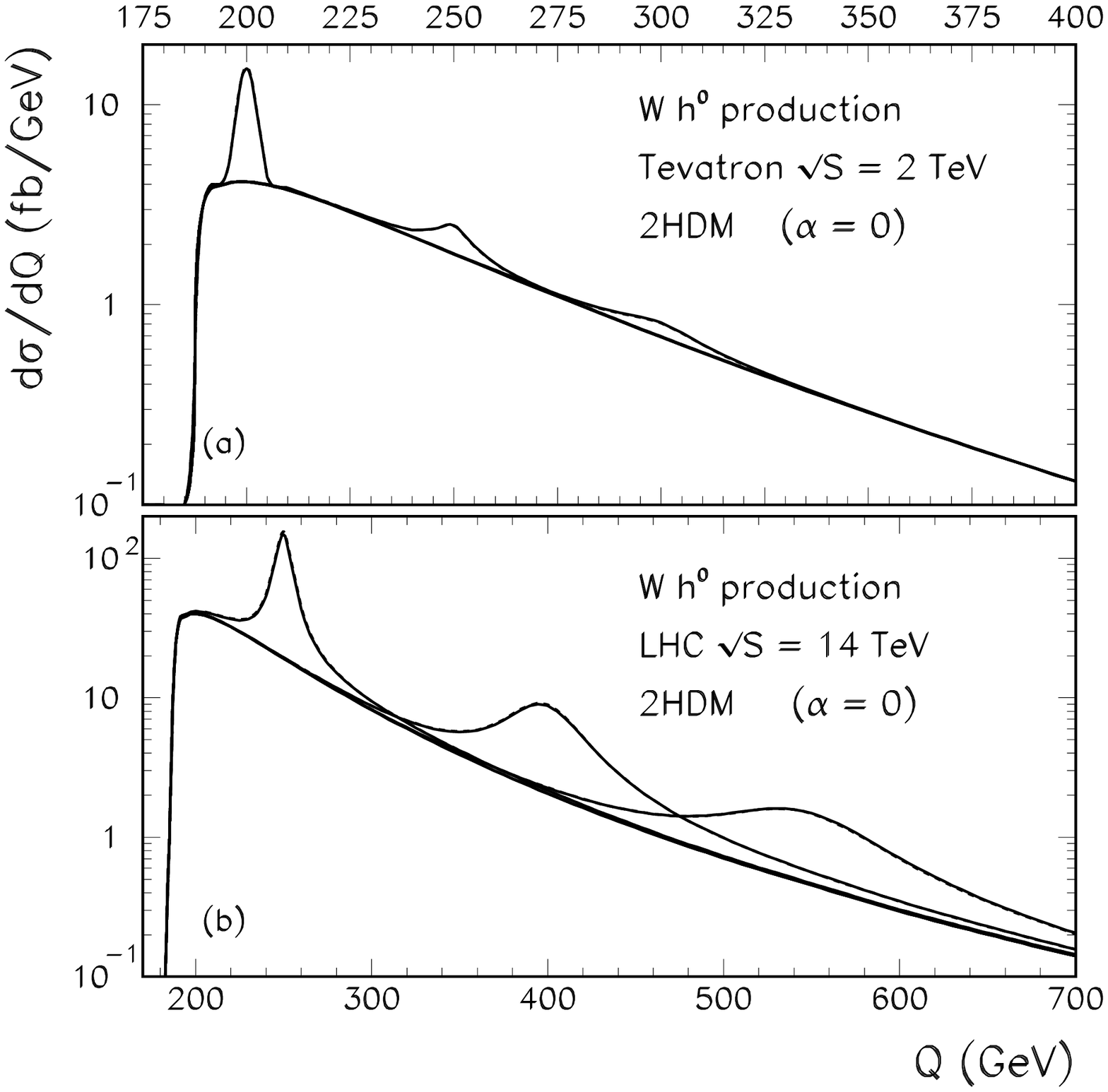}
\end{tabular}
\end{center}
\vspace{-0.5cm}
\caption{
Invariant mass distributions of $W^+$-$h^0$ and $W^-$-$h^0$ pairs from
$\phi^\pm$ ($s$-channel resonance) and 
$W^{\pm *}$ ($s$-channel non-resonance) decays at the Tevatron
Run-II, and at the LHC, for the 2HDM with Higgs mixing angles
$\alpha=0$. We show the signal for $m_{\phi}=200,~250$ and 300\,GeV at
the Tevatron (a), and for $m_{\phi}=250,~400$ and 550\,GeV at the LHC (b). The
solid curves show the results of the NLO calculation, and the dashed
ones of the LO analysis. 
}
\label{fig:DsDQ-Wh}
\end{figure}

Before concluding this section, we discuss how to generalize the above
results to the generic 2HDM (called type-III \cite{2HDM3}), 
in which the two Higgs doublets $\Phi_1$
and $\Phi_2$ couple to both up- and down-type quarks and the {\it ad hoc}
discrete symmetry \cite{GW} is not imposed.
The flavor-mixing Yukawa couplings in this model can be conveniently
formulated under a proper basis of Higgs doublets so that
{\small $~<\hspace*{-1.2mm}\Phi_1\hspace*{-1.2mm}>=(0,v/\sq2 )^T$} and
{\small $<\hspace*{-1.2mm}\Phi_2\hspace*{-1.2mm}>=(0,0)^T$.}
Thus, the diagonalization of the fermion mass matrix also 
diagonalizes the Yukawa couplings of $\Phi_1$,  
and all the flavor-mixing effects are generated by Yukawa couplings 
($\widehat{Y}^U_{ij}$ and $\widehat{Y}^D_{ij}$) 
of $\Phi_2$ which exhibit a natural hierarchy under the 
ansatz \cite{ansatz,2HDM3}
\be
\dis \widehat{Y}_{ij}^{U,D}=\xi_{ij}^{U,D}
{\sqrt{m_im_j}}/
{<\hspace*{-1.2mm}\Phi_1\hspace*{-1.2mm}>}
\label{eq:ansatz}
\ee
with {\small $~\xi_{ij}^{U,D}\sim {\cal O}(1)$}.~
This ansatz highly suppresses the flavor-mixings among light quarks
and identifies the largest mixing coupling as the one from the
$t$-$c$ or $c$-$t$ transition. 
A recent renormalization group analysis~\cite{RG-2HDM3}
shows that such a suppression persists at the high energy scales.
The relevant Yukawa interactions involving the charged Higgs bosons
$H^\pm$ are \cite{HY}:
\be 
\ba{ll}
& {\cal L}_Y^{CC} = \\[2mm]
& ~ H^+\left[\overline{t_R}~(\widehat{Y}_U^\dag V)_{tb}~b_L
        -\overline{t_L}~(V\widehat{Y}_D)_{tb}~b_R  
        ~+\overline{c_R}~(\widehat{Y}_U^\dag V)_{cb}~b_L
        -\overline{c_L}~(V\widehat{Y}_D)_{cb}~b_R~\right]+{\rm h.c.}
\\[2mm]
& ~ \simeq~ H^+\left[\overline{t_R}~\widehat{Y}^{U\ast}_{tt}~b_L
               +\overline{c_R}~\widehat{Y}^{U\ast}_{tc}~b_L\right]
               +{\rm h.c.}+({\rm small~terms})\,,
\ea
\label{eq:Hcb}
\ee
where $ ~\widehat{Y}^U_{tt}=\xi^U_{tt}\times (\sqrt{2}m_t/v)\simeq \xi^U_{tt}$,
and  $~\widehat{Y}^U_{tc}=\xi^U_{tc}\times (\sqrt{2m_tm_c}/v)
                  \simeq  \xi^U_{tc}\times 9\%~$,
in which $~\xi^U_{tc}\sim {\cal O}(1)$ is allowed by 
the current low energy data \cite{2HDM3,Laura}.
As a result, the Yukawa counter term in Fig.~1d involves both $\delta m_t$ and
$\delta m_c$. Consequently, we need to replace the NLO quantity $\Omega$ 
in the finite part of the Yukawa counter term 
[cf. the definition below (\ref{eq:NLO})] by 
\be
\dis\Omega ({\rm 2HDM}) = 3\ln \left[m_\phi^2/(m_tm_c)\right]+4~, 
\label{eq:omega-2HDM}
\ee
for the type-III 2HDM.
In the relevant $\phi^\pm$-$c$-$b$ coupling of this 2HDM, we note that, 
similar to the case of the TopC model,
only the right-handed charm is involved \cite{HY}, i.e.,
\be
{\cal C}_L^{tb}={\cal C}_L^{cb}= 0,~~~~
{\cal C}_R^{tb}=\xi_{tt}^U(\sqrt{2}m_t/v),~~~
{\cal C}_R^{cb}\simeq \xi^U_{tc}\times 9\% .
\label{eq:choice0}
\ee
where the parameters ($\xi^U_{tt},\,\xi_{tc}^U$) are expected to be naturally
around ${\cal O}(1)$. We have examined the possible constraint of $\xi^U_{tt}$
from the current $R_b$ data and found that the typical values of
$\xi^U_{tt}\!\sim\! 1.0\!-\!1.5$ are allowed for 
$m_{H^\pm}\gtrsim 200$\,GeV (cf. Fig.\,\ref{fig:Rb}c).\footnote{
Our calculation of $R_b$ in the 2HDM-III is consistent with those
in Ref.\,\cite{Rb2HDM} and Ref.\,\cite{Laura} (though a larger 
$\xi_{tt}^U$-value was chosen in the numerical analysis of 
Ref.\,\cite{Laura}).}\,
The production cross section of $H^\pm$ in this 2HDM can be obtained by 
rescaling the result of the TopC model according to the ratio of the 
coupling-square
$\left[{\cal C}^{tc}_R({\rm 2HDM})/{\cal C}_R^{tc}({\rm TopC})\right]^2
\!\sim\! \left[0.09\xi_{tc}^U/0.2{\cal C}_R^{tb}({\rm TopC})\right]^2$ (which is
about $1/7$ for $\xi_{tc}^U=1.5$ and
the charged scalar mass around $400$\,GeV).

Finally, we note that there are three neutral Higgs bosons 
in the 2HDM, the CP-even scalars ($h^0,~H^0$) 
and the CP-odd pseudo-scalar $A^0$. The mass diagonalization
for $h^0$ and $H^0$ induces the Higgs mixing angle $\alpha$. 
The low energy constraints on this model require \cite{2HDM3,Laura}
$~m_h,\,m_H \leq m_{H^\pm} \leq m_A~$ or 
$~m_A \leq m_{H^\pm} \leq m_h,\,m_H $.~   
For the case of
$m_{H^\pm} > m_{h^0} + M_W$, the $H^\pm\to W^\pm h^0$ decay channel is also
open. Taking, for example,  
$\alpha =0$ and $(m_h,\,m_A)=(120,1200)$\,GeV, we find from
Fig.~\ref{fig:GammaBr}b that the $tb$ and $Wh^0$ decay 
modes are complementary at low and
high mass regions of the charged Higgs boson $H^\pm$. 
In Figs.~\ref{fig:DsDQ}b-c and \ref{fig:DsDQ-LHC}b-c,
we plot the invariant mass distributions of $t\bbar$ 
and $\tbar b$ pairs from $H^\pm$ (signal) and $W^{\pm\ast}$ (background)
decays in the 2HDM  at the 2\,TeV Tevatron and the 14\,TeV LHC,
with the typical choice of the parameters: 
$~\left(\xi^U_{tt},\,\xi^U_{tc}\right)=(1.5,\,1.5)$ in  Eq.~(\ref{eq:choice0}), 
$\left(m_h,\,m_A\right)=\left(120,\,1200\right)$\,GeV, and $\alpha=0$ or $\pi/2$.
[A larger value of $\xi^U_{tt}$ will
simultaneously increase (reduce) the BR of $tb$ ($Wh^0$) mode.]
We see that, due to a smaller $c$-$b$-$H^\pm$ coupling  
[cf. (\ref{eq:choice0})], it is hard to detect such a charged Higgs boson
with mass $m_{H^\pm}>250$\,GeV at the Tevatron Run-II. We then examine
the potential of the LHC for the high mass range of $H^\pm$. Similar
plots are shown in Figs.~\ref{fig:DsDQ}b-c for 
$\alpha =0$ and $\alpha =\pi/2$, respectively. 
When $\cos\alpha$ is large (e.g., $\alpha =0$),
the branching ratio of the $tb$-channel decreases as $m_{H^\pm}$ 
increases (cf. Fig.~\ref{fig:GammaBr}b),  so that the LHC does not 
significantly improve the probe of the large $m_{H^\pm}$, 
range via the single-top mode (cf. Fig.~\ref{fig:DsDQ-LHC}b).
In this case,  the $W^\pm h^0$ channel, however,
becomes important for large $m_{H^\pm}$, as shown in Fig.~\ref{fig:DsDQ-Wh}
(cf. Fig.~\ref{fig:GammaBr}b, for its decay branching ratios) 
since the $H^\pm$-$W^\mp$-$h^0$
coupling is proportional to $\cos \alpha$ \cite{2HDM3}.
On the other hand, for the parameter space with
small $\cos\alpha$ (e.g., $\alpha =\pi /2$), the $W^\pm h^0$ channel 
is suppressed so that the single-top mode is important even for large 
mass region of $H^\pm$.\footnote{Note that the $H^\pm$-$W^\mp$-$H^0$
coupling is proportional to $\sin\alpha$ and is thus enhanced for
small $\cos\alpha$. In this case, the $WH^0$ mode may be important
provided that $H^0$ is relatively light. We will not further elaborate
this point here since it largely depends on the mass of $H^0$.}\,
This is illustrated in Fig.~\ref{fig:DsDQ-LHC}c at the LHC for
$\alpha =\pi/2$. In order to probe the whole parameter space and
larger $m_{H^\pm}$, it is important to study both $tb$ 
and $Wh^0$ (or $WH^0$) channels.

\section{Generalization to Neutral (Pseudo-)Scalar \\ Production 
via $b\bbar$ Fusion \label{sec:Generalization}} 

The QCD corrections are universal so that the generalization to 
the production of neutral scalar or pseudo-scalar $\phi^0$ via
the $b\bbar$ fusion is straightforward,
i.e., we only need to replace (\ref{eq:omega-2HDM}) by 
\be
\dis\Omega (\phi^0 b\bbar ) = 3\ln \left[m_\phi^2/m_b^2\right]+4\,, 
\label{eq:omega-Hbb}
\ee
in which $m_\phi$ is the mass of $\phi^0$.
The finite piece of the Yukawa renormalization
[cf. the quantity $\Omega$ in (\ref{eq:delta_mt})]
is scheme-dependent.
We can always define the $\phi^0$-$b$-$\bbar$ Yukawa coupling as 
$\sqrt{2}m_b/v$ times an enhancement factor $K$ so that the Yukawa
counter term is generated by $\delta m_b/m_b$.\footnote{
This specific definition works even if the Yukawa coupling is not related to 
any quark mass. For instance,
the bottom Yukawa couplings of the $b$-Higgs and
$b$-pion in the TopC model \cite{TopC,hbb} are independent of quark masses
because the $b$-Higgs does not develop vev.} 
After resumming the leading logarithmic terms,
$\left[\alpha_s \ln (m_\phi^2/m_b^2)\right]^n$, via the renormalization
group technique, the net effect of the Yukawa renormalization is to
change the Yukawa coupling or the related quark-mass into the corresponding
$\overline{\rm MS}$ running coupling or mass, as discussed in the previous
section.

The $b\bbar$ decay branching ratios of the neutral Higgs bosons in 
the MSSM with large $\tanb$
are almost equal to one \cite{Hdecay}. The same is
true for the $b$-Higgs or $b$-pion in the TopC model \cite{TopC}.
It has been shown that at the Tevatron,
the $b\bbar$ di-jet final states can be properly identified \cite{CDF_bb}.
The same technique developed for studying the resonance of the
coloron or techni-$\rho$ in the $b\bbar$ decay mode \cite{CDF_bb} can
also be applied to the search of the neutral Higgs bosons 
with large bottom Yukawa coupling.
When the neutral scalar or pseudo-scalar $\phi^0$ 
is relatively heavy, e.g., in the range of ${\cal O}(250-1000)$~GeV, 
the QCD di-jet backgrounds can be effectively removed by requiring the two
$b$-jets to be tagged with large transverse momenta ($P_T$) because 
the $P_T$ of each $b$-jet from the $\phi^0$ decay 
is typically at the order of $m_\phi/2$.
Hence, this process can provide complementary information to that 
obtained from studying 
the $\phi^0 b\bbar$ associate production \cite{Jack,hbb,hbx}.

We first consider the production of the
neutral Higgs boson $\phi^0$, which can be either $A^0$, $h^0$, or $H^0$,
in the MSSM with large $\tanb$, where the corresponding Yukawa couplings
to $b\bbar$ and $\tau^+\tau^-$ 
are enhanced relative to that of the SM since 
$y_{D}/y_{D}^{\rm SM}$ is equal to $\tanb$, $-\sin\alpha/\cos\beta$, or
$\cos\alpha/\cos\beta$, respectively, at the tree-level.
In the large $\tanb$ region,
the MSSM neutral Higgs bosons dominantly decay into $b\bbar$ and
$\tau^+\tau^-$ final states,
which can be detected at the hadron colliders. In comparison with
the recent studies on the $\phi^0b\bbar$ \cite{hbb} 
and $\phi^0\tau^+\tau^-$ \cite{htautau} 
associate production, we expect the inclusive $\phi^0$
production via the $b\bbar$-fusion would be more useful for $m_\phi$ being
relatively heavy (e.g., $m_\phi \geq 200-300$~GeV) 
because of the much larger phase space as well as
a better suppression of the backgrounds in the high $P_T$ region. 
The total LO and NLO cross sections 
for the inclusive production process $pp,p\bar{p}\to A^0X$ at the
Tevatron and the LHC are shown in Figs.~\ref{fig:Sigma_MSSMbb}a and b, 
in parallel to Figs.~\ref{fig:FigSigma_FixMt.eps} and \ref{fig:Sigma1}
for the case of charged top-pion production.
Here, we have chosen \,$\tanb =40$\, for illustration. The cross sections
at other values of $\tanb$ can be obtained by multiplying the scaling factor
$\, \left(\tanb /40\right)^2 \,$.
From Fig.~\ref{fig:Sigma_MSSMbb}a, we see a significant improvement from
the pure LO results (dash-dotted curves) by 
resumming over the large logarithms of $m_\phi^2/m_b^2$ into the running
Yukawa coupling.  The good agreement
between the LO results with running Yukawa coupling and the NLO results
is due to a non-trivial, and process-dependent,
cancellation between the individual ${\cal O}(\alpha_s)$ contributions 
of the $b\bbar$ and $bg$ sub-processes.
In contrast to the production of the charged top-pion or Higgs boson
via the initial state $c\bbar$ or $\cbar b$ partons, the neutral Higgs 
boson production involves the $b\bar{b}$ parton densities. 
The $K$-factors for the ratios of the NLO versus LO cross sections of
$p\bar{p}/pp\to A^0X$ are presented in Fig.~\ref{fig:K-A0} for the MSSM 
with $\tanb =40$. The main difference is due to 
the fact that the individual contribution
by the ${\cal O}(\alpha_s)$ $bg$-fusion becomes more negative as compared to
the case of the charged top-pion production shown in Fig.~\ref{fig:InitKFac}. 
This makes the overall $K$-factor of the NLO versus LO cross sections
range from about $-(16$$\sim$17)\% to +5\% at the Tevatron and the LHC.
In parallel to Table~1 and Fig.~\ref{fig:PDF}, 
we have examined the uncertainties of the CTEQ4 PDFs
for the $A^0$-production at the Tevatron and the LHC, and the results are
summarized in Table~\ref{tb:PDFS2} and Fig.~\ref{fig:PDF-bb}).  

The transverse momentum ($Q_T$) distributions of $A^0$,
produced at the upgraded Tevatron and at the LHC, are shown in
Fig.~\ref{fig:QT-MSSM} for various $A^0$ masses ($m_A$) with $\tan \beta = 40$.
The solid curves are the result of the multiple soft-gluon resummation,
and the dashed ones are from the ${\cal O}(\alpha_s)$ calculation.
The shape of these transverse momentum distributions is similar to that
of the charged top-pion (cf. Fig.~\ref{fig:QTDistn}). 
The fixed order distributions are singular as
$Q_T \to 0$, while the resummed ones have a maximum at some finite $Q_T$
and vanish at $Q_T = 0$. When $Q_T$ becomes large, of the order of $m_A$, 
the resummed curves merge into the fixed order ones. The average resummed
$Q_T$ varies between 25 and 30 (40 and 60) GeV in the 
mass range of $m_A$ from 200 to 300 (250 to 550)\,GeV
at the Tevatron (LHC).

We also note that for large $\tanb$,
the SUSY correction to the running $\phi^0$-$b$-$\bbar$ Yukawa coupling
is significant \cite{large-tanb} and can be
included in a way similar to our recent analysis of the $\phi^0b\bbar$
associate production~\cite{hbb}.  
To illustrate the SUSY correction to the $b$-Yukawa coupling, we choose all 
MSSM soft-breaking parameters as $500$~GeV, and the Higgs mixing parameter
$\mu = \pm 500$~GeV. Depending on the sign of $\mu$, the SUSY correction 
to the $\phi^0$-$b$-${\bar b}$ coupling can either take the same sign as 
the QCD correction or have an opposite sign \cite{hbb}.
In Fig.~\ref{fig:Sigma_MSSMbb}c, the solid curves represent the NLO
cross sections with QCD correction alone, while the results including the 
SUSY corrections to the running bottom Yukawa coupling are shown for
$\mu =+500$\,GeV (upper dashed curves) and $\mu =-500$\,GeV (lower dashed
curves). As shown, these partial SUSY corrections can change the cross
sections by about a factor of 2.
The above results are for the inclusive production of the CP-odd Higgs 
boson $A^0$ in the MSSM. Similar results can be easily obtained for the
other neutral Higgs bosons ($h^0$ and $H^0$) by properly rescaling
the coupling strength. We also note that in the large $\tanb$ region,
there is always a good mass-degeneracy between either $h^0$ and $A^0$ (in 
the low mass region with $m_A \lae 120$\,GeV) or $H^0$ and $A^0$ (in
the high mass region with $m_A \gae 120$\,GeV), as shown in Figs.~10 and 11
of Ref.~\cite{hbb}.
\begin{table}[p]
\begin{center}
\begin{tabular}{c  r r r r  r r r r}
\hline\hline
&&&&&&&&\\[-0.2cm]
Collider & 
\multicolumn{4}{c}{Upgraded Tevatron (2\,TeV)} 
& \multicolumn{4}{c}{LHC (14\,TeV)} \\
[0.15cm] \cline{1-9} &&&&&&&&\\[-0.2cm]
Process$\backslash$PDF 
&   4A1 &    4M &   4A5 &   4HJ &   4A1 &    4M &   4A5 &   4HJ \\
[0.15cm]\hline\hline &&&&&&&&\\[-0.2cm]
&  2020 &  1900 &  1660 &  1920 & 18100 & 19800 & 16600 & 17900 \\
&&&&&&&&\\[-0.3cm] LO                  
&   166 &   153 &   129 &   163 &  1520 &  1440 &  1280 &  1440 \\
&&&&&&&&\\[-0.3cm]
&  19.9 &  18.2 &  15.0 &  21.7 &   258 &   238 &   206 &   238 \\
[0.2cm]\hline &&&&&&&&\\[-0.2cm]
&  1810 &  1780 &  1620 &  1800 & 17100 & 17400 & 16700 & 17500 \\
&&&&&&&&\\[-0.3cm] NLO                 
&   160 &   154 &   134 &   164 &  1520 &  1470 &  1350 &  1470 \\
&&&&&&&&\\[-0.3cm]
&  20.3 &  19.3 &  16.4 &  22.9 &   265 &   250 &   222 &   251 \\
[0.2cm]\hline &&&&&&&&\\[-0.2cm]
&  3040 &  2900 &  2590 &  2930 & 25400 & 25400 & 24100 & 25600 \\
&&&&&&&&\\[-0.3cm] $q\bar{q}\to\phi^0X$
&   253 &   237 &   203 &   251 &  2140 &  2050 &  1850 &  2050 \\
&&&&&&&&\\[-0.3cm]
&  31.0 &  28.8 &  24.0 &  33.8 &   364 &   339 &   298 &   340 \\
[0.2cm]\hline &&&&&&&&\\[-0.2cm]
&$-$1230&$-$1120&$-$970&$-$1130&$-$8320&$-$8010&$-$7370&$-$8050\\
&&&&&&&&\\[-0.3cm] $qg\to\phi^0X$      
&$-$92.9&$-$83.1&$-$69.0&$-$87.5& $-$623& $-$575& $-$505& $-$574\\
&&&&&&&&\\[-0.3cm]
&$-$10.6&$-$9.42&$-$7.59&$-$10.9& $-$98.8&$-$88.8&$-$75.8&$-$88.7\\
[0.2cm]\hline\hline
\end{tabular}
\end{center}
\caption{ 
Cross sections in fb for neutral Higgs boson production in the MSSM 
with $\tanb =40$, at the upgraded Tevatron and the LHC, are shown
for four different CTEQ4 PDFs. 
They are separately given for the LO and NLO processes, 
and for the $b\bar{b}\to A^0X$ and $bg\to A^0X$ sub-processes. 
For the upgraded Tevatron the 
top number is for $m_{A} = 200$ GeV, 
the middle is for $m_{A} = 300$ GeV, and
the lowest is for $m_{A} = 400$ GeV.
For the LHC the 
top number is for $m_{A} = 400$ GeV, 
the middle is for $m_{A} = 700$ GeV, and
the lowest is for $m_{A} = 1$ TeV.
}
\label{tb:PDFS2}
\end{table}
\begin{figure}[p]
\vspace{-1cm}
\begin{center}
\begin{tabular}{c}
\epsfysize=14.0cm
\epsfbox{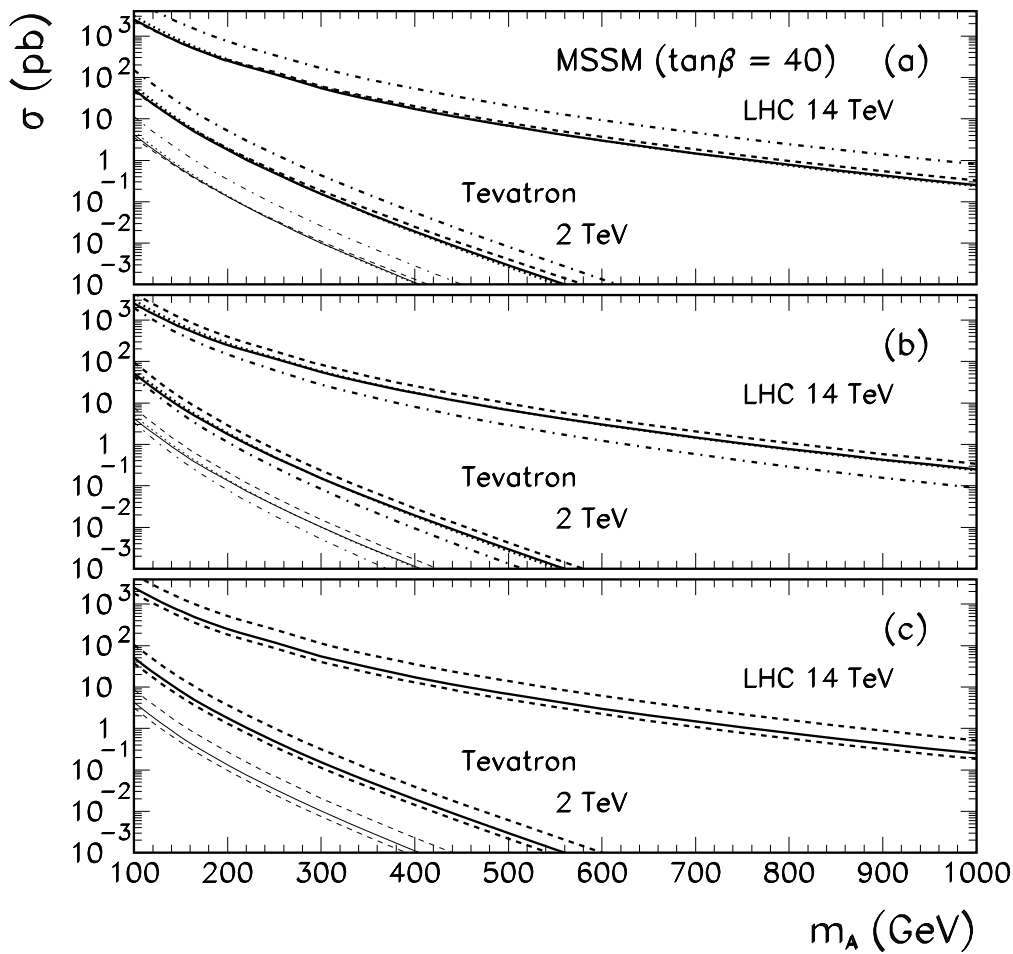}
\end{tabular}
\end{center}
\vspace{-0.5cm}
\caption{ 
LO and NLO cross sections for the neutral Higgs $A^0$ 
production in the MSSM with $\tanb =40$,
at the Tevatron and the LHC. 
(a) For each collider we show the NLO cross sections with
the resummed running Yukawa coupling (solid) and with one-loop
Yukawa coupling (dashed), as well as the
LO cross sections with resummed running Yukawa coupling (dotted) and 
with tree-level Yukawa coupling (dash-dotted).
(b) The NLO (solid), the $b\bar b$ (dashed) and $bg$
(dash-dotted) sub-contributions, and the LO (dotted) contributions are shown.
Since the $bg$ cross sections are negative, they are multiplied by
$-1$ in the plot. The cross sections at $\sqrt{S} = 1.8$~TeV 
(thin set of lowest curves)
are multiplied by 0.1 to avoid overlap with the $\sqrt{S} = 2$ TeV curves.   
(c) The NLO cross sections
with QCD running Yukawa coupling (solid curves) and those with additional
SUSY correction to the running coupling are shown 
(upper dashed lines for the Higgs-mixing parameter $\mu =+500$\,GeV 
and lower dashed lines for $\mu =-500$\,GeV).
}
\label{fig:Sigma_MSSMbb}
\end{figure}
%
%
\begin{figure}[t]
\begin{center}
\vspace{-1cm}
\begin{tabular}{c}
\epsfysize=10cm
\epsfbox{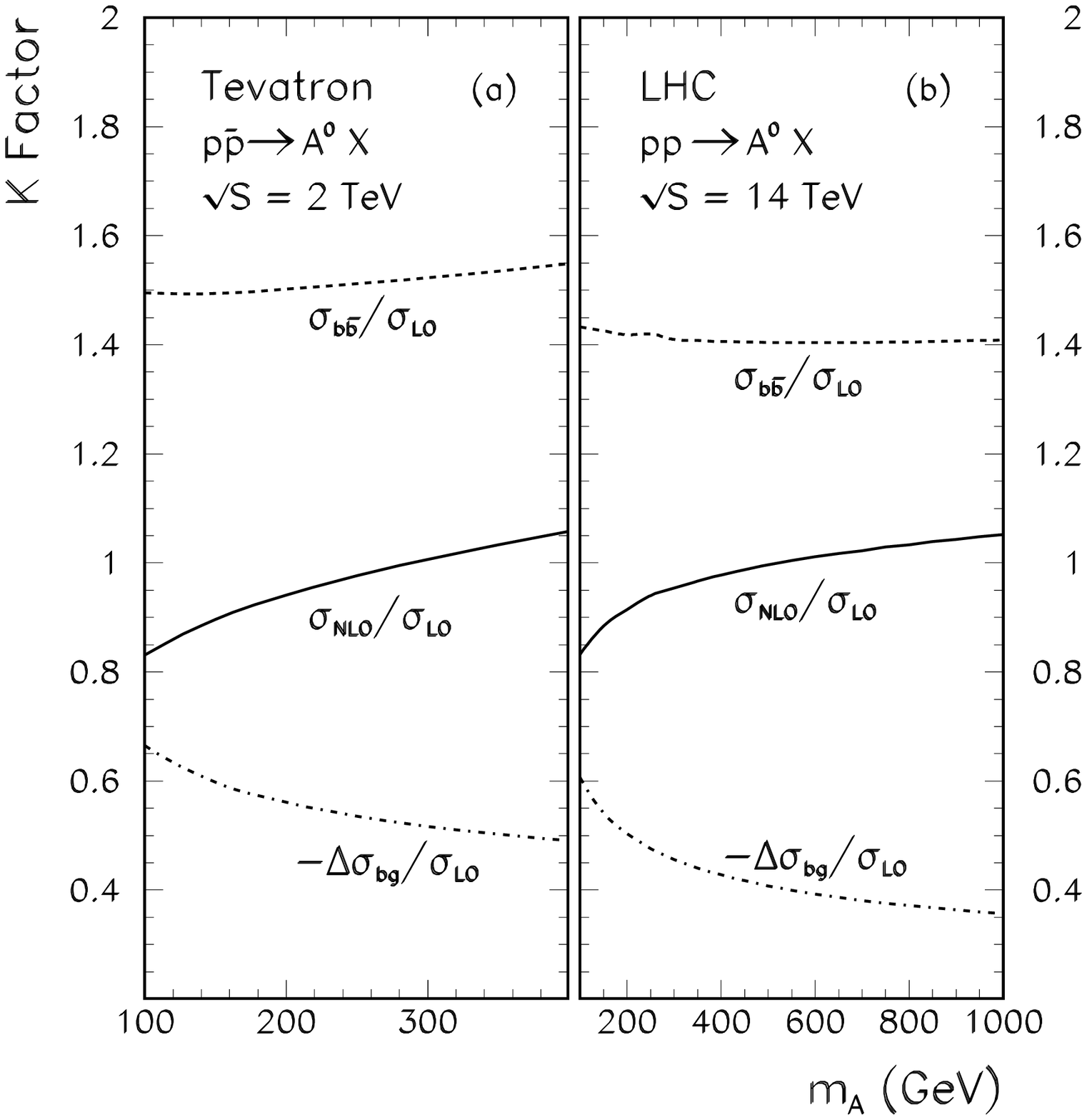}
\vspace{-0.5cm}
\end{tabular}
\end{center}
\caption{ 
The $K$-factors for the $A^0$ production in the MSSM with 
$\tanb =40$ are shown for
the NLO ($K=\sigma_{\rm NLO}/\sigma_{\rm LO}$, solid lines),
$b\bar{b}$ ($K=\sigma_{\bb}/\sigma_{\rm LO}=
(\sigma_{\rm LO}+\Delta\sigma_{\bb})/\sigma_{\rm LO}$, 
dashed lines), and $bg$ ($K=-\Delta\sigma_{bg}/\sigma_{\rm LO}$, 
dash-dotted lines) contributions, 
at the upgraded Tevatron (a) and the LHC (b).
}
\label{fig:K-A0}
\end{figure}
\begin{figure}[t]
\vspace{-1cm}
\begin{center}
\begin{tabular}{c}
\epsfysize=10cm
\epsfbox{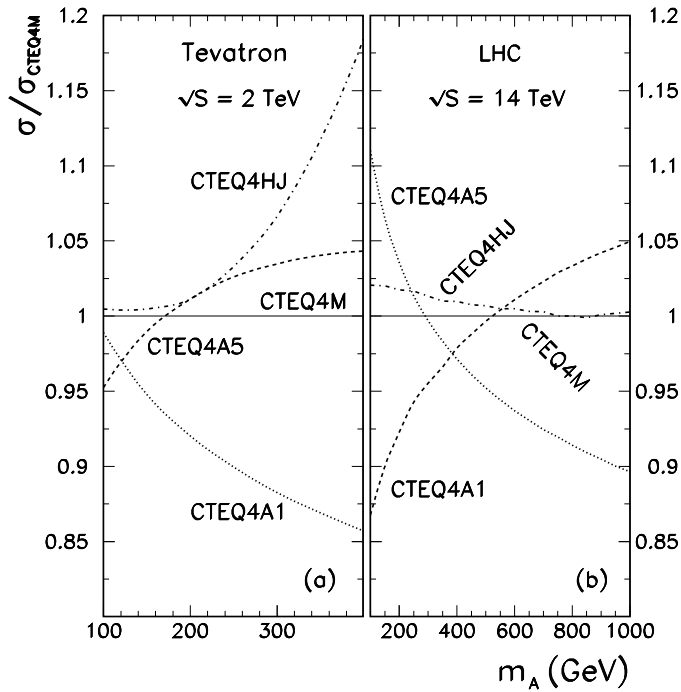}
\end{tabular}
\end{center}
\vspace{-0.5cm}
\caption{
The ratios of NLO cross sections computed by four different sets of
CTEQ4 PDFs relative to that by the CTEQ4M for neutral pseudo-scalar
($A^0$) production in the MSSM with $\tanb =40$, at the upgraded
Tevatron (a) and the LHC (b). 
}
\label{fig:PDF-bb}
\end{figure}
\begin{figure}[t]
\vspace{-1cm}
\begin{center}
\begin{tabular}{c}
\epsfysize=10cm
\epsfbox{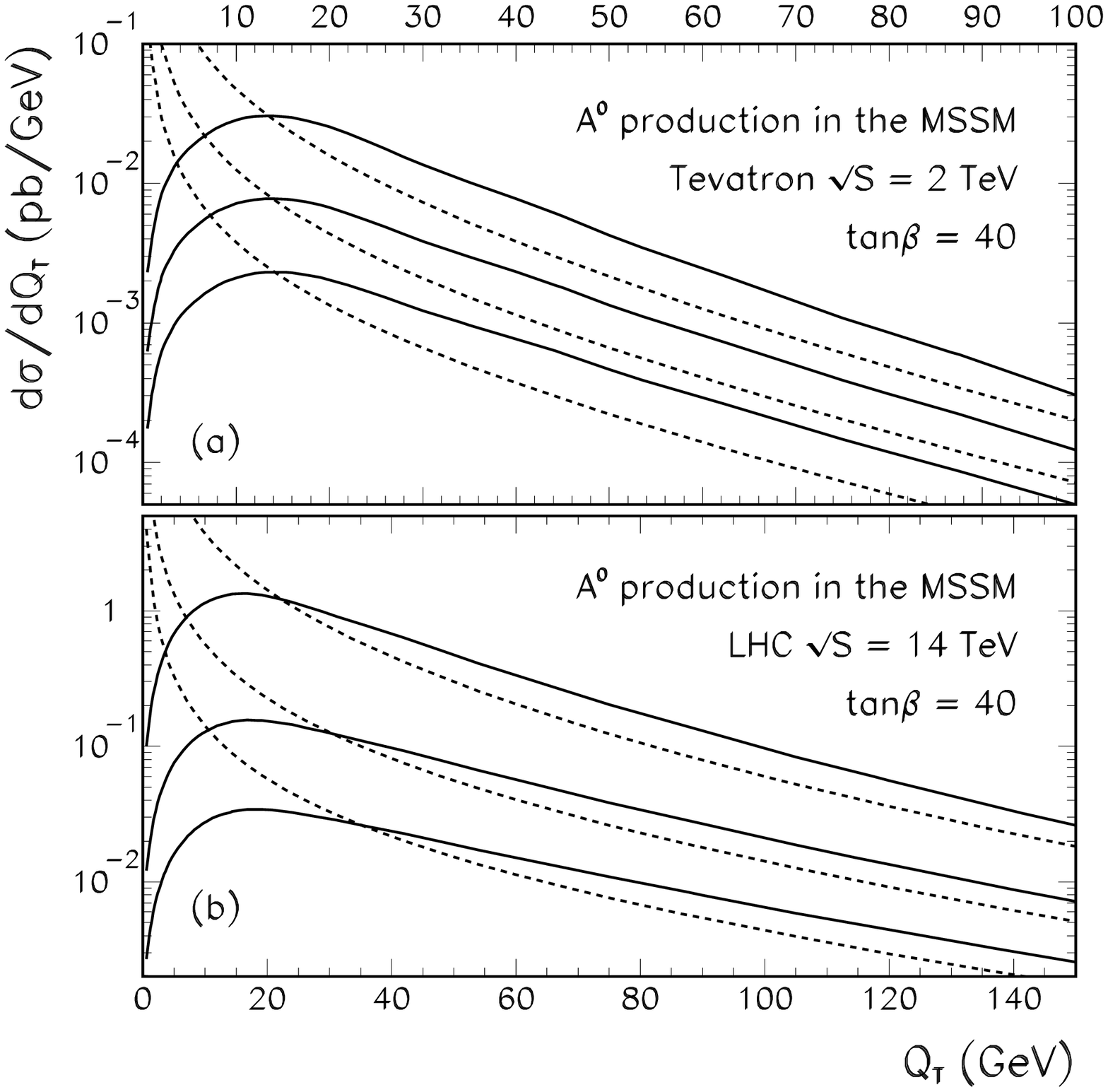}
\end{tabular}
\end{center}
\vspace{-0.5cm}
\caption{
Transverse momentum distributions of pseudo-scalar $A^0$ produced
via hadronic collisions, calculated in the MSSM with $\tanb =40$. 
The resummed (solid) and ${\cal O}(\alpha _s)$ (dashed) curves 
are shown for $m_A =$ 200, 250, and
300\,GeV at the upgraded Tevatron (a), and 
for $m_A =$ 250, 400, and 550\,GeV at the LHC (b).
}
\label{fig:QT-MSSM}
\end{figure}

We then consider the large bottom Yukawa coupling of
the neutral $b$-Higgs ($h_b^0$) and $b$-pion
($\pi_b^0$) in the TopC model \cite{TopC,TopC2,hbb}.
The new strong $U(1)$ force in this model is attractive in the 
$<\hspace*{-0.15cm}\tbar t\hspace*{-0.15cm}>$ channel but repulsive 
in the $<\hspace*{-0.15cm}\bbar b\hspace*{-0.15cm}>$ channel. Thus, 
the top but not the bottom acquires dynamical mass from the vacuum. 
This makes the  $t$-Yukawa coupling ($y_t$) super-critical
while the $b$-Yukawa coupling ($y_b$) sub-critical, at the TopC breaking
scale $\Lambda$, i.e., 
~$y_b(\Lambda ){~\lae ~} y_{\rm crit}
\hspace*{-1mm}=\hspace*{-1mm}\sqrt{{8\pi^2}/{3}} 
{~\lae ~} y_t(\Lambda ) $~,   which requires
$y_b$ being close to $y_t$ and thus naturally large. Our recent
renormalization group analysis \cite{hbb} shows that the relation
$y_b(\mu ){\sim } y_t(\mu )$ holds well at any scale $\mu$ below $\Lambda$.
For the current numerical analysis, we shall choose a typical value of 
~$y_b(m_t)\simeq y_t(m_t)\approx 3$,~ i.e., 
~$|{\cal C}_L^{bb}|=|{\cal C}_R^{bb}|\simeq 3/\sqrt{2}.~$
In Fig.~\ref{fig:Sigma-bHiggs}, we plot the production cross sections of 
$h^0_b$ or $\pi_b^0$ at the Tevatron and the LHC.  This is similar to the
charged top-pion production in Fig.~\ref{fig:Sigma1}, except the non-trivial
differences in the Yukawa couplings (due to the different tree-level values 
and the running behaviors) and the charm versus bottom parton luminosities.
\begin{figure}[t]
\vspace{-1cm}
\begin{center}
\begin{tabular}{c}
\epsfysize=10cm
\epsfbox{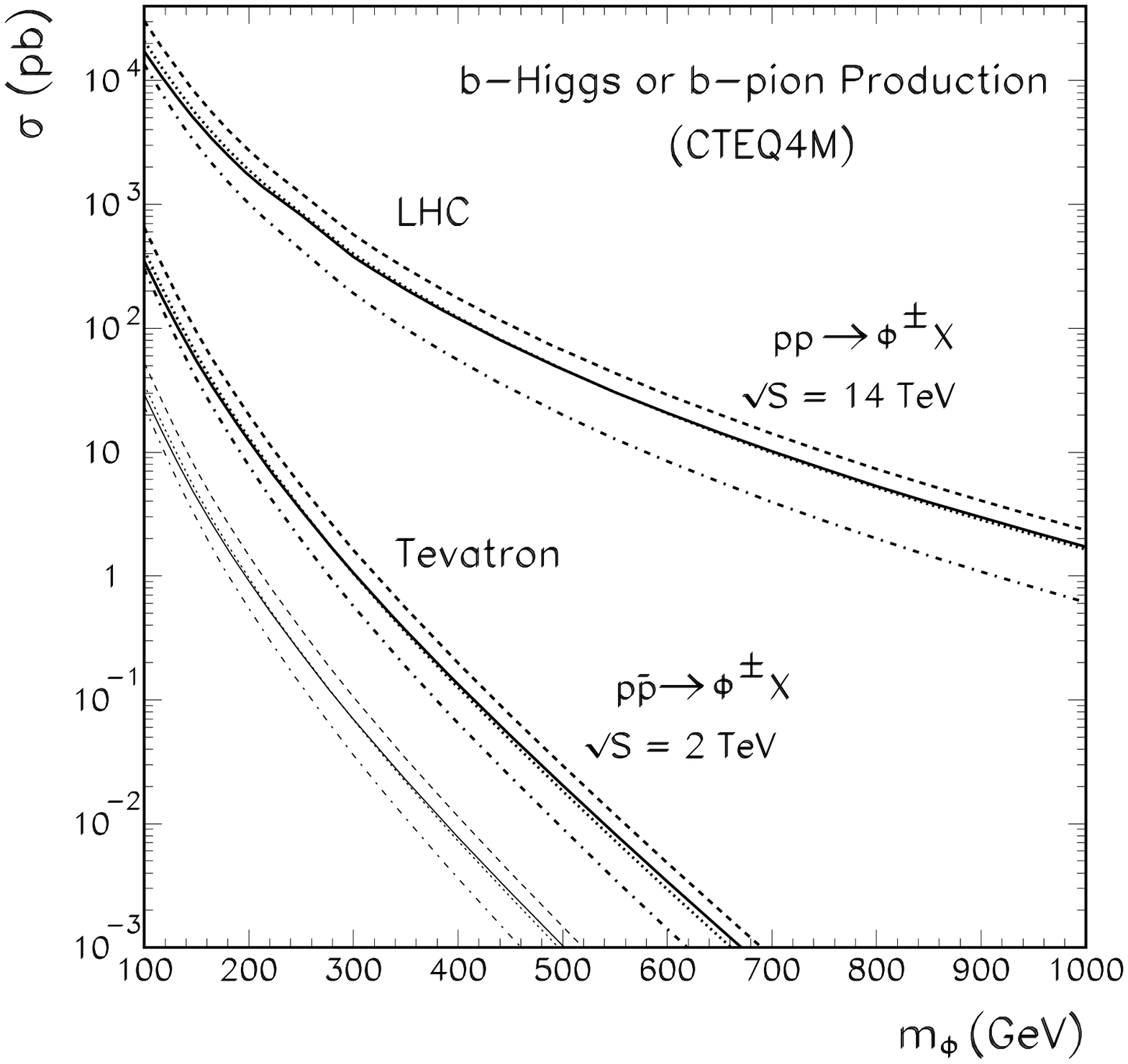}
\end{tabular}
\end{center}
\vspace{-0.5cm}
\caption{
Cross sections for the neutral $b$-pion $\pi_b^0$ or $b$-Higgs $h_b^0$
production via the $b\bbar$-fusion in the TopC model at the Tevatron and
the LHC. The NLO (solid), the $q\bar{q}'$ (dashed) and $qg$
(dash-dotted) sub-contributions, and the LO (dotted) contributions with
resummed running Yukawa coupling are shown. Since the $qg$ cross
sections are negative, they are multiplied by $-1$ in the plot. The cross
sections at $\sqrt{S} = 1.8$~TeV 
(thin set of lowest curves)
are multiplied by 0.1 to avoid overlap with the $\sqrt{S} = 2$ TeV curves. 
}
\label{fig:Sigma-bHiggs}
\end{figure}

\section{Summary of the Analytic Results \label{sec:SummaryOfAnalytic}}

In this Section, we present the individual NLO parton cross sections
computed at $D=4-2\epsilon$ dimensions.
We note that, unlike the usual Drell-Yan type processes, the one-loop 
virtual contributions (cf. Fig.~1b-d) 
are not ultraviolet (UV) finite unless the new
counter term from Yukawa coupling (related to the quark-mass renormalization,
cf. Fig.~1e) is included.

\subsection{Partonic processes $c\bbar\to\phi^+X$}

The spin- and color-averaged amplitude-square for the 
$c\bbar\to\phi^+g$ process is
\be
\overline{|{\cal M}|^2}=\dis
\f{2\pi C_F}{3}\alpha_s\left(|\CL|^2+|\CR|^2\right)\mu^{2\epsilon}
\left[(1-\epsilon)\left(\f{\widehat{t}}{\widehat{u}}+
\f{\widehat{u}}{\widehat{t}}+2\right)+
2\f{\widehat{s}\,m_\phi^2}{\widehat{t}\,\widehat{u}} \right].
\label{eq:M2qq}
\ee

The individual contributions (from the virtual loop and real gluon emission)
to the NLO partonic cross section are:
\be
\ba{ll}
\Delta\widehat{\sigma}^{\rm virtual}_{\rm loop} \hspace*{-2mm}
&=~\dis\widehat{\sigma}_0\f{\alpha_sC_F}{2\pi}
 \left(\f{4\pi\mu^2}{Q^2}\right)^{\epsilon}
 \f{\Gamma (1-\epsilon )}{\Gamma (1-2\epsilon )}
 \left[ -\f{2}{\epsilon^2}
    +\f{2\pi^2}{3}-2\right]\delta (1-\tauhat ),
\\[5mm]
\Delta\widehat{\sigma}^{\rm virtual}_{\rm count} \hspace*{-2mm}
&=~\dis\widehat{\sigma}_0\f{\alpha_sC_F}{2\pi}
 \left(4\pi\right)^{\epsilon}
 \f{\Gamma (1-\epsilon )}{\Gamma (1-2\epsilon )}
 \left[ -\f{3}{\epsilon}-\Omega\right]\delta (1-\tauhat ),
\\[5mm]
\Delta\widehat{\sigma}^{\rm real}_{c\bbar} \hspace*{-2mm}
&=~\dis\widehat{\sigma}_0\f{\alpha_sC_F}{2\pi}
\left(\f{4\pi\mu^2}{Q^2}\right)^{\epsilon}
 \f{\Gamma (1-\epsilon )}{\Gamma (1-2\epsilon )}
\left[ \f{2}{\epsilon^2}\delta (1-\tauhat )
+\f{3}{\epsilon}\delta (1-\tauhat )
-\f{2}{\epsilon}P^{(1)}_{q\leftarrow q}(\tauhat )C_F^{-1}
\right.\\[5mm]
&~~~\left. \dis
+                                                                                                                                                                  4\left(1+\tauhat^2\right)\left(\f{\ln (1-\tauhat )}{1-\tauhat}\right)_+
-2\f{1+\tauhat ^2}{1-\tauhat }\ln\tauhat +2(1-\tauhat ) \right],
\\[5mm]
\dis P^{(1)}_{q\leftarrow q}(\tauhat ) \hspace*{-2mm}
&=~\dis C_F\left(\f{1+\tauhat^2}{1-\tauhat}\right)_+ 
~=~C_F\left[\f{1+\tauhat^2}{\left(1-\tauhat\right)_+}+\f{3}{2}\delta (1-\tauhat )\right],
\ea
\label{eq:A1}
\ee
where the standard plus prescription $(\cdots )_+$ is given by
\be
\dis\int^1_0 d\alpha\, \xi (\alpha )\left[\chi (\alpha )\right]_+ \,=\,
\int_0^1 d\alpha\,\chi (\alpha )\left[\xi(\alpha )-\xi (1)\right].
\label{eq:plus}
\ee
In (\ref{eq:A1}),
the infrared $\dis\f{1}{\epsilon^2}$ poles cancel between 
$\Delta\widehat{\sigma}^{\rm virtual}_{\rm loop}$ and 
$\Delta\widehat{\sigma}^{\rm real}_{c\bbar}$.
The term $\Delta\widehat{\sigma}^{\rm virtual}_{\rm loop}$ 
from the virtual loop actually
contains two types of $\dis\f{1}{\epsilon}$ poles inside $[\cdots ]$ : 
$\dis\f{3}{\epsilon_{UV}}+\f{3}{\epsilon_{IR}}~$ 
 with $\epsilon_{UV}
=-\epsilon_{IR}\equiv \epsilon =(4-D)/2 >0$. Also, the 
$-\dis\f{3}{\epsilon}$ pole inside the Yukawa counter-term contribution
$\Delta\widehat{\sigma}^{\rm virtual}_{\rm count}$ is ultraviolet 
while the $+\dis\f{3}{\epsilon}$ pole inside
$\Delta\widehat{\sigma}^{\rm real}_{c\bbar}$ is infrared (IR). 
We see that the contribution 
$\Delta\widehat{\sigma}^{\rm virtual}_{\rm count}$ 
from the counter term of the Yukawa coupling is
crucial for cancelling the UV divergence from 
$\Delta\widehat{\sigma}^{\rm virtual}_{\rm loop}$ (which is absent in
the usual Drell-Yan type processes), while  
the soft $\dis\f{1}{\epsilon}$ divergences between
$\Delta\widehat{\sigma}^{\rm virtual}_{\rm loop}$
and $\Delta\widehat{\sigma}^{\rm real}_{c\bbar}$ cancel. Finally, 
the $\dis\f{1}{\epsilon}$ collinear singularity inside 
$\Delta\widehat{\sigma}^{\rm real}_{c\bbar}$  will be absorbed into
the re-definition of the PDF via the quark-quark transition function 
$P^{(1)}_{q\leftarrow q}(\tauhat )$. All the finite terms are
summarized in Eq.~(\ref{eq:NLO}).

\subsection{Partonic processes $gc,g\bbar\to\phi^+X$}

The spin- and color-averaged amplitude-square for the 
$gc,g\bbar\to\phi^+X$ process is
\be
\overline{|{\cal M}|^2}=\dis
\f{\pi \alpha_s}{3(1-\epsilon )}\left(|\CL|^2+|\CR|^2\right)\mu^{2\epsilon}
\left[(1-\epsilon )\left(\f{\widehat{s}}{-\widehat{t}} 
   +\f{-\widehat{t}}{\widehat{s}} -2\right)
   -2\f{\widehat{u}\,m_\phi^2}{\widehat{s}\,\widehat{t}}\right].
\label{eq:M2gq}
\ee

The ${\cal O}(\alpha_s)$ partonic cross section for 
the quark-gluon fusions is given by:
\be
\ba{ll}
\Delta\widehat{\sigma}^{\rm real}_{cg,\bar{b}g} \hspace*{-2mm}
& =\,
\dis\widehat{\sigma}_0\f{\alpha_sC_F}{2\pi}
\left(\f{4\pi\mu^2}{Q^2}\right)^{\epsilon}\left[
\left( -\f{1}{\epsilon}
 \f{\Gamma (1-\epsilon )}{\Gamma (1-2\epsilon )}
  +\ln\f{(1-\tauhat )^2}{\tauhat} \right)
P^{(1)}_{q\leftarrow g}(\tauhat )
\right. \\[4.6mm]
& ~~~~~~~~~~~~~~~~~~~~~~~~~~~~~ 
\left. \dis +\f{1}{4}(-3+7\tauhat )(1-\tauhat ) \right],
\\[4.6mm]
P^{(1)}_{q\leftarrow g}(\tauhat ) \hspace*{-2mm}
& =\,\dis \f{1}{2}\left[\tauhat^2+(1-\tauhat )^2 \right],
\ea
\label{eq:A2}
\ee
where it is clear that the collinear $\dis\f{1}{\epsilon}$ singularity
will be absorbed into the re-definition of the PDF via the gluon-splitting
function $P^{(1)}_{q\leftarrow g}(\tauhat )$. The final result is finite
and is given in Eq.~(\ref{eq:NLO}).

\section{Conclusions}

In summary, we have presented the complete 
${\cal O}(\alpha_s)$ QCD corrections
to the charged scalar or pseudo-scalar production via the
partonic heavy quark fusion process at hadron colliders. 
We found that 
the overall NLO corrections to the $p\bar{p}/pp \to \phi^\pm$
processes are positive for $m_\phi$ above $\sim$150\,(200)~GeV
and lie below $\sim$15\,(10)\% for the
Tevatron (LHC) in the relevant range of $m_\phi$
(cf.  Fig.~\ref{fig:InitKFac}). 
The inclusion of the NLO contributions thus
justifies and improves our recent LO analysis \cite{HY}.
The uncertainties of the NLO rates
due to the different PDFs are systematically examined and are found to be
around $20\%$ (cf. Table~1 and Fig.~\ref{fig:PDF}).  
The QCD resummation to include the effects of multiple soft-gluon 
radiation is also performed,
which provides a better prediction of the transverse momentum ($Q_T$)
distribution of the scalar $\phi^{0,\pm}$, 
and is important for extracting the experimental signals 
(cf. Fig.~\ref{fig:QTDistn}).
We confirm that the Tevatron Run-II (with a $2$~fb$^{-1}$ 
integrated luminosity) is able to explore the natural 
mass range of the top-pions up to about 300--350\,GeV
in the TopC model \cite{TopC,TopC2}.  On the other hand,
due to a possibly smaller $\phi^\pm$-$b$-$c$ coupling in the
2HDM, we show that to probe the charged Higgs boson with mass
above 200\,GeV in this model may
require a high luminosity Tevatron 
(with a $10-30$~fb$^{-1}$ integrated luminosity). 
The LHC can further probe the charged Higgs 
boson of the 2HDM up to about ${\cal O}(1)\,$TeV via the single-top and
$W^\pm h^0$ (or $W^\pm H^0$) production.  
The complementary roles of the $tb$ and $W^\pm h^0$ 
channels in the different regions of the Higgs mass and the Higgs
mixing angle $\alpha$ are demonstrated.   We have also analyzed a
direct extension of our NLO results to the neutral 
(pseudo-)scalar production via the $b\bbar$-fusion for the 
neutral Higgs bosons $(A^0,h^0,H^0)$ in the MSSM with large $\tanb$,
and for the neutral $b$-pion ($\pi_b^0$) or $b$-Higgs ($h_b^0$)
in the TopC model with $U(1)$-tilted large bottom Yukawa coupling.
In comparison with the $\phi^0 b\bbar$ associate production \cite{hbb},
this inclusive $\phi^0$-production mechanism provides a complementary 
probe for a neutral Higgs boson (with relatively large mass),
whose decay products,
e.g., in the $b\bbar$ or $\tau\tau$ channel, typically have
high transverse momenta ($\sim m_\phi /2$) and
can be effectively detected \cite{CDF_bb}.
This is particularly helpful for the discovery reach of the Tevatron. 
Further detailed  Monte Carlo analyses at the detector level 
should be carried out to finally conclude the sensitivity of 
the Tevatron Run-II and the LHC via this process. 
 
We mention that a recent paper \cite{note-added}
also studied the QCD corrections for the neutral
Higgs production $b\bbar \to H^0$ within the SM,  
and partially overlaps with our work
as the pure NLO QCD correction is concerned. 
The overlapped part is in general agreement with ours except  
that we determine the counter term of the 
Yukawa coupling (expressed in terms of the relevant quark
mass) by the on-shell scheme (cf. Refs.~\cite{Htb-NLO,CSL}) 
while Ref. \cite{note-added} used $\overline{\rm MS}$ scheme. 
After resumming the leading logarithms into
the running mass or Yukawa coupling, the two results coincide.
Note that the apparent large ${\cal O}(\alphas )$ correction derived in 
Ref.~\cite{note-added} is due to the fact that it only includes
the contribution from the $b\bar{b}$ sub-process, which is 
part of our complete ${\cal O}(\alphas )$ contribution.
The inclusion of the NLO contribution from the $bg$ sub-process,
which turns out to be negative and partially cancels the $b {\bar b}$ 
contribution, yields a typical size of ${\cal O}(\alphas )$ correction 
to the production rate of a neutral Higgs boson produced via 
heavy quark fusion. The $bg$ sub-process is identified as 
${\cal O}(1/\ln[m_H/m_b])$ instead of ${\cal O}(\alpha_s)$ correction in 
Ref.~\cite{note-added}.



\appendix
\def\Tblcutstab
{
\begin{table}[t] 
\begin{center}
\begin{tabular}{c||cccc}
Process & Acceptance Cuts   & $p_T$ Cuts & $\Delta R$ Cut & $\Delta M$ 
Cut\\ \hline
$\phibb$  & 4923            & 1936       & 1389       & 1389       \\
$\Zbb$  & 1432              & 580        & 357        & 357        \\
$\bbbb$ & $5.1 \times 10^4$ & 3760       & 1368       & 1284       \\
$\bbjj$ & $1.2 \times 10^7$ & $1.5 \times 10^6$ & 
$6.3 \times 10^5$ & $5.9 \times 10^5$ \\
\end{tabular}
\end{center} 
\vskip 0.08in 
\caption{The signal and background events for 2~\ifb
of Tevatron data, assuming
$m_\phi = 100$\,GeV, $2 \Delta m_\phi = 26$\,GeV, and $K = 40$
after imposing the acceptance cuts, $p_T$ cuts,
and reconstructed $m_\phi$ cuts described in the text.
(A $k$-factor of 2 is included in both the signal and the background 
rates.)}
\label{cutstab}
\end{table}
}

\def\Tblptcutstab
{
\begin{table}[t] 
\begin{center}
\begin{tabular}{c||cc}
$m_\phi$ (GeV) & $p^{(1)}_T$ Cut & $p^{(2)}_T$ Cut \\ \hline
75             & 35              & 25              \\
100            & 50              & 30              \\
125            & 60              & 35              \\
150            & 70              & 45              \\
175            & 85              & 55              \\
200            & 90              & 60              \\
250            & 125             & 80              \\
300            & 150             & 105             \\
350            & 175             & 190             \\
400            & 190             & 120             \\
500            & 245             & 160             \\
800            & 390             & 260             \\
1000           & 500             & 320             \\
\end{tabular}
\end{center} 
\vskip 0.08in 
\caption{The optimal $p^{(1)}_T$ and $p^{(2)}_T$
cuts for isolating a Higgs boson
of mass $m_\phi$ from the QCD $\bbbb$ background.}
\label{ptcutstab}
\end{table}
}

\def\Tblbtagtab
{
\begin{table}[t] 
\begin{center}
\begin{tabular}{c||ccc}
~Process     & 2 or more $b$-tags & 3 
or more $b$-tags & 4 $b$-tags  \\ \hline
$\phibb$     & 1139              & 660           & 180              \\
$\Zbb$       & 293               & 170           & 46               \\
$\bbbb$      & 1054              & 610           & 166              \\
$\bbjj$      & $1.2 \times 10^5$ & 2141          & 4   
             \\ \hline
Significance & 3.3               & 12.21         & 12.25            \\
\end{tabular}
\end{center} 
\vskip 0.08in 
\caption{The signal and background events for 2~\ifb
of Tevatron data, assuming
$m_\phi = 100$\,GeV, $2 \Delta m_\phi = 26$\,GeV, and $K = 40$ 
for two or more,
three or more, or four $b$-tags, 
and the resulting significance of the signal. }
\label{btagtab}
\end{table}
}

\def\Tbleventnumtab
{
\begin{table}[t] 
\begin{center}
\begin{tabular}{c||cccc}
~$m_\phi$~(GeV)& \multicolumn{2}{c}{Tevatron}  & \multicolumn{2}{c}{LHC} \\
               & $N_S$        & $N_B$     & $N_S$       & $N_B$     \\ \hline
 75            & 583          & 640       & 3.4 $\times 10^6$ &
 4.8 $\times 10^6$  \\
 100           & 180          & 216       & 2.0 $\times 10^6$ &
 3.0 $\times 10^6$  \\
 150           & 58           & 92        & 9.2 $\times 10^5$ &
 1.2 $\times 10^6$  \\
 200           & 17           & 31        & 4.2 $\times 10^5$  &
 5.6 $\times 10^5$  \\
 300           & 1.3          & 2.1       & 83000       & 70000     \\
 500           &              &           & 12000       & 5700      \\
 800           &              &           & 1500        & 406       \\
 1000          &              &           & 407         & 70        \\
\end{tabular}
\end{center} 
\vskip 0.08in 
\caption{The signal ($N_S$) and background ($N_B$) event numbers 
for a 2~\ifb of Tevatron data and a 100~\ifb of LHC data, for various
values of $m_\phi$, after applying the cuts described in the
text, and requiring 4 $b$-tags. An enhancement of $K = m_t / m_b
= 40$ is assumed for the signal, though the numbers may be simply
scaled for any $K_{\rm new}$ by multiplying by $(K_{\rm new}/40)^2$.}
\label{eventnumtab}
\end{table}
}

\def\TblHcoupling
{
\begin{table}[t] 
\begin{center}
\begin{tabular}{c ccc}
\hline\hline \\[-0.25cm]
Higgs &  
$A\hspace*{1.7cm}$  &  
$h\hspace*{1.7cm}$  &  
$H\hspace*{1.7cm}$  \\ [0.25cm]
\hline \\[-0.25cm]
$y_{U}/y_{U}^{\rm SM}$ & 
$\cotb\hspace*{1.5cm}$ & 
$\cosa /\sinb \hspace*{1.5cm}$ & 
$\sina /\sinb\hspace*{1.5cm}$ \\ [0.25cm]
$y_{D}/y_{D}^{\rm SM}$ & 
$\tanb\hspace*{1.5cm}$ &  
$-\sina /\cosb \hspace*{1.5cm}$ & 
$\cosa /\cosb\hspace*{1.5cm}$\\[0.25cm]
$~~g_{\phi VV}/y_{\phi VV}^{\rm SM}~~$ & 
$0\hspace*{1.5cm}$ &  
$\sinba\hspace*{1.5cm}$ & 
$\cosba\hspace*{1.5cm}$\\ [0.2cm]
\hline\hline
\end{tabular}
\end{center} 
\vskip 0.08in 
\caption{
Comparison of the neutral MSSM Higgs couplings to up-type ($U=u,c,t$) 
and down-type ($D=d,s,b; e,\mu , \tau$) fermions and 
to the gauge-boson ($V=W,Z$) pairs. The ratios to the 
corresponding SM couplings are shown, which
are determined by angles $\beta$ and $\alpha$ at the tree-level.
}
\label{Hcoupling}
\end{table}
}


\def\Figzbbfeynfig
{
\begin{figure}
\centerline{\hbox{
\psfig{figure=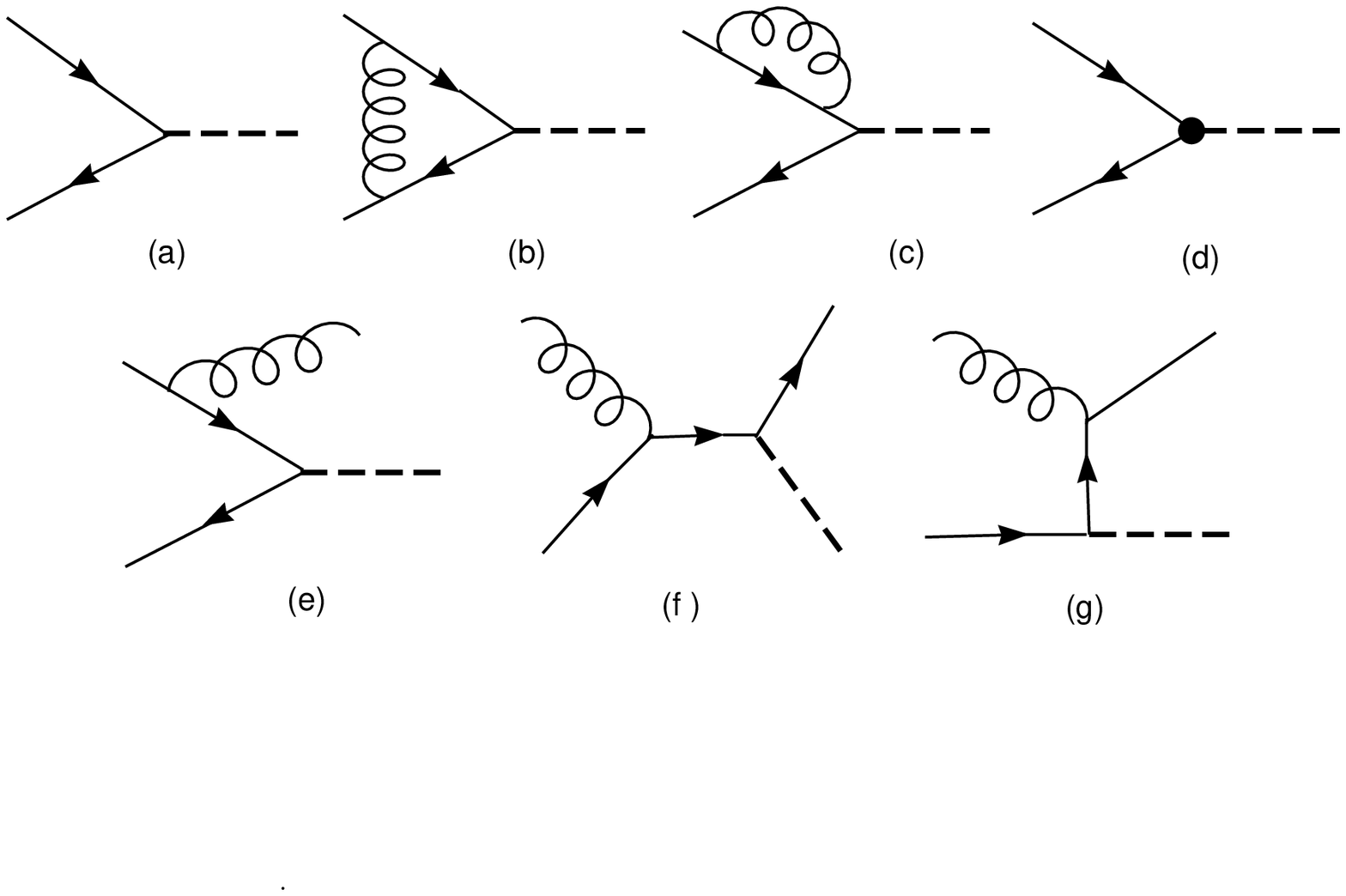,height=2.0in}}}
\vskip 0.1 in
\caption{Representative leading order Feynman diagrams for 
$\phibb$ production
at a hadron collider.  The decay $\phi \to b \bar{b}$ is not shown.}
\label{sigfeynfig}
\end{figure}
\begin{figure}
\centerline{\hbox{
\psfig{figure=\Figdir/fig2.ps,height=2.0in}}}
\vskip 0.1 in
\caption{Representative Feynman diagrams for leading 
order $\Zbb$ production
at a hadron collider.  The decay $Z \to b \bar{b}$ is not shown.}
\label{zbbfeynfig}
\end{figure}
}

\def\Figbbjjfeynfig
{
\begin{figure}
\centerline{\hbox{
\psfig{figure=\Figdir/fig3.ps,height=2.3in}}}
\vskip 0.1 in
\caption{Representative leading order Feynman diagrams for QCD
$\bbbb$ production at a hadron collider.}
\label{bbbbfeynfig}
\end{figure}
\begin{figure}
\centerline{\hbox{
\psfig{figure=\Figdir/fig4.ps,height=4.0in}}}
\vskip 0.1 in
\caption{Representative leading order Feynman diagrams for 
QCD $\bbjj$ production at a hadron collider.}
\label{bbjjfeynfig}
\end{figure}
}

\def\FigYsMs
{
\begin{figure}[t]
\vspace*{-1cm}
\centerline{\hbox{\psfig{figure=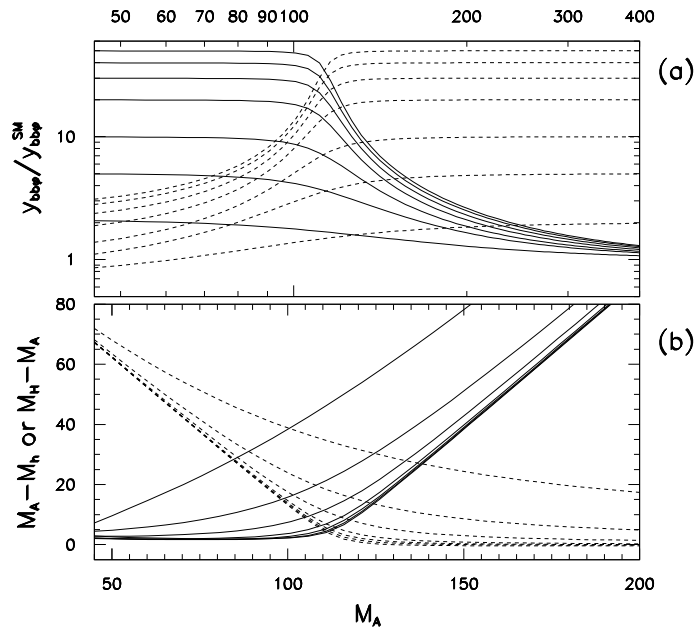,height=10cm,width=10cm}}}
\vspace*{-.5cm}
\caption{
Bottom Yukawa couplings to the MSSM Higgs bosons and the 
mass differences, $m_A - m_h$ and $m_H - m_A$,
as a function of $m_A$ for $\tan \beta$ values:
2.0, 5.0, 10.0, 20.0, 30.0, 40., 50.0. 
In~(a), $y_{bbh}$ is in solid and $y_{bbH}$ is in dashed,
and $\tan \beta$ decreases from top to bottom curves.
In~(b), $m_A-m_h$ is in solid, $m_H-m_A$ is in dashed, and 
$\tan \beta$ increases from the top to bottom curves. 
Here, all the
SUSY soft-breaking mass parameters are chosen to be 500~GeV.
}
\label{fig:YsMs}
\end{figure}
}

\def\FigYsMsTwo
{
\begin{figure*}[t]
\vspace*{-1cm}
\centerline{\hbox{
\psfig{figure=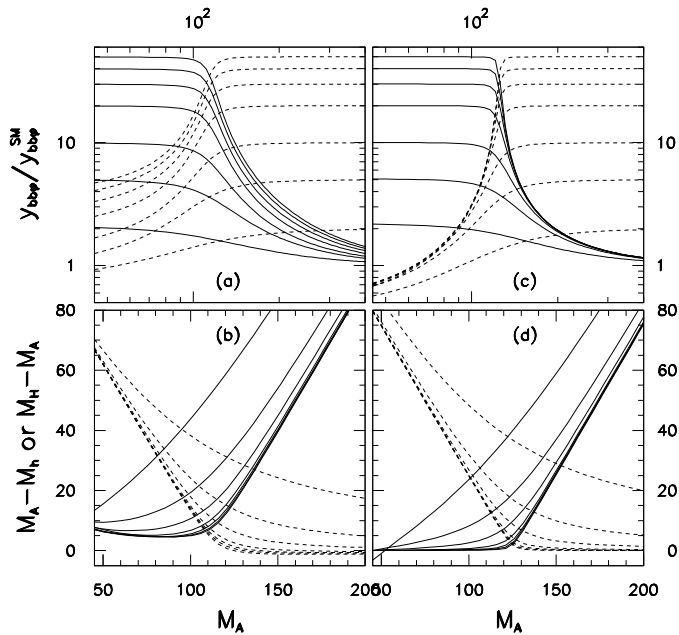,height=10cm,width=10cm}}}
\vspace*{-.5cm}
\caption{
The same as the previous figure, but in (a)-(b), we change  
the right-handed stop mass to 200~GeV, and in (c)-(d), we use the
``LEP~II Scan A2'' set of SUSY parameters.
}
\label{fig:YsMs2}
\end{figure*}
}

\def\FigExclusion
{
\begin{figure*}[p]
\begin{center}
\begin{tabular}{cc} 
\psfig{figure=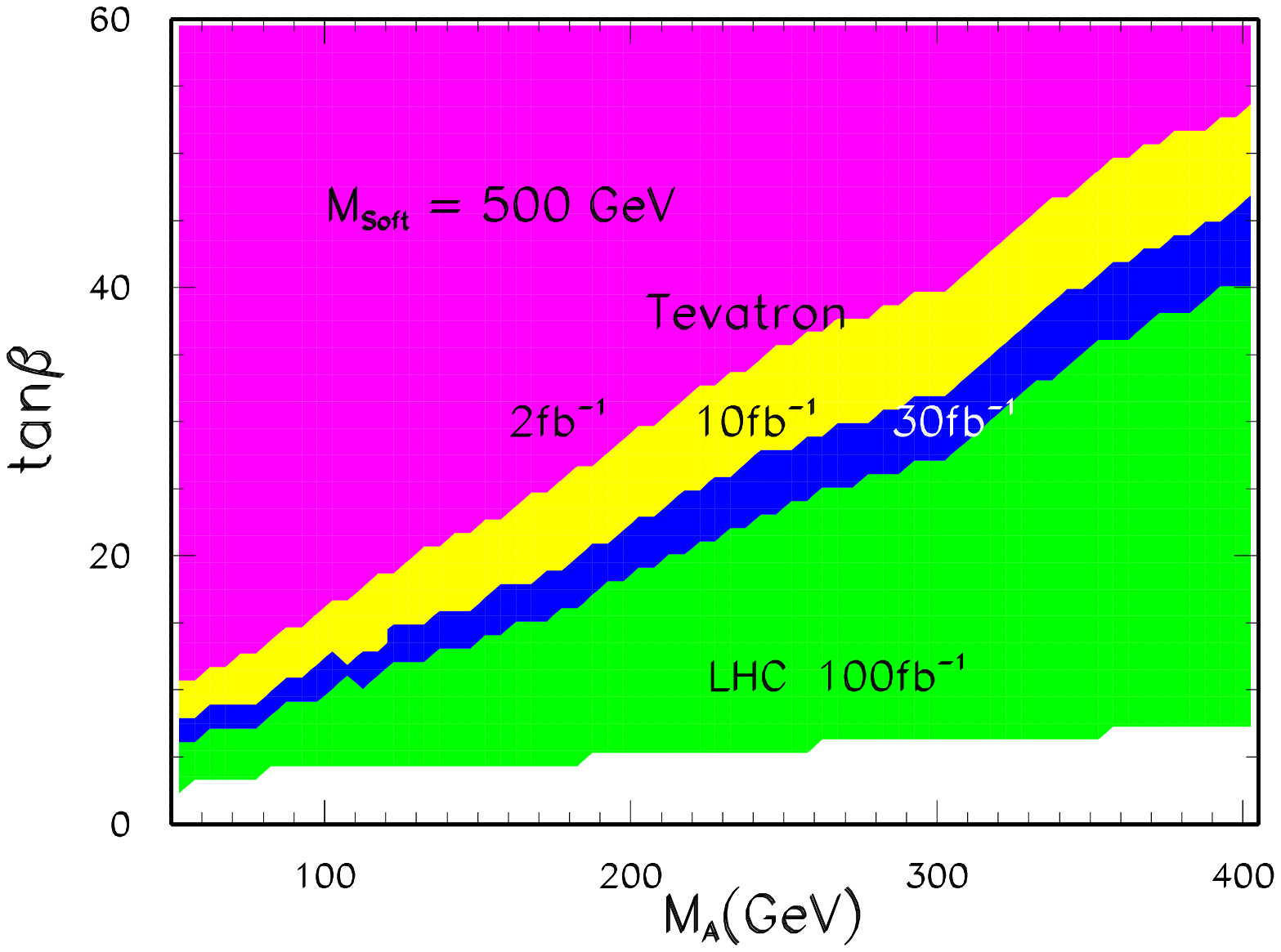,height=3.8in} 
\\[-3.4in]\hskip 4.7in (a)\\[3.1in] \\[-2.0cm] 
\psfig{figure=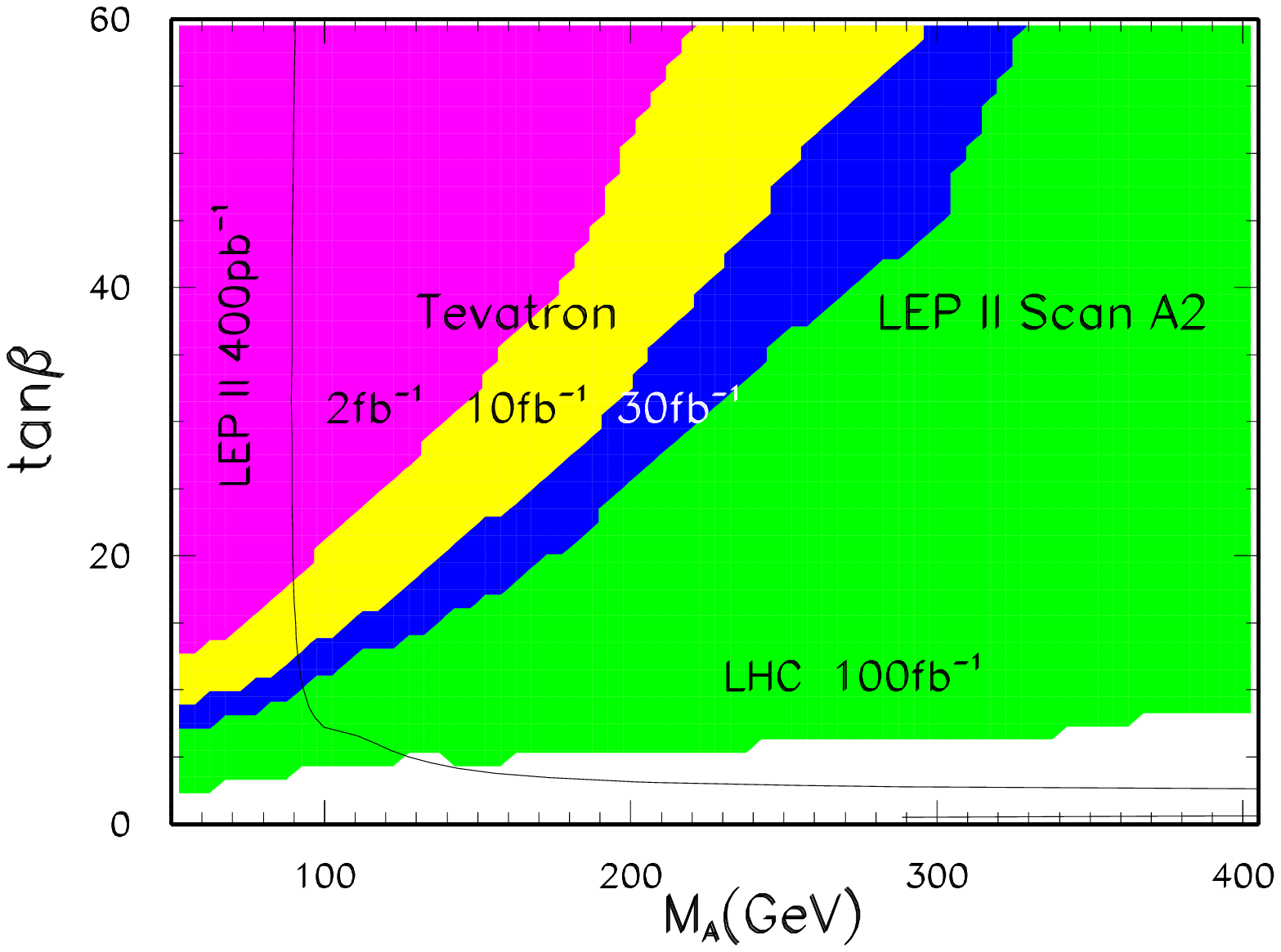,height=3.8in} 
\\[-3.4in]\hskip 4.7in (b)\\[3.1in]
\end{tabular}
\end{center}
\vspace*{-0.5cm}
\caption{
$95\%$~C.L. exclusion contours in the 
$m_A$-$\tan\beta$ plane of the MSSM.
The areas above the four boundaries are excluded for the Tevatron Run~II
with the indicated luminosities,
and for the LHC with an integrated luminosity of 100 fb$^{-1}$.
The soft SUSY breaking parameters were chosen uniformly to be 500 GeV
in Fig.~(a), while the inputs of the ``LEP~II Scan~A2'' are used for
the Fig.~(b) in which LEP~II excludes the left area of the solid curve. 
}
\label{fig:Exclusion}
\end{figure*}
}

\def\FigExclusionLHC
{
\begin{figure*}[t]
\epsfysize=10cm \centerline{\epsfbox{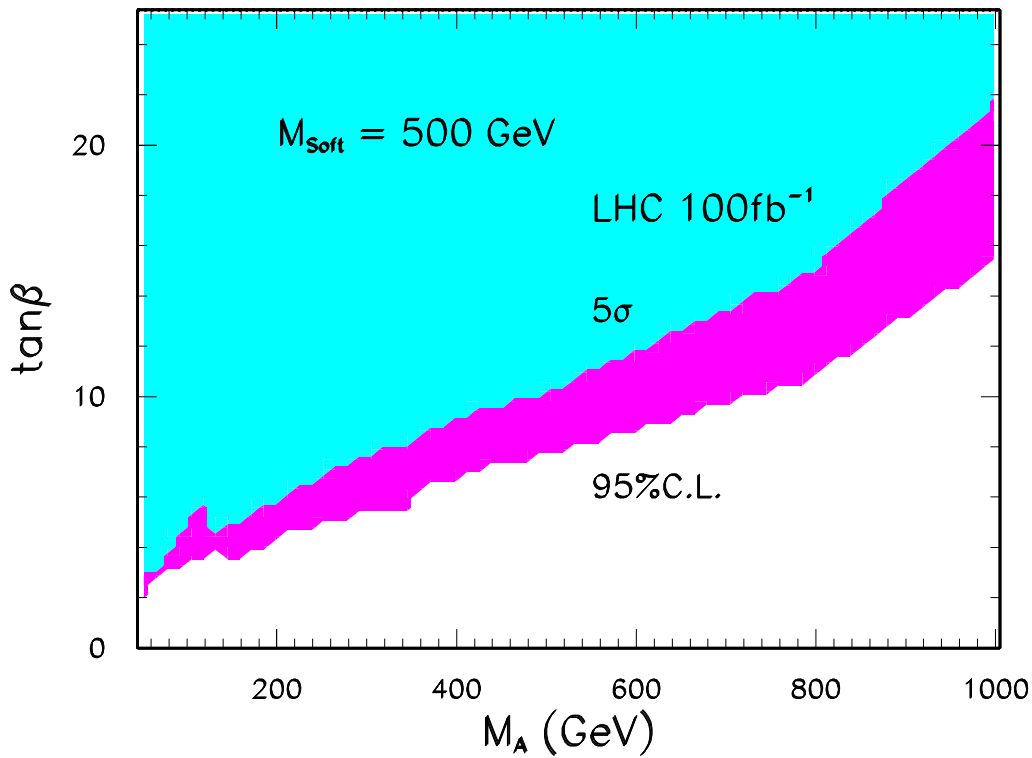}}
\vskip 0.1 in
\caption{
Discovery and exclusion contours in the 
$m_A$-$\tan\beta$ plane of the MSSM 
for the LHC with an integrated luminosity of 100 fb$^{-1}$.
The area above the lower boundary is excluded at $95\%$~C.L., while
the upper boundary is the $5 \sigma$ discovery contour.
The soft SUSY breaking parameters were chosen uniformly to be 500 GeV.
}
\label{fig:ExclusionLHC}
\end{figure*}
}

\def\FigExclusionTwo
{
\begin{figure*}[p]
\begin{center}
\begin{tabular}{cc} 
\psfig{figure=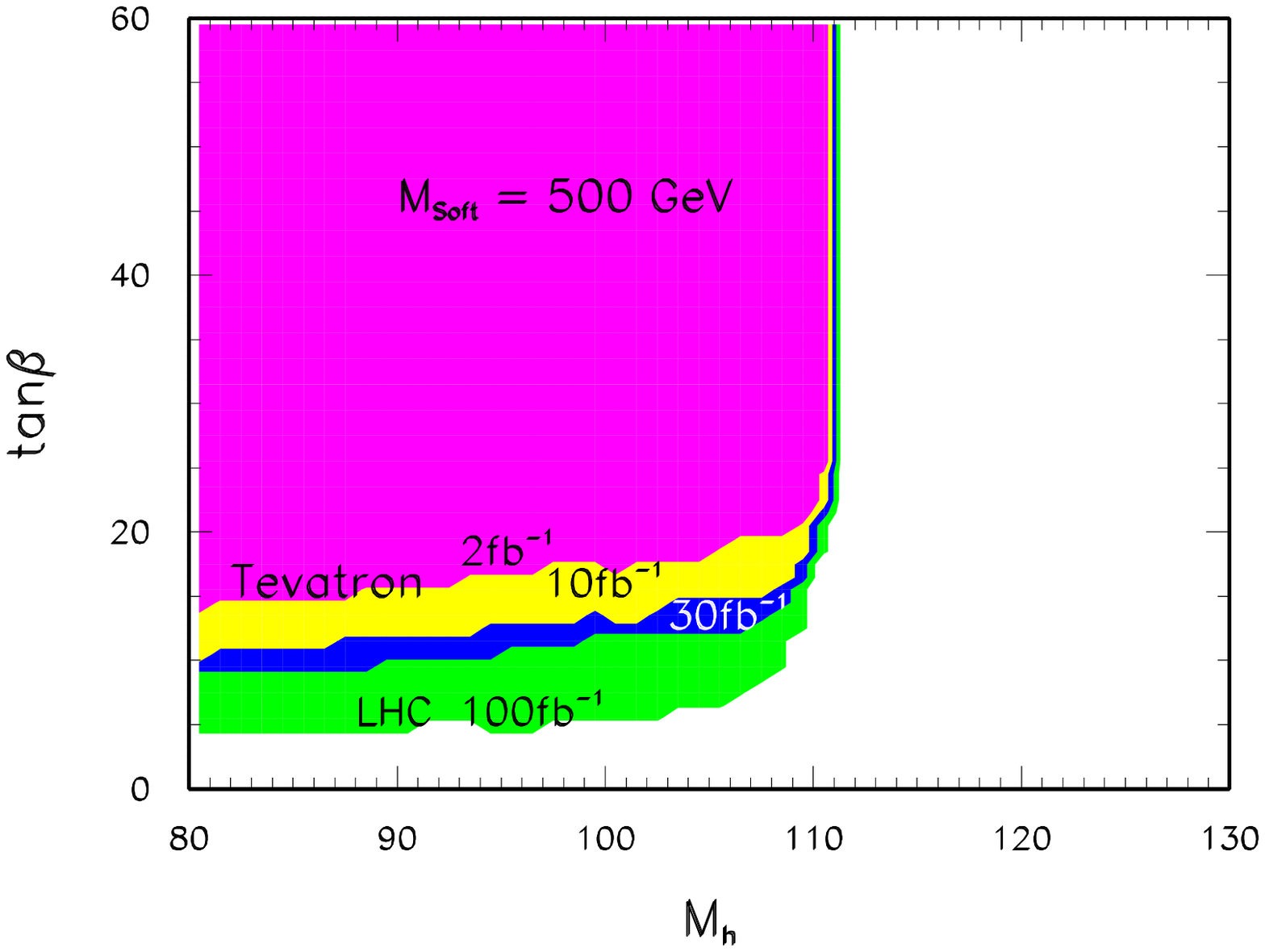,height=3.8in} 
\\[-3.4in]\hskip 4.7in (a)\\[3.1in] \\[-1.6cm] 
\psfig{figure=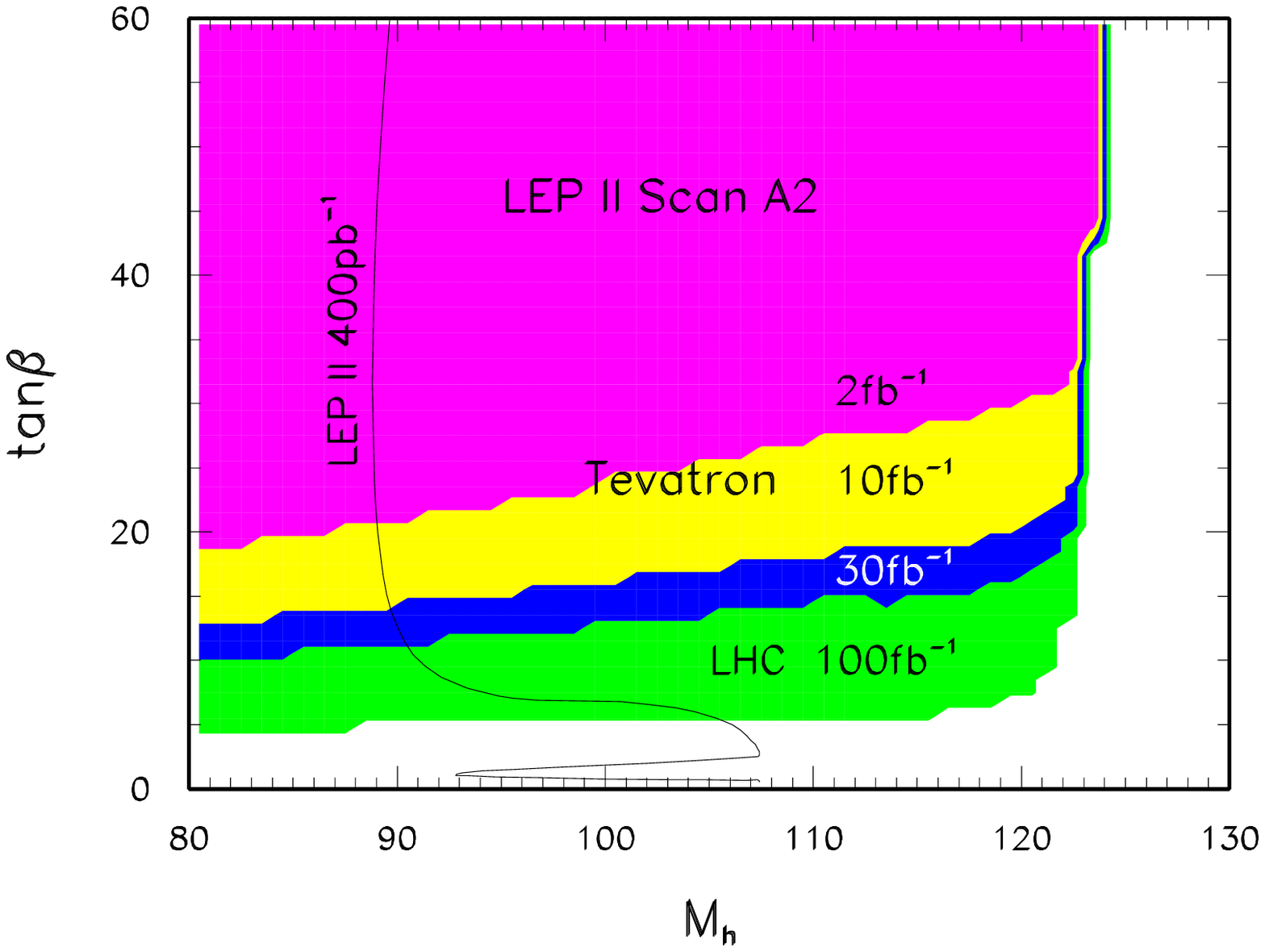,height=3.8in} 
\\[-3.4in]\hskip 4.7in (b)\\[3.1in]
\end{tabular}
\end{center}
\vspace*{-0.5cm}
\caption{
$95\%$~C.L. exclusion contours in the $m_h$-$\tan\beta$ plane of the MSSM.
The areas above the four boundaries are excluded for the Tevatron Run~II
with the indicated luminosities,
and for the LHC with an integrated luminosity of 100 fb$^{-1}$.
LEP~II can exclude the area on the left-hand side of the solid 
curve in the lower plot. 
}
\label{fig:Exclusion2}
\end{figure*}
}

\def\FigExclusionThree
{
\begin{figure*}[p]
\begin{center}
\begin{tabular}{cc} 
\psfig{figure=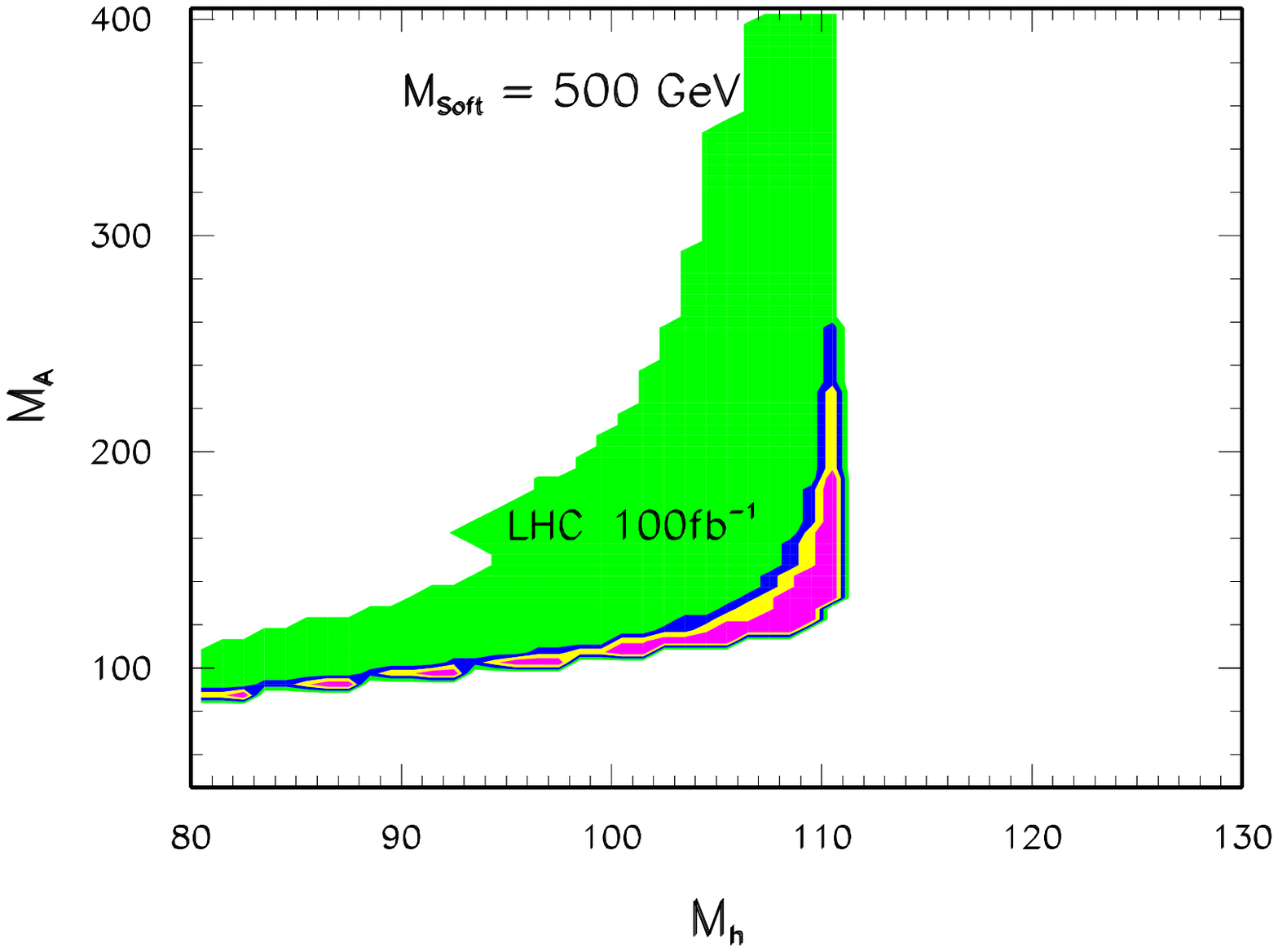,height=3.8in} 
\\[-3.4in]\hskip 4.7in (a)\\[3.1in] \\[-1.6cm] 
\psfig{figure=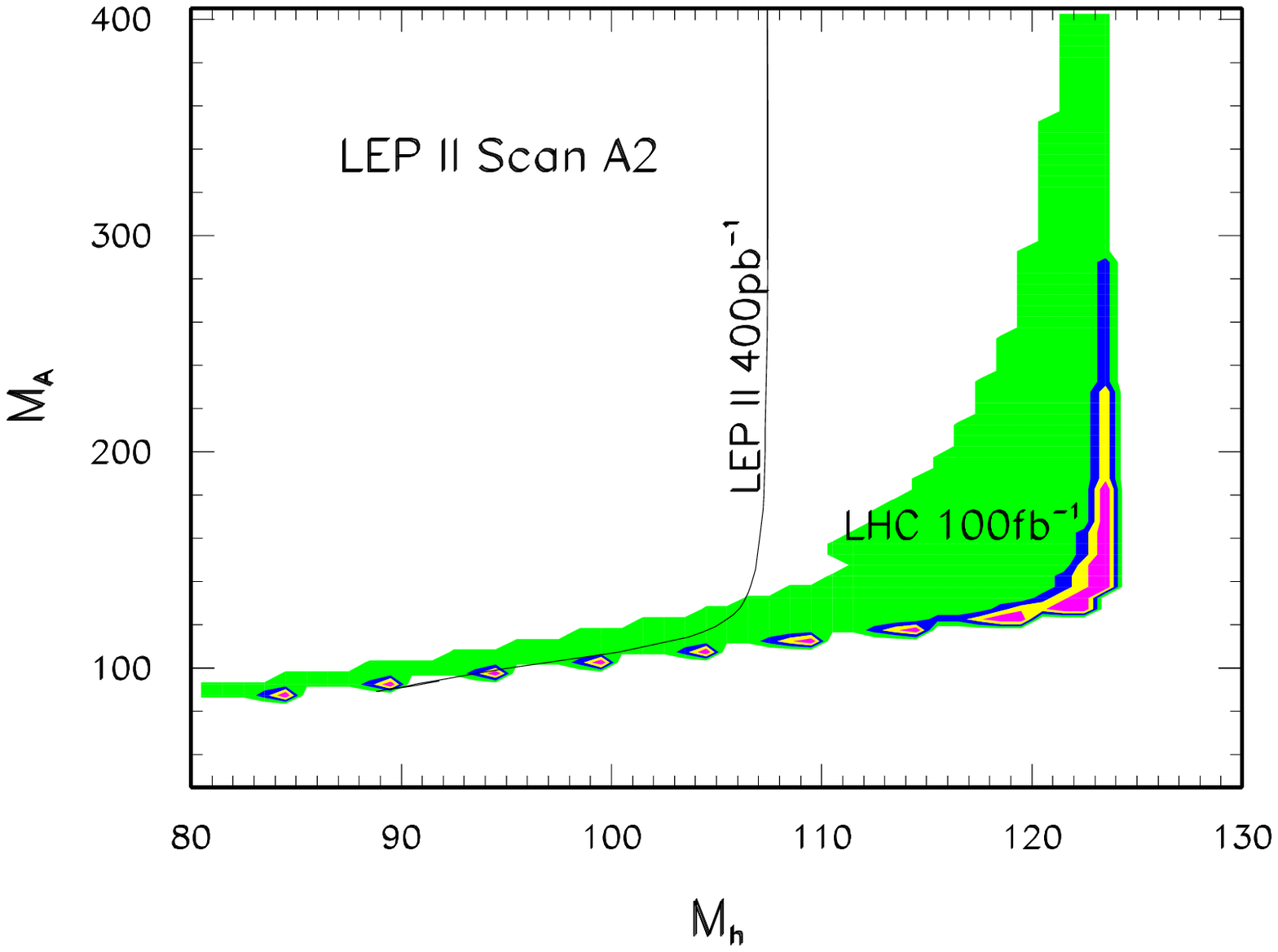,height=3.8in} 
\\[-3.4in]\hskip 4.7in (b)\\[3.1in]
\end{tabular}
\end{center}
\vspace*{-0.5cm}
\caption{
$95\%$ ~C.L. exclusion contours in the $m_A$-$m_h$ plane of the MSSM.
The shaded areas indicate the excluded regions for the Tevatron Run~II
with integrated luminosities, 2, 10, 30 fb$^{-1}$,
and for the LHC with an integrated luminosity of 100 fb$^{-1}$,
as those in the previous figures.
LEP~II can exclude the area on the left-hand side of the
solid curve in the lower plot. 
}
\label{fig:Exclusion3}
\end{figure*}
}

\def\FigmbRun2L
{
\begin{figure*}[t]
\vspace*{-1cm}
\epsfysize=10cm \centerline{\epsfbox{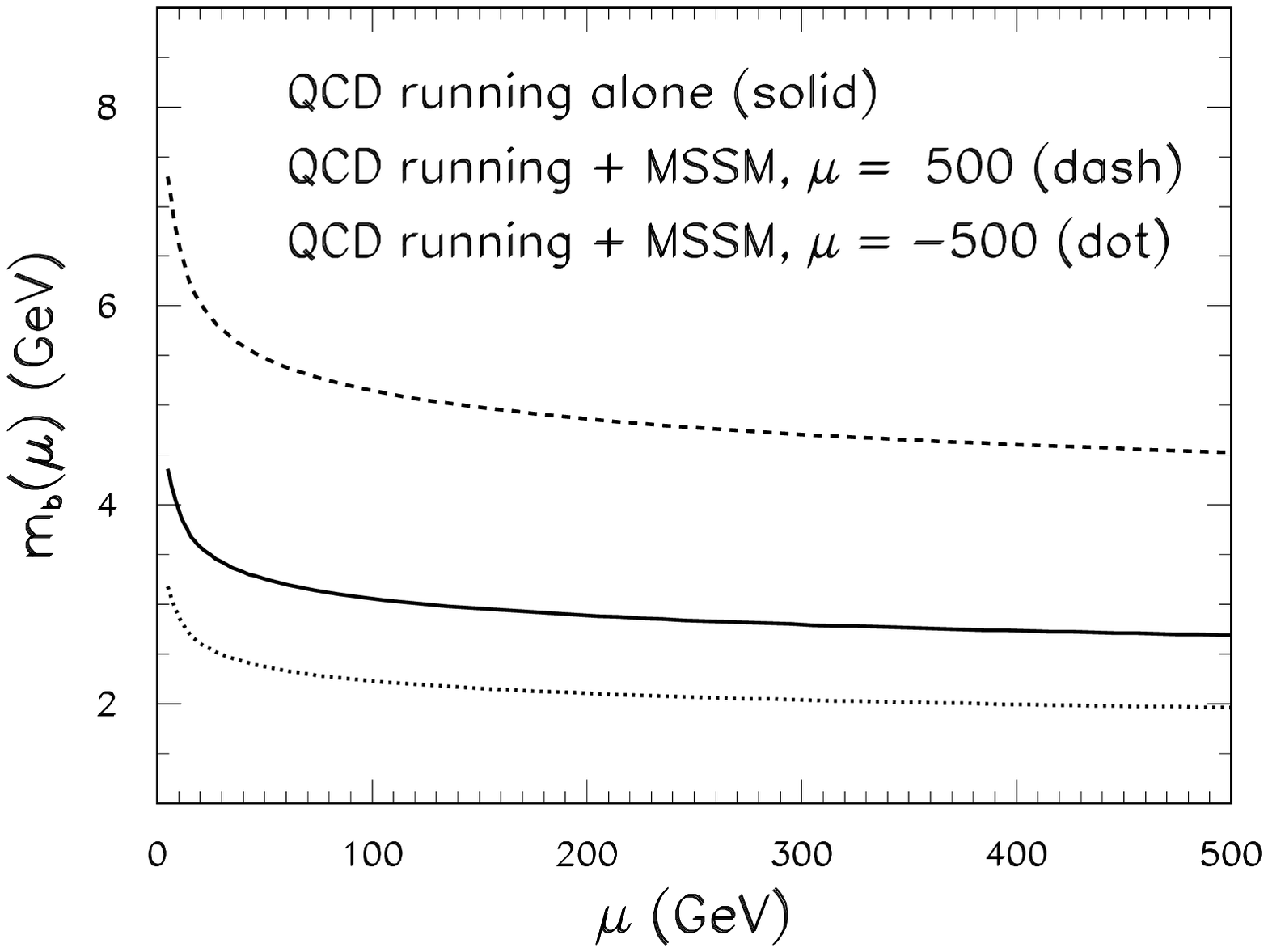}}
\vskip 0.1 in
\vspace*{-.75cm}
\caption{
The running of the bottom quark mass $m_b(\mu_R)$ 
as a function of the renormalization scale $\mu_R$.  
The solid curve shows the QCD evolution alone. 
The dashed curve further includes the supersymmetric corrections to the 
``effective'' running mass, for $\tan \beta = 30$. 
All soft SUSY breaking parameters have been fixed as $500$~GeV.
The dotted curve includes the SUSY corrections but with the sign
of the Higgs-mixing parameter $\mu$ flipped.
}
\label{fig:mbRun2L}
\end{figure*}
}



\chapter{Probing Higgs Bosons with Large Bottom Yukawa Coupling
at Hadron Colliders}


\section{Introduction}

A major task for all future high energy colliders is to
determine the electroweak symmetry breaking (EWSB) mechanism 
for generating the $(W^\pm ,Z^0)$ masses and the mechanism for the 
fermion mass generation \cite{hhunt}. 
Whether the two mechanisms are correlated or not is an interesting 
and yet to be determined issue. Given the large top quark mass 
($m_t =175.6\pm 5.5$~GeV \cite{topmass}), as high as the EWSB scale,
it has been speculated that the top quark may play a special role
for the EWSB.  One of such ideas is that
some new strong dynamics may involve a composite Higgs
sector to generate the EWSB and
to provide a dynamical origin for the top quark 
mass generation ({\it e.g.,} 
the top-condensate/top-color models \cite{TopCrev}).
Another idea is realized in the supersymmetric
theories in which the EWSB is driven radiatively by the large top quark
Yukawa coupling \cite{SUSY-rad}.

In the minimal standard model (SM) there is only 
one Higgs doublet, which 
leaves a physical neutral scalar boson 
as the remnant of the spontaneous EWSB.
The Yukawa couplings of the SM Higgs are determined from 
the relevant SM fermion masses divided by the vacuum expectation
value (vev), $v\simeq 246$~GeV. 
Thus, aside from the coupling to the heavy top 
quark, all the other SM Yukawa couplings are highly suppressed,
independent of the Higgs boson mass.
For the top-condensate/topcolor type of models \cite{TopCrev}, with a
composite Higgs sector, the new strong dynamics associated with 
the top sector
plays a crucial role for generating the large top mass and (possibly) 
the $W,Z$ boson
masses. As to be discussed below, in this scenario,
the interactions of the top-bottom sector with the 
composite Higgs bosons are different from that in the SM.
Due to the infrared quasi-fixed-point structure \cite{IRQFP}
and the particular boundary conditions at the compositeness scale, the
bottom Yukawa coupling to the relevant Higgs scalar is naturally 
as large as that of the top in such models. 
In the minimal supersymmetric extension of the SM (MSSM) 
\cite{mssm}, there are two Higgs doublets, whose mass
spectrum includes two neutral scalars ($h$ and $H$), 
a pseudoscalar ($A$) and a pair of charged scalars $(H^\pm)$.
The MSSM Higgs sector contains two free parameters 
which are traditionally chosen
as the ratio of the two Higgs vev's ($\tan\beta = v_u/v_d$) 
and the pseudoscalar mass ($m_A$).
A distinct feature of the MSSM is that in the large $\tan\beta$ region
the Higgs Yukawa couplings to all the down-type fermions are 
enhanced by either $\tan\beta$ or $1/\cos\beta$.  
Among the down-type fermions, the bottom quark has the largest mass and so
the largest Yukawa coupling. Thus, it represents a likely place where new 
physics could reveal itself experimentally.
This common feature, the large bottom Yukawa coupling relative
to that of the SM, present in both types of the (conceptually 
quite distinct) models discussed above, serves as the theoretical
motivation for our analysis.

Since the third family $b$ quark, as the weak isospin
partner of the top, can  have large 
Yukawa coupling with the Higgs 
scalar(s) in both composite and supersymmetric models, 
we recently proposed \cite{bbbb_prl}
to use the $b$ quark as a probe of possible non-standard
dynamics in Higgs and top sectors.
In fact, because of the light $b$ mass ($\simeq 5$~GeV)
relative to that of the top ($\simeq 175$~GeV),
the production of Higgs boson associated with $b$ quarks
($\ppbar /pp \to \phibb \to \bbbb$) may be experimentally
accessible at the Fermilab Tevatron\footnote{
A $p \bar{p}$ collider with $\sqrt{s} = 2$\, TeV.} 
or the CERN Large Hadron Collider (LHC)\footnote{
A $p p$ collider with $\sqrt{s} = 14$\, TeV.}, even though the large
top mass could render associated Higgs production
with top quarks ($\ppbar /pp \to \phi t \bar{t}$) infeasible.
As we will show, this makes it possible for the Run~II of the Tevatron
and the LHC to confirm the various models in which the $b$-quark
Yukawa coupling is naturally enhanced relative to the SM prediction.
However, if the $\phi b\bar{b}$ signal is not found,
the Tevatron and the LHC can impose stringent constraints 
on the models with either a composite or a supersymmetric Higgs 
sector, in which the Yukawa coupling of the Higgs boson(s) and bottom
quark is expected to be large.
In \cite{dai}, this reaction was explored at
the LHC and the Tevatron\footnote{
In \cite{dai}, a $1.8$~TeV $p\bar{p}$ Tevatron was assumed with an
integrated luminosity of 30~fb$^{-1}$, and
the squark mixings of the MSSM were neglected.} 
to probe the supersymmetry (SUSY) parameters of the MSSM.  
The conclusion
was that, with efficient $b$-tagging, useful information concerning
the MSSM could be extracted from either the LHC or a high luminosity
Tevatron.  In this work, we expand upon earlier results
\cite{bbbb_prl} in which it was concluded that even the
much lower integrated luminosity of the Tevatron Run~II can provide 
useful information through this reaction, provided an optimized search
strategy is employed.  
This analysis is based on the model-independent results in 
Ref.~\cite{Balazs-Cruz-he-Tait-Yuan}, where the relevant backgrounds 
were calculated and an effective
search strategy to extract a signal from the backgrounds were presented.

In what follows, 
after deriving the exclusion contours on the $m_A$-$\tanb$
plane of the MSSM, we analyze its implication on the 
supergravity\cite{susyrev} and gauge-mediated\cite{GMSB} models
of soft SUSY breaking that naturally predict a large $\tan\beta$.
The comparison of our Tevatron Run~II results with the LEP~II bounds 
(from the $Zh$ and $hA$ channels) is also presented,
illustrating the complementarity of our analysis to other 
existing Higgs search strategies. 

\section{Constraints On MSSM Parameters and Implications for 
Models of Soft-breaking of SUSY}

Supersymmetry (SUSY) is one of the most natural extensions of the SM, 
mainly because of its ability to solve the hierarchy problem,
as well as for its capacity to imitate the current experimental 
success of the SM, despite the plethora of
the introduced new particles and free parameters \cite{susyrev}. 
The Minimal Supersymmetric SM (MSSM) \cite{mssm}
requires a two Higgs doublet extension of the SM
\cite{hhunt} together with the corresponding supersymmetric partners.  
The model includes all renormalizable interactions that respect  
the standard gauge group $SU(3)_C\otimes SU(2)_L\otimes U(1)_Y$ and
supersymmetry. In order to prevent potentially dangerous
baryon and lepton number violating interactions, invariance under
a discrete $R$-parity\footnote{The $R$-parity is
defined in such a way that SM particles are
even under $R$ and their superpartners are odd.} is also required. 
To be compatible with data, supersymmetry has to be broken.
The breaking of SUSY is parametrized by the general set of
soft-breaking terms, which, in principle, should be deduced from a
specific underlying model for SUSY breaking, such as the
supergravity \cite{susyrev} and gauge-mediated \cite{GMSB} models.
In this section, we discuss the potential of
the Tevatron and the LHC to test MSSM via measuring the production 
rate of $\phibb$ mode. We shall also discuss the implication of
this result on various supergravity and gauge-mediated 
models of soft SUSY-breaking.

\subsection{Bottom Yukawa Couplings and the MSSM Higgs Sector} 

\TblHcoupling
In the MSSM, the Higgs couplings to the SM fermions and gauge bosons
involve two new free parameters at the 
tree-level, which are the vacuum angle 
$\beta (\equiv \arctan v_u/v_d)$ and the Higgs mixing angle $\alpha$.
These couplings are shown in Table~\ref{Hcoupling}. 
We see that the MSSM Higgs
couplings to the gauge boson pairs 
are always suppressed relative to that of 
the SM, while their couplings to 
the down(up)-type fermions are enhanced in
the large (small) $\tanb$ region. These enhanced Yukawa couplings are of
great phenomenological importance for the Higgs 
detection and especially for probing
the associated new dynamics in the top-bottom sector.  
Alternatively, we can choose $\tanb$ and the pseudoscalar mass $m_A$ 
as two free parameters. 
Then, at the one-loop level, $\alpha$ can be calculated from
\begin{equation}
\dis\tan 2\alpha 
   =\tan 2\beta\left(m_A^2+m_Z^2\right)
               \left[m_A^2-m_Z^2+\f{\epsilon_t}{\cos 2\beta}\right]^{-1}~,
\label{eq:MSSM-alpha}
\end{equation}
with $~\alpha\in \left(-\pi /2, 0\right)$.~ Here the parameter $\epsilon_t$
represents the dominant top and stop loop corrections which depend on
the fourth power of the top mass $m_t$ and the logarithm of the 
stop mass $M^2_{\tilde t}$:  
\begin{equation}
\dis\epsilon_t= \frac{3 G_F m^4_t}{\sqrt{2}\pi^2 \sin^2 \beta}
\log\left(\f{M^2_{\tilde t}}{m^2_t}+1\right)~.
\label{eq:MSSM-alpha1}
\end{equation}
Note that for large $\tanb$, the bottom and sbottom loop corrections
can also be important. Hence, in our numerical analysis below 
we have included the complete
radiative corrections with full mixing in the stop and sbottom sectors,
and the renormalization group improvements are also adopted.\footnote{ 
We have used the HDECAY program \cite{hdecay} to compute the Higgs
masses, couplings and decay branching ratios.}
As shown in Table~\ref{Hcoupling}, the $A$-$b$-$\bar b$
coupling has no explicit $\alpha$ dependence.
The bottom Yukawa couplings $y_{bbh}$ and $y_{bbH}$ are 
$\alpha$- and $\beta$-dependent, their magnitudes relative to the SM 
prediction are displayed in Fig.~\ref{fig:YsMs}a 
as a function of $m_A$ for various $\tanb$ values. It shows that for
$m_A$ above $\sim 110$~GeV, the $h$-$b$-$\bar b$
coupling quickly decreases, approaching to the
SM value for all $\tanb$, while the $H$-$b$-$\bar b$
coupling increases for large $\tanb$. 
Therefore, we expect that in the large $\tanb$ region, the production
rate of $Ab\bar b$ or $hb\bar b$ can be large for small $m_A$, while
the rate of $Ab\bar b$ or $Hb\bar b$ are enhanced for large $m_A$.   
Whether the signals of the two Higgs scalars 
($A$ and $h$ or $A$ and $H$) can be experimentally resolved as 
two separate signals 
({\it e.g.,} two bumps in the $b \bar b$ invariant mass distribution)
depends on their mass degeneracy.
\FigYsMs 

The MSSM Higgs boson mass spectrum
 can be determined by taking the second derivative on
Higgs effective potential with respect to the Higgs fields. 
At tree-level, the resulting Higgs masses obey the relations:
$m_h \leq m_Z \cos 2\beta$, $m_Z \leq m_H$, 
$m_h\leq m_A \leq m_H$, and $m_W \leq m_{H^\pm}$.
However, these relations are substantially modified 
by radiative corrections \cite{himrc}. 
Including the important contributions from top and stop
loops, the masses of $h$ and $H$ can be written as: 
\begin{eqnarray}
&& m^2_{h,H} = 
 \frac{1}{2} \left\{ \left(M^2+\epsilon_t\right) \mp 
\right. \nonumber \\ && ~~~ \left. 
\left[ \left(M^2+\epsilon_t\right)^2
-4\epsilon_t\left(m_Z^2\cos^2\beta +m_A^2\sin^2\beta\right)
-4m_Z^2m_A^2\cos^22\beta 
\right]^{1\over 2} \right\},
\end{eqnarray}
where $M^2\equiv m_Z^2+m_A^2$ and the parameter $\epsilon_t$ is defined in 
(\ref{eq:MSSM-alpha1}).
To analyze the Higgs mass degeneracies, we plot the mass differences
$m_A$-$m_h$ and $m_H$-$m_A$ in Fig.~\ref{fig:YsMs}b 
using the complete radiative corrections to the Higgs mass spectrum 
\cite{hdecay}.  We see that for the large $\tanb$ values, 
the pseudoscalar $A$ is about degenerate in mass
with the lighter neutral scalar $h$ below $\sim 120$~GeV and with the 
heavier neutral $H$ above $\sim 120$~GeV. This degeneracy indicates that
the $\phibb$ signal from the MSSM generally contains two
mass-degenerate scalars with similar couplings, and thus
results in a stronger bound on $\tanb$ by about a factor of $\sqrt{2}$.

\FigYsMsTwo
Finally, we note that in Fig.~\ref{fig:YsMs}, all 
soft-breaking mass parameters were chosen to be 500~GeV. 
Various choices of SUSY soft-breaking parameters 
typically affect these quantities
by about 10\%-30$\%$. To illustrate these effects, we plot in 
Fig.~\ref{fig:YsMs2} the same quantities, but
changing the right-handed stop mass to
$M_{\tilde t}=200$~GeV in (a-b), and in (c-d) we use the
``LEP2~II Scan~A2'' set of SUSY parameters\footnote{
The parameters $m_0$ and $M_2$ are fixed at $1$~TeV, $\mu$ is
chosen to be $-100$~GeV and $m_t=175$~GeV. The scalar trilinear
coupling $A_i$ is fixed at $\sqrt{6}$~TeV, corresponding to
the maximal left- and right-handed top-squark mixing. Detailed 
prescription about this set of parameter scan can be found in 
Ref.~\cite{madison-LEP2,FNAL-LEP2}.}
for comparison. 

Because the MSSM predicts a large bottom quark Yukawa coupling 
for large $\tanb$, and the mass of the lightest neutral scalar has
to be less than $\sim 130$\,GeV, we expect that
the Tevatron and the LHC can test this model via measuring 
the $\phibb$ production rate.
In the following, we shall discuss the range of the
$m_A$-$\tanb$ plane that can be explored at various colliders.
Some models of SUSY soft-breaking predict a large $\tanb$ with
light Higgs scalar(s), and thus predict a large 
$\phibb$ rate. Without observing such a signal, one can impose a 
stringent constrain on the model. 
This will also be discussed below.

\subsection{Constraints	on MSSM from $\phibb$
production at Tevatron and LHC}

To use the model-independent result of $K_{min}$ obtained in 
Ref.~\cite{Balazs-Cruz-he-Tait-Yuan} 
to constrain the  $m_A$-$\tanb$ plane in the MSSM, 
one needs to calculate the SUSY Higgs boson masses, 
decay branching ratios, and their couplings to
the bottom quark for a given set of the soft breaking parameters.  
In the following numerical analysis, we use the HDECAY code to include 
the full mixings in the stop/sbottom sector with QCD and electroweak 
radiative corrections \cite{hdecay}.
For simplicity, we assume that the superpartners are all heavy enough 
so that the decays of the Higgs bosons into them are forbidden.
Under this assumption, we find that the 
decay branching ratio of $h \to b
\bar{b}$ is close to one for the relevant region of the parameter space.

As explained above, we combine signals from more than one scalar boson
provided their masses differ by less than $\Delta m_{exp}$, which
is the maximum of the experimental mass resolution (cf. footnote-7)
and the natural decay width of the scalar boson.
Since the results of Ref.~\cite{Balazs-Cruz-he-Tait-Yuan} 
are given in terms of the
minimal enhancement factor $K_{\min}$ defined in Eq.(1) of that reference,
we need to convert them into exclusion bounds in the $m_A$-$\tanb$
plane of the MSSM, in case that a signal is not found.
The bound on $\tan\beta_{\min}$
(with the possible mass degeneracy included)
can be derived from that on $K_{\min}$ (for a single scalar) by 
requiring
\begin{equation}
\tan^2\hspace*{-0.15cm}\beta~ {\rm BR}(A\to b\bbar) 
+\hspace*{-0.15cm}\sum_{\phi =h,H}\hspace*{-0.2cm}
\dis\theta \left(\Delta m_{\rm exp} -|\Delta M_{A\phi}|\right)
\dis\left(\f{y_b^\phi}{y_b^{\rm SM}}\right)^2{\rm BR}(\phi\to b\bbar)
\geq K_{\min}^2 ~,
\label{eq:MSSM-condition}
\end{equation}
where  $y_b^{\rm SM}$ and $y_b^\phi$ denote the $b$ 
quark Yukawa coupling 
in the SM and the MSSM (with $\phi =h$ or $H$),  respectively.
Inside the argument of the $\theta$-function,  
$\Delta M_{A\phi}$ is the mass difference
between $A$ and $\phi$.
Thus, the equality sign in the above relation 
determines the minimal value
$\tan\beta_{\min}$ for each given $K_{\min}$. 

\FigExclusion
To estimate the exclusion regions in the $m_A$-$\tanb$ plane, 
a set of soft breaking parameters has to be chosen, which 
should be compatible with the current data from the 
LEP~II and the Tevatron experiments, while not much larger than 1 TeV.
For simplicity, we choose 
all the soft SUSY breaking parameters 
(and the Higgs mixing parameter-$\mu$)
to be 500 GeV as our ``default''
values, i.e., $M_{\rm soft}=500$~GeV.
In Fig.~\ref{fig:Exclusion}a, we show the 
$95\%$~C.L. exclusion contours in the $m_A$-$\tan\beta$ plane 
derived from the measurement of 
$p\bar{p}/pp \to \phibb \to \bbbb$, using this ``default'' set
of SUSY parameters.
The areas above the four boundaries are 
excluded for the Tevatron Run~II 
with the indicated luminosities, 
and for the LHC with an integrated luminosity of 100 fb$^{-1}$.
Needless to say, different choice of SUSY parameters, such as the 
mass and the mixing of the top squarks and the value (and sign) of the
parameter $\mu$, would modify this result. 
To compare the potential
of the Tevatron and the LHC in constraining 
the MSSM parameters via
the $\phibb$ ($\phi =h,A,H$) production 
to that of the LEP~II experiments
via $Z\phi$ and $hA$ production, we consider one of the ``benchmark''
parameter scans discussed in 
\cite{madison-LEP2,FNAL-LEP2}, which is called the
``LEP II Scan~A2''$^{12}$ set. 
For this set of SUSY parameters, the 
LEP~II exclusion contour ~\cite{madison-LEP2,FNAL-LEP2} 
at the 95\% C.L. is displayed in 
Fig.~\ref{fig:Exclusion}b,
for a center-of-mass energy of 
$200$~GeV and an integrated luminosity of
$100$~pb$^{-1}$ per LEP~II experiment.
As shown in Fig.~\ref{fig:Exclusion}b, 
the Tevatron Run~II result, in comparison with the LEP~II result,
can already cover a substantial region
of the parameter space with only a 2 fb$^{-1}$ luminosity.
Thus, detecting the $\phibb$ signal at hadron colliders
can effectively probe the MSSM Higgs sector,
especially for models with large $\tan \beta$ values.
Furthermore, for $m_A \gae 100$ GeV, 
Tevatron Run~II is complementary with LEP~II,
because the latter is not sensitive to that region of parameter space.
The LHC can further probe the MSSM down to $\tanb \sim$ 7~(15) 
for  $m_A < 400~(1000)$ GeV.
This is also shown in Fig.~\ref{fig:ExclusionLHC}
using the ``default'' set of SUSY parameters,
in which 
the region above the upper curve is the discovery contour
at the $5 \sigma$ level
for the LHC with an integrated luminosity of 100 fb$^{-1}$,
and the area above the lower curve can be excluded at $95\%$~C.L.,
if a signal is not found. 
\FigExclusionLHC

\FigExclusionTwo
For completeness, we also present the exclusion contours 
in the $m_h$-$\tanb$ and $m_A$-$\tanb$
planes for both the ``default" and the ``LEP II Scan A2'' sets
of SUSY parameters. 
They are shown in Figs.~\ref{fig:Exclusion2} and 
\ref{fig:Exclusion3}.
Again, we see that the Tevatron Run II and the LHC bounds sensitively
and complementarily cover the MSSM parameter space in contrast with
the LEP II results.
We have also studied the bounds with the ``LEP II Scan A1''  inputs
\cite{madison-LEP2,FNAL-LEP2} and found that the exclusion 
contours from the $\phibb$ production are similar to those with 
``LEP II Scan A2'' inputs.
The most noticeable difference is that the theoretically allowed 
range for $m_h$ becomes smaller by about 10\,GeV in the ``Scan A1'' set,
as compared to the ``Scan A2'' inputs.
\FigExclusionThree

Even though our analyses, described above and in 
Ref.~\cite{Balazs-Cruz-he-Tait-Yuan}
are quite different from that of Ref.~\cite{dai}, the final bounds 
at the LHC happen to be in qualitative agreement. 
We also note that our bounds on the $m_A$-$\tanb$ plane
improve considerably the one obtained in Ref. \cite{dressetal},
in which  
the $p{\bar p}\to \phibb \to \tau^+ \tau^- b \bar b$ production
rate at the Tevatron Run~I data
was compared to the MSSM prediction.
Though, we
do not choose to explicitly present projected results for the Tevatron
Run~I data, we encourage our experimental colleagues to pursue this 
analysis
on the existing Run~I data sample, as it seems likely that one could
obtain useful information even with the lower luminosity and collider 
energy of Run~I as well as a somewhat lower $b$-tagging efficiency.

\FigmbRun2L
Before concluding this subsection,
we remark upon the effects on our bounds from the
possible radiative corrections to the
$\phibb$ production process.\footnote{This point has also been recently
discussed in Ref.~\cite{stev}.}  
As mentioned in Sec.~II of Ref.~\cite{Balazs-Cruz-he-Tait-Yuan}, 
one of the dominant correction is from the next-to-leading
order (NLO) QCD loops, which are not currently available
for the $\phibb$ signal and background cross sections. 
However, aside from the QCD corrections to the
$\phibb$ vertex (part of that can be simply 
included into the running of the
$\phibb$ Yukawa coupling or the running $b$-mass), there are 
pentagon loops formed by the virtual gluons 
radiated from an initial state quark (gluon) and re-absorbed by the 
final state $b$ quark with the $\phibb$ vertex included in the loop.
Such QCD corrections are not factorizable so 
that a consistent improvement
of our results is impossible before a full NLO QCD analysis is 
completed\footnote{Such a full NLO QCD calculation is beyond the scope
of our current study. A systematic calculation for this is in 
progress \cite{spira}.}. 
Putting aside the complexity of the full NLO QCD
corrections, we briefly comment upon how the
radiative corrections to the running $\phibb$ Yukawa coupling
affect our final bounds.
At the one-loop level, the
relation between the pole quark mass $m_q^{\rm pol}$
and the $\overline{\rm MS}$ QCD running mass at the scale 
$\mu =m_q^{\rm pol}$ is:
\be
m_q(m_q^{\rm pol}) = 
m_q^{\rm pol}\left[1+
\dis\f{4\alpha_s(m_q^{\rm pol})}{3\pi}\right]^{-1}~.
\ee
When running upward to any scale $\mu$,
\be
m_q(\mu )=m_q(m_q^{\rm pol})\dis\f{c\left[\alpha_s(\mu )/\pi\right]}
                        {c\left[\alpha_s(m_q^{\rm pol})/\pi\right]}~,
\label{eq:running}
\ee
where (cf. Ref.~\cite{hdecay})
\begin{eqnarray}
&& c[x] = (23x/6)^{12/23}\left[1+1.175x\right] ~~ {\rm for}~~ 
m_b^{\rm pol}<\mu <m_t^{\rm pol}, 
\nonumber \\
&& c(x)=(7x/2)^{4/7}\left[1+1.398x\right] ~~ {\rm for} ~~ 
\mu >m_t^{\rm pol}. \nonumber
\end{eqnarray}
Numerically,  $m_t(m_t^{\rm pol})\simeq 166$\,GeV and 
$m_b(m_t^{\rm pol})\simeq 3$\,GeV.
This QCD correction, Eq.~ (\ref{eq:running}),
alone will reduce the running mass $m_b(\mu )$ by about $40\%$ 
from the scale $\mu = m_b^{\rm pole}\simeq 5$~GeV 
up to the weak scale of ${\cal O}(200)$~GeV (cf. the
solid curve of Fig.~\ref{fig:mbRun2L}).
This is however not the full story. The complexity comes from the
finite SUSY threshold correction in the large $\tanb$ region
which are potentially large \cite{mbrun2,mbrun,mbrun3}. 
As shown in Ref.~\cite{mbrun}, the dominant one-loop
SUSY correction contributes to running $b$-mass a finite term so that
$m_b(\mu )$ at the SUSY scale 
$\mu \equiv \mu_R = M_{\rm soft}$ is multiplied by
a constant factor $~1/[1+\Delta_b({\rm SUSY})]~$, which 
appears as a common factor in the bottom Yukawa couplings
of all three neutral Higgs bosons.
For large $\tanb$,
$\Delta_b({\rm SUSY})$ contains the following $\tanb$-enhanced terms
from sbottom-gluino and stop-chargino loops\footnote{We
thank K.~Matchev for discussing his published results in Ref.~\cite{mbrun},  
and to him and W.A.~Bardeen for discussing 
the issue of the running $b$-mass. 
Our convention of the MSSM Higgs parameter $\mu$ 
differs from that of Ref.~\cite{mbrun} by a minus sign.},
\begin{equation}
\begin{array}{l}
\Delta_b({\rm SUSY})~=~ 
\dis\left(\f{\Delta m_b}{m_b}\right)^{\tilde{b}\tilde{g}}
   +\left(\f{\Delta m_b}{m_b}\right)^{\tilde{t}\tilde{\chi}}\\[0.45cm]
=\dis\f{-\mu\tanb}{16\pi^2}\left\{\f{8}{3}g_3^2m_{\tilde{g}}
F\left(m_{\tilde{b}}, m_{\tilde{b}}, m_{\tilde{g}} \right)+
\left[ y_tA_t F\left(m_{\tilde{t}}, m_{\tilde{t}}, \mu \right) -
g_2^2M_2 F\left(m_{\tilde{t}}, m_2, \mu \right)  
\right]\right\},
\end{array}
\label{eq:mbrun}
\end{equation}
with the function $F$ defined as:
$$
F(\sqrt{x},\sqrt{y},\sqrt{z})
=-\dis\f{xy\ln x/y+yz\ln y/z+zx\ln z/x}{(x-y)(y-z)(z-x)} ~.
$$
In these equations, the MSSM Higgs parameter 
$\mu$ should not be confused with 
the usual renormalization scale $\mu_R$.
In (\ref{eq:mbrun}), we have assumed, for simplicity,  mass 
degeneracy for the top and bottom squarks, respectively.
Eq.~(\ref{eq:mbrun}) shows that the SUSY correction to the running $m_b$
is proportional to $\tanb$ and $\mu$. Thus, this correction is 
enhanced for large $\tanb$ and non-negligible in comparison with
the QCD corrections. Also changing the sign of $\mu$ will
vary the sign of the whole correction $\Delta_b({\rm SUSY})$
and implies that the SUSY correction can either increase or decrease
the running $b$-mass at the energy scale around of ${\cal O}(M_{\rm soft})$.
Normally, when
defining the running coupling using the Collins-Wilczek-Zee (CWZ)
scheme \cite{CWZ}, only the $\mu_R$-dependent contributions are included
while all the $\mu_R$-independent terms are absorbed 
into the corresponding Wilson
coefficient functions. 
However, since the $\mu_R$-independent contribution
$\Delta_b({\rm SUSY})$ is not small for 
large $\tanb$ and a full NLO SUSY
calculation is not yet available, 
we include $\Delta_b({\rm SUSY})$ to 
define an ``effective'' running coupling/mass of $b$-quark
even below the SUSY threshold scale $M_{\rm soft}$.
This could give a rough estimate on the large
SUSY loop corrections from the $\phi$-$b$-$\bar b$ vertex. 
Obviously, when $\mu_R$ is much smaller than
$M_{\rm soft}$, the CWZ scheme should be used. 
Hence,
in Fig.~\ref{fig:mbRun2L}, we only show the ``effective'' running mass
$m_b(\mu )$ down to about $100$~GeV which is the relevant energy scale
and the mass scale ($\sim M_{\rm Higgs}$) considered in this work.
Fig.~\ref{fig:mbRun2L}  
illustrates that due to the SUSY correction the 
``effective'' running $b$ mass at
the weak scale can be either larger or smaller than the 
SM QCD running value ($\sim 3$~GeV), depending on the choice of
the sign of $\mu$ parameter (and also other soft-breaking parameters
such as the trilinear coupling $A_t$ and masses of the gluino, gaugino,
stop and sbottom).  
For the ``default'' set of SUSY parameters used in our analysis,
all the soft-breaking parameters are set to $500$~GeV for simplicity.
It happens to be the case that
the SM QCD and SUSY corrections nearly cancel each other so that
the ``effective'' running mass $m_b$ is very close to the 
pole mass value ($\sim 5$~GeV) for the scale above $\sim 100$~GeV 
(cf. upper curve of Fig.~\ref{fig:mbRun2L}).
For comparison, in our analysis using 
the ``LEP~II Scan~A2'' set of SUSY parameters, 
the SM QCD and SUSY corrections do not cancel and tend
to reduce the ``effective'' running $b$-mass or the 
Yukawa coupling $y_b(\mu )$
which results in a weaker bound for the Tevatron
Run~II and the LHC, 
as shown in Fig.~\ref{fig:Exclusion}b.\footnote{
A detailed analysis of these effects at the Tevatron Run~II is
currently underway \cite{stev}.} 
The difference in the exclusion contours shown in 
Figs.~\ref{fig:Exclusion}a and~\ref{fig:Exclusion}b,
derived from the measurement
of the $\phibb$ production rate at hadron colliders,
is mainly due to the difference in the ``effective'' 
running coupling or mass (including the QCD and
SUSY corrections), as described above.\footnote{
From Fig.~\ref{fig:Exclusion}, it is also clear that, 
for the LHC bounds, the SUSY correction has much less 
impact since the relevant $\tanb$ values become much lower,
around of ${\cal O}(2-15)$.}~
We therefore conclude that a full NLO QCD calculation 
is important for a consistent improvement
of our current analysis.

\subsection{Interpretation of results for
Models of soft breaking parameters}

The MSSM allows for a very general set of soft SUSY-breaking
terms and thus is specified by a large 
number of free parameters ($\simeq 124$ \cite{mssm}), 
though only a complicated subset
of this parameter space is consistent with all current
experimental results.  It is therefore important to understand
the mechanism of supersymmetry breaking (which presumably
occurs at a high energy scale \cite{kanet}) and to predict
the soft parameters at the weak scale from an underlying model.
Many alternative ideas about how supersymmetry might be broken, and how
this will result in the low energy soft breaking parameters exist in the
literature, including the supergravity inspired (SUGRA) models and
gauge-mediated SUSY breaking (GMSB) models.  In this section we
examine the sensitivity of the $\phibb$ process to probe 
a few models of SUSY breaking, concentrating for the most part on
the popular SUGRA and GMSB models.  However, there are also 
other interesting ideas to which the $\phibb$ process may provide
interesting information,
because these models naturally prefer a large $\tan \beta$.  
A few examples include the SO(10) 
grand unification theories \cite{sotengut}
(SO(10) GUTs); the infrared fixed-point scenario
\cite{irfixedp}; and also a scenario with compositeness,
the ``more minimal supersymmetric SM" \cite{moremssm}.

\subsubsection{Supergravity Models with large $\tan\beta$}

The supergravity inspired (SUGRA) models \cite{sugrarev}
incorporate gravity in a natural manner, and solve the problem 
of SUSY breaking through the introduction of a hidden sector, 
which breaks SUSY at a very high scale [$\sim {\cal O}(10^{11})$ GeV] 
and interacts with the MSSM fields only gravitationally.
This model offers an exciting glimpse into the possible connection
between the heavy top quark and the EWSB by allowing radiative breaking
of the electroweak symmetry, in which the large top quark Yukawa
coupling can drive one of the Higgs masses negative at energies
$\sim m_Z$.  In the limit of large $\tan \beta$ the bottom and tau
Yukawa couplings can also play an important role \cite{dressnoji}.

Under the assumption that the gravitational
interactions are flavor-blind, this model
determines the entire SUSY spectrum
in terms of five free parameters (at the high energy scale of the
SUSY breaking) including a common scalar mass
($\tilde m_0$), a common gaugino mass ($M$), a common scalar
tri-linear interaction term ($A$), $\tan \beta$, and the sign of the
Higgs mixing parameter (${\rm sgn}(\mu)$).  The weak scale particle
spectrum can then be determined by using the renormalization group
analysis to run the sparticle masses from the high scale to the weak
scale.

Though large $\tan \beta$ is not required by the minimal SUGRA model,
it can naturally be accommodated, as demonstrated in 
\cite{dressnoji,sugrtbg},   where it was found that
a large $\tan \beta$ also generally requires that the pseudoscalar
Higgs mass be light ($m_A \leq$ 200 GeV for $\tan \beta \geq 30$),
because the enhanced $b$ and $\tau$ Yukawa couplings act through the
renormalization group equations to reduce the down-type Higgs mass
term at the weak scale, thus resulting in a light Higgs spectrum.
Since the importance of the $b$ and $\tau$ effects in the RG
analysis increases
with larger $\tan \beta$, as $\tan \beta$ increases the resulting
$m_A$ decreases, making the large $\tan \beta$ scenario in the
SUGRA model particularly easy to probe through
the $\phibb$ process.
From the limits on the $m_A$-$\tan \beta$ plane derived above,
it thus seems likely that from the data of the
Tevatron Run II with 2 \ifb of integrated luminosity, a large
portion of the minimal SUGRA model with $\tan \beta \geq 20$
may be excluded.

\subsubsection{Gauge-mediated SUSY Breaking Models with 
Large $\tan\beta$}

Models with GMSB break SUSY at a scale which is typically much lower
than that present in the SUGRA models.  
The supersymmetry is generally
broken in a hidden sector which directly couples to a set of messenger
chiral superfields.  This induces a difference in mass between 
the fermion and
scalar components of the messenger fields, which in turn generates
masses for the gauginos and sfermions of the MSSM fields via loops
involving the ordinary gauge interactions \cite{dinetal,gmmrev}.
A generic feature of this scenario is that because of the relatively
low scale of SUSY breaking, the gravitino acquires a much smaller mass
than in the SUGRA scenarios, and is generally the lightest
supersymmetric partner (LSP).
While specific models of GMSB vary as to their assumptions and relevant
parameters, generally what must be assumed is the field content of the
messenger sector (including transformation properties under the gauge
group and number of multiplets in the theory) and the scale at which
SUSY is broken in the hidden sector.

The minimal GMSB models can also result in a radiative breaking of the
EWSB, through the renormalization group evolution of the Higgs masses
(driven by the large top Yukawa coupling)
from the effective SUSY breaking scale to the weak scale.  As in 
the SUGRA model case, this
evolution can drive the mass term of the up-type Higgs
negative at the weak scale, thus breaking the electroweak symmetry.  
In fact, because the effective SUSY breaking
scale in a GMSB model is typically much lower than in the SUGRA model
(and thus closer to the weak scale),
in order for the proper EWSB to occur, it was demonstrated in 
\cite{babuetal} that a large $\tan \beta$ is required (about $30$-$40$).
However, because the effective SUSY breaking scale is typically
much lower than in the SUGRA model, the large effects of the $b$ and
$\tau$ on the Higgs mass running do not reduce the Higgs spectrum to the
degree that occurs in the SUGRA model with large $\tan \beta$,
and thus result in a heavier $A$ with mass
of about 400 GeV.  So, this
particular model would only be explored through the $\phibb$ process
at the LHC.  However, more general analyses of the GMSB scenario 
\cite{baggetal,baertal,ratsarid,borzumati},
introducing more degrees of freedom in the messenger
sector than the minimal model, can allow for 
large $\tan \beta$ and relax
$m_A$ to be as low as about 200 GeV.  
Thus, through the $\phibb$ production these more general 
models can be first 
probed at the Tevatron for the relevant mass range, 
and then largely explored at the LHC.

\section{Conclusions}

It remains a challenging task to determine the underlying 
dynamics of the 
electroweak symmetry breaking and the flavor symmetry breaking.
Either fundamental or
composite Higgs boson(s) may play a central role in the 
mass generation of the weak gauge bosons and the fermions.  
The heavy top quark, with a mass on
the same order as the scale of the electroweak symmetry breaking, 
suggests that the top quark may play 
a special role in the mechanism of mass generation.  
In this work, we have shown that in the typical 
models of this type, the bottom quark,
 as the weak isospin partner of the 
top quark, can also participate in the dynamics of 
mass generation, and serves as an effective probe of the possible new 
physics associated with the Higgs and top sectors.
 
We have presented a model-independent analysis on Higgs boson
production in
association with bottom quarks, via the reaction
$\ppbar/pp \to \phibb \to \bbbb$, at the Tevatron Run~II
and the LHC. 
Using the complete tree level calculation with an estimated QCD
$k$-factor of 2, we derive the exclusion contour for the enhancement
factor (in the coupling of $\phibb$ relative to that of the SM)
versus the Higgs mass $m_\phi$ at the $95\%$~C.L., assuming 
a signal is not found.

We apply these results to analyze the constraints on the parameter
space of the MSSM (in the large $\tanb$ region) 
with naturally large bottom Yukawa couplings. 
To confirm the MSSM, it is necessary to 
detect all the predicted neutral 
Higgs bosons $h,~H,~A$ and the charged scalars $H^\pm$.
From LEP~II, depending on the choice 
of the MSSM soft-breaking parameters,
the current $95\%$~C.L. bounds on the masses of the MSSM
Higgs bosons are about 70~GeV for both the $CP$-even scalar $h$ and
the $CP$-odd scalar $A$ \cite{himlim}. It can be improved at LEP~II
with higher luminosity and maximal energy, but the bounds on the
Higgs masses will not be much larger than $\sim m_Z$ for an   
arbitrary $\tanb$ value. The $Wh$ and $WH$ associated production at
the Tevatron Run~II can further improve these bounds, if a signal
is not observed. At the LHC, a large portion of parameter space can be 
tested via $pp\to t\bar t +h(\to \gamma \gamma )+X$, and
$pp\to h(\to ZZ^\ast )+X$, etc. \cite{kunstz,gunorr}. 
A future high energy $e^+e^-$ collider will fully test
the MSSM Higgs sector \cite{eehiggs,eehiggs2} through the
reactions $ e^+ e^- \to Z+h(H),\, A+h(H), \, H^+ H^-$, etc. 
In this work, we demonstrate that studying the $\phibb$
channel at hadron colliders can further improve our knowledge on MSSM.
The exclusion contours on the $m_A$-$\tanb$ plane of the MSSM
shows that the Tevatron and the LHC 
are sensitive to a large portion of the parameter space
via this mode.
It therefore provides a complementary   
probe of the MSSM Higgs sector in comparison with that from LEP~II.
The implications of these bounds in the parameter space
on both the supergravity and the gauge-mediated SUSY
breaking models are further discussed.
We find that the $\phibb$ process can effectively test 
the  models with either scheme of the SUSY soft-breaking
in the large $\tanb$ scenario.

In conclusion, the stringent constraints obtained by
studying the associated $\phibb$ production at hadron colliders
for the supersymmetric
models show the utility of this search mode.
However, much work remains to be done.  For example, we
have found that a $b$-trigger is essential for this analysis.
In fact, the CDF group at the Fermilab has demonstrated that
it is possible to detect events with four or more jets and 
two or more $b$-tags \cite{cdfwh},
which can significantly be improved with 
the implementation of $b$-trigger at the Run~II of the Tevatron
\cite{run2btag}.
However, the large QCD 4-jet background rate at the LHC can
potentially make triggering on the 
$b {\bar b} \phi(\rightarrow b {\bar b})$ 
events difficult, though 
it is expected that $b$-trigger would becomes more efficient for a 
heavier (pseudo-)scalar $\phi$. (This is because the $b$-jets from 
the decay of a heavier $\phi$ are more energetic, and 
its QCD background rate drops rapidly as a function of the 
transverse momentum of the triggered jet.)
We hope that the interesting results
afforded by studying this channel will stimulate interest in
working on these problems in our experimental colleagues.

We have found that the $\phibb$ process 
complements Higgs searches in other channels, and 
thus it is expected
that experimental searches for this signature at the Tevatron Run~II
(and possibly beyond) and the CERN LHC will provide interesting
and important information about the mechanism of 
the electroweak symmetry
breaking and the fermion mass generation.


\chapter{``Intrinsic $k_T$'' and Soft Gluon Resummation \label{sec:In1}}



\section{Introduction}

For many years, it has been a common practice to ``smear'' the transverse
momentum ($p_T$) distribution of the Drell-Yan pairs or the direct photons
calculated in perturbative QCD to obtain a better agreement
between the theoretical predictions and the experimental data. The argument
of smearing the $p_T$ distribution is that any parton inside a hadron
generally has a non-zero transverse momentum which is called the ``intrinsic
transverse momentum'' (or ``intrinsic $k_T$'') of the parton. The concept of
intrinsic transverse momentum evolved in the literature from the covariant
parton model of Landshoff, Polkinghorne and Short \cite{Landshoff} to the 
$k_T$ dependent parton distributions of Collins and Soper \cite{Collins80}. 
Intrinsic $k_T$ emerges from the Fermi motion of partons within the hadron.
Since the Fermi motion is uncorrelated with the direction of the
acceleration, it leads to momentum fluctuations in the transverse
directions. By definition, this ab initio transverse momentum has a
non-perturbative nature and typically is of the order of $\Lambda _{QCD}$,
i.e. a few hundred MeV. It was however found that the amount of ``smearing''
needed to make the theoretical prediction of the $p_T$ distribution agree
with data in some cases could be as large as a few GeV which is much larger
than $\Lambda _{QCD}$. To understand this, we need the full armory of QCD
including both the perturbative and the non-perturbative part. In the low 
$p_T$ region, which is the region where a perturbative calculation will
deviate the most from experimental data, one needs to take into account the
effects of multiple (soft and collinear) gluon radiation from both the
initial and the final states. For extremely low $p_T$ regions, the
non-perturbative effects of the QCD theory, e.g. effects due to higher twist
operators, should also be included when comparing with the measurements. In
this work we make use of the soft gluon resummation formalism developed by
Collins and Soper to estimate the amount of smearing needed to apply onto a
perturbative QCD calculation so that its prediction would agrees with that
from a resummation calculation.

For the Drell-Yan process, in which only the initial state involves QCD
colors, Collins, Soper and Sterman (CSS) gave a set of comprehensive formula
to account for the effects on the $p_T$ distribution of the lepton pair due
to the multiple soft (and collinear) gluon emissions in the initial state.
In the CSS formalism, the perturbative part of the multiple gluon radiation
effects is summarized by the perturbative Sudakov form factor and some
Wilson coefficients (i.e. $C$-functions); the non-perturbative part of QCD
physics which is relevant to the low $p_T$ distribution is parametrized by
some non-perturbative functions which at the present time can only be fitted
to some existing Drell-Yan data. The moral is that if the CSS formalism is
the correct parametrization of the multiple gluon effects, then the same
form of the non-perturbative functions (fitted to particular sets of data)
should also be applied to the other Drell-Yan data at different energies of
the colliders and/or invariant masses of the lepton pairs.

The effects due to initial state gluon radiations should be universal to any
process involving the same kinds of incoming partons with similar QCD color
structures in the events. 
As we saw earlier, the CSS formalism describes well the transverse momentum
distribution of $W^{\pm}$, $Z$-bosons and diphotons at the Tevatron.
Although the direct photon process does not
have the same kinds of incoming partons or color structures as those
contributing to the Drell-Yan process, we expect that the dominant effects
on the low $p_T$ distribution of the direct photon would come from the
initial state radiation whose non-perturbative contributions are more or
less the same as those measured in the Drell-Yan process. As far as we know,
there is no proof for this statement. It is only our educated guess that the
non-perturbative behavior of the partons inside a hadron would be similar.
Of course, the calculable perturbative Sudakov form factors would
distinguish an initial gluon parton from an initial quark (or anti-quark)
parton, and should be properly included in the CSS formalism for calculating
the $p_T$ distributions. Although we lack theoretical proof, there are
indications from phenomenology which seem to confirm the need of the
resummation of the initial state QCD radiation for the direct photon process
for low $p_T$. Monte Carlo simulations (using Bauer-Reno's MC or PYTHIA)
indicate that initial state radiation improves the finite order QCD\
results. There are also efforts to derive CSS type formulas  with partial
success.

In this work, we do not intend to do a complete calculation on the $p_T$
distribution of the direct photon using the CSS formalism, which as
described in the above, involves both the perturbative Sudakov form factor
and the non-perturbative functions. We only apply the non-perturbative part
of the CSS formalism to explain the amount of $p_T$-smearing needed to make
a ``smeared'' next-to-leading order (NLO) 
QCD perturbative calculation agree with
the experimental data of the direct photon in its transverse momentum
distribution. We show that for the low energy fixed-target experiments, the
non-perturbative part of the CSS resummation formalism alone can explain the
required amount of low-$p_T$-smearing. However, to explain the high energy
collider data, such as those at the Tevatron, the perturbative part of the
Sudakov form factor should also be included in the calculation. We defer the
complete calculation, which separately includes the effects from the
perturbative Sudakov form factor associated with $qq$ ($q{\bar q}$) or $gq$ (%
$g{\bar q}$) initiated direct photon production processes, to our future
work.

\section{Smearing $p_T$ Distributions of Direct Photons}

\indent

In this section we show that it is possible to resolve the discrepancy
between the theoretical prediction and the experimental data of the $p_T$
distribution of the direct photon measured in fixed target experiments. This
is achieved by applying the CSS resummation formalism and by including only
the contribution from the non-perturbative functions of the formalism, which
(in the direct photon context) has just one free parameter to be determined
by experimental data. Our approach is to assume the approximative
universality (i.e. insensitivity to the flavors of the active partons) of
the non-perturbative function in the CSS resummation. Hence, the same
non-perturbative function that is used to describe the Drell-Yan process can
also be used for the direct photon production to improve the next-to-leading
order predictions. What follows in this section is rather intuitive and is
not backed up by full theoretical understanding or precise calculations.

It is a fact that the transverse momentum distribution of the photons
produced directly at hadron-hadron collisions deviates systematically from
the experimental measurement. The NLO QCD calculations \cite{Aurenche} give
less steep curves than the experimental data. A recent global analysis of
the problem \cite{Huston}, examining both fixed target and collider
experiments, concluded that neither new parton distributions nor new
fragmentation functions can resolve the problem, since the deviation occurs
at different $x$ values in experiments with different energies. In the above
study the authors also exclude the possibility that the deviation can be
entirely accounted for the scale choice in the calculations.

It was shown in Ref.~\cite{Huston} that convoluting the theoretical $p_T$
distribution with a Gaussian resolution function can raise the cross section
at the low $p_T$ end so that it agrees with the experimental data better.
This is because smearing a falling distribution by a Gaussian can steepen
its original shape. Writing the smeared $p_T$ distribution as 
\begin{eqnarray}\label{ConEqu}
{d\sigma \over dp_T} = {1 \over \Delta \sqrt{\pi/2} }
\int_{-\infty}^{\infty} dp_T' 
{d\sigma^{NLO} \over dp_T'} ~ {\rm e}^{-(p_T - p_T')^2/2 \Delta^2}
\end{eqnarray}
defines $\Delta $ as the quantitative measure of the smearing. The amount of
the smearing needed differs for different center of mass energies (see the
first and second columns of Table \ref{TBLSmear}).

There are several puzzling details here. First, there is no theoretical
reason to introduce an ad hoc smearing of the NLO results. Secondly, the
amount of the intrinsic $k_T$ that the partons acquire because of the Fermi
motion within the hadron is in the order of $\Lambda _{QCD}$, which is a few
hundred MeV. The amount of smearing that brings theory and experiment to
agreement is more than that. Finally, even if one invents a mechanism to
introduce some type of a smearing, one has to find a smearing function that
has a dependence on $S$ which is the square of the center of mass energy of
the hadron-hadron collision, and make sure that the smearing does not ruin
the agreement at higher $p_T$'s.

To overcome all the above difficulties we propose the application of the
non-perturbative function of the CSS resummation as a smearing device. This
simple idea seems to be plausible because the non-perturbative function has
a Gaussian form and depends on the center of mass energy of the process.
Although we do not have a proof, after noticing that the non-perturbative
function acts like a smearing factor at the Drell-Yan process (for details
see the next section), we may as well try it at the direct photon process.

The non-perturbative function is usually
parametrized in a simple Gaussian form: 
\begin{eqnarray}
  W^{NP} (Q_T,Q,Q_0) = {1 \over 2 G_2(Q,Q_0)}
  \exp{\left\{ -Q_T^2/4 G_2(Q,Q_0) \right\}},
  \end{eqnarray}
with $G_2(Q,Q_0) = g_1 + g_2 \ln( {Q / 2 Q_0} )$, $g_1 = 0.11\,{\rm GeV}^2$, 
$g_2 = 0.58\,{\rm GeV}^2$ and $Q_0 = 1.6\,{\rm GeV}$. We can easily check
whether the amount of the smearing from the non-perturbative function comes
close to what phenomenology demands. First we substitute $\sqrt{S x^2}$ for $%
Q$ in $G_2(Q,Q_0)$. This substitution is inspired by the relation $Q = \sqrt{%
S x_A x_B} = \sqrt{\hat{s}}$ that holds in the case of the Drell-Yan
process, with $\sqrt{S}$ being the center of mass energy of the colliding
hadrons, and $x_A$, $x_B$ the longitudinal momentum fractions of the
incoming partons. Since these latter quantities are not measured at the
direct photon production we take $x$ as a free parameter and let data
determine its value. The best fit to the fixed target experimental data
gives $x = 0.3$. %
%
%
Now we can calculate the numerical value of $G_2(Q,Q_0)$ for the six
different experiment. The results are shown in the last column of Table \ref
{TBLSmear}. 
\begin{table}
\begin{center}
\begin{tabular}{ cccc }
\hline \hline \\[-3mm]
 Experiment & $\sqrt{S}$ (GeV) & $\Delta$ (GeV) & 
$\sqrt{2 G (Q,Q_0)}$  (GeV)\\
\hline \\[-3mm]
 WA70       & ~~23         & ${\ \lower-1.2pt\vbox{\hbox{\rlap{$<$}\lower5pt\vbox{\hbox{$\sim$}}}}\ } 1$
                                            & 1.0 \\
 UA6        & ~~24         & ${\ \lower-1.2pt\vbox{\hbox{\rlap{$<$}\lower5pt\vbox{\hbox{$\sim$}}}}\ } 1$
                                            & 1.1 \\
 E706       & ~~31         & 1-2            & 1.2 \\
 R806       & ~~63         & 2-3            & 1.5 \\
 UA2        & ~630         & 3-4            & 2.2 \\
 CDF        & 1800         & 3-4            & 2.5 \\
\hline \hline
\end{tabular}
\end{center}
\caption{The amount of the smearing needed 
and the amount provided by the non-perturbative function
to make theory agree with experiment
for different direct photon experiments.
Values of $\Delta$ are taken from Ref.~[4].
}
\label{TBLSmear}
\end{table}

We see at first sight that the non-perturbative function provides about the
right amount of smearing needed to bring the theory prediction and the
experimental data in a better agreement. Examining Table \ref{TBLSmear} more
closely, we observe that the smearing from the non-perturbative function
falls short as the center of mass energy of the collider increases. Hence,
for the fixed target experiments at a higher energy and for the collider
experiments, the contribution from the non-perturbative function alone
cannot account for all the amount of smearing needed.

This can also be seen from Fig.~\ref{FigDP}. Substituting the
non-perturbative function into the convolution equation (\ref{ConEqu}), the
theoretical distributions can be rendered close to the experimental ones as
it is shown in Fig.~\ref{FigDP}. (Among the fixed target experiments, UA6 $p 
\overline{p}$ data are not shown, but they have the same characteristics as
UA6 $pp$ data.) The NLO QCD curve is shown in short dashes. The long dashed
curve is the convolution of the theory with the non-perturbative function.
As anticipated, the agreement between the data and the theoretical
prediction improves in low the $p_T$ region after including the smearing
effects for all the low energy fixed target experiments. However, the
convolution (i.e. the $p_T$-smearing) is less effective for the collider
data (not shown). As discussed above, at collider energies the perturbative
Sudakov form factor arising from resumming over the large logs which are due
to the initial state radiation will be important besides the
non-perturbative effects, and it is not included in the present study. 
\begin{figure*}[t]
\begin{center} 
\begin{tabular}{c}
\epsfysize=10.0cm
\epsffile{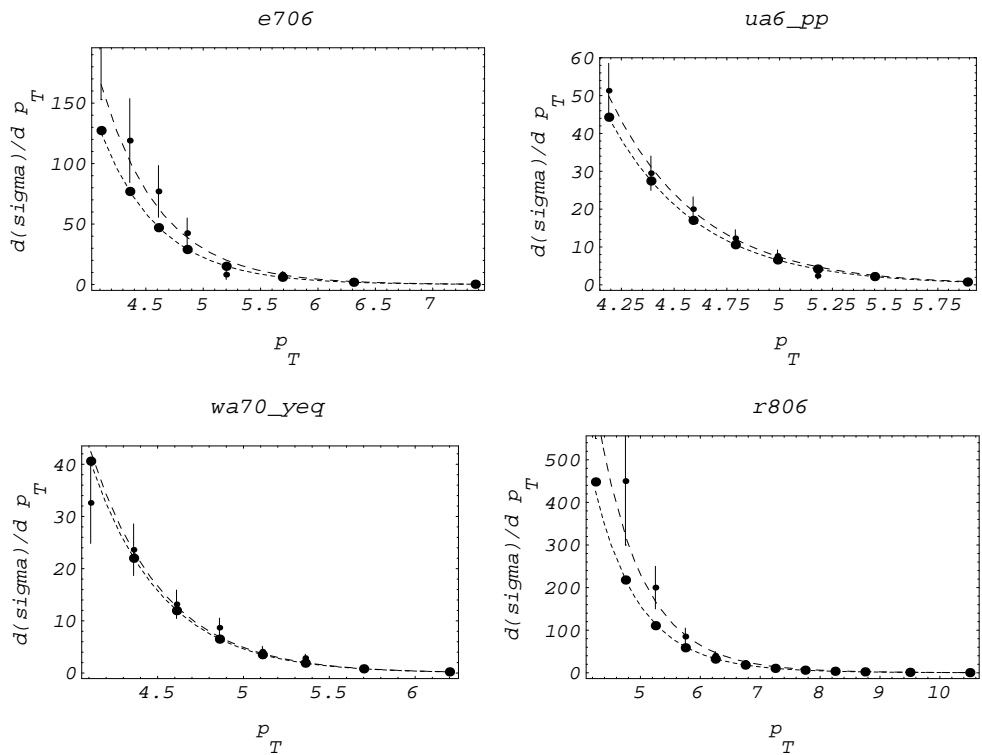}
\end{tabular} 
\caption{
Experiment numbers are listed at the top of each plot. Short dashed line: NLO 
transverse momentum distribution. Long dashed line: NLO theory convoluted with 
the Fourier transform of the non-perturbative function used in the Drell-Yan 
resummation formalism.}
\label{FigDP}
\end{center}
\end{figure*}

In conclusion, the above results suggest that there is an important
similarity between the Drell-Yan and the direct photon processes. The roots
of this correspondence are found in the initial state radiation, which is
present in both cases.

\section{$p_T$-Smearing and Soft gluon Resummation}

\indent

In this section we demonstrate that ``$p_T$-smearing'' naturally arises
within the resummation formalism of Collins, Soper and Sterman 
\cite{CS,CSS,Collins-Soper-Sterman}. 
It is the same type of smearing that is needed to make
the theoretical prediction agree with the experimental data in $p_T$
distribution of the direct photon produced in fixed target experiments.

The canonical example of the CSS formalism is the hadronic production of
electroweak vector bosons. 
In the Drell-Yan type of process the kinematical variables of the vector
boson $V$ (real or virtual) are selected to be its invariant mass $Q$,
rapidity $y$, transverse momentum $Q_T$, and azimuthal angle $\phi_V$,
measured in the laboratory frame. The kinematics of the leptons from the
decay of the vector boson can be described by the polar angle $\theta$ and
the azimuthal angle $\phi$, defined in the Collins-Soper frame \cite{CSFrame}%
. The fully differential cross section of the vector boson production and
decay is given by the resummation formula: 
\begin{eqnarray} & &
  \left( { d \sigma(AB \rightarrow V(\rightarrow {l {\bar l'}}) X )
  \over dQ^2 \, dy \, dQ^2_T \, d\phi_V \, d\cos{\theta} \, d\phi}
  \right)_{res} =
  {1 \over 96 \pi^2 S} \,
  {Q^2 \over (Q^2 - M_V^2)^2 + M_V^2 \Gamma_V^2}
  \nonumber \\  & & ~~
  \times \bigg\{ {1\over (2 \pi)^2}
  \int_{}^{} d^2 b \, e^{i {\vec Q_T} \cdot {\vec b}} \,
  \sum_{j,k}{\widetilde{W}_{jk} (b_*,Q,x_A,x_B, \theta, \phi)} \,
  W^{NP}_{jk} (b,Q,x_A,x_B)
  \nonumber \\  & & ~~~~
  + ~ Y(Q_T,Q,x_A,x_B, \theta, \phi) \bigg\}.
  \label{ResFor}
  \end{eqnarray}
Here 
\begin{eqnarray}
  \widetilde{W}_{jk} (b,Q,x_A,x_B, \theta, \phi)  =
  \widetilde{w}_{jk} (x_A,x_B, \theta, \phi)
  \exp{\left\{ -S(b,Q) \right\}},
  \label{WTwi}
  \end{eqnarray}
with the Sudakov exponent 
\begin{eqnarray} & &
  S(b,Q) =
  \int_{b_0^2/b^2}^{Q^2}
  {d {\bar \mu}^2\over {\bar \mu}^2}
       \left[ \ln\left({Q^2\over {\bar \mu}^2}\right)
        A\big(\alpha_s({\bar \mu})\big) +
        B\big(\alpha_s({\bar \mu})\big)
       \right],
  \label{SudExp}
  \end{eqnarray}
and $x_{A,B} = \sqrt{Q} \exp{\left( \pm y \right)}$. In the above
expressions $j$ represents quark flavors and $\bar{k}$ stands for anti-quark
flavors. 
The definition of $\widetilde{w}$ together with further details can be found
in Chapter \ref{ch:Resummation}.

In Eq.~(\ref{ResFor}), the impact parameter $b$ is to be integrated from 0
to $\infty$. However, for $b$ larger than a certain $b_{max}$, corresponding
to energy scales less than $1/b_{max}$, the QCD coupling $\alpha_s$ becomes
so large that a perturbative calculation is no longer reliable. To avoid
this, $\widetilde{W}$ is evaluated at $b_*$, with 
\begin{eqnarray}
  b_* = {b \over \sqrt{1+(b/b_{max})^2} }
  \label{bStar}
  \end{eqnarray}
such that $b_*$ never exceeds $b_{max}$. Thus $b_{max}$ is the theoretical
scale that separates perturbative and non-perturbative physics in the
resummation formalism. Just like other theoretical scales (e.g.
renormalization or factorization scales), $b_{max}$ defines a somewhat
arbitrary separation. Choosing $b_{max} = 1/2 $ GeV$^{-1}$ (as usual)
instead of $1/\Lambda_{QCD}$ shuffles parts of the low $Q_T$ perturbative
physics related to the initial state radiation into the non-perturbative
function. Thus the non-perturbative function contains more than just the
intrinsic partonic $k_T$. $W^{NP}$ is needed in the formalism in order to
suppress the contribution of $\widetilde{W}_{jk}$ from the large $b$ (small $%
Q_T$) region and to parametrize our ignorance of the non-perturbative
phenomenon there. Furthermore, since physics cannot depend on $b_{max}$, $%
W^{NP}$ carries an implicit $b_{max}$ dependence through the
non-perturbative parameters $g_i$ defined below. Generally $W^{NP}$ has the
structure 
  \begin{eqnarray} &
  W^{NP}_{jk} (b,Q,Q_0,x_A,x_B) = 
  \exp{\left\{ - S^{NP}(b,Q,Q_0,x_A,x_B) \right\}} , 
  \label{WNP}
  \end{eqnarray}
with the non-perturbative exponent 
  \begin{eqnarray} &
  S^{NP}(b,Q,Q_0,x_A,x_B) =
   h_1(b) \ln \left( Q^2\over Q^2_0 \right) +
     h_{j/A}(x_A,b) + h_{{\bar k}/B}(x_B,b).
  \label{SNPh}
  \end{eqnarray}
Here $h_1$, $h_{j/A}$ and $h_{{\bar k}/B}$ cannot be calculated using
perturbation theory, they must be determined using experimental data. The
form of the non-perturbative exponent $S^{NP}$ is inspired
by the form of the Sudakov form factor which in turn is derived by
renormalization group arguments discussed in Chapter~\ref{ch:Resummation}.
Since for large values of
the impact parameter the non-perturbative function has to cancel the
contribution of $\widetilde{W}$, the structure of the non-perturbative
exponent is determined by the Sudakov exponent, that is, by perturbative
resummation physics.

We restrict our attention to the small $Q_T$ region, where 
resummation effects may be important and are detectable. In the $Q_T^2 \ll
Q^2$ region the $Y$-term in Eq.~(\ref{ResFor}) is negligible. 
Dropping the $Y$ piece and rewriting $\widetilde{W}_{jk}$ in a form of a
Fourier transform%
\footnote{
We use the same symbol for the Fourier transformed function,
and we distinguish the Fourier transformed quantity
only by displaying the different argument.} we obtain 
\begin{eqnarray} &&
  \left( { d \sigma(AB \rightarrow V(\rightarrow {l {\bar l'}}) X )
  \over dQ^2 \, dy \, dQ^2_T \, d\phi_V \, d\cos{\theta} \, d\phi}
  \right)_{res - Y} =
  {1 \over 96 \pi^2 S} \,
  {Q^2 \over (Q^2 - M_V^2)^2 + M_V^2 \Gamma_V^2}
  \nonumber \\  && ~~
  \times \bigg\{ {1\over (2 \pi)^2}
  \int_{}^{} d^2 b \, e^{i {\vec Q_T} \cdot {\vec b}} \,
  \sum_{j,k}
  \left(  {1 \over 2 \pi} 
     \int_{}^{} d^2 Q_T' \, e^{-i {\vec Q_T'} \cdot {\vec b}} \,
    {\widetilde{W}_{jk} (Q_T',Q,x_A,x_B, \theta, \phi)} \right)
  \nonumber \\  && ~~~~~~~~~~~~~~~~~~~~~~~~~~~~~~~
  \times ~ W^{NP}_{jk} (b,Q,x_A,x_B) \bigg\}
  \label{}
  \end{eqnarray}
The $b$ integral can formally be carried out, resulting in the appearance of
the inverse Fourier transform of the non-perturbative function. Introducing
the convolution notation for the remaining integral, the first part of the
resummation formula in $Q_T$ space simplifies to 
\begin{eqnarray} &&
  \left( { d \sigma(AB \rightarrow V(\rightarrow {l {\bar l'}}) X )
  \over dQ^2 \, dy \, dQ^2_T \, d\phi_V \, d\cos{\theta} \, d\phi}
  \right)_{res - Y} =
  {1 \over 96 \pi^2 S} \,
  {Q^2 \over (Q^2 - M_V^2)^2 + M_V^2 \Gamma_V^2}
  \nonumber \\  && ~~
  \times {1\over (2 \pi)^2}
  \left( \widetilde{W}_{jk}
  \otimes ~ W^{NP}_{jk} \right) (Q_T,Q,x_A,x_B,\theta,\phi)
  \label{ResForQ}
  \end{eqnarray}
where $\otimes$ denotes the convolution and is defined by 
\begin{eqnarray} &&
  \left( \widetilde{W}_{jk}
  \otimes ~ W^{NP}_{jk} \right) (Q_T,Q,x_A,x_B,\theta,\phi) =
  \nonumber \\  && ~~~~
  \sum_{j,k}
  \int_{}^{} d^2 Q_T' \,
  \widetilde{W}_{jk} (Q_T',Q,x_A,x_B, \theta, \phi)
  \, W^{NP}_{jk} (Q_T-Q_T',Q,x_A,x_B).
  \label{Convol}
  \end{eqnarray}
Eq.~(\ref{ResForQ}) expresses that the resummed cross section at low $Q_T$
region is given by the convolution of the exponentiated singular part of the
hard scattering cross section with the non-perturbative function. It can be
viewed as if the hard cross section were smeared by the non-perturbative
function.

The non-perturbative function is approximated with the flavor independent,
Gaussian parametrization \cite{Ladinsky-Yuan,Davies} 
\begin{eqnarray} &&
  S^{NP} (b,Q,Q_0,x_A,x_B) = 
  G_1 (x_A,x_B) b + G_2 (Q,Q_0) b^2  ~~~ {\rm with}
  \nonumber \\ &&
  G_1 = g_1 g_3  \ln{(100 x_A x_B)} ~~~ \rm{and} ~~~ 
  G_2 = g_1 + g_2 \ln\left( {Q \over 2 Q_0} \right)
  \label{SNPg}
  \end{eqnarray}
Recent analysis gives $g_1 = 0.11\,{\rm GeV}^2$, $g_2 = 0.58\,{\rm GeV}^2$ , 
$g_3 = -1.5\,{\rm GeV}^{-1}$ and $Q_0 = 1.6\,{\rm GeV}$ \cite{Ladinsky-Yuan}. 
The Fourier transform of the above non-perturbative function 
(for simplicity taken without the linear term) is 
\begin{eqnarray}
  W^{NP} (Q_T,Q,Q_0) = {1 \over 2 G_2(Q,Q_0)}
  \exp{\left\{ -Q_T^2/4 G_2(Q,Q_0) \right\}}.
  \label{FNPQ}
  \end{eqnarray}
This is the form that we used to smear the direct photon NLO distribution.

The resummation formula (\ref{ResForQ}) indicates that the effect of
resumming the initial state soft gluon radiations is twofold. First, it
exponentiates contributions that are singular as $Q_T\rightarrow 0$ and
therefore contain large logs ($\ln(Q/Q_T)$) of the finite order cross
section, into the Sudakov form factor $\exp (-S)$. Second, it smears the
above perturbative Sudakov form factor with the non-perturbative function $%
W^{NP}$. The amount of the smearing provided by the non-perturbative
function is larger than the amount of the intrinsic $k_T$ because the
non-perturbative function contains parts of the perturbative multiple soft
gluon emission effects in the low $Q_T$ region. Note that the separation
between the perturbative Sudakov form factor and the non-perturbative
function depends on the arbitrary parameter $b_{max}$ used in the CSS
formalism. Hence, the non-perturbative function also contains some
perturbative physics.

We expect a similar underlying mechanism of the initial state radiation to
be present in the direct photon process. That would alter the shape of the
transverse momentum distribution of the direct photon in the low $p_T$
region. This is proven in the agreement between the experimental data and
the predicted low-$p_T$ distribution after including the smearing effects
due to the non-perturbative part of the CSS resummation formalism. For the
direct photon process, however, the color flow of the initial state partons
is different from that for the Drell-Yan process. This slightly complicates
the above resummation results and affects the non-perturbative function
given above. Even though, for example, the value of the $g_i$ 
parameters may change, the form of the non-perturbative function should be
the same. Our phenomenological parameter $x$ partly accounts for these
changes. 

At low energies, due to lack of phase space, the multi gluon initial state
radiation is suppressed. Mathematically this means that the contribution of
the perturbative Sudakov form factor is less important. In the low energy
region, expansion of $\widetilde{W}$ in term of $\alpha _s$ can be
approximated by the first few terms, since $\ln(Q/Q_T)$ will not be large.
(The two characteristic scales $Q_T$ and $Q$ are of the same order in
magnitude.) The first few terms in the expansion of $\widetilde{W}$ are the
singular parts of the fixed order cross section, which dominate the low $Q_T$
behavior. In the CSS resummation formalism, this singular part is smeared by
the non-perturbative function. This suggests that one can improve the
theoretical prediction of a fixed order perturbative calculation at the low $%
p_T$ region by smearing it with the non-perturbative function described
above. This is indeed what we see in Table 1 for low energy fixed target
data.

According to the resummation formula (\ref{ResFor}) the effect of the
non-perturbative function on $\widetilde{W}$ is larger at the low $Q_T$
(large $b$) end. As $Q_T$ increases ($b$ decreases) the non-perturbative
function approaches unity. The Sudakov form factor, on the other hand,
increases its contribution with increasing $Q_T$ (decreasing $b$), which
becomes evident after expanding it in $\alpha _s$:
  \begin{eqnarray}
S(b,Q)=\frac{\alpha _s}\pi \ln \left( \frac{bQ}{b_0}\right) \left[ A_1\ln
\left( \frac{bQ}{b_0}\right) +B_1\right] +{\cal O}(\alpha _s^2). 
  \end{eqnarray}
This means that at higher direct photon $p_T$ region the effect of the
Sudakov factor would also be important. This effect might just cancel the
smearing effect of the non-perturbative function leaving the higher $p_T$
data virtually unchanged, which would explain why $k_T$ smearing has to be
performed only in the lowest $p_T$ regions.

For data from experiments with higher center of mass energies, the initial
state multiple gluon radiation effects are important. This also means that
the perturbative Sudakov form factor is important. In this case, the full
resummation calculations should be carried out to describe the $p_T$
distribution of the direct photon produced at high energy hadron colliders,
such as the Tevatron. For a rough estimate of the effect of a Sudakov type
of form factor at high energy direct photon experiments we turn to the
resummation formula again. We observe that writing (\ref{ResFor}) into a
form of a convolution can be repeated grouping terms in a slightly different
way. Taking the Sudakov factor out from $\widetilde{W}$ and putting it into $%
W^{NP}$, the argument that leads to the convolution form (\ref{ResForQ}) can
be repeated without a change. Now that the smearing function contains the
Sudakov part we ask the question, whether this effective smearing function
has a broader width, that is whether it provides an additional smearing what
the collider experiments need. To answer this we have to evaluate the (two
dimensional) Fourier transform of the non-perturbative function multiplied
with the Sudakov factor evaluated at $b_{*}$. The simplest way of doing this
is to perform the integrations numerically. The result agrees with our
expectations. The effective smearing function has an approximate Gaussian
shape and a width $\Delta _{eff}$ larger than of the non-perturbative
function, with the increase depending on the center of mass energy. We can
interpret this result as an indication that the missing $p_T$ smearing may
well come from the Sudakov form factor.

To further illustrate the power of the Fourier transformed form of the
resummation formula, we attempt some simple estimates on observables that
are related to the transverse momentum of the $W$-boson in hadron
collisions. Consider the average transverse momentum $\left\langle
Q_T\right\rangle $ of the weak vector boson produced in hadron collisions.
Inserting the Fourier transform (\ref{FNPQ}) of the non-perturbative
function into formula (\ref{ResForQ}) the only $Q_T$ dependence of the right
hand side comes from the exponential term of the non-perturbative function $%
\exp ({-(Q_T-Q_T^{\prime })^2/4G)}$. Since the largest contribution in the $%
Q_T^{\prime }$ integral comes from the region where $Q_T^{\prime }\sim Q_T$
we approximate the $Q_T^{\prime }{}^2$ term by $Q_T^2$. To calculate the
average transverse momentum we make use of the integral 
\begin{eqnarray}
  \int_{0}^{+ \infty} d Q_T' \, Q_T' 
  \exp\left\{ Q_T' Q_T/2 G_2 \right\} = {4 G_2^2 \over Q_T^2}.
  \label{integral}
  \end{eqnarray}
Under the above approximations we obtain that the average $Q_T$ is
proportional to $G_2(Q,Q_0)$. 
(A more elaborate numerical calculation supports this.) The result is given
on Fig.~\ref{MPlot}a. We find that the average $Q_T$ grows as log of $Q$.

\input epsf.tex 
\begin{figure*}[t]
\begin{center} 
\begin{tabular}{cc}
\epsfysize=5.3cm
$\!\!\!\!\!\!\!\!\!\!\!\!\!\!\!\!$
\epsffile{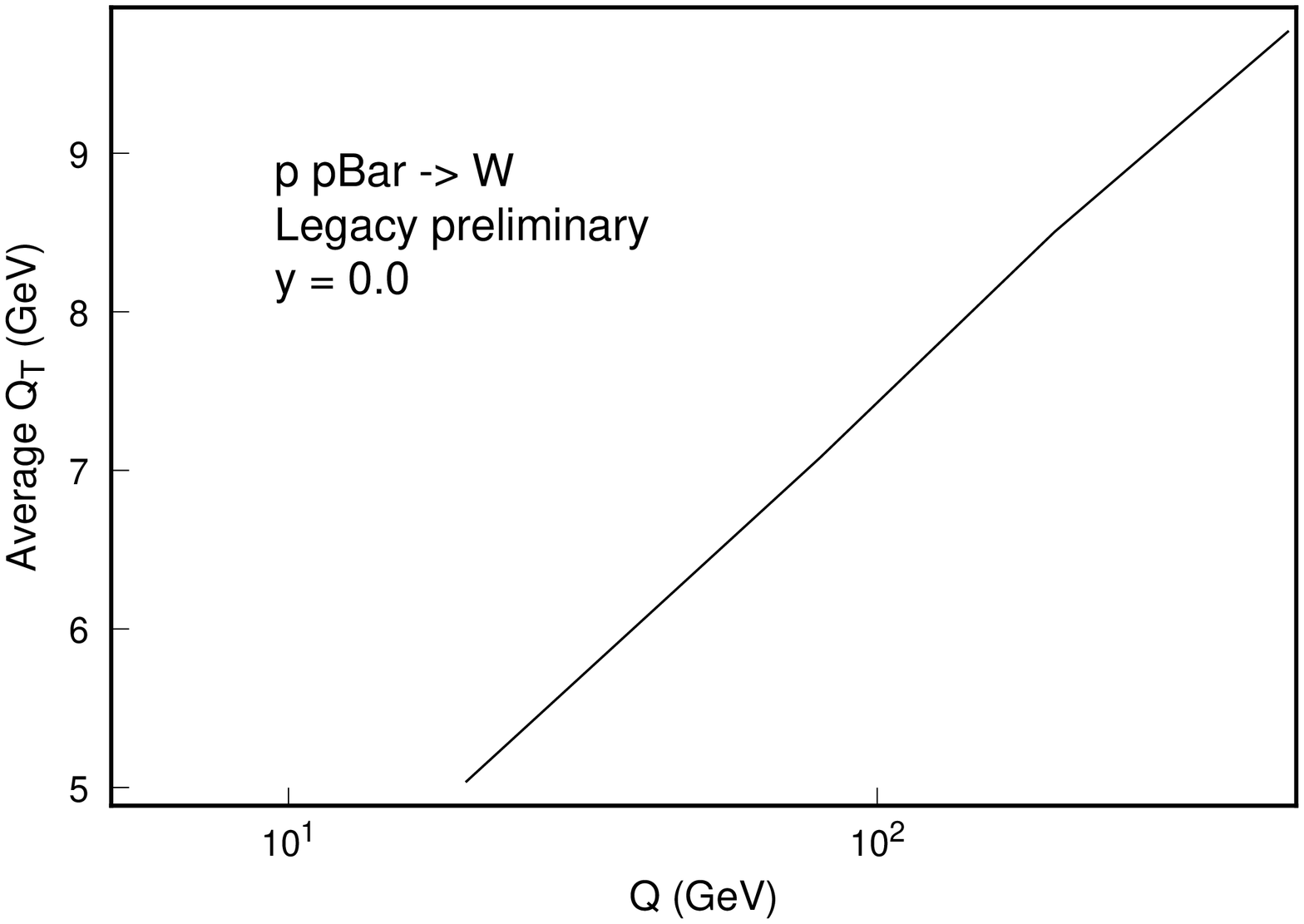} & 
$\!\!\!\!\!\!\!\!\!\!\!$
\epsfysize=5.3cm
\epsffile{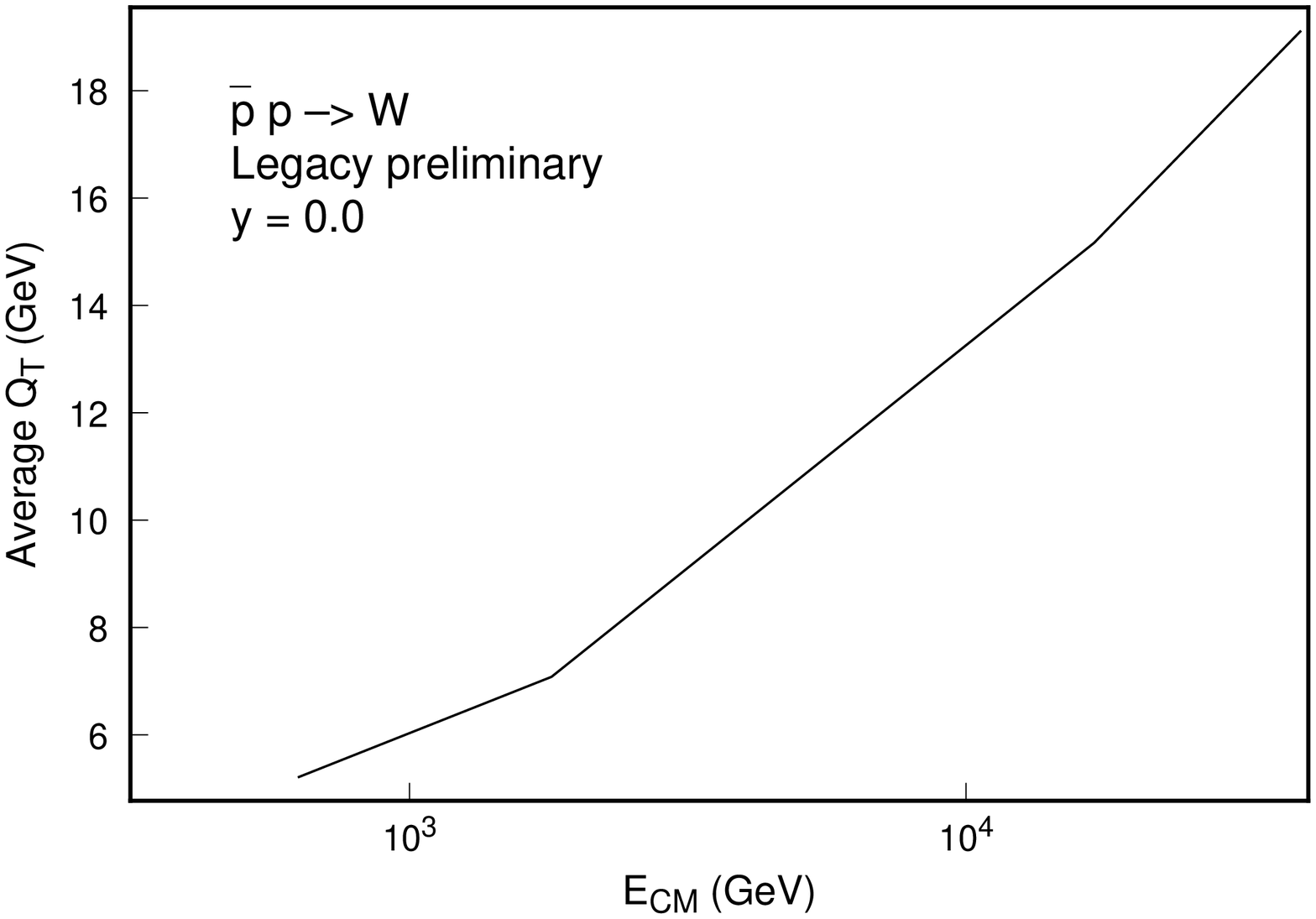}
\end{tabular} 
\caption{
a) The invariant mass versus the average $W^\pm$ boson 
transverse momentum in the resummed calculation. 
b) The center of mass energy versus the average $W^\pm$ boson
transverse momentum in the resummed calculation.
}
\label{MPlot}
\end{center}
\end{figure*}

Using the first order equality $Q=\sqrt{Sx_Ax_B}$ we can obtain a similar
conclusion for the dependence of the average transverse momentum on the
center of mass energy of the hadron colliders. Of course, the $S$ dependence
will not be purely logarithmic, since $\widetilde{W}_{jk}$ (and a more
thorough parametrization of the non-perturbative function) makes it more
complicated. Nevertheless, as Fig.~\ref{MPlot}b shows, even the $S$
dependence of $\left\langle Q_T\right\rangle $ is not very far from being
logarithmic. This result differs from what one would infer from a finite
order perturbative calculation which would predict that $\left\langle
Q_T\right\rangle $ is proportional to $E_{cm}=\sqrt{S}$.

In conclusion, we have shown that the discrepancy between the theoretical
prediction and the experimental data of the $p_T$ distribution of the direct
photon can be resolved for the fixed target experiments after smearing the
NLO calculation by the non-perturbative function that is suggested by the
CSS resummation formalism. To improve the phenomenological method described
in this work for high energy collider experiments, one needs to include the
effects on $p_T$ distribution of direct photons from the perturbative
Sudakov form factor in the CSS resummation formalism.


%




\singlespacing

\end{document}